\documentclass[aps,pra,twocolumn,amsfonts,superscriptaddress,longbibliography,nofootinbib]{revtex4-1}

\usepackage{macros}

\begin{document}

\title{Non-perturbative methodologies for low-dimensional
  strongly-correlated systems: From non-abelian bosonization to truncated spectrum methods}

\author{Andrew J. A. James}
\affiliation{London Centre for Nanotechnology, University College London, Gordon Street, London WC1H 0AH, United Kingdom}

\author{Robert M. Konik}
\affiliation{Condensed Matter Physics and Materials Science Division, Brookhaven National Laboratory, Upton, NY 11973-5000, USA} 

\author{Philippe Lecheminant}
\affiliation{Laboratoire de Physique Th\'eorique et Mod\'elisation, CNRS UMR 8089, Universit\'e de Cergy-Pontoise, Site de Saint-Martin, 2 avenue Adolphe Chauvin, 95302 Cergy-Pontoise Cedex, France}

\author{Neil J. Robinson}
\affiliation{Condensed Matter Physics and Materials Science Division, Brookhaven National Laboratory, Upton, NY 11973-5000, USA} 

\author{Alexei M. Tsvelik}
\affiliation{Condensed Matter Physics and Materials Science Division, Brookhaven National Laboratory, Upton, NY 11973-5000, USA} 

\date{\today}

\begin{abstract} 
We review two important non-perturbative approaches for extracting the
physics of low-dimensional strongly correlated quantum
systems. Firstly, we
start by providing a comprehensive review of non-Abelian bosonization. This includes an introduction to the basic elements of conformal field theory as applied to systems with a current algebra, and we orient the reader by presenting a number of applications of non-Abelian bosonization to models with large symmetries. We then tie this technique into recent advances in the ability of cold atomic systems to realize complex symmetries. Secondly, we discuss truncated spectrum methods for the numerical study of systems in one and two dimensions. For one-dimensional systems we provide the reader with considerable insight into the methodology by reviewing canonical applications of the technique to the Ising model (and its variants) and the sine-Gordon model. Following this we review recent work on the development of renormalization groups, both numerical and analytical, that alleviate the effects of truncating the spectrum. Using these technologies, we consider a number of applications to one-dimensional systems: properties of carbon nanotubes, quenches in the Lieb-Liniger model, 1+1D quantum chromodynamics, as well as Landau-Ginzburg theories. In the final part we move our attention to consider truncated spectrum methods applied to two-dimensional systems. This involves combining truncated spectrum methods with matrix product state algorithms. We describe applications of this method to two-dimensional systems of free fermions and the quantum Ising model, including their non-equilibrium dynamics.

$ $ 


 \end{abstract}
\maketitle

\tableofcontents


\section{Introduction}

Quantum systems have been under intense investigations for well over a century, following the pioneering work of Max Planck at the very beginning of the 20th century~\cite{PlanckAnnPhys01}. With the establishment of the new quantum mechanics a number of important and well-known results flowed forth in quick succession: blackbody radiation~\cite{PlanckAnnPhys01}, the photoelectric effect~\cite{EinsteinAnnPhys05}, predictions for the energy levels of the electrons in the hydrogen atom~\cite{BohrPhilMag13}, and so on (see, e.g., Refs.~\cite{HeisenbergBook,DiracBook,FeynmanLecturesVol3}).  

In the 1920s many-body quantum systems came under an increasing amount of attention. Once Wolfgang Pauli introduced the exclusion principle~\cite{PauliZPhys25a,PauliZPhys25b} it was realized that many-particle correlations might lead to fundamentally new physics. Paradigmatic models, such as the Ising model~\cite{IsingZPhys25} and the Heisenberg model~\cite{HeisenbergZPhys28,BetheZPhys31} were established, and Schr\"odinger developed his wave equation for quantum mechanics~\cite{SchrodingerPR26}. Dirac emphasized the application of Schr\"odinger's formalism to many-electron problems~\cite{DiracProcRoySocA29}, and shortly after Hylleraas~\cite{HylleraasZPhys29} presented an approximate solution of the helium atom via a variational wavefunction. This simple calculation showed much of the power of quantum theory, predicting the ground state energy of helium to within one half of one percent of its measured value. 

Despite the successful description of the helium atom, it was also apparent that interactions present a significant challenge. In the case of helium, one is dealing with a `simple' few-body problem and even here an exact result is not known. For computing properties of helium it is fortunate that the Coulomb interaction is weak\footnote{The weakness of the Coulomb interaction is controlled by the value of the fine structure constant \[ \alpha = \frac{e^2}{4\pi \epsilon_0 \hbar c}  \approx \frac{1}{137} \ll 1. \]} and perturbative techniques give reasonable results. On the other hand, when we have a many-particle problem in which interactions are not weak, there is \textit{a priori} no obvious route towards solving the problem.  Furthermore, careful study of the hydrogen atom revealed that interactions can lead to subtleties in even the apparently trivial case of the two-body problem. This is perhaps best exemplified by the 1947 experiments of Lamb and Rutherford, where a shift in the energy between the $2S$ and $2P$ orbitals of hydrogen was observed~\cite{LambPR47}. This so-called Lamb shift was not predicted by the exact solution of the Dirac equation for hydrogen~\cite{GordonZPhys28,DarwinProcRoySocA28}, and was explained shortly afterwards by Bethe, who computed the electron self-energy in the two orbitals and showed that they differ~\cite{BethePR47}.

So, even in the case of few-body problems, it is clear that interactions are challenging in the theory of quantum systems. Moving towards the many particle problem, it becomes important to develop a systematic understanding of the effect of interactions. At first blush, such an aim may appear hopeless -- our eventual goal is to describe the behavior of macroscopic ($\sim10^{23}$) numbers of interacting particles. From experimental observations, we already know that depending on the precise details of the system, we can realize a plethora of phases of matter with strikingly different physical properties. Whilst for the case of weak interactions (or another small parameters) one can apply the extensive framework of perturbative quantum field theory (see, e.g., Refs.~\cite{PeskinSchroeder,ZinnJustinBook,SrednickiBook,TsvelikBook,AltlandAndSimons,MussardoBook}), in the absence of a small parameter (so-called \textit{strongly correlated systems}) one must develop non-perturbative techniques. This is perhaps one of the grandest challenges of modern theoretical physics. 

In pursuit of non-perturbative techniques to attack strongly correlated problems, we turn our attention towards low-dimensional quantum systems. At first glance, it is not obvious that this is the easiest regime to consider: particles confined to move on a line must scatter in order to move past one another. As a result, strong correlations and collective phenomena rule the roost in low dimensional quantum systems. Yet despite this, a number of exact results and methods peculiar to low-dimensions exist, and these help guide the way.

Relatively early in the development of quantum mechanics, two important advances in the study of many-body systems occurred. Firstly, Jordan and Wigner suggested the transformation which establishes a relationship between fermionic and bosonic one-dimensional quantum systems~\cite{JordanZPhys28}. Secondly,  Bethe presented his now famous ansatz for the eigenstates of the one-dimensional isotropic Heisenberg model~\cite{BetheZPhys31} -- a truly strongly correlated system in which no small parameter exists for perturbative expansions. 

These two important results existed in isolation for almost 30 years before an explosion of results for integrable 1+1-dimensional quantum models and closely related 2+0-dimensional statistical mechanics models, beginning in the late 1950s: the Heisenberg XXZ chain~\cite{OrbachPR58,WalkerPR59,YangPR66a,YangPR66b,YangPR66c}, the six-vertex model~\cite{SutherlandPRL67,BaxterStudApplMath71}, the eight-vertex model~\cite{BaxterAnnPhys72,BaxterAnnPhys73a,BaxterAnnPhys73b,BaxterAnnPhys73c}, the Lieb-Liniger model~\cite{LiebPR63a,LiebPR63b}, the massive Thirring model~\cite{BergknoffPRL79,BergknoffPRD79}, the sine-Gordon model~\cite{SklyaninTheorMathPhys79}, the Gross-Neveu model~\cite{AndreiPRL79,AndreiPhysLettB80}, and the $SU(2)$-Thirring model~\cite{BelavinPhysLettB79}. Whilst integrable models form a set of measure zero in the space of all models, they provide a valuable starting point for understanding strongly correlated systems and they include a number of models of experimental interest (see, for example, Refs.~\cite{HofferberthNature07,ColdeaScience10,LakePRL13,MourigalNaturePhys13,GuanRMP13}).  

Further to developments in integrable models, in the mid-1970s there were parallel developments in the condensed matter and high-energy communities on the formal one-to-one correspondence between fermionic and bosonic models in 1+1D~\cite{MattisJMathPhys74,LutherPRB74,ColemanPRD75,MandelstamPRD75}. This formalized the links between interacting fermion and boson systems, as had already been realized with the noninteracting Tomanaga-Luttinger liquid~\cite{TomonagaPTP50,MattisJMathPhys65,LuttingerJMathPhys63}, which extended early works by Bloch on describing the electron gas through its sound waves~\cite{BlochZPhys33,BlochHelvPhysActa34}. By exploiting this correspondence between fermionic and bosonic theories, through a toolbox now known as \textit{bosonization and refermionization}, the door was opened to studying nonintegrable strongly correlated problems~\cite{HaldaneJPhysC81,GiamarchiBook,GNTBook,TsvelikBook,FradkinBook}. This framework remains at the forefront of understanding of various exotic phenomena, including the well-known spin-charge separation~\cite{HaldaneJPhysC81,KimPRL96,ClaessenPRL02,AuslaenderScience05,KimNatPhys06,HilkerArxiv17}. 

As well as analytical approaches, based upon integrability and bosonization, there are a number of powerful numerical techniques that shed light on the properties of low-dimensional strongly correlated quantum systems. Exact diagonalization~\cite{NoackAIPConfProc05,ZhangEurJPhys10} is a useful tool for one-dimensional models with small local Hilbert spaces (such as spin-1/2 chains) allowing access to the eigenstates of moderately large systems (up to $\sim30$ sites for full and $\sim40$ sites for iterative diagonalization of a spin-1/2 chain). Hamiltonian truncation methods can pivot the power of exact diagonalization to tackle problems with larger Hilbert spaces: Wilson's numerical renormalization group (NRG)~\cite{NoackAIPConfProc05,wilson1975the} and the truncated space approach (TSA)~\cite{yurov1990truncated,yurov1991truncated} both embrace the philosophy of the renormalization group to work with restricted Hilbert spaces. Beyond exact diagonalization, there is a proliferation of techniques based upon matrix product states and their tensor network generalizations (see the reviews~\cite{VerstraeteAdvPhys08,OrusAnnPhys14}), which includes the ubiquitous density matrix renormalization group (DMRG) algorithm~\cite{white1992density,WhitePRB93,NoackAIPConfProc05,SchollwockAnnPhys11}. For finite temperature properties and large systems, quantum Monte Carlo (QMC)~\cite{FoulkesRMP01,HochkeppelChapter09} remains at the forefront of available methods.

Despite this diverse range of methods, there is a never-ending demand to advance and extend the non-perturbative techniques available to us. In recent years this has been driven by the desire to meet fascinating new experimental challenges, such as describing materials with large and complex symmetries (such as transition metal~\cite{TokuraScience00,WangNatureNano12} and rare-earth~\cite{StewartRMP84,ColemanArticle07} compounds) and understanding ground-breaking studies of cold atomic gases with enlarged symmetries~\cite{WuPRL03,WuModPhysLettB06,CazalillaNewJPhys09,GorshkovNaturePhys10,CazalillaRPP14,CapponiAnnPhys16}. We have already seen that integrability can be a useful tool on this road, but it by no means exhausts the problems which need to be addressed. Indeed, in higher spatial dimensions integrability has little to directly say at all. In this review we will present a number of techniques, some partially based upon integrability, some partially based upon matrix product states, which have been developed in an attempt to overcome some of the challenges of the field and address some of the experimentally relevant questions. 

\subsection{Overview}

We will first discuss non-Abelian bosonization and its application to systems with complicated symmetries. In the course of our discussion, we will make explicit the links to recent studies of condensed matter systems with large symmetries (such as spin and orbital degeneracy), as well as experiments on cold atomic gases with symmetries hard to realize in the solid state (such as $SU(N)$ spin symmetry). Following this, we will review the truncated space approach (TSA). Using exact information from integrability or conformal field theory, this method allows one to compute the low-energy excitation spectrum and correlation functions of perturbed integrable models (and not necessarily weakly perturbed). At its base the TSA is a numerical approach, and its realm of applicability can be greatly extended with powerful renormalization group improvements. 

Two-dimensional quantum systems can be even richer than their one-dimensional counterparts, and there exist few methods which can accurately decipher their properties. In the third technique that we review, we attack a number of two-dimensional problems by combining data from integrability with matrix product state based numerics. With such methods it is possible to glue together one-dimensional integrable sub-units to form large two-dimensional arrays, which we then study for several example systems. By following such a path, we will show that certain two-dimensional systems and their critical points can be studied. 

Throughout the review, we have tried to keep our use of acronyms to a minimum; nevertheless, we provide a glossary of those that we do use at the end of the main body of the review. 

The theory of strongly correlated low-dimensional quantum systems is a vast and rapidly advancing field. As a result, there are topics too numerous to name that we do not have space to cover. However, in relation to the topics of focus of this review, it would be remiss of us not to mention a few particular examples.

(1) Recent experimental advances in the field of ultracold atoms have stimulated a huge theoretical effort to understand the non-equilibrium dynamics of low-dimensional quantum systems. Issues at the core of understanding quantum mechanics are being addressed, with the aim of addressing even basic questions such as: Does thermalization emerge from unitary time-evolution? How do conservation laws modify the dynamics of a system? Can non-equilibrium systems relax to states with properties very different to those accessible in equilibrium? How does one describe non-equilibrium steady states in which there are finite flows of currents? An introduction to some of the theoretical techniques of this field can be found in the recent review articles~\cite{GogolinReview15,DAlessioReview15,EsslerReview16,CalabreseReview16,CazalillaReview16,BernardReview16,CauxReview16,VidmarReview16,LangenReview16,ProsenReview16,VasseurReview16,DeLucaReview16} and references therein.  

(2) As well as the non-equilibrium dynamics, over the past decade there have been significant advances in the computation of equilibrium dynamical correlation functions. It is well known that Abelian bosonization (e.g., the Luttinger liquid) fails to capture the correct physics of dynamical correlation functions at finite frequency and momentum -- in part due to the linearization of the spectrum, which only applies in the vicinity of the Fermi points. To resolve this problem, the non-linear Luttinger liquid formalism~\cite{ImambekovScience2009,ImambekovRMP12} was developed, in which Abelian bosonization is modified to include mobile impurities which allow one to capture the correct finite frequency and momentum behavior. Combined with information from integrability, exact results can be obtained for threshold singularities (see, e.g., Refs.~\cite{PereiraPRL08,EsslerPRB10,AdityaPRB12,TiegelPRB16,VenessPRB16}) and the real-time dynamics~\cite{SeabraPRB14}. 

(3) Integrability is an important tool and cornerstone of both the previously mentioned topics. In itself, there have been significant advances in studying integrable quantum systems, from the development of efficient numerical routines for computing correlation functions (such as {\sc abacus}~\cite{CauxJMathPhys09}) to new analytical results for matrix elements in multi-component models~\cite{PozsgayJPhysA12,BelliardJStatMech12,BelliardJStatMech13,PakuliakNuclPhysB14,PakuliakNuclPhysB15,PakuliakJPhysA15,PakuliakSIGMA15,HutsalyukNuclPhysB16}. One of the most beautiful mathematical results has been the development of the correspondence between integrable models (e.g., thermodynamic Bethe ansatz) and ordinary differential equations, see for example the review article~\cite{DoreyJPhysA07} and references therein. 

(4) There have also been significant advances in the study of critical theories in higher dimensions, spurred on by the development of the numerical conformal bootstrap~\cite{RychkovArxiv11,SimmonsDuffinArxiv16,PolandNatPhys16}. This has allowed for important quantities, such as the critical exponents, to be computed to extremely high accuracy in physically interesting systems, such as the three-dimensional Ising model~\cite{ElShowkPRD12,El-Showk2014}. 

(5) Also on the numerical methods front, there have been recent interesting developments in the application of machine learning methods to strongly correlated systems. This includes attempts to describe strongly correlated states of matter~\cite{ArsenaultPRB14,MentaArxiv14,ChngArxiv16,CarrasquillaArxiv16,BroeckerArxiv16,CarleoArxiv16} and suggest new materials~\cite{KusneSciRep14,KalininNatureMat15,GhiringhelliPRL15}.

\section{Non-Abelian bosonization}
\label{Sec:NonAbelianBosonization}

\subsection{Background} 

\subsubsection{Motivation}
In physics it is frequently the case that making the right choice of variables dramatically simplifies the problem, allowing the solution to be grasped. In the field of condensed matter, we are often dealing with electrons and so the original variables are fermionic fields. In many problems of interest, these fields are strongly interacting: the associated excitations of these fields become incoherent and extracting the physics of the problem becomes muddied. It is then that we seek new variables, whose excitations are coherent, in which the physics is more transparent. Bosonization, the topic of this section of the review, provides us with one such reformulation: the problem is expressed in terms of collective variables which are bosonic or even fermionic, but different to the original fields ~\cite{StoneBosonization,CFTBook,GNTBook,TsvelikBook,FradkinBook,GiamarchiBook}. Such a formulation in many cases significantly simplifies the problem, helping us to find the solution and understand the physics. 

Non-Abelian bosonization, much like its Abelian counterpart (see Appendix~\ref{App:AbelianBosonization} for a brief discussion) is a mathematical procedure that establishes a formal equivalence between fermionic and bosonic versions of the same model in 1+1 dimensions. Our discussion of non-Abelian bosonization will be applications driven\footnote{The technique itself has been reviewed before, see Refs.~\cite{CFTBook,GNTBook,TsvelikBook,FradkinBook,MudryBook} for some prominent examples.} -- technical aspects will be explained in the context of models that exhibit new and interesting physics. In particular, our focus will be on models with complicated symmetries that may emerge, for example, when orbital degrees of freedom must be taken into consideration in an electronic system. Examples of such systems include transition metal~\cite{TokuraScience00,WangNatureNano12} and rare-earth compounds~\cite{StewartRMP84,ColemanArticle07}, as well as many cold atomic gas systems~\cite{WuPRL03,WuModPhysLettB06,CazalillaNewJPhys09,GorshkovNaturePhys10,CazalillaRPP14,CapponiAnnPhys16}. Our main focus will be on such systems in the vicinity of a quantum critical point (QCP): the quantum aspect of the problem is enhanced close to a QCP, and models with complicated symmetries will be described by highly entangled, strongly correlated states in this regime~\cite{VidalPRL03}. 

At the very core of non-Abelian bosonization is a mathematical theorem~\cite{WittenCommMathPhys84,KnizhikNuclPhysB84,CFTBook}: the Hamiltonian of non-interacting massless fermions in (1+1) dimensions that transform according to some symmetry group can be written as the sum of Wess-Zumino-Novikov-Witten (WZNW) models. Whilst at first glance such a reformulation looks rather complicated, the fact that each WZNW model commutes with the others allows us to treat each symmetry sector independently (this is reminiscent of spin-charge separation in Abelian bosonization, see Appendix~\ref{App:AbelianBosonization}) and often makes the problem tractable. The reformulation also enables us to incorporate various interactions, and occasionally (if we are lucky!) the problem can turn out to be exactly solvable, or at least amenable to approximate methods. 

\subsubsection{Applications of non-Abelian bosonization}

As we have mentioned in the previous section, our discussion of non-Abelian bosonization will be focused upon applications in condensed matter and cold atom systems with complicated (e.g., high) symmetry. This is, of course, not the only scenario in which one can apply non-Abelian bosonization; in this section, we give (a certainly  incomplete!)  list of other applications which we do not have space to cover. 

1. \textit{Spin chains and ladders.---} There is an extensive literature on applications of non-Abelian bosonization to spin chains and ladders, see the text books~\cite{GNTBook,TsvelikBook}. The manifest realization of non-Abelian symmetries serves to make the physics much more transparent, as was shown by the seminal early works of Polyakov and Wiegmann~\cite{PolyakovPhysLettB84}, Affleck~\cite{AffleckPRL85,AffleckNuclPhysB86}, and Affleck and Haldane~\cite{AffleckPRB87}.

2. \textit{The Kondo problem and generalizations.---} Non-Abelian bosonization is a standard tool for attacking the Kondo problem, starting from the work of Fradkin and collaborators~\cite{FradkinNuclPhysB89} and subsequent works by Affleck and Ludwig~\cite{AffleckNuclPhysB90,AffleckLudwigNuclPhysB91a,AffleckLudwigNuclPhysB91b,LudwigAffleckPRL91}, much of which is reviewed in Ref.~\cite{AffleckActaPhysPolon95}. Generalizations of the Kondo problem to multiple channels~\cite{AffleckPRB95,AffleckActaPhysPolon95,AndreiPRB00}, cluster impurities~\cite{IngersentPRL05,FerreroJPhysCondMatt07} or to the Kondo lattice~\cite{FujimotoJPSJ94} can also be treated. 

3. \textit{Disordered fermions.---} Problems featuring disorder have also been the subject of intense study with non-Abelian bosonization. These include: Dirac fermions in a random non-Abelian gauge potential~\cite{BernardArxiv95,CauxNuclPhysB96,MudryNuclPhysB96,CauxPRL98,CauxNuclPhysB98,CauxPRL98b,BhaseenNuclPhysB01}, disordered d-wave superconductors~\cite{NersesyanPRL94,NersesyanNuclPhysB95,AltlandPhysRep02}, non-Hermitian theories with random mass terms~\cite{GuruswamyNuclPhysB00}, and random potentials related to percolation transitions~\cite{LudwigPRB94}.

4. \textit{Quantum Hall transitions and edge states.---} Non-Abelian bosonization also has various applications to the quantum Hall effect. These include relations to transitions between quantum Hall states~\cite{LudwigPRB94,AffleckNuclPhysB86} and the description of quantum Hall edge states~\cite{WenPRL90,WenPRB90,StoneAnnPhys91,WenIntJModPhysB92,WenAdvPhys95}. More recently,  non-Abelian bosonization has been extensively used in the coupled-wire construction~\cite{KanePRL02} of two-dimensional non-Abelian fractional quantum Hall states and chiral-spin liquid phases, where one starts from an array of one-dimensional fermionic or bosonic wires~\cite{TeoPRB14,MengPRB15,GorohovskyPRB15,HuangPRB16,lecheminant2016lattice,huang2016coupled,fuji2016non}.

5. \textit{Quantum chromodynamics in 1+1-dimensions and Quark-Gluon plasma in 1+3-dimensions.---} Outside the realm of condensed matter physics, non-Abelian bosonization is a powerful tool in high energy physics, including for the description of toy models of quantum chromodynamics, see for example Refs.~\cite{GepnerNuclPhysB85,AffleckNuclPhysB86b,FrishmanPhysRep93,AzariaPRD16}, and realistic models of dense quark-gluon plasma~\cite{KojoPRD10}.

\subsubsection{This section of the Review}

The path for our discussion is as follows: we will begin by introducing non-Abelian bosonization in quite some detail, starting from the basic idea of linearizing the dispersion of a one-dimension quantum system, and moving on to discuss the current algebra, the conformal embedding theorem, the diagonalization of WZNW models, and the Lagrangian formulation. To supplement this discourse, we provide brief introductions to Abelian bosonization and conformal field theory (CFT) in Appendices~\ref{App:AbelianBosonization}~and~\ref{App:CFT}, where we summarize some useful basic concepts. 

In Sec.~\ref{Sec:ExamplesNonAbelian} we move on to discuss a number of examples of non-Abelian bosonization motivated by applications to materials of current interest in solid state experiments, such as transition metal and rare earth compounds. The electrons in these models carry both spin and orbital degrees of freedom, leading to complicated symmetries such as $U(1) \times SU(k) \times SU(N)$ or $U(1) \times Sp(2N)$. Here we will discuss some truly exotic physics, including topological phases and emergent parafermions. We follow this with Sec.~\ref{Sec:ColdAtomsExamples}, where applications of non-Abelian bosonization to cold atomic gases will be covered. 

\subsection{Linearizing the dispersion}
\label{Sec:Linearization}
To begin, let us briefly recap the standard field theoretical approach to $(1+1)$-dimensional quantum systems, which starts with linearizing the dispersion~\cite{GNTBook,TsvelikBook,GiamarchiBook}. In our discussion of non-Abelian bosonization, we will assume that non-interacting fermions have a linear spectrum, which is a valid point of view for states sufficiently close to the Fermi points in a condensed matter system. The formal transition from a quadratic theory to a linear dispersion is achieved by writing the fermion fields as a combination of a fast (oscillatory) exponent and slow right- and left-moving fields $R,L$:  
\be
\psi(x) = \re^{i k_Fx}R(x) +\re^{-i k_F x}L(x), \label{psi}
\ee
where $k_F$ is the Fermi wave vector (we work in units where $\hbar = 1$).\footnote{In doing the expansion~\fr{psi} we neglect the presence of higher harmonics, which may arise as a result of, e.g., interactions.} Substituting~\fr{psi} into the non-interacting Hamiltonian with a quadratic dispersion relation we obtain 
\bea
H &=& \frac{1}{2m}\int \rd x\, \psi^\dagger \Big(- \p_x^2 - k_F^2\Big)\psi, \nn
&\approx& i v_F\int \rd x\Big(-R^\dagger \p_x R + L^\dagger \p_xL\Big), \label{linearHam}
\eea
where $v_F = k_F/m$ is the Fermi velocity. In obtaining~\fr{linearHam} we have neglected terms that are oscillatory (which are suppressed by the integration over $x$) and second derivatives of the slow fields, which are assumed to be small (hence the name ``slow''). It is clear that the linearization procedure will not capture the correct physics for all energies and momentum: a cut-off energy $\Lambda \sim k_F^2/2m$ (the Fermi energy) for the theory is introduced to account for this. Under this linearization procedure, low energy non-relativistic one-dimensional fermions are transformed into relativistic Dirac ones; this emergent Lorentz symmetry plays a very important role in the theory of strongly correlated one-dimensional (1D) systems~\cite{TsvelikBook,GNTBook}.

The Dirac Hamiltonian~\fr{linearHam} will serve as a starting point for the remainder of our discussions of non-Abelian bosonization. The introduction of local degrees of freedom (e.g., higher symmetry) does not change the discussion: consider left- and right-moving fermion fields $L_{j\alpha}$, $R_{j\alpha}$ that carry both orbital ($j=1,\ldots,k$) and spin~($\alpha=1,\ldots,N$) indices. The fields are governed by the Dirac Hamiltonian (cf. Eq.~\fr{linearHam})
\be
H = i v_F\sum_{j=1}^k\sum_{\alpha=1}^N\int \rd x\Big(-R^\dagger _{j\alpha}\p_x R_{j\alpha} + L^\dagger _{j\alpha}\p_xL_{j\alpha}\Big), \label{Dirac}
\ee
and obey the standard anti-commutation relations 
\be
\begin{split}
\Big\{R^\dagger _{j\alpha}(x),R^{\phantom\dagger}_{j'\beta}(y)\Big\} &=\delta_{jj'}\delta_{\alpha\beta}\delta(x-y),\\
\Big\{L^\dagger _{j\alpha}(x),L^{\phantom\dagger}_{j'\beta}(y)\Big\} &=  \delta_{jj'}\delta_{\alpha\beta}\delta(x-y), \\
\Big\{R_{j\alpha}(x),L_{j'\beta}(y)\Big\} &= 0.
\end{split} \label{comm}
\ee
Herein, we will set the Fermi velocity $v_F=1$ and measure energy in appropriate units. 

\subsection{The Kac-Moody algebra}
Let us now consider one of the most fundamental concepts of low-dimensional quantum physics, the Kac-Moody algebra~\cite{KacMathUSSAIzv68,MoodyJAlgebra68}, and discuss its central role in non-Abelian bosonization. 

\subsubsection{Current Operators}
We consider the Hamiltonian~\fr{Dirac} where the fermions carry both orbital ($j=1,\ldots,k$) and spin ($\alpha=1,\ldots,N$) indices. We define the \textit{current operators} 
\be
\begin{split}
& J^a_R = R^\dagger (I\otimes s^a)R,\qquad a=1,\ldots,N^2-1,\\
& F^a_R = R^\dagger (t^a\otimes I)R,\qquad a=1,\ldots,k^2-1,
\end{split} \label{JF}
\ee
with identical definitions for left-moving currents with $R \to L$. In Eqs.~\fr{JF} we use the convenient short hand notation
\be
R^\dagger (t^a \otimes s^b) R = \sum_{j,j'=1}^k \sum_{\alpha,\beta=1}^N R^\dagger_{j\alpha} t^a_{jj'} s^b_{\alpha\beta}R^{\phantom\dagger}_{j'\beta},  \label{shorthand}
\ee
while $I$ is the unit matrix, $s^a$ are the generators of the $\mathfrak{su}(N)$ algebra associated with the local spin degrees of freedom, and $t^b$ are the generators of the $\mathfrak{su}(k)$ algebra associated with the local orbital degrees of freedom. The generators of the $\mathfrak{su}(N)$ algebra are normalized according to 
\be
\mbox{Tr}( s^a s^b ) = \frac{1}{2}\delta_{ab},\qquad [s^a,s^b] = \sum_c i f^{abc}s^c, \label{generators}
\ee
where $\delta_{ab}$ is the Kronecker delta and $f^{abc}$ are the structure constants of the Lie algebra (see, e.g., Ref.~\cite{FultonHarris}).\footnote{For the case of $N=2$, the generators of the $\mathfrak{su}(2)$ algebra in this normalization are $s^a = \s^a/2$, with $\s^a$ the Pauli matrices. The structure constants are $f^{abc} = \epsilon^{abc}$, where $\epsilon^{abc}$ is the Levi-Civita symbol.} Similar relations hold for the generators $t^a$ of the $\mathfrak{su}(k)$ algebra.

\subsubsection{Commutation relations}
The anti-commutation relations~\fr{comm} imply that currents with different chirality ($R$ or $L$) or from different groups ($SU(N)$ or $SU(k)$) commute. Currents which have the same chirality and group structure compose the \textit{Kac-Moody algebra}~\cite{KacMathUSSAIzv68,MoodyJAlgebra68}. For the currents featuring the generators of the $\mathfrak{su}(N)$ algebra, we have ($\ell=R,L=0,1$)
\be
[J^a_\ell(x), J^b_\ell(y)] = i f^{abc}J^c_\ell(x)\delta(x-y) - (-1)^\ell \frac{i k}{4\pi}\delta'(x-y)\delta_{ab}, \label{kac}
\ee
where summation over repeated indices is implied (henceforth we adopt this convention) and $\delta'(x)$ is the derivative of the Dirac delta function. 

The current $J^a$ that satisfies~\fr{kac} with $f^{abc}$ the structure constants of the $\mathfrak{su}(N)$ algebra is called an $SU(N)_k$ current, where $k$ is called the `level'.\footnote{In the mathematics literature, $k$ is known as the `central extension' of the Kac-Moody algebra~\cite{KatsJFunctAnalAppl74,LepowskyCommunMathPhys78,FrenkelPNAS80}.} It follows from the definition~\fr{JF} that $F^a$ is an $SU(k)_N$ current.\footnote{This should be read as ``an $SU(k)$ level $N$ current''.} 

The final term on the right-hand side of Eq.~\fr{kac} is often called the anomalous commutator or the Schwinger term.\footnote{It is intimately related to the presence of a quantum anomaly, see for example Refs.~\cite{GotoProgTheorPhys55,SchwingerPRL59,ColemanPR69}.} It can be derived in a straightforward manner: recall that commutation in a field theory is defined inside of a time-order correlation function. For two operators, $A(x)$ and $B(y)$, the commutators is defined as~\cite{GotoProgTheorPhys55,SchwingerPRL59}
\bea
&&\lla [A(x),B(y)]...\rra \nn
&&= \lim_{\tau \rightarrow 0^+ }\lla \Big[ A(\tau,x)B(0,y)- A(-\tau,x)B(0,y) \Big] ...\rra, \label{anom}
\eea
where the ellipses denote any other fields present in the correlation function. Replacing $A(\tau,x)$ and $B(0,y)$ in Eq.~\fr{anom} with the expressions for the $SU(N)_k$ currents 
\bea
A(\tau,x) &=&  R^\dagger_{j\alpha}(\tau,x) s^a_{\alpha\beta}R^{\phantom\dagger}_{j\beta}(\tau,x),\\
B(0,y) &=&  R^\dagger_{j'\gamma}(0,y) s^b_{\gamma\delta}R^{\phantom\dagger}_{j'\delta}(0,y),
\eea
and using the well-known result for the correlation function of the fermion fields~\cite{CFTBook} 
\be
\lla R^{\phantom\dagger}_{j\alpha}(\tau,x) R_{l\beta}^\dagger(\tau',x')\rra = \frac{1}{2\pi}\frac{\delta_{jl}\delta_{\alpha\beta}}{(\tau-\tau') - i (x-x')},
\ee
we obtain the anomalous commutator 
\bea
&& \lla 
\Big[ R^\dagger _{j\alpha}(x) s^a_{\alpha\beta}R^{\phantom\dagger}_{j\beta}(x) , R^\dagger _{j'\gamma}(y) s^b_{\gamma\delta}R^{\phantom\dagger}_{j'\delta}(y) \Big] \rra \nn
&&=  \frac{k\delta_{ab}}{2}\lim_{\tau \rightarrow 0^+} \frac{1}{4\pi^2}\bigg\{\frac{1}{[\tau-i(x-y)]^2} - \frac{1}{[\tau+i(x-y)]^2}\bigg\}\ , \nn
&&= \frac{k\delta_{ab}}{8\pi^2}\p_x\bigg(\frac{1}{x-y+i 0^+}-\frac{1}{x-y-i 0^+}\bigg)\nn
&&= -\frac{i k}{4\pi}\delta'(x-y)\delta_{ab}.
\eea

\subsubsection{Fourier Components}
It will often be convenient to work with the Fourier components of the current operators, where one assumes the system of fermions is placed in a box of length $l$ with periodic boundary conditions,
\be\label{modes}
J^a(x) = \frac{1}{l}\sum_{n=-\infty}^{\infty} \re^{-2\pi i nx/l}J^a_n.
\ee
In terms of the Fourier components $J^a_n$, the Kac-Moody algebra is
\be
[J_n^a,J_m^b] = i f^{abc}J^c_{n+m} + \frac{nk}{2}\delta_{n+m,0}\delta_{ab}. \label{kacFourier}
\ee
It is clear that the zeroth component of the currents constitutes a subalgebra 
\be
[J_0^a,J_0^b] = i f^{abc}J^c_{0},
\ee
that is isomorphic to the global algebra~\fr{kacFourier}. 

\subsection{Conformal embedding and the Sugawara Hamiltonian}
\label{Sec:ConformalEmbedding}

We now turn our attention to another important concept that is at the core of non-Abelian bosonization: the theorem that non-interacting fermions that transform according to some symmetry in (1+1)-dimensions can be written as a sum of WZNW models~\cite{KnizhikNuclPhysB84}. As the theory of non-interacting massless Dirac fermions in (1+1)-dimensions possesses conformal symmetry~\cite{BPZJStatPhys84,BPZNuclPhysB84}, this theorem is often called \textit{conformal embedding}~\cite{CFTBook}. On a basic level the conformal embedding defines a set of \textit{fractionalization rules} for breaking up the free fermion Hamiltonian in terms of Hamiltonians of different critical models that commute with one-another. 

To illustrate the conformal embedding, we consider the Hamiltonian $H$ defined in Eq.~\fr{Dirac}. The fermions possess both orbital ($j=1,\ldots,k$) and spin ($\alpha = 1,\ldots,N$) indices, so the Hamiltonian has the unitary group symmetry $U(1)\times SU(N) \times SU(k)$. The conformal embedding for $H$ takes the form
\be
H = H[U(1)] + W[SU(N);k] + W[SU(k);N], \label{embedding}
\ee
where $W[G;k]$ is the WZNW Hamiltonian for the group $G$ at level $k$, which can be written in Sugawara form~\cite{SugawaraPR68}
\bea
&& W[SU(N);k]\nn
&&~~ = \frac{2\pi}{N+k}\int_0^l \rd x \Big(:J_R^aJ_R^a: +:J_L^aJ_L^a: \Big), \nn
&&~~ = \frac{2\pi}{l(N+k)}\Big[J_{d,0}^aJ_{d,0}^a +2\sum_{n>0}J^a_{d,-n}J^a_{d,n}\Big], \label{sugawara}
\eea
where $J^a$ are the $SU(N)_k$ currents and $d=R,L$. Normal ordering of an operator (denoted by colons) is defined such that Fourier components with $n > 0$ annihilate the vacuum~\cite{FateevSovJNuclPhys86}. The $U(1)$ Hamiltonian in~\fr{embedding} is the Gaussian model, which may also be expressed in the Sugawara form~\cite{SugawaraPR68}
\be
H[U(1)] = \frac{\pi}{Nk} \int \rd x(:j_R^2: + :j_L^2:),
\ee
with $U(1)$ currents defined by 
\be
j_R = :R^\dagger _{j\alpha}R^{\phantom\dagger}_{j\alpha}:, \quad j_L = : L^\dagger_{j\alpha} L^{\phantom\dagger}_{j\alpha}:. \label{u1cur}
\ee

The conformal embedding~\fr{embedding} is, essentially, a field theory analogue of the decomposition of kinetic energy into radial and angular motion in classical mechanics: 
\be
\frac{m{\bf v}^2}{2} = \frac{m(\dot r)^2}{2} + \frac{{\bf L}^2}{2mr^2}, \label{mech}
\ee
where the first term on the right-hand side would correspond to the Gaussian theory. 

The most important point to take away from the conformal embedding~\fr{embedding} is that all three Hamiltonians on the right-hand side commute with one-another. This means that each symmetry sector can be treated separately -- in many cases this leads to substantial simplifications in calculations. The reader may be familiar with a similar phenomenon in Abelian bosonization: spin-charge separation~\cite{GNTBook,GiamarchiBook}.\footnote{See Appendix~\ref{App:AbelianBosonization} for one such example of this phenomenon.} Also in analogy to the Abelian case, interactions that include solely Kac-Moody current operators of a given group do not violate the conformal embedding~\fr{embedding}, often allowing for their treatment. In terms of the mechanical analogy~\fr{mech}, this is similar to the simplifications that occur when working with a radially symmetric potential (for example). In the examples and discussions below we will extensively use this feature of the theory. 

Analogies between non-Abelian and Abelian bosonization cannot always be drawn. One prominent example of this is to consider the problem of bosonization on the level of operators. The situation here is more nuanced: it well known (see Appendix~\ref{App:AbelianBosonization} for a discussion) that Abelian bosonization allows one to express fermionic operators (including chiral ones, such as the $L,R$ fermions) as sums or products of local operators acting in the chiral sectors of the Gaussian model (e.g., the free boson). Consider, for example, a single species of massless fermion: the bosonization rules states the fermion operators can be written in terms of vertex functions (exponentials) of the chiral bosonic field~\cite{CFTBook,GNTBook,GiamarchiBook}
\be
R= \frac{1}{\sqrt{2\pi a_0}}\re^{i\varphi},\qquad L= \frac{1}{\sqrt{2\pi a_0}}\re^{-i\bar\varphi}, \label{bosid}
\ee
where $a_0$ is the lattice constant, and the bosonic fields are governed by the actions
\be
\begin{split}
S_R &= \frac{1}{4\pi}\int \rd\tau\rd x\ \p_x\varphi(i\p_{\tau} +\p_x)\varphi,\\
S_L &= \frac{1}{4\pi}\int \rd\tau\rd x\ \p_x\bar\varphi(-i\p_{\tau} +\p_x)\bar\varphi.
\end{split} \label{BosonAction}
\ee
The convenient separation~\fr{bosid} of the operators into chiral sectors is \textit{not a universal property of CFTs}. In fact, this can be seen even in the simplest CFT: the critical Ising model~\cite{CFTBook}!\footnote{We discuss this case in detail in Appendix~\ref{App:CFT}.} In general, multi-point correlation functions of CFTs cannot be factorized into products of holomorphic functions (as would be implied by~\fr{bosid}), but are instead expressed in terms of \textit{sums of products of holomorphic functions}~\cite{CFTBook}
\bea
&&\la A(z_1,\bar z_1)\ldots A(z_N,\bar z_N)\ra\nn
&& ~~~= \sum_j C_j F_j(z_1,\ldots,z_N)\bar F_j(\bar z_1,\ldots,\bar z_N),
\eea
where $z= \tau - i x$ and $\bar z=\tau + i x$. The holomorphic functions $F,\bar F$ are called \textit{conformal blocks} and the coefficients, $C_j$ are fixed by the requirement that the entire correlation function is single valued~\cite{CFTBook}. With this in mind, it is generally not possible to speak about the factorization of operators in theories such as the WZNW model, where instead one can only speak of the factorization of conformal blocks. We will discuss this further below, in cases where we deal with perturbations of fermionic models.

\subsubsection{Diagonalization of the Sugawara Hamiltonian}
\label{Sec:SugawaraDiag}
Let us return to the Sugawara Hamiltonian~\fr{sugawara}. This appears to be rather complicated, so it is perhaps natural to think that the conformal embedding~\fr{embedding} is not terribly useful. Fortunately, things are not so bad: it is relatively straightforward to diagonalize the Sugawara Hamiltonian~\fr{sugawara}. 

Firstly, we should remember that~\fr{sugawara} is formed from two commuting pieces, which describe the left- and right-moving excitations
\bea
&&W[G;k] = H_R + H_L, \\
&& H_d = \frac{2\pi}{l(k+c_v)}\Big[J_{d,0}^aJ_{d,0}^a +2\sum_{n>0}J^a_{d,-n}J^a_{d,n}\Big]. \label{HR}
\eea
Here we have written the Hamiltonian in a more general form in terms of $c_v$, the quadratic Casimir in the adjoint representation~\cite{FultonHarris}
\be
f_{abc}f_{\bar abc} = c_v\delta_{a\bar a}. \label{casimir}
\ee
The overall separation of the Hamiltonian into chiral parts is reasonable: after all, the Hamiltonian describes a sub-sector of the theory of non-interacting massless Dirac fermions~\fr{Dirac} where, indeed, right- and left-movers are independent. In fact, this decomposition of the Hilbert space is a general property of CFTs~\cite{CFTBook,BelavinNuclPhysB84} and it allows us to discuss the left- and right-moving sectors independently.

Secondly, we can construct the lowest eigenstates of~\fr{HR} by starting with the vacuum states $|h\ra$, which are defined as the states which are annihilated by the positive Fourier components of the currents:
\be
J^a_n |h\ra = 0, \quad n > 0. 
\ee
The lowest eigenstates $|h\ra$ are then solutions of the Hamiltonian of a quantum spinning top
\be
H_{\rm top} = \frac{2\pi}{l(k+c_v)}J_{0}^aJ_{0}^a, \qquad [J_0^a,J_0^b] = i f^{abc}J^c_{0}. \label{top}
\ee
The eigenvalues of the states $|h\ra$ are proportional to the quadratic Casimir invariants $c_2[h]$ of the group;\footnote{Consider a representation $h$ of a group with generators $T^a[h]$. The quadratic Casimir operator is $\hat C_2[h] = T^a[h] T^a[h]$. This commutes with every element of the algebra, so it follows from Schur's Lemma~\cite{FultonHarris} that $\hat C_2[h] = c_2[h]I$, where $c_2[h]$ is a number known as the quadratic Casimir invariant.} focusing on the case of the $SU(N)$ group, we have 
\be
E[h] - E_0= \frac{2\pi}{l} \frac{c_2[h]}{N+k}. \label{topeig}
\ee
For the simple case of $N=2$, the states realize irreducible representations of $SU(2)$ and the associated quadratic Casimir invariants are numbered by the eigenvalues of the total spin operator, taking the familiar form $c_2[j] = j(j+1)$ with $j = 1/2, 1, 3/2, \ldots$~\cite{FateevSovJNuclPhys86}. The lowest energy states are degenerate, being characterized by both the total angular momentum $j$ and its projection $j^z = -j,-j+1,\ldots,j$: we denote each of these states by $|j,j^z\ra$. All other eigenstates are constructed by acting upon these states with the negative Fourier components of the Kac-Moody currents 
\be
J_{-n_1}^{a_1}\ldots J_{-n_p}^{a_p}|j,j^z\ra, \label{states}
\ee
where $n_q$ are positive integers. In the $SU(2)_k$ WZNW model these states have eigenvalues~\cite{TsvelikBook}
\be
E - E_0 = \frac{2\pi}{l} \Bigg[\frac{c_2[j]}{2+k} + \sum_{q=1}^p n_q\Bigg].
\ee
Thus we have knowledge of the eigenstates and eigenvalues of the Sugawara Hamiltonian.

\subsubsection{The central charge}
As the WZNW model is a CFT, another important characteristic is the value of the central charge $c$~\cite{CFTBook}. In a (1+1)-dimensional CFT with dispersion relation $\omega = v|k|$, the value of the central charge is related to the specific heat $C_v$ for a fixed volume $l$ at temperature $T$: 
\be
\frac{C_v}{l} = \frac{\pi c}{6v}T\, .
\ee
The central charge also appears in many other contexts, including the finite-size scaling of the free energy~\cite{BlotePRL86,AffleckPRL86}, the finite-size scaling of the entanglement entropy~\cite{CalabreseJPhysA09}, and the algebra and operator product expansion obeyed by the stress-energy tensor~\cite{BPZJStatPhys84,BelavinNuclPhysB84,FriedanPRL84}.\footnote{For more details about this, and CFTs in general, we provide some useful results in Appendix~\ref{App:CFT}.} In the WZNW model for the group $G$ at level $k$, the central charge is given by~\cite{KnizhikNuclPhysB84}
\bea
c= \frac{kD_G}{k+c_v}, \label{WZNWc}
\eea
where $D_G$ is the number of the generators of the algebra of the group $G$ and $c_v$ is the quadratic Casimir in the adjoint representation~\fr{casimir}. For the $SU(N)$ group, $D_{SU(N)}= N^2-1$ and $c_v = N$.

The central charge provides a useful check of the validity of a given conformal embedding: the central charge of the original Hamiltonian and the conformal embedding should be equal. Consider an example: there are $Nk$ species of free fermions in the Hamiltonian~\fr{Dirac} and hence the central charge is $c=Nk$. Using Eq.~\fr{WZNWc}, the sum of central charges of the WZNW models in the conformal embedding~\fr{embedding} is
\be
1 + \frac{k(N^2-1)}{N+k} + \frac{N(k^2-1)}{k+N} = Nk,
\ee
and hence the central charge of~\fr{embedding} is consistent with that of~\fr{Dirac}. 

\subsubsection{The conformal dimensions of primary fields}\label{hwstates}
In field theory there is a one-to-one correspondence between operators in the theory and eigenstates of the Hamiltonian~\cite{CFTBook}. This is established through the Lehmann expansion of the two-point correlation functions
\be
\la {\cal O}(\tau,x){\cal O}^\dagger (0,0)\ra = \sum_n\re^{(-E_n\tau + i P_nx)} |\la n|{\cal O}(0,0)|0\ra|^2\ ,
\ee
where the sum is performed over the complete set of eigenstates $|n\ra$ of the Hamiltonian. In a CFT this correspondence between operators and eigenstates significantly simplifies: two-point correlation functions of primary fields are fixed solely by conformal invariance~\cite{CFTBook}. For an operator ${\cal O}$ with conformal dimensions~$\Delta,\bar\Delta$ the two-point correlation functions in a cylinder geometry (with circumference $r$) are~\cite{CFTBook}
\bea
&&\la {\cal O}_{\Delta,\bar\Delta}(\tau,x){\cal O}_{\Delta,\bar\Delta}^\dagger (0,0)\ra \nn  
&& = \bigg\{\frac{\pi}{r \sinh[\frac{\pi}{r}(v\tau -i x)]}\bigg\}^{2\Delta}\bigg\{\frac{\pi}{r\sinh[\frac{\pi}{r}(v\tau +i x)]}\bigg\}^{2\bar\Delta}.\nn
\eea
Expanding this correlation function for large $\tau$ and $x$, and comparing to the Lehmann expansion, we obtain
\be
E_n - E_0= \frac{2\pi v}{r}(\Delta + \bar\Delta), ~~ P_n = \frac{2\pi }{r}(\Delta - \bar\Delta).
\ee
In WZNW models these formulae establish a correspondence between the primary fields of the theory and the eigenstates of the quantum spinning top Hamiltonian~\fr{top}. Specifically, in the $SU(N)$ WZNW model primary fields transform as tensors with respect to the $SU(N)$ group and are labelled by its representation $h$; primary fields transforming according to the $h$-representation thus have conformal dimensions (cf. Eq.~\fr{topeig})
\be
\Delta[h] = \bar\Delta[h] = \frac{c_2[h]}{k+c_v}, \label{cftDim}
\ee
with $c_2[h]$ the quadratic Casimir invariant of the representation $h$ of $SU(N)$. 

Higher representations can be obtained by arranging tensor products of lower representations, see~\cite{FultonHarris}. In analogy, one may hope to generate primary fields in higher representations through fusing fields from the fundamental representation. Indeed this is the case, with some caveats: for a WZNW model at a given level $k$, the fusion process will terminate at a certain representation, with further fusing of primary fields leading not to new primary fields, but instead descendants~\cite{CFTBook}. For example, in the $SU(2)_k$ WZNW model there are only primary fields with $j \leq k/2$~\cite{CFTBook}. 

It should also be noted that Eq.~\fr{states} implies that states with non-zero $n_q$ are created through the fusion of current operators with primary fields (which are in one-to-one correspondence with $|h\ra$). This is just another way of saying that the corresponding fields are descendants of the corresponding primaries. 

\subsection{The Wess-Zumino-Novikov-Witten Lagrangian}
\label{WZNWLagrangian}
It will be useful to have a Lagrangian formulation of the WZNW model. The action for the Sugawara Hamiltonian $W[G;k]$,~\fr{sugawara}, is given by~\cite{PolyakovPhysLettB83,PolyakovPhysLettB84,DiVecchiaCERN84,WittenCommMathPhys84,KnizhikNuclPhysB84,CFTBook}
\begin{eqnarray}
S&=& kW(g)\label{WZaction},\\
W(g) &=& \frac{1}{8\pi} \int \rd^2 x \; {\rm Tr}\Big(\partial^{\mu} g^{-1} \partial_{\mu} g\Big) + \Gamma(g), \label{defW}
\end{eqnarray}
where $g$ is a matrix from the fundamental representation of the Lie group $G$ and $\Gamma(g)$ is the famous WZNW topological term~\cite{WittenCommMathPhys84,WittenNuclPhysB83}
\be
\Gamma(g) = \frac{-i}{12\pi}   \int_B \rd^3 y \; \epsilon^{\alpha \beta \gamma}  {\rm Tr}(g^{-1} \partial_{\alpha} g g^{-1} \partial_{\beta} g g^{-1} \partial_{\gamma} g),
\label{WZNW}
\ee
where $y_i$ ($i=1,2,3$) are the coordinates of the three-dimensional ball whose two-dimensional boundary is identified with the space-time~\cite{WittenCommMathPhys84} and $\partial_\alpha \equiv \partial_{y_\alpha}$. 

An important (and rather remarkable) identity for the action~\fr{WZaction} acting on a product of fields $hg$ is~\cite{PolyakovPhysLettB83,DiVecchiaCERN84}
\bw
\bea
W(hg) &=& W(g) + W(h)+  \frac{1}{4\pi}\int \rd\tau\rd x\ \mbox{Tr}\Big[g^{-1}(\p_{\tau} -i\p_x) g h(\p_{\tau}+i \p_x)h^{-1}\Big]. \label{id}
\eea
\ew
This can be thought of as a generalization of the simple identity 
\be
[\p_{\mu}(\phi +\chi)]^2 = (\p_{\mu}\phi)^2 + 2\p_{\mu}\phi\p_{\mu}\chi + (\p_{\mu}\chi)^2,
\ee
which one can check by direction substitution of two simple $SU(2)$ matrices, $g = \exp(\frac{i}{2}\s^z\phi)$ and $h=\exp(\frac{i}{2}\s^z\chi)$, into Eq.~\fr{id}. 

\subsection{Operator correspondence between bosonic and fermionic sectors}
\label{Sec:OperatorCorrespondence}
We have already mentioned (in Sec.~\ref{Sec:ConformalEmbedding}) that the operator correspondence between the fermionic and bosonic theories in non-Abelian bosonization is more nuanced than in the Abelian case (cf. Appendix~\ref{App:AbelianBosonization}). The simplest identities concern the Kac-Moody currents. The currents for the group $G$ at level $k$ are related to the matrix field $g$ (which is in the fundamental representation of $G$ and governed by the WZNW action) through~\cite{WittenCommMathPhys84,DiVecchiaCERN84,KnizhikNuclPhysB84}
\be
\begin{split}
J_R &=- \frac{k}{4\pi}g(\p_{\tau} + i\p_x)g^{-1},\\
 J_L &= \frac{k}{4\pi}g(\p_{\tau} - i\p_x)g^{-1} \ .
 \end{split}
 \label{currents}
\ee

While currents from different symmetry sectors do not talk to one another (as they commute), this is not true for other simple fermionic operators. Take, for example, the conformal embedding~\fr{embedding} and consider generic fermion bilinears $R^\dagger_{j\alpha}L_{l\beta}$. These will feature matrix fields $g$, $U$ from the fundamental representations of $SU(N)$ and $SU(k)$: 
\be
\begin{split}
&R^\dagger _{j\alpha}L^{\phantom\dagger}_{l\beta} = \bigg\{\re^{i \Phi\sqrt{1/Nk}} g_{\alpha\beta}U_{jl}\bigg\}, \\
&j ,l = 1,...,k; \quad \alpha,\beta = 1,...,N.  
\end{split} \label{RL}
\ee
The curly brackets $\{\ldots\}$ denote that this identity is \textit{not valid in the operator sense, but applies at the level of conformal blocks}. To be precise, $N$-point correlation functions of the fermion bilinear~\fr{RL} can be constructed from $N$-point conformal blocks of the primary fields of $SU(N)_k$ and $SU(k)_N$ WZNW models and $U(1)$ bosonic vertex functions.

In order for the identity~\fr{RL} to be valid, it must be the case that the scaling dimensions of the operators of the left- and right-hand sides are equal. Substituting the values for the quadratic Casimir invariant in the fundamental representation of $SU(N)$, $c_2[h] = (N^2-1)/2N$, into Eq.~\fr{cftDim} we find 
\be
\frac{1}{2} = \frac{1}{2Nk} + \frac{N-1/N}{2(N+k)} + \frac{k-1/k}{2(N+k)},
\ee
which is valid for all $N,k$ as required. 

\subsubsection{The primary field in the adjoint representation}
An operator that we will frequently encounter (and we will discuss it in detail below) and that has a simple operator correspondence is the primary field in the adjoint representation
\be
\Phi_{\rm adj}^{ab}(x) = :\mbox{Tr}[ t^a g(x+\epsilon) t^b g^{-1}(x)]:\ .\label{primaryadj}
\ee
In models with complicated symmetries, such an operator is often symmetry-allowed and so generically appears in the low-energy field theory description. Examples of this scenario include the low-energy theories of $SU(2n)$ two-leg spin ladders~\cite{LecheminantPRB15}, two-orbital $SU(N)$ cold atomic Fermi gases~\cite{BoisPRB15}, and certain Kondo models~\cite{AkhanjeePRB13}. We will return to some of these examples later. 

In the WZNW model for group $G_k$, the conformal dimension of this operator is~\cite{KnizhikNuclPhysB84}
\be
\Delta_{\rm adj} =\bar\Delta_{\rm adj} = \frac{c_v}{k+c_v}. \label{adjoint}
\ee

\section{Some examples of non-Abelian bosonization} 
\label{Sec:ExamplesNonAbelian}

In this section, we will discuss the application of non-Abelian bosonization in several conformal embedding schemes.\footnote{We note that there are two ways in which to write the conformal embedding. Firstly, as in Eq.~\fr{emb2}, it is presented as a direct sum ($\oplus$) of symmetry groups, which can be thought of as applying at the level of the Hamiltonian or the stress-energy tensor of the theory. Alternatively, as in Sec.~\ref{Sec:k4}, it can be presented in terms of the product ($\times$), which is extremely useful for understanding the correspondence at the level of the fields appearing within the equivalent theories. We will use both conventions where appropriate.} These include the case discussed above~\fr{embedding}
\be
U(Nk)_1 = U(1) \oplus SU(N)_k\oplus SU(k)_N, \label{emb2}
\ee
and two other cases~\cite{AltschulerNuclPhysB89}: 
\bea
O(4nk)_1 &=& Sp(2n)_k\oplus Sp(2k)_n, \\
SU(2)_N &=& U(1)\oplus {\mathbb{Z}}_N,
\eea
where ${\mathbb{Z}}_N$ denotes the conformal theory of ${\mathbb{Z}}_N$ parafermions~\cite{ZamolodchikovJETP85}. Applications to cold atomic gases will be considered in detail in the subsequent section. 

\subsection{$SU(2)\times SU(k)$ model and its perturbations}
\label{Sec:SU2SUk}
Let us begin from a lattice model. Consider electrons with orbital indices $n=1,\ldots,k$ and spin index $\alpha =\, \up,\,\dn$ hopping on a one-dimensional lattice of $L$ sites and interacting via Hubbard and Hund's interactions
\bea
H &=& -t \sum_{j=1}^L\sum_{n=1}^k \sum_{\alpha=\up,\dn} \Big[c^\dagger _{n\alpha}(j+1)c^{\phantom\dagger}_{n\alpha}(j) +{\rm H.c.}\Big]\nn
&& + \sum_{j=1}^L \Big[ U n(j)n(j) - J {\bf S}(j)\cdot{\bf S}(j) \Big]. \label{orb}
\eea
Here $c^\dagger_{n\alpha}(j)$ is the creation operator for a spin-$\alpha$ electron in orbital $n$ of the $j$th lattice site, and we define the number and spin operators
\bea
n(j) &=& \sum_{n=1}^k \sum_{\alpha = \up,\dn} c^\dagger _{n\alpha}(j)c^{\phantom\dagger}_{n\alpha}(j), \\
S^a(j) &=& \sum_{n=1}^k \sum_{\alpha,\beta} c^\dagger _{n\alpha}(j) s^a_{\alpha\beta}c^{\phantom\dagger}_{n\beta} (j),
\eea
where $s^a = \s^a/2$ with $\s^a$ the Pauli matrices. 

\subsubsection{Applications of the model}

The Hamiltonian~\fr{orb} is particularly simple, taking into account onsite Hubbard and Hund's interactions for electrons with both orbital and spin degrees of freedom. As a result,~\fr{orb} and closely related models\footnote{For example, those with a modified band structure due to more complicated hopping terms, often input directly from density functional theory calculations.} have been well-studied in higher spatial dimensions, with various application to condensed matter systems. The model~\fr{orb} with $k=3$ at $1/3$ filling has been studied on the Bethe lattice~\cite{StadlerPRL15} using dynamical mean field theory (DMFT) to gain insight into spin-orbital separation in Hund's metals. The case with $k=3$ has also been studied in three spatial dimensions using DMFT~\cite{YinPRB12} in an attempt to explain the unusual frequency-dependence of the optical conductivity in iron-chalcogenide and ruthenate superconductors. A closely related three-dimensional model (with band structure from density functional theory (DFT)) with $k=5$ has been studied with slave boson mean field theory and DMFT~\cite{GiovanettiPRB15} as a description of the insulating iron selenide La$_2$O$_3$Fe$_2$Se$_2$. 

\subsubsection{Low-energy effective theory at weak coupling}
We will focus on the weak coupling limit, $t \gg |U|,|J|$, and we expand the fermionic fields in the vicinity of the Fermi points~\fr{psi}. We obtain the $U(1)\times SU(2) \times SU(k)$-invariant chiral Gross-Neveu model~\cite{GrossPRD74} with the most general symmetry allowed current-current interaction. The Hamiltonian density reads
\bea  
{\cal H} &=& -i R^\dagger _{j\s}\p_xR^{\phantom\dagger}_{j\s} + i L^\dagger _{j \s}\p_x L^{\phantom\dagger}_{j \s} 
+ g_cR^\dagger _{j \s}R^{\phantom\dagger}_{j \s}L^\dagger _{j' \s'}L^{\phantom\dagger}_{j' \s'}\nn
&& + g_{o}[R^\dagger (t^a\otimes I)R][ L^\dagger (t^a\otimes I)L]\nn
&& + g_{so}[R^\dagger (t^a\otimes s^b)R][ L^\dagger (t^a\otimes s^b)L]\nn 
&& + g_s[R^\dagger (I\otimes s^a)R][ L^\dagger (I\otimes s^a)L], \label{ModelF}
\eea
where $s^a$ ($a=1,2,3$) and $t^a$ ($a=1,...,k^2-1$) are generators of the $\mathfrak{su}(2)$ and $\mathfrak{su}(k)$ Lie algebras, respectively.\footnote{We remind the reader that normalization conventions are defined in Eqs.~\fr{generators}.}

In writing~\fr{ModelF} we have neglected two classes of interaction terms. 
\begin{enumerate}
\item Those terms which are completely chiral, such as 
\be 
[R^\dagger (t^a \otimes I) R][R^\dagger (t^a \otimes I) R].
\ee
\item Those terms which are not completely chiral, but carry net chirality, such as
\be
[ R^\dagger (t^a \otimes I )R][R^\dagger (t^a \otimes I)L].
\ee
\end{enumerate}
Neglecting such terms is justified in the following manner. In the first case, the generated terms describe forward scattering and, to leading order, generate a mode-dependent renormalization of the Fermi velocity $v_F \to \tilde v_{j\sigma}$, which we neglect for weak coupling. In the second case, these terms appear with oscillatory factors and hence are suppressed by integration over $x$ in the Hamiltonian.

This model has two integrable points. At one of them, the symmetry of the low-energy theory is extended to $U(1)\times SU(2k)$~\cite{KonikNuclPhysB15} and the interaction term can be written in the compact form 
\be
V = g_s\Big(R^\dagger _{j\s}L^{\phantom\dagger}_{j\s}\Big)\Big(L^\dagger_{p\s'}R^{\phantom\dagger}_{p\s'}\Big) + g_c\Big(R^\dagger _{j\s}R^{\phantom\dagger}_{j\s}\Big)\Big(L^\dagger _{p\s'}L^{\phantom\dagger}_{p\s'}\Big). 
\ee
This case is well understood  ---it is described by the highly-symmetric $SU(2k)$ Gross-Neveu model--- so we will mostly be interested in the case where integrability is broken. A renormalization group (RG) analysis of the model~\fr{ModelF} suggests that the $SU(2k)$ symmetry is restored in the strong coupling regime -- we will comment in more detail on this case in the following. 

The other integrable point corresponds to $g_{so} = 0$, where one can apply the conformal embedding~\fr{embedding} so that the model~\fr{ModelF} is written as the sum of three independent WZNW models perturbed by current-current interactions:
\bea
 {\cal H} = &\Big[& \frac{2\pi}{k+2}\Big(:J_R^aJ_R^a: +:J_L^aJ_L^a: \Big) + g_sJ_R^aJ_L^a\Big] \nonumber\\
&+& \Big[\frac{2\pi}{k+2}\Big(:F_R^aF_R^a: +:F_L^aF_L^a: \Big) + g_oF_R^aF_L^a\Big] \nn
&+&  \Big[\frac{\pi}{2k}\Big(:j_Rj_R: +:j_Lj_L: \Big) + g_cj_Rj_L\Big], \label{wzwo}
\eea
where $J^a$ [$F^a$] are the $SU(2)_k$ [$SU(k)_2$] currents~\fr{JF}, and $j_{R,L}$ are the $U(1)$ currents~\fr{u1cur}. Each of the symmetry sectors of the model~\fr{wzwo} are WZNW models written in the form ($J$ should replaced by $F$ or $j$ as appropriate)
\be
{\cal H}[G_k] = \frac{2\pi}{c_v+k}\Big(:J_R^aJ_R^a: +:J_L^aJ_L^a: \Big) + g J_R^aJ_L^a,\label{pertWZNW}
\ee
with $G_k = SU(2)_k,\, SU(k)_2,\, U(1)$ to be explicit. Each of these models~\fr{pertWZNW} are integrable and exactly solvable~\cite{TsvelikJETP87,SmirnovIntJModPhysA94}. From such an analysis, it is known that when the interaction parameter $g$ is positive, excitations are massive and have non-Abelian statistics. On the other hand, when $g < 0$ the interaction scales to zero under the RG, and the low-energy excitations of the model are gapless -- this is the case for the $U(1)$ charge sector of theory.

In the following, we will focus on the case with $g_s < 0$, $g_o > 0$, and we treat the model~\fr{ModelF} with a small cross-coupling interaction $g_{so}$, which can then be thought of as a perturbation about the $SU(2)_k$ WZNW critical point. We will find that this perturbation is relevant (in the RG sense), and as a result the spectrum of low-energy excitations is very different in the low-symmetry case to the spectrum of the highly-symmetric $SU(2k)$ Gross-Neveu model. 

\subsubsection{Renormalization group and low-energy projection}

We adopt the standard approach to low-energy effective field theories, starting with the RG equations~\cite{WilsonPRB71,WegnerPRB72,WegnerPRB72b}.\footnote{See Ref.~\cite{BalentsPRB96} for an example of the RG applied to a simple one-dimensional system, the two-leg Hubbard ladder, using the operator product expansion.} Strong predictions have been made from such analyses, in particular it has been argued that in some simple models~\cite{BalentsPRB96,LinPRB97,AssarafPRL04} the largest possible symmetry is restored (in our case, this would be the $U(1)\times SU(2k)$ symmetry of the integrable point). The reliability of such approaches is not entirely evident -- for models with more than one coupling constant, the Gell-Mann-Low function is universal only at first loop (but see the discussion of Ref.~\cite{KonikPRB02}). Beyond this, it is expected that the details of the RG flow depend upon the regularization scheme and so forth. Keeping these points in mind, the RG equations at first loop for~\fr{ModelF} are
\be
\begin{split}
\dot g_o &= \frac{k}{2} g_o^2 + \frac{3k}{32} g_{so}^2, \\
\dot g_{so} &= \frac{k^2 - 4}{4k} g_{so}^2 + g_{so}(2g_s + k g_o), \\
\dot g_s &= g_{s}^2 + \frac{k^2 - 1}{4k^2} g_{so}^2, 
\end{split}
 \label{RG1}
\ee
where the dots denote derivatives with respect to $\xi = 1/2\pi \ln(\Lambda/|E|)$, where $E$ is the energy and  $\Lambda$ is the momentum cutoff.  

As we mentioned previously, we focus on the case with bare couplings $g_o(0) > 0$ and $g_s(0) < 0$. When integrability is preserved ($g_{so} = 0$), the RG equations simplify $\dot g_o = k g_o^2/2$, $\dot g_s = g_s^2$. The current-current interaction in the spin sector scales to zero $g_s \to 0$ under the RG flow and the sector is gapless. On the contrary, the orbital sector flows to strong coupling $g_o \to O(1)$ and the excitations in the sector are massive. The RG flow is cut-off at the \textit{RG scale}, $\xi_o =(1/g_o(0)-1)/k \approx 1/g_o(0)k$. 

In the non-integrable case ($g_{so} \neq 0$) a marginally relevant perturbation is added to the theory. If we assume that the coupling $g_{so}$ is much smaller than $g_s, g_o$ for the whole RG flow (that is $g_{so}(\xi) \ll g_{s}(\xi), g_o(\xi)$ up to $\xi = \xi_o$), we can extract the renormalized spin orbit coupling parameter from the RG equations
\be
g_{so}(\xi_o) \approx \frac{g_{so}(0)}{g_o(0) + 2|g_s(0)|/k}. 
\ee
Notice that such an assumption is valid if the bare coupling $g_{so}(0)$ is sufficiently small. Consistent with this assumption, herein we take $|g_{so}(\xi_o)| \ll 1$ and treat the spin-orbit current-current interaction as a perturbation. 

\subsubsection{The  $SU(2)_k$ WZNW model perturbed by the adjoint operator}
\label{Sec:SU2Kadj}

We now focus on formulating a low-energy effective description of the model at energies smaller than the orbital gap. This is done by projecting the spin-orbit term $g_{so}$ onto the ground state of the perturbed $SU(k)_2$ WZNW theory. In Refs.~\cite{AkhanjeePRB13,KonikNuclPhysB15}, it was argued that the resulting perturbation is described in terms of the primary field of the $SU(2)_k$ WZNW model in the adjoint representation, $\Phi^{ab}_{\rm adj}$ introduced in Eq.~\fr{primaryadj}. The main argument for this was based upon the following observations:
\begin{enumerate}[label=(\roman*)]
\item The scaling dimension of the spin-orbit coupling $g_{so}$ term is $2$.
\item The perturbing operator should be represented as a product of conformal blocks of the $SU(k)_2$ and $SU(2)_k$ primary fields. 
\item The primary fields in the adjoint representation of the $SU(k)_2$ and $SU(2)_k$ theories have scaling dimensions~\cite{AkhanjeePRB13} (cf. Eqs.~\fr{adjoint})
\be
d_{\rm adj}[SU(k)_2] = \frac{2k}{k+2}, \quad d_{\rm adj}[SU(2)_k] = \frac{4}{k+2}. \nonumber
\ee 
Hence the product of the primary fields in the adjoint representation of the two sectors produces an operator with the correct scaling dimension. 
\item The orbital sector of the theory flows to strong coupling and becomes gapped. On the vacuum the only operator which has a non-zero average is the trace of the adjoint field ${\rm Tr}\, \Phi_{\rm adj}[SU(k)_2]$~\cite{KonikNuclPhysB15}. 
\item After integrating out high-energy degrees of the freedom, the local operator ${\rm Tr}\, \Phi_{\rm adj}[SU(2)_k]$ will be present in the spin sector of theory, emerging from the entire product of the conformal blocks. 
\end{enumerate}

So, to describe the low-energy spin sector of the theory we have a $SU(2)_k$ WZNW model perturbed by the primary field in the adjoint representation
\bea
S&=& kW(g) + \lambda \sum_{a=1}^3\int \rd^2x\, {\rm Tr}[\s^a g\s^a g^\dagger ]\label{pertaction},
\eea
where $W(g)$ is the WZNW Lagrangian defined in Eq.~\fr{defW}. The mass scale in the orbital $SU(k)_2$ sector, $M_o \sim g_o^{1/k}\exp(-2\pi/kg_o)$, plays the role of the ultra-violet cut-off in this theory.

In order to relate the action~\fr{pertaction} to the original fermionic model~\fr{ModelF}, we require that 
\be
\lambda \sim g_{so}(\xi_0)\Big\la \mbox{Tr}\, \Phi_{\rm adj}[SU(k)_2]\Big\ra. \label{lambdascale}
\ee
This statement is a little problematic: Eq.~\fr{lambdascale} is not well-defined as the ground state of~\fr{pertWZNW} is degenerate and the expectation value can take multiple values. For the purposes of the following, we will treat $\lambda$ as an arbitrary parameter, which can take either sign. In the physical realization~\fr{ModelF}, we argue that the system will choose the ground state that maximizes the energy gap in the spin sector, and hence~\fr{lambdascale} is fine. 

In the low-energy effective action~\fr{pertaction}, the perturbing operator is strongly relevant with scaling dimension $d = 4/(k+2)$. As a result, it generates a characteristic energy scale 
\be
\Lambda_{so} \sim |\lambda|^{1/(2-d)}M_o. \label{RG}
\ee
Before we consider the case of general $k$, we will first discuss two particularly simple examples when there are two or four orbitals per site. 

\subsubsection{Simple case (i) $k=2$} \label{Sec:casei}
For $k=2$ orbitals per site, the model is equivalent to three massive Majorana fermions~\cite{GNTBook,FateevSovJNuclPhys86} with masses $\Lambda_{so} \sim \lambda$. This follows from a relation between the $SU(2)_2$ currents and products of Majorana fermions~\cite{FateevSovJNuclPhys86}
\be
J^a_R = -\frac{i}{2} \epsilon^{abc}\chi_R^b \chi_R^c, ~~ F^a_R = -\frac{i}{2}\epsilon^{abc} \xi^b_R \xi^c_R. 
\ee
As a result of this relation, an equivalent reformulation of the model~\fr{ModelF} with $k=2$ is 
\bea
{\cal H} &=& \frac{i}{2}(\chi^a_L\p_x\chi_L^a -\chi^a_R\p_x\chi_R^a + \xi^a_L\p_x\xi_L^a-\xi^a_R\p_x\xi_R^a) \nn
&& + \frac12 \sum_{a>b} \Big[ g_s(\chi_R^a\chi_L^a)(\chi_R^b\chi_L^b)+ g_o(\xi_R^a\xi_L^a)(\xi_R^b\xi_L^b) \Big] \nn
&& + \sum_{a,b} 2g_{so}(\chi_R^a\chi_L^a)(\xi_R^b\xi_L^b). \label{k2}
\eea
When $g_{so}=0$, the averages $\la \chi_R^a \chi_L^a \ra$ and $\la \xi^a_R \xi^a_L \ra$ do not have a definite sign. It is also apparent that the sign of the averages should not depend on the sign of coupling $g_{so}$, and so the system should choose signs self-consistently. 

When the model~\fr{k2} with $g_s < 0$, $g_0 > 0$ is perturbed by the spin-orbit coupling $g_{so}$ the low-energy effective theory is formed from a triplet of massive Majorana fermions $\chi^a$ with a current-current interaction. This interaction can lead to the creation of bound states of the fermions, see for example Ref.~\cite{FrishmanArxiv16}. 

\subsubsection{Simple case (ii) $k=4$}
\label{Sec:k4}
An additional case of interest is $k=4$, where the conformal embedding is $SU(4)_2\times SU(2)_4$. This case is special because the central charges of each of the two WZNW models are integers ($c=5$ and $c=2$ respectively). This indicates that they can be bosonized using Abelian bosonization.  

In particular, the action~\fr{pertaction} for the $SU(2)_4$ spin sector can be reformulated in terms of two bosonic fields $\phi_{1,2}$~\cite{FateevSovJNuclPhys86,FabrizioPRB94}
\bea
S = \int \rd^2x \left[\frac{1}{8\pi}\sum_{a=1,2}(\p_{\mu}\phi_a)^2 + \lambda \sum_{i=1}^3\cos\Big(e_a^{(i)}\phi_a\Big)\right],\nn \label{k4} 
\eea
with
\be
\Big({\bf e}^{(i)}\Big)^2 = 2/3, \quad \Big({\bf e}^{(i)}{\bf e}^{(j)}\Big) = - 1/3.\nonumber
\ee

Similarly, when perturbed by the trace of primary field in the adjoint representation, the $SU(4)_2$ orbital part of the WZNW action can be expressed in terms of  six bosonic fields $\theta_a$ ($a=1,\ldots,6$)~\cite{LecheminantPRB15}
\be
S = \int \rd^2x \left\{ \frac{1}{8\pi} \sum_{a=1}^6(\p_{\mu}\theta_a)^2 + \bar\lambda\sum_{a>b}\cos\left[\sqrt{2/3}(\theta_a - \theta_b)\right]\right\}. \label{c5}
\ee
One of the fields is redundant since in the perturbed $SU(4)_2$ theory, the non-linear part of the action does not depend upon the center of mass field $\theta_0 \equiv \sum\theta_a$, which can be factored out as a Gaussian theory.

The form ~\fr{c5} is convenient for refermionization:
\bea
S = \int \rd^2x &\Big[& R^+_a(\p_\tau - i\p_x)R^{\phantom\dagger}_a + L^+_a(\p_\tau + i\p_x)L^{\phantom\dagger}_a \nonumber\\
&& + g_0R^+_aR^{\phantom\dagger}_aL^+_bL^{\phantom\dagger}_b - gR^+_aL^{\phantom\dagger}_aL^+_bR^{\phantom\dagger}_b \Big], 
\eea
where $g_0 = 2\pi/3$ is chosen to change the compactification radius of the fields and $g \sim \bar\lambda$. The fermionized action can be more convenient for numerical calculations and for the application of the $1/N$-expansion.   

The model~\fr{k4} is related to the low-energy effective theory for the four channel Kondo model (e.g., spin-half electrons with four-orbital degrees of freedom coupled to a spin-1/2 impurity), which shares its description with a spin-half impurity coupled to a spin-one Fermi gas~\cite{FabrizioPRB94}. In the case of the impurity model, the nonlinear terms in~\fr{k4} are located at a single spatial point. 
\subsubsection{The semi-classical limit: $k\gg1$}
\label{Sec:SemiClassics}
Having discussed two simple cases, let us now return to general values for the number of orbitals $k$. Focusing on the case when $k\gg1$, we can treat the action~\fr{pertaction} semi-classically. To do so, we use the identity
\be
\sum_{a=1}^3 {\rm Tr}[\s^a g\s^a g^\dagger ] = 2\, {\rm Tr}[g] {\rm Tr}[g^\dagger ] - 2, \label{semiclass}
\ee
and then we parameterize $g$, the $SU(2)$ matrix, by 
\be
g = n_0\hat I + i \s^an_a~~{\rm with}~~ n_0^2 + {\bf n}^2 =1. \label{parameterization}
\ee
As the Pauli matrices are traceless, we see that the perturbation is of the form
\be
V = \tilde \lambda n_0^2 = \tilde \lambda (1 - \bf{n}^2 ).
\ee

When $\tilde\lambda < 0$, the low-energy theory describes three weakly interacting vector bosons governed by the Lagrangian density
\be
{\cal L}_{\rm eff} = \frac{k}{4\pi}(\p_{\mu}{\bf n})^2 + |\tilde \lambda|{\bf n}^2 +...\,,
\ee
where the ellipses denote higher order terms, such as interactions. The higher order terms are suppressed with increasing $k$, as can be seen by rescaling the fields $n^a \to n^a/\sqrt{k}$. Due to the degeneracy in expanding about either $n_0 = \pm 1$, the ground state is formed from two degenerate massive triplets of vector bosons, which are $SU(k)$ singlets. The mass of the excitations (ignoring renormalization due to interactions) is $M_{\rm tr,-} \sim \sqrt{|\tilde\lambda|/k}$, which is also the mass envisaged from the RG considerations, see Eq.~\fr{RG}. Furthermore, the scaling of the mass $M_{\rm tr,-}$ with $k$ is supported by truncated conformal space approach (TCSA) numerical calculations~\cite{KonikNuclPhysB15}, which also show that when $k>3$ there exist bound states of the vector bosons. 

For $\tilde \lambda > 0$ the situation is more interesting. For energies below $\sqrt{\tilde\lambda}$, the field component $n_0$ is suppressed and ${\bf n}$ becomes the unit vector (that is ${\bf n}\cdot{\bf n} = 1$). As a consequence the WZNW term in the action~\fr{WZNW} becomes a topological term~\cite{AffleckPRB87,AffleckNuclPhysB88,SheltonPRB96}:
\be
\Gamma(i\s^an^a) = \frac{i}{8}\int \rd^2x \epsilon_{\mu\nu}\Big({\bf n}\cdot[\p_{\mu}{\bf n}\times\p_{\nu}{\bf n}]\Big) \equiv i\pi\Theta, \label{thetaterm}
\ee
where $\Theta$ is an integer. This can be interpreted as the number of points $(x,\tau)$ mapped to identical values of 
\be
{\bf n}(x,\tau) = (\cos\theta,\sin\theta\cos\psi,\sin\theta\sin\psi),
\ee
for the transformation $\theta(x,\tau),\, \psi(x,\tau)$~\cite{TsvelikBook}.

For the particular case under consideration~\fr{pertaction}, the topological term appears with coefficient $k\pi$, so it contributes non-trivially to the action only when $k$ is odd: 
\be
S = \frac{k}{4\pi}\int \rd^2x (\p_{\mu}{\bf n})^2 + i\pi k\Theta, ~~ {\bf n}^2=1. \label{nlsm}
\ee
This model is exactly solvable~\cite{ZamoldchikovAnnPhys79,WiegmannPhysLettB85,WiegmannJETPLett85,ZamolodchikovNuclPhysB92}; for $k$ even the triplet of vector bosons is gapped with mass 
\be
M_{\rm tr,+} \sim k {\tilde\lambda}^{1/2} \exp(- k/2). \label{scale}
\ee
This structure agrees with the result for $k=2$ (see Sec.~\ref{Sec:casei}) and suggests that the same may be valid for any even $k$. However, it worth keeping in mind that the small scale~\fr{scale} is much smaller than the RG scale~\fr{RG}. For the case of odd $k$, the mass scale~\fr{scale} marks a crossover to a basin of attraction described by the critical $SU(2)_1$ WZNW model~\cite{FateevPhysLettB91}.

We see that the mass $M_{\rm tr,+} < M_{\rm tr,-}$ and as a consequence, the system with $\tilde \lambda > 0$ has a greater ground state energy. This means that the fermionic model~\fr{ModelF} will energetically favor $\lambda < 0$ (recall that Eq.~\fr{lambdascale} is a little problematic due to the degenerate ground states, so the sign of $\lambda$ is not given \textit{a priori} in our analysis). This assertion is supported by the TCSA calculations of Ref.~\cite{KonikNuclPhysB15}. It is worth noting, however, that for small $k=3,4$ the difference between the ground state energy in the two phases ($\lambda < 0$ or $\lambda > 0$) is a small fraction of the mass $M_{\rm tr,-}^2$ per unit cell, and hence the $\lambda > 0$ phase should be thought of as being metastable. 

\subsubsection{Comparing two limits: correlation functions and quasi-long-range order}
Let us now compare the maximally symmetric point and the non-integrable case considered above in terms of their correlation functions and the quasi-long-range order. 

\begin{enumerate}
\item \textit{The maximally symmetric $U(1)\times SU(2k)$ limit.} 
\item[] This is realized when the bare couplings $g_o, \, g_s,\, g_{so}$ are positive and of the same order: under the one-loop RG flow, the symmetry is restored in the strong coupling limit. The low-energy theory is the $SU(2k)$ chiral Gross-Neveu model, whose spectrum of excitations is well known~\cite{GrossPRD74} and consists of gapless $U(1)$ collective modes and massive excitations in the $SU(2k)$ sector with masses~\cite{AndreiPhysLettB80}
\be
M_j = M_1 \frac{\sin(\pi j / 2k)}{\sin(\pi/2k)}, \quad j = 1,\ldots,2k-1. \label{MjAndrei}
\ee  
These excitations belong to multiplets which transform according to a representation described by a Young tableau consisting of a single column of $j$ boxes. In the low-energy limit, single fermions are incoherent, made up of both a $U(1)$ collective excitation and an $SU(2k)$ excitation with smallest mass, $j=1$. 

\item \textit{The non-symmetric $U(1)\times SU(2) \times SU(k)$ limit.}
\item[] This is the case that we have described above, where the coupling $g_{so}$ is small such that spin and orbital sectors of the model are weakly coupled at high energies. Due to the electron carrying charge, spin, and orbital indices, the single electron excitation is much higher in energy than the collective modes of the spin sector. 
\end{enumerate}

In the non-symmetric case, the low-energy collective spin modes can be seen in spectral functions of fermion bilinears which are orbital singlets. There are two such operators
\be
O_{2k_F} = R^\dagger (I\otimes I)L,\quad S^a_{2k_F} = R^\dagger (I\otimes s^a)L,
\ee
which correspond to $2k_F$ charge density wave (CDW) and spin density wave (SDW) order parameters, respectively. The operators can be expressed as products of conformal blocks of the $SU(k)_2$ and $SU(2)_k$ primary fields, multiplied by a vertex operator of the bosonic field $\Phi$ associated with the charge degree of freedom, $\exp(i\sqrt{2\pi/k}\Phi)$. The charge boson is governed by a Gaussian action $S_c = \frac12 \int \rd^2 x (\p_\mu \Phi)^2$. At low-energies, the order parameters can be replaced by
\bea
O_{2k_F} &=& A\re^{i\sqrt{2\pi/k}\Phi}\mbox{Tr}(g),\nn
S^a_{2k_F} &=& A'\re^{i\sqrt{2\pi/k}\Phi}\mbox{Tr}(is^ag)\ ,
\eea
where $g$ is a matrix in the fundamental representation of $SU(2)$, cf. Sec.~\ref{WZNWLagrangian}. 

The ground state energy for the model~\fr{pertaction} is lower when $\lambda < 0$, and as a result the CDW order forms with ${\rm Tr}(g) \sim n_0$. Within the ground state the ${\mathbb{Z}}_2$ symmetry between $n_0 = \pm 1$ is broken and $n_0$ acquires a finite average. The large distance $x \gg M_{\rm tr, -}^{-1}$ asymptotics for the two-point function of the CDW order parameter are 
\be
\left\la O_{2k_F}(\tau,x)O^\dagger _{2k_F}(0,0)\right\ra =\frac{Z}{(\tau^2 +x^2)^{1/2k}}. \label{O2kF}
\ee
This follows simply from the correlation function of the bosonic exponents and ${\rm Tr}(g)$ developing a finite average. $Z$ can be estimated by recalling that the operator ${\rm Tr}(g)$ has power law correlations at intermediate distances $M_o^{-1} \ll |x| \ll M_{\rm tr,-}^{-1}$ and scaling dimension $3/2(k+2)$. Hence
\be
Z \sim \left(\frac{M_{\rm tr,-}}{M_o}\right)^{3/2(k+2)}.
\ee

So, we see that the ground state in the non-symmetric case has $2k_F$ CDW quasi-long-range order, which is not too different from the high symmetry case. One important difference, however, appears when examining the spin-spin correlation functions: the asymptotics of these look very different in the two cases. In the high symmetry $SU(2k)$ Gross-Neveu model, there is both the ubiquitous $U(1)$ charge excitation continuum and a continuum of single particle excitations~\cite{GrossPRD74}, which in the present case is dominated by the triplet modes
\be
\lla {\bf S}_{2k_F}(\tau,x)\cdot{\bf S}_{-2k_F}(0,0)\rra \sim \frac{K_0(M_1\sqrt{\tau^2+x^2})}{(\tau^2 +x^2)^{1/2k}} + \ldots, \label{spinsusc}
\ee
where $K_0(x)$ is the modified Bessel function of the second kind, and the ellipses denote higher order terms corresponding to emission of more than one massive particle. The threshold energy for the spin spectral function is $M_1$, a much smaller energy scale than the threshold of the particle-hole continuum.

Beyond the spin sectors of the two limits of model~\fr{ModelF}, another difference that emerges is that the non-symmetric limit has an orbital dynamical susceptibility that is drastically different from the spin one~\fr{spinsusc}, as a result of the large mass for orbital excitations, $M_o \gg M_{tr}$. Spectral functions of operators that involve the emission of orbital excitations thus have large spectral gaps, unlike the symmetric case. 

\subsubsection{Alternative quasi-long-range order}
As we have discussed, the phase with $n_0 = 0$ is not the ground state of model~\fr{ModelF}, but for small couplings $g_{so}$ it is close in energy to the ground state. One can then speculate what will happen if this phase is stabilized as the ground state by the presence of additional interactions not included within our model~\fr{ModelF}. In the state with $n_0 = 0$, the long distance $|x| \gg M_1^{-1}$ asymptotics of the two-point function of the SDW order parameter are
\bea
\la {\bf S}_{2k_F}(\tau,x)\cdot{\bf S}_{-2k_F}(0,0)\ra &\sim& (\tau^2 +x^2)^{-\frac{1}{2k}}\nn
&&\times{\cal F} \Big(M_1(\tau^2 +x^2)^{1/2}\Big), \quad
\eea
where ${\cal F}$ is the correlation function of the unit vector fields ${\bf n}$ in the $O(3)$ nonlinear sigma model~\fr{nlsm}~\cite{TsvelikBook}. When the number of orbitals $k$ is odd ${\cal F}(y) \sim 1/y$ is a power law, whilst for even $k$ it decays exponentially. At intermediate distances, $M_o^{-1} < |x| < M^{-1}_1$, the correlation functions for $k$ even or odd are indistinguishable. 

Due to the behavior of ${\cal F}$ the SDW susceptibility (cf. Eq.~\fr{spinsusc}) is singular only when $k$ is odd. However, there is quasi-long-range order in the spin sector for even $k$, but this is associated with a higher harmonic of the SDW, the $4k_F$ component, whose operator is 
\be
{\bf S}_{2k_F}^2 \sim \re^{i\sqrt{8\pi/k}\Phi}.
\ee
This behavior is rather reminiscent of the under-doped phase of the high-$T_c$ cuprate superconductors~\cite{FradkinRMP15}. 

\subsection{Generalization from $SU(2)$ to $SU(N)$}

We have focused on the model~\fr{ModelF} which possesses $U(1)\times SU(2) \times SU(k)$ symmetry. The above results are easily generalized to the case with higher spin symmetry, replacing $SU(2)$ by $SU(N)$. The low-energy theory in the spin sector will be of the same form as~\fr{pertaction}, with $SU(2)_k$ Kac-Moody currents replaced with $SU(N)_k$ currents. Let us briefly discuss some results in this case. 

\subsubsection{The large $k$ limit}
\label{SUNsemiclassic}
We once again consider the semi-classical limit with $k\gg1$, where it is easy to see that some new features emerge from the enlarged spin symmetry. To begin we parameterize the $SU(N)$ matrix $g$ in the following manner:
\be
g = U^\dagger \Lambda U, \quad 
\Lambda = \mbox{diag}(\re^{i\alpha_1},\ldots,\re^{i\alpha_N}),
\ee
where $U$ is an $N\times N$ matrix containing $N(N-1)$ real parameters, and
\be
\sum_{j=1}^N\alpha_j = 0\,{\rm mod}(2\pi)\, .
\ee
It is then easy to see that the semi-classical limit of the perturbing operator~\fr{semiclass} becomes 
\be
{\rm Tr}(g){\rm Tr}(g^\dagger) = 2  \sum_{k>l}\cos(\alpha_k - \alpha_l). 
\ee

When the coupling $\lambda < 0$, the lowest energy state (e.g., the vacuum) maximizes the value of ${\rm Tr}(g)$. This is achieved through fixing 
\be
\alpha_i = \alpha_j = \frac{2\pi m}{N}, \qquad m=1,\ldots,N, \label{alp}
\ee
and the matrix $g$ can then be approximated by
\be
g = \re^{2\pi i m/N}\Big[(1- b_a b_a/2)\hat I+ i b_a r^a +O(b^4)\Big],
\ee
where $r^a$ are the generators of the $\mathfrak{su}(N)$ Lie algebra ($a=1,\ldots,N^2-1$). Under this parameterization, the quadratic part of the action becomes
\be
S \approx \int \rd^2 x\, \Big( k( \p_{\mu} b_a )^2 + \tilde c\, b_a b_a \Big).
\ee
This is a theory of $N^2-1$ massive bosons $b_a$, each of which corresponds to a generator of $\mathfrak{su}(N)$.

With knowledge of the ground state and low-energy action in place, we can infer that the model has two types of excitations: (i) kinks that interpolate between the $N$ degenerate ground states (corresponding to $m=1,\ldots,N$ in~\fr{alp}); (ii) small fluctuations about the ground states. The latter excitations will transform according to the adjoint representation of $SU(N)$. This picture constitutes a straightforward generalization of the $N=2$ case. We note that it may so happen that the adjoint particles become unstable at small $k$ and disappear from the spectrum, but this is obviously beyond the semi-classical analysis. 

When the coupling $\lambda > 0$ and $N=2n$, the vacuum energy is minimized by 
\be
\Lambda = {\rm diag}(\underbrace{1,\ldots,1}_{n~{\rm times}},\underbrace{-1,\ldots,-1}_{n~{\rm times}})\, .
\ee
The $SU(N)$ matrix $g$ then becomes
\be
g = i Q\re^{2\pi i l/N},~~{\rm with}~~Q^2 =I,\,~{\rm Tr}\, Q =0. \label{semiclassQ}
\ee
with $l = 1,\ldots, N$ corresponding to different ground states, as in the $\lambda < 0$ case. 

When the phase factor in Eq.~\fr{semiclassQ} is absent, the low-energy model would correspond to the Grassmanian sigma model on the $U(2n)/[U(n)\times U(n)]$ manifold ($N=2n$)~\cite{AffleckNuclPhysB88}. In that case, the WZNW term is topological, $k \Gamma = i \pi \Theta k$, cf. Eq.~\fr{thetaterm}. For $N>2$, the theory is in a gapped phase with a broken discrete (${\mathbb{Z}}_2$) symmetry~\cite{AffleckNuclPhysB88}. In the limit of $n\to0$, such a model describes the integer quantum Hall effect~\cite{LevinePRL83,LevineNuclPhysB84a,LevineNuclPhysB84b,LevineNuclPhysB84c,PruiskenNuclPhysB84}, whilst for $n\to1$ it becomes the well-known $O(3)$ nonlinear sigma model with a topological term~\cite{AffleckNuclPhysB86,AffleckPRB87}. The existence of a critical point for odd values of the topological term is firmly established in these two cases: for $n\to 0$ the universality class of the critical point is unknown, whilst for $n = 1$ it is $SU(2)_1$~\cite{AffleckPRB87}. 

\subsubsection{Special case: $k=N$}
In the special case $N=k$ the model~\fr{pertaction} is exactly solvable: the central charge of the model is $c = (N^2-1)/2$ (cf. Eq.~\fr{WZNWc}) and the $SU(N)_N$ WZNW model is equivalent to the model of $N^2-1$ massless Majorana fermions. In this case, the operator corresponding to the primary field in the adjoint representation takes a particularly simple form -- it is the Majorana mass term. This has two consequences: (i) there are no kink excitations (as the ground state is now non-degenerate); (ii) the sign of $\lambda$ does not make a difference to the spectrum for $k=N$, although it certainly affects correlation functions of the fields. 

\subsection{$Sp(2N)$ model: Competition between superconductivity and charge density wave order}

Let us now turn our attention to a different model of spin-$1/2$ fermions with orbital degeneracy ($N$ orbitals), governed by the Hamiltonian 
\bea
H &=& -t \sum_{j=1}^L \sum_{n=1}^{N} \sum_{\alpha = \up,\dn} \Big[ c^\dagger_{n\alpha}(j+1)c^{\phantom\dagger}_{n\alpha}(j) + {\rm H.c.}\Big] \nn
&& + V \sum_j \sum_{\alpha,\beta,\gamma,\delta} \Big[c^\dagger_{n\alpha}(j)\epsilon_{\alpha\beta} c^\dagger_{n\beta}(j)\Big]\Big[ c^{\phantom\dagger}_{m\gamma}(j) \epsilon_{\gamma\delta} c^{\phantom\dagger}_{m\delta}(j)\Big] \nn
&& + U \sum_j n(j)n(j), \label{Model2}
\eea
where $\epsilon_{\alpha\beta}$ is the Levi-Civita symbol, $n(j) = \sum_{m=1}^{N}\sum_{\alpha = \up,\dn} c^\dagger_{m\alpha}(j)c_{m\alpha}(j)$ is the number operator on each site $j$. The fermions interact via an onsite Hubbard interaction $U$ and a pairing interaction $V$. We will consider the model far from half-filling, such that umklapp processes are negligible. 

\subsubsection{Low-energy effective theory at weak coupling}

As with the previous case, we consider the weak-coupling limit $|U|,|V| \ll t$, and proceed by linearizing the spectrum [see, e.g.,~\fr{psi}]. The left- and right-moving fermionic fields are governed by the following Hamiltonian density
\bea
 {\cal H} &=& -i R^\dagger _{n\alpha}\p_xR^{\phantom\dagger}_{n\alpha} + i L^\dagger _{n\alpha}\p_x L^{\phantom\dagger}_{n\alpha}   -  g_{cdw}O^{\phantom\dagger}_{cdw}O^\dagger _{cdw}\nn
 &&  - g_{sc}O^{\phantom\dagger}_{sc} O^\dagger _{sc} + g_c R^\dagger _{n\alpha}R^{\phantom\dagger}_{n\alpha}L^\dagger _{m\beta}L^{\phantom\dagger}_{m\beta},   \label{Model1}
\eea
where we have explicitly written the interaction terms as products of the charge density wave (CDW) and superconducting (SC) order parameters
\be
 O_{cdw} = \Big(R^\dagger _{n\alpha}L^{\phantom\dagger}_{n\alpha}\Big),\quad O_{sc} = \Big(R^\dagger _{n\alpha}\epsilon_{\alpha\beta}L^\dagger _{n\beta}\Big),
\ee
and the interaction parameters are given in terms of the parameters of the microscopic model~\fr{Model2} by
\be
g_{cdw} = -2U,\quad g_{sc} = -4V,\quad g_c = 2U.
\ee
As usual, we have neglected terms which carry net chirality [see the discussion following Eq.~\fr{ModelF}]. 

The model~\fr{Model1} explicitly features both the SC pairing and CDW order parameters in its interaction terms. This feature of the model leads to direct competition between these two types of order: when $g_{cdw}, \, g_{sc} > 0$ both the interaction terms are relevant in the RG sense, whilst $g_c$ is always marginal. Depending on which coupling is larger, $g_{cdw}$ or $g_{sc}$, the dominant fluctuations at low-energies are of either CDW or SC type. 

Although the competition between SC and CDW order is a feature of the theory for $N=1$, we will be interested in its generalization to higher numbers of orbitals $N$. In part, our interest in the multi-orbital case stems from the facts that such a model may possess an enlarged $Sp(2N)$ symplectic group symmetry~\cite{LecheminantPRL05,LecheminantNuclPhysB08}, which we will discuss further below. 

\subsubsection{The symplectic group $Sp(2N)$} 
\label{Sec:Sp2N}

The symplectic group $Sp(2N)$ is a subgroup of the special unitary group $SU(2N)$. Its generators $T^a$ change sign under
\be
\Omega (T^a)^T \Omega = - T^a, 
\ee
where the $2N\times 2N$ matrix $\Omega$ is
\be
\Omega = \left(\begin{array}{ccc} 0 && I_N \\ -I_N & & 0 \end{array} \right), \label{Omega}
\ee
with $I_N$ the $N\times N$ unit matrix. 

For our discussion, it will be useful to label each element of the matrix by a pair of numbers $(\alpha,n)$ with $\alpha = \pm 1$ and $n=1,\ldots,N$.\footnote{In the fermionic model, these indices will correspond to spin and orbital quantum numbers, respectively.} Under such a relabeling, the matrix $\Omega$ acts as the antisymmetric tensor on the greek indices and trivially on the roman indices: $\Omega = \epsilon \otimes I_N$, cf.~\fr{Omega}.

With this parameterization of the indices, the completeness relation for the generators of the $\mathfrak{sp}(2N)$ Lie algebra read
\bea
&&(T^a)_{(\alpha,m)(\beta,n)}(T^a)_{(\gamma,o)(\delta,p)}\nn
&&\qquad = \delta_{mp}\delta_{no} \delta_{\alpha\delta}\delta_{\beta\gamma} -  \delta_{mo}\delta_{np} \epsilon_{\alpha\gamma}\epsilon_{\beta\delta}.  \label{completeness}
\eea

\subsubsection{Applications of the model}

The Hamiltonian~\fr{Model2} can be seen as a straightforward generalization (to fermions) of the model introduced in Ref.~\cite{HoPRL98} to describe the behavior of higher spin (e.g., spinor) bosonic gases studied in cold atomic gases. In particular, it is special case of another model, describing the most general Hamiltonian for fermions with half-integer spin $F$ with point-like interaction (the general model contains $F+1/2$ parameters, rather than the two present in our model~\cite{LecheminantPRL05,LecheminantNuclPhysB08}). 

This simplified model serves as a starting point for understanding the physics of many systems with higher spin and orbital degeneracy. This includes ultra-cold gases of fermions, such as $^6$Li, $^{40}$K, and $^{173}$Yb, where unusual superfluid phases are expected to occur~\cite{HoPRL99,HonerkampPRL04,HonerkampPRB04,PaananenPRA06,HePRA06,PaananenPRA07,CherngPRL07}, including superfluids composed from molecular bound states of the constituent fermions.  This is discussed further in Sec.~\ref{Sec:ColdAtomsExamples}. As an aside, we also note that the type of pairing in~\fr{Model2} has been studied in the context of frustrated spin-$1/2$ quantum magnets in two spatial dimensions~\cite{ReadPRL91}.
Recently it has been found that the enlarged $Sp(2N)$ symmetry also emerges in the model of an interacting metallic wire in a strong longitudinal field~\cite{BulmashArxiv16}.

The model~\fr{Model2} has been studied using both RG and CFT techniques~\cite{LecheminantPRL05,LecheminantNuclPhysB08}. Related models with higher orbital number have availed themselves of additional approximate techniques, such as the $1/N$ expansion -- see, for example, Ref.~\cite{FlintNatPhys08}.

\subsubsection{The quantum critical point between CDW and SC phases} 
\label{Sec:cdwsc}
A good starting point for our discussion of model~\fr{Model1} is the special case of $g_{cdw} = g_{sc} \equiv g_o$, where the model acquires an enlarged $Sp(2N)$ symmetry and is integrable. The interaction terms can be written in terms of the $Sp(2N)_1$ currents\footnote{This can easily been seen from the completeness relation~\fr{completeness} which implies
\be
\Big(R^\dagger T^a R\Big) \Big(L^\dagger T^a L\Big) = -{\cal O}_{cdw}{\cal O}^\dagger _{cdw} - {\cal O}_{sc}{\cal O}_{sc}^\dagger\, . \nonumber
\ee
}
\be
J^a = R^\dagger T^a R, ~~  \bar J^a = L^\dagger T^a L, \label{sp2Ncurrents}
\ee
and the $SU(2)_N$ currents 
\be
\begin{split}
j^z &= R^\dagger_{n\alpha} R^{\phantom\dagger}_{n\alpha},~~ j^+ = R^\dagger_{n\alpha}\epsilon_{\alpha\beta} R^\dagger_{n\beta},~~j^- = (j^+)^\dagger, \\ 
\bar j^z &= L^\dagger_{n\alpha} L^{\phantom\dagger}_{n\alpha},~~\bar j^+ = L^\dagger_{n\alpha}\epsilon_{\alpha\beta} L^\dagger_{n\beta},~~\bar j^- = (\bar j^+)^\dagger,
\end{split}  \label{su2currents}
\ee
as the sum of two commuting WZNW models perturbed by current-current interactions
\bea
{\cal H} &=& {\cal H}_{Sp} + {\cal H}_{SU}, \label{wzwspsu} \\
{\cal H}_{Sp} &=& \frac{2\pi}{2N+1} \Big(:J^a J^a: + :\bar J^a \bar J^a: \Big) + g_o J^{a}\bar J^{a},\qquad \label{sp1}\\
{\cal H}_{SU} &=& \frac{2\pi}{2+N}\Big(:j^a j^a: + :\bar j^a \bar j^a: \Big) + g_c j^z\bar{j}^z. \label{wzwsu}
\eea
This is a realization of the conformal embedding~\cite{CFTBook}
\be
O(4Nk)_1 = Sp(2N)_k\oplus Sp(2k)_N, 
\ee
with $k=1$
\be
O(4N)_1 = Sp(2N)_1\oplus SU(2)_N. \label{embedding1}
\ee
As the interaction terms in~\fr{wzwspsu} preserve the structure of the conformal embedding~\fr{embedding1} (e.g., they do not couple different symmetry sectors), the model remains integrable~\cite{TsvelikJETP87,OgievetskyNuclPhysB87,SmirnovIntJModPhysA94}. 

If the perturbing current-current interaction in the $Sp(2N)_1$ `orbital' sector of the theory~\fr{sp1} is relevant, it generates a gap in the spectrum and the low-energy orbital excitations are non-Abelian anyons~\cite{TsvelikPRL14} with masses 
\be
M_n = M \sin\left[ \frac{\pi n}{2(N+1)}\right], ~~ n=1,\ldots,N.
\ee
The anyons are formed from the kinks that interpolate between the different ground states of the model~\fr{sp1} (cf. Sec.~\ref{SUNsemiclassic}) and parafermion zero modes that reside upon these kinks. On the other hand, the $SU(2)_N$ sector of the model~\fr{wzwsu} is gapless, and hence the model describes a QCP between phases with CDW and SC quasi-long-range order. 

The orbital excitations are not static -- the kink and its accompanying parafermion zero mode can propagate through the system; the case where such excitations move slowly was considered in Ref.~\cite{TsvelikPRL14}. In the following we depart from the symmetric $g_{cdw} = g_{sc}$ limit, so that the ground state degeneracy in the orbital sector is lifted by the presence of an external perturbation.

\subsubsection{Away from the symmetric limit: emergent integrability, ${\mathbb{Z}}_N$ parafermions, and competing orders} 
\label{Sec:HubStrat}

We now want to consider small deviations from the symmetric point, caused by inequality of the coupling parameters. To undertake such a study, we need to identify and treat the most relevant operator that arises from such a deviation. To do so, we will first consider the symmetric model and decouple the interaction term with a Hubbard-Stratonovich transformation. The saddle point of the resulting theory will suggest a natural order parameter matrix, which combines both SC and CDW order parameters. In this basis, the perturbing operator will be quite obvious. Our results will coincide with those found in Refs.~\cite{LecheminantPRL05,LecheminantNuclPhysB08} for the perturbing operator, despite taking a rather different route.

To begin, we perform a Hubbard-Stratonovich transformation on the interaction term described by the coupled $Sp(2N)_1$ currents~\fr{sp2Ncurrents}:
\bea
{\cal H}_{\rm int,o} &=& \frac{|\Delta_1|^2 + |\Delta_2|^2}{g_o} \nn
&& + \Big(\Delta_1 R^\dagger _{n\alpha}L^{\phantom\dagger}_{n\alpha} + {\rm H.c.}\Big)\nn
&& + \Big(\Delta_2 R^\dagger _{n\alpha}\epsilon_{\alpha\beta}L^\dagger _{n\beta} + {\rm H.c.}\Big)\, ,\label{HS} 
\eea
where $\Delta_{1,2}$ are the scalar auxiliary fields introduced by the Hubbard-Stratonovich transformation. At the weak-coupling $g_o \ll 1$ saddle point, the auxiliary fields can be approximated by
\be
\Delta_a = |\Delta|z_a, ~~ \sum_a |z_a|^2 =1.
\ee
We obtain 
\be
{\cal H}_{\rm int,o} =  \frac{|\Delta|^2}{g_o} -i \frac{|\Delta|}{2}\Big(\bar\Psi_R \hat G\Psi_L - \bar\Psi_L\hat G^\dagger \Psi_R\Big),
\ee
with $\bar \Psi_L = (L_{n\up}^\dagger,~-L^{\phantom\dagger}_{n\dn},~ -L^\dagger_{n\dn},~L^{\phantom\dagger}_{n\up})$, $\bar \Psi = \Psi^\dagger$, and 
\be
\hat G = \left( \begin{array}{cc} \hat g & 0\\ 0 & \hat g \end{array} \right), \quad
\hat g = i\left( \begin{array}{cc} z_1 & z_2\\ z_2^* & - z_1^* \end{array}\right).  \label{g}
\ee
Integrating out the fermionic fields $L,R$ we recover the $SU(2)_N$ WZNW model, as required. The order parameter combines both the CDW and SC order parameters. 

Now, we consider the term that arises when the couplings for the CDW and SC order parameters are slightly different, $g_{cdw} - g_{sc} = \delta g_o$. In the theory after the Hubbard-Stratonovich transformation, this gives rise to the perturbation 
\be
V = \frac{\delta g_o}{g_o^2}|\Delta|^2 \Big(|z_1|^2 - |z_2|^2\Big) = \lambda \Phi_{zz}^{\rm adj}, \label{relevant}
\ee
where $\Phi_{ab}^{adj}$ is the $SU(2)_N$ primary field in the adjoint representation (cf. Sec.~\ref{Sec:OperatorCorrespondence}) and $\lambda \sim \delta g_o$ is its coupling constant. Notice that this situation is different to that considered in the previous sections: the perturbation contains only one component of the adjoint field (which is a $3\times 3$ matrix). 

The perturbing term~\fr{relevant} is relevant, but does not break integrability. To see this, we can use an additional conformal embedding~\cite{CFTBook,ZamolodchikovJETP85}
\be
SU(2)_N = U(1)\oplus {\mathbb{Z}}_N, \label{sun2embed}
\ee
to rewrite the action of the orbital sector
\be
W(\hat g) + \lambda\Phi_{zz}^{\rm adj} = \frac{N}{4\pi}(\p_{\mu}\phi)^2 + A[{\mathbb{Z}}_N] + \lambda\Phi_{zz}^{\rm adj}, \label{ZN}
\ee
where we denote the Lagrangian of the $SU(2)_N$ WZNW model by $W(\hat g)$, $\phi$ is the field associated with the $U(1)$ part of the embedding~\fr{sun2embed}, and $A[{{\mathbb{Z}}_N}]$ is the Lagrangian for critical ${\mathbb{Z}}_N$ parafermions~\cite{ZamolodchikovJETP85,FateevPhysLettB91}. A detailed discussion of parafermions will follow in the next section. 

The perturbing field $\Phi_{zz}^{\rm adj}$ is described by the thermal operator of the ${\mathbb{Z}}_N$ theory; such a perturbation was shown to be integrable in Refs.~\cite{TsvelikJETP88,FateevIntJModPhysA91}. For any sign of the coupling $\lambda$ in~\fr{ZN}, the orbital excitations are massive, and their scattering is governed by the diagonal $S$-matrix~\cite{FateevIntJModPhysA91}
\be
[S(\theta)]_{a,b}^{\bar a,\bar b} = \frac{\sinh(\theta/2 + i\pi/N)}{\sinh(\theta/2 - i\pi/N)}\delta_a^{\bar a}\delta_b^{\bar b}\, . \label{Sm}
\ee
The masses of the orbital excitations are given by~\cite{FateevIntJModPhysA91}
\bea
&& m_n = m\frac{\sin(\pi n/N)}{\sin(\pi/N)}, ~~ n = 1,2,\ldots,N-1, \label{masses}\\
&& m \sim M \left| \frac{g_{cdw}- g_{sc}}{g_{cdw}+g_{sc}}\right|^{1/(2-d_{\rm adj})}, \nonumber
\eea
which follows from the pole structure of the $S$-matrix~\fr{Sm}. 

The conformal embedding of the $SU(2)_N$ theory in terms of $U(1)$ and ${\mathbb{Z}}_N$ degrees of freedom~\fr{sun2embed} suggests that the components of the order parameter matrix $\hat g$~\fr{g} can be expressed in terms of the primary fields of the $U(1)$ and ${\mathbb{Z}}_N$ theories
\be
z_1 \sim \s \re^{i\phi}, \qquad z_2 \sim \mu\re^{i\theta}, \label{z1z2}
\ee
where $\theta$ is a $U(1)$ field dual to $\phi$, while $\s,\mu$ are the order and disorder parameters of the ${\mathbb{Z}}_N$ parafermion model~\cite{ZamolodchikovJETP85}. Each of the bosonic exponents in~\fr{z1z2} has scaling dimension $1/2N$. As already mentioned, the excitation spectrum of the perturbed model~\fr{ZN} does not depend on the sign of the coupling of the perturbation, $\lambda$, but it does affect the vacuum averages of operators. Depending on the sign of $\lambda$, either the order or disorder parameter of the ${\mathbb{Z}}_N$ parafermions acquires a finite vacuum average.\footnote{A similar scenario in the Ising field theory may be familiar to the reader~\cite{CFTBook,IsingBook}.} These two scenarios would correspond to either CDW ($\la \s \ra$ finite) or SC ($\la \mu \ra$ finite) phases with the corresponding quasi-long-range order.

\subsubsection{Common features of the $SU(2)\times SU(k)$ model~\fr{ModelF} and the $Sp(2N)$ model~\fr{Model1}} 
Model~\fr{Model1} and model~\fr{ModelF} have different symmetries, but despite this they have certain physical features in common. In the limit on which we have focused, where the different symmetry sectors are weakly coupled, the lowest massive modes describe collective bosonic degrees of the freedom. In the $SU(2) \times SU(k)$ symmetric model~\fr{ModelF}, the $2k_F$ CDW ground state has almost coherent spin triplet excitations~\fr{spinsusc}. Likewise, for the $Sp(2N)$ model~\fr{Model1}, in the CDW phase where $\la\s\ra \neq 0$ the spectral function of the SC order parameter is\bea
\la O_{sc}(\tau,x)&&O^\dagger _{sc}(0,0)\ra \sim \la z_2(\tau,x)z_2^*(0,0)\ra \nn
&&\sim \la \re^{i\theta(\tau,x)}\re^{-i\theta(0,0)}\ra\la \mu(\tau,x)\mu^*(0,0)\ra\nn
&&\sim \frac{K_0(m_1\sqrt{\tau^2 +x^2})}{(\tau^2 + x^2)^{1/2N}} + \ldots, \label{scsc}
\eea
where the ellipses denote terms involving emission of higher numbers of excitations.\footnote{Similarly, in the SC phase ($\la \mu \ra \neq 0$), the spectral function of the CDW order parameter behaves as Eq.~\fr{scsc}.} Once again, we see that we have quasi-long-range order as a result of the condensation of a bosonic degree of freedom and almost coherent excitations associated with a competing order parameter.

\subsection{Parafermions and their zero modes}
\label{Sec:ParafermZeroModes}

In the previous section, we have seen how parafermions can emerge from a model which possesses $SU(2)_N$ symmetry. In this section, we will discuss parafermions in detail and show some interesting results about their zero modes.

\subsubsection{General motivation for studying parafermions}

The holy grail of topological quantum computing are non-Abelian anyons~\cite{KitaevAnnPhys03,FreedmanBAMS03,BondersonPRL08,NayakRMP08} -- excitations of a system which have non-trivial braiding statistics,\footnote{In CFT this is realized through the braiding of conformal blocks~\cite{BelavinNuclPhysB84,KnizhikNuclPhysB84,MooreNuclPhysB91}.} and whose permutation transforms between two different ground states which are locally indistinguishable~\cite{BaisNuclPhysB80,GoldinJMathPhys81,GoldinPRL85}. Such excitations are perhaps the most exotic known to man, so an obvious question to ask is do such excitations arise in physically meaningful systems, or are they simply a theorist's dream? Fortunately, nature seems to be on the physicists' side; anyons naturally arise in some particularly simple models of Majorana fermions, related to spin chains and (unconventional) superconductors~\cite{KitaevAnnPhys03} (see also the discussion of Ref.~[\onlinecite{NayakRMP08}]), as well as in the exotic setting of fractional quantum Hall states~\cite{MooreNuclPhysB91,NayakRMP08}. Such realizations may have already been achieved in experiments~\cite{MourikScience12,WillettRepProgPhys13}. 

A simple generalization of the Majorana fermion, promoting the ${\mathbb{Z}}_2$ symmetry to ${Z}_N$ ($N>2$), is the ${Z}_N$ parafermion~\cite{GreenPR53,FradkinNuclPhysB80,ZamolodchikovJETP85,AliceaAnnRevCMP16}. These excitations have multi-valued correlation functions~\cite{GreenPR53,FradkinNuclPhysB80,ZamolodchikovJETP85}, reflecting their intrinsically non-local nature; this makes them a great resource for information storage in quantum computation -- local perturbations, such as disorder, should not effect such an excitation. Of course, one has to figure out how to obtain and manipulate the parafermions, with numerous schemes having been proposed in recent years~\cite{NersesyanEPL11,MongPRX14,TsvelikPRL14,ZhuangPRB15,AliceaAnnRevCMP16}. As part of such studies, it is important to consider what happens when many anyons are brought together -- they may interact and the ground state degeneracy may be lifted, leading to restrictions on real-world devices -- developing an understanding of multi-anyon systems is an interesting subject in its own right (see, for example, Refs.~\cite{TrebstPRL08,GilsPRB13,FinchNJP13,JermynPRB14}). 

In models composed of Majorana fermions, anyon excitations generally reside on soliton-like ``kinks'' which interpolate between ground states with different topological properties (see Ref.~\cite{MongPRX14} for one such example). However, ${Z}_N$ parafermions with $N>2$ are interacting objects~\cite{FateevPhysLettB91}\footnote{In fact, non-interacting parafermions have a non-Hermitian Hamiltonian, yielding complex energy eigenvalues~\cite{FendleyJPhysA14}!}  -- considering a case where the state is inhomogeneous (containing kinks between different ground states) is then exceedingly difficult. The existence of anyons residing on the kinks was demonstrated for $N=3$ in Ref.~\cite{MongPRX14}, a special case in which Abelian bosonization can be applied. 

In this section, we discuss how to define parafermions in terms of the $SU(2)_N$ WZNW model currents, and we discuss a simple Hermitian fermionic model which contains anyon parafermion zero modes.

\subsubsection{Parafermions from the $SU(2)_N$ currents}
\label{Sec:parafermSU2N}
In relation to the material previously covered in the review, the easiest way for us to define parafermions is to `gauge away' the $U(1)$ subsector of the $\mathfrak{su}(2)_N$ Kac-Moody algebra -- this is easily achieve by factoring out the $U(1)$ part of the corresponding $SU(2)_N$ currents~\fr{su2currents} 
\bea
j^z &=& -i \sqrt{\frac{N}{2\pi}} \p_z \varphi,~~j^+ = \frac{\sqrt{N}}{2\pi} e^{i\sqrt{8\pi/N}\varphi}\psi,~~j^- = (j^+)^\dagger, \nn \label{left} \\
\bar j^z &=& i\sqrt{\frac{N}{2\pi}}\p_{\bar z}\bar\varphi,~~\bar j^+ = \frac{\sqrt N}{2\pi}\re^{-i\sqrt{8\pi/N}\bar\varphi}\bar\psi^\dagger \ , \bar j^- = (\bar j^+)^\dagger, \nn \label{right}
\eea
where we have defined the right/left moving parafermion fields $\psi,\bar\psi$. The $U(1)$ part of the current is described by the chiral components $\varphi, \bar\varphi$ of the bosonic field $\Phi$, which is governed by the Gaussian action  
\be
S = \frac{1}{2}\int \rd^2x \Big(\p_{\mu}\Phi\Big)^2.
\label{FreeBosonAction}
\ee
This construction essentially defines the conformal embedding that we used in the previous section, see Eq.~\fr{sun2embed}. 

From the expressions for the currents~\fr{left}--\fr{right}, one can compute the multi-point correlation functions of the parafermion fields, revealing their non-trivial braiding statistics when $N>2$. At the critical point, we have the two-point correlation functions 			
\be
\begin{split}
& \lla \psi(z)\psi^\dagger (0)\rra \sim z^{-2(N-1)/N}, \\
& \lla \bar\psi(\bar z)\bar\psi^\dagger (0)\rra \sim {\bar z}^{-2(N-1)/N}. 
\end{split}
\label{para2}
\ee
The $2n$-point correlation functions are  
\bea
&&\lla\psi(1)...\psi(n)\psi^\dagger (n+1)...\psi^\dagger (2n)\rra \nn
&&~= \lla j^+ (1)...j^+ (n)j^-(n+1)...j^-(2n)\rra \nn
&&~~~\times \prod_{i<j\leq n}z_{ij}^{-2/N}\prod_{n<i<j\leq 2n}z_{ij}^{-2/N}\prod_{i,j\leq n}z_{i,j+n}^{2/N}\ ,
\label{paran}
\eea
which reveal that for $N>2$ parafermions do not satisfy Wick's theorem. We note that for $N=2$, the parafermion field is real, describing Majorana fermions $\psi = \psi^\dagger$.

It is, of course, possible to introduce a mass term for ${\mathbb{Z}}_N$ parafermions by modifying the action as
\be
S = A[{\mathbb{Z}}_N] - \lambda\int \rd^2x[\psi\bar\psi + \psi^\dagger \bar\psi^\dagger ], \label{parafermmass}
\ee
where $A[{\mathbb{Z}}_N]$ is the critical parafermion action. For $N>2$ this is an interacting theory, but it remains integrable~\cite{FateevPhysLettB91}. The presence of such a term modifies the long distance asymptotics of correlation functions, such as~\fr{para2}--\fr{paran}, but the essential properties of the parafermions, such as their braiding statistics, remain.  

\subsubsection{A fermionic model with parafermion bound states}

Our aim is to consider a simple Hermitian model of fermions which has parafermionic zero-energy anyon modes on the boundary between topologically different states. Our analysis will use non-Abelian bosonization to construct a low-energy effective theory, and subsequently we use the integrability of this theory to provide supporting evidence for the presence of parafermion zero modes on the ``kinks'' which interpolate between topologically different ground states.\footnote{The arguments for the presence of parafermion modes can be generalized to other simple Lie groups, such as $SU(N)_k$ (see, e.g.~\cite{CastroAlvaredoNuclPhysB00}).} We also construct an effective theory that describes a finite density of such excitations, and through its exact solution study how the parafermions interact. 

The starting point for our study is a model of fermions, carrying both spin $\alpha = \up,\dn$ and orbital $k = 1,\ldots,N$ degrees of freedom, with Hamiltonian
\bea
{\cal H}_f &=&i(- R^\dagger _{k\alpha}\p_xR_{k\alpha} + L^\dagger _{k\alpha}\p_xL_{k\alpha}) \nn
&& + g_{\parallel} j^z\bar j^z + \frac{g_{\perp}}{2}\left(j^+ \bar j^- + j^-\bar j^\dagger \right). \label{ferm}
\eea
The interaction term is written directly in terms of the $j^a,\bar j^a$ $SU(2)_N$ Kac-Moody currents. This is an anisotropic version of the spin sector of the model considered in Sec.~\ref{Sec:SU2SUk}, cf. Eq.~\fr{pertWZNW}, and as such it has many features in common with it. The model is exactly solvable for generic values of $g_{\parallel}, g_\perp$~\cite{JerezPRB98}, but was first solved at the isotropic point $g_{\parallel} = g_\perp$ via the Bethe ansatz~\cite{TsvelikJETP87}. Our discussion will rely on the Bethe ansatz solution of this model, and we will discuss this in some detail.  

We begin by using the conformal embedding~\fr{embedding} to separate the kinetic term of~\fr{ferm} into the $U(1)$ charge field and a critical $SU(2)_N$ WZNW model. The non-Abelian sector is then described by the WZNW model perturbed by anisotropic current-current interactions, described by the Hamiltonian density 
\bea
{\cal H} &=& \frac{2\pi}{N+2}\left(:j^aj^a: + :\bar j^a\bar j^a:\right) \nn
&& + g_{\parallel} j^z\bar j^z + \frac{g_{\perp}}{2}\left(j^+ \bar j^- + j^-\bar j^+ \right). \label{wznwH}
\eea

We use formulae~\fr{left}--\fr{right} to express the theory in terms of an additional $U(1)$ field $\Phi$ and $\mathbb{Z}_N$ parafermions. The Lagrangian density for the $SU(2)_N$ sector is then 
\be
{\cal L} = \frac{1}{2}(\p_{\mu}\Phi)^2 + A[{\mathbb{Z}}_N] - \lambda \left(e^{i\beta\Phi}\psi\bar\psi + {\rm H.c.}\right), 
\label{boson} 
\ee
where the coupling $\lambda \sim Ng_{\perp}$, and $\beta$ is related to $g_{\parallel}$ -- for small values of $g_{\parallel}$ it satisfies $\beta^2 = (1+ Ng_{\parallel}/\pi)^{-1}$. 

From the exact solution of the model, we know that when $g_\parallel > 0$ the model is massive, and its excitations are solitons and antisolitons. The scattering S-matrix for the soliton is a tensor product of the XXZ and restricted solid on solid (RSOS) S-matrices~\cite{SmirnovIntJModPhysA94}. For sufficiently strong interactions $g_\parallel$, there exist soliton-antisoliton bound states, but these will not be of interest to us here. The exact solution of the model will reveal that each soliton/antisoliton carries a parafermion zero mode which endows it with non-Abelian statistics.  

Our model for parafermions~\fr{boson} features a coupling between the $U(1)$ field $\Phi$ and the mass term for the parafermions. This is reminiscent of a theory
\be
S = A[{\mathbb{Z}}_N] + \int \rd^2x\, \lambda(x) \Big[ \psi\bar\psi + \psi^\dagger\bar\psi^\dagger\Big], \label{varmass}
\ee
with a spatially dependent mass term.\footnote{We remind the reader that in the $N=2$ Majorana fermion theory, there will be zero energy modes localized at the positions $\{x_i\}$ where $\lambda(x_i) = 0$. There, the mass term interpolates between topologically trivial (a conventional insulator of massive fermions) and non-trivially phases. In the topological phase, exponentially localized Majorana fermions will appear on an edge, as can be easily proved by solving the field equations, see e.g., Ref.~\cite{LecheminantPRB02}.} Instead, in the theory~\fr{boson}, the role of a spatially dependent mass term is played by a dynamical field $\exp[i\beta\Phi(x)]$. Importantly, this field changes sign with a soliton configuration of the field $\Phi(x)$, so one can use~\fr{boson} as a substitute for the model~\fr{varmass} with a coordinate-dependent mass gap providing certain requirements are met. Firstly, the soliton configurations of the field $\Phi$ should be slow, in order to be considered quasi-static. Secondly, the solitons should be far from one another, on average. We will extract a more precise criteria from the exact solution. Thirdly, quantum fluctuations of the dynamical field $\exp[i\beta\Phi(x)]$ should be small in order that it can mimic a static $\lambda(x)$ in~\fr{varmass} -- this essentially requires a small value of $\beta$. 

The requirements on the dynamical field can be met in the following manner. We apply a magnetic field $H$ to our system of fermions, this couples to the $U(1)$ bosonic sector of~\fr{boson}. The applied field breaks the symmetry between soliton and antisolitons. We focus on a field strength that is slightly below the soliton mass $M$ threshold, such that 
\bea
T \ll M-H \ll M,\label{limit}
\eea
where $T$ is the temperature. In this limit, the system is described by a dilute gas of thermally excited solitons accompanied by a negligible number of antisolitons. The velocity of the solitons is 
\be
\sqrt {\la v^2\ra} = \sqrt{2T/M} \ll 1,
\ee
whilst the average soliton density is 
\be
n \sim e^{-(M-H)/T}  \ll 1.
\ee
As a result, this can be thought of as a gas of slow solitons which are undisturbed, due to the exponentially long collision time scale $\tau \sim \exp[(M-H)/T]$. 

\subsubsection{Bethe ansatz solution of the model~\fr{boson}}
Having established that in the limit~\fr{limit} we have a gas of quasi-static solitons, we now use the thermodynamic Bethe ansatz (TBA) to establish that the solitons can carry parafermion zero modes. The TBA equations for the soliton sector of the theory in the limit~\fr{limit} can be extracted from Ref.~\cite{TsvelikPRB95}, for example. They are part of a more general system of equations which can contain massive soliton-antisoliton bound states (see Ref.~\cite{TsvelikJETP87b}), but these are irrelevant for the current discussion. 

The free energy $F$ of the model~\fr{boson} in the limit~\fr{limit} is 
\bea
\frac{F}{L} = - TM\int \frac{\rd \theta}{2\pi}\cosh\theta \ln\Big(1 + e^{\epsilon_N(\theta)/T}\Big), \label{F}
\eea  
where $L$ is the system size, and the function $\epsilon_N(\theta)$ is determined from the system of non-linear integral equations
\bw
\bea
&&\frac{\epsilon_j}{T} = s*\ln\left(1+e^{\epsilon_{j-1}/T}\right)\left(1+e^{\epsilon_{j+1}/T}\right) + s*\ln\left(1+ e^{\epsilon_N/T}\right)\delta_{j,N-1}, \qquad j = 1,\ldots,N-1, \label{BA2} \\
&&\frac{\epsilon_N}{T} - K * \ln\left(1 + e^{\epsilon_N/T}\right) = - \frac{M}{T} \cosh\theta + \frac{H}{T}  + s*\ln\left(1+ e^{\epsilon_{N-1}/T}\right) + O(e^{- H/T}), \label{BA1}
\eea
\ew
where we define the convolution $a\ast b(x)$ as
\be
a\ast b(x) = \int_{-\infty}^\infty \rd y\, a(x-y)b(y),
\ee 
and the kernels
\bea
s(x) &=& \frac{1}{\pi \cosh(x)}, \nn
K(\omega) &=& \frac{\sinh\left[\pi(\xi -1)\frac{\omega}{2}\right]}{2\cosh\left(\frac{\pi\omega}{2}\right) \sinh\left(\frac{\pi\xi\omega}{2}\right)}, \quad \frac{1}{\xi} = \frac{8\pi}{N\beta^2} -1,\nonumber
\eea
with $\xi = 1 /( 8\pi/N\beta^2 - 1)$. 

As we are interested in the limit~\fr{limit}, a first approximation is to replace the quasi-energies $\epsilon_j$ ($j=1,\ldots,N-1$) by their (constant) asymptotic values. Then, the corresponding integral equations~\fr{BA2} become algebraic, with the solution 
\be
1 + e^{\epsilon_j/T} = \frac{\sin^2\Big(\frac{\pi(j+1)}{N+2}\Big)}{\sin^2\Big(\frac{\pi}{N+2}\Big)}.
\ee
Substituting this into (\ref{F}) we obtain the following expression for the free energy: 
\bea
\frac{F}{L} &=& - 2T \cos\left(\frac{\pi}{N+2}\right)\int \frac{\rd p}{2\pi} e^{- \frac{M-H}{T} - \frac{p^2}{2MT}} \nn
&& + O\Big(e^{-\frac{2(M-H)}{T}} \Big),\label{F1}
\eea
This is simply the free energy of an ideal gas of particles with mass $M$ and chemical potential $H$. The prefactor $Q = 2\cos(\pi/(N+2))$ arises from the degeneracy of the ${\cal N}$ particles with a given energy: this degeneracy is equal to $Q^{\cal N}$ and arises from the parafermionic zero modes which live on the solitons. Notice that $Q$ is not an integer -- this is a direct indicator that the zero modes attached to the solitons do not commute. For $N=2$, we find the well-known result for the dimension of the Hilbert space with ${\cal N}$ Majorana fermions: $D(2)_{\cal N} = 2^{{\cal N}/2}$ (this is simply the dimensionality of the Clifford algebra representation of ${\cal N}$ gamma matrices). In the case of $N=3$, the obtained dimensionality coincides with the large ${\cal N}$ asymptotic of the Fibonacci numbers:
\bea
D(3)_{\cal N} &=& \frac{{\phi}^{\cal N} - (-\phi)^{-{\cal N}}}{\sqrt 5}, \nn
\phi &=& 2\cos\left(\frac{\pi}{5}\right) = \frac{1+\sqrt 5}{2}, \nonumber
\eea
where $\phi$ is the golden ratio. 

As we have mentioned,~\fr{F1} describes an ideal gas of anyons (solitons + parafermions) when we neglect the next order terms (the first order term corresponds to the leading term in the soliton density expansion). Taking into account higher order terms, from~\fr{BA2} one can move towards equations for the interacting anyonic gas. Importantly, interactions will lift the ground state degeneracy. 

At low temperatures, we can invert the matrix kernel in Eq.~\fr{BA2} to obtain equations where the kernel acts on terms which vanish in the $T\to0$ limit
\bw
\bea
&&T\ln\Big(1+e^{\epsilon_j/T}\Big) - T{\cal A}_{jk}\ln\Big(1+e^{-\epsilon_k/T}\Big) =  {\cal A}_{j,N-1}*s*T\ln\Big(1+e^{\epsilon_N/T}\Big),\qquad j,k = 1,\ldots,N-1,  \label{BA3} \\
&&{\cal A}_{jk}(\omega) = 2\coth\left(\frac{\pi\omega}{2}\right) \frac{\sinh\left(\frac{\pi \omega}{2}[N-\mbox{max}(j,k)]\right)\sinh\left(\frac{\pi\omega}{2}\mbox{min}(j,k)\right)}{\sinh(\frac{N\pi\omega}{2})}.\nonumber
\eea
\ew 
At low temperatures $T\ll M$, the distribution on the right hand side (RHS) of Eq.~\fr{BA3} is sharp -- we approximate it by a delta function 
\be
{\cal A}_{j,N-1}*s*T\ln(1+e^{\epsilon_N(\theta)/T})\approx n_{\rm s}(T){\cal A}_{j,N-1}*s(\theta), \label{BA4} 
\ee
where $n_{\rm s}$ is the number of solitons. Substituting~\fr{BA4} into the right hand side of~\fr{BA3}, the TBA equations look very similar to those for the ferromagnetic XXZ model with $n_{\rm s}$ sites and anisotropy $\gamma = \pi/N$~\cite{TakahashiProgTheorPhys71,GaudinPRL71,TakahashiProgTheorPhys72}. One difference is that there is an additional restriction on the solutions of~\fr{BA3}, where solutions with rapidities shifted by $i\pi/2$ are forbidden -- such a set of TBA equations in fact describe the critical restricted solid-on-solid model~\cite{BazhanovPTPS90,ReshetikhinCMP90} with conformal charge $c = 2(N-1)/(N+2)$.   

The right hand side of Eqs.~\fr{BA4} is proportional to the soliton $n_{\rm s}$, which reflects that the excitation bandwidth of the interacting anyon gas is proportional to the average distance between the solitons $\ell \sim n^{-1}_{\rm s}$. This is an interesting result, as it contradicts the naive expectation that the bandwidth is proportional to the overlap of the zero mode wavefunctions (which will be exponentially small in $M\ell$). Instead, we have a scenario where the bandwidth is related to the collision time for the mobile solitons. 

So, having started from a model of electrons with orbital degeneracy~\fr{ferm}, we have used non-Abelian bosonization and the conformal embedding to obtain an integrable low-energy effective theory of parafermions coupled to a $U(1)$ boson~\fr{boson}. Under a certain set of physically reasonable requirements, the coupling between the $U(1)$ boson and the parafermion acts like a spatially varying mass term for the parafermion (cf. the action~\fr{varmass}). Such a term interpolates between topologically-distinguishable ground states when the bosonic field has a soliton configuration. The TBA equations suggest that on these solitons reside zero-energy parafermions -- the bound state of the soliton-parafermion can be thought of as a non-Abelian anyon. In the case when there is a finite density of solitons, these anyons interact and the ground state degeneracy is lifted.

In the next part of the review we will explore some physical manifestations of non-Abelian bosonization, in the setting of ultra cold atomic gas experiments.

\section{Applications to cold atoms physics} 
\label{Sec:ColdAtomsExamples}

\subsection{High-symmetry and cold atoms}
\label{Sec:HighSymColdAtoms}

Thanks to the high level of control of interactions and lattice geometries, recent experimental progress in trapped ultracold atomic gases provides a great opportunity to explore the physics of strong correlations~\cite{LewensteinBook,BlochRMP08,BlochNatPhys12}. The effect of spin degeneracy can also be probed in these systems as the total angular momentum $F$ of the atom, which includes both electron and nuclear spins, can be larger than 1/2 (resulting in $2F + 1$ hyperfine states). In optical traps the $2F + 1$ components are degenerate, and as a result novel and interesting fermionic phases may be stabilized~\cite{WuPRL03,HonerkampPRL04,LecheminantPRL05,WuModPhysLettB06,RappPRL07,CazalillaNewJPhys09,GorshkovNaturePhys10,HermelePRB11,CazalillaRPP14,CapponiAnnPhys16,NatafPRL16}.  

In the low-energy limit, the interaction between two half-integer hyperfine spin-$F$ fermionic atoms is governed by $s$-wave scattering processes. For $SU(2)$ rotationally invariant problems, the interaction Hamiltonian $H_{\rm int}$ takes the general form~\cite{HoPRL98,OhmiJPSJ98}:
\begin{eqnarray}
H_{\rm int} &=& \int \rd^3 r  \sum_{f=0,2,\ldots}^{2F-1} g_{f} {\cal P}_{f}({\bf r})  \nonumber \\
&=& \int \rd^3 r \sum_{f=0,2,\ldots}^{2F-1} g_{f}
\sum_{m= -f}^{f}  P^{\dagger}_{fm}({\bf r})  P^{\phantom\dagger}_{fm}({\bf r}) ,
\label{F-F-scattering}
\end{eqnarray}
where the total hyperfine spin $f$ should be antisymmetric according to Pauli's principle and thus $f=0,2,\ldots, 2F-1$. In Eq.~(\ref{F-F-scattering}), ${\cal P}_{f}$ is the projection operator onto the total spin-$f$ sector and the coupling constants $g_{f}$ are related to the corresponding $s$-wave scattering lengths $a_{f}$~\cite{LewensteinBook,BlochRMP08}. For instance, in the ground state of ${}^{173}$Yb (respectively ${}^{40}$K),  we have $F=5/2$ (respectively  $F=9/2$) and thus three (respectively five) independent coupling constants. In Eq.~(\ref{F-F-scattering}), the pairing operators $P^{\dagger}_{f m}$ are defined through the underlying Clebsch-Gordan coefficients:
\begin{equation}
 P^{\dagger}_{f m} ({\bf r}) = \sum_{\alpha \beta} \langle fm |F,F;\alpha
\beta\rangle c^{\dagger}_{\alpha} ({\bf r}) c^{\dagger}_{\beta} ({\bf r}),
\label{sumruleClebsch}
\end{equation}  
with $c^{\dagger}_{\alpha}({\bf r})$ ($\alpha = 1,\ldots, 2F+1$) being the fermion creation operators corresponding to the $2F+1$ hyperfine atomic states. The operators $P^\dagger_{fm}$ are also related to the density operator $n ({\bf r}) = \sum_{\alpha}  c^{\dagger}_{\alpha}({\bf r})  c_{\alpha}({\bf r})$ through the identity:
\begin{equation}
\sum_{f,m} P^{\dagger}_{fm}({\bf r})  P^{\phantom\dagger}_{fm} ({\bf r}) = n^2 ({\bf r}) .
\label{sumruleClebsch2}
\end{equation}

The $SU(2)$ symmetry of the atom-atom interaction~(\ref{F-F-scattering}) can be enlarged by fine-tuning the scattering lengths. In this respect, when the couplings $g_{f} = g$ do not depend on $f$ (i.e., $a_{0}=a_{2}=\cdots=a_{2F-1}$) one has
\be
H_{\rm int}{\Big|}_{g_f = g}  = g  \int \rd^3 r \,n^2({\bf r}),\label{HintUN}
\ee
which follows from Eq.~(\ref{sumruleClebsch2}). As the density operator is invariant under the transformation $c_{\alpha} \rightarrow U_{\alpha \beta} c_{\beta}$, with $U$ being a unitary matrix, the two-body interaction~(\ref{F-F-scattering}), cf. Eq.~\fr{HintUN}, enjoys an enlarged $U(N)=U(1) \times SU(N)$ continuous symmetry with $N=2F+1$. While the $U(1)$ symmetry accounts for the conservation of the total number of atoms,\footnote{Herein we use the terminology of the $U(1)$ `charge' symmetry, as is the jargon of condensed matter physics.} the non-trivial part is the $SU(N)$ symmetry which acts in the hyperfine spin subspace.

Such a symmetry enlargement may seem academic but remarkably enough, the fine-tuning of all scattering lengths \textit{is indeed possible} for a specific class of atoms: those where the total \textit{electron} angular momentum $J=0$ vanishes, such as in the $^1S_0$ ground state of alkaline earth and ytterbium atoms~\cite{CazalillaNewJPhys09,GorshkovNaturePhys10,CazalillaRPP14}. Then, for $J=0$ the hyperfine state depends solely on the nuclear spin\footnote{The total degeneracy $N$ satisfies $N=2F+1=2I+1$.} $I$ and the nuclear-spin-dependent variation of the scattering lengths is estimated to be smaller than $\sim 10^{-9}$  from perturbation theory~\cite{GorshkovNaturePhys10}. Recent experiments with $^{87}$Sr ($N=10$) and  $^{173}$Yb ($N=6$) atoms have measured the scattering lengths and indeed confirm the presence of the $SU(N)$ symmetry in the hyperfine spin space~\cite{ZhangScience14,ScazzaNatPhys14,CappelliniPRL14}. The cooling of such atoms below the quantum degeneracy temperature has been achieved for $^{87}$Sr, $^{171}$Yb and $^{173}$Yb, with $I= 9/2$, $I= 1/2$ and $I=5/2$ respectively~\cite{DeSalvoPRL10,TeyPRA10,FukuharaPRL07,TaiePRL10,TaieNatPhys12}. With these examples, the experimental exploration of exotic physics associated with fermions possessing $SU(N)$ hyperfine spin symmetry (where $N$ can be as large as 10) can be undertaken. Also of particular interest to us is the fact that these systems can also be confined to a one-dimensional geometry, see e.g., Ref.~\cite{PaganoNatPhys14}.

Besides the $SU(N)$ hyperfine spin symmetry, one can also find another extended symmetry by fine-tuning of the scattering lengths: $a_{2} = \ldots = a_{2F-1} \ne a_0$. Here, the atom-atom interaction~(\ref{F-F-scattering}) has two independent coupling constants, $g_0$ and $g_{2}$, and it can be rewritten by means of Eq.~(\ref{sumruleClebsch}) as:
\be
H_{\rm int} = \int \rd^3 r  \left[  g_2 n^2 ({\bf r}) + (g_0 - g_2)  P^{\dagger}_{00} ({\bf r}) P_{00}^{\phantom\dagger} ({\bf r}) \right],
\label{Sp2Nmodelatom}
\ee
where $P_{00} ({\bf r}) $ is the spin-$F$ singlet [e.g., Bardeen-Cooper-Schrieffer~(BCS)] pairing operator:
\begin{eqnarray}
P^{\dagger}_{00} ({\bf r})  &=& \frac{1}{\sqrt{N}}
c^{\dagger}_{\alpha}  ({\bf r}) {\cal J}_{\alpha \beta} c^{\dagger}_{\beta}({\bf r})  \nonumber \\
&=& -  \frac{1}{\sqrt{N}} \sum_{\alpha} \left(-1\right)^{\alpha}
c^{\dagger}_{\alpha} ({\bf r})  c^{\dagger}_{2F+2-\alpha}({\bf r}) ,
\label{Sp2Nsingletpairing}
\end{eqnarray}
where the $N \times N$ matrix ${\cal J}$ (with $N=2F+1=2n$) is the natural generalization of the familiar antisymmetric tensor $\epsilon = i \sigma_2$ to half-integer (hyperfine) spin $F > 1/2$. The interaction~(\ref{Sp2Nmodelatom}) enjoys an extended continuous symmetry as the singlet-pairing operator~(\ref{Sp2Nsingletpairing}) is invariant under the $Sp(2n)$ group~\cite{WuPRB05}.\footnote{$Sp(2n)$ consists of unitary matrices ${\cal U}$ that satisfy ${\cal U}^* {\cal J} {\cal U}^{\dagger} = {\cal J}$. See also Sec.~\ref{Sec:Sp2N}} In the $F=1/2$ case ($N=2n=2$) it reduces to the $SU(2)$ Hubbard model since $SU(2) \simeq Sp(2)$. Interestingly, in the $F=3/2$ case (i.e., $N=2n=4$) there is no need for fine-tuning and the original model~(\ref{F-F-scattering}) enjoys an exact $Sp(4)$ symmetry, which is locally isomorphic to $SO(5)$~\cite{WuPRL03,WuModPhysLettB06}.

In this part, we will review the physics of one-dimensional fermionic cold atoms with enlarged $Sp(N)$ and $SU(N)$ continuous (hyperfine) spin symmetries that can be investigated with non-Abelian bosonization, introduced in Sec.~\ref{Sec:NonAbelianBosonization}~and~\ref{Sec:ExamplesNonAbelian}.

\subsection{$Sp(2n)$ ultracold fermions: a low-energy approach}
We first consider the interacting Hamiltonian~(\ref{Sp2Nmodelatom}) with $Sp(2n)$ symmetry and load the underlying atoms into a one-dimensional optical lattice, resulting in the following lattice model~\cite{WuPRB05,LecheminantPRL05,LecheminantNuclPhysB08}:
\bea
H &=& -t \sum_{i,\alpha} \Big[c^{\dagger}_{\alpha}(i) c^{\phantom\dagger}_{\alpha}(i+1) + {\rm H.c.} \Big]\nn
&& + \frac{U}{2} \sum_i n(i)^2 + V \sum_i P^{\dagger}_{00}(i) P^{\phantom\dagger}_{00}(i),
\label{hubbardS}
\eea
where $c^{\dagger}_{\alpha}(i)$ ($\alpha = 1,\ldots, 2F+1=2n$) is the fermion creation operator on site $i$ for spin state~$\alpha$, $n(i) = \sum_{\alpha}  c^{\dagger}_{\alpha}(i) c_{\alpha}(i)$ is the lattice density operator, and the singlet-pairing operator $P^{\dagger}_{00}(i)$ on site $i$ is given by Eq.~(\ref{Sp2Nsingletpairing}). The continuous symmetry of model~(\ref{hubbardS}) is $U(1) \times Sp(2n)$, with the $U(1)$ part being the standard $U(1)$ charge symmetry:
\be
c_{\alpha}(j) \rightarrow e^{i \theta} c_{\alpha}(j).
\ee
In the following we will focus on two particular cases: (i) incommensurate filling; (ii) half-filling of the lattice, and we will investigate the competition between the density and singlet pairing operators in the weak-coupling regime, $ |U|,|V|  \ll t$.

\subsubsection{Molecular Luttinger liquids and $\mathbb{Z}_n$ quantum criticality}
Using the continuum description of the lattice fermionic operators $c_{\alpha}(i)$ in terms of left and right-moving Dirac fermions $L_{\alpha}$, $R_{\alpha}$ (see Eq.~\fr{psi}), the non-interacting part of the Hamiltonian in Eq.~(\ref{hubbardS}) is described by the Hamiltonian density:
\begin{equation}
{\cal H}_0 = - i \left( R_{\alpha}^{\dagger} \partial_x R_{\alpha}^{\phantom\dagger}
- L_{\alpha}^{\dagger} \partial_x L_{\alpha}^{\phantom\dagger}  \right),
\label{contfreeham}
\end{equation}
where the Fermi velocity has been set to unity. As in Sec.~\ref{Sec:ConformalEmbedding}, we use non-Abelian bosonization and introduce a $U(1)$ charge boson $\Phi_c$ and its dual $\Theta_c$,
\begin{eqnarray}
:R^{\dagger}_{\alpha} R_{\alpha}: &=& \sqrt{n/2\pi} \; \partial_x \big(\Phi_c - \Theta_c\big),  \nonumber \\
:L^{\dagger}_{\alpha} L_{\alpha} : &=& \sqrt{n/2\pi} \;  \partial_x \big(\Phi_c + \Theta_c\big), \nonumber 
\end{eqnarray}
and chiral $SU(2n)_1$ currents:
\be
J^A_{R} = R_{\alpha}^{\dagger} T^A_{\alpha \beta} R^{\phantom\dagger}_{\beta},  \qquad
J^A_{L} = L_{\alpha}^{\dagger} T^A_{\alpha \beta} L^{\phantom\dagger}_{\beta}, 
\label{chargeconserquant}
\ee
where $T^A$ ($A= 1, \ldots, 4 n^2 -1$) are the generators of $SU(2n)$ in the fundamental representation normalized such that $\mbox{Tr} (T^A T^B) = \delta^{AB}/2$, as in Sec.~\ref{Sec:NonAbelianBosonization}. 
The Sugawara form of the non-interacting Hamiltonian (\ref{contfreeham}) then reads:
\bea
{\cal H}_0 &=& 
\frac{1}{2} \left[ \left(\partial_x \Phi_c \right)^2 +  \left(\partial_x \Theta_c \right)^2 \right]\nn
&& + \frac{2\pi}{2n + 1} \left[ : J^A_{R} J^A_{R}: + : J^A_{L} J^A_{L}: \right].
\label{contfreehambis}
\eea
Since the continuous symmetry group of the interaction of Eq.~(\ref{hubbardS}) is generically $Sp(2n)$, we need  to introduce the currents $J^a_{R(L)}, a=1,...,n(2n+1)$ which  generate the $Sp(2n)_1$ CFT with central charge $c=n(2n+1)/(n+2)$:
\begin{equation}
J^a_{R} = R_{\alpha}^{\dagger} T^a_{\alpha \beta} R^{\phantom\dagger}_{\beta}, \qquad J^a_{L} = L_{\alpha}^{\dagger} T^a_{\alpha \beta} L^{\phantom\dagger}_{\beta} ,
\label{sp2ncur}
\end{equation}
$T^a$ being the generators of $Sp(2n)$ in the fundamental representation and normalized such that: $\mbox{Tr} (T^a T^b) = \delta^{ab}/2$. The remaining currents, i.e., the $SU(2n)_1/Sp(2n)_1$ currents, are denoted by: $ J_L^i = L_{\alpha}^\dag T^i_{\alpha \beta} L_{\beta}$ ($i=1, \ldots, 2 n^2 - n  -1$) with a similar definition for the right currents.

With these definitions, the low-energy effective Hamiltonian of model~(\ref{hubbardS}) can be derived for \textit{incommensurate filling}:
\begin{eqnarray}
  \mathcal{H} &=&  \mathcal{H}_c +  \mathcal{H}_s, \; \;  [{\cal H}_c, {\cal H}_s] = 0,  \label{Speffectiveinteracting}\\
 \mathcal{H}_c &=& \frac{v_c}{2} \left[\frac{1}{K_c}
\left(\partial_x \Phi_c \right)^2 + K_c \left(\partial_x \Theta_c \right)^2 \right] \nn
 \mathcal{H}_s &=& \frac{2\pi v_s}{2n + 1} \left[ : J^A_{R} J^A_{R}: + : J^A_{L} J^A_{L}: 
\right]  + g_1 J_R^a J_L^a + g_2 J_R^iJ_L^i ,\nonumber
\end{eqnarray}
where $v_c$ (respectively $v_s$) is the charge (respectively spin) velocity and $K_c$ denotes the Luttinger parameter. The continuum limit gives the identification: $g_1 = - 2 a_0 (2V + n U)/n$, $g_2 = 2 a_0 (2V - n U)/n$ (with $a_0$ being the lattice spacing). In the $F=3/2$ case, i.e., $N=4$, one can express the non-Abelian part of model~(\ref{Speffectiveinteracting}) in a more transparent basis\footnote{In this case, it is also possible to investigate the physical properties of $F=3/2$ cold atoms by Abelian bosonization as in Ref.~\cite{WuPRL05}.} by exploiting the equivalence $SU(4)_1 \sim SO(6)_1$. By introducing six Majorana fermions $\xi^{i}_{R,L}$ $(i=0,\ldots,5)$ as in the study of the two-leg spin-1/2 ladder with four-spin exchange interactions~\cite{AzariaPRL99}, the interacting part of model~(\ref{Speffectiveinteracting}) simplifies to
\begin{equation}
{\cal H}^{\rm int}_s =  \lambda_{1} (\xi^{i}_R \xi^{i}_L)^2 + \lambda_{2} \xi^{0}_{R} \xi^{0}_{L} \xi^{i}_R \xi^{i}_L,
\label{spin3demirefer}
\end{equation}
which turns out to be exactly solvable~\cite{ControzziPRL06}. The one-loop RG equations for model~(\ref{Speffectiveinteracting}) in the general $n$ case have been discussed in Refs.~\cite{LecheminantPRL05,LecheminantNuclPhysB08}.  

\paragraph{Molecular Luttinger liquids.}
A first spin gap phase is stabilized when $U < 0$ and $V > n U /2$, where the long-distance physics of the RG flow is governed by the symmetric ray $g_{1} = g_{2} = \tilde g  > 0$. The low-energy (infrared) Hamiltonian for the (hyperfine) spin degrees of freedom then takes the form of a chiral $SU(2n)$ Gross-Neveu model~\cite{GrossPRD74}:
\begin{equation}
{\cal H}_{{\rm s}}^{{\rm int} *} = \tilde g \Big(J^a_{R} J^a_{L} + J^i_{R} J^i_{L} \Big) = \tilde g J^A_R J^A_L .
\label{su2NGN}
\end{equation}
This is an integrable massive field theory with the low-energy spectrum~\fr{MjAndrei} for $\tilde g >0$~\cite{AndreiPhysLettB80}. The spin degrees of freedom are thus fully gapped and a $c=1$ critical phase is formed, stemming from the gapless charge degrees of freedom described by the bosonic field $\Phi_c$ in Eq.~(\ref{Speffectiveinteracting}). It is then natural to expect the emergence of a gapless $2k_F$ CDW phase due to the presence of the dynamical $SU(2n)$ symmetry enlargement. The corresponding lattice order parameter is $n_{2k_F}(j) = \sum_{\alpha} e^{i 2 k_F x} c^{\dagger}_{\alpha} (j) c^{\phantom\dagger}_{\alpha}(j)$, ($x= ja_0$) which has the continuum limit
\begin{equation}
n_{2k_F} =  R_{\alpha}^\dagger L_{\alpha} \sim \exp \left( i \sqrt{2 \pi /n} \Phi_c \right) {\rm Tr} \; g , 
\label{2kFCDW}
\end{equation}
where the non-Abelian bosonization identity~\fr{RL} has been used and $g$ is the $SU(2n)_1$ WZNW primary field with scaling dimension $(2n-1)/2n$. In the ground state of the chiral $SU(2n)$ Gross-Neveu model~(\ref{su2NGN}), we have $\langle {\rm Tr} \, g \rangle \ne 0$, and as a result the equal-time density-density correlation function is
 \begin{equation}
 \langle n (x)  n (0) \rangle \simeq 
 A  \cos\big(2 k_F x\big) x^{-K_c/n}  - \frac{n K_c}{\pi ^2 x^2},
\label{2kfdensitycorrel}
\end{equation}
where $A$ is a non-universal amplitude. 

The restoration of the $SU(2n)$ symmetry in the low-energy limit means that there is no pairing instability to compete with the $2k_F$ CDW, in stark contrast to the $F=1/2$ ($n=1$) case~\cite{GNTBook,GiamarchiBook}. Indeed, a general superconducting pairing operator $c^{\dagger}_{\alpha} (i) c^{\dagger}_{ \beta} (i)$ is not a singlet under the $SU(2n)$ symmetry when $n>1$. In the continuum limit, its hyperfine spin part cannot sustain a non-zero expectation value in the gapped $SU(2n)$  invariant model~(\ref{su2NGN}). However, we may consider a molecular superfluid instability made of $2n$ fermions: $M_i = \prod_{\alpha=1}^{2n} c^{\dagger}_{\alpha} (i) $ which is now a singlet under the $SU(2n)$ symmetry.  Its equal-time correlation function can be determined in the $SU(2n)$ restored phase~(\ref{su2NGN})~\cite{LecheminantPRL05,LecheminantNuclPhysB08}:
\begin{equation}
 \big\langle M^{\dagger} (x)  M  (0) \big\rangle \sim  x^{- n/K_c} .
\label{2Ntets}
\end{equation}
We thus see that $2k_F$ CDW and molecular superfluid instabilities compete. In particular, a dominant molecular superfluid instability requires $K_c > n$. It was shown numerically in Refs.~\cite{CapponiPRB07,CapponiPRA08,RouxEPJB09} that such a scenario can be achieved for local (on-site) attractive interactions in the low-density regime, signaling the emergence of a molecular Luttinger liquid phase. This is characterized by the formation of bound states of $N$ fermions (analogous to baryons in high-energy physics) and the suppression of Cooper pairs. A related phase has already been stabilized in other one-dimensional systems~\cite{BurovskiPRL09,OrsoPRL10,RouxPRA11,DalmontePRL11,AzariaArxiv16}.

\paragraph{BCS singlet pairing phase.}
A second spin-gapped phase arises in the model~\fr{hubbardS} when $V<0$ and $V < n U/2$. The RG flow is now attracted towards the asymptote $g_{1} = - g_{2} = \tilde g > 0$.  The low-energy Hamiltonian for the (hyperfine) spin degrees of freedom again takes the form of a chiral $SU(2n)$ Gross-Neveu model:
\begin{equation}
{\cal H}_{{\rm s}}^{{\rm int} *} = \tilde g \Big( J^a_{R} J^a_{L} - J^i_{R} J^i_{L} \Big)  = \tilde g {\tilde J}^A_R {\tilde J}^A_L ,
\label{su2NGNdual}
\end{equation}
where the duality transformation~\cite{BoulatNuclPhysB09}
\begin{equation}
{\tilde L}_{\alpha} = {\cal J}_{\alpha \beta} L^{\dagger}_{\beta}, \; \;
{\tilde R}_{\alpha}= R_{\alpha}
\label{dualityfer}
\end{equation}
has been applied to the Dirac fermions, resulting in new $SU(2n)_1$ currents ${\tilde J}^A_{L,R}$.

Besides the opening of a spectral gap, it is clear the model possesses a hidden enlarged symmetry at low energy, which we denote as $\widetilde{SU}(2n)$ symmetry, which is generated by the currents ${\tilde J}^A_{R,L}$. The physical properties of the phase can be inferred from the transformation of the CDW order parameter~(\ref{2kFCDW}) and the BCS singlet pairing operator under~(\ref{dualityfer})
\begin{eqnarray}
n_{2k_F} &\rightarrow&  {\cal J}_{\alpha \beta} {\tilde R}_{\alpha}^\dagger  {\tilde L}^{\dagger}_{\beta} \nonumber \\
P_{00}^{\dagger} &\sim& R_{\alpha}^\dagger {\cal J}_{\alpha \beta}  L^{\dagger}_{\beta} 
\rightarrow {\tilde R}_{\alpha}^\dagger {\tilde L}_{\alpha}^{\phantom\dagger}.
\label{dualityorderparameter}
\end{eqnarray}
We see that $n_{2k_F}$ is no longer a singlet under the hidden $\widetilde{SU}(2n)$ symmetry, and as a result it is short-range (e.g., it has exponentially decaying correlation functions) in the $c=1$ critical phase described by the $\widetilde{SU}(2n)$ Gross-Neveu model~(\ref{su2NGNdual}). On the other hand, the BCS singlet pairing operator $P_{00}^\dagger$ is now $\widetilde{SU}(2n)$ invariant and exhibits power-law decay of the equal-time correlation function:
\begin{equation}
\Big\langle P^{\dagger}_{00} (x) P_{00} (0) \Big\rangle \sim x^{- 1/(nK_c)} .
\label{bcscor}
\end{equation}
As a result, the leading instability of the critical $c=1$ phase is BCS singlet pairing. 

\paragraph{$\mathbb{Z}_n$ quantum criticality.}
The nature of the quantum phase transition (QPT) between the two discussed phases, a spin-gapped CDW and a BCS phase, is governed by discrete soft modes. This can be revealed through a conformal embedding approach: the coset $SU(2n)_1/Sp(2n)_1$ CFT has central charge $c= 2n - 1 - n(2n+1)/(n+2) = 2(n-1)/(n+2)$. This is identical to the central charge of the ${\mathbb{Z}}_n$ parafermionic CFT~\cite{ZamolodchikovJETP85}, and this signals that the $SU(2n)_1/Sp(2n)_1$ CFT is equivalent to the ${\mathbb{Z}}_n$ CFT~\cite{AltschulerNuclPhysB89}. The latter CFT describes self-dual critical points of the two-dimensional ${\mathbb{Z}}_n$ generalization of the Ising model. 

In the vicinity of the QPT, the $Sp(2n)$ degrees of freedom have a large spectral gap, and the remaining ${\mathbb{Z}}_n$ degrees of freedom are governed by the effective action
\begin{equation}
{\cal S}_{\rm eff} = A[{\mathbb{Z}}_n]  + \lambda \int \rd^2 x \, \epsilon_1(x) , 
\label{efftrans}
\end{equation}
where $A[{\mathbb{Z}}_n] $ stands for the action of the ${\mathbb{Z}}_n$ CFT and $\epsilon_1$ is the first ${\mathbb{Z}}_n$ thermal operator with scaling dimension $d_{\epsilon} = 4/(n+2)$. Model~(\ref{efftrans}) is a massive integrable deformation of the ${\mathbb{Z}}_n$ CFT for both signs of the coupling constant $\lambda$~\cite{FateevIntJModPhysA91}. The order parameters of the CDW and BCS phases can be expressed in terms of the ${\mathbb{Z}}_n$  fields~\cite{LecheminantNuclPhysB08}:
\begin{eqnarray}
n_{2k_F} &\sim&   \exp \left( i \sqrt{2 \pi /n} \Phi_c \right) \mu_1, 
 \nonumber \\
P_{00}^{\dagger} &\sim&  \exp \left( i \sqrt{2 \pi /n} \Theta_c \right) \sigma_1,
\label{orderparametersZN}
\end{eqnarray}
where $ \mu_1$ ($\sigma_1$) is the disorder (order) parameter of the ${\mathbb{Z}}_n$ CFT with scaling dimension $(n-1)/n(n+2)$. The identification (\ref{orderparametersZN}) is closely related to the one~\fr{z1z2} obtained within the Hubbard-Stratonovich approach of Sec.~\ref{Sec:HubStrat}. One thus observes that the CDW phase corresponds to $\lambda > 0$ where the $\mathbb{Z}_n$ degrees of freedom are disordered, whilst the BCS phase corresponds to $\lambda < 0$ with ordered $\mathbb{Z}_n$ degrees of freedom. In the $n=2,3$ cases, the QPT between these two phases is universal and belongs to the Ising ($n=2$) and three-state Potts ($n=3$) universality classes, respectively. This has been confirmed numerically in the $F=3/2$ ($n=2$) case~\cite{CapponiPRA08,RouxEPJB09}. 

When $n \ge 4$, the second ${\mathbb{Z}}_n$ thermal operator $\epsilon_2$ with scaling dimension $12/(n+2)$ is generated in the ${\mathbb{Z}}_n$ sector. The field theory capturing the QPT between the CDW and BCS phases is then 
\begin{equation}
{\cal S}_{\rm transition} = A[{\mathbb{Z}}_n]   + {\tilde \lambda} \int d^2 x \, \epsilon_2 \left(x\right) , 
\label{efftransbis}
\end{equation}
which is also an integrable deformation of the ${\mathbb{Z}}_n$ CFT~\cite{FateevIntJModPhysA91}. The nature of the phase transition now depends on the sign of the coupling constant ${\tilde \lambda}$~\cite{FateevIntJModPhysA91}. When ${\tilde \lambda}  <0$, the field theory~\fr{efftransbis} is massive and there is a first-order QPT between the two phases. On the other hand, for ${\tilde \lambda}   > 0$ the action~\fr{efftransbis} flows under the RG to a $c=1$ phase. As a result, the critical theory at the QCP is described by a $c=1$ gapless theory for the $\mathbb{Z}_n$ degrees of freedom and a $c=1$ gapless theory for the decoupled charge theory, resulting in an overall $c=2$ theory at the QPT. 

\subsubsection{Haldane-charge insulator}
So far we have concentrated on the case with incommensurate filling. We now turn our attention to half-filling; we consider the $U(1)\times Sp(2n)$ model~\fr{hubbardS} with one atom per site ($k_F = \pi/2 a_0$). At half-filling there is no expectation of spin-charge separation (as found in~\fr{Speffectiveinteracting}) for $n > 1$;  the low-energy properties of~\fr{hubbardS} can be investigated via the conformal embedding~\cite{AltschulerNuclPhysB89} :
\be
SO(4n)_1 \sim SU(2)_n \times Sp(2n)_1, \label{embeddingso4n}
\ee
where the $SO(4n)$ group is the maximal continuous symmetry of the $2n$ Dirac fermions~(\ref{contfreeham}) of the non-interacting limit.\footnote{This can be simply realized by decomposing the Dirac fermions into their real components, i.e. in the Majorana basis.} The $SU(2)_n$ currents of the embedding~(\ref{embeddingso4n}) are~\cite{NonnePRB11}:
\be
\mathfrak{J}_L^{\dagger} = \frac{1}{2} \; L_\alpha^\dagger {\cal J}_{\alpha\beta} L_\beta^\dagger, \qquad
\mathfrak{J}_L^z = \frac{1}{2}  \; :L_{\alpha}^\dagger L_{\alpha}^{\phantom\dagger}: ,   \label{su2Ncurrentssp}
\ee
with a similar definition for the right currents. At half-filling, we need also to introduce umklapp terms which are built from
\begin{equation}
       J_L^{i+}
    = L_\alpha^\dagger \tilde{T}^i_{\alpha\beta} L_\beta^\dagger, \qquad
    J_R^{i+}
    = R_\alpha^\dagger \tilde{T}^i_{\alpha\beta} R_\beta^\dagger ,
    \label{LEpseudocurrents}
\end{equation}
where the generators $\tilde{T}^i_{\alpha\beta}$ ($i= 1, \ldots, 2n^2-n-1$), together with ${\cal J}_{\alpha\beta}$,  form the set of antisymmetric generators of $SU(2n)$. The interacting part of the low-energy Hamiltonian of model~(\ref{hubbardS}) at half-filling is then
\begin{eqnarray}
  \mathcal{H}_{\mathrm{int}}&=&
  g_1 J_R^aJ_L^a + g_2 J_R^iJ_L^i+g_3 \mathfrak{J}_R^z\mathfrak{J}_L^z\nonumber\\
  &&+\frac{g_4}{2}(J_R^{i+} J_L^{i-}+\mathrm{H.c.}) \nn
  &&+ \frac{g_5}{2} (\mathfrak{J}_R^{+}\mathfrak{J}_L^{-}+\mathrm{H.c.}) .
  \label{LEeffectiveinteracting}
\end{eqnarray}

A detailed RG analysis of~\fr{LEeffectiveinteracting} has been presented in Ref.~\cite{NonnePRB11}. Two phases exhibiting dynamical symmetry enlargement were found, with low-energy properties governed by the $SO(4n)$ Gross-Neveu model up to duality symmetries~\cite{BoulatNuclPhysB09}. The $SO(4n)$ Gross-Neveu model is a massive integrable field theory, whose mass spectrum is known exactly~\cite{ZamoldchikovAnnPhys79,KarowskiNuclPhysB81}. The excitation spectrum consists of elementary fermions of mass $m$, bound states of these fermions, and kinks. 
The bound states have masses ($n > 1$)
\begin{equation}
m_p = m\; \frac{\sin \left( \frac{\pi p}{2\left(2 n - 1\right)} \right)}{\sin \left( \frac{\pi }{2  \left(2n - 1\right)} \right)},
 \label{masspecO4N}
\end{equation}
with $p=2, \ldots, 2 n - 2$, while the mass of the kinks is
\begin{equation}
m_{\rm kinks} = \frac{m}{2 \sin \left( \frac{\pi }{2  \left(2n - 1\right)} \right)} .
 \label{kinkO4N}
\end{equation}
The two symmetry enlarged phases are fully gapped Mott insulators with leading instabilities of the CDW and Spin-Peierls (bond-ordering) type. The phases are two-fold degenerate as a result of spontaneous breaking of the one-site translation symmetry.

Interestingly, the RG analysis reveals that there is another Mott insulator phase which displays no symmetry enlargement. In stark contrast to the previous two phases, the $Sp(2n)_1$ current-current perturbation with coupling constant $g_1$ in~(\ref{LEeffectiveinteracting}) reaches the strong-coupling regime before the others. Integrating out the resulting massive $Sp(2n)$ degrees of freedom, the low-energy field theory of the third Mott insulator phase is expressed in terms of the $SU(2)_n$ fields of the conformal embedding~\fr{embeddingso4n}~\cite{NonnePRB11}:
\begin{equation}
\mathcal{H}_{\mathrm{int}} \simeq  \lambda {\rm Tr}\, {\Phi}^{(1)},
\label{heffsu2}
\end{equation}
where $\lambda >0$ and ${\Phi}^{(1)}$ is the spin-1 (or adjoint) primary field of the $SU(2)_n$ CFT with scaling dimension $4/(n+2)$ (see the discussion of Sec.~\ref{Sec:SU2Kadj}). 

The effective Hamiltonian~(\ref{heffsu2}) is directly related to the low-energy theory of the spin-$n/2$ $SU(2)$ Heisenberg chain derived by Affleck and Haldane in Ref.~\cite{AffleckPRB87}. As we have reviewed above, the model~(\ref{heffsu2}) has a spectral gap when $n$ is even, while it flows under the RG to the $SU(2)_1$ CFT when $n$ is odd. This phenomenology is in full agreement with Haldane's conjecture for the Heisenberg spin chain~\cite{HaldanePhysLettA83,HaldanePRL83}. In the case of $n$ even, there is a gapped non-degenerate Mott insulating phase with properties similar to the spin-one Haldane phase. However, in the definitions~\fr{su2Ncurrentssp} we see that the $SU(2)_n$ currents are singlets with respect to the hyperfine spin $Sp(2n)$ symmetry, and as a result they depend only upon the charge degrees of freedom. In this respect, the fully gapped Mott insulating phase for even $n$ is a Haldane-charge insulator~\cite{NonnePRB10,NonnePRB11}. 

The emergence of this exotic insulating phase can also be understood from the strong coupling limit of the lattice Hamiltonian~\fr{hubbardS}~\cite{NonnePRB10}. For strong attractive $U$ and $V = NU/2$, the model~\fr{hubbardS} becomes equivalent to a (pseudo) spin-$n$ antiferromagnetic Heisenberg chain~\cite{NonnePRB10,NonnePRB10a,NonnePRB11}: 
\begin{equation}
	{\cal H}_{\mathrm{eff}} = J_{\mathrm{eff}} \sum_i \Vec{\cal S}_i \cdot \Vec{\cal S}_{i+1},
\label{HaldaneSU2Ham}
\end{equation}
with $J_{\mathrm{eff}} =\frac{4t^2}{n(2n+1)|U|}$ and pseudo spin operators which carry charge and are $Sp(2n)$ spin singlets
\be
{\cal S}^{\dagger}_i = \sqrt{n/2} \; P^{\dagger}_{00}(i)\qquad {\cal   S}^{z}_i = \frac12 [n(i) - n ].\label{spinop.eq}
\ee
These operators satisfy the $SU(2)$ commutation relations with $\Vec{\cal S}_i^2 = n(n+2)/4$. They generalize the $\eta$-pairing operators introduced by Yang for the half-filled spin-1/2 (i.e., $n=1$) Hubbard model~\cite{YangPRL89} or those introduced by Anderson in his study of BCS superconductivity~\cite{AndersonPR58}. 

The even/odd dichotomy revealed by the RG analysis can also be simply explained within the strong-coupling framework. For the case of even $n$, the pseudo spin is integer and the Haldane-charge insulator phase is formed. On the other hand, when $n$ is odd, the pseudo spin is half-integer and a metallic (i.e., gapless) phase is stabilized. This is in complete analogy with Haldane's conjecture for spin chains, where here the underlying spin $\Vec{\cal S}$ is non-magnetic and carries charge. For this reason, the even/odd behavior was coined the ``Haldane-charge conjecture'' in Ref.~\cite{NonnePRB10}.

\subsection{$SU(N)$ ultracold fermions}

Let us now turn our attention to the low-energy properties of ultracold alkaline earth and ytterbium fermions atoms loaded into a 1D optical lattice. For atoms in the $^1S_0$ (i.e., $g$) state, the lattice Hamiltonian is a generalization of the well-known Fermi-Hubbard model where the hyperfine spin degrees of freedom enjoy $SU(N)$ rotational invariance~\cite{CapponiAnnPhys16}
\be
H = - t \sum_{i}\sum_{\alpha=1}^{N}  \Big[c_{\alpha}^\dag(i) c^{\phantom\dagger}_{\alpha}(i+1) + \text{H.c.}\Big]
+ \frac{U}{2}  \sum_{i} n(i)^2 ,
\label{HubbardSUN}
\ee 
where $\alpha = 1, \ldots, N=2I+1$ now describes the nuclear spin states of the underlying atoms, as discussed in Sec.~\ref{Sec:HighSymColdAtoms}. This model is invariant under the global charge $U(1)$ symmetry 
\be 
c_{\alpha}(j) \mapsto e^{i \theta} c_{\alpha}(j)
\ee
and the $SU(N)$ symmetry: 
\be
c_{\alpha}(j) \mapsto {\cal U}_{\alpha \beta}  c_{\beta}(j)
\ee
with ${\cal U}$ being an $SU(N)$ matrix. As a result, the continuous symmetry group of the Hamiltonian~(\ref{HubbardSUN}) is $U(N) = U(1) \times SU(N)$. When $N=2$, model~(\ref{HubbardSUN}) is exactly solvable by means of the Bethe ansatz~\cite{LiebPRL68,HubbardBook}. However, for $N>2$ the Hamiltonian~(\ref{HubbardSUN}) is not integrable for arbitrary $U$ and filling $n$. In the absence of a lattice, the model is again integrable and its properties have been described in Ref.~\cite{GuanRMP13}.

\subsubsection{Mott transition}
The continuum description of model~(\ref{HubbardSUN}) was studied by Affleck in Refs.~\cite{AffleckNuclPhysB86,AffleckNuclPhysB88}. At incommensurate filling, there is spin-charge separation and the Hamiltonian density decomposes into two commuting parts, $ [{\cal H}_c, {\cal H}_s] = 0$:
\begin{eqnarray}
  \mathcal{H} &=&  \mathcal{H}_c +  \mathcal{H}_s, \nn
 \mathcal{H}_c &=& \frac{v_c}{2} \left[\frac{1}{K_c}
\left(\partial_x \Phi_c \right)^2 + K_c \left(\partial_x \Theta_c \right)^2 \right] \nn
 \mathcal{H}_s &=& \frac{2\pi v_s}{N + 1} \left[ : J^a_{R} J^a_{R}: + : J^a_{L} J^a_{L}: 
\right]  + g J_R^{a} J_L^{a},
\label{Hsuncont}
\end{eqnarray}
where $v_c$ ($v_s$) is the charge (spin) velocity, $K_c$ the Luttinger parameter, and
$g = - 2 a_0 U$. In the Hamiltonian density~(\ref{Hsuncont}), $J^a_{R,L}$ ($a= 1, \ldots, N^2 -1$) are the chiral $SU(N)_1$ currents defined in Eq.~(\ref{chargeconserquant}) in terms of the Dirac fermions of the non-interacting Hamiltonian~(\ref{contfreeham}). For a repulsive interaction $U > 0$, which is the case in the experiments of Ref.~\cite{PaganoNatPhys14}, the interaction in the spin sector is marginally irrelevant and thus scales to zero in the far infrared (low energy) limit. As a result, for incommensurate filling and repulsive interaction, all modes are gapless and the central charge is $c=N$. In this respect, a $N$-component metallic Luttinger liquid phase emerges with $2k_{F}$ CDW oscillations and non-universal power-law exponents in the density-density and $SU(N)$ spin-spin correlation functions~\cite{AffleckNuclPhysB86,AffleckNuclPhysB88,AssarafPRB99,ManmanaPRA11}. 

The most interesting situation to consider is commensurate filling with one atom per site, such that $k_F = \pi/Na_0$. In contrast to the $N=2$ case, the umklapp term is always strongly irrelevant for $N>2$ for sufficiently small $U$~\cite{AffleckNuclPhysB88,AssarafPRB99}. The leading umklapp contribution, which is in fact generated at higher order in perturbation theory, leads to a sine-Gordon model for the charge degree of freedom~\cite{AssarafPRB99}
\begin{equation}
 {\cal H}_c = \frac{v_c}{2} \left[\frac{1}{K_c}
\left(\partial_x \Phi_c \right)^2 + K_c \left(\partial_x \Theta_c \right)^2 \right] 
+ g_c \cos( \sqrt{4 \pi N} \Phi_c ).
 \label{SGumklapp}
\end{equation}
Here the sine-Gordon potential term has scaling dimension $NK_c$, and as a result it becomes a relevant perturbation when $NK_c < 2$ and a Mott insulating phase emerges. For the spin-1/2 Fermi Hubbard model ($N=2$) one has $K_c < 1$ for arbitrarily small repulsive interactions  and a charge gap opens, leading to a Mott insulating phase with a single gapless spin mode~\cite{GNTBook,GiamarchiBook,HubbardBook}. For fermions with $SU(N)$ spins with $N>2$, the sine-Gordon term in~\fr{SGumklapp} is irrelevant at small $U$, with a Mott transition occurring at finite $U = U_{\rm c} \neq 0$. On the Mott insulator side of the transition, one expects gapless $SU(N)_1$ spin modes described by a $c = N-1$ CFT~\cite{AssarafPRB99}. The Mott transition has been numerically investigated for $N=3,4$ with QMC and DMRG~\cite{AssarafPRB99,ManmanaPRA11,BuchtaPRB07}. The existence of a Mott transition at finite $U$ was reported in Refs.~\cite{AssarafPRB99,ManmanaPRA11}, while the DMRG calculations of Ref.~\cite{BuchtaPRB07} concluded that $U_{\text{c}} = 0$ for all $N \geq 2$. The latter DMRG results~\cite{BuchtaPRB07} are in strong disagreement with the irrelevance of the umklapp term for weak $U$ and $N > 2$  found in Refs.~\cite{AffleckNuclPhysB88,AssarafPRB99}. They also disagree with more recent DMRG results from Manmana \textit{et al.}~\cite{ManmanaPRA11}, where the Mott transition was identified through the minima of the fidelity susceptibility.

In the large $U$ limit and with one atom per site, the physical properties of the model~\fr{HubbardSUN} are governed by the $N-1$ gapless spin modes. In direct analogy with the $N=2$ Hubbard model~\cite{HubbardBook}, \fr{HubbardSUN} reduces to the $SU(N)$ Heisenberg antiferromagnetic spin chain, i.e., the Sutherland model~\cite{SutherlandPRB75}:
\begin{equation}
{\cal H}_{\rm spin} = J \sum_i \sum_{a=1}^{N^{2}-1} S_{i}^{a} S_{i+1}^{a}  ,
\label{sutherland}
\end{equation}
where $J = 4 t^2/U$ is the antiferromagnetic spin exchange and $S_{i}^{a}$ is the spin operator at site $i$ which transforms in the fundamental representation of the $SU(N)$ group. Model~(\ref{sutherland}) can be solved exactly by means of the Bethe ansatz~\cite{SutherlandPRB75}. The low-energy spectrum is gapless with $N-1$ relativistic modes, each of which has the same velocity $v_{s} =  \pi J/N$. The critical theory is the $SU(N)_1$ CFT perturbed by a marginally irrelevant current-current interaction (cf. Eq.~\fr{Hsuncont})
\begin{equation}
  \mathcal{H}_{\rm spin} = \frac{2\pi v_s}{N + 1} \Big[ : J^a_{R} J^a_{R}: + : J^a_{L} J^a_{L}: 
\Big]  + g J_R^{a} J_L^{a} .
\label{Hcontsutherland}
\end{equation}
In the low-energy limit, $SU(N)_1$ quantum critical behavior with central charge $c=N-1$ is stabilized and the marginally irrelevant current-current interaction leads to logarithmic corrections in correlation functions~\cite{AffleckJPhysA89,MajumdarJPhysA02}. In turn, the continuum $SU(N)$ spin operator can be expressed in terms of the gapless fields~\cite{AffleckNuclPhysB86,AffleckNuclPhysB88}
\begin{equation}
 S^{a}_j \sim J^a_{L} +  J^a_{R} +  e^{2 i j \pi/N} \lambda {\rm Tr} ( g T^{a}) + \mathrm{H.c.},
\label{Spincontsutherland}
\end{equation}
where $g$ is the $SU(N)_1$ WZNW primary field and $\lambda$ is a non-universal constant which depends on the gapped charge degrees of freedom.\footnote{An analogous non-universal factor that depends upon the charge degrees of freedom appears in the Abelian bosonization identities of spin operators.} The low-lying gapless excitations of the Sutherland model occur in pairs with individual dispersion relations covering a fraction of the Brillouin zone~\cite{JohannesonNuclPhysB86}. The elementary excitations of the model are then a generalization of the spinons of the spin-1/2 Heisenberg chain and carry fractional quantum numbers. They display  fractional statistics with angle $\theta = \pi/N$ and transform in the conjugate ${\bf {\bar N}}$ representation of the $SU(N)$ group~\cite{BouwknegtNuclPhysB96, SchurichtPRB06}. In this respect, they may be viewed as an analogue of antiquarks in quantum chromodynamics.

\subsubsection{Mott insulating phases}
\label{Sec:MIphases}
The nature of the $SU(N)$ Mott insulating phases for other commensurate fillings can also be investigated in the weak-coupling regime by means of bosonization~\cite{SzirmaiPRB08}. Alternatively, one can directly consider the large $U$ limit and study the leading relevant perturbation, which describes the departure from the $SU(N)_1$ fixed point~\cite{AffleckNuclPhysB88}.

In this respect, let us consider a filling of $m/N$ ($m=1,\ldots,N-1$) where a Mott insulator with $m$ atoms per site is formed. In the large-$U$ limit, it is described by the $SU(N)$ spin chain Hamiltonian~(\ref{sutherland}) where the spin operators transform under the antisymmetric $m$-tensor representation of $SU(N)$ with the Young tableau: 
\begin{equation}
\text{\scriptsize $m$~times} \left\{ 
\yng(1,1,1,1)
 \right.  \; .
 \label{antisymrep}
\end{equation}
A field theory analysis of this problem can be obtained by finding the leading relevant perturbation to the $SU(N)_1$ CFT which obeys all the symmetries of the lattice model. 

The $SU(N)_1$ CFT has $N-1$ primary fields $\Phi_m$ ($m=1,\ldots,N-1$) which transform in the antisymmetric representation~(\ref{antisymrep}) of the $SU(N)$ group. Their scaling dimensions can be determined as a result of Eq.~\fr{cftDim}: $d_m = m(N-m)/N$. From the point of view of the $SU(N)_1$ WZNW model and its field $g$, the primary field $\Phi_m$ can be obtained through $m$ fusions of $g$ with itself~\cite{CFTBook}. 

An important symmetry constraint stems from the one-site translation symmetry of the lattice model, which acts as follows on the WZNW field $g$ for the filling $m/N$ (i.e,. $k_F = \pi m/Na_0$)~\cite{AffleckNuclPhysB88}:
 \be
 g \rightarrow  e^{2 i \pi m /N}  g  .
 \label{transantisymrep}
\ee
As a result, when $m$ and $N$ have no common divisor, a relevant $SU(N)_1$ primary field is not allowed by one-step translational symmetry~\fr{transantisymrep}.\footnote{This can easily be seen from the fact that $\Phi_m$ ($m=1,\ldots,N-1$) is constructed from fusing $g$ with itself $m$ times. Then under one-site translation the primary field must transform as $\Phi_m \to e^{2i \pi m^2/N} \Phi_m$ which cannot equal $\Phi_m$ if $m$ and $N$ have no common divisors. Thus all primary fields $\Phi_m$ are symmetry forbidden.} With no symmetry allowed relevant perturbations, a gapless $SU(N)_1$ WZNW  QCP emerges, as in the Sutherland model~\fr{sutherland}. On the other hand, when $N$ is divisible by $m$ ($N=mp$) the primary field $\Phi_p$ is symmetry allowed and may appear in the low-energy theory. The low-energy effective Hamiltonian density then reads
\be
{\cal H}^{N = mp}_{\rm eff} \simeq  \frac{2\pi v_s}{N + 1} \Big[ : J^a_{R} J^a_{R}: + : J^a_{L} J^a_{L}: 
\Big]  +  \kappa  \Big( {\rm Tr} g \Big)^p + \mathrm{H.c.} 
\label{antisymHameff}
\ee
The added operator is strongly relevant when $p(N-p)/N < 2$, i.e., when $N(m-1) < 2 m^2$. Then, a fully gapped bond-ordered phase emerges with a $p$-fold degenerate ground state with spontaneously broken one-site translation symmetry. A paradigmatic example of this is the self-conjugate representation where $m=N/2$ with $N$ even. There a fully gapped dimerized phase is stabilized, corresponding to the Mott insulating phase of the half-filled $SU(2m)$ Fermi-Hubbard model~\fr{HubbardSUN}~\cite{SzirmaiPRB08,AffleckNuclPhysB91,OnufrievPRB99,AssarafPRL04,BoisPRB15}. On the other hand, when  $N(m-1) > 2 m^2$ the $SU(N)_1$ QCP is realized. Recently these predictions have been checked to high accuracy by variational QMC calculations~\cite{DufourPRB15}.  

\subsubsection{Orbital effects}

Beyond the existence of the $SU(N)$ symmetry, another interesting aspect of alkaline-earth-like atoms stems from the fact that one can incorporate an additional orbital degree of freedom into the system~\cite{GorshkovNaturePhys10}. Indeed by considering the metastable ${}^{3}P_{0}$ state `$e$' of alkaline earth atoms, the interplay between orbital and $SU(N)$ symmetries can be investigated. A paradigmatic model for this competition in one dimension is the $g-e$ $SU(N)$ model which is defined by the Hamiltonian~\cite{GorshkovNaturePhys10,CapponiAnnPhys16}
\bea
 H_{g\text{-}e} &=& 
  -  t \sum_{i}  \sum_{m=g,e}  \sum_{\alpha=1}^{N} 
   \Big[c_{m\alpha}^\dag(i) c_{m\alpha}^{\phantom\dagger}(i+1)  + \text{H.c.}\Big]  
   \nn
&&
 + \frac{U}{2} \sum_{i} n(i)^2  +V \sum_i n_{g}(i) n_{e}(i)\nn
 &&
 + V_{\text{ex}}  \sum_{i,\alpha,\beta} 
  c_{g\alpha}^\dag(i) c_{e\beta}^\dag(i) 
  c_{g\beta}^{\phantom\dagger}(i) c_{e\alpha}^{\phantom\dagger}(i) ,
\label{gemodel}
\eea
where the index $\alpha$ labels the $SU(N)$ nuclear-spin states (as before) and $m=g,e$ labels the two atomic states of alkaline earth atoms (${}^{1}S_{0}$ and ${}^{3}P_{0}$, respectively). In Eq.~(\ref{gemodel}), $n_{m,i}$ denotes the density of the $m=g,e$ fermions at each site: $n_{m}(i) = \sum_{\alpha=1}^{N}c^{\dagger}_{m\alpha}(i)c_{m\alpha}(i)$. 

On top of the $U(N)$ symmetry discussed for the previous model, the $g-e$ model~(\ref{gemodel}) is invariant under a $U(1)_{o}$ orbital symmetry:
\begin{equation}
c_{g\alpha}(j) \mapsto e^{i \theta_{\text{o}}} c_{g\alpha}(j) , \qquad
c_{e\alpha}(j) \mapsto e^{-i \theta_{\text{o}}} c_{e\alpha}(j),
\label{orbital-U1}
\end{equation}
which reflects the fact that the total fermion numbers for $g$ and $e$ are conserved separately.

The continuum description of the $g-e$  model~(\ref{gemodel}) can be derived as before by introducing $2N$ left-right moving Dirac fermions $L_{m\alpha}$ and $R_{m\alpha}$. The resulting low-energy approach has been investigated for incommensurate filling~\cite{SzirmaiPRB13,BoisPRB16,SzirmaiArxiv16} and at half-filling~\cite{NonneEPL13,BoisPRB15}. In this respect, non-Abelian bosonization can be used alongside the conformal embedding~\fr{emb2} of Sec.~\ref{Sec:ExamplesNonAbelian}. The conformal embedding for the $g-e$ model reads
\begin{equation}
U(2N) \rightarrow U(1)_{\text{c}} \times SU(2)_N   \times  SU(N)_2.
\label{conformalembgemodel}
\end{equation}
The non-Abelian left currents, $J_\text{L}^a$ and $j_L^a$, for the $SU(N)_2$ (nuclear) spin and $SU(2)_N$ orbital sectors are defined as
\be
J_\text{L}^a = L_{n\alpha}^\dagger T^a_{\alpha\beta} L^{\phantom\dagger}_{n\beta} \qquad  j_{\text{L}}^i = \frac{1}{2} L_{m\alpha}^\dagger \sigma^i_{mn} L^{\phantom\dagger}_{n\alpha},
\label{defalkacurrents}
\end{equation}
with $T^a$ ($a=1,\ldots,N^{2}-1$) and $\sigma^i$ ($i=x,y,z$) being, respectively, the $SU(N)$ generators and the Pauli matrices. 

At half-filling, several Mott insulating phases have been found within the bosonization approach~\cite{BoisPRB15}. Perhaps the most interesting is a symmetry-protected topological phase which occurs in the large $U$ regime~\cite{CapponiAnnPhys16,NonneEPL13,BoisPRB15,DuivenvoordenPRB13,TanimotoArxiv15,RoyArxiv15} with a symmetry based on the $PSU(2n)= SU(2n)/{\mathbb{Z}}_{2n}$ projective unitary group. 

When $V_{\rm ex}>0$ and $N=2n$, the strong-coupling (large $U$) Hamiltonian is given by~\cite{NonneEPL13,BoisPRB15}:
\begin{equation}
\mathcal{H}_{\text{eff}}  = J
\sum_{a=1}^{N^{2}-1} \mathcal{S}_{i}^{a}\mathcal{S}_{i+1}^{a}   \; ,
\label{stronHamGorshkov}
\end{equation}
where the $SU(2n)$ spin operators transform under the  self-conjugate representation with the Young tableau:
\begin{equation}
\text{\scriptsize $n$~times} \left\{
\yng(2,2,2) \right. .
\label{irrepSPT}
\end{equation}
When $n=1$, model~\fr{stronHamGorshkov} is the spin-1 Heisenberg chain with the Haldane phase as a ground state. In the general $n$ case, on general grounds, it is expected that model~\fr{stronHamGorshkov} in the representation~\fr{irrepSPT} displays a non-degenerate fully gapped phase~\cite{AffleckLettMathPhys86,RachelPRB09}. In this respect, a MPS state has been constructed to describe the properties of the ground state~\cite{NonneEPL13,BoisPRB15,TanimotoArxiv15,RoyArxiv15}. The latter is $SU(2n)$-symmetric and featureless in the bulk  and has exponentially decaying spin-spin correlation functions with a very short correlation length. The hallmark of this phase is the existence of edge states which transform in the antisymmetric  self-conjugate representation with dimension $N!/[(N/2)!]^{2}$ of the $SU(2n)$ group. The topological nature of this phase can also be revealed by numerical investigation of the entanglement spectrum: for $N=4$ this exhibits a six-fold degeneracy~\cite{TanimotoArxiv15}, which marks the emergence of a symmetry-protected topological phase with edge state that transform under the projective representation of the $PSU(4)$ group~\cite{CapponiAnnPhys16,DuivenvoordenPRB13}. 

The symmetry-protected topological phase, found as the ground state of the spin model~\fr{stronHamGorshkov}, is the natural generalization of the spin-1 Haldane phase to the $PSU(2n)$ group. DMRG calculations for the $g-e$  model~(\ref{gemodel}) show that this $PSU(2n)$ Haldane phase occurs in the large-$U$ regime~\cite{NonneEPL13,BoisPRB15}. In contrast to the $n=1$ case, it is not adiabatically connected to the weak-coupling regime where one finds a spin-Peierls phase with bond-order. A QPT occurs at finite interaction strength, which can be inferred from the conformal embedding~\fr{conformalembgemodel}. In the vicinity of the QPT, the $U(1)$ charge and $SU(2)_N$ orbital degrees of freedom have large spectral gaps. The low-energy degrees of freedom, which control the nature of the phase transition, are the $SU(N)_2$ nuclear spin states. In order to discuss the QPT, we then need to identify the primary fields of the $SU(N)_2$ CFT which are strongly relevant and allowed by the symmetry of the underlying lattice model~\fr{gemodel} (cf. the discussion of the previous section). The $SU(N)_2$  CFT has $N(N+1)/2$ primary fields and they are obtained by fusion of the WZNW $g$ field, which has scaling dimension $(N^2-1)/N(N+1)$. Under the one-site translation symmetry of the lattice Hamiltonian, the WZNW field $g$ transforms as: $g \rightarrow -g$ at half-filling. There are three possible relevant perturbations that can be considered as a result, which may describe the theory in the vicinity of the QCP between the $PSU(2n)$ Haldane phase and the spin-Peierls phase: 
\begin{equation}
\begin{split}
\Phi_{\pm} &\sim  \left( \mbox{Tr}  \;  g\right)^2 \pm   \mbox{Tr}  \;  g^2  + \text{H.c.} \\
\Phi_{\rm adj} &\sim  |\mbox{Tr}  \; g|^2 .
\end{split}
\label{SUN2primaries}
\end{equation} 
The scaling dimensions of these perturbations are, respectively, 
\be
\begin{split}
d_{+} &=  \frac{2(N-1)}{N}, \\ d_{-} &= \frac{2(N +1)(N - 2)}{N(N+2)},\\ d_{\rm adj} &= \frac{2N}{(N+2)}.
\end{split}
\ee
For even $N=2n$, the general low-energy effective field theory that governs the properties of the model in the 
vicinity of the QCP is 
\begin{eqnarray}
&& {\cal S}_{\rm QCP} =  {\cal S}[SU(N)_2; g]\nn
&& +  \int \rd^2 x \; \left[ {\tilde \lambda}_+ \Phi_{+} +  {\tilde \lambda}_{-} \Phi_{-}
+ {\tilde \lambda}_{\rm adj} \Phi_{\rm adj} \right]. 
\label{actionQCP}
\end{eqnarray}
Following the semiclassical approximation of Ref.~\cite{AffleckNuclPhysB88},  one can show that the $PSU(2n)$ Haldane phase and the spin-Peierls phase appear in the space of parameters of the action (\ref{actionQCP}). When ${\tilde \lambda}_{\pm,{\rm adj}} <0$, minimization of the perturbation leads to $g = \pm I$ ($N$ even) and one has $ \langle {\rm Tr} g \rangle \ne 0$. The ground state is two-fold degenerate as a consequence of the spontaneously broken translation symmetry ($g \rightarrow - g$). This corresponds to the spin-Peierls phase identified in the weak-coupling limit~\cite{NonneEPL13,BoisPRB15}. On the strong coupling side,  for instance when ${\tilde \lambda}_+  =  {\tilde \lambda}_- > 0$ and ${\tilde \lambda}_{\rm adj}  >0$, the semiclassical analysis now gives an $SU(N)$ matrix of the form:
\begin{equation}
g = U_0\, {\rm diag}(\underbrace{i,\ldots,i}_{n~{\rm times}},\underbrace{-i,\ldots,-i}_{n~{\rm times}})\, U^{\dagger}_0 ,
\label{Grassmanmat}
\end{equation}
with $U_0$ being a unitary matrix. The ground state is now non-degenerate and invariant under one-site translations as $g \rightarrow -g$ can be absorbed in a redefinition of $U_0$. The resulting effective field theory is known to be the Grassmannian sigma model on the $U(N)/[U(N/2)\times U(N/2)]$ manifold with a $\theta = 2 \pi$ topological theta term~\cite{AffleckNuclPhysB88}. The latter is known to be massive and describes the semiclassical field theory of the $SU(N)$ Heisenberg spin chain in self-conjugate representations~\fr{irrepSPT}, see Ref.~\cite{ReadNuclPhysB89}. Thus, we conclude that the QCP between the spin-Peierls and $PSU(2n)$ Haldane phases should belong to the $SU(N)_2$ WZNW universality class as predicted in Ref.~\cite{NonneEPL13}. In the special $N=2$ case, one recovers  the 
well-known $SU(2)_2$ quantum critical behavior of the integrable Babujian-Takhtajan model~\cite{TakhtajanPhysLettA82,BabujianPhysLettA82} which is the QCP between the Haldane and dimerized phases~\cite{TsvelikPRB90}.
 
\subsection{$SU(N)$ two-leg spin ladder}

\subsubsection{Introduction}

Two-leg spin ladders have been a focus of much theoretical and experimental work over more than two decades. This strong interest stems from the desire to understand the crossover between one and higher dimensions, as well as being motivated by their experimental realizations~\cite{GNTBook,GiamarchiBook}. These simple magnetic quantum systems might also be employed as quantum simulators for fundamental theories of particle and many-body physics~\cite{WardJPhysCondMatt13,LakeNatPhys10}. In this respect, the problem of confinement of fractional quantum number excitations can be investigated in a simple two-leg spin ladder which consists of two spin-1/2 antiferromagnetic Heisenberg chains coupled by an interchain spin-exchange interaction. Gapless fractional spin-1/2 excitations (spinons) of individual chains turns out to be confined into gapped spin-1 (triplon) excitations even by an infinitesimal interchain coupling~\cite{SheltonPRB96}.

One generalization of this confinement problem in two-leg spin ladders is to consider spins where the internal symmetry group is enlarged to $SU(N)$. Such problems can be experimentally investigated by considering ultracold alkaline earth or ytterbium atoms loaded into a double-well optical lattice, with the lattice Hamiltonian
\begin{eqnarray}
{\cal H} &=&  J_{\parallel} \sum_i \sum_{a=1}^{N^{2}-1} \left( S^{a}_{1,i}  S^{a}_{1,i+1} + S^{a}_{2,i}  S^{a}_{2,i+1} \right)
\nonumber \\
&+&   J_{\perp} \sum_i \sum_{a=1}^{N^{2}-1} S^{a}_{1,i}  S^{a}_{2,i}  ,
\label{2leg}
\end{eqnarray}
where $S^{a}_{l,i}$ ($a =1, \ldots, N^2 -1$) denote the $SU(N)$ spin operators, which transform in the fundamental representation of the $SU(N)$ group, on the $i$-th site of the chain (leg) and the index $l=1,2$ stands for a leg of the ladder. The intrachain and interchain spin exchange interactions are antiferromagnetic for applications to ultracold alkaline earth or ytterbium atoms. When $J_{\perp} = 0$, the Hamiltonian~\fr{2leg} describes two decoupled $SU(N)$ Sutherland models, with quantum critical behavior in the $SU(N)_1$ WZNW universality class as reviewed above. The elementary gapless excitations of the model are the generalization of the spinons of the spin-1/2 Heisenberg chain with fractional statistics with angle $\theta = \pi/N$~\cite{BouwknegtNuclPhysB96,SchurichtPRB06}. The two-leg $SU(N)$ spin ladder~\fr{2leg} is thus a paradigmatic model for studying the confinement or deconfinement of these excitations with fractional quantum numbers upon switching on an antiferromagnetic interchain spin-exchange ($J_{\perp} > 0$). 

\subsubsection{The strong-coupling limit}

Some insights into this problem can be gained by considering the strong-coupling regime $J_{\perp}  \gg J_{\parallel}$ where the Hamiltonian~\fr{2leg} reduces to a single $SU(N)$ spin chain model~\fr{sutherland}. The underlying $SU(N)$ spin operator now transforms in the antisymmetric representation described by the Young's tableau
\be
\yng(1,1)\, .
\ee

As discussed in Sec.~\ref{Sec:MIphases}, when $N$ is odd, an $SU(N)_1$ quantum critical behavior is expected and the $SU(N)$ spinons are still deconfined, in contrast to the case with $N=2$. For example, when $N=3$ the strong-coupling limit of the ladder~\fr{2leg} again gives a Sutherland model~\fr{sutherland}, where the $SU(3)$ spin operators belong to the conjugate ${\bar {\bf 3}}$ representation. The $SU(3)$ spinon excitations are gapless and incoherent, and they transform in the {\bf 3} representation of the $SU(3)$ group, not the ${\bar {\bf 3}}$ representation as is the case in the limit $J_\perp = 0$. 

When $N$ is even (i.e., $N=2n$), the strong-coupling Hamiltonian is described by the effective field theory~\fr{antisymHameff} with $p=n$. A fully gapped phase with an $n$-fold degenerate ground-state is stabilized when the perturbation~\fr{antisymHameff} is relevant, i.e., for $n \le 4$ ($N=4,6$).\footnote{In the $N=8$ case, the perturbation of Eq.~\fr{antisymHameff} is marginal and the sign of the coupling constant $\kappa$ is important. However, the latter cannot be fixed within the symmetry argument that leads to Eq.~\fr{antisymHameff}. The variational QMC calculations of Ref. \cite{DufourPRB15} found critical behavior in the $SU(8)_1$ universality class.} The spinons now correspond to the gapped domain walls between the $\mathbb{Z}_n$ degenerate ground states. Finally, when $N$ is even and $N > 8$, the interacting part of model~\fr{antisymHameff} is strongly irrelevant and $SU(N)_1$ quantum criticality is restored in the strong-coupling limit, leading to gapless deconfined spinon excitations.

\subsubsection{Weak-coupling approach}
We now consider the opposite limit of weak-coupling, $J_{\perp}  \ll J_{\parallel}$, and investigate the low-energy physics of the two-leg ladder~\fr{2leg}. Using Eq.~\fr{Spincontsutherland}, the $SU(N)$ operators in the continuum limit are described by: 
\begin{equation}
\frac{S^{a}_{l,j}}{a_0} \simeq J^{a}_{l L} (x) +  J^{a}_{l R} (x) + \mbox{e}^{ i 2  \pi x /Na_0} \lambda   
\; {\rm Tr} \Big( g_l (x)   T^a \Big)  +  \mathrm{H.c.}  ,
\label{spinop}
\end{equation}
where $x = j a_0$ and $J^{a}_{l L,R}$ are the left and right $SU(N)_1$ currents. In Eq.~\fr{spinop}, $g_l$ is the $SU(N)_1$ WZNW field with scaling dimension $(N-1)/N$ corresponding to the $l$-th chain. The continuum limit of the two-leg $SU(N)$ spin ladder is then described by the action~\cite{LecheminantPRB15}:
\bea
{\cal S} &=& {\cal S}[SU(N)_1; g_1] +   {\cal S}[SU(N)_1; g_2] \nonumber \\
&&+   \int \rd^2x \; \Big[ \lambda_1   \mbox{Tr}(g_1g_2^+) + \lambda_2 \mbox{Tr}g_1\mbox{Tr}g_2^+
+ \mathrm{H.c.} \Big],\nn
\label{cftactionladder}
\eea
where ${\cal S}[SU(N)_k; g_l]$ denotes the action of the $SU(N)_k$ WZNW model for the $l$-th chain and $\lambda_1 = J_{\perp} \lambda^2 /2$, $\lambda_2 = - \lambda_1/N$. Model~(\ref{cftactionladder})  thus describes two $SU(N)_1$ WZNW models perturbed by two strongly relevant operators with the same scaling dimension, $2(N-1)/N  < 2$. 

\paragraph{Field theory strong coupling approach.}
Under the RG, the perturbations of~\fr{cftactionladder} flow towards strong coupling, where one can undertake a strong coupling approach when $|\lambda_1| \gg |\lambda_2|$, i.e., $N \gg 1$. Minimizing the $\lambda_1$ term ($\lambda_1 > 0$) in the action~\fr{cftactionladder} gives $g_1 = -g_2 = g$ when $N$ is even, or $g_1 = e^{\pm i 2\pi/N} g_2 = g$ when $N$ is odd. In both cases, the WZNW topological term in Eq.~\fr{cftactionladder} is doubled and the low-energy effective action is: 
\begin{eqnarray}
{\cal S}_{\rm eff} &=&  {\cal S}[SU(N)_2; g] + 2 {\tilde \lambda}_2  \int \rd^2 x \;  |\mbox{Tr}  \; g|^2
\nonumber \\
 &=&  {\cal S}[SU(N)_2; g] +  {\tilde \lambda}_2  \int \rd^2 x  \;  {\rm Tr}  \;  \Phi_{\rm adj}  ,
\label{strongcoupladderCFT}
\end{eqnarray}
where $\Phi_{\rm adj}$ is the $SU(N)_2$ primary field transforming in the adjoint representation with scaling dimension $2N/(N+2)$. In Eq.~\fr{strongcoupladderCFT}, we have ${\tilde \lambda}_2 >0$ for all $N$ and $J_{\perp} >0$. The perturbation present in Eq.~\fr{strongcoupladderCFT} is strongly relevant and a mass gap may open. However, in Ref.~\cite{LecheminantNuclPhysB15} it was argued that  the effective action~(\ref{strongcoupladderCFT}) displays a massless flow to $SU(N)_1$ when $N$ is odd and ${\tilde \lambda}_2 >0$. The result is then in perfect agreement with the conclusion obtained from the direct strong-coupling limit of the lattice model~\fr{2leg}. When $N$ is even, there is no known result for the infrared limit of the action~\fr{strongcoupladderCFT} with ${\tilde \lambda}_2 >0$ except when $N=2$. As reviewed in Sec.~\ref{Sec:SU2Kadj}, the perturbation is integrable in that case and a mass gap opens.

\paragraph{Conformal embedding approach.}
To shed light on the possible phases,  a conformal embedding analysis based on the symmetries of model~(\ref{cftactionladder}) can be performed. In the decoupled limit, the CFT governing the low-energy properties of model~(\ref{cftactionladder}) is $SU(N)_1 \times SU(N)_1$. However, when $J_{\perp} \ne 0$ the continuous symmetry group is reduced to $SU(N)$, making it more natural to consider the following conformal embedding~\cite{CFTBook}:
\begin{equation}
SU(N)_1 \times SU(N)_1 \sim SU(N)_2 \times  \mathbb{Z}_N .
\label{embeddingSUN1}
\end{equation}
The action, Eq.~\fr{cftactionladder}, can be expressed in terms of the fields of this conformal embedding~\cite{LecheminantPRB15}:
\begin{eqnarray}
 {\cal S} &=& {\cal S}[SU(N)_2; g] 
 + A[{\mathbb{Z}}_n]      -  {\tilde g} \int \rd^2 x    \;  \left( \psi  {\bar \psi}  + \mathrm{H.c.} \right)  
 \nonumber \\
&&+ \lambda_2 \int \rd^2 x   \, {\rm Tr}\, \Phi_{\rm adj} \big( \sigma_2 + \sigma_2^{\dagger} \big) ,
\label{cftnewbasis}
\end{eqnarray}
with ${\tilde g} = - N a_0 J_{\perp} \lambda^2  /8\pi^2$ and  $\lambda_2 = -  a_0 J_{\perp} \lambda^2 /N$. In Eq.~\fr{cftnewbasis}, $\psi , {\bar \psi}$ stand for the first parafermion currents with conformal weights $\Delta, {\bar \Delta} = (N-1)/N$ which generate the $\mathbb{Z}_N$ CFT and $\sigma_2$  denotes the second spin field with scaling dimension $2(N-2)/N(N+2)$~\cite{ZamolodchikovJETP85}.

The effective field theory~(\ref{cftnewbasis}) contains two different sectors, the $SU(N)$ singlet sector described by the $\mathbb{Z}_N$ parafermions, and the magnetic one which depends on the $SU(N)$ degrees of freedom. The main difference between $SU(2)$ and $SU(N>2)$ cases stems from the fact that for $N=2$ there is no $\sigma_2$ spin field which couples the two sectors of the theory~\cite{SheltonPRB96}. In that case, model~\fr{cftnewbasis} separates into two parts which can be expressed in terms of four massive Majorana fermions. This describes the non-denegerate gapped phases for both signs of $J_{\perp}$ when $N=2$~\cite{SheltonPRB96}. The situation is much more involved in the $N>2$ case due to the coupling of the magnetic and singlet sectors in Eq.~\fr{cftnewbasis}. 

The low-energy properties of model~\fr{cftnewbasis} can be deduced by exploiting the integrability of the $\mathbb{Z}_N$ parafermionic model of Fateev~\cite{FateevIntJModPhysA91,FateevPhysLettB91}:
\begin{eqnarray}
 {\cal S}_{\rm Fateev} = A[\mathbb{Z}_n]  - {\tilde g} \int \rd^2 x \;  \big( \psi  {\bar \psi}  + \mathrm{H.c.} \big)  .
\label{fateevmodel}
\end{eqnarray}
This model was already introduced in Sec.~\ref{Sec:ParafermZeroModes} in the context of $\mathbb{Z}_N$ parafermionic zero mode. 

The low-energy properties of this integrable model depend upon the parity of $N$. When $N$ is even a spectral gap is generated for the $\mathbb{Z}_N$ modes for both positive and negative coupling ${\tilde g}$. The analysis for the $SU(N)$ modes of Eq.~\fr{cftnewbasis} takes place in Ref.~\cite{LecheminantPRB15}: a spin-gapped phase with a $N/2$-fold degenerate ground-state was predicted for $J_{\perp} > 0$. When $N=4$, this gives rise to a plaquette phase with a two-fold degenerate ground-state which has been identified numerically~\cite{vandenBosschePRL01,SU3ladderpreprint}.

On the other hand, when $N$ is odd the  low-energy properties of~\fr{fateevmodel} depend on the sign of the coupling ${\tilde g}$. Interestingly, for $J_{\perp}>0$ (i.e., ${\tilde g} < 0$) the model~\fr{fateevmodel} displays an integrable massless RG flow from the $\mathbb{Z}_N$ ultraviolet fixed point to the infrared one governed by the minimal model ${\cal M}_{N+1}$ CFT with central charge $c=1 - 6/(N+2)(N+1)$~\cite{FateevPhysLettB91}. In the simplest case of $N=3$, the resulting massless degrees of freedom are described by the $c=7/10$ tricritical Ising model (TIM) CFT. As discussed in Ref.~\cite{SU3ladderpreprint}, the low-energy limit of model~\fr{cftnewbasis} for $N=3$ can be written in terms of the TIM$\times$SU(3)$_2$ CFTs as:
 \be
  {\cal S} = {\cal S}[SU(3)_2; g] + {\cal S}_{{\rm TIM}}
+ \kappa \int \rd^2x   \,  \epsilon_{\rm TIM} {\rm Tr} \, \Phi_{\rm adj} ,
\label{cftnewbasisIR}
\ee
where ${\cal S}_{{\rm TIM}}$ is the action of the TIM CFT and $\epsilon_{\rm TIM}$ is the thermal operator with scaling dimension $1/5$ of the TIM CFT~\cite{CFTBook}. The interacting part of model~\fr{cftnewbasisIR} is a strongly relevant perturbation with scaling dimension $7/5 < 2$, which opens a mass gap for the TIM degrees of freedom. By a simple mean-field decoupling of the $SU(3)_2$ and TIM sectors, a fully gapped trimerized phase is revealed when the interchain spin-exchange is weak, $J_{\perp}>0$~\cite{SU3ladderpreprint}. The latter phase has a three-fold degenerate ground state and spontaneously breaks one-site translation symmetry. Taking into account the conclusion of the strong-coupling analysis, with the emergence of the gapless $c=2$ phase of the Sutherland model, a quantum phase transition should occur for an intermediate $J_{\perp}$. This transition has recently been numerically identified and it seems to be a generic feature of the two-leg $SU(N)$ spin ladder~\fr{2leg} when $N$ is odd~\cite{SU3ladderpreprint}. These results pave the way for its experimental observation in the context of ultracold alkaline earth or ytterbium atoms loaded into a double-well optical lattice.

\subsection{Summary: non-Abelian bosonization}

In the previous three sections, we have introduced non-Abelian bosonization and applied it to a number of interesting scenarios, both in condensed matter (Sec.~\ref{Sec:ExamplesNonAbelian}) and cold atom systems (Sec.~\ref{Sec:ColdAtomsExamples}). We focussed on models with extended symmetries, including electrons that carry both spin and orbital indices, as well as cold atoms with multiple internal states (such as hyperfine levels). In each case, we have applied the conformal embedding to split the non-interacting part of the model into separate WZNW models for each symmetry sector. We then introduced interactions that preserve the symmetry structure, allowing us to study each symmetry sector separately and use appropriate tools (such as integrability, the renormalization group, and semi-classical analyses) to understand the low-energy phase diagram and correlation functions. 

In our example applications, we have seen some truly exotic physics: from competing CDW and SC  orders (as is well known in higher dimensions, such as the high-$T_c$ cuprates) to non-Abelian anyons -- excitations of low-dimensional systems that do not conform to the usual fermion/boson classification of particles. There is the tantalizing possibility that such exotic excitations may arise in the $\mathbb{Z}_n$ quantum phase transition that occurs between the spin-gapped CDW and the BCS phase of $Sp(2n)$ cold atoms. $SU(N)$ cold atoms and spin ladders have also been revealed as rich sources of intriguing physics, including phase transitions in the $SU(N)_2$ universality class and low-energy theories described by the well-studied minimal model conformal field theories. 

Nevertheless, despite its wide-ranging applications and versatility, non-Abelian bosonization can only get one so far. Scenarios where one applies the RG are controlled only when interactions are weak; semi-classical analyses apply in some large $N$ limit; away from these, it is necessary to use other tools and techniques. Indeed, we have already mentioned some of these in our discussions; numerical approaches such as DMRG~\cite{white1992density} play a large role in our understanding of physics away from analytically tractable limits. Studies using the TCSA have been cited throughout the previous sections for providing supporting evidence to non-Abelian bosonization analyses, or interpreting results away from analytically tractable limits. 

Now we turn our attention away from non-Abelian bosonization (although we will use it in some of our discussions) to numerical methods for tackling low-dimensional strongly correlated systems. In the next four sections we will introduce and discuss truncated space approaches for studying low-dimensional field theories. These methods, and their numerical renormalization group extensions, will then be used to study a number of interesting applications, including some of the theories that we have discussed above, such as perturbed WZNW models and the tricritical Ising model.  

\section{Beyond integrability I: truncated space approach (TSA)}
\label{Sec:TSA}

\subsection{Introduction to the approach}
The aim of this section is to present a comprehensive methodology, the truncated space approach, that permits the study of perturbations of integrable and conformal models in one spatial dimension.  While such unperturbed models form a relatively small (i.e. a measure zero) subset of all one-dimensional theories, they are remarkably well distributed throughout `theory' space.  Thus if one is able to study their perturbed variants, one will be able to understand the physics of much of this space.  

The TSA methodology was first developed by V. Yurov and Al. Zamolodchikov in two papers, one treating perturbations of the scaling Yang-Lee model~\cite{yurov1990truncated}, and one treating the critical Ising model perturbed by a magnetic field~\cite{yurov1991truncated}. These initial two papers sparked a sustained period of work on perturbed (both unitary and non-unitary) conformal minimal models where the TSA was used to elucidate a wide variety of the properties of these models: see, for example, Refs.~\cite{delfino1996non,MongPRX14,dorey1998excited,klassen1992spectral,klassen1991relation,pozsgay2008formII,pozsgay2008formI,guida1997vacuum,klassen1992kinks,kausch1997relation,lassig1991finite,takacs1997new,ahn2005nlie,koubek1992scattering,lepori2008particle,martins1991scaling,colomo1992s,martins1991constructing,lencses2014excited,mussardo2009effective,ellem1998thermodynamic,ellem2002excited,lencses2015confinement,toth2007study,lassig1991exact,mossa2008analytic,pozsgay2014exact,pozsgay2015exact,dorey1996excited,bazhanov1997quantum,von1991critical,zamolodchikov2002scaling,delfino2006decay,fioravanti2000universal,lassig1991scaling}. Beyond this work on conformal minimal models, the TSA has also been used to study variants of sine-Gordon models~\cite{feverati1999non,feverati1998truncated,feverati1998scaling,bajnok2004susy,bajnok2002finite,toth2004nonperturbative,feher2011sine,feher2012sine,palmai2013diagonal,takacs2011determining,palmai2015edge,buccheri2014finite,bajnok2001nonperturbative,bajnok2000k,takacs2006double,pozsgay2006characterization,takacs2010form}.  These papers all concern perturbed $c=1$ compact free bosons with the notable exception of Ref. \cite{bajnok2004susy} which considered a perturbation of the $c=3/2$ supersymmetric generalization of sine-Gordon and is the first paper to consider a model where the underlying unperturbed theory had $c>1$.    
It also has been used extensively to study perturbations of conformal theories with boundaries~\cite{dorey1998tba,dorey2000g,dorey2001one,bajnok2001boundary,bajnok2002finite,kormos2010one,kormos2009defect,kormos2008boundary,dorey2000finite,kormos2006boundary,lencses2011breather,bajnok2002spectrum,takacs2012excited,kormos2007boundary}.

The vast majority of the early works using TSA studied perturbations of theories with central charge no greater than one. However, more recently the TSA has been used to study more complicated cases, including multi-boson theories~\cite{konik2015predicting,konik2011exciton} as well as perturbed WZNW theories~\cite{beria2013truncated,AzariaPRD16,konik2015studying,AzariaPRD16}. In part the study of more complicated theories has become possible due to the development of renormalization group techniques, both numerical and analytical, that alleviate the consequences of the truncation in the TSA~\cite{konik2007numerical,feverati2008renormalization,watts2012renormalisation,giokas2011renormalisation,lencses2014excited}. In the past few years the TSA has been used to study a number of non-traditional models and quantities, including Landau-Ginsburg theories~\cite{rychkov2015hamiltonian,coser2014truncated,rychkov2016hamiltonian,bajnok2016truncated,Elias-Miro2016renormalized}, the fractional quantum Hall effect~\cite{MongPRX14,wu2014braiding,wu2015matrix,zaletel2012exact}, entanglement
properties~\cite{palmai2014excited,palmai2016entanglement}, quantum chromodynamics~\cite{katz2014solution,AzariaPRD16}, non-equilibrium dynamics~\cite{rakovszky2016hamiltonian,palmai2015edge,caux2012constructing,brandino2015glimmers}, as well as the properties of non-relativistic continuum field theories (such as the Lieb-Liniger model)~\cite{caux2012constructing,brandino2015glimmers}. Perhaps the most exciting direction for the TSA in recent research is its extension to higher dimensional theories~
\cite{hogervorst2015truncated,katz2016conformal}.

In all of the cases, the basic problem the TSA treats is easy enough to state.  The TSA enables the study of a Hamiltonian of the following form:
\begin{equation}
H = H_{\rm known} + \lambda V_{\rm pert}.
\end{equation}
Here $H_{\rm known}$ is either an integrable or conformal theory, and $V_{\rm pert}$ is some perturbing operator, which need not be of a form that renders the full Hamiltonian $H$ integrable or exactly solvable in any fashion. $H_{\rm known}$ is a ``known'' theory in the sense that we have a complete understanding of its spectrum and matrix elements {\it in finite volume}. 

We will see that the size of the system is a control parameter for the TSA, the varying of which allows us to explore different regimes of the theory, from the deep UV to far IR. In conformal theories, understanding the spectrum in finite volume poses no difficulty~\cite{CFTBook}, whilst in an integrable model understanding the spectrum in infinite volume typically allows one to understand the spectrum in 
finite volume.
However, unlike the conformal case, in integrable theories there is some additional work involved, inasmuch as we have to solve Bethe-type quantization relations that are present for an integrable system in finite volume.

\begin{figure}[h]
\includegraphics[width=0.4\textwidth]{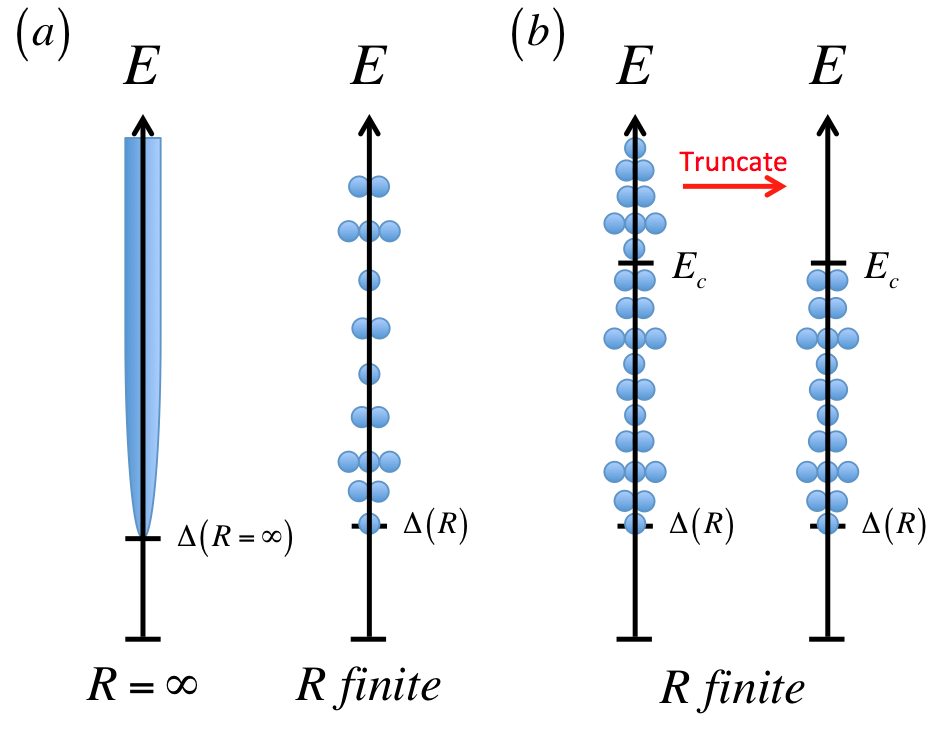}
\caption{(a) A schematic depiction of the spectrum of $H_{\rm known}$ in the infinite volume (left) and the finite volume (right). In the infinite volume, there is a continuum of states, whilst in the finite volume the spectrum is discrete (and possibly with finite degeneracy). (b) A cartoon illustration of the TSA procedure; a cutoff energy $E_c$ is introduced and states in the spectrum of $H_{\rm known}$ above this energy are discarded.}
\label{Fig:Spectrum}
\end{figure}

For the purpose of the TSA, it is important that we understand the spectrum in the finite volume (as opposed to the infinite volume) as here the spectrum is discrete (we will shortly see why this is important). We do note, however, that there are conformal theories that possess a continuous spectra even in finite volume, for example theories involving non-compact bosons. There are ways to treat perturbations of such conformal theories (amounting to correctly handling the bosonic zero mode), but for now we will restrict our attention to theories whose spectrum is discrete for finite volume. We will denote this spectrum by $\{|E_i\rangle\}_{i=1}^{\infty}$, and portray it schematically in Fig.~\ref{Fig:Spectrum}.

Having knowledge of the spectrum, the next ingredient that we require is an understanding of the matrix elements of the perturbing operator relative to unperturbed basis. That is, we need to know 
\begin{equation}
\langle E_i |V_{\rm pert} | E_j\rangle.
\end{equation}
If the theory is a conformal theory, such matrix elements are readily computable. For example, the states $|E_i\rangle$ will (at least) have a representation as a sum of products of the Virasoro generators, $L_{-n}$, acting on some highest weight state $|\Delta,\bar \Delta\rangle$,\footnote{A brief discussion of the Virasoro algebra and its generators is given in Appendix~\ref{App:CFT}; a detailed discussion can be found in, e.g., Ref.~\cite{CFTBook}.}  i.e.
\begin{equation}
|E_i\rangle = \sum_j c_j \prod^{M_j}_{k_j=1} L_{-n_{k_j}}\prod^{\bar M_j}_{\bar k_j=1} L_{-n_{\bar k_{j}}}|\Delta,\bar \Delta\rangle,
\end{equation}
with $n_{k_j},n_{\bar k_{j}}>0$.
As we know how the Virasoro generators $L_{-n_j}$, $L_{-n_{\bar j}}$ commute with the perturbation $V_{\rm pert}$, as well as how they commute with one
another, we are able to compute $\langle E_i |V_{\rm pert} | E_j\rangle$ in principle. In practice, we may need to compute these commutators numerically. For continuum relativistic integrable models, such matrix elements can be computed {\it in infinite volume} via the form factor bootstrap, 
i.e. Ref.~\cite{smirnov1992form}.  Under the bootstrap, they are computable by applying
analyticity constraints based on the two-particle $S$-matrix, crossing symmetry, and
unitarity.  
For states in an integrable model with a relatively small number of particles, the matrix elements take on a tractable form. For matrix elements involving states with many particles, the matrix elements can be formidable and, while analytic expressions are available, they are typically not easily evaluated. There however exceptions -- typically integrable theories with so-called diagonal $S$-matrices have matrix elements that are far more accessible.  Like with the spectrum,
having the matrix elements in infinite volume gives one the ability to write down the matrix
elements in finite volume, although here the path from infinite to finite volume is much more
involved~\cite{pozsgay2008formI,pozsgay2008formII,konik2007numerical}.

Supposing that we have full knowledge of both the unperturbed spectrum and the matrix elements of the perturbing operator, we can represent the full Hamiltonian in matrix form:
\begin{widetext}
\begin{equation}
H = \begin{bmatrix}
E_1 +\lambda\langle E_1|V_{\rm pert}|E_1\rangle  & \lambda\langle E_1|V_{\rm pert}|E_2\rangle  & \lambda\langle E_1|V_{\rm pert}|E_3\rangle  & \dots   \\
\lambda\langle E_2|V_{\rm pert}|E_1\rangle  &  E_2 +\lambda\langle E_2|V_{\rm pert}|E_2\rangle  & \lambda\langle E_2|V_{\rm pert}|E_3\rangle & \dots  \\
\lambda\langle E_3|V_{\rm pert}|E_1\rangle  &  \lambda\langle E_3|V_{\rm pert}|E_2\rangle  & E_3 + \lambda\langle E_3|V_{\rm pert}|E_3\rangle & \dots  \\
\vdots & \vdots & \vdots & \ddots \\
\end{bmatrix}.
\end{equation}
\end{widetext}
As it stands $H$ is an infinite dimensional matrix.  So what to do?  The most crude thing we can imagine doing is simple truncating the space of states in energy.  All states whose unperturbed energy exceeds a cutoff, $E_c$, we toss away, as pictured in Fig.~\ref{Fig:Spectrum}(b). This leaves us with a finite number of states (say $N$) and a truncated Hamiltonian matrix, $H_N$ that is finite:
\begin{widetext}
\begin{equation}
H_N = \begin{bmatrix}
E_1 +\lambda\langle E_1|V_{\rm pert}|E_1\rangle  & \lambda\langle E_1|V_{\rm pert}|E_2\rangle  & \lambda\langle E_1|V_{\rm pert}|E_3\rangle  & \dots & \lambda\langle E_1|V_{\rm pert}|E_N\rangle  \\
\lambda\langle E_2|V_{\rm pert}|E_1\rangle  &  E_2 +\lambda\langle E_2|V_{\rm pert}|E_2\rangle  & \lambda\langle E_2|V_{\rm pert}|E_3\rangle & \dots & \lambda\langle E_2|V_{\rm pert}|E_N\rangle \\
\lambda\langle E_3|V_{\rm pert}|E_1\rangle  &  \lambda\langle E_3|V_{\rm pert}|E_2\rangle  & E_3 + \lambda\langle E_3|V_{\rm pert}|E_3\rangle & \dots & \lambda\langle E_3|V_{\rm pert}|E_N\rangle \\
\vdots & \vdots & \vdots & \ddots \\
\lambda\langle E_N|V_{\rm pert}|E_1\rangle  &  \lambda\langle E_N|V_{\rm pert}|E_2\rangle  & \lambda\langle E_N|V_{\rm pert}|E_3\rangle & \dots & E_N + \lambda\langle E_N|V_{\rm pert}|E_N\rangle \\
\end{bmatrix}
\end{equation}
\end{widetext}
This Hamiltonian we can easily diagonalize (e.g., numerically) and extract the spectrum.

In this crude truncation scheme, we simply ignore the effects of the unperturbed high energy Hilbert space; this works remarkably well for a surprisingly large number of cases! 
We will now consider three of them: i) the continuum limit of the transverse field quantum Ising model perturbed by a longitudinal field; ii) the tricritical Ising model, a conformal minimal model,
perturbed by its energy operator; and iii) a compact free boson perturbed by the cosine of the boson, i.e. the sine-Gordon model. The essential reason why the truncation may not strongly affect the results is found in the relevancy (in the RG sense) of the perturbing operator. A strongly relevant perturbing operator will not strongly mix the low and high energy Hilbert spaces of the unperturbed theory and so the truncation goes unfelt in the low energy sector of the full theory. This is not to say that the states in the low-energy sector are not mixed strongly amongst themselves: indeed, they are. In this procedure we {\it are not doing something akin to perturbation theory!}

\subsection{The transverse field Ising model perturbed by a longitudinal field}
\label{Sec:QuantumIsing}
As the first presented example of the TSA, we will consider the quantum Ising model perturbed by a longitudinal field.  Our presentation is modeled after the treatment of Ref.~\cite{fonseca2003ising}.  Ref.~\cite{fonseca2003ising} represents
the first time that the TSA was used to study perturbations of a massive integrable model.

\begin{figure}
\includegraphics[width=0.4\textwidth]{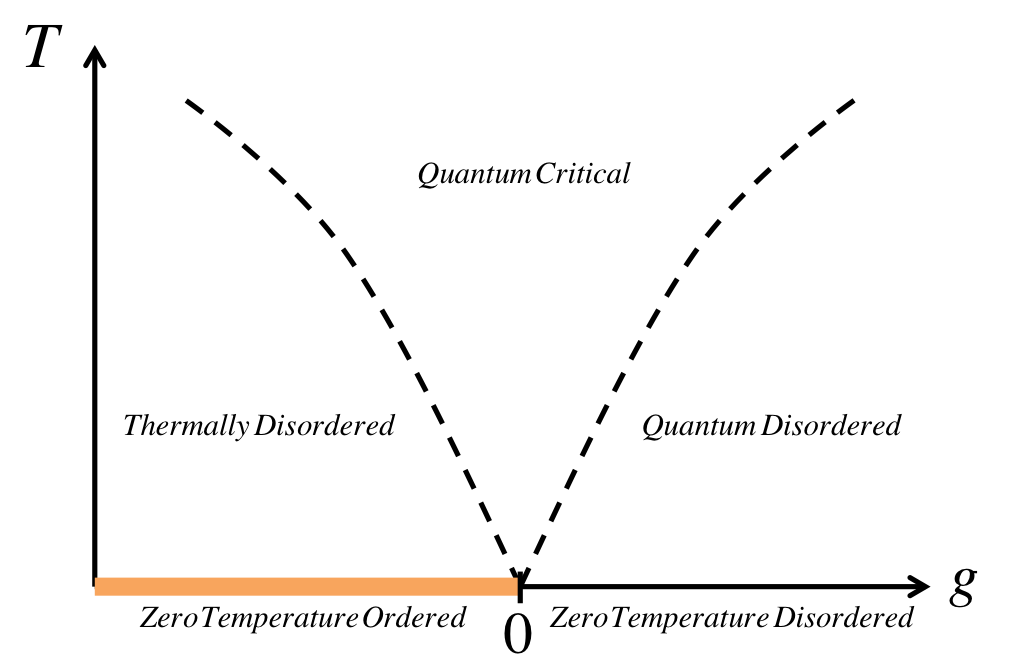}
\caption{Phase diagram of the one-dimensional quantum Ising model with $h=0$. Order is only possible at $T=0$, with $g=0$ separating an ordered phase ($J<0$ ferromagnet, $J>0$ antiferromagnet, shown as a solid orange bar) from a disordered phase.}
\label{fig:Ising_phase_diagram}
\end{figure}

On a one dimensional lattice, this theory has a Hamiltonian given by
\begin{equation}
H = \sum_i \bigg(J\sigma^z_i\sigma^z_{i+1}+J(g+1)\sigma^x_i + h\sigma^z_i\bigg).
\end{equation}
For $h=0$ the theory has the phase diagram pictured in Fig.~\ref{fig:Ising_phase_diagram}.  For $g<0$ the system is in its ordered phase, i.e. at $T=0$, $\sigma^z$ has an expectation value.  For $g>0$, the system is instead in its disordered phase, and $\langle \sigma^z \rangle = 0$.

To apply the TSA to this model, we need to first take its continuum limit. In this limit, the theory is equivalent to a theory of a Majorana fermion (see, for example,~\cite{IsingBook,CFTBook}).   The Hamiltonian then takes the form
\begin{eqnarray}\label{continuumH}
H = iv_F \int^R_0 \rd x &\bigg(&\bar\psi (x)\partial_x \bar\psi(x) - \psi (x)\partial_x\psi(x) \cr\cr
&& + im\bar\psi(x)\psi(x) + h\sigma^z(x)\bigg),
\label{eq:majorana_longitudinal}
\end{eqnarray}
where the various parameters of this continuum Hamiltonian can be expressed in terms of their lattice counterparts via
\be
v_F \sim Ja; \quad m \sim gJ; \quad h \sim ha^{-7/8}.
\ee
Here $a$ is the lattice spacing, $\psi$ and $\bar\psi$ are left and right moving Majorana Fermi fields, and $R$ is the system size. Herein we set $v_F=1$, but it is straightforward to restore it on dimensional grounds in any quantity. We will largely follow the conventions of Ref.~\cite{fonseca2003ising}.

In order to apply the TSA, we need both the spectrum of the model unperturbed by $h\sigma^z$ and matrix elements of the perturbing field, $\sigma^z$, with regards to this basis. First we consider the spectrum.  The spectrum consists of two sectors, Ramond and Neveu-Schwarz. In the Ramond sector, the Fermi fields obey periodic boundary conditions, i.e. 
\be
\psi(x+R) = \psi(x),\quad\bar\psi(x+R) = \bar\psi(x),
\ee
and hence have mode expansions given by 
\begin{eqnarray}
\psi(x) &=& i\sum_{n\in Z} \left(\frac{\epsilon(k_n)-k_n}{R\epsilon(k_n)}\right)^{\frac12}\left(e^{ik_nx}a_{k_n}  -e^{-ik_nx}a^\dagger_{k_n} \right);\nn
\bar\psi(x) &=& \sum_{n\in Z} \bigg(\frac{\epsilon(k_n)+k_n}{R\epsilon(k_n)}\bigg)^{\frac12}\left(e^{ik_nx}a_{k_n}+ e^{-ik_nx}a^\dagger_{k_n}\right),\nonumber
\end{eqnarray}
where $k_n = 2\pi n/R$ and $\epsilon(k) = \sqrt{k^2+m^2}$.  The modes $a,a^\dagger$ satisfy the anti-commutation
relations
\begin{equation}
\big\{a_k,a^\dagger_{k'}\big\} = \delta_{k,k'}. \label{anticomm}
\end{equation}

On the other hand, in the Neveu-Schwarz sector the Fermi fields obey anti-periodic boundary conditions, i.e.
\begin{eqnarray}
\psi(x+R) = -\psi(x),\quad\bar\psi(x+R) = -\bar\psi(x),
\end{eqnarray}
and consequently have mode expansions with half-integer moding:
\begin{eqnarray}
\psi(x) &=& i\sum_{m\in Z+\frac12} \bigg(\frac{\epsilon(q_m)-q_m}{R\epsilon(q_m)}\bigg)^{\frac12}\left(e^{iq_mx}a_{q_m} -e^{-iq_mx}a^\dagger_{q_m}\right),\nn
\bar\psi(x) &=& \sum_{m\in Z+\frac12} \bigg(\frac{\epsilon(q_m)+q_m}{R\epsilon(q_m)}\bigg)^{\frac12}\left( e^{-iq_mx}a^\dagger_{q_m} + e^{iq_mx}a_{q_m}\right). \nonumber
\end{eqnarray}
The half-integer modes $a_q,a^\dagger_q$ satisfy the same anti-commutation relations as their integer moded counterparts, Eq.~\fr{anticomm}.

States spanning the complete Hilbert space can be constructed from the modes. However, the Hilbert space of the model differs depending on whether the Ising chain is in its ordered ($m<0$) or disordered ($m>0$) phase. In the ordered phase, there are two near degenerate vacua (differing up to $e^{-mR}$ corrections), with one lying in the Ramond (${\cal R}$)  sector, $|{\cal R}\rangle$, and one in the Neveu-Schwarz (${\cal NS}$) sector, $|{\cal NS}\rangle$. In the ordered phase, the Hilbert space is spanned by states with an even number of modes built above these two vacua: 
\bea
a^\dagger_{k_1}a^\dagger_{k_2}&\cdots& a^\dagger_{k_{2N}}|{\cal R}\rangle, \qquad k_{n_i} = \frac{2\pi n_i}{R},~~n_i\in Z,\cr\cr
a^\dagger_{q_1}a^\dagger_{q_2}&\cdots& a^\dagger_{q_{2M}}|{\cal NS}\rangle, ~~~q_{m_i} = \frac{2\pi m_i}{R},~~m_i\in Z+\frac{1}{2}. \nonumber
\eea
This is a simple realization of the physical requirement that there must be an even number of domain walls: the fermions represent domain walls between ordered segments of the chain, and in the presence of periodic boundary conditions there must be an even number of domain walls (and hence an even number of fermions). The energy of these states is given by
\bea
E_{{\cal R}}(k_1,\cdots,k_{2N}) &=& \sum_{i=1}^{2N} \epsilon(k_i) + \frac{Rm^2}{8\pi}\log \big(a^2m^2\big),\nn
E_{{\cal NS}}(q_1,\cdots,q_{2M}) &=& \sum_{i=1}^{2M} \epsilon(q_i) + \frac{Rm^2}{8\pi}\log \big(a^2m^2\big), \nonumber
\eea
respectively, where the constant term is Onsager's singularity~\cite{onsager1944crystal}.

In the disordered phase, there is a unique vacuum state that is found in the ${\cal NS}$ sector, $|{\cal NS}\rangle$. As with the ordered phase, excited states in the ${\cal NS}$ sector are built from even numbers of excitations above this vacua. On the other hand, in the ${\cal R}$ sector, states are now built from odd numbers of excitations. As a result, we have the following states in the Hilbert space for the disordered phase: 
\begin{eqnarray}
a^\dagger_{k_1}a^\dagger_{k_2}&\cdots& a^\dagger_{k_{2N+1}}|{\cal R}\rangle, ~~~k_{n_i} = \frac{2\pi n_i}{R},~~n_i\in Z;\cr\cr
a^\dagger_{q_1}a^\dagger_{q_2}&\cdots& a^\dagger_{q_{2M}}|{\cal NS}\rangle, ~~~q_{m_i} = \frac{2\pi m_i}{R},~~m_i\in Z+\frac{1}{2}. \nonumber
\end{eqnarray}
The construction of this state space reflects, in part, the conditions that perturbing by $\sigma^z$ places on the theory. The operator $\sigma^z(x+R)=\sigma^z(x)$ is periodic and connects the two sectors, ${\cal R}$ and ${\cal NS}$, through $\langle {\cal NS}|\sigma^z(0)| {\cal R}\rangle \neq 0$. However, because $\langle {\cal NS}|\sigma^z(0)| {\cal NS}\rangle = \langle {\cal R}|\sigma^z(0)| {\cal R}\rangle = 0$, the half-integer modes of the ${\cal NS}$ sector must only appear in even numbers. In the disordered phase, the ${\cal R}$ sector is inequivalent to the ${\cal NS}$ sector~\cite{IsingBook}, and so it must involve odd numbers of (integer-moded) fermions. Such modes can exist in the disordered phase as the fermions do not correspond to domain walls (as they do in the ordered phase). 

With the spectrum in hand, we now consider the matrix elements of the perturbing operators on these states. The (unperturbed) quantum Ising model is good in this way, as it is one of the few theories where all the matrix elements can be written down explicitly in a rather simple form. We will not discuss how these matrix elements are (analytically) arrived at in detail but merely state them.  The reader can however find
derivations in an appendix of Ref.~\cite{fonseca2003ising} as well as via a lattice formulation of the
problem in Ref.~\cite{bugrii2001correlation}.

The matrix elements take the general form (regardless of phase, and remembering that $\sigma^z$ connects only states in different ${\cal NS}$ and ${\cal R}$ sectors)
\begin{widetext}
\begin{eqnarray}
\langle {\cal NS}| a_{q_l}\cdots a_{q_1}|\sigma^z(0)|a^\dagger_{k_1}\cdots a^\dagger_{k_n}|{\cal R}\rangle = S(R)\prod^l_{j=1}\tilde{g}(\theta_{q_j}) \times \prod^n_{i=1}g(\theta_{k_i})\times
F_{l,n}(\theta_{q_1},\cdots ,\theta_{q_l}|\theta_{k_1},\cdots,\theta_{k_n}),
\label{matrixelsigmaz}
\label{eq:TFIM_ME}
\end{eqnarray}
where the variable $\theta$
\begin{equation}
\theta_{q_i} = \sinh^{-1}\left(\frac{q_i}{\absval{m}}\right), \quad \theta_{k_j} = \sinh^{-1}\left(\frac{k_j}{\absval{m}}\right),
\end{equation}
is a convenient parameterization of the momenta and $g(\theta) = \exp[\kappa(\theta)]/\sqrt{\absval{m}\!R\cosh(\theta)}$ and $\tilde{g}(\theta) = \exp[-\kappa(\theta)]/\sqrt{\absval{m}\!R\cosh(\theta)}$, are normalization factors related to working in finite volume, $R$.  The factor $S(R)$, first derived by Subir Sachdev in Ref.~\cite{sachdev1996universal}, is close to $1$ for $\absval{m}\!R$ much greater than $1$; we will write the expression nonetheless to emphasize the remarkable fact that the matrix elements of $\sigma^z$ are known exactly for any volume $R$: 
\begin{eqnarray}
S(R) = \exp\bigg(\frac{(mR)^2}{2}\int^\infty_{-\infty}\frac{\text{d}\theta_1}{2\pi}\frac{\text{d}\theta_2}{2\pi}
\frac{\sinh\theta_1\sinh\theta_2}{\sinh\big(mR\cosh\theta_1\big)\sinh\big(mR\cosh\theta_2\big)}
\log \left|\coth\bigg(\frac{\theta_1-\theta_2}{2}\bigg)\right|\bigg).
\end{eqnarray}
The factor $\kappa(\theta)$ is given by
\begin{eqnarray}
\kappa(\theta)=\int^\infty_{-\infty}\frac{\text{d}\theta'}{2\pi} \frac{1}{\cosh(\theta-\theta')} \log\left[\frac{1-e^{-\absval{m}\! R \cosh \theta'}}{1+e^{-\absval{m}\! R \cosh \theta'}}\right].
\end{eqnarray}
It now remains to specify the function $F_{l,n}$ that carries the non-trivial dependence on the modes' momenta:
\begin{eqnarray}
F_{l,n}(\theta_{q_1},\cdots ,\theta_{q_l}|\theta_{k_1},\cdots,\theta_{k_n}) = i^{\frac{l+n}{2}}\bar\sigma \prod^n_{i<j}\tanh\bigg(\frac{\theta_{k_i}-\theta_{k_j}}{2}\bigg)
\prod^l_{r<s}\tanh\bigg(\frac{\theta_{q_r}-\theta_{q_s}}{2}\bigg)\prod^n_i\prod^l_r\coth\bigg(\frac{\theta_{q_r}-\theta_{k_i}}{2}\bigg).
\end{eqnarray}
\end{widetext}
Here $\bar\sigma = s\absval{m}^{1/8}$, where $s$ can be given in terms of $A$, Glaisher's constant,\footnote{$A$ itself is related to the Riemann zeta function through $\zeta'(1) = 1/12- \ln (A)$.} by 
$s = 2^{1/12}e^{-1/8}A^{3/2} = 1.35783834\ldots$.

\subsection{The Expected Spectrum: From Mesons to $E_8$}

\begin{figure}
\includegraphics[width=0.45\textwidth]{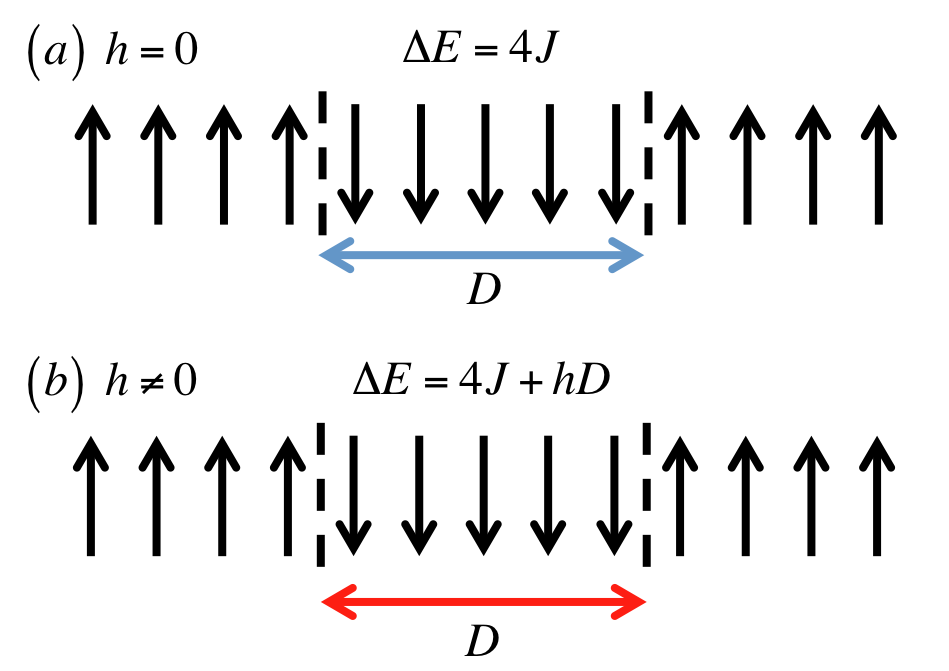}
\caption{(a) A sketch of two domain walls (dashed lines) in an ordered background of the quantum Ising chain.  We see that the spins between the domain walls are overturned relative to the system's overall order. (b) In the presence of a longitudinal magnetic field, the domain walls are linearly confined (energy cost $\Delta E$ grows with domain size $D$). }
\label{domainwalls}
\end{figure}

Before applying the TSA to the quantum Ising Hamiltonian, we first review the spectrum of the model. For the purpose of this discussion, we will restrict ourselves to the ordered phase of the model, i.e., $m \geq 0$. We first consider the unperturbed $h=0$ limit; here the fundamental excitations of the theory are free fermions. In the spin-chain description, these free fermions are domain walls in an ordered background, see Fig.~\ref{domainwalls}.  As we work with periodic boundary conditions, the domain walls necessarily come in pairs, and in the region between the domain walls the spins are overturned. When $h=0$, the size $D$ of this overturned region can be arbitrary. 

However, let us now consider adding $h\neq0$: the region of overturned spins now costs an energy state is proportional to $hD$. As a result, the domain walls become confined (in the two-particle problem, two domain wall fermions experience an interaction potential that grows linearly in their separation). The energies of such confined states were first analyzed by McCoy and Wu in Ref.~\cite{mccoy1978two}.  
The behaviour of the model in the $(h,T)$ was first discussed in the context of a TSA computation
in Ref.~\cite{delfino1996non}.  In the continuum limit (in which we work), the energies of the confined states can be computed using an elegant Bethe-Salpeter analysis, as described in Ref.~\cite{fonseca2003ising}.\footnote{For an extension of this Bethe-Salpeter analysis to other models exhibiting confinement,
such as the three state Potts model, 
see Refs.~\cite{rutkevich2009twokink,rutkevich2015baryon}.  The predictions in these papers were subsequently
verified in Ref.~\cite{lencses2015confinement} using the TSA.}
In this later analysis, the assumed wavefunction of the confined states has only a two-fermion contribution; taking a page from QCD, these states can be thought of as `mesons' and the underlying fermions `quarks'. The energies, $E_i$, of the mesons ($i=1,2,3,\ldots$) are given in terms of the zeroes, $z_i$, of the Airy function, ${\rm Ai}(y)$:
\be
E_i = 2m +mt^2\Big(-z_i + \delta_2 t^2 + \delta_4 t^4 + O(t^6)\Big), \label{BSanalysis}
\ee
where
\bea
t &=& \bigg(\frac{2\bar\sigma^2h}{m^2}\bigg)^{1/3}, \quad \delta_2 = -\frac{\mu z_i^2}{5},\nn
\delta_4 &=& \bigg(\frac{84\mu^2}{350} - \frac{2\mu^2}{25} - \frac{\nu}{7}\bigg)z_i^3 - \bigg(\frac{2\mu^2}{5} - \frac{4\nu}{7} + \frac{\rho}{2}\bigg),\nn
\mu &=& \frac{1}{4};\quad\nu=\frac{1}{8};\quad\rho=\frac{1}{2}. \nonumber
\eea
These energies are computed in the limit that $h \ll m$. This nominally infinite ($i=1,2,3,\ldots$) sequence of energies of the `bound' states must, for $m\neq0$, be understood to terminant. Physically, the reason for this is simple: when a bound state with energy $E_n$ crosses the two-particle threshold (i.e., $E_n > 2E_1$), the bound state has a decay channel (into two lower energy bound states) and hence is unstable and not a true (e.g., long-lived) excitation of the system. 

From Eq.~\fr{BSanalysis}, it is clear to see that as $m\to0$ something special occurs. At $m=0$ and $h\neq0$, the continuum quantum Ising model becomes integrable; it has been shown to have a sequence of non-trivial conserved quantities (beyond those of energy and momentum). Through a remarkable application of the $S$-matrix bootstrap~\cite{zamolodchikov1989integrals}, knowledge of the Lorentz spin of these conserved charges was exploited to deduce the full spectrum of the model with $m=0$. The spectrum consists of eight excitations whose mass ratios correspond to the ratios of the components $S_i$ of the Perron-Frobenius vector of the Cartan matrix of the $E_8$ Lie algebra.\footnote{To then say that the $E_8$ symmetry is not explicitly manifest in the quantum Ising chain is an understatement.} The excitation energies at zero momentum are given by~\cite{zamolodchikov1989integrals}
\begin{eqnarray}
m_1 &=& C h^{8/15};\cr\cr
m_2 &=& \frac{1}{2}\Big(1 + \sqrt{5}\Big)m_1=1.61803\ldots m_1;\nn
m_3 &=& 2\cos\bigg(\frac{\pi}{30}\bigg)m_1=1.98904\ldots m_1;\nn
m_4 &=& 2\cos\bigg(\frac{7\pi}{30}\bigg)m_2=2.40486\ldots m_1;\nn
m_5 &=& 2\cos\bigg(\frac{2\pi}{15}\bigg)m_2=2.95629\ldots m_1;\nn
m_6 &=& 2\cos\bigg(\frac{\pi}{30}\bigg)m_2=3.21834\ldots m_1;\nn
m_7 &=& 4\cos\bigg(\frac{\pi}{5}\bigg)\cos\bigg(\frac{7\pi}{30}\bigg)m_2=3.89115\ldots m_1;\nn
m_8 &=& 4\cos\bigg(\frac{\pi}{5}\bigg)\cos\bigg(\frac{2\pi}{15}\bigg)m_2=4.78338\ldots m_1.\nonumber
\end{eqnarray}
The first energy, $m_1$, gives the fundamental excitation scale for the system. By dimensional analysis, it is a function of $h^{8/15}$ and the dimensionless proportionality constant, $C$, can be determined exactly~\cite{fateev1994exact}:
\begin{eqnarray}
C &=& \frac{4\sin(\frac{\pi}{5})\Gamma(\frac{1}{5})}{\Gamma(\frac{2}{3})\Gamma(\frac{8}{15})}
\bigg(\frac{4\pi^2\Gamma(\frac{3}{4})\Gamma^2(\frac{13}{16})}{\Gamma(\frac{1}{4})\Gamma^2(\frac{3}{16})}\bigg)^{\frac{4}{5}}\cr\cr
&=& 4.40490\ldots.
\end{eqnarray}  
In this integrable limit, there exist {\it stable} excitations with energies that exceed the two-particle threshold, i.e. $m_{4,5,6,7,8}>2m_1$. The stability of these excitations is guaranteed by the existence of the non-trivial conserved quantities.

These excitations have been observed in the quasi-1D Ising spin chain, CoNb$_2$O$_6$~\cite{coldea2010quantum,morris2014hierarchy}. In Ref.~\cite{coldea2010quantum}, the first five stable mesons were observed by measuring the spin-spin response (dynamical structure factor) through inelastic neutron scattering. There, by adjusting an applied transverse magnetic field, the spin chain was able to be tuned towards its critical point (i.e., $m=0$) where the $E_8$ spectrum emerges. Due to both the smallness of certain matrix elements in the $E_8$ Ising theory, as well as the finite value of $m$ and the integrability breaking terms that will necessarily be present in a real material, they were able to clearly observe only the first two of the eight $E_8$ excitations. Nevertheless, as one tunes towards the critical point, the emergence of the golden ratio $m_2/m_1 = (1+\sqrt5)/2$ is clearly observed. In later terahertz spectroscopic measurements~\cite{morris2014hierarchy} carried out at zero transverse field, the increased energy resolution of the technique allowed the observation of all nine stable mesons below the two-particle threshold. 

\subsection{TSA Results}

We now move to discussing the TSA results for this model.  We will walk the reader through different aspects of how the TSA data typically presents itself; we start with the behavior of the ground state energy.

\subsubsection{TSA Raw Data and Behavior of Ground State}

\begin{figure}
\includegraphics[width=0.45\textwidth]{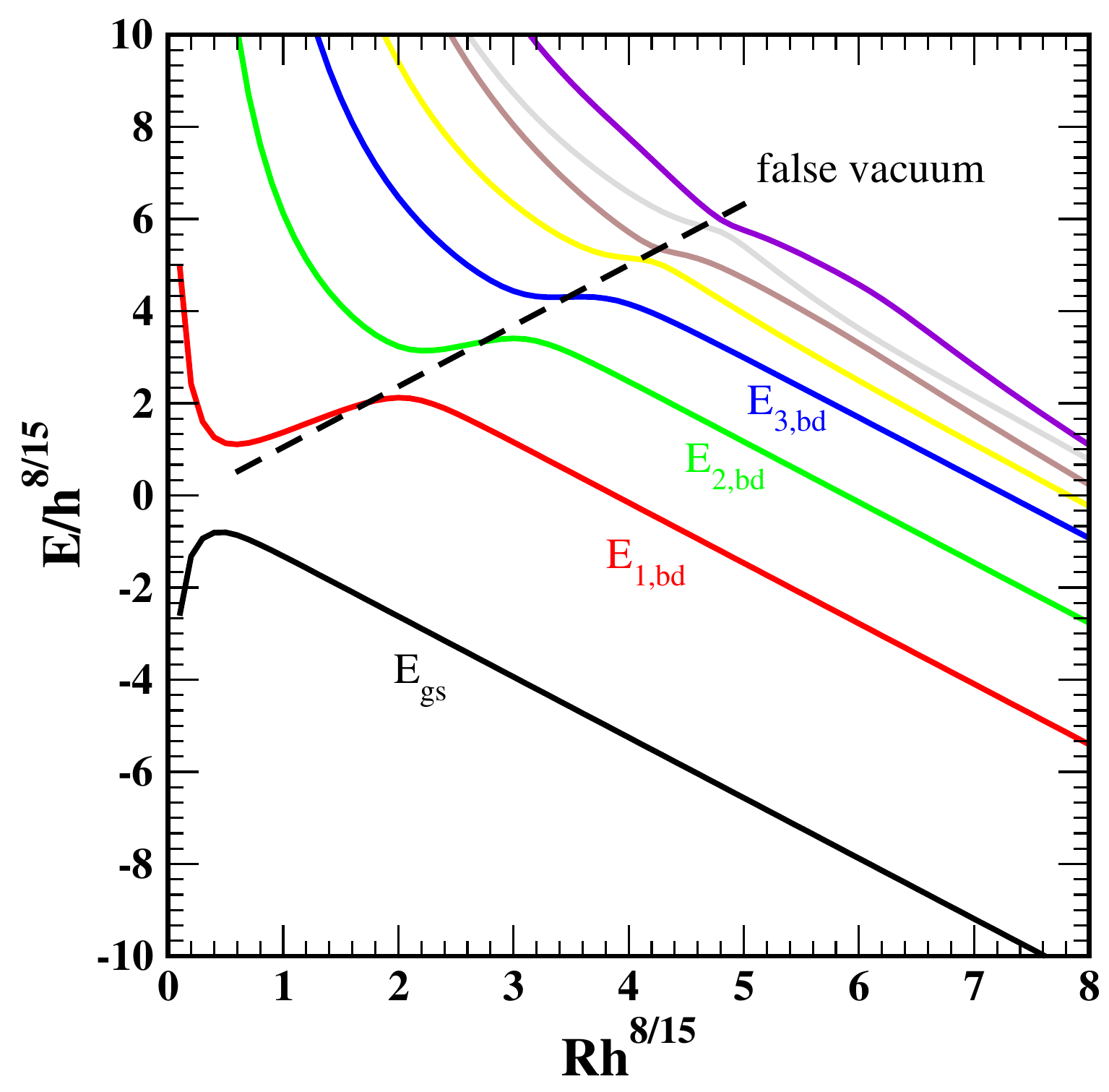}
\caption{Raw TSA data for the lowest lying energy levels of the Hamiltonian in Eq.~\fr{continuumH} for $h=(2m)^{15/8}$, $m=1$, and a cutoff of $N=RE_c/(2\pi)=30$ plotted against the dimensionless system size, $Rh^{8/15}$.  The presented data focuses on the zero-momentum (ground state) sector.  One sees that the energy levels all roughly have a constant negative slope.  The dashed black line that has a positive slope corresponds to a false vacuum state (equal to one of the linear combinations, $|NS\rangle +{\rm sign}(h)|R\rangle$ -- see text).}
\label{TSAfig1}
\end{figure}

In Fig.~\ref{TSAfig1} we plot the lowest lying set of energies coming from the TSA (the parameters chosen for the computation are given in the figure caption).  We see that for sufficiently large system size $R$ all of these energies decrease linearly with increasing $R$ (for smaller values of $R$, the energies evolve into their unperturbed ($h=0$) forms). This linear decrease reflects the negative energy density of the ground state, i.e. 
\be
E_{gs} = -f(m,h) R,
\ee
where $f(m,h)$ can be written as~\cite{fonseca2003ising}:
\begin{equation}
f(m,h) = \frac{m^2}{8\pi}\log(m^2a^2) + m^2\Phi(mh^{-8/15}),
\end{equation}
where $\Phi(\eta)$ is a universal scaling function. In the presence of a magnetic field, $h\gg m^{15/8}$, the dominant contribution to the ground state comes from a particular linear combination of the near degenerate ${\cal R}$ and ${\cal NS}$ vacua,
\be
|E_{gs}\rangle \sim |{\cal NS}\rangle -{\rm sign}(h) |{\cal R}\rangle + \ldots, 
\ee
where the ellipses denote states with finite fermion number. The ground state energy $E_{gs}$ then reduces to 
\be
E_{gs}\sim -2\bar\sigma m^{1/8}h R.
\ee
That is, the ground state represents spins aligned in a direction anti-parallel to the applied magnetic field. 

However, what of the state with its spins parallel to the applied field? In the infinite volume ($R=\infty$), this state would have infinite energy and hence would not exist in the theory. In the finite volume (where we work when using the TSA), however, this state indeed exists. It has finite positive energy, $2\bar\sigma m^{1/8}hR \equiv -E_{gs}$ (at least for $h \gg m^{15/8}$) and in terms of the unperturbed basis is given roughly by 
\begin{eqnarray}
|E_{\rm false~vac.}\rangle &\sim& |{\cal NS}\rangle + {\rm sign}(h) |{\cal R}\rangle.
\end{eqnarray}
The presence of this false vacuum state in the TSA data, Fig.~\ref{TSAfig1}, can be inferred by regions where the energy of a particular state increases with system size $R$ (say the second excited state between $R=1$ and $R=2$). As we are typically interested in low-energy excitations about the true vacuum, we have to be sure to note mistake a state that is the false vacuum (or an excitation about the false vacuum) for one that is of direct interest. This is always an issue in models where the unperturbed Hamiltonian has a discrete spontaneous (near-) symmetry breaking where the perturbation explicitly breaks this symmetry. 

\subsubsection{Behavior of Excited States}
\begin{figure}
\includegraphics[width=0.45\textwidth]{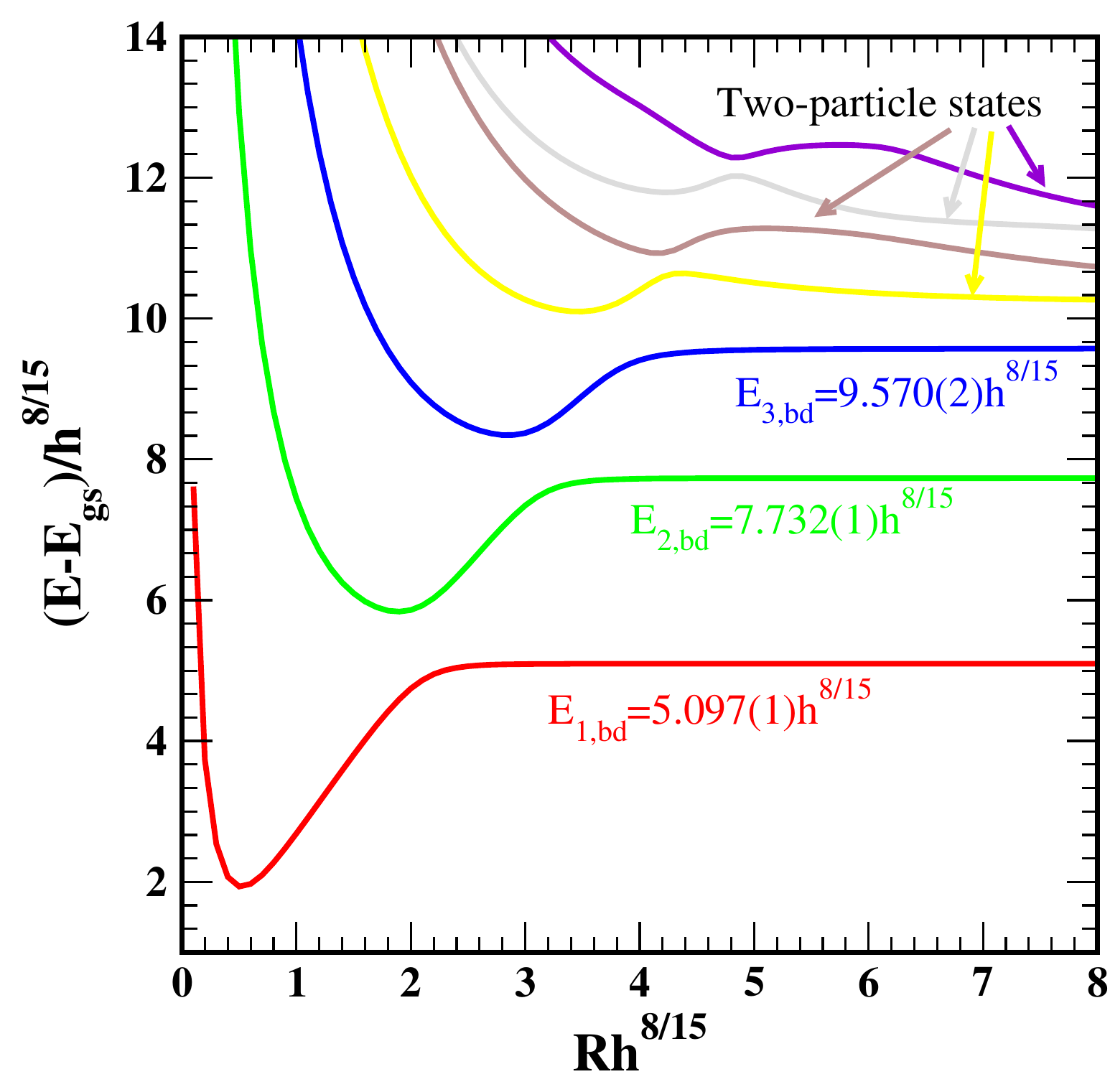}
\caption{The same data as presented in Fig.~\ref{TSAfig1} with the ground state energy subtracted.}
\label{TSAfig2}
\end{figure}

We now consider the behavior of the excited states.  In Fig.~\ref{TSAfig2} we plot the excited state energies relative to the ground state energy.  We now see clearly that there are regions in the finite volume $R$ where the lowest excited states are unchanging. This region in $R$ is the region in which we want to work within the TSA. We furthermore see that the energies can be determined with relatively high precision (with small errors in the fourth significant digit). We note that the data presented here is taken outside the region of validity of the Bethe-Salpeter analysis~\fr{BSanalysis} (as $h=(2m)^{15/8}$). In the final subsection we will, however, make a more detailed comparison between the predicted energies from Bethe-Salpeter and those of the TSA analysis.

One point to stress here is that different excitations have different ``stability regions'': the first bound state, $E_{1,\rm bd}$, becomes stable after the dimensionless system size $Rh^{8/15}$ exceeds $2.4$, while the third meson state $E_{3,\rm bd}$ stabilizes when $Rh^{8/15}>4.5$. For the particular choice of $m$ and $h$ presented, only the first three mesons are stable.  States higher in energy coming from the TSA represent multi-meson states (marked as two-particle states in Fig.~\ref{TSAfig2}). We will discuss the behavior of two-particle states in what is to come -- however, roughly speaking, their energy as a function of $R$ should behave as
\begin{equation}\label{2particleEn}
E_{\rm 2-particle~state} = E_{i,\rm bd} + E_{j,\rm bd} + \frac{\alpha}{R^2},
\end{equation}
that is, its energy should be the sum of the energies of two different bound states plus a term going as $1/R^2$ that indicates the two mesons may have finite (and opposite) momentum. This will be true for sufficiently large system sizes $R$; at smaller values of $R$ we can see regions in Fig.~\ref{TSAfig2} where the energy is constant. These regions represent finite volume resonances in the model that correspond to metastable mesons. The existence of such states in certain regions of $R$ again requires the TSA data to be treated with interpretational care (see, for example, Ref.~\cite{pozsgay2006characterization}).

\subsubsection{Cutoff Dependence of Excited State Energies}

\begin{figure}
\includegraphics[width=0.45\textwidth]{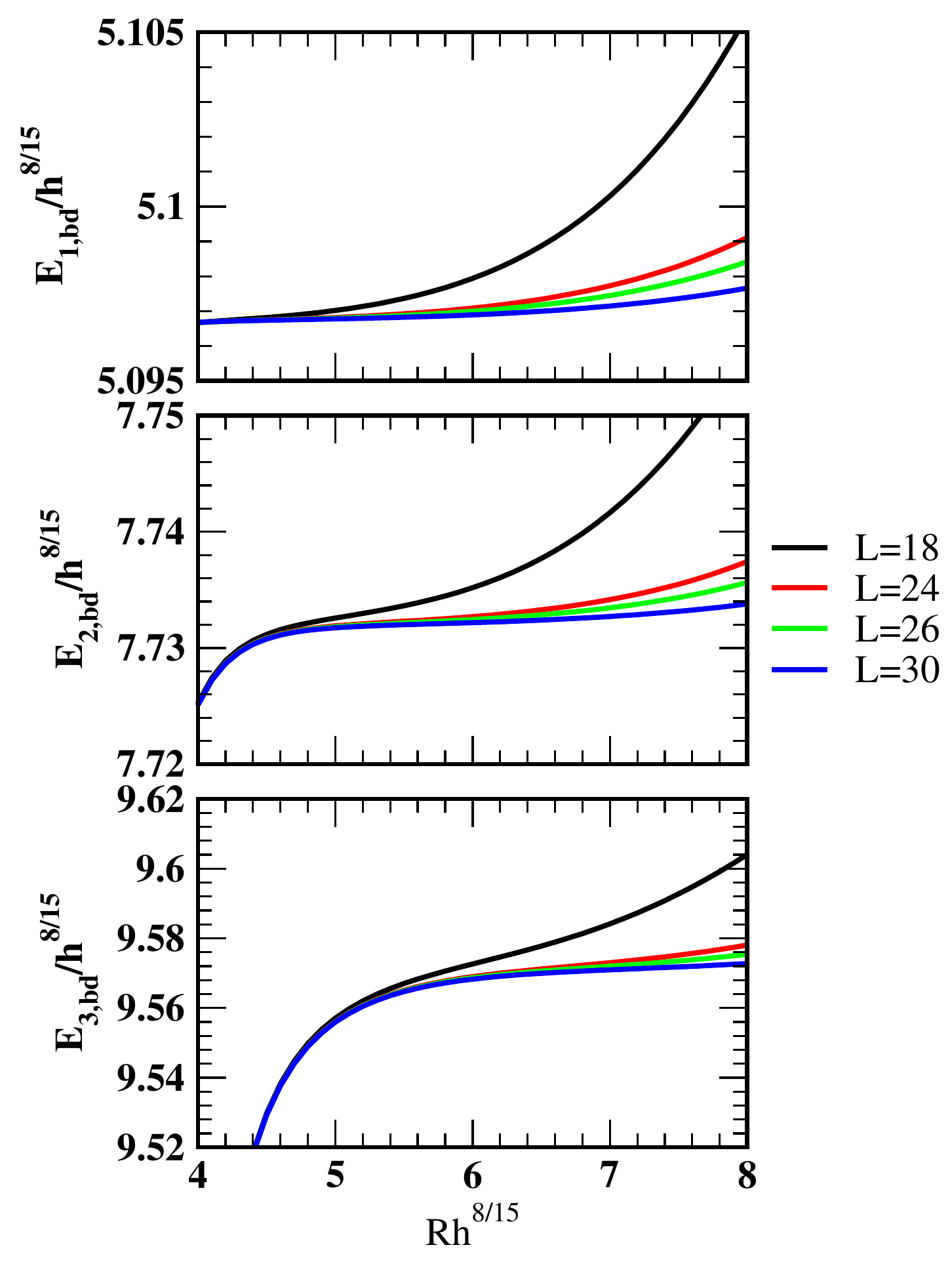}
\caption{Energies for the first, second, and third (upper to lower panels) vs. the dimensionless system size $Rh^{8/15}$ for four different values of the cutoff $L\equiv N=E_cR/(2\pi)$.}
\label{TSAfig3}
\end{figure}

One of the key aspects in analyzing TSA data is to understand the effects of the cutoff. Typically we work at fixed dimensionless cutoff $N=E_cR/(2\pi)$.  But this means that as $R$ increases the cutoff $E_c$ is decreasing.  Thus, at large $R$ we expect to see deviations in the data from results based on the absence of a cutoff.  We make a study of such deviations in Fig.~\ref{TSAfig3} for the first three mesons excitation energies.

In Fig.~\ref{TSAfig3} we plot the meson energies for four different values of the dimensionless cutoff. In each case, the cutoff effects are manifest at large $R$ through upward deviations in the energies.  Cutoff effects are more pronounced for states higher in energy; for the first meson, the lowest cutoff, $N=18$, leads to a $0.17\%$ error in the meson energy at $Rh^{8/15}=8$.  However the third meson at this same value of $R$ has an error of $0.29\%$.  More crucially, one can see that at any given cutoff the first meson has a wider region where its energy is (almost) independent of $R$ than the third meson. Even at the highest value of the cutoff $N=30$ employed in the TSA, its energy does not see a true plateau in $R$.

In general, this demonstrates a general need to account for the effects of cutoffs in TSA data.  Reducing these effects forms the major thrust of Sec.~\ref{Sec:TSA_Cutoff}.

\subsubsection{Evolution of Spectrum with $m$}

\begin{figure}
\includegraphics[width=0.45\textwidth]{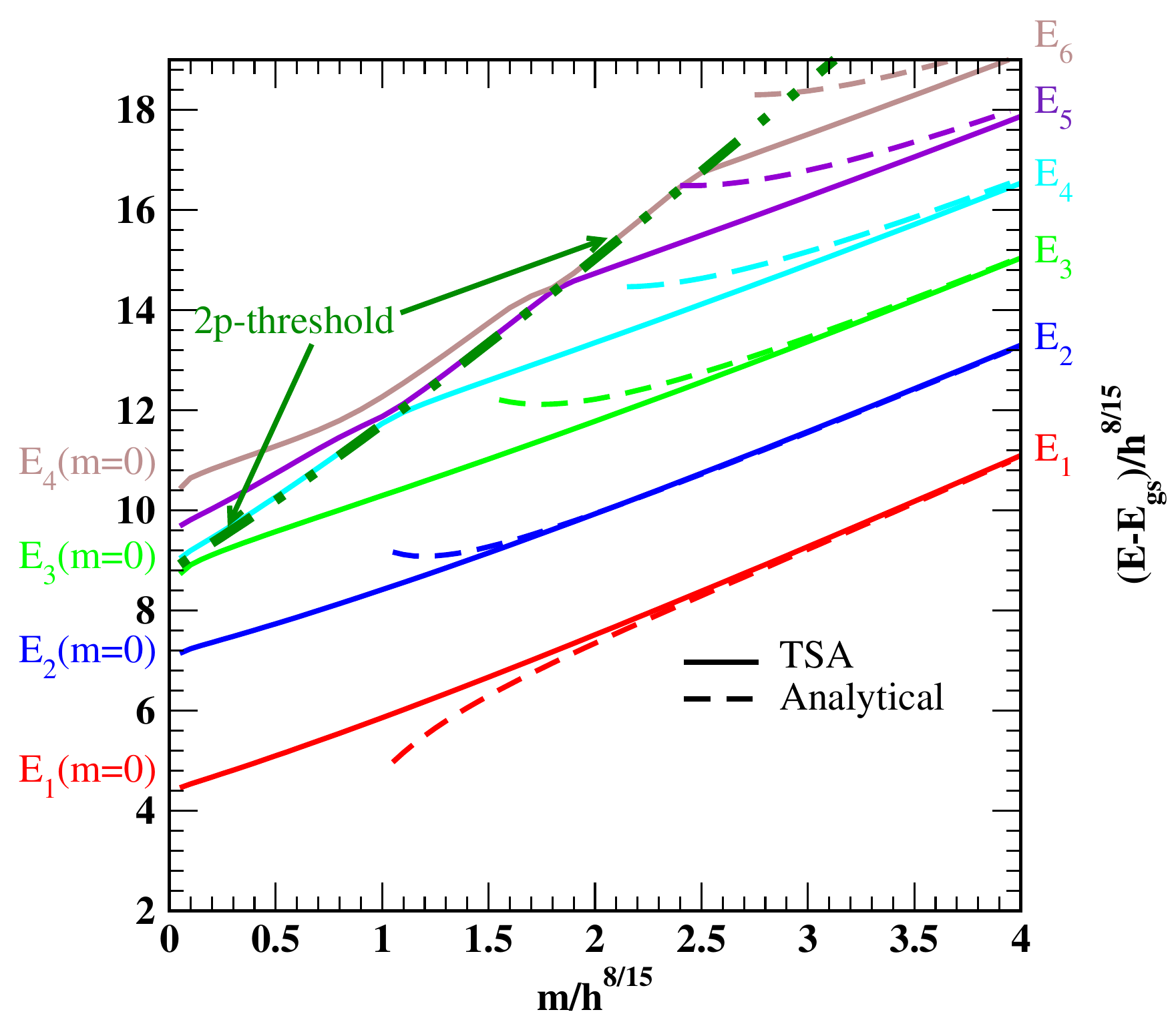}
\caption{The bound state energies as a function of the original fermion mass (scaled to be dimensionless). Energies evolve from their value in the integrable E8 limit (at $m=0$) to their values as predicted by the Bethe-Salpeter analysis, Eq.~\fr{BSanalysis}, as given by the dashed lines.}
\label{TSAfig4}
\end{figure}

The final piece of analysis that we will perform for the quantum Ising model in the presence of a longitudinal magnetic field is the evolution of the spectrum as one increases $m$, the mass of the unperturbed fermions, from $0$.  As we have explained, at $m=0$ the full Hamiltonian is integrable and has a spectrum consisting of eight stable excitations whose energies are related to the $E_8$ Lie algebra.  At finite $m$, integrability is broken and we expect the $E_8$ spectrum to evolve into one composed of bound domain walls, the mesons.  We can observe this evolution in Fig.~\ref{TSAfig4}.

Figure~\ref{TSAfig4} presents the six lowest lying energies as a function of the dimensionless scaling parameter, $\eta = m/h^{8/15}$.  At $\eta=4$, we see that all six excitations can be treated as meson bound states: we have plotted the energies of mesons as predicted in Eq.~\fr{BSanalysis} and we see that we obtain good agreement.\footnote{This is to be expected: large $\eta$ corresponds to small $h$, where the Bethe-Salpeter analysis of Ref.~\cite{fonseca2003ising} is expected to be most robust.} We know that these states are indeed stable mesons because they lie below the two-particle threshold (dashed-dotted green line in Fig.~\ref{TSAfig4}).  As $\eta$ decreases the two-particle threshold decreases more rapidly than the meson energies with the result that certain mesons cease to meet the stability criterion of being below threshold\footnote{For a comprehensive analysis of the decay of mesons that find themselves
above the two-particle threshold, see Ref.~\cite{delfino2006decay}.}. Once above, they become two-particle states; as we have indicated above, this evolution is complicated by working at finite $R$.  A meson state that finds itself above the two-particle threshold only becomes a two-particle state at sufficiently large $R$, and there may be regions in $R$ where the meson is metastable.

By the time we reach $\eta$ small but still finite, only three of the mesons remain below the two-particle threshold.  These three mesons are the first three excitations of the $E_8$ spectrum. Above the two-particle threshold, the first two mesons have evolved into two-particle (e.g., two-meson) excitations, whilst the third excitation above the two-particle threshold corresponds to the four excitation, $m_4$ of the $E_8$ spectrum.

One might be wondering if it is possible to see all eight of the excitations of the $E_8$ spectrum using the TSA. Indeed, one can, however the massive basis that we have employed here is suboptimal for doing so as the expression for the matrix elements of the spin operator~\fr{matrixelsigmaz} requires one to work at finite $m$ (that is, with finite integrability breaking).  It is possible to instead work with a massless basis from the start, as was done in Ref.~\cite{yurov1991truncated,delfino1996non}. In the massless limit, cutoff effects are remarkably small, and using a basis of just $39$ states (a cutoff of $N=10$), estimates good to a few percent of the $E_8$ mass spectrum for the first five excitations were obtained.  With a massless basis, using the cutoffs that the data
in this section were computed under (up to $N=30$), all eight excitations are readily found. The massless basis is, however, less intuitive and closed form expressions for general matrix elements are not available, unlike the massive case. Nevertheless, in the next two examples of applying the TSA, we will focus on perturbations about a massless conformal field theory.

\subsection{Tricritical Ising perturbed by the energy operator}

We now consider applying the TSA to a conformal field theory perturbed by a relevant operator. This class of problems form the widest range of problems studied using the TSA.\footnote{Indeed, it is often referred to as truncated conformal space approach (TCSA) in this setting.} We will focus on a particular subclass of such theories here: perturbed conformal minimal models (see, e.g., Ref.~\cite{CFTBook,MussardoBook} for further information about minimal models). Specifically, we will consider a moderately non-trivial example: the tricritical Ising model perturbed by its leading energy operator $\epsilon$: 
\begin{equation}
H = H_{\rm tricritical~Ising} + g_2\int^R_0 \rd x \epsilon (x).
\end{equation}
All of the TSA results presented in this section are computed using the TruSpace code~\cite{TruSpace}, 
developed in part by one of the authors. TruSpace is able to study generic relevant perturbations of generic 
conformal minimal models.  Making this code publicly available recalls the practice of
G. Mussardo and M. Lassig, 
two of the first scientists to employ
the TSA following its introduction by V. Yurov and Al. Zamolodchikov,
making their code \cite{lassig1991hilbert} available to the community.

The tricritical Ising model and its perturbations were one of the earliest targets of the TSA~\cite{lassig1991scaling}, since its introduction by Yurov and Zamolodchikov. There have been extensive follow-on studies where the TSA was used to elucidate various aspects of the tricritical Ising model~\cite{lassig1991scaling,fioravanti2000universal1,fioravanti2000universal,lepori2008particle,mossa2008analytic}. Furthermore, it is a good example for describing the capabilities of the TSA because some of its perturbations lead to integrable models whose properties have also been well studied~\cite{mussardo1987ramond,christe1990integrable,fendley1990second,henkel1990mass,zamolodchikov1991tricritical,zamolodchikov1991thermodynamic,colomo1992s,acerbi1996form,guida1998tricritical,ellem1998thermodynamic,fioravanti2000universal1,fioravanti2000universal}.

\subsubsection{Overview of the tricritical Ising model}

\begin{figure}
\includegraphics[width=0.4\textwidth]{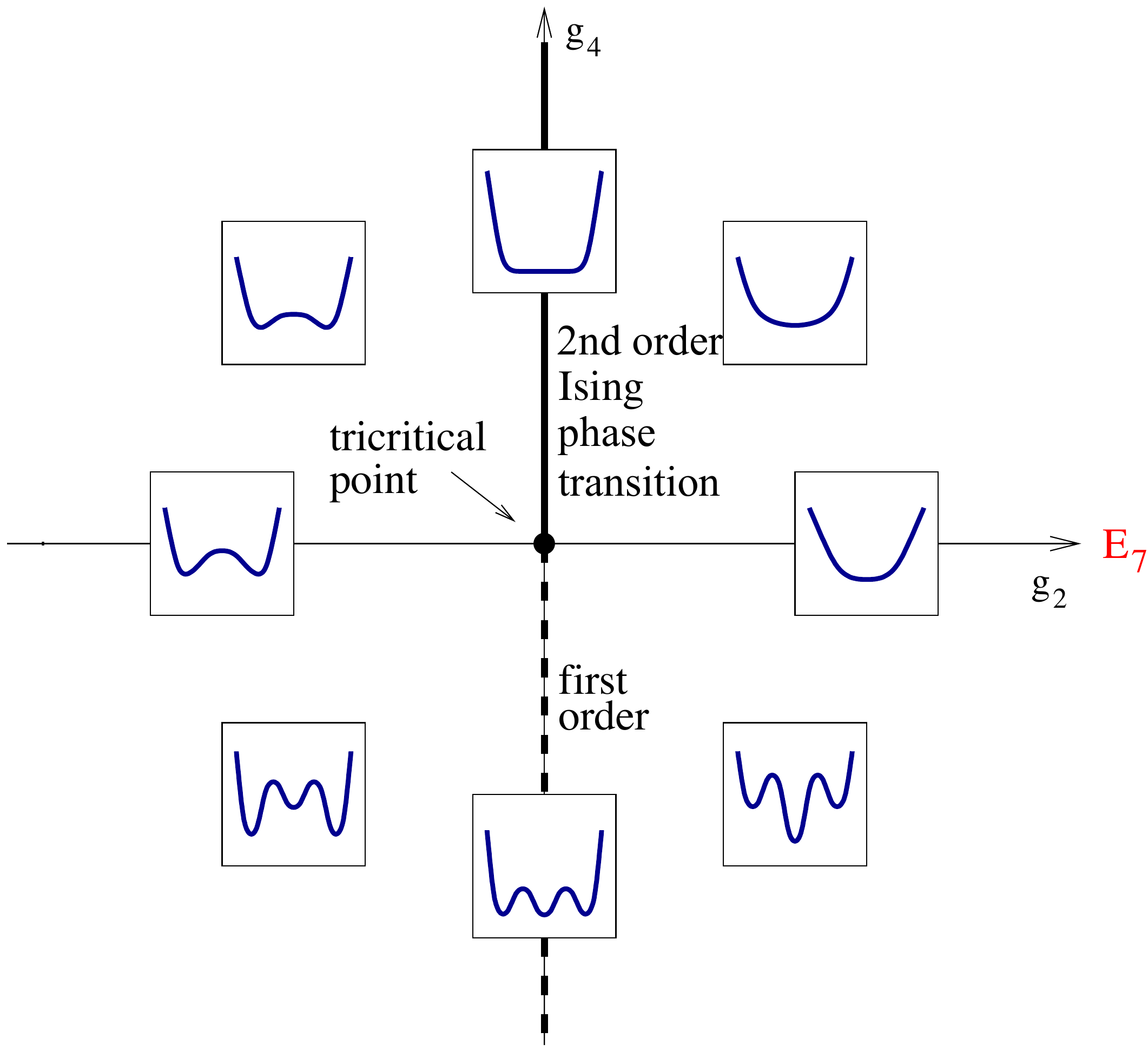}
\caption{The phase diagram in the $g_2-g_4$ plane for the Landau-Ginzburg representation of the tricritical Ising model including the vacuum structure of the theory in the different quadrants.  
Crossing the positive $g_4$ axis (bold solid line) leads to a second order phase transition (in the Ising universality class) between an ordered and disordered phase.  Crossing the negative $g_4$ axis (dashed line) leads to a first order phase transition. The model has a hidden $E_7$ symmetry along the line $g_4=0$. The tricritical point of the theory is found at $g_2=g_4=0$.  This figure is adapted from Ref.~\cite{lepori2008particle}.}
\label{phasediagram}
\end{figure}

The tricritical Ising model has a number of realizations.  It can be written as a two-dimensional classical statistical mechanics model of an (classical) Ising model with vacancies.  Here the Hamiltonian is 
\begin{equation}\label{classical2d}
H = -\sum_{\langle ij\rangle}(K+\sigma_i\sigma_j)t_it_j - \mu\sum_i t_i;
\end{equation}
where $\sigma_i = \pm1$ are the standard Ising variables at site~$i$ of a two-dimensional square lattice and $t_i=0,1$ indicates whether the site is vacant ($t_i=0$) or not ($t_i=1)$. The energy of a pair of nearest-neighbor aligned spins is $K+1$ while those that are anti-aligned have energy $K-1$. $\mu$ is a chemical potential which determines the number of vacancies in the system.  As a function of the three parameters $(\beta,K,\mu)$ this model is known to have a tricritical point where a line of second order phase transitions terminates \cite{blume1971ising}.

Another useful representation of the tricritical Ising model is the $\Phi^6$ Landau-Ginzburg (LG) theory with an action of the form~\cite{zamolodchikov1986conformal}
\begin{equation}
S = \int \rd^2x \bigg(\frac{1}{8\pi}\partial^\mu\phi\partial_\mu\phi + g_2\phi^2 + g_4\phi^4 + g_6\phi^6\bigg).
\end{equation}
This representation is useful inasmuch as one can readily understand the vacuum structure of the theory and hence the possible excitations, including those that are kink-like.  At the tricritical point, $g_2=g_4=0$. In our application of the TSA to this model, we are going to be interested in turning on a finite $g_2$, in part because this perturbation is integrable -- hence we will present analytical results to which we will compare the TSA analysis. However, in Fig.~\ref{phasediagram} we present all different the phases in the $g_2-g_4$ plane and their associated vacuum structures in the LG framework.

Finally, there is also a representation as a quantum spin chain, in terms of the spin-1 Blume-Capel model~\cite{gefen1981phase,alcaraz1985hamiltonian,balbao1987operator,von1990off}.  
The spin chain has the Hamiltonian
\begin{equation}\label{spinchain}
H = \xi\sum^L_{i=1}\bigg(\alpha(S_i^x)^2 + \beta S^z_i +\gamma (S^z_i)^2 - S^x_i S^x_{i+1}\bigg),
\end{equation}
where $S^z_i$ and $S^x_i$ are spin-one operators acting on site $i$ of an $L$ site lattice:
\begin{eqnarray}
S^z_i &=& \begin{pmatrix} 
1 & 0 & 0 \\
0 & 0 & 0 \\
0 & 0 & -1 \\
\end{pmatrix};
S^x_i = \frac{1}{\sqrt{2}}\begin{pmatrix} 
0 & 1 & 0 \\
1 & 0 & 1 \\
0 & 1 & 0 \\
\end{pmatrix}.
\end{eqnarray}
The spin chain can be tuned to its critical point by setting~\cite{von1990off,alcaraz1985hamiltonian,von1986conformal}
\be
\alpha = 0.910207(4);~\beta=0.415685(6),~\xi^{-1}=0.56557(50).
\ee
For this choice of normalization of the Hamiltonian, the level spacing in a given Verma model is `conformal-like', given by $2\pi/L$. 

This model has a $\mathbb{Z}_2$ symmetry with an associated charge, $Q$:
\begin{eqnarray}
Q &=& \sum_i \begin{pmatrix} 
0 & 0 & 0 \\
0 & 1 & 0 \\
0 & 0 & 0 \\
\end{pmatrix}_i.
\end{eqnarray}
Under such a $\mathbb{Z}_2$ transformation $S^x_i$ is an odd operator, i.e. $e^{i\pi Q}S^x_ie^{-i\pi Q}=-S^x_i$.  The spectrum of the model correspondingly has an even and an odd sector. The action of the $\mathbb{Z}_2$ symmetry in the classical 2D lattice model is to take $\sigma_i \rightarrow -\sigma_i$, while in the LG representation it is $\phi \rightarrow -\phi$.

\subsubsection{Conformal field theory description of the tricritical Ising model}

The CFT that corresponds to the tricritical Ising model is the second in the series of unitary conformal minimal models (it is the first in the sequence of unitary minimal models with $N=1$ supersymmetry). It has central charge $c=7/10$ and has six scaling operators. Four of the operators are even under $Q$ while the remaining two are odd.  The four even operators include the identity operator, $I$, and three ``energy-like'' operators, $\epsilon,t,\epsilon''$. In the classical statistical mechanics picture, these three operators correspond to the three different terms in the Hamiltonian~\fr{classical2d}; in the LG picture $\epsilon=\phi^2$, $t=\phi^4$, and $\epsilon''=\phi^6$. Alternatively, in the quantum spin chain representation, we have direct expressions for $\epsilon$ and $t$ in terms of the spin variables~\cite{von1990off}
\begin{eqnarray}
\epsilon &=& \sum^L_{i=1} \Big(\sin(\theta) (S^x_i)^2+\cos(\theta)S^z_i \Big);\cr\cr
t &=& \sum^L_{i=1} \Big(-\cos(\theta) (S^x_i)^2 - \sin(\theta)S^z_i\Big) ,
\end{eqnarray}
with $\theta = \tan^{-1}(2.224)$. The two odd operators, $\sigma$ and $\sigma'$,  are related to the $\sigma_i$ degrees of freedom in the classical lattice picture while in the LG formulation they are given by $\sigma = \phi$ and $\sigma'=\phi^3$. The scaling dimensions of these operators, their action under $Q$, and their representation in the LG formalism are summarized in Table~\ref{table_tc_ops}.
\begin{table}\label{table_tc_ops}
\centering
\begin{tabular}{|c |c |c |c |}
\hline\hline
CFT  & scaling & LG & action \\
operator & dimension, $\Delta+\bar \Delta$ & representation & under Q\\
\hline\hline
$I$ & 0 & $I$ & even \\
$\epsilon$ & 1/5  & $\phi^2$ & even \\
$\epsilon''$ & 3 & $\phi^6$ & even \\
$t$ & 6/5 & $\phi^4$ & even \\
$\sigma$ & 3/40 & $\phi$ & odd \\
$\sigma'$ & 7/8 & $\phi^3$ & odd \\
\hline 
\hline
\end{tabular}
\caption{The six scaling operators of the tricritical Ising theory together with their scaling dimensions, their Landau-Ginzburg (LG) representation, and their action under $Q$, the $\mathbb{Z}_2$ symmetry transformation. $I$ is the identity operator, $\epsilon$ is the leading energy operator, $\epsilon''$ is the sub-leading energy operator, $t$ is the vacancy density, $\sigma$ is the spin operator (proportional to $S^x$ on the lattice), and finally $\sigma'$ is the sub-leading spin operator.}
\end{table}

Knowledge of the scaling operators is key to being able to writing down the unperturbed spectrum of the theory, one of the two requirements for applying the TSA to a model. For every scaling operator there is a highest weight state $|\Delta,\bar \Delta\rangle$ where the total scaling dimension of the operator is given by $\Delta +\bar \Delta$. The highest weight states are formed (on the plane) by the action of the operator field at $z=0$ on the vacuum\footnote{This can also be pictured as a state at time $t=-\infty$ by applying the conformal transformation that maps the plane to a cylinder.}
\begin{equation}
|\Delta_i,\bar\Delta_i\rangle \equiv \phi^{\rm plane}_{\Delta_i,\bar\Delta_i}(0)|0\rangle .
\end{equation}
The full Hilbert space is then spanned by the list of states formed by the Virasoro generators acting on the highest weight states:
\begin{equation}\label{basicstates}
\prod^{M}_{j=1} L_{-n_j}\prod^{\bar M}_{\bar j=1}\bar L_{-n_{\bar j}}|\Delta_i,\bar\Delta_i\rangle,
\end{equation}
where $i=1,\ldots,6$ label the six scaling operators, and $n_{j},n_{\bar j} > 0$. The energy and momentum of such a state is given by 
\begin{eqnarray}
E &=& \frac{2\pi}{R}\Bigg(c+ \Delta_i +\bar\Delta_i+ \sum^{M}_{j=1} n_j + \sum^{\bar M}_{\bar j=1} n_{\bar j}\Bigg), \\
P &=& \sum^{M}_{j=1} n_j - \sum^{\bar M}_{\bar j=1} n_{\bar j},
\end{eqnarray}
where here the central charge is $c=7/10$. Due to the translational invariance of the Hamiltonians in which we are interested, we can perform the TSA computations in subsectors with fixed values of the momentum $P$. 

One technical, but important and unfortunate, point is that not all states of the form~\fr{basicstates} are linearly independent. One has to remove so-called ``null states'' from this list. This is most easily done numerically through computing the Gram matrix (the matrix of the overlaps of such states), diagonalizing it, and dropping linear combinations of such states which have zero eigenvalues. This leaves one with a set of states formed from linear combinations of states of the form~\fr{basicstates} that are orthonormal and form a complete basis. This procedure is numerically implemented in the TruSpace code~\cite{TruSpace}.  We do note however that Ref.~\cite{alba2011on} has suggested a means to generate a complete basis analytically.

The next requirement for applying the TSA to perturbations of the tricritical Ising model is the ability to compute matrix elements of the perturbing field. Thus we are forced to compute matrix elements of the perturbing field between two states $|1\rangle$ and $|2\rangle$ of the form found in Eq.~\fr{basicstates}:
\begin{eqnarray}
\int^R_0 \rd x \langle 1|\Phi_{\rm pert}(x)|2\rangle &=& \delta_{P_1,P_2}R \langle 1|\Phi_{\rm pert}(0)|2\rangle. \quad
\end{eqnarray}
The integral over space enforces that the momenta, $P_1, P_2$, of the two states be equal in order for the matrix element to be non-zero. To compute this matrix element, two ingredients are needed: (i) the commutation relations of the Virasoro modes,
\begin{eqnarray}
[L_n,L_m] &=& (n-m)L_{n+m} + n(n^2-1)\frac{c}{12}\delta_{n+m,0};\cr\cr
[\bar L_n,\bar L_m] &=& (n-m)\bar L_{n+m} + n(n^2-1)\frac{c}{12}\delta_{n+m,0};\cr\cr
[\bar L_n, L_{m}] &=& 0, \nonumber
\end{eqnarray}
as well as (ii) the commutation relationship of the Virasoro modes with the perturbing field itself:
\begin{eqnarray}
[L_n-L_0,\Phi_{\rm pert}(0)] &=& n\Delta_{\Phi_{\rm pert}}\Phi_{\rm pert},\cr\cr
[\bar L_n-\bar L_0,\Phi_{\rm pert}(0)] &=& n\Delta_{\Phi_{\rm pert}}\Phi_{\rm pert},\nonumber
\end{eqnarray}
We see that this latter commutation relationship with $\Phi_{\rm pert}$ is completely determined by the field's scaling dimension, $\Delta_{\Phi_{\rm pert}}$.

With these two ingredients in hand, all the matrix elements can be reduced to those involving two highest weight states with the perturbing field. These matrix elements are no more than the structure constants of the theory associated with the three point functions, i.e.
\begin{equation}
\langle \Delta_1,\bar\Delta_1 | \Phi_{\rm pert}(0)|\Delta_2,\bar\Delta_2\rangle \equiv \bigg(\frac{2\pi}{R}\bigg)^{\Delta_{\Phi_{\rm pert}}+\bar\Delta_{\Phi_{\rm pert}}}C_{\Phi_1\Phi_{\rm pert}\Phi_2},
\end{equation}
where the associated three point function on the plane is given by
\bw
\begin{eqnarray}
\langle \Phi_{1}(x) \Phi_{\Delta_{\rm pert}}(y) \Phi_{2}(z)\rangle &=&  \frac{C_{\Phi_1\Phi_{\rm pert}\Phi_2}}{|x-y|^{2\Delta_1+2\Delta_{\rm pert}-2\Delta_2} |x-z|^{2\Delta_1-2\Delta_{\rm pert}+2\Delta_2}|y-z|^{-2\Delta_1+2\Delta_{\rm pert}+2\Delta_2}},
\end{eqnarray}
\ew
where here we have assumed that $\Delta=\bar\Delta$ in all cases.
For conformal minimal models the structure constants are in general available~\cite{dotsenko1985four}. In Table~\ref{struct_const} we explicitly list the structure constants for the perturbation that we are interested in, the leading energy operator $\epsilon$.

\begin{table}[h]
\centering
\vskip 20pt
\begin{tabular}{|c ||c |c |c |c |c |c |}
\hline
\diaghead(5,-4){\hskip 25pt }%
       {$\Phi_1$}{$\Phi_2$}& $I$ & $\epsilon$ & $\epsilon''$ & t & $\sigma$ & $\sigma'$ \\
\hline
\hline
$I$  &  ~0~    &  ~1~     &  0   &  0    & 0   & 0  \\
\hline
$\epsilon$  &    1    &   0     &  0   &  $c$    & 0   & 0  \\
\hline
$\epsilon''$ &   0    &     0    &  0  & $3/7$   & 0     & 0  \\
\hline
$t$  &  0    &  $c$   &  $3/7$     &  0   & 0  & 0 \\
\hline
$\sigma$  &  0    & 0   &  0     &  0   & $3c/2$ & $1/2$ \\
\hline
$\sigma'$  &  0    & 0   &  0     &  0   & $1/2$ & 0 \\
\hline 
\hline
\end{tabular}
\caption{The structure constants $C_{\Phi_1\epsilon\Phi_2}$ involving the leading energy operator $\epsilon$.  Here the constant $c$ is given by $c=\frac{2}{3}\sqrt{\frac{\Gamma(4/5)\Gamma^3(2/5)}{\Gamma(1/5)\Gamma^3(3/5)}}$.}
\label{struct_const}
\end{table}

\subsubsection{The $E_7$ spectrum of $H_{\rm tricritical~Ising} + g_2\int^R_0 \rd x \epsilon(x)$}

Before analyzing the numerical data coming from the TSA, we first discuss the available analytic results for this model.  As this model is integrable~\cite{zamolodchikov1989integrable,christe1990integrable,fateev1990conformal}, these are considerable.  As illustrated in Fig.~\ref{phasediagram}, finite $g_2$ drives the model into a massive phase, which may possess either order $\langle 0|\sigma |0\rangle \neq 0$ ($g_2<0$) or disorder $\langle 0|\sigma |0\rangle = 0$ ($g_2>0$). Remarkably, in both cases the massive spectrum is related to the $E_7$ Lie algebra, with the ratio of the masses in the spectrum being  equal to the ratios of the components of the $E_7$ Perron-Frobenius eigenvector associated with the $E_7$ Cartan matrix (in much the same way as the spectrum of the critical quantum Ising model perturbed by the spin operator is related to the $E_8$ algebra). The spectrum consists of 7 particles with masses~\cite{fateev1990conformal,christe1990integrable}
\begin{eqnarray}\label{E7spec}
m_1 &=& Cg_2^{5/9} \equiv \frac{2\Gamma(\frac{2}{9})}{\Gamma(\frac{2}{3})\Gamma(\frac{5}{9})}
\bigg(\frac{4\pi^2\Gamma(\frac{2}{5})\Gamma^3(\frac{4}{5})}{\Gamma^3(\frac{1}{3})\Gamma(\frac{3}{5})}\bigg)^{5/18}g_2^{5/9} \nn
&=& 3.74537\ldots g_2^{5/9};\\
m_2 &=& 2\cos\bigg(\frac{5\pi}{18}\bigg)m_1=1.28557\ldots m_1, \cr\cr
m_3 &=& 2\cos\bigg(\frac{\pi}{9}\bigg)m_1=1.87938\ldots m_1,\cr\cr
m_4 &=& 2\cos\bigg(\frac{\pi}{18}\bigg)m_1 =1.96961\ldots m_1,\cr\cr
m_5 &=& 4\cos\bigg(\frac{\pi}{18}\bigg)\cos\bigg(\frac{5\pi}{18}\bigg)m_1=2.53208\ldots m_1,\cr\cr
m_6 &=& 4\cos\bigg(\frac{2\pi}{9}\bigg)\cos\bigg(\frac{\pi}{9}\bigg)m_1=2.87938\ldots m_1,\cr\cr
m_7 &=& 4\cos\bigg(\frac{\pi}{18}\bigg)\cos\bigg(\frac{\pi}{9}\bigg)m_1 =3.70166\ldots m_1. \nonumber
\end{eqnarray}
Here, in Eq.~\fr{E7spec}, we have given the relation of the fundamental mass scale, $m_1$, in terms of the strength of the perturbation $g_2$~\cite{fateev1994exact}. These excitations have definite parity under $Q$: $m_2,m_4,m_5$ and $m_7$ are even excitations, while $m_1,m_3$, and $m_6$ are odd. The masses are the same for both signs of the coupling $g_2$ because, as in the Ising model, there is a Kramers-Wannier duality that maps the ordered phase onto the disordered \cite{lassig1991scaling,fioravanti2000universal,lepori2008particle,fioravanti2000universal1}, akin to that of the standard Ising model~\cite{IsingBook}. While the masses are the same, the nature of the excitations are different. In the ordered phase some of the excitations (the odd ones) are kinks, i.e. they interpolate between the two available vacua.  We will, for the sake of convenience, only present TSA data for the disordered sector.

Beyond the masses, the scattering matrices of the theory are known. As the theory remains integrable in the presence of the perturbing operator $\epsilon$, all scattering in the theory is encoded in the two-body $S$-matrices. These can be expressed most compactly as follows. If $A^\dagger_i(\theta)$ creates a particle with mass $m_i$, energy $E$ and momentum $p$ described by  
\begin{equation}
p= m_i \sinh (\theta), ~~ E= m_i\cosh(\theta ),
\end{equation}
the scattering matrices are defined by the generalized commutation relations,
\be
\label{defn_S}
A^\dagger_i(\theta_i)A^\dagger_j(\theta_j)= S_{ij}(\theta_i-\theta_j) A^\dagger_j(\theta_j)A^\dagger_i(\theta_i).
\ee
Here the rapidity, $\theta$, parameterization of the energy-momentum of a particle is convenient as it leads to an $S$-matrix that depends on the difference of the particles' rapidities (as dictated by Lorentz invariance). We have written down a simplified form of the $S$-matrix where there are no processes that interchange particle species; this follows from each of the $E_7$ masses being different: integrability together with kinematic constraints forbid such processes. The full list of the $S$-matrices are available in Refs.~\cite{fateev1990conformal,christe1990integrable,acerbi1996form}. To analyze the TSA data, we will only need one $S$-matrix, $S_{11}(\theta)$,
which is given by
\begin{eqnarray}\label{S11}
S_{11}(\theta) &=&  f_2(\theta)f_{10}(\theta);\\
f_a(\theta) &=& \frac{\tanh(\frac{1}{2}+\frac{i\pi a}{18})}{\tanh(\frac{1}{2}-\frac{i\pi a}{18})}.\nonumber
\end{eqnarray}

Within the TSA framework, we can do more than compute the particle spectra of a theory. For example, matrix elements of the various operators in the theory can also be computed. We will show such computations in the sections that follow: we will consider matrix elements that involve zero, one, and two-particle states. As a particular operator to consider, we will focus our attention on the leading energy operator, $\epsilon$, the perturbation itself. As the perturbation of a critical theory, this operator is closely related to the trace of the stress energy tensor
\begin{equation}
\Theta_T(x) = 2\pi g_2 (2-2\Delta_\epsilon)\epsilon(x). \label{setensor}
\end{equation}
The vacuum expectation value of $\Theta_T(x)$, $\langle 0| \Theta_T(x)|0\rangle$, together with its one particle-matrix elements, 
$\langle 0| \Theta_T(x) A^\dagger_i(\theta=0)|0\rangle$ can be computed exactly from the integrability of the $E_7$ theory. 
The matrix elements are given in Table~\ref{1p_me_tab}.

\begin{table}[t]
\centering
\vskip 20pt
\begin{tabular}{|c |c |}
\hline
 State, $|i\rangle$ & $F^{\Theta_T}_i=\langle 0|\Theta_T(0) |i\rangle/m_1^2$  \\
\hline
\hline
$|0\rangle$ &  1.18388\ldots    \\
\hline
$A^\dagger_2(\theta)|0\rangle$  &    0.9604936853\ldots\\
\hline
$A^\dagger_4(\theta)|0\rangle$  &    0.4500141924\ldots \\
\hline 
\hline
\end{tabular}
\caption{The exact matrix elements of the trace of the stress energy tensor involving the vacuum and the first two even one-particle states.}
\label{1p_me_tab}
\end{table}

Beyond the one-particle matrix elements, the TSA can also access two-particle matrix elements. Matching the TSA data onto the analytics is more involved because of the need to take into account non-trivial finite-size effects on the matrix elements. To illustrate this matching, we will
make a detailed study of the two-particle matrix element
\be
\langle 0|\Theta_T(0)A^\dagger_1(\theta_1)A^\dagger_1(\theta_2)|0\rangle.
\ee
Its analytic form in the infinite volume is given by~\cite{acerbi1996form} 
\bw
\be
\label{2p11}
\langle 0|\Theta_T(0)A^\dagger_1(\theta_1)A^\dagger_1(\theta_2)|0\rangle = - 2\pi i\, m_1^2 \sinh\bigg(\frac{\theta_{12}}{2}\bigg) \frac{F_{11,\rm min}(\theta_{12})}{D_{11}(\theta_{12})},\ \ 
\ee
with $\theta_{12} = \theta_1 - \theta_2$ and the functions 
\bea 
&& F_{11,\rm min}(\theta) = G\bigg(\frac{10}{18},\theta\bigg)G\bigg(\frac{2}{18},\theta\bigg),\qquad
 G(a,\theta) \!=\! \exp\Bigg[ \!2\!\int^\infty_0  \frac{\rd x}{x}\frac{\cosh\Big(x \big(a-\frac{1}{2}\big)\Big) \sin^2\Big(\frac{x (i\pi-\theta)}{2\pi}\Big)}{\cosh\big(\frac{x}{2}\big)\sinh(x)}\Bigg],\nn
&& D_{11}(\theta ) = P\bigg(\frac{10}{18},\theta\bigg)P\bigg(\frac{2}{18},\theta\bigg), \qquad P(a,\theta) =\frac{\cos(\pi a) - \cosh(\theta)}{2\cos^2(\pi a)}.
\end{eqnarray}
\ew
One arrives at this form through a non-trivial application of the form factor bootstrap, a set of analytic constraints based on Lorentz invariance and consistency with the scattering matrix that the matrix elements must satisfy~\cite{acerbi1996form,smirnov1992form}. A detailed explanation of the bootstrap can be found in a number of different references. See, for example, Refs.~\cite{smirnov1992form,essler2005review}.
We will show in the sections that follow how to connect this form factor in infinite volume
to their finite volume counterparts available through the TSA.  The theory on how to make this connection was first worked out in generality in Refs.~\cite{pozsgay2008formI,pozsgay2008formII}.

\subsubsection{TSA analysis of $E_7$ spectrum}

\begin{figure}
\includegraphics[width=0.42\textwidth]{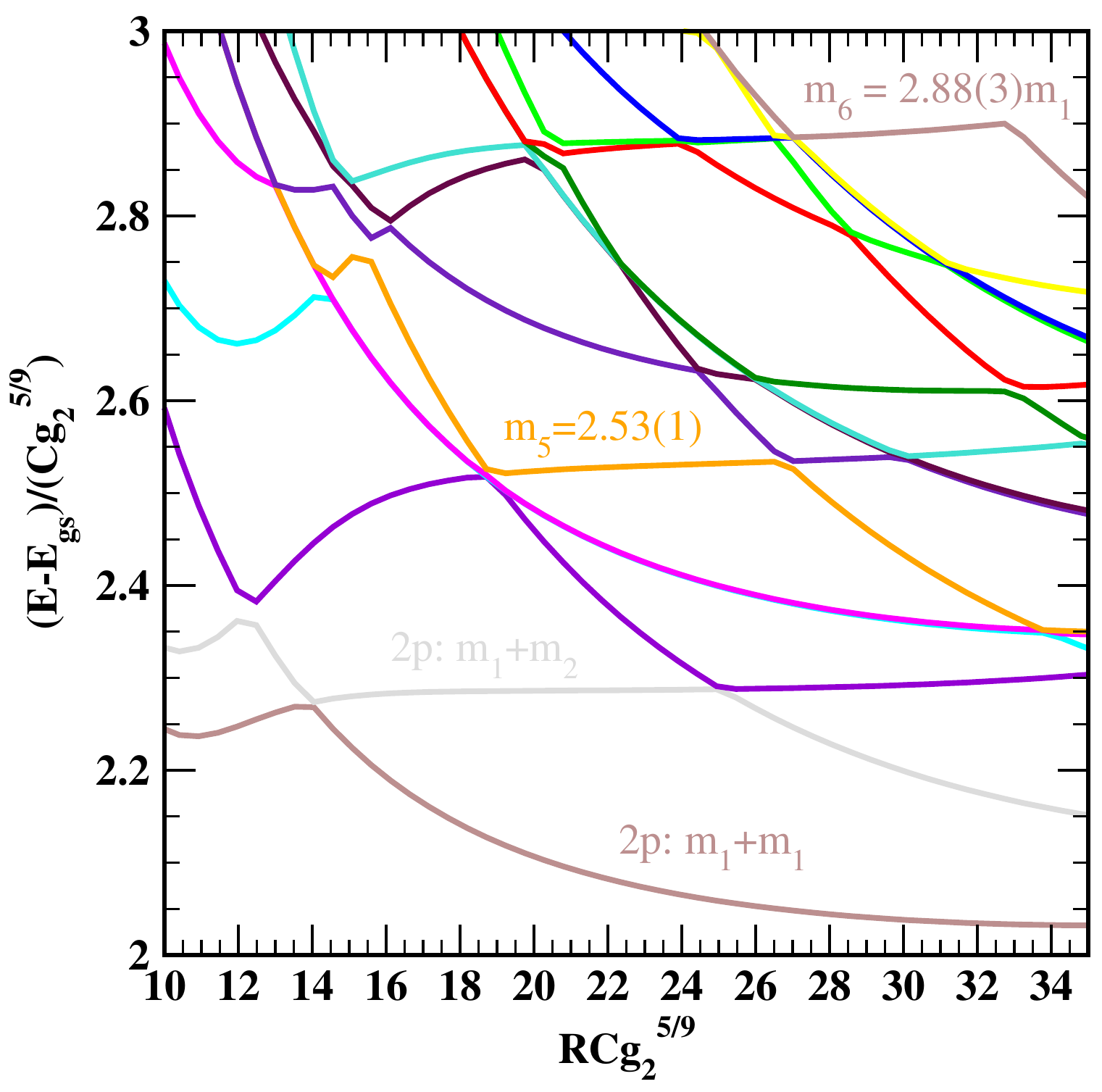}
\includegraphics[width=0.42\textwidth]{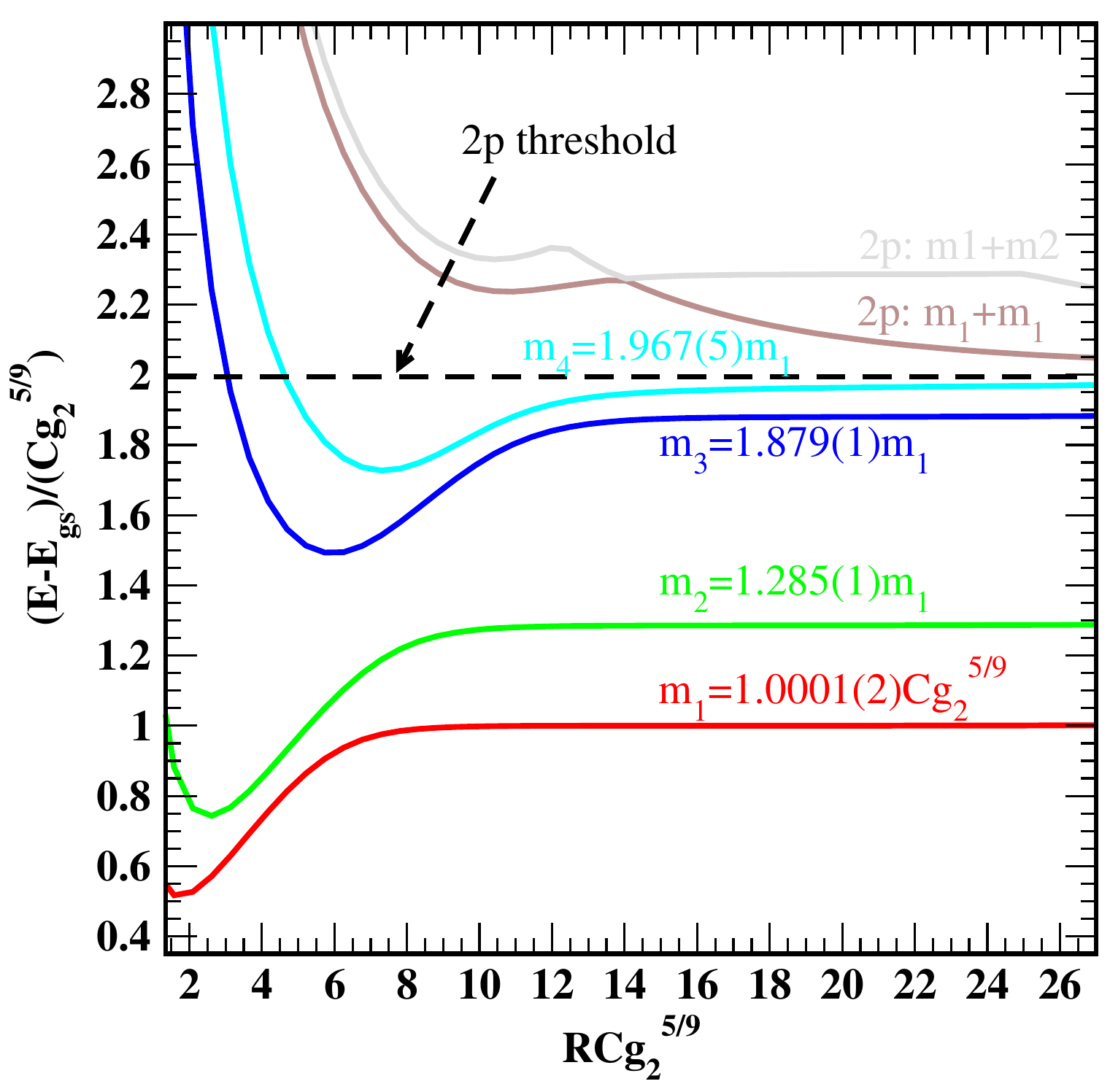}
\caption{The dimensionless excited state energies ($E_i/Cg_2^{5/9}$) as a function of the dimensionless system size, $RCg_2^{5/9}$, computed with the TSA. Lower panel: The first six excited states consisting of four single particle states and two two-particle states. Upper panel: The next fifteen excited states including the fifth and sixth single particle states.  For the single particles, we give the best estimate, including error, of their masses [to be compared with the exact masses in Eq.~\fr{E7spec}]. The dimensionless cutoff used for this computation is $N=22$ and involves 8810 states from all 6 Verma modules (here we do not work individually in the even and odd sectors of the theory).}  
\label{TSAfig5}
\end{figure}

Let us now turn to presenting the TSA analysis of the $E_7$ model. This spectrum was first
computed using the TSA in Ref.~\cite{lassig1991scaling}.  We first consider the model's low-lying excitation spectrum, which we present in Fig.~\ref{TSAfig5}. We have plotted the data in terms of dimensionless energies, $E_i/m_1=E/(Cg_2^{5/9})$ and dimensionless volume $RCg_2^{5/9}$ where the constant $C$ is defined in Eq.~\fr{E7spec}.  While this data has been computed at $g_2=1/(2\pi)$, the data for any choice of $g_2$ will rescale on to these curves (provided the same cutoff is employed) once recast in these dimensionless variables.

In the lower figure we present the energies of the lowest six excited states (with the ground state energy subtracted). We see that we obtain the expected masses of the first four one-particle states to better than 0.5\%.  We note, however, that the excitations have different stability regions. The first excited state, $m_1$, has a much wider range in $R$ where it equals its infinite volume value than the $m_4$ excitation. The fifth excited state (the brown curve for $RCg_2^{5/9}>14$) is a two-particle state consisting of two $m_1$ particles. It is easy to identify as such because of its $1/R^2$ decay to a value of $2m_1$. This $1/R^2$ term in its energy comes from the constituent particles of the state having finite (and opposite) momentum.\footnote{The two particles have opposite momentum as we work in the zero-momentum sector of the theory.} The sixth state in the lower panel (grey curve) is also a two-particle state, but is formed from one $m_1$-particle and one $m_2$-particle. The curve is flat, because unlike the two-particle state of two $m_1$ particles, here the $m_1$ and $m_2$ particles both have zero-momentum. As the two particles are different, they are allowed to have their other quantum numbers, such as momentum, equal. In general, there is typically a Fermi-like exclusion principle in the allowed quantum numbers for the constituent particles in multi-particle states. We will turn to a more detailed analysis of the energy of the two-particle states shortly.

In the upper panel of Fig.~\ref{TSAfig5} we present fifteen excited states whose energies are greater than the two-particle threshold, $2m_1$.  Among these fifteen are the single-particle $m_5$
and $m_6$ excited states.  We see that the accuracy at which we obtain these energies is less than the first four, at roughly 1\%. We also see that the spectrum is populated by a variety of two-particle excitations; at the bottom of this panel are the same two two-particle states that appear in the lower panel. However, as we consider a larger range of $R$, we see that the grey curve for $RCg_2^{5/9} > 26$ is a 2-$m_1$ state (it is such because
it experiences the characteristic $1/R^2$ decay in energy of a two-particle state when the two particles each carry finite momentum), while for $14<RCg_2^{5/9} < 26$ it is a $m_1$-$m_2$ state. 

\begin{figure}
\includegraphics[width=0.45\textwidth]{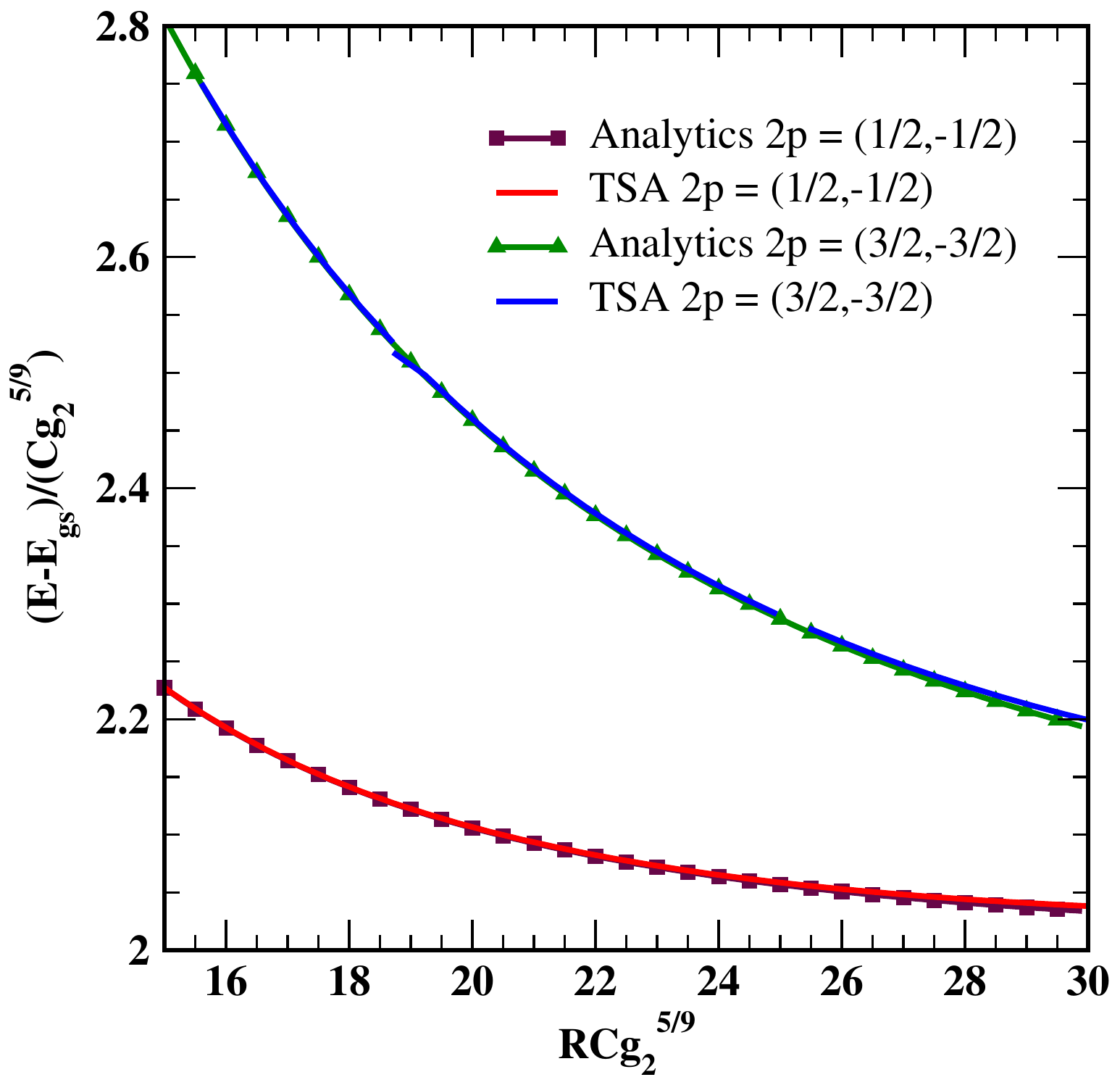}
\caption{Energies of the first two two-particle states as a function of the dimensionless system size. We compare our TSA data to an analytic calculation where the energies are determined by solving the two-particle quantization condition, see Eq.~\fr{quant}.}
\label{TSAfig6}
\end{figure}

In the region $RCg_2^{5/9}>26$, we thus have two different 2-$m_1$ states. How do these differ? They differ in terms of the momentum quantum numbers assigned to each of the particles. We can make this notion more precise. The momentum of each particle is subject to a quantization rule; this rule recognizes that the two particles in the state interact via their $S$-matrix (Eq.~\fr{S11}):
\begin{eqnarray}
1 &=& e^{Rm_1\sinh(\theta_1)}S_{11}(\theta_1-\theta_2);\cr\cr
1 &=& e^{Rm_1\sinh(\theta_2)}S_{11}(\theta_2-\theta_1).
\end{eqnarray}
Here $p_1=m_1\sinh(\theta_1)$ and $p_2=m_1\sinh(\theta_2)$ are the two momenta of the particles. If $S_{11}$ is trivial (i.e. $S_{11}=1$), these quantization conditions are those of free particles, i.e. $p_i = 2\pi n_i/R$.  They are derived by taking the particle `around the world': in doing so the wavefunction picks up both a geometric phase proportional to the system size and a phase due to the interaction of the two particles, which is encoded in $S_{11}$. In logarithmic form, these quantization relations can be written as
\begin{eqnarray}
2\pi n_1 &=& Y(\theta_1,\theta_2), \quad 2\pi n_2 = Y(\theta_2,\theta_1),\label{quant}
\end{eqnarray}
where $n_1,n_2$ are integers or half-integers forming the quantum numbers that describe the state, while
\begin{align}
Y(\theta_1,\theta_2) =& Rm_1\sinh(\theta_1) + \frac{1}{i}\log\big(S_{11}(\theta_1-\theta_2)\big).\label{quantY}
\end{align}
The $\log$ of the $S$-matrix can be written as
\begin{align}
 \frac{1}{i}\log\big(S_{11}(\theta)\big) =& \frac{1}{i} \log f\bigg(\theta,\frac{2}{18}\bigg) +  \frac{1}{i}\log f\bigg(\theta,\frac{10}{18}\bigg),\nonumber \\
 \frac{1}{i}\log f(\theta,a) =& -2\tan^{-1}\bigg(\frac{\tanh(\frac{\theta}{2})}{\tan(\frac{\pi a}{36})}\bigg)\nonumber \\
& \quad -2\tan^{-1}\bigg[\tanh\bigg(\frac{\theta}{2}\bigg)\tan\bigg(\frac{\pi a}{36}\bigg)\bigg]. \nonumber
\end{align}
The total momentum of the state is $\frac{2\pi}{R}(n_1+n_2)$ and as we work in the zero momentum sector, $n_1+n_2=0$.\footnote{As we mentioned previous, $n_1 \neq n_2$ for two identical particles, as the quantization conditions satisfy a Pauli-like exclusion principle.} These equations can be readily solved for $\theta_1,\theta_2$.  The energy of the state is then
\begin{equation}
E_{2-\rm particle} = m_1\cosh(\theta_1) + m_1\cosh(\theta_2).
\end{equation}
Using this analysis, in Fig.~\ref{TSAfig6} we plot the analytically computed energies against those derived from the TSA for two two-particle states, one corresponding to $(n_1,n_2)=(1/2,-1/2)$ and one corresponding to $(n_1,n_2)=(3/2,-3/2)$. The $n_i$ are half-integers because two-particle states reside in the Neveu-Schwarz sector for the disordered phase of the model (i.e. $g_2>0$), as in the ordinary Ising model. We see that there is excellent agreement between the TSA numerics and the analytical result.

\subsubsection{TSA analysis of matrix elements of the stress-energy tensor $\Theta_T$ for the $E_7$ spectrum}
\label{IsingE7me}

\begin{figure}
\includegraphics[width=0.45\textwidth]{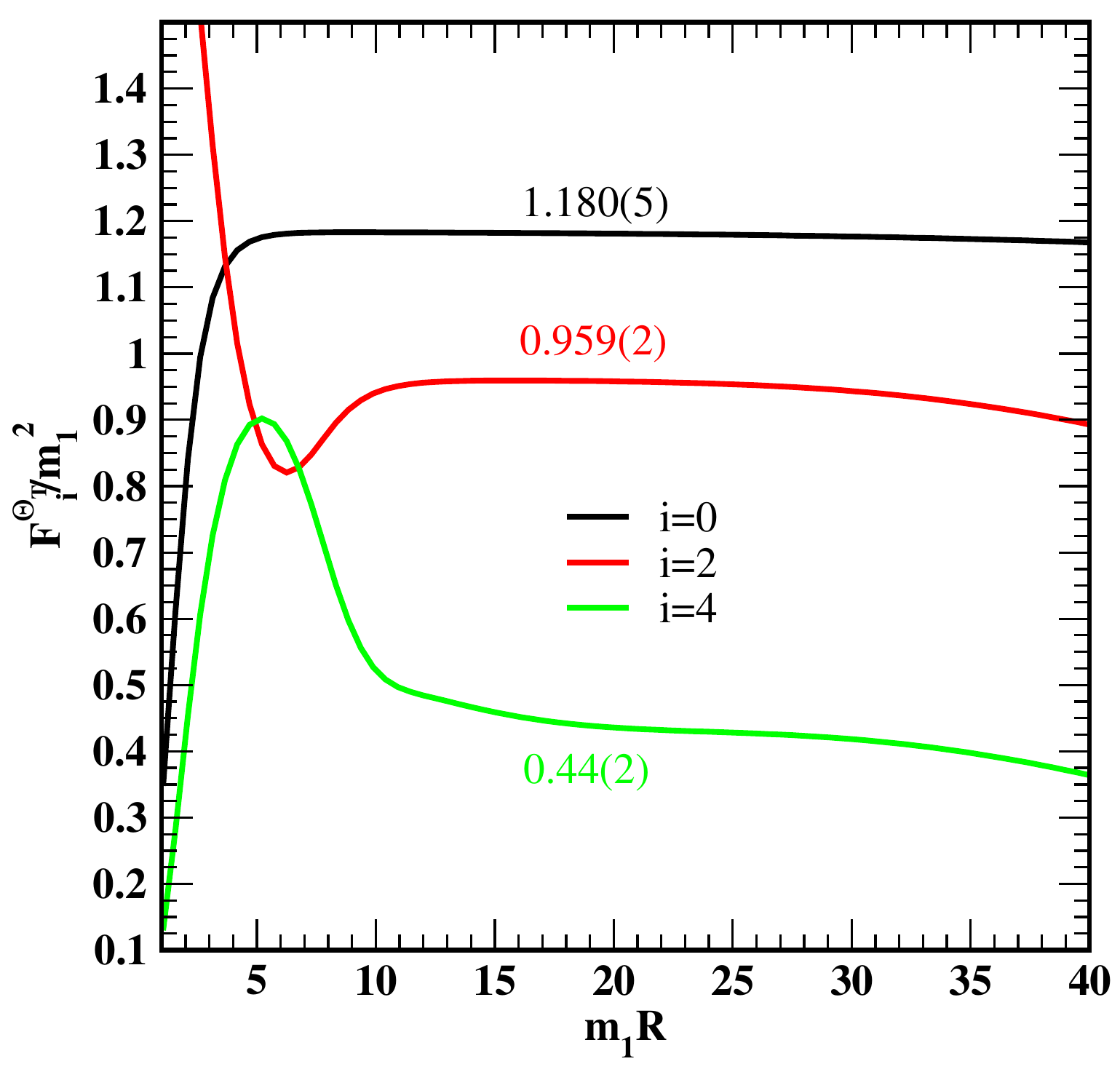}
\caption{The matrix elements of the stress energy tensor (proportional to the leading energy operator) between the vacuum and the first three even one-particle states as a function of the dimensionless system size $Rh^{8/15}$.  For the one-particle matrix elements we have adjusted, according to Eq.~\fr{factor}, the TSA values to accord to their infinite volume counterparts.  The cutoff used here is again $N=22$.}
\label{TSAfig7}
\end{figure}

Having considered the $E_7$ energies using the TSA, we now turn our attention to the matrix elements of the stress-energy tensor $\Theta_T$~\fr{setensor}. In Fig.~\ref{TSAfig7} we first consider the vacuum expectation value (VEV) of $\Theta_T$, as well as the first two one-particle matrix elements involving even single-particle states (i.e. $m_2$ and $m_4$). For the VEV, $F^{\Theta_T}_{0}\equiv \langle 0 |\Theta_T (0)|0\rangle$, we see that the matrix elements plateau over an intermediate range of system sizes, $R$. It is in this plateau region that we want to compare the value of the VEV to its infinite volume value presented in Table~\ref{1p_me_tab}. For small systems, the VEV is dominated by finite size effects (for a analysis of such effects in the context of critical Ising perturbed by the spin operator see Ref.~\cite{pozsgay2008Luscher}). On the other hand, at large values of $R$, the VEV begins to change its value as cutoff effects appear. The intermediate region of $R$ is then the sweet spot.

To compare the one-particle matrix elements, 
\begin{equation}
F^{\Theta_T}_i \equiv \langle 0|\Theta_T (0)A^\dagger_i(\theta)|0\rangle,
\end{equation}
computed using the TSA to the infinite volume versions in Table~\ref{1p_me_tab}, one must take into account the different normalizations of the particle states assumed in the two cases.  This amounts to the finite and infinite volume matrix elements differing by a factor of $(m_iR\cosh(\theta))^{1/2}$ (where $\theta$ is the rapidity of the particle):
\begin{eqnarray}\label{factor}
F^{\Theta_T}_{i}\Big\vert_{R=\infty} =\sqrt{m_i R\cosh(\theta)} F^{\Theta_T}_{i}\Big\vert_R.
\end{eqnarray}
This scaling has been performed for the data presented in Fig. \ref{TSAfig7}. We again see that there is a region of $R$ where a plateau exists. However, compared to the VEV, this region is smaller (and at least for 
$\langle 0|\Theta_T (0)A^\dagger_4(0)|0\rangle$ could be said to not strictly exist at all). 
To expand the plateau region we can employ a higher cutoff, $N$. However as we will see, the 
plateau only expands slowly with increasing $N$; in the next section we will consider renormalization 
group strategies to maneuver around this difficulty without paying a heavy numerical cost.

\begin{figure}
\includegraphics[width=0.45\textwidth]{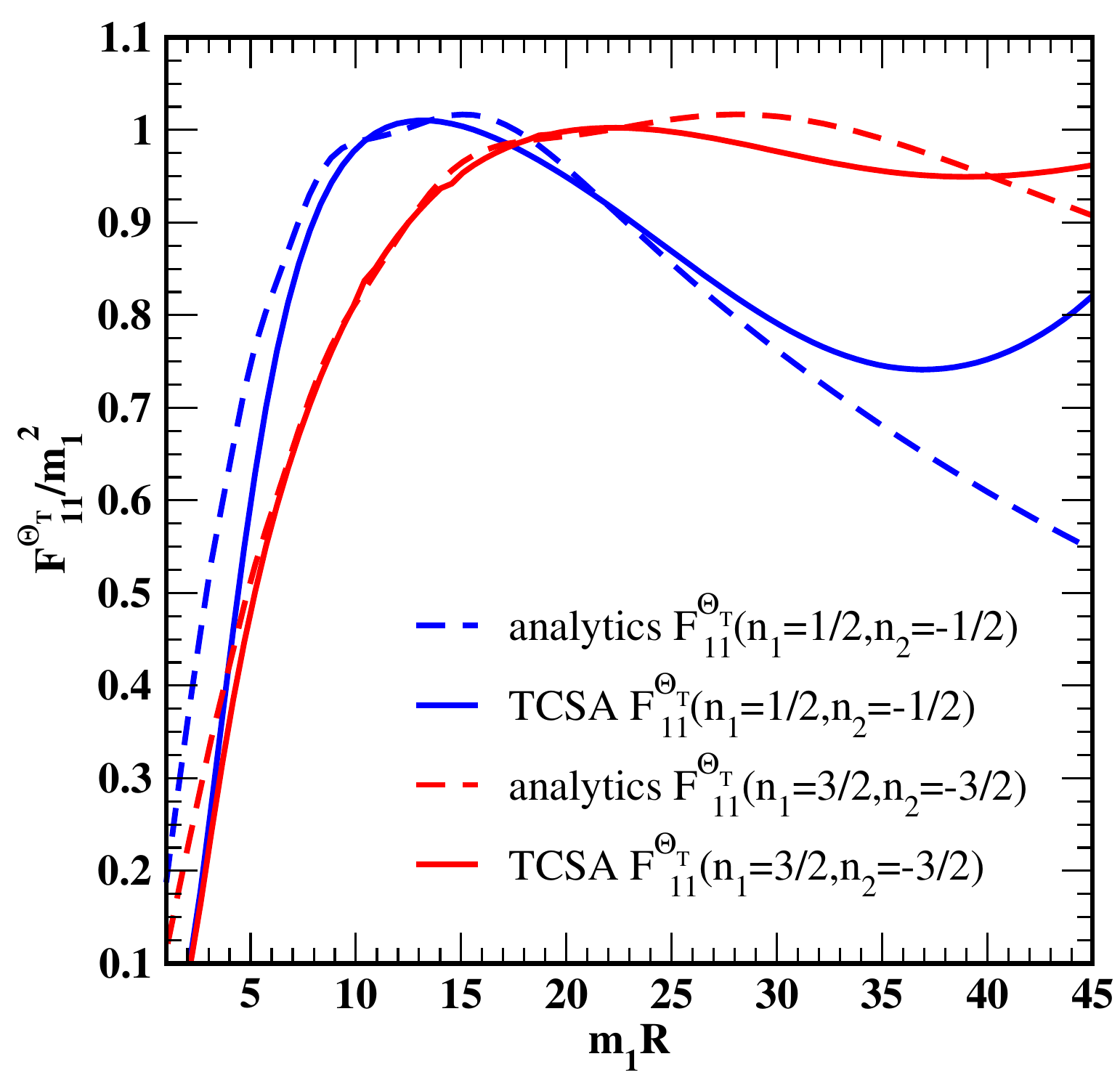}
\caption{The infinite volume matrix elements of the stress energy tensor between the vacuum and the first two two-particle states as a function of the dimensionless system size $Rm_1$.  Both the analytic bootstrap result for the matrix element, Ref.~\cite{acerbi1996form}, as well as the value inferred from the TSA using the relation in Eq.~\fr{2quant} are shown.}
\label{TSAfig8}
\end{figure}

Let us now consider the two-particle matrix elements. In particular, we focus upon
\begin{equation}
F_{11}^{\Theta_T}(\theta_1,\theta_2) \equiv \langle 0|\Theta_T(0)A^\dagger_1(\theta_1)A^\dagger_1(\theta_2)|0\rangle.
\end{equation}
Here the comparison between the TSA numerics and the analytics from the form-factor bootstrap is considerably richer, because the matrix elements have a genuine dependence on the rapidities $\theta$ of the constituent particles, see Eq.~\fr{2p11}. If we plot a two-particle matrix element against $R$, we are in fact plotting the matrix element against the center of mass momentum-energy 
\begin{equation}
E_{\rm c.o.m}\equiv (E_1-E_2)^2-(p_1-p_2)^2 = 2m_1^2(1-\cosh(\theta_1-\theta_2))
\end{equation}
because as we vary $R$ we vary $\theta_1-\theta_2$ via quantization relation~\fr{quant}. Thus unlike the one-particle matrix elements, we expect a plot vs. $R$ for two-particle matrix elements to reveal their non-trivial energy-momentum dependence. We see precisely this in Fig.~\ref{TSAfig8} where we plot the matrix elements for two different 2-$m_1$ states [$(n_1,n_2)=(1/2,-1/2),(3/2,-3/2)$] against the system size. 

Comparison of the TSA data  with the infinite volume bootstrap result, Eq.~\fr{2p11}, is more complicated for a number of reasons. At small $R$ where $E_{\rm c.o.m.}$ is large, we expect to encounter finite size effects due to the smallness $R$. We clearly see these deviations in Fig.~\ref{TSAfig8}. As with the one-particle matrix elements, we also expect large $R$ to be dominated by cutoff effects. Indeed, we see sharp deviations between the TSA numerics and the analytics in Fig.~\ref{TSAfig8} for $m_1R>30$.

One subtlety here is that in order to compare the TSA numerics with the bootstrap, we need to again take into account the different state normalizations. For two-particle states this normalization is not merely the product of two $\sqrt{m_1R\cosh(\theta)}$ factors, but takes into account that the normalization is affected by the interaction between the particles. If we define $\rho_2(\theta_1,\theta_2)$ through 
\begin{eqnarray}
\rho_2(\theta_1,\theta_2) &=& {\rm Det}
\begin{bmatrix}
\partial_{\theta_1} Y(\theta_1,\theta_2)& \partial_{\theta_2} Y(\theta_1,\theta_2)  \\
\partial_{\theta_1} Y(\theta_2,\theta_1)& \partial_{\theta_2} Y(\theta_2,\theta_1)  \\
\end{bmatrix},
\end{eqnarray}
where $Y(\theta_1,\theta_2)$ is defined in Eq.~\fr{quantY}, then the finite volume and infinite volume matrix elements are related via 
\be
\label{2quant}
F_{11}^{\Theta_T}(\theta_1,\theta_2)\Big\vert_{R=\infty} = \rho_2 ^{1/2} (\theta_1,\theta_2)F_{11}^{\Theta_T}(\theta_1,\theta_2) \Big\vert_{R}.
\ee
In the absence of interactions, i.e. $S_{11}=1$, we have $\rho_2(\theta_1,\theta_2) = m_1^2R^2\cosh(\theta_1)\cosh(\theta_2)$. 
The relation in Eq.~\fr{2quant} was used in Ref.~\cite{konik2007numerical} to obtain agreement between the TSA data and the analytical result, for a 
two-particle matrix element of the spin operator for the critical Ising model perturbed by a magnetic field.
However, the differing normalizations of matrix elements in finite and infinite volume (together with how to 
handle disconnected terms in matrix elements) were first elaborated upon comprehensively for general matrix elements in general 
integrable models in two papers, Refs.~\cite{pozsgay2008formI,pozsgay2008formII}.

\subsection{Applying the TSA to the sine-Gordon model}
We now turn to the last detailed example we use to illustrate the TSA: perturbations of a compact free boson by vertex operators, 
with a particular focus on the sine-Gordon model.  Systems represented by such perturbations are ubiquitous in low dimensional 
strongly correlated condensed matter and cold atomic systems (see, for example, Refs.~\cite{GiamarchiBook,GNTBook,TsvelikBook} 
for discussions of applications to condensed matter systems; one example of a cold atoms scenario can be found in Ref.~\cite{gritsev2007linear}). 
Compact free bosons, $\Theta$, in themselves describe a remarkable number of one-dimensional systems including Heisenberg 
spin chains, doped one-dimensional Hubbard models, Luttinger liquids, metallic carbon nanotubes, 
etc~\cite{GiamarchiBook,GNTBook,TsvelikBook,MussardoBook}. The various relevant perturbations of such systems 
typically take the form of `vertex operators',\footnote{This terminology comes from high energy physics, and has been adopted by the condensed matter physics community working on one-dimensional quantum systems.} i.e. $e^{i\alpha\Theta}$ or $e^{i\alpha\Phi}$, 
where $\Phi$ is the boson dual to $\Theta$. The ability to study all such systems using the TSA makes it an extremely versatile tool.  
As a specific example, we will consider the sine-Gordon model, as it is integrable~\cite{zamolodchikov1979factorized,izergin1981lattice} 
and there exists a large body of results computed analytically exploiting this integrability. These will then provide a benchmark to 
compare TSA numerical results against.  While we present TSA data here on the sine-Gordon 
generated specifically for this review,
the sine-Gordon was first studied using the TSA in a set of three papers by G. Feverati, F. Ravanini, and G. Tak\'acs
(Refs.~\cite{feverati1999non,feverati1998truncated,feverati1998scaling}).

The sine-Gordon model has an action given by
\begin{align}
S =\int \!\rd x \rd t \Big(\frac{1}{8\pi}& \partial^\mu\Theta(x,t)\partial_\mu\Theta(x,t)\nonumber\\
&\quad - \lambda \cos \big(\beta\Theta(x,t)\big) \Big).
\end{align}
The model has a $U(1)$ current given by $j^\mu = \epsilon^{\mu\nu}\partial_\nu \Theta(x,t)$, whose form is independent of the cosine perturbation. The excitation spectrum of the sine-Gordon model consists of two solitons with $U(1)$ charge $\pm 1$ whose classical counterpart are ``kinks'' in the field that interpolate between two different adjacent minima of the cosine potential. These solitons have mass $m$ given by~\cite{zamolodchikov1995mass}
\begin{eqnarray}
m &=& C_{sg}\lambda^{\frac{1}{2-\beta^2}}, \label{mass-coupling}\\
C_{sg} &=& \frac{2\Gamma\Big(\frac{\xi}{2}\Big)}{\sqrt{\pi}\Gamma\Big(\frac{1}{2}+\frac{\xi}{2}\Big)}
\left( \frac{\pi\Gamma\Big(1-\frac{\beta^2}{2}\Big)}{2\Gamma\Big(\frac{\beta^2}{2}\Big)} \right)^{\frac{1}{2-\beta^2}}\\
\xi &=& \frac{\beta^2}{2-\beta^2}. .
\end{eqnarray}
For $\beta < 1$, there exist bound states of solitons, known as breathers. For a given $\beta$, it is known that there are $n = 1, \ldots, [1/\xi]$ (here $[O]$ denotes the integer part of $O$) such bound state excitations with masses 
\be
m_n = 2m \sin\bigg(\frac{\pi n \xi}{2}\bigg).
\ee
To illustrate the use of the TSA, we will focus on the specific value of $\beta^2=1/2$. Here the spectrum is two solitons and two breathers. One of the breathers is degenerate with the two solitons. In fact, at this point in phase space ($\beta^2=1/2$) the sine-Gordon model has an $SU(2)$ symmetry, with the two solitons and breather of mass $m$ forming a spin-one representation of $SU(2)$. The remaining breather, of mass $\sqrt{3}m$, is a singlet under this $SU(2)$ symmetry. Additionally, for this value of $\beta$ the sine-Gordon model is equivalent to the $SU(2)_1$ WZNW model perturbed by the trace of the fundamental WZNW field. 

As with the $E_7$ perturbed tricritical Ising theory, we can write down the two-particle scattering matrices for these excitations. Defining the $S$-matrices that appear in the generalized commutation relations (see, e.g., Eqs.~\fr{defn_S},~\fr{S11}, for the analog of this in the $E_7$ deformation of the tricritical Ising model as studied in the previous section) for the creation operators of the four $\beta^2=1/2$ sine-Gordon excitations (we label these by $s,\bar s, b_1, b_2$ for the soliton, antisoliton, and the two breathers) 
\begin{equation}
\begin{array}{c||c|c}
 & ~~{\rm mass}~~ & ~U(1)~{\rm charge} \\
 \hline
 A^\dagger_s(\theta) & m & +1 \\
 A^\dagger_{\bar s}(\theta) & m & -1 \\
 A^\dagger_{b_1}(\theta) & m & 0 \\
 A^\dagger_{b_2}(\theta) & \sqrt{3}m & 0 
 \end{array}
 \label{sGexc}
 \end{equation}
 we then have
\begin{eqnarray}
S_0(\theta) &\equiv& S_{ss}(\theta) = S_{s\bar s}(\theta) = S_{sb_1}(\theta) = S_{\bar s b_1}(\theta) \nn
&=& \frac{\sinh(\theta) + i\sin(\frac{\pi}{3})}{\sinh(\theta) -i\sin(\frac{\pi}{3})} \nn
S_{sb_2}(\theta) &=& S_{\bar sb_2}(\theta) = S_{b_1b_2}(\theta) \nn
&=& S_0\bigg(\theta + i\frac{\pi}{6}\bigg)S_0\bigg(\theta -i\frac{\pi}{6}\bigg);\nn
S_{b_2b_2}(\theta) &=& \Big(S_0(\theta )\Big)^3. \nonumber
\end{eqnarray}
These $S$-matrices will be necessary in understanding the finite size corrections to both the ground state and excited state energies.

\subsubsection{Overview of the massless compact boson}
To study the sine-Gordon model using the TSA we need to specify the spectrum of the unperturbed compact boson as well as how to compute matrix elements of vertex operators relative to this basis. This case is considerably easier than conformal minimal models (such as the tricritical Ising model considered in the previous section) as we do not have to worry about null states. 

Let us first consider the spectrum. To delineate it, it is useful to consider the mode expansion of the boson~\cite{CFTBook}
\begin{eqnarray}
\Theta (x,t) &=& \Theta_0 + \frac{4\pi}{R}\Pi_0 t + \frac{2\pi m}{\beta R}x \cr\
&& + i\sum_{l\neq 0} \frac{1}{l}\Big(a_l e^{\frac{2\pi i l}{R}(x-t)}-\bar a_{-l}e^{\frac{2\pi i l}{R}(x+t)} \Big). \qquad
\end{eqnarray}
This mode expansion assumes the boson has compactification radius $2\pi/\beta$, i.e. $\Theta(x+R,t) = \Theta(x,t) + \frac{2\pi}{\beta} m$, 
where $m$ denotes the winding number, which is related to the $U(1)$ charge of the sector. The operator $\Theta_0$ 
is the `center of mass' of the Bose field and $\Pi_0$ is its conjugate momentum, which has permitted values $n\beta$, with integer $n$. 
These obey the commutator $[\Theta_0,\Pi_0] = i$. 

The bosonic Hilbert space emerges from an infinite set of highest weight states marked by the bosonic winding number and the value of conjugate momentum
\begin{eqnarray}
|n,m\rangle = e^{in\beta \Theta(0) + i\frac{m}{2\beta} \Phi(0)}|0\rangle.
\end{eqnarray}
These highest weight states $|n,m\ra$ are defined by acting with vertex operators involving the boson and its dual on the vacuum $|0\rangle$. The dual boson, $\Phi$, can be defined via the relation
\begin{equation}
\partial_x\Theta(x,t) = \partial_t \Phi (x,t).
\end{equation}
The full Hilbert space is then recovered by the acting with the right and left moving modes ($a_n$ and $\bar a_n$) of the field on the highest weight states:
\begin{equation}\label{bosonic_states}
|s\rangle = \prod^M_{j=1}a_{k_j}\prod^{\bar M}_{\bar j=1}\bar a_{k_{\bar j}}|n,m\rangle.
\end{equation}
The energy and momentum of such a state is
\begin{eqnarray}
E_s &=& \frac{2\pi}{R}\bigg(n^2\beta^2 + \frac{m^2}{4\beta^2}+\sum^M_{j=1} k_j + \sum^{\bar M}_{\bar{j}=1} k_{\bar j}  -\frac{1}{12}\bigg),\nn
P_s &=& \frac{2\pi}{R}\bigg((n-m) + \sum^M_{j=1} k_j - \sum^{\bar M}_{\bar{j}=1} k_{\bar j}\bigg). \nonumber
\end{eqnarray}
The $1/12$ term in $E_s$ reflects the fact that the vacuum energy in the conformal limit on the cylinder does not vanish if it is assumed to be zero on the plane. The $a_n/\bar a_{n}$ satisfy the following commutation relations:
\begin{eqnarray}
[ a_n,a_{m}] &=& n\delta_{n+m,0};\cr\cr
[ \bar a_n,\bar a_{m}] &=& n\delta_{n+m,0};\cr\cr
[ a_n,\bar a_{m}] &=& 0.
\end{eqnarray}
These commutators, together with the relation governing commuting the modes with vertex operators
\begin{equation}
[ a_n, e^{i\beta \Theta(0)} ] = -\beta e^{i\beta \Theta(0)},
\end{equation}
allow one to compute generic matrix elements of the states (Eq.~\fr{bosonic_states}) with the vertex operators appearing in the sine-Gordon Hamiltonian.  

\subsubsection{sine-Gordon ground state energy at $\beta^2=1/2$}

\begin{figure}
\includegraphics[width=0.4\textwidth]{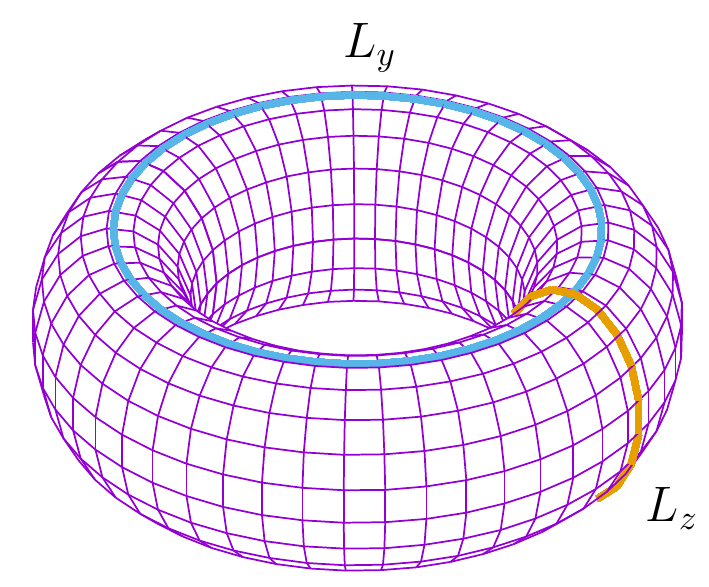}
\caption{A sketch of the toroidal space-time employed in the TSA when (imaginary) time is periodic.  By viewing the two periods of torus (here denoted by $L_{y}$ and $L_z$) as alternatively space or time, we can derive an expression for the finite size correction to the ground state energy.}
\label{torus}
\end{figure}

We now turn to the TSA results for the ground state energy.  Analytically the ground state energy can be characterized completely and is given by
\begin{eqnarray}
E_{gs} &=& -\epsilon_{\rm bulk}R + E_{\rm TBA}(R).
\end{eqnarray}
The bulk contribution to the ground state energy, $E_{gs}$, is given by~\cite{zamolodchikov1979factorized,baxter1982exactly}
\begin{equation}
\epsilon_{\rm bulk} = -\frac{m^2}{4}\tan\bigg(\frac{\pi\xi}{2}\bigg).
\end{equation}
The contribution $E_{\rm TBA}$ takes the form of a coupled set of integral equations known as the thermodynamic Bethe ansatz (TBA) equations involving the $S$-matrices of the various excitations in the model~\cite{zamolodchikov1990thermodynamic}:
\bw
\begin{eqnarray}\label{fullTBA}
E_{\rm TBA}(R) &=& 
-\int^\infty_{-\infty} \rd\theta \sum_{i=s,\bar s,b_1}\log(1+e^{-R\epsilon_s (\theta)})m\cosh(\theta) -\int^\infty_{-\infty} \rd\theta \log(1+e^{-R\epsilon_{b_2} (\theta)})\sqrt{3}m\cosh(\theta),\\
\epsilon_i (\theta) &=& 
m\cosh(\theta) - \sum_{j=s,\bar s,b_1}\int^\infty_{-\infty} \rd\theta \log\Big(1+e^{-R\epsilon_j (\theta)}\Big)K_{ij}(\theta-\theta')
-\int^\infty_{-\infty} \rd\theta \log\Big(1+e^{-R\epsilon_{b_2} (\theta)}\Big)K_{ib_2}(\theta-\theta'), \cr\cr
\epsilon_{b_2} (\theta) &=& 
\sqrt{3}m\cosh(\theta) - \sum_{j=s,\bar s,b_1}\int^\infty_{-\infty} \rd\theta \log(1+e^{-R\epsilon_j (\theta)})K_{b_2j}(\theta-\theta') 
-\int^\infty_{-\infty} \rd\theta \log(1+e^{-R\epsilon_{b_2} (\theta)})K_{b_2b_2}(\theta-\theta'),\cr\cr
K_{kl}(\theta) &=& \frac{1}{i}\partial_\theta\log S_{kl}(\theta),\nonumber
\end{eqnarray}
where $i=s,\bar s, b_1$ in the second equation. For large system sizes, $R$, this considerably simplifies and we obtain
\begin{eqnarray}\label{ETBA}
E_{\rm TBA}(R) &=& 
-3m\int^\infty_{-\infty} \frac{d\theta }{2\pi}\cosh(\theta) e^{-mR\cosh(\theta)} -\sqrt{3}m\int^\infty_{-\infty} \frac{d\theta }{2\pi}\cosh(\theta) e^{-\sqrt{3}mR\cosh(\theta)}
+ O(e^{-2mR}).
\end{eqnarray}
\ew

The large $R$ expression can be understood in a simple intuitive manner. To do so, let us imagine that we are working in imaginary (periodic) time and so our space-time is a torus (see Fig.~\ref{torus}).  The torus has two periods, $L_y$ and $L_z$, with $m^{-1} \ll R=L_z \ll L_y$. First we suppose $L_z$ corresponds to space and $L_y$ to time (i.e. $L_y = T^{-1}$ where $T$ is the temperature). The partition function for the system in this picture is 
\begin{equation}
Z = e^{-E_{gs}(R)L_y} + \cdots
\end{equation}
where the ellipses denote terms that are suppressed by working at low temperature in a gapped system. Alternatively, we can take the view that $L_z=R$ is the inverse temperature and $L_y$ is the volume of space. Here, the computation of the partition function must take into account states beyond the ground state:
\begin{eqnarray}
Z &=& e^{-E_{gs}(L_y)R}\Bigg(1 + \sum_{i=s,\bar s, b_1,b_2}\sum_{p} e^{-R(p^2+m_i^2)^{1/2}}\Bigg) \cr\cr
&& + {\rm two~particle~contributions}.
\end{eqnarray}
However, because the inverse temperature $R$ in this picture is such that $Rm \gg 1$, higher particle contributions to the partition function are suppressed and can be ignored. 

Now, if we compare the logarithm of the two different ways of computing the partition function of the system, we see that
\bw
\begin{eqnarray}
E_{gs}(R)L_y &=& E_{gs}(L_y)R + \sum_{i=s,\bar s, b_1,b_2}\sum_p e^{-R(p^2+m_i^2)^{1/2}} + \ldots, \nn
&=& E_{gs}(L_y)R +  \sum_{i=s,\bar s, b_1,b_2} L_ym_i\int^\infty_{-\infty}\frac{\rd\theta}{2\pi}\cosh(\theta) e^{-R(m_i^2\sinh^2(\theta)+m_i^2)^{1/2}} + O(e^{-2mR}).
\end{eqnarray}
\ew
In the second line, the factor $L_ym_i\cosh(\theta)/(2\pi)$ appears through making the sum over modes with momenta $2\pi n/L_y$ an integral. The ground state energy $E_{gs}(L_y)$ in the picture with $L_y$ space must be proportional to $L_y$ (i.e. the vacuum must have a uniform energy density)
\begin{equation}
E_{gs}(L_y) = \epsilon_{\rm bulk} L_y.
\end{equation}
Hence we see how rather simple considerations recover the asymptotically large $R$ form of $E_{\rm TBA}(R)$, see Eq.~\fr{ETBA}. We also see how to interpret the large $R$ correction to the ground state energy: at finite $R$, the vacuum is modified by virtual processes where excitations are created, travel around the system, and are then annihilated. These virtual processes are suppressed exponentially in the size of the system.

\begin{figure}
\includegraphics[width=0.45\textwidth]{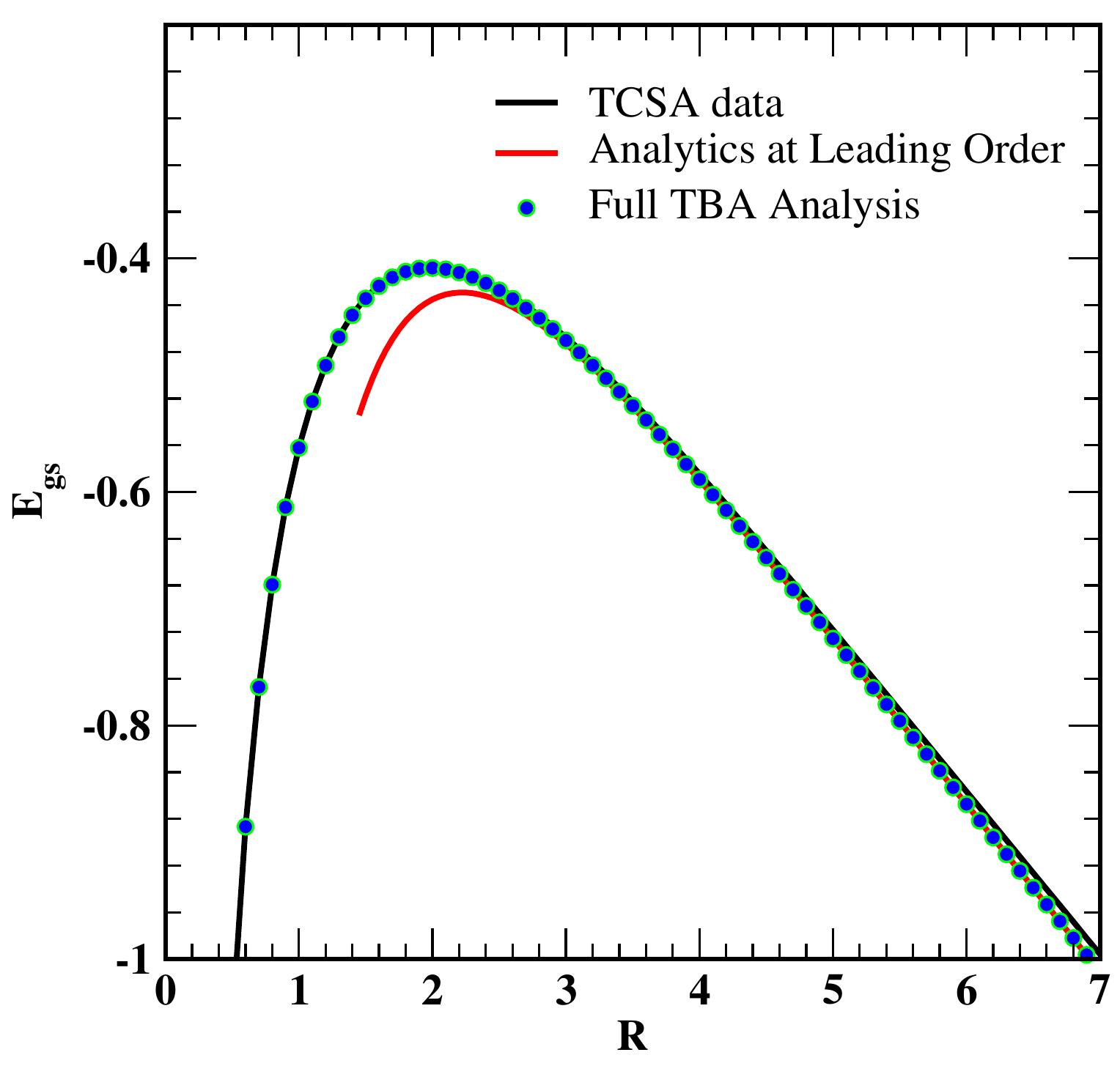}
\caption{The ground state energy as a function of system size for the sine-Gordon model at $\beta^2=1/2$.  TSA data is compared to analytic computations, both the full TBA analysis [see Eq.~\fr{fullTBA}] as well as the analytic expression for the ground state energy valid at large $R$, see Eq.~\fr{ETBA}.  For the TSA data we have chosen a value of the coupling $\lambda$ such that $m=1$ (using Eq.~\fr{mass-coupling}). The truncation level employed for the TSA is $N=18$ and involved 6917 states in the charge 0 ground state sector.}
\label{TSAfig9}
\end{figure}

With these considerations out of the way, we now can consider how the analytic expressions for the ground state energy of the sine-Gordon model compare to the TSA data. In Fig.~\ref{TSAfig9} we show the ground state energy computed using the TSA alongside the full TBA expression and the asymptotic form~\fr{ETBA}. We see that the full analytic expression agrees well over the entire range of $R$ shown in the figure. However, for the very largest values of $R$ shown, deviations can begin to be seen, reflecting cutoff effects in the TSA. The asymptotic form of the energy only begins to agree with the TSA data for $R>2.5$. For small systems the TSA ground state energy must return to its conformal value,
\begin{equation}
E_{gs}\big(R\ll m^{-1}\big) = -\frac{2\pi}{R}c + \ldots,
\end{equation}
where, for a boson, the central charge $c = 1$. For Eq.~\fr{fullTBA} to agree with the TSA data, the complicated non-linear integral equations for $E_{\rm TBA}(R)$ must reduce to 
\begin{equation}
E_{\rm TBA}(Rm\ll 1) = \epsilon_{\rm bulk} R - \frac{2\pi}{R}c + \ldots
\end{equation}
This was established in Ref.~\cite{zamolodchikov1990thermodynamic}.

\subsubsection{sine-Gordon excitations at $\beta^2=1/2$}

\begin{figure}
\includegraphics[width=0.45\textwidth]{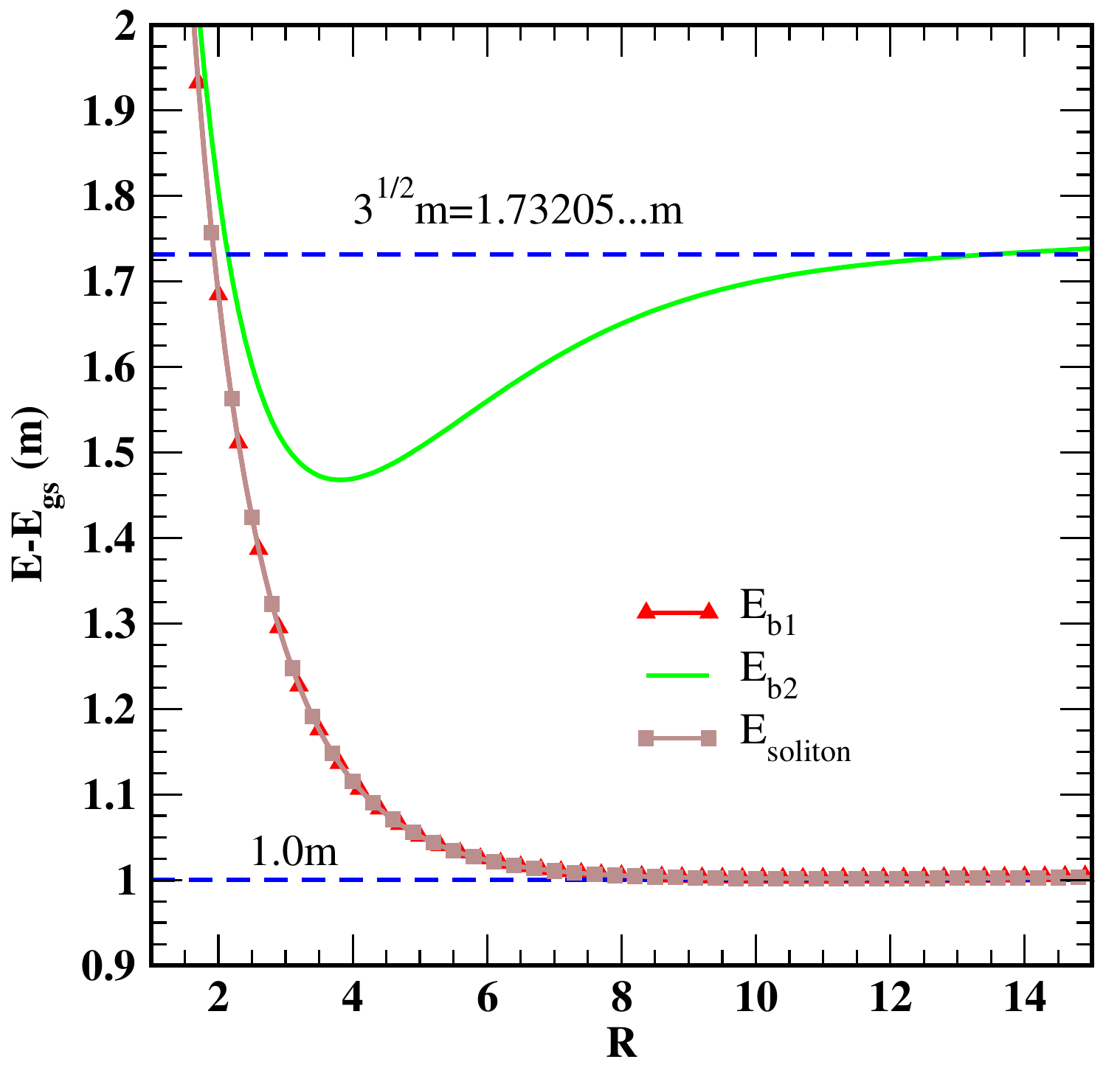}
\caption{The single particle excited state energies as a function of system size for the sine-Gordon model at $\beta^2=1/2$. The truncation level employed for the TSA was again $N=18$ and involved 6917 states in the charge 0 ground state sector and 7917 states in the charge $\pm 1$ sectors. The different charge sectors of the theory were treated separately so as to reduce the numerical burden.}
\label{TSAfig10}
\end{figure}

Having analyzed the ground state energy of the sine-Gordon model, we now turn to its one-particle excitations. There are four: two solitons and one breather of mass $m$ and one breather of mass $\sqrt{3}m$ [see Eq.~\fr{sGexc}].  We plot the excitation energies in Fig.~\ref{TSAfig10}, where we see the masses of the two solitons and the first breather are degenerate (numerically they agree to five significant digits) for all system sizes and obtain the expected value of $m=1$ (we have chosen the coupling constant through Eq.~\fr{mass-coupling} such that $m=1$) for $R>8$. The TSA prediction of the mass of the second breather however shows marked finite size corrections and only approaches its expected value of $\sqrt{3}$ for a value of $R\approx12$.  Even here, it overshoots the expected value for larger values of $R$, which indicates significant cutoff effects in the TSA data.

\begin{figure}
\includegraphics[width=0.35\textwidth]{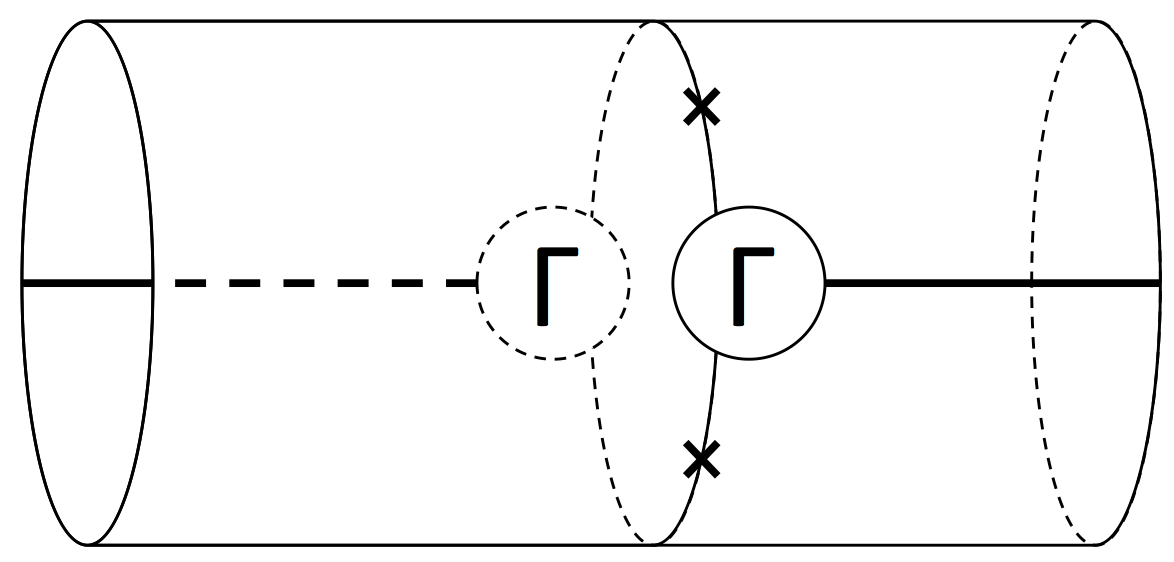}
\caption{The process that leads to the so-called $\mu$-term contributions to the finite size corrections to the excitation energies,
whereby an excitation experiences a virtual decay with amplitude $\Gamma$ into two other particles.  These two excitations
then travel around the system and recombine into the original excitation.  Adapted from Ref.~\cite{yurov1991truncated}.}
\label{mu-term}
\end{figure}

We will now analyze the finite size corrections to the energy of the second breather, $b_2$, in more detail. Finite size corrections come in two flavors. The first results from a particle undergoing virtual decay into two other particles and subsequently reforming after having travelled around the system. This is sketched schematically in Fig.~\ref{mu-term}. The $b_2$ breather has two possible decay channels, as it can be a bound state of a soliton-anti-soliton as well as a bound state of two $b_1$ breathers. The first process is proportional to $(\Gamma_{s\bar s}^{b_2})^2$ while the second is proportional to $(\Gamma_{b_1b_1}^{b_2})^2$, where $(\Gamma^{b_2}_{ab})^2$ is the imaginary part of the residue in the $S$-matrix $S_{ab}\big(\theta=iu_{ab}^{b_2}\big)$ at the value of imaginary rapidity that corresponds to $a,b$ forming a $b_2$ bound state:
\begin{equation}
m^2_{b_2}=m^2_{a}+m^2_{b} + 2m_am_b\cos\Big(u_{ab}^{b_2}\Big).
\end{equation}
For both the considered processes $u^{b_2}_{ab}=\pi/3$ and $(\Gamma^{b_2}_{ab})^2=2\sqrt{3}$. 

The full expression for this finite size correction to the mass of the excitation is then
\begin{eqnarray}
\Delta_\mu m_{b_2} &=& \sum_{(a,b)=(s,\bar s),(\bar s,s),(b_1,b_1)}(\Gamma^{b_2}_{ab})^2\mu_{b_2,ab} e^{-\mu_{b_2,ab}R}\cr\cr
&=& 3\sqrt{3}e^{-\frac{mR}{2}}. 
\end{eqnarray}
Here $\mu_{b_2,ab} = \frac{m_am_b}{m_{b_2}}\sin(u^{b_2}_{ab}) = 1/2$.  The $\mu$ index affixed to the correction ($\Delta_\mu m$) indicates that this term is often referred to as a ``$\mu$-term'' correction to the mass~\cite{klassen1991on}. We see that this correction is exponentially suppressed in the system size $R$. 

\begin{figure}
\includegraphics[width=0.35\textwidth]{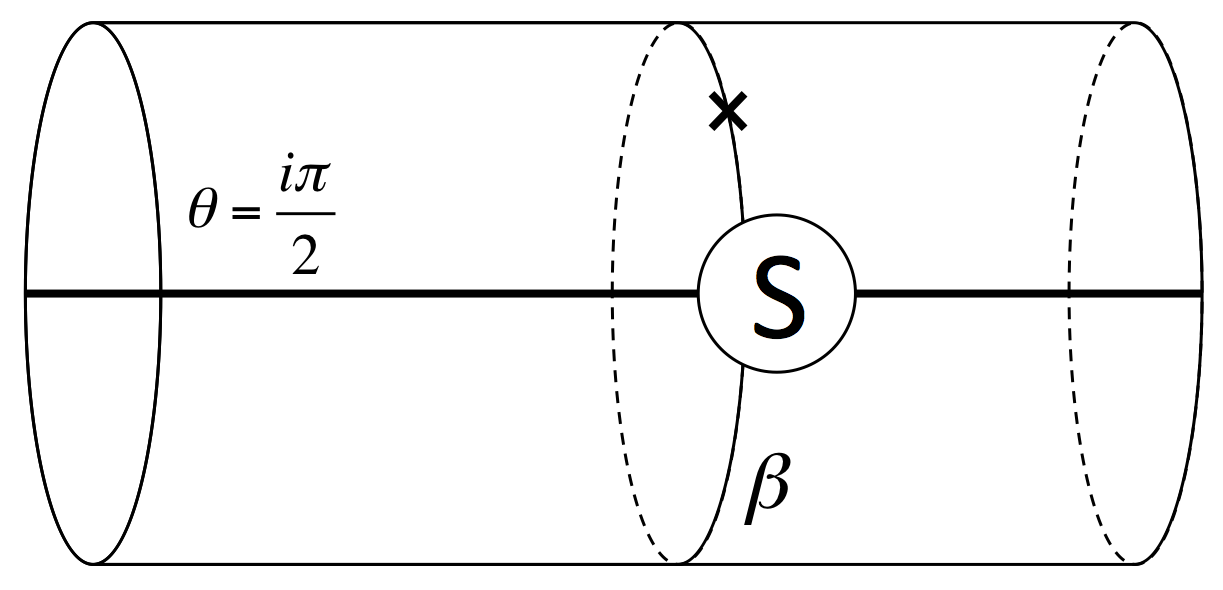}
\caption{The process that leads to the so-called $F$-term contributions to the finite size corrections to the excitation energies.
Here an excitation interacts via its scattering amplitude $S$ with particles that have been virtually created from the vacuum. Adapted from Ref.~\cite{yurov1991truncated}.}
\label{Fterm}
\end{figure}

The second type of finite size correction is illustrated in Fig.~\ref{Fterm} and is known as an $F$-term \cite{klassen1991on}. This correction arises because, in the finite volume particles are virtually
created from the background and can interact with a real excitation, thus altering its mass.  This correction takes the form
\begin{eqnarray}
\Delta_F m_{b_2} &=& -\sum_{a}{}^{'} \mathcal{P}\int^\infty_{-\infty} \frac{d\theta}{2\pi} e^{-m_aR\cosh(\theta )}m_aR\cosh(\theta )\cr\cr
&& \hskip .4in \times\Big(S_{ab_2}(\theta+i\pi/2)-1\Big),
\end{eqnarray}
where $\mathcal{P}$ indicates the principle value of the integral should be taken and the prime on the sum $\sum^{'}_a$ indicates that the $S$-matrix $S_{ab_2}(\theta + i\pi/2)$ should not have a multiple pole for real $\theta$~\cite{klassen1991on}. The only value of $a$ for which this is true is $a=b_2$ and hence the $F$-term is of order $e^{-\sqrt{3}mR}$ and is much smaller than $\Delta_{\mu} m_{b_2}$. As a result, we will ignore it. In general, the problem of determining the error one is making in including only the $\mu$-terms and $F$-terms is complex; a full discussion can be found in Ref.~\cite{klassen1991on}.  

\begin{figure}
\includegraphics[width=0.45\textwidth]{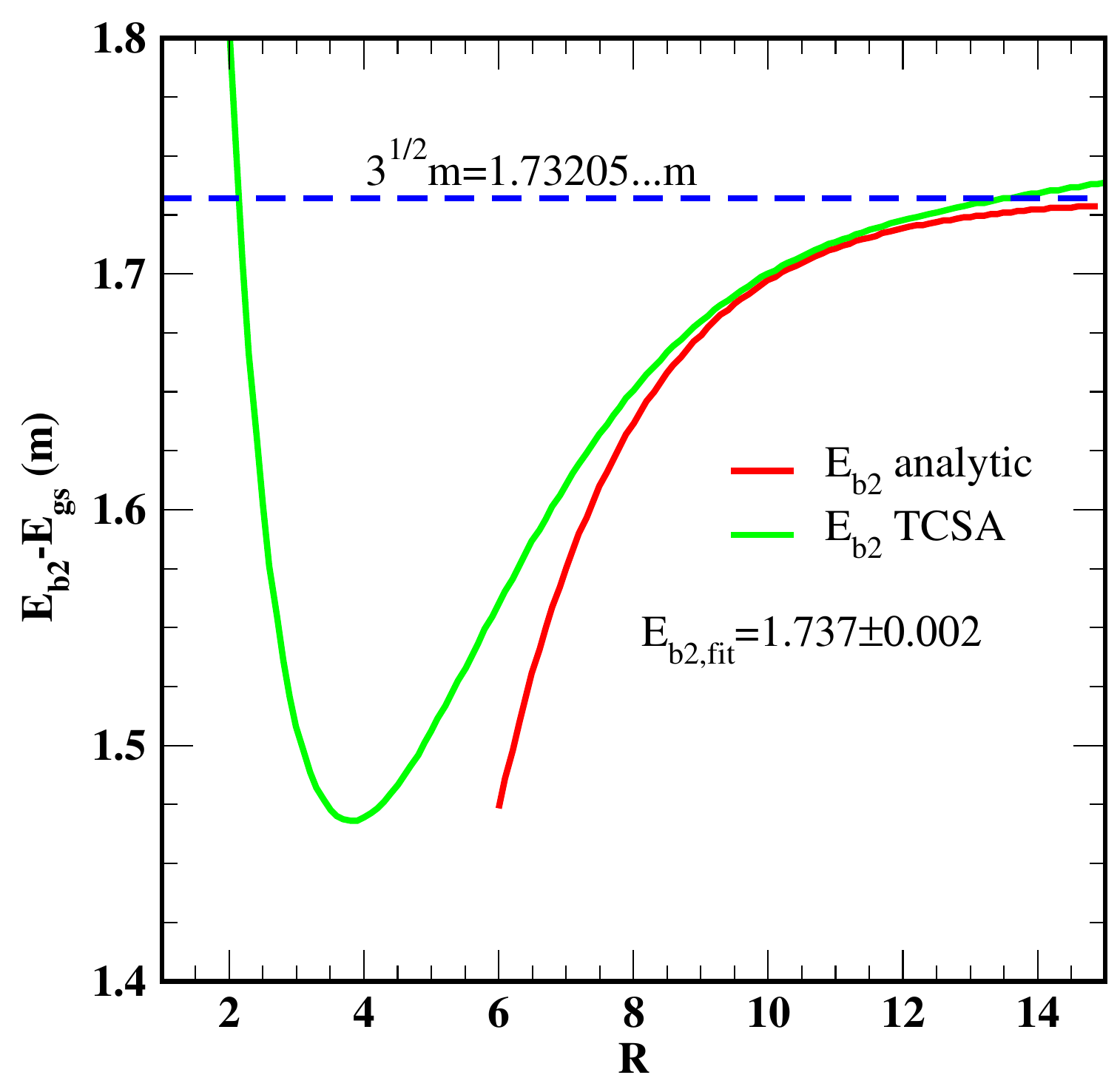}
\caption{An expanded view of the TSA energy of the second breather for the sine-Gordon model at $\beta^2=1/2$.   Here we
plot the leading order finite size corrections to its mass vs the TSA data.  If we fit this leading order form to the data treating
the $m_{b_2}$ as a fitting parameter we find $m_{b_2}=1.737(5)$ where the primary source of the uncertainty comes from
different choices of the fitting region.}
\label{TSAfig11}
\end{figure}

In Fig.~\ref{TSAfig11} we show the TSA results for the second breather with the analytic prediction for its energy including the $\mu$-term; we see the agreement is reasonable. If we treat $m_{b_2}$ as a fitting parameter and fit the analytic expression against the TSA data we obtain $m_{b_2} = 1.737(5)$, a roughly 0.2\% error from its true value of $\sqrt{3}$.

\section{Removing the effects of the cutoff from the TSA}
\label{Sec:TSA_Cutoff}

In this section we consider strategies for removing the effects of the cutoff on TSA computations. These will go in two directions: one primarily numerical \cite{konik2007numerical}, and one analytical \cite{feverati2008renormalization,watts2012renormalisation,giokas2011renormalisation,lencses2014excited,hogervorst2015truncated}.  We first consider the numerical approach.

\subsection{The numerical renormalization group and TSA}\label{nrgalgorithm}

\begin{figure*}
\includegraphics[width=0.7\textwidth]{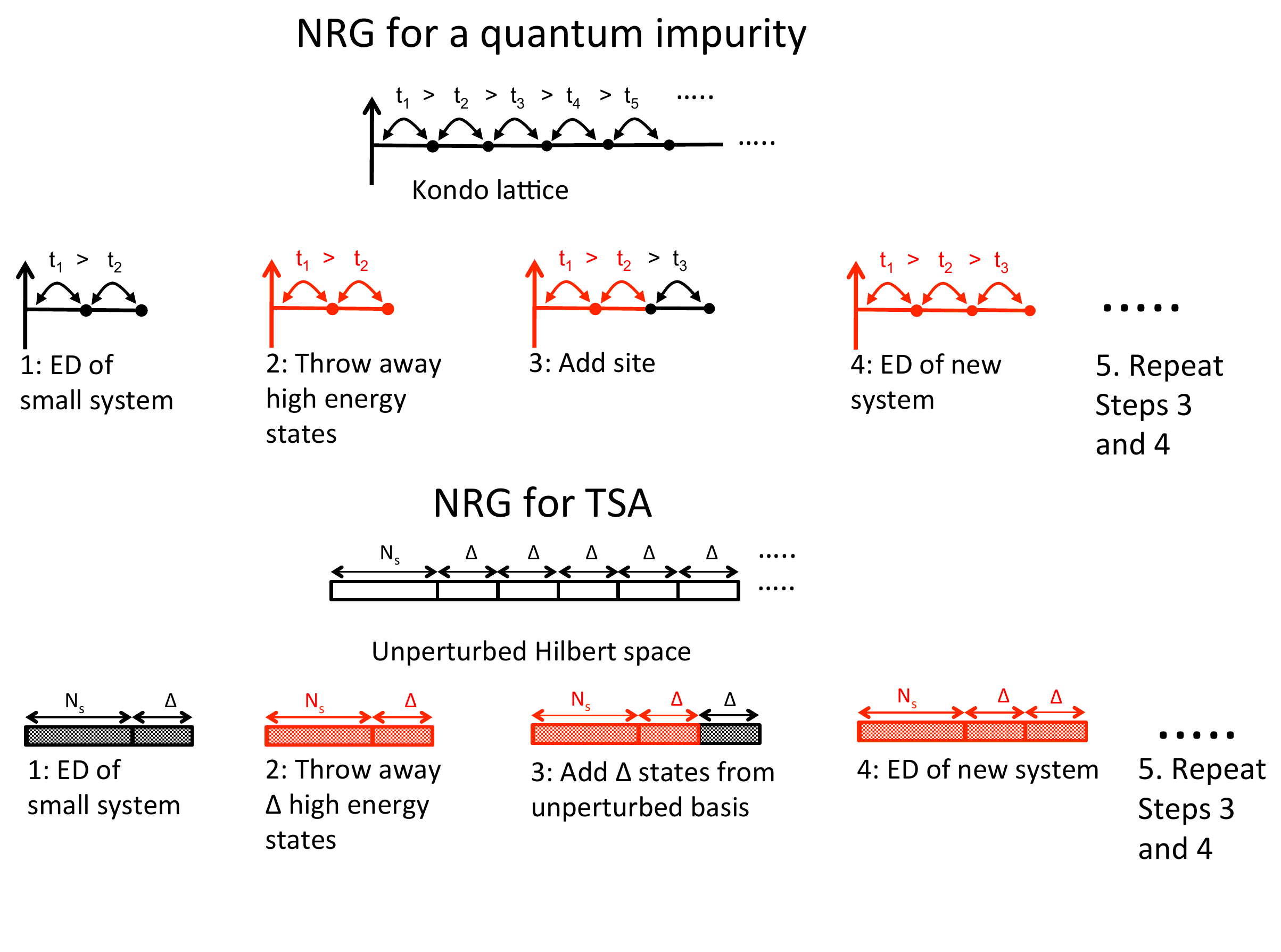}
\caption{A schematic representation of the NRG algorithm for quantum impurity problems and for TSA.}
\label{nrg_algorithm}
\end{figure*}

The first strategy~\cite{konik2007numerical} we employ to alleviate the effects of the cutoff is based on the numerical renormalization group (NRG), as developed by Kenneth Wilson~\cite{wilson1975the}. The NRG is a tremendously successful technique (see~\cite{bulla2008numerical}) for the study of generic quantum impurity problems.  It is based on the realization
that there is a hierarchy of energy scales in a quantum impurity problem that can be exploited to find solutions. The NRG for quantum impurity problems comes in two parts. In the first part the quantum impurity problem, that of a single localized spin or electron interacting with a fermionic bath, is mapped to an equivalent lattice model.  For a spin interacting with a Fermi sea, this lattice model takes the form
\be
H_{\rm Kondo} =- J{\bar S}\cdot c^\dagger_{1\sigma}{\bar \sigma}_{\sigma\sigma '}c_{1\sigma'}
+ \sum_{i=1,\sigma}^\infty t_i c^\dagger_{i\sigma}c_{i+1\sigma},
\label{WilsonHam}
\ee
where $\bar \sigma = (\sigma_x,\sigma_y,\sigma_z)$ are the Pauli matrices. This model consists of a spin-1/2 degree of freedom, $\bar S$, living at the end of a half-infinite lattice. The spin interacts with electrons that are able to hop along the lattice. The hierarchy of scales arises because the hopping parameters, $t_i$, decrease in strength the further one goes along the lattice.  Roughly speaking $t_i$ behaves as $\Lambda^{-i}$ where $\Lambda > 1$ is a parameter that arises in a logarithmic discretization of the Fermi sea of electrons. The mapping of the quantum impurity problem to this lattice model involves various non-trivial details~\cite{wilson1975the}. However these are not relevant for our purposes and we will suppose we begin with a Hamiltonian of the form in Eq.~\fr{WilsonHam}.

Wilson's key insight was that this Hamiltonian can be diagonalized in a set of iterative steps. Because the hopping parameters fall off as one moves away from the impurity, the electrons living on
sites close to the spin impurity interact most strongly with it. If we are interested in finding the ground state energy, a first (crude) approximation would be to truncate the lattice to a small finite number of lattice sites.  In this way we capture the portion of the Hamiltonian with the couplings that are largest in magnitude.  After this truncation, the size of the Hilbert space of lattice is small and the energies can easily be found through a numerical exact diagonalization (ED). This is the first step of Wilson's numerical renormalization group, and is shown as step~1 in the top part of Fig.~\ref{nrg_algorithm}.

In the second step we begin to account for the sites that we threw away in the first step. We take the states ($N_s+\Delta_1$ in total) obtained with the ED in step 1, $\{|E^{1}_s\rangle\}_{s=1}^{N_s}$ and order them by energy:
\begin{equation}\label{first_en}
E^{1}_1 < E^{1}_2 < \cdots < E^{1}_{N_s}. 
\end{equation}
As we are interested in low energy properties of the model, we only keep $N_s$ of these states, throwing away the remainder. This is step 2 in Fig.~\ref{nrg_algorithm}. In step 3, we add a site to the truncated lattice. The Hilbert space of this augmented finite lattice then consists of a tensor product of states in Eq.~\fr{first_en} with the states of the added lattice site:
\be
|e^{2}_{ij}\rangle = |E^{1}_{i}\rangle \otimes |i\rangle
\ee
In the third step of Fig.~\ref{nrg_algorithm}, we represent the tensor structure of the Hilbert space by picturing the initially truncated lattice in red and the added site in black. This new Hilbert state has $N_s+\Delta_2$ states. In step 4 we diagonalize this new problem, obtaining a new set of eigenstates with energies that we have ordered:
\begin{equation}\label{second_en}
E^{2}_1 < E^{2}_2 < \cdots < E^{2}_{N_s}. 
\end{equation}
We can then proceed to repeat the procedure: we throw away the $\Delta_2$ most energetic states, add a site to obtain an expanded basis, reformulate the Hamiltonian in this new basis, and perform an ED.  Because we keep only $N_s$ states from each diagonalization, the size of Hilbert space for each ED does not grow.  Yet the iterative procedure allows sites far away from the spin impurity to influence its physics.  It works because the sequence of EDs take into account the portion of the lattice with the largest hoppings first, leaving weaker couplings to later in the procedure.

\begin{table*}[ht]
\centering
\begin{tabular}{|c ||c |c |c |c |c|}
\hline\hline
 & ED & $N_s=500,\Delta=250$ & $N_s=500,\Delta=500$ & $N_s=1000,\Delta=500$ & $N_s=1500,\Delta=500$ \\
\hline\hline
$R=10$ & &  & & & \\
\hline
$E_{gs}$ &-1.41824  & -1.41813 &-1.41814  &-1.41817 &-1.41819 \\
$E_{b1}$ &1.00126 & 1.00122 &1.00123  &1.00123 &1.00124 \\
$E_{b2}$ & 1.69959 & 1.69955 & 1.69956  &1.69956 & 1.69957\\
\hline
$R=15$ & &  & & & \\
\hline
$E_{gs}$ & -2.10451 &-2.10350&-2.10361  &-2.10387 &-2.10404 \\
$E_{b1}$ & 1.0018&1.00173& 1.00174 & 1.00172& 1.00172 \\
$E_{b2}$ & 1.73396&  1.73440& 1.73434 & 1.73410& 1.73401\\
\hline
\end{tabular}
\caption{Effects of the choice of $N_{s}$ and $\Delta$ upon the TSA+NRG results for the low lying energies of the sine-Gordon
model at $\beta^2=1/2$ at two different values of $R$.  The overall NRG cutoff used here is $N=22$.  
The number of states falling below this cutoff is 39279.  We compare the different NRG
results with an exact diagonalization (ED) which serves as a reference point.}
\label{NRG_data}
\end{table*}

While Wilson had to map the spin impurity interacting with a Fermi sea to a half-line lattice with varying hoppings, we in a sense start in this position when we perform the TSA on a Hamiltonian
$H_0$ perturbed by a relevant operator.  A relevant perturbation will not strongly mix high and low energy states (although states about a given energy will be mixed strongly).  We can then imagine applying the same iterative procedure employed by Wilson.  We begin with states in our unperturbed Hilbert space, ordered by their energies: 
\be
|0\rangle,|1\rangle, \ldots .
\ee
As a first step we take the $N_s+\Delta$ of the states lowest in energy.  We form the full Hamiltonian (both $H_0$ and perturbation) in this truncated basis of $N_s+\Delta$ states. Just as with the ordinary TSA, we numerically diagonalize the problem and obtain the $N_s+\Delta$ eigenvalues, $E^1_1,\cdots ,E^1_{N_s+\Delta}$, and eigenvectors, $|E\rangle^1_1,\cdots ,|E\rangle^1_{N_s+\Delta}$.  This step of the TSA+NRG is pictured in the bottom part of Fig.~\ref{nrg_algorithm}. As Wilson did for the NRG in the quantum impurity problem, in the next step (step 2), we order these $N_s+\Delta$ states in ascending order of their energies and toss away the top $\Delta$ states. We express the remaining eigenvectors in terms of the unperturbed basis, $\{|m\rangle\}$ as follows:
\begin{equation}
|E\rangle^1_k = A^1_{km}|m\rangle + B^1_{km'}|m'+N_s\rangle; ~~~~k=1,\cdots,N_s,
\end{equation}
where $A^1$ is an $N_s\times N_s$ matrix and $B^1$ is an $N_s\times \Delta$ matrix and repeated indices are summed on. To these $N_s$ eigenstates we add the next $\Delta$ states from the unperturbed theory, 
\be
|m+\Delta+1\rangle,\cdots,|m+2\Delta\rangle.
\ee  
This leaves us again with a truncated Hilbert space of $N_s + \Delta$ states (step 3). We reform the Hamiltonian in this new basis and then re-diagonalize to extract a new set of $N_s+\Delta$ 
energies, $E^2_1,\cdots ,E^2_{N_s+\Delta}$, and eigenstates, $|E\rangle^2_1,\cdots ,|E\rangle^2_{N_s+\Delta}$.  We again order the eigenstates in energy and toss away the top most $\Delta$ states.  If we re-express the remaining eigenstates, $|E\rangle^2_k$ in terms of the original conformal basis we obtain
\begin{eqnarray}
|E\rangle^2_k &=& A^2_{km}|E\rangle^1_m + B^2_{km}|m+N_s+\Delta\rangle\cr\cr
&=& \big(A^2A^1\big)_{km}|m\rangle + \big(A^2B^1\big)_{km'}|m'+N_s\rangle\cr\cr
&& + B^2_{km'}|m'+N_s+\Delta\rangle,
\end{eqnarray}
where again $A^2$ is an $N_s\times N_s$ matrix and $B^2$ is an $N_s\times \Delta$ matrix.  Here the index $m$ runs from $1$ to $N_s$ while the index $m'$ runs from $1$ to $\Delta$.

We can repeat this procedure {\em ad libitum}: we order the new set of $N_s+\Delta$ states, toss away the topmost $\Delta$ states, add the next $\Delta$ unperturbed states from the CFT, reform the Hamiltonian, re-diagonalize, etc. In this way, we allow the higher energy states of the unperturbed CFT to mix in with eigenstates of the full (but truncated) theory.  As we keep $N_s+\Delta$ fixed at each step, the associated computational problem grows only as the square of the total number of states kept [owing to the need to manipulate states (albeit only $N_s+\Delta$ of them) which are expressed in terms of an ever growing basis as the NRG proceeds].  

At the $n$th-iteration, the eigenstates have the form
\begin{eqnarray}
|E\rangle^n_k &=& \big(A^n\cdots A^1\big)_{km}|m\rangle + \big(A^n\cdots A^2B^1\big)_{km'}|m+N_s\rangle \cr\cr
&& \hskip 0in +
\big(A^n\cdots A^3 B^2\big)_{km'}|m'+N_s+\Delta\rangle + \cdots \cr\cr 
&& + \big(A^n B^{n-1}\big)_{km'}|m'+N_s+(n-2)\Delta\rangle\cr\cr
&& + B^n_{km'}|m'+N_s+(n-1)\Delta\rangle.
\end{eqnarray}
We see that each term in the above sum has a matrix product state form.  The approximation encoded in the NRG is then one where we study a Hamiltonian arrived at by projecting the original Hamiltonian onto a space composed of matrix product states of the above form. 

What we are in effect are doing is constructing a variational wavefunction for the true eigenstates of the full Hamiltonian.  The quality of the variational ansatz is controlled by $N_s$ (the number
of states kept at the end of each step) and $\Delta$ (the number of states thrown away in each step). The larger $N_s$ and $\Delta$ are, the more accurate the results of the NRG procedure.  Typically we have found that taking $N_s \sim 1000-4000$ and $\Delta \sim 500-1000$ leads to robust results (answers that are accurate to 3-4 significant digits).\footnote{Although we must stress that the exact values of $(N_s,\Delta)$ needed to produce a given accuracy from the NRG+TSA is a model dependent statement.}

As an example, in Table~\ref{NRG_data} we present data for the ground state and first two excited energies of the sine-Gordon model at $\beta^2=1/2$, in its charge zero sector, as computed using the NRG for different values of $(N_s,\Delta)$.  We allow the NRG to proceed so that states with energies below a (dimensionless) cutoff of $N=22$ are accounted for (39279 in total).  We compare our NRG results to a straight ED for this same cutoff.  For the smallest pair of values $(N_s,\Delta)=(500,250)$ we obtain 4 significant digit agreement between the NRG and the ED. For the largest $(N_s,\Delta)=(1500,500)$ this has improved to 5 significant digit agreement. Here we have limited ourselves to a relatively small cutoff ($N=22$ with 39279 states in the
Hilbert space) so that we could compare it to an ED.  In principle the NRG can go to cutoffs far higher (up to $10^6$ states)~\cite{konik2011exciton,konik2015predicting,brandino2015glimmers,caux2012constructing}.

While the NRG allows one to go to higher cutoffs than the plain TSA, it can only do so much.  The size of the Hilbert space grows exponentially with the cutoff, which then puts a limit on the cutoff that the NRG can be pushed to.  However as a natural output of the NRG algorithm, we obtain the flow of the energy of a state as a function of cutoff.  Provided the cutoff is large enough then this flow can be described by a one-loop-like RG equation~\cite{konik2007numerical}
\begin{equation}\label{nrg_rg}
\frac{d\Delta E}{d\log N } = -\alpha \Delta E,
\end{equation}
where $\Delta E = E(N)-E(N=\infty)$ is the deviation of the energy from its value in a theory with no cutoff. The numerical NRG data can then be fit to this equation, allowing us to obtain a value for $E(N=\infty)$, in effect removing the cutoff entirely.  There are two issues in using this procedure: 1) knowing if you are in a regime where this one-loop equation is valid; 2) the value of $\alpha$. In general, 1) is a difficult question to answer as the subleading terms to Eq.~\fr{nrg_rg} will depend on the particular model. We will see this more clearly in the next section, where we will see that subleading terms arise from operator product expansions (OPEs) of the perturbing operator with itself. Practically, however, we have found that provided the NRG procedure is within 5\% of the exact value, the one-loop extrapolation encoded in Eq.~\fr{nrg_rg} can reduce the error to well below 1\%.

The second question concerns the appropriate value of $\alpha$.  For the energies, second order perturbation theory relates $\alpha$ to the scaling dimension of the perturbing operator.  However for perturbation theory to be accurate we need $\lambda (R/N)^{2-2\Delta_{\pp}}$ to be a small parameter, where $\Delta_{\pp}$ is the chiral scaling dimension of the perturbing operator and $\lambda$ is the strength of the perturbation.  We may well be forced to work at system sizes $R$ where this parameter is not small (i.e., where higher order terms in the perturbation theory are not sufficiently small). In these cases, we have found that taking $\alpha =1$ as a heuristic leads to robust results~\cite{konik2007numerical}.

\subsubsection{Applying the NRG and its extrapolations to the ground state energy for the sine-Gordon model at $\beta^2=1/2$}

\begin{figure}
\includegraphics[width=0.45\textwidth]{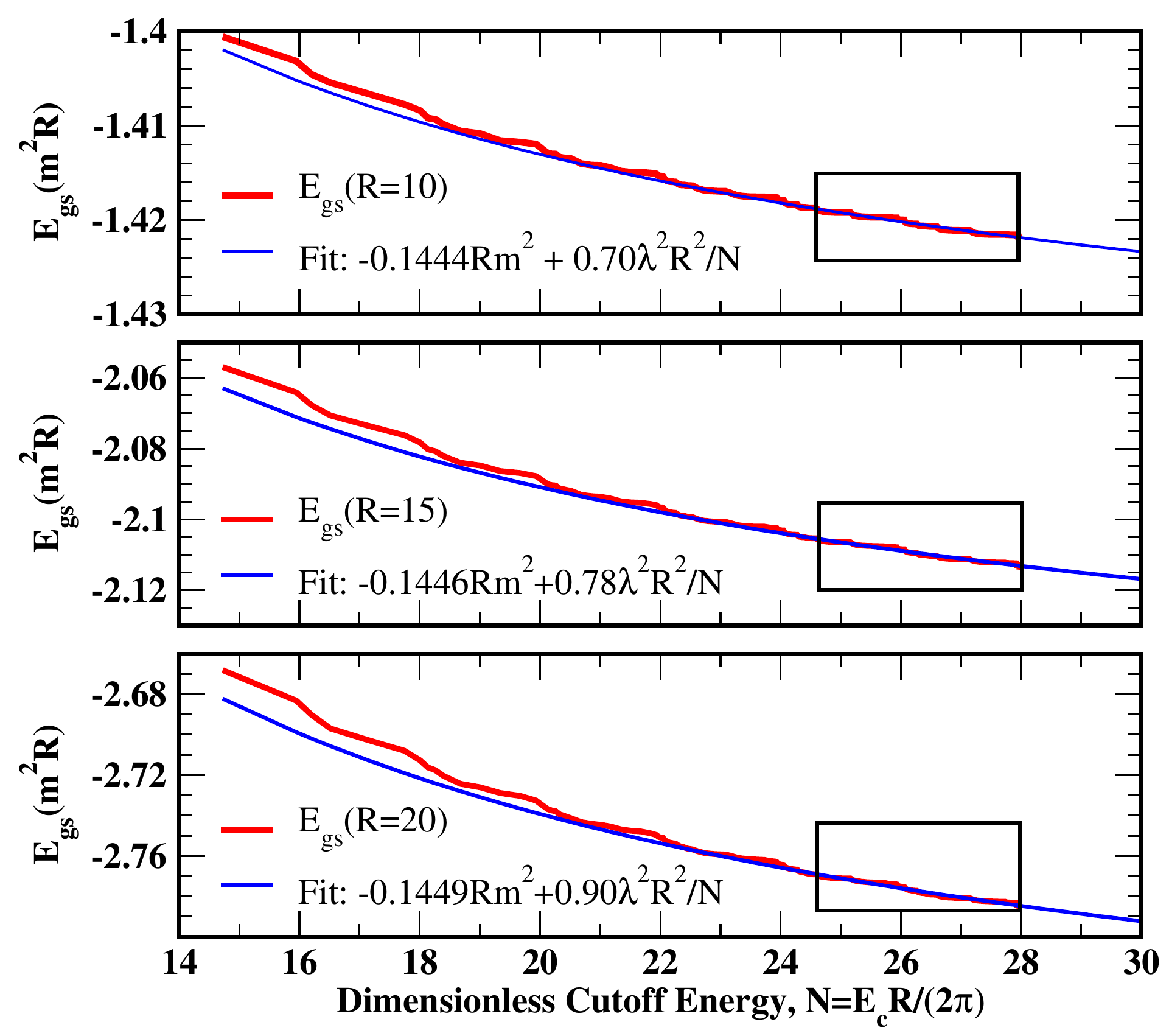}
\caption{The ground state energy at three different values of the system size, R ($R=10,15,20$, upper to lower), computed for the sine-Gordon model at $\beta^2=1/2$ as determined using the NRG. The energies are plotted as a function of the cutoff energy (effectively the NRG step). In red are presented the raw NRG data and in blue the fits to this data.  The black squares mark the region over which the fit is made.  The exact value of the ground state energy is $E_{{\rm gs},\beta^2=1/2}  = -\frac{1}{4}\tan\big(\frac{\pi}{3}\big)m^2R=-0.144338m^2R$.}
\label{SGNRG}
\end{figure}

We will now consider some specific examples where we apply the NRG together with the extrapolation encoded in Eq.~\fr{nrg_rg}. In Fig. \ref{SGNRG} we plot the evolution of the ground state energy as a function of the cutoff $N$ for three different system sizes (one in each panel).  We run the NRG to a cutoff energy of $N=28$.  We then use the region of energy flow marked by the boxes in each panel to extrapolate the energies to $N=\infty$.  For these extrapolations, we use the value of $\alpha$ determined from the leading order term (corresponding to the identity operator in the OPE of the $\cos(\beta\Theta)$ with itself) arising in second order perturbation theory [see Eq.~\fr{dH2}].  With this value of $\alpha$, the analytic fitting form is dictated to be
\begin{equation}
E_{gs,\beta^2=\frac12}(N) = E_{gs,\beta^2=\frac12}(N=\infty) + \frac{\gamma\lambda^2R^2}{N}.
\end{equation}
Nominally the perturbation theory of the next section gives $\gamma=1/2$.  However, here we treat $\gamma$ as a fitting parameter and find $\gamma=0.70$ at $R=10$ to $\gamma=0.90$ at $R=20$. That the value of $\gamma$ overshoots the analytically expected number indicates that at these values of $R$ higher order terms are non-negligible. In allowing $\gamma$ to be a fitting parameter, we see that the energies so extrapolated to $N=\infty$ agree well (within three to four significant digits) with the exact answer, $E_{gs,\beta^2=1/2}= -0.144338...m^2R$.  This also provides heuristic insight into why taking $\alpha=1$ in~\cite{konik2007numerical} produced accurate extrapolations: by assuming the approach of $E(N)$ to its $N=\infty$ counterpart was slower than predicted by second order perturbation theory, it mimicked the effects of higher order terms. In general, the second order corrections lead to underestimates in the energies as we will see in greater detail in the next section. 

\subsubsection{Applying the NRG and its extrapolations to the matrix elements of the $E_7$ deformation of the tricritical Ising model}

We have discussed heretofore the application of the NRG and its one-loop extrapolations to the problem of eliminating the cutoff dependence to TSA energies.  This approach is not simply restricted to energies, but works for matrix elements of operators as well.  The one-loop equation in Eq.~\fr{nrg_rg} has a similar form as before, but now arises from first order perturbation theory.  The coefficient $\alpha$ in the equation differs from that appropriate to the energies; from first order perturbation theory $\alpha$ will depend upon the scaling dimension of the perturbing operator, the operator whose matrix element is being considered, as well as operators appearing in the OPE of the two.

\begin{figure}
\includegraphics[width=0.45\textwidth]{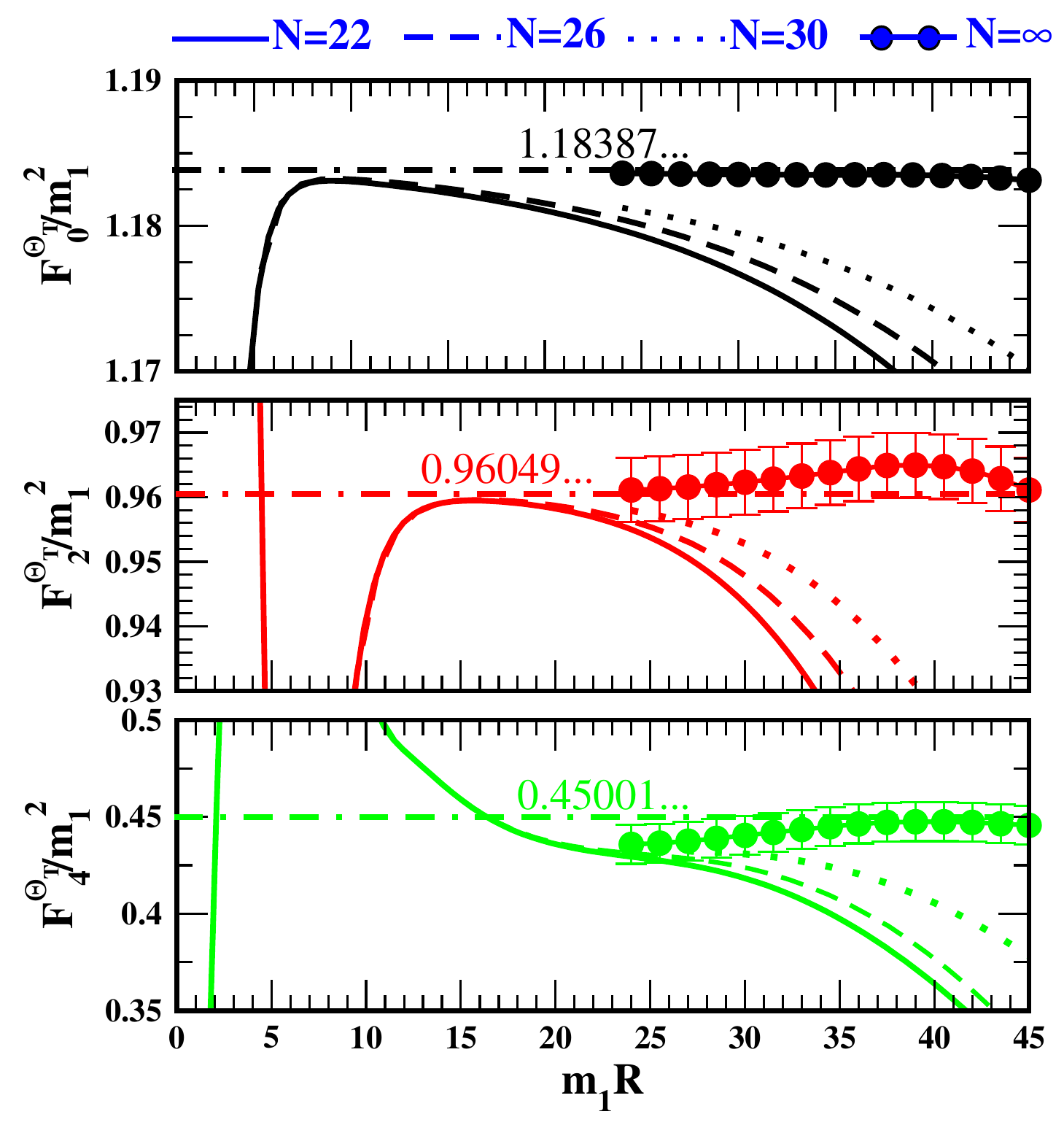}
\caption{TCSA data for the matrix elements $F^{\Theta_T}_i$ (with $i=0,2,4$) in the tricritical Ising model perturbed by the leading energy perturbation as a function of the system size, $R$. The upper panel ($i=0$) is the vacuum expectation value of the stress-energy tensor, $\Theta_T$, whilst $i=2,4$ describes the matrix elements of the operator with the first and second even excitations. Three values of the cutoff, $N=22,26,30$, are shown alongside extrapolations to the $N=\infty$ limit (see text). The extrapolated values have error bars indicating the uncertainty arising from different choices of the energy region over which one performs the extrapolation. The dot-dashed horizontal lines denote the exact values of the matrix elements.}
\label{isingE7MENRG}
\end{figure}

To study how to apply the NRG to the computation of matrix elements we consider the VEV, two one-particle matrix elements, and a single two-particle matrix element of the stress energy tensor in the $E_7$ deformation of the tricritical Ising model (cf. the discussion of the previous section). We first analyze the VEV and one-particle states in Fig.~\ref{isingE7MENRG}. There we plot the values of matrix elements as a function of the system size for different cutoffs under the NRG, as well as an extrapolated value using a one-loop RG equation. As we explained in Sec.~\ref{IsingE7me}, all four of the matrix elements should nominally be independent of $R$.  However, at small $R$ we have significant finite size corrections and at large $R$ there are cutoff effects.  We thus want to see how the large $R$ TSA+NRG results for the matrix elements deviate from their (analytically) expected values. In Fig.~\ref{isingE7MENRG} these values are shown as dot-dashed lines; we see that as we increase the cutoff, the NRG values of the matrix elements bend upwards towards their exact value. We also see that this approach is relatively slow, although we obtain much better results when we extrapolate the NRG results using the one-loop equation:
\be 
\label{rg_me}
\frac{d \delta\langle 0|\Theta_T(0)|i\rangle}{d\log N } = -\alpha \delta\langle 0|\Theta_T(0)|i\rangle, 
\ee
with $\alpha=2-4\Delta_{\epsilon}-2\Delta_I=8/5$.  In all three cases we see that the extrapolated value [using Eq.~\fr{rg_me}] greatly improves the agreement between the numerics and the expected analytical value of the matrix elements; in particular, it all but eliminates any dependence on system size at large $R$.

\begin{figure}
\includegraphics[width=0.45\textwidth]{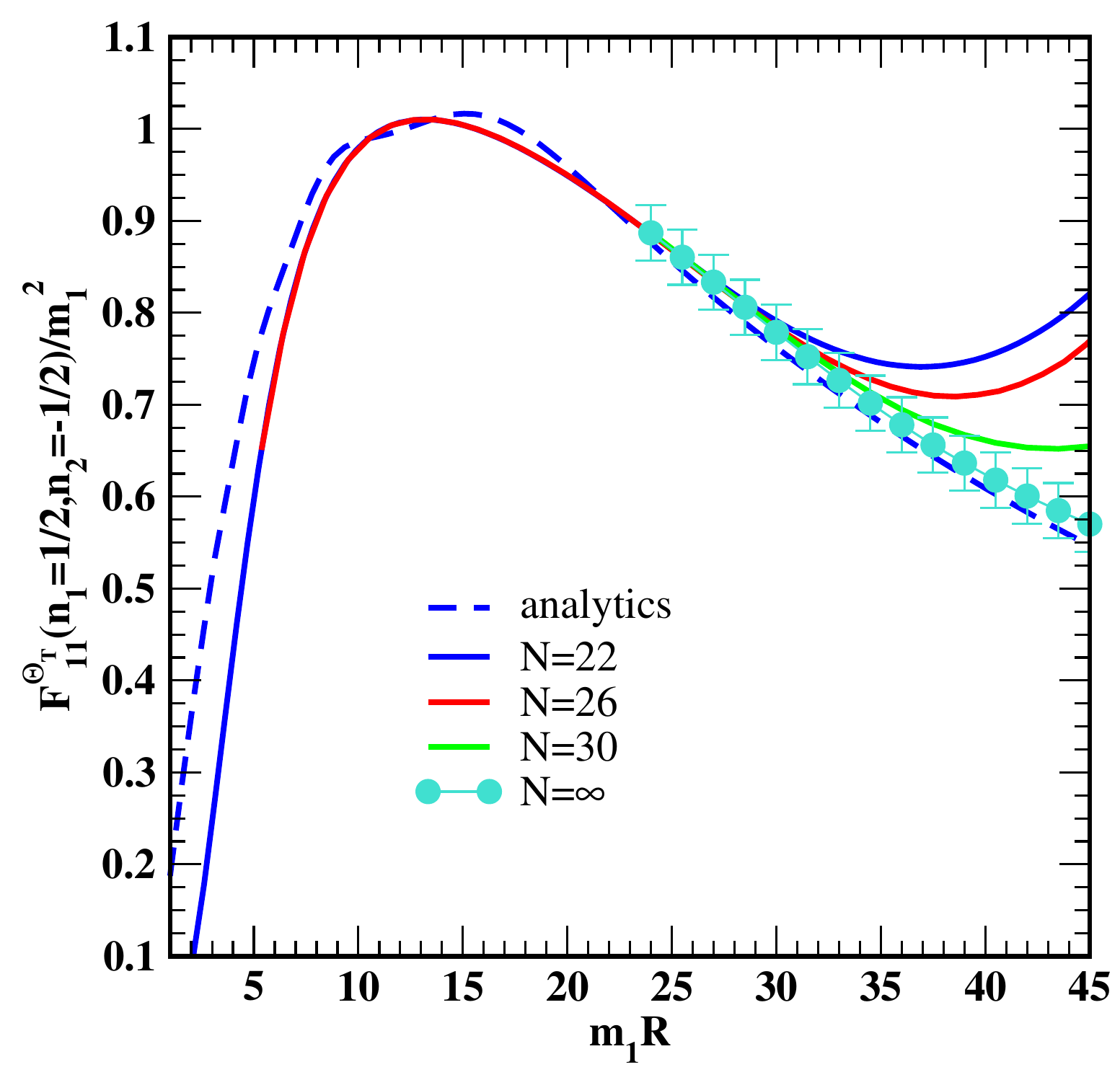}
\caption{TSA data for the two-particle matrix element of the trace of the stress-energy for the tricritical Ising model perturbed by the leading energy perturbation as a function of the system size, $R$.  Data is presented for different TSA cutoffs, $N$, and we plot the extrapolated value for the matrix element at $N=\infty$. To compare, we also present the matrix element's exact analytical value computed using the form factor bootstrap (dashed line). At small values of the system size $R$ the agreement between the numerics and the analytical result is imperfect due to finite size effects not taken into account in the analytical treatment. For large systems, there are deviations due to cutoff effects in the TSA, but we see that the extrapolated $N=\infty$ values agree well with the analytical result.}
\label{Ising2pMENRG}
\end{figure}

We find much the same behavior for the large R behavior of two-particle matrix elements. In Fig.~\ref{Ising2pMENRG} we see that at large $R$ the TSA+NRG values of the matrix elements deviate from the exact value. As the cutoff is increased, the TSA+NRG curve approaches the exact curve, but only slowly. Once again extrapolating the value of the matrix element to $N=\infty$, we find that the result agrees (within the error of the extrapolation) with the exact curve.

The improvement that using the TSA+NRG+extrapolation offers is clearly more significant for matrix elements than for the ground state energies.  This is largely due to the greater sensitivity of matrix elements to cutoff effects.

\subsection{An analytical renormalization group for the TSA}\label{analyticrg}

In the previous section we considered how to adapt the NRG (first developed to study quantum impurity problem) to alleviate the effects of the cutoff in TSA computations. In this section we turn our attention to a complementary analytical approach by which we can remove the effects of the cutoff.
The idea behind this was first introduced in \cite{feverati2008renormalization} and then further elaborated on in \cite{watts2012renormalisation,giokas2011renormalisation,lencses2014excited,hogervorst2015truncated}.

To see how we can do this, we follow the discussion in Ref.~\cite{hogervorst2015truncated}. We first take our Hilbert space, $\Ha$, and divide it into two parts: $\Ha$ = $\Hl$ $\otimes$ $\Hh$.  Here $\Hl$ consists of all states whose unperturbed energies are equal to or lower than $\C$, while $\Hh$ consists of all states whose unperturbed energies are greater than $\C$. Then, we can write our Hamiltonian in the following manner:
\begin{equation}\label{e1}
H = \begin{bmatrix}
H_{ll} & H_{lh} \\
H_{hl} & H_{hh} 
\end{bmatrix},
\end{equation}
where $H_{ij}$ ($i,j=h,l$) corresponds to the Hamiltonian matrix restricted to the two subdivisions of the Hilbert space. If we have an eigenstate 
\be
\begin{bmatrix}
c_l \\
c_h 
\end{bmatrix},
\ee
with energy $E$, we can write the Schr\"odinger equation as
\begin{eqnarray}\label{e2}
H_{ll}c_l + H_{lh}c_h &=& Ec_l,\cr\cr
H_{hl}c_l + H_{hh}c_h &=& Ec_h .
\end{eqnarray}
By eliminating $c_h$ from the above set of equations, we have 
\begin{equation}\label{e3}
\bigg(H_{ll} + H_{lh}\frac{1}{E-H_{hh}}H_{hl}\bigg)c_l = (H_{ll} + \delta H )c_l = Ec_l.
\end{equation}
In doing so, we have reformulated the eigenvalue problem in terms of coefficients of states that live in the low energy Hilbert space alone. The Hamiltonians we are studying take the form $H = H_0 + \lambda V$ where $V = \int^R_0 \rd x\, \pp(x)$.  We can then expand $\delta H$ in powers of $\lambda$, giving
\begin{eqnarray}\label{e4}
\delta H &=& -\lambda^2 V_{lh} \frac{1}{H_0-E} V_{hl} \cr\cr
&& + \lambda^3 V_{lh}\frac{1}{H_0-E}V_{hh}\frac{1}{H_0-E}V_{hl} + O(\lambda^4).~~
\end{eqnarray}
Introducing the (imaginary) time dependence of operators in the interaction picture,
\be
{\cal O}(\tau) = e^{H_0\tau}{\cal O}(0)e^{-H_0\tau},
\ee
we can rewrite Eq.~\fr{e4} as 
\begin{eqnarray}\label{delH}
\delta H &=& -\lambda^2 \sum_{c\in \Hh} \int^\infty_0 \rd\tau\, e^{(E-H_0)\tau} V(\tau)|c\rangle\langle c| V(0) + O(\lambda^3) \cr\cr
&\equiv &
\delta H_2 +  O(\lambda^3).
\end{eqnarray}
Evaluating the matrix elements of $\delta H_2$ with respect to the states $|a\ra,\ |b\ra$ in the unperturbed basis, we end up evaluating expressions of the form
\bea
D_{ab}(\tau) = \int^R_0 dx_1 dx_2 \sum_{c\in\Hh}&& \langle a|\pp(x_1,\tau)|c\rangle \nn
&& \times \langle c|\pp(x_2,0)|b\rangle.
\eea
The key to readily computing this quantity is the use of OPEs.

The OPE of the perturbing field (assuming a spacetime geometry of an infinite cylinder) can be written as
\bw
\begin{eqnarray}\label{e6}
\pp(x_1,\tau)\pp(x_2,0) &=& \sum_\varphi \bigg(\frac{R}{2\pi}\bigg)^{2\Delta_\varphi-4\Delta_\pp} C_{\varphi\pp\pp}|z_1-z_2|^{-4\Delta_\pp+2\Delta_\varphi}|z_1|^{2\Delta_\pp}|z_2|^{2\Delta_\pp-2\Delta_\varphi}\varphi(x_2,0) + \ldots, 
\end{eqnarray}
\end{widetext}
where 
\be
z_1 = e^{-\frac{2\pi}{R}(\tau + ix_1)}, \quad z_2 = e^{-\frac{2\pi}{R}ix_2}, \quad \bar z_i = z_i^*, 
\ee
and $\Delta_\pp$/$\Delta_\varphi$ is the chiral scaling dimension of $\pp$/$\varphi$, 
$C_{\varphi\pp\pp}$ is the structure constant for $\langle\varphi\pp\pp\rangle$, and the ellipses 
denote less singular terms in the OPE.  
We assume here that the left, $\Delta_{\cal O}$, and right, $\bar\Delta_{\cal O}$ scaling dimensions
for all operators ${\cal O}$ are the same.

However, the OPE~\fr{e6} is not quite what we want; instead, we would like to evaluate a modified OPE
\be
\sum_{c\in\Hh}\pp(x_1,\tau)|c\rangle\langle c|\pp(x_2,0), \label{resOPE}
\ee
where a partial resolution of the identity involving high energy states has been inserted between the fields. To see how one can evaluate such a quantity, we ask the following question: what is the temporal dependence of $D_{ab}(\tau)$? We see that $D_{ab}(\tau)$ can be rewritten as
\begin{eqnarray}
D_{ab}(\tau) &=& \sum_{c\in\Hh} e^{-(E_c-E_a)\tau} \cr\cr
&&\times \int^R_0 \rd x_1 \rd x_2  \pp_{ac}(x_1,0)\pp_{bc}(x_2,0), \quad
\end{eqnarray}
where $\pp_{ac} = \langle a|\pp |c\rangle$. We also see that $D_{ab}(\tau)$ only involves terms $e^{-E_c\tau}$ with $E_c > \Lambda \equiv 2\pi N/R$, where we have introduced
$N$ as a dimensionless cutoff. Thus our strategy will be to evaluate $D(\tau)$ using the original, unrestricted OPE and then throw away terms involving powers of $e^{-2\pi\tau/R}$ smaller than $N$. Identification of these powers will be possible through the Taylor series expansion:
\begin{equation}\label{e7}
(1-z)^{-a} = \sum^\infty_{n=0} \frac{1}{n!}\frac{\Gamma(a+n)}{\Gamma(a)}z^n \equiv \sum_{n=0}^\infty S(n,a)z^n.
\end{equation}
With this at hand, we are able to write $D(\tau)$ as
\bw
\begin{eqnarray}\label{e8}
D_{ab}(\tau) &=& \int^R_0 \rd x_1 \rd x_2 \sum_{c\in\Hh} \langle a| \pp(x_1,\tau)|c\rangle\langle c|\pp(x_2,0)|b\rangle, \nn
&=& \sum_{\varphi} \bigg(\frac{R}{2\pi}\bigg)^{2\Delta_\varphi-4\Delta_\pp} C_{\varphi\pp\pp}
\sum_{2n>N}S^2(n,2\Delta_\pp-\Delta_\varphi) e^{-\frac{2\pi (2n+2\Delta_\pp)\tau}{R}}
 R\int^R_0dx_2\langle a|\varphi(x_2,0) |b\rangle.
\end{eqnarray}
In the above, we see that our sum is restricted so that $2n > N$, which corresponds to our restricted OPE~\fr{resOPE}.

Taking the form~\fr{e8} for $D_{ab}(\tau)$, and using it to evaluate $(\delta H_{2})_{ab}$ we obtain
\begin{eqnarray}\label{delHab}
(\delta H_2)_{ab} & = & -\lambda^2  R^2\sum_\varphi \bigg(\frac{R}{2\pi}\bigg)^{2\Delta_\varphi-4\Delta_\pp}C_{\varphi\pp\pp}
 \int^\infty_0 \rd\tau e^{\tau(E-E_a)}\delta_{P_a,P_b}\langle a|\varphi(0,0)|b\rangle
 \sum_{2n>N}\!\!S^2\bigg(n,2\Delta_\pp-\Delta_\varphi\bigg) e^{-\frac{2\pi (2n+2\Delta_\pp)\tau}{R}}\cr\cr
&=& -\lambda^2 R^2 \sum_\varphi \bigg(\frac{R}{2\pi}\bigg)^{2\Delta_\varphi - 4\Delta_{\pp}} C_{\varphi\pp\pp}
\sum_{2n>N}S^2\bigg(n,2\Delta_\pp-\Delta_\varphi\bigg) \delta_{P_a,P_b}
\bigg(\frac{1}{E_a-E+\frac{2\pi}{R}(2n+2\Delta_\pp)}\bigg) \langle a|\varphi(0,0)|b\rangle,\nn
\end{eqnarray}
\ew
where $\delta_{P_a,P_b}$ indicates that the momentum of states $|a\rangle$ and $|b\rangle$ must be the same.  Thus, we have succeeded in writing $\delta H_2$ as a sum over single fields.   As written, we can compute the correction to the eigenstate $\sum_{a\in\Hl} c_a |a\rangle$ with energy $E$ via
\be
\delta E = \sum_{a,b\in \Hl} c_a c_b (\delta H_2)_{ab}.
\ee
However if we approximate $\delta H_2$ by dropping the dependence on $E-E_a$ (which is weak provided $\Lambda \gg E$), we can use $\delta H_2$ much more expeditiously. We can study the theory $H_0 + \lambda V_{ll} + \delta H_2$ equipped with the cutoff $\Lambda$. This theory is no harder to diagonalize than the original.  However, by doing so we find the the entire low-lying spectrum of the theory, $H_0 + \lambda V$ without cutoff, making relative errors of $O(\lambda (\frac{R}{N})^{2-4\Delta_\pp}),\ O(\frac{E}{\Lambda})$. The reader may note that matrix $(\delta H_2)_{ab}$ is not symmetric in $a,b$.  This is a consequence of our choice in Eq.~\fr{delH}, where manifest Hermiticity is lost, together with dropping less singular terms in the OPE. If instead of Eq.~(\ref{delH}), we represent $\delta H_2$ via
\be
\delta H_2 = -\lambda^2 \sum_{c\in \Hh} \int^\infty_0 \rd\tau e^{E\tau} V(\tau)|c\rangle\langle c| V(0) e^{-H_0\tau},
\ee
and take the OPE of the perturbing field with itself to read (this essentially involves a different choice of subleading terms in the OPE)
\bw
\be
\pp(x_1,0)\pp(x_2,-\tau) = \sum_\varphi \bigg(\frac{R}{2\pi}\bigg)^{2\Delta_\varphi-4\Delta_\pp} C_{\varphi\pp\pp}
 |z_1-z_2|^{-4\Delta_\pp+2\Delta_\varphi}|z_1|^{2\Delta_\pp}|z_2|^{2\Delta_\pp-2\Delta_\varphi}\varphi(x_1,0) + \ldots,
\ee
with $z_1 = e^{-i\frac{2\pi}{R}x_1}$, $z_2 = e^{-\frac{2\pi}{R}(-\tau+ix_2)}$, we obtain instead 
\be\label{delHabp}
(\delta H'_2)_{ab}
= -\lambda^2 R^2 \sum_\varphi \bigg(\frac{R}{2\pi}\bigg)^{2\Delta_\varphi - 4\Delta_{\pp}} C_{\varphi\pp\pp}
\sum_{2n>N}S^2\bigg(n,2\Delta_\pp-\Delta_\varphi\bigg) \delta_{P_a,P_b}
\bigg(\frac{1}{E_b-E+\frac{2\pi}{R}(2n+2\Delta_\pp)}\bigg) \langle a|\varphi(0,0)|b\rangle.
\ee
This is the same expression as in Eq.~(\ref{delHab}) with $E_a$ replaced with $E_b$ (we mark $(\delta H_2)_{ab}$ with a prime to indicate this swap). This rewriting of $\delta H_2$ then opens up the possibility of studying the fully Hermitian Hamiltonian $H_0 + \lambda V_{ll} + \frac{1}{2}(\delta H_2 + \delta H'_2)$ as a model that is free of cutoff effects [to $O(\lambda^2)$]. where instead of dropping $E-E_a,E-E_b$ in $\delta H_2$ and $\delta H'_2$ respectively, we replace them with $H_{ll}-H_0=\lambda V$.  This would remove $O(\frac{E}{\Lambda})$ errors we were making by ignoring the $E-E_a$ dependence previously of $\delta (H_2)_{ab}$.

\subsection{Third order contributions to $\delta H$}
One can also consider the third order contribution to $\delta H$. This can be put in a similar form to Eq.~\fr{delH} 
\be
\delta H_3 = \sum_{c,d \in H_h} \int^\infty_0 \rd\tau_1 \int^0_{-\infty} \rd\tau_2 e^{(E-H_0)\tau_1} V(\tau_1 )|c\rangle
 \langle c|V_{hh}(0)|d\rangle\langle d| V(\tau_2) e^{-(E-H_0)\tau_2}.
\ee
Manipulations, along the same lines as those for $\delta H_2$, allow us to rewrite this in the form
\begin{eqnarray}
\delta H_3 &=& \lambda^3 R^3 \sum_{\varphi,\varphi'}C_{\varphi\pp\pp}C_{\varphi'\varphi\pp}\bigg(\frac{R}{2\pi}\bigg)^{2+2\Delta_{\varphi'}-6\Delta_\pp}
\sum_{n,m > N/2}S^2\bigg(n,2\Delta_\pp-\Delta_\varphi\bigg) S^2\bigg(n,\Delta_\varphi+\Delta_\pp-\Delta_\varphi'\bigg)\cr\cr
&& \times \frac{1}{\frac{R}{2\pi}(H_0-E)+2\Delta_\pp + 2n}\varphi'(0,0) \frac{1}{\frac{R}{2\pi}(H_0-E)+(4\Delta_\varphi + 2n+2m)}.
\end{eqnarray}
\ew
Now, provided we once again drop the $(H_0-E)$ dependence in the denominators, we can add it to the Hamiltonian as a Hermitian `counter term' to eliminate errors of $O(\lambda^3)$ arising from our use of the cutoff. As at the end of the last section, if we are willing to play with the representation of $\delta H_3$ (as an integral over time dependent operators, as well as
the exact form of the OPE), we can arrive at a Hermitian form for $\delta H_3$ where the $(H_0-E)$ terms in the denominators are kept.

\subsection{Dependence on cutoff}

Despite writing explicit forms for $\delta H_2$ and $\delta H_3$, we have not exhibited their dependence on the dimensionless cutoff $N$. To do so, we note that for large $n$, $S(n,a)$ is given by
\begin{equation}
S(n,a) \sim n^{a-1}.
\end{equation}
As a result, $\delta H_2$ has the form
\begin{eqnarray}\label{dH2}
\delta H_2 &\sim& \lambda R^{1-2\Delta_\pp} \times \lambda\bigg(\frac{R}{N}\bigg)^{2-2\Delta_\pp} \cr\cr
&& \times \sum_{\varphi} C_{\varphi\pp\pp}N^{2\Delta_\pp-2\Dvp} 
R^{2\Dvp}\langle \varphi\rangle,
\end{eqnarray}
where the last term, $R^{2\Dvp}\langle \varphi\rangle$, is a dimensionless $O(1)$ number. The size of $\delta H_2$ is controlled by a factor, $\lambda R^{1-2\Delta_\pp}$, that sets the energy scale of the correction multiplied by a dimensionless factor, $\lambda(\frac{R}{N})^{2-2\Delta_\pp}$, that characterizes how convergent the perturbation theory (in $\lambda$) is for a particular system size, $R$, as well as the dimensionless cutoff $N$. There is an additional cutoff dependence, $N^{2\Delta_\pp-2\Dvp}$, whose exact effect depends on the relevancy of the perturbing operator relative to the operators, $\varphi$, that appear in the OPE of $\pp$ with itself.

Similarly we see that $\delta H_3$ is of the order
\begin{eqnarray}
\delta H_3 &\sim& \lambda R^{1-2\Delta_\pp} \times \lambda^2\bigg(\frac{R}{N}\bigg)^{4-4\Delta_\pp} \cr\cr
&& \hskip -.5in \times \sum_{\varphi,\varphi'}C_{\varphi\pp\pp}C_{\varphi'\varphi\pp}
N^{2\Delta_\pp-2\Delta_{\varphi'}} R^{2\Delta_{\varphi'}}\langle \varphi'\rangle.
\end{eqnarray}
This can continued to the $n$th order contribution:
\begin{equation}
\delta H_n \sim \lambda R^{1-2\Delta_\pp} \times \lambda^{n-1}\bigg(\frac{R}{N}\bigg)^{2n-2-2(n-1)\Delta_\pp},
\end{equation}
where here we have dropped the dependency on the dimensional factor arising from the operators that appear in the OPE of $\pp$ with itself.

\subsection{Examples of perturbative improvement upon the TSA+NRG}

\begin{figure}
\includegraphics[width=0.45\textwidth]{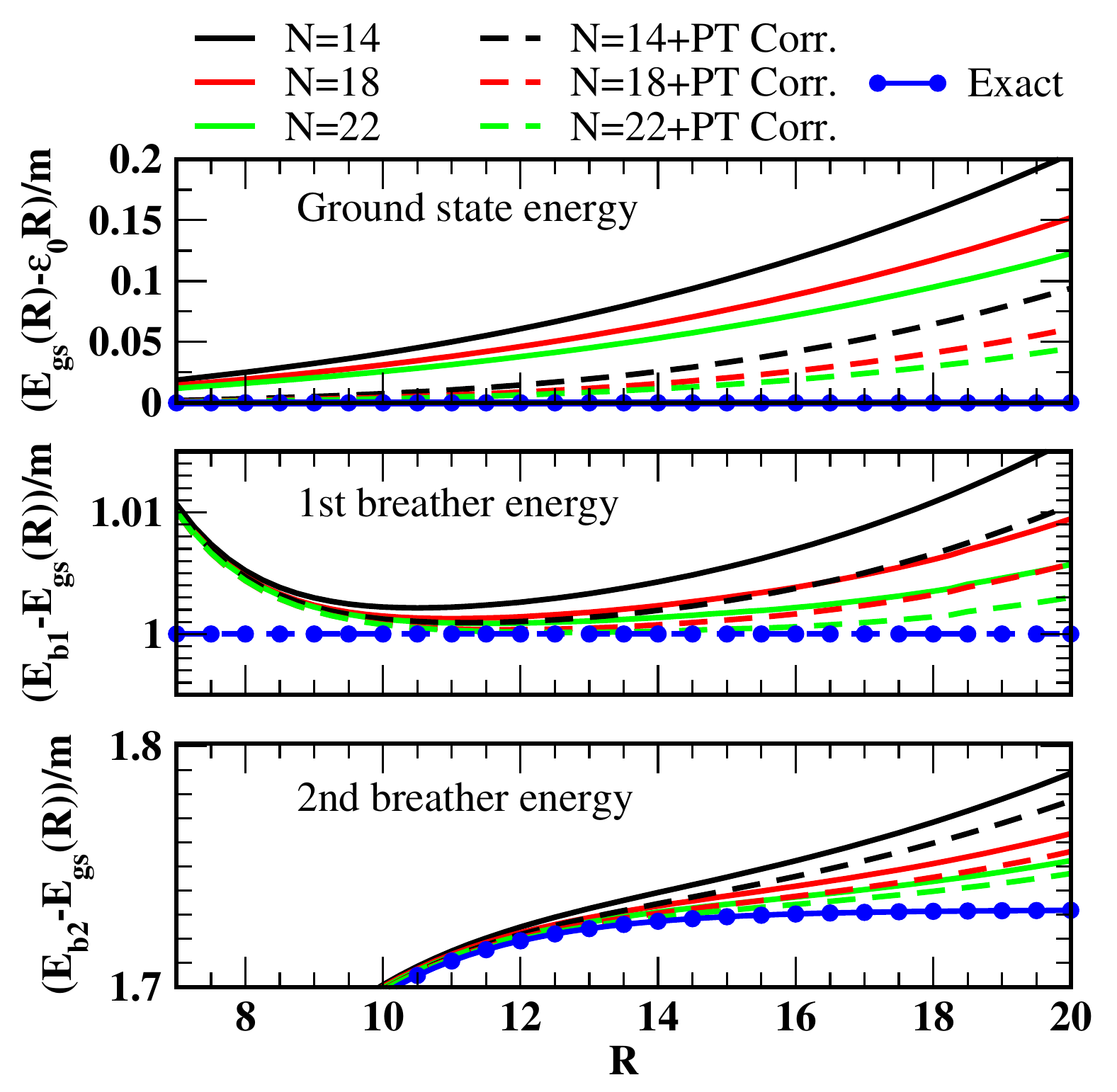}
\caption{The perturbative energy corrections for the sine-Gordon model at $\beta^2=1/2$. In the three panels we show TSA energy data as a function of system size, $R$, both without (solid lines) and with (dashed lines) the perturbative correction accounting for the effects of the cutoff.  The top, middle, and bottom panels consider the ground state energy, the first breather energy, and the second breather energy, respectively.  The blue line with circle symbols is the exact energy of these states at a given R.  The data is given for three different dimensionless cutoffs, $N=14, 18$, and $22$.}
\label{sg_pt}
\end{figure}

We will now consider two examples where Eq.~\fr{delHab} is used to improve upon the raw TSA+NRG results.  In the first example, we study the corrections to the energies of the ground state and first two excited state energies for the sine-Gordon model at $\beta^2=1/2$. In Fig.~\ref{sg_pt} we plot these energies as a function of system size for a number of different cutoffs.  The solid lines are the energies arrived at using the TSA+NRG and the dashed lines are those energies plus the correction coming from Eq.~\ref{delHab}.  We also plot the exact values (i.e. without cutoff) of the energies (blue line with circle symbols) for comparison.  While the correction coming from Eq.~\fr{delHab} in all cases improves the answer, it does not completely eliminate the effects of the cutoff.  This is not surprising: Eq.~\fr{delHab} is perturbative in nature and so clearly we see that $O(\lambda^3)$ (and above) corrections are non-negligible. We also see that these higher order corrections grow in importance with system size $R$.  Again this reflects the fact that Eq.~\fr{delHab} works best when the perturbing operator $\pp$ leaves the theory close to its conformal UV fixed point, i.e. when $R$ is small.

\begin{figure}
\includegraphics[width=0.45\textwidth]{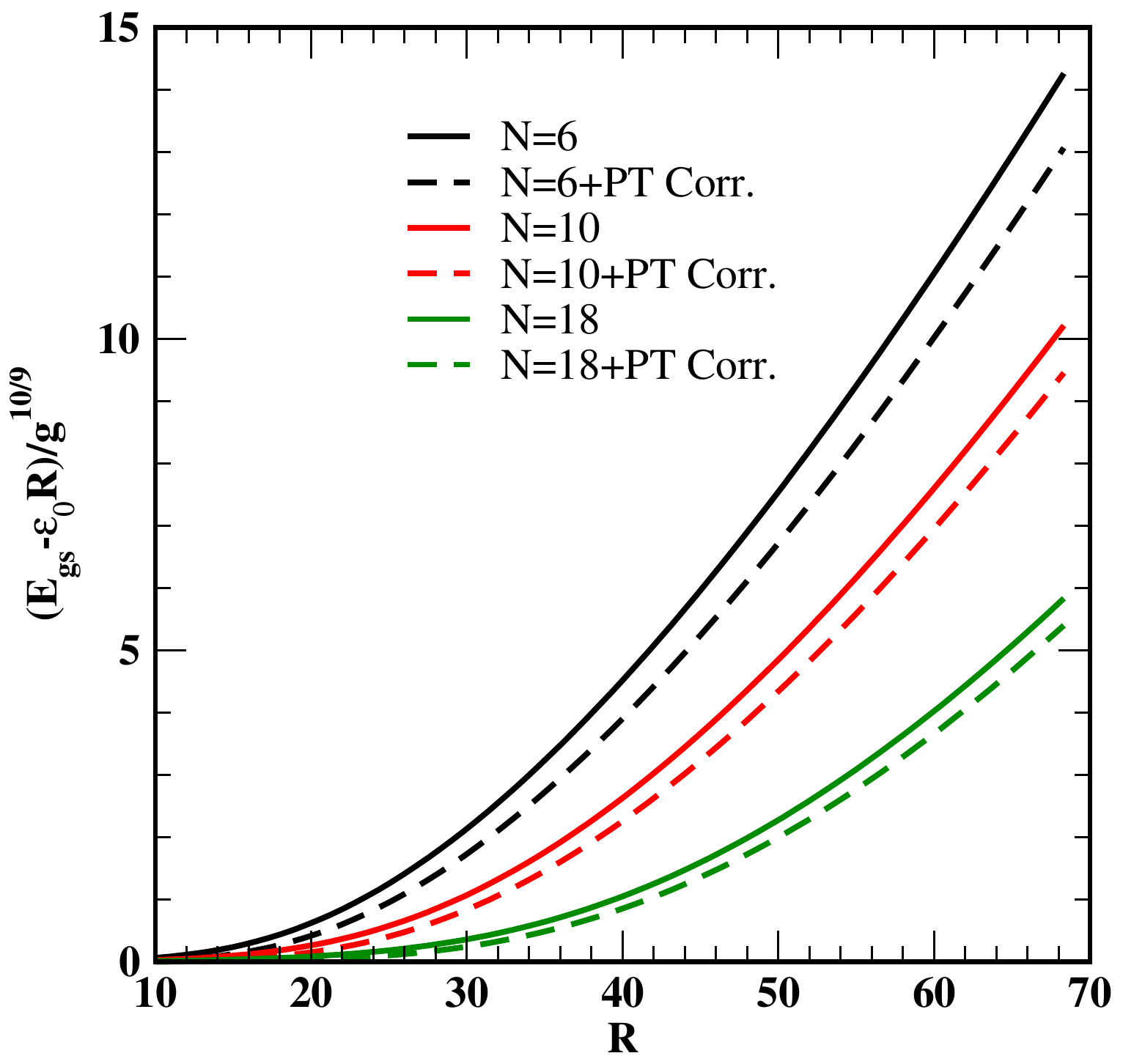}
\caption{The perturbative energy corrections for the tricritical Ising model perturbed by the leading energy perturbation.  We present data for the ground state energy (with its exact analytical
value subtracted off) for three different values of the dimensionless cutoff, $N=6, 10,$ and $18$.  The uncorrected TSA data is shown with solid lines while the corrected data is presented with dashed lines.}
\label{tricriticalIsing_pt}
\end{figure}

This behavior is not particular to the sine-Gordon model. It can also be seen in Fig. \ref{tricriticalIsing_pt} where we plot the correction Eq.~\fr{delHab} makes to the ground state energy of the tricritical Ising theory perturbed by the leading energy operator. The correction, while improving the results towards the exact value of the ground state energy, is both small as well as relatively smaller than that seen for the sine-Gordon model (where the correction term makes up roughly half the distance between the raw TSA+NRG data and the exact value of the ground state energy).

\subsection{Perturbative corrections for matrix elements}
 
When discussing the analytical derivation of corrections, we have focused on the perturbative corrections to the energy levels. We can also analyze such corrections for matrix elements; we will not do so in detail here, but will give the dependence on cutoff $N$ and system size $R$ that first order perturbation theory predicts. If we consider the matrix element
\begin{equation}
M_{ab}^{\cal O}=\langle a | {\cal O}(0)|b\rangle, 
\end{equation}
of an operator ${\cal O}$ between two states $|a\rangle$ and $|b\rangle$, then the correction to the matrix element that comes from taking into account states above the cutoff to first order in perturbation theory is 
\begin{eqnarray}\label{me_pert}
\delta M_{ab}^{\cal O} &\sim& \lambda \sum_{\varphi} C_{\varphi {\cal O}\phi_p} R^{2+2\Delta_\varphi-2\Delta_{\cal O}-2\Delta_{\phi_p}}\cr\cr
&& \times N^{-2-2\Delta_\varphi+2\Delta_{\cal O}+2\Delta_{\phi_p}} \langle a|\varphi(0)|b\rangle .
\end{eqnarray}
Here we see that the correction involves a sum over operators that appear in the operator product expansion of ${\cal O}$ with the perturbing field $\pp$. In the previous section we have used this scaling form in combination with the numerical renormalization group to dramatically improve the predictions of the values of matrix elements by the TSA.
Eq.~(\ref{me_pert}) was worked out in detail in Ref.~\cite{szecsenyi2013one}, but the dependence on $N$ alone 
can be deduced by a simple scaling analysis \cite{konik2007numerical} provided the OPE 
of ${\cal O}$ and $\phi_p$ is known.

\subsection{Resummation of higher order terms: Development of an RG equation}

So far we have only considered the leading order perturbative corrections to the energies and matrix elements coming from introducing a cutoff.  However, it is possible to resum this perturbation theory~\cite{watts2012renormalisation,giokas2011renormalisation,feverati2008renormalization} by deriving a one-loop RG equation.  To see how, suppose for the sake of simplicity that the only field appearing in the OPE of $\pp$ with itself is $\pp$, i.e. $C_{\phi_p\phi_p\varphi} = \delta_{\varphi,\phi_p}C_{\phi_p\phi_p\phi_p}$.  Then matrix elements of the correction Hamiltonian take the form [see Eq.~\fr{delHab}]
\begin{equation}
(\delta H_2)_{ab}= -\alpha\lambda^2 (2\pi)^{2\Delta_{\phi_p}}R^{1-2\Delta_{\phi_p}}\frac{R^{2-2\Delta_{\phi_p}}}{N^{2-2\Delta_{\phi_p}}}D_{ab,\phi_p},
\end{equation}
where $\alpha$ and $D_{ab,\phi_p}$ are dimensionless constants and $D_{ab,\phi_p}$ depends on the states $|a\rangle$ and $|b\rangle$.  Here we have assumed the denominators in Eq.~\fr{delHab} can be approximated by
\begin{equation}\label{approx}
\frac{1}{E_a-E+\frac{2\pi}{R}(2n+2\Delta_{\phi_p})} \rightarrow \frac{R}{4\pi n}.
\end{equation}
This is a valid approximation (in the sense that the correction terms for the approximation are suppressed by a power of the inverse cutoff) if the dimensionless energy $R E/2 \pi$ is far below the cutoff.

Now compare this matrix element with the matrix elements of the original Hamiltonian:
\begin{equation}
(H_{\rm pert})_{ab} = \lambda R^{1-2\Delta_p}(2\pi)^{2\Delta_p}D_{ab,\phi_p}.
\end{equation}
Here it looks like the effect of accounting for the states above the cutoff in second order perturbation theory is equivalent to simply replacing the coupling $\lambda$ in the original Hamiltonian by
\begin{equation}
\lambda \rightarrow \lambda - \alpha\lambda^2 \frac{R^{2-2\Delta_p}}{N^{2-2\Delta_p}}.
\end{equation}

Instead of asking what all the states above the cutoff $N$ contribute, we can instead consider only the contributions arising from states within an energy shell $[N,N+1]$
\begin{eqnarray}
(\delta H_2)_{ab}(N) - (\delta H_2)_{ab}(N+1) &=&\cr\cr
&& \hskip -1.95in  -\alpha \lambda^2 R^{3-4\Delta_p}(2\pi)^{2\Delta_p}(2-2\Delta_p)N^{2\Delta_p-3}D_{ab,\phi_p}.
\end{eqnarray}
So, if we want to compute energies in a theory with cutoff $N+1$ and coupling $\lambda_{N+1}$, we can simulate a theory with cutoff $N$ and simply choose the coupling $\lambda_N$ given by 
\begin{equation}
\lambda_{N} = \lambda_{N+1} - \alpha(2-2\Delta_p)\lambda^2_{N+1}\frac{R^{2-2\Delta_p}}{N^{3-2\Delta_p}}.
\end{equation}
In which case, we expect to obtain identical results, up to $\lambda^3$ errors. This statement can be written as a differential equation:
\begin{equation}
\frac{d\lambda_N}{dN} = \alpha(2-2\Delta_p)\lambda^2_{N}\frac{R^{2-2\Delta_p}}{N^{3-2\Delta_p}}.
\end{equation}
Integrating this equation, we find 
\begin{equation}
\lambda_N = \frac{\lambda_\infty}{1+\alpha\lambda_\infty R^{2-2\Delta_p}N^{2\Delta_p-2}}
\end{equation}
Physically, this should be understood in the following manner. A theory with cutoff $N$ and coupling $\lambda_N$ is \textit{equivalent to a theory with no cutoff and coupling $\lambda_\infty$}. This is a rather powerful formula, inasmuch as it allows us to compute contributions from very high energy states with finite numerical effort.

\begin{figure}
\includegraphics[width=0.45\textwidth]{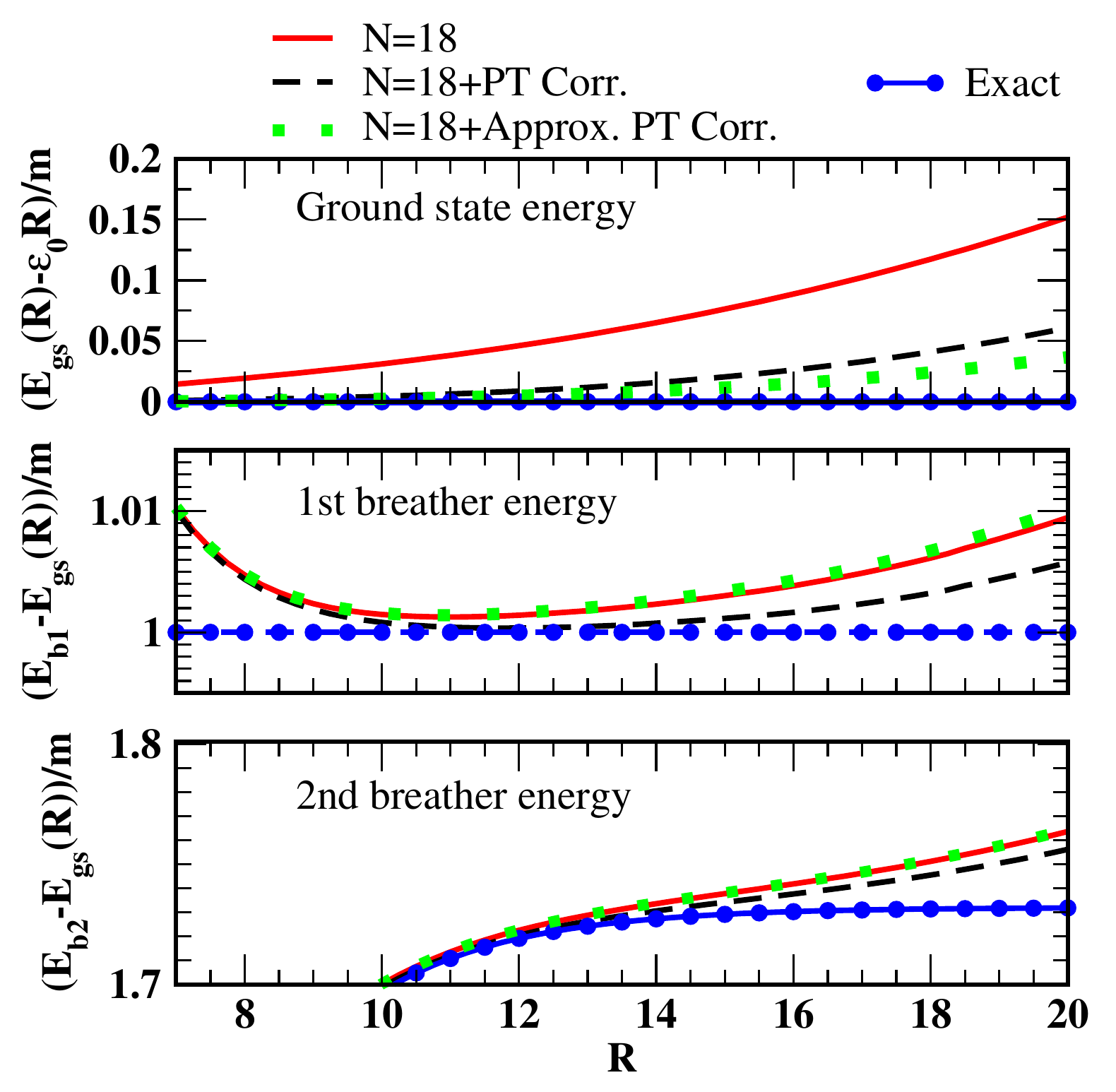}
\caption{The approximate perturbative energy corrections for the sine-Gordon model at $\beta^2=1/2$ where the state energy dependence is ignored, i.e. the denominator in Eq.~\fr{dH2} is simplified by dropping $E_a-E$. In the three panels we show TSA energy data for the cutoff $N=18$ as a function of system size, $R$, for:
i) the TSA data without the energy correction (red solid line); 
ii) the TSA data with the exact perturbative correction accounting for the effects of the cutoff (black dashed line); and 
iii) the TSA data with approximate perturbative correction (green dotted line).  
As in Fig.~\ref{sg_pt}, the top, middle, and bottom panels consider the ground state energy, the first breather energy, and the second breather energy respectively.  The blue line with circle
symbols is the exact energy of these states at a given R. We see that the approximate correction does slightly better in improving the ground state energy to its exact value in comparison to the exact correction with the full energy denominator.  However for the two breathers, the approximate correction does considerably worse than its exact counterpart.}
\label{sg_pt1}
\end{figure}

Having said this, a note of caution is needed. A key approximation in the development of this RG equation is the rewriting of the denominator in Eq.~\fr{approx}. The quality of this approximation is, however, a model dependent statement. We give an an example of this in Fig.~\ref{sg_pt1}; we recompute the perturbative corrections to the low lying energy states coming from Eq.~\fr{delHab} by making the approximation in Eq.~\fr{approx}.  We see that the improvement in the ground state energy due to adding in the perturbative correction is largely unchanged by making this approximation.  However,  this approximation renders the corrections to the first two excited states coming from Eq.~\fr{delHab} negligible, whereas before the improvement was
notable (compare Fig.~\ref{sg_pt} to Fig.~\ref{sg_pt1}).  The detailed reason for this is that in sine-Gordon model the operator in the sum $\sum_\varphi$ dominating Eq.~\fr{delHab} is the identity operator. In making the approximation in Eq.~\fr{approx}, the correction term due to the identity operator is then the same for all states, leaving the relative energies of the ground state
and excited states unchanged.

\section{Other improvements on the TSA}\label{otherimprovements}

\subsection{Sweeping}\label{sweeping}

The RG improvements on the TSA described in the previous section are mostly geared towards extracting low energy information. However, there will be occasions where knowledge of the low energy states together with their matrix elements is insufficient and where information for
states with extensive energy is needed.  An important example of where such information might be needed is the description of a system after a quantum quench (see, e.g., Refs.~\cite{GogolinReview15,DAlessioReview15,EsslerReview16,CalabreseReview16,CazalillaReview16,BernardReview16,CauxReview16,VidmarReview16,LangenReview16,ProsenReview16,VasseurReview16,DeLucaReview16}).  Suppose one initializes a system in the ground state, $|gs\rangle$, of a Hamiltonian, $H$. Then, at time $t=0$, one non-trivially changes the Hamiltonian $H$ to $H'$ such that $|gs\rangle$ is no longer an eigenstate of the system.  The old ground state is now some non-trivial linear combination of the new eigenstates of $H'$, $|E'_s\rangle$:
\begin{equation}
|gs\rangle = \sum_s c_s |E'_s\rangle
\end{equation}
The time-evolution of the state post-quench is easily described in this representation:
\begin{equation}
|gs(t)\rangle = \sum_s c_se^{-itE'_s} |E'_s\rangle
\end{equation}
If the quench injects sufficient energy into the system, the coefficients of the expansion $c_s$ will have non-zero weight at energies $E'_s$ mid-spectrum.  Hence in order to describe the post-quench dynamics of the system, we need to understand states, $|E'_s\rangle$, that lie mid-spectrum.

In order to obtain mid-spectrum states in the fully perturbed system with good accuracy we add a sweeping procedure to the NRG, which is not dissimilar to the finite volume algorithm in DMRG~\cite{white1992density}.  Or equally good, this procedure may be considered as an extended set of Jacobi transformations done to diagonalize a symmetric matrix (extended in the sense that we are zeroing blocks not individual elements of a matrix). While this technique was first developed to understand level spacing statistics in perturbed conformal field theories~\cite{brandino2010energy}, it was crucial in work for the study of quenches of the Lieb-Liniger model where a one-body integrability breaking perturbation was added~\cite{brandino2015glimmers}. 

The procedure works as follows.  Suppose we have completed the NRG procedure described in Sec.~\ref{nrgalgorithm} for $M$ iterations.  In doing so,  we run through the first $N+M\Delta$ states of the conformal basis. We now begin anew, but instead of working with $N+M\Delta$ conformal states we work with a basis formed from the eigenstates generated in the first NRG procedure. This basis is given by
\begin{eqnarray}
&&|E\rangle^M_1,\ldots,|E\rangle^M_N,|E\rangle^1_{N+1},\ldots,|E\rangle^1_{N+\Delta},\cr\cr
&& |E\rangle^2_{N+1},\ldots |E\rangle^2_{N+\Delta},\ldots
|E\rangle^{M-1}_{N+1},\ldots,|E\rangle^{M-1}_{N+\Delta},\cr\cr
&&|E\rangle^M_{N+1},\ldots,|E\rangle^M_{N+\Delta}.
\end{eqnarray}
We see that this basis is formed from taking the first $N$ states coming from the last NRG iteration (and so the best guess we have at the $N$ lowest energy states in the theory), followed by the $M\times\Delta$ states we discarded in the $M$ iterations of the NRG (those with upper indices $1,\ldots,M$ and lower indices $N+1,\ldots,N+\Delta$).  This basis is much closer to the true eigenstates of the system than the initial conformal basis was. Using this basis, we repeat the NRG iterations; this set of iterations constitutes a single sweep.  We typically find that the eigenenergies converge rapidly after only a handful of sweeps.  

\begin{figure}
\includegraphics[width=0.45\textwidth]{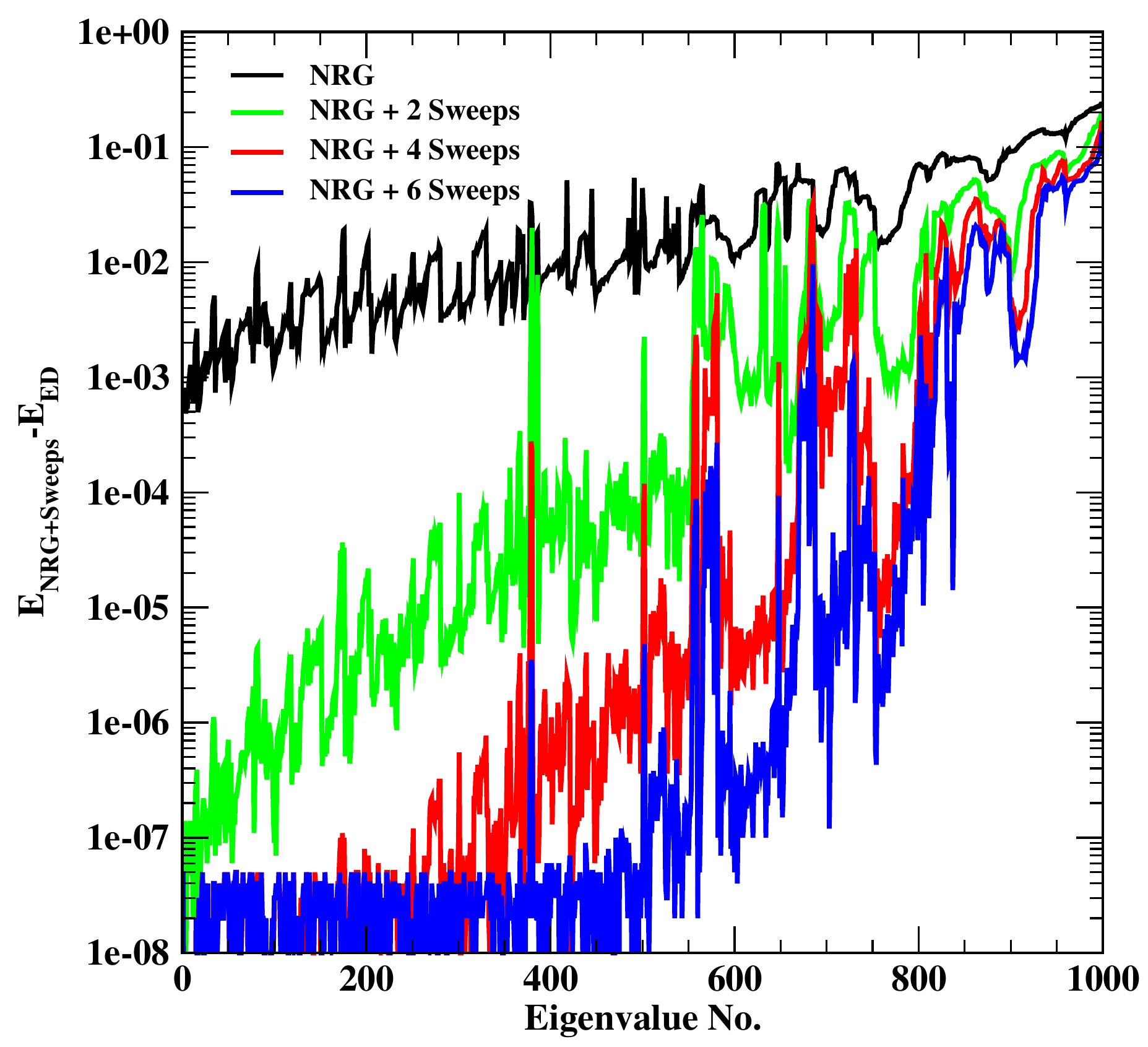}
\caption{The evolution of the first 1000 energies of the sine-Gordon model at $\beta =1/2$ and $R=15$
after successive sweeps.}
\label{Reiterfig}
\end{figure}

To test this procedure we compared the results of an exact diagonalization for the first 1000 eigenenergies of the $\beta^2=1/2$ sine-Gordon model with a cutoff of $N=22$ (with this cutoff there were $39279$ states in Hilbert space) with the same 1000 eigenenergies as computed using the NRG where $N=1000$ and $\Delta=500$ + different numbers of sweeps. This is shown in Fig.~\ref{Reiterfig}. With no sweeps, we see that the lowest lying eigenenergies from the NRG agree with those of the ED at the $10^{-3}$ level (the eigenvalues themselves are roughly $O(1)$).  However we see that by the 500th eigenvalue, this disagreement is at the $10^{-2}$ level.  If we now perform two sweeps, we see a dramatic improvement between the ED and NRG+sweeps.  Up to the 500th eigenvalue, we obtain agreement that is no worse than $10^{-4}$.  If we perform six sweeps, we find the agreement improves to $10^{-8}$.  Further sweeps do not improve on this (although they do increase the agreement for higher energy eigenstates).  This floor of $10^{-8}$ is likely a result of inherent numerical noise.  To reduce this would require greater precision numerics than that offered by C++ doubles.

\subsection{Metrics for the Hilbert space other than energy}\label{othermetric}

A fundamental assumption of the TSA and the NRG is that low energy states in the unperturbed eigenbasis of $H_0$ are the most important for determining the physics of the low energy states in the full theory, $H_0 + \lambda V_{\rm pert}$. In turn this is tied in to the assumption that $V_{\rm pert}$ is a relevant operator. However it is worthwhile to question just how valid this assumption is in one particular case.  

\begin{figure}
\includegraphics[width=0.45\textwidth]{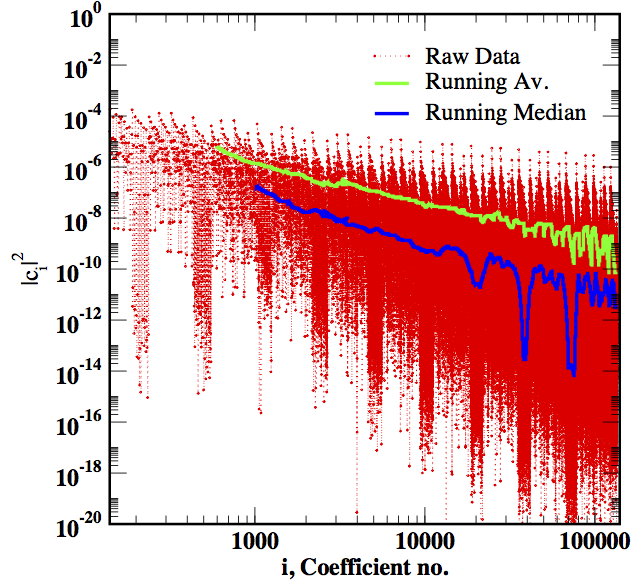}
\caption{The numerical values of the expansion coefficients of the ground state of the sine-Gordon model at $\beta^2 =1/2$ and $R=10$ in terms of the basis of the massless compact boson. The majority of basis states make only a small contribution to the ground state eigenstate.}
\label{coeff}
\end{figure}

We do so for the ground state of the sine-Gordon model at $\beta^2=1/2, R=10$. We compute the ground state using the NRG+TSA with $(N_s,\Delta)=(1000,500)$ and dimensionless cutoff $N=26$. In this example, a basis of 135,901 unperturbed eigenstates of the massless boson is employed. In Fig.~\ref{coeff} we present the coefficients of expansion, $\{c_i\}$, for the ground state coming from this computation, i.e.
\be
|E_{gs}\rangle = \sum_i c_i |i\rangle_{massless~boson}.
\ee
We see two trends in the size of these coefficients. Firstly, we see a general trend (as determined by either the mean or the median) where the size of the coefficients decay exponential with the unperturbed energy of the state (we roughly expect that the unperturbed energy of the $i$th state $|i\rangle_{massless~boson}$ ranked in energy to behave as $E_i \sim \log(i)$ as the size of the Hilbert space grows exponentially in energy).  However while this is the general trend, there is tremendous scatter in the magnitudes of the coefficients.  For any given energy range there are a set of states that are orders of magnitude more important than the mean/median and similarly a set of states whose coefficients are orders of magnitude smaller than the mean/median.  It would greatly improve numerically efficiency if the latter could be identified before any calculation was done and excluded from the computation.

While there is no general principle by which this can be done (at least none of which we know), there is a practical way of accomplishing this aim~\cite{caux2012constructing,brandino2015glimmers,konik2011exciton,konik2015predicting}. There is a heuristic expectation that the high energy states, $|\rm high~en.\rangle$, that are important for low-energy properties will have a matrix element of appreciable magnitude with the ground state, i.e. the matrix element
\be
\langle {\rm high~en.}|V_{\rm pert}|{\rm g.s.}\rangle,
\ee
will be large. Here the relevant ground state $|\rm g.s.\rangle$  is \textit{not} the unperturbed one but the full ground state.  However if one has to compute the full ground state of the theory {\it before} determining which high energy states are important, one is no further ahead. Crucially, however, one can compute a (rough) approximate ground state of the full theory using, e.g., a low cutoff with little numerical effort. With the rough approximant at hand, one computes the above set of matrix elements for all unperturbed states up to a much higher energy cutoff. Ordering the unperturbed eigenstate by the size of $\langle {\rm high~en.}|V_{\rm pert}|{\rm g.s.}\rangle$, rather than by energy, provides a much more efficient way of moving through the Hilbert space of the theory.

\begin{figure}
\includegraphics[width=0.45\textwidth]{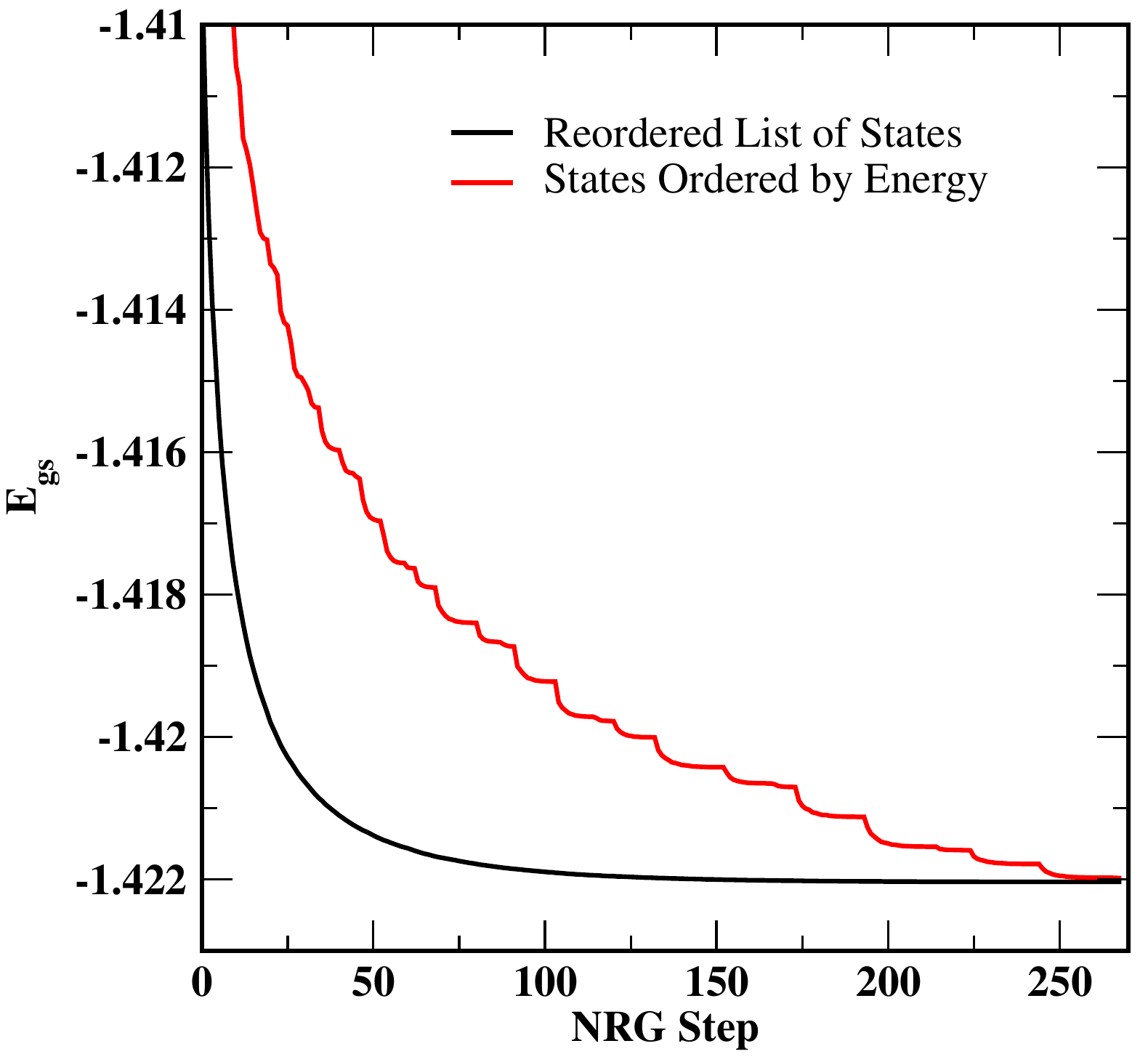}
\caption{The ground state energy as a function of NRG step size for both the basis ordered in terms of energy (the traditional TSA approach) and the reordered states. In the second case, the basis is reordered according to the size of the overlap of the perturbation ($\cos(\beta\phi)$) between a state and the three lowest energy states (which is obtained through a TSA computation with a much smaller cutoff, consisting of roughly 2000 states). This reordering captures, in some sense, the relative importance of a state. The NRG converges much more quickly to its final answer with the reordered basis than with the basis ordered in terms of energy. }
\label{reorder_nrg}
\end{figure}

To illustrate this, we present the ground state energy of the sine-Gordon model at $\beta^2=1/2$ under the TSA+NRG with the conformal basis of states ordered by energy (i.e., the traditional TSA) and by matrix element overlap (as described above) in Fig.~\ref{reorder_nrg}. In particular, we consider the overlap matrix elements for the first three low-lying states (not just the ground state) of the theory computed with much lower cutoff (corresponding to roughly 2000 states, as opposed to the $\approx135,000$ states considered with cutoff $N=26$). We see that the TSA+NRG with the reordered set of states converges much more quickly to the final answer than the TSA+NRG with the original energy ordering of the states. This reflects that the reordering by the size 
of matrix elements does a good job of identifying the states of greatest importance. In fact with the reordered set of states, we could terminate the TSA+NRG at step 100, obtaining the same answer (to the fourth significant digit) as the TSA+NRG performed with the list of states ordered by energy.

Using lists of states ordered by metrics other than energy was important in TSA studies of the Lieb-Liniger model~\cite{caux2012constructing,brandino2015glimmers} and in TSA studies of the single-particle sector of gapped carbon nanotubes~\cite{konik2011exciton,konik2015predicting}. In both cases, one had to work at cutoffs corresponding to truncated Hilbert spaces with sizes of order $10^6$.  By reordering the states with this method one was left, typically, with a space of `important' states of size $10^5$.

\section{Applications of the TSA}
\label{Sec:TSAapplications}

To close our discussion of the TSA, in this final section we briefly consider a variety of relatively recent applications of the TSA. 
But reader be warned: these applications are not a comprehensive summary, rather they reflect the particular interests of the authors.

\subsection{Semiconducting carbon nanotubes}

We begin with an application that involves direct comparison with experimental data.  One consequence of making contact with experimental data here was the need to take into account the effects of the TSA cutoff in a non-trivial way.  In everything we have discussed before, we have been wanting to remove the effects of having a cutoff in the theory.  But real experimental systems always have a finite bandwidth and so a finite cutoff.  Thus in this work we faced the challenge of connecting a TSA cutoff with a physical bandwidth.
 
Even beyond questions of the role of the cutoff, the study of semiconducting carbon nanotubes in Refs.~\cite{konik2011exciton,konik2015predicting} represent a non-trivial application of the TSA. These quasi-one dimensional quantum systems, can be shown to be equivalent to a generalized sine-Gordon model of four bosons~\cite{levitov2003narrow,konik2011exciton}, a much more complicated theory than typically studied with the TSA:
\begin{eqnarray}
H &=& H_0 + H_{\rm gap},\nn
H_0 &=& \int \rd x \sum_{\substack{i=c+,c-,\\~~s-,s+}} \frac{v_i}{8\pi}\Big(K_i(\partial_x\Theta_i)^2 + K_i^{-1}(\partial_x \Phi_i)^2\Big),\nn
H_{\rm gap} &=& \int \rd x\, \frac{4\tilde\Delta_0}{\pi}\bigg[\prod_{i}\cos\bigg(\frac{\Theta_i}{2}\bigg)+\prod_{i}\sin\bigg(\frac{\Theta_i}{2}\bigg)\bigg].\nonumber
\end{eqnarray}
The four bosons arise from the Abelian bosonization of the fermions in a particular subband of the nanotube (see, for example, Appendix~\ref{App:AbelianBosonization}). In each subband the electron has a four-fold degeneracy, arising from the two spin and two valley degrees of freedom; this leads to four bosons. Whilst the electronic degrees of freedom carry charge, spin, valley and chirality quantum numbers, the bosonic fields in the Hamiltonian represent particular linear combinations of these; in particular, $\Theta_{c+}$ is the total charge boson. 

\begin{figure}
\includegraphics[width=0.45\textwidth]{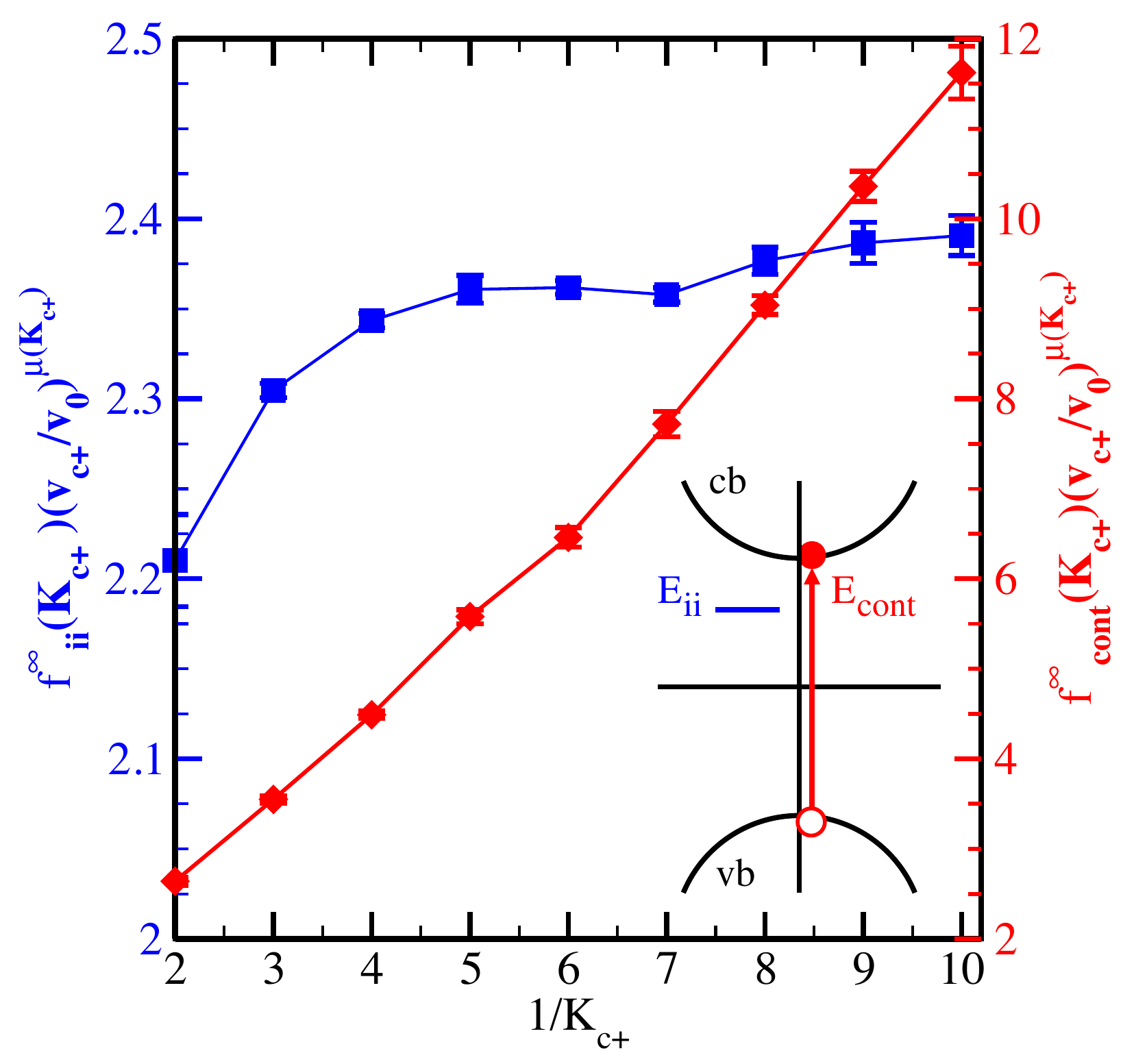}
\caption{The scaling functions in the limit of infinite bandwidth for the $E_{\rm ii,exc}$ excitons (excitons formed as a bound state of a hole in the $i$th 
valence subband and a particle in the $i$th conduction subband) and the particle-hole continuum, $E_{cont}$. At $K_{c+}$ = 1 (the noninteracting point) 
these functions converge to the value 2. Inset: Sketch of $E_{\rm ii,exc}$ and $E_{\rm cont}$ excitations. This figure is adapted from Ref.~\cite{konik2015predicting}.}
\label{exc_scaling}
\end{figure}

The electronic interactions are characterized by $K_i$, the Luttinger parameters for each of the bosons. The long-range Coulomb interaction that is present between charges in the nanotube strongly renormalizes the total charge Luttinger parameter, $K_{c+}$, whilst the remaining three remain close to one. $K_{c+}$ can be determined from various parameters of the tube 
\begin{equation}
K_{c+} = \bigg(1+ \frac{8e^2}{\pi\kappa\hbar v_0}\Big[-\log (k_{\rm min}R_{\rm tube}) + c_0\Big]\bigg)^{-1/2}.
\end{equation}
This expression for $K_{c+}$ takes into account all of the key parameters of the tube: 
(i) $\kappa$ is the dielectric constant of the medium surrounding the tube and is the factor that determines most strongly the effective strength of the Coulomb interaction, i.e., how much $K_{c+}$ deviates from 1;
(ii) $k_{\rm min}$ is the minimum allowed wave vector in the tube, which is necessarily larger than $2\pi/L$ (where $L$ is the length of the tube). In principle $k_{\rm min}$ can be much larger than this minimal scale, say, on the order of the inverse mean free path in the tube;
(iii) $c_0$ is an $O(1)$ constant that depends on the wrapping vector [the vector (n,m) that identifies how a graphene sheet is rolled up to form a particular tube], and has been derived  in Refs.~\cite{egger1998correlated,degottardi2009transverse}; and
(iv) $R_{\rm tube}$ is the radius of the tube.

Typically in carbon nanotubes, the total charge Luttinger parameter is strongly renormalized with $K_{c+}$ taking values in the range of $\sim 0.2$. We also note that because the Luttinger parameters for each of the bosons is different, their velocities are also different as $v_{i} = v_0/K_i$, where $v_0$ is bare Fermi velocity in the subband. 

The coupling $\tilde \Delta_0$ in $H_{\rm gap}$ is a function of the bare gap $\Delta_0$ of the subband through the relation
\begin{equation}
\tilde \Delta_0 = \Delta_0 \left(\frac{\Lambda_{\rm tube}}{v_{c+}}\right)^{(1-K_{c+})/4},
\end{equation}
where $\Lambda_{\rm tube}$ is the effective bandwidth of the tube (not to be mistaken for the TSA cutoff).

One key aim of the work~\cite{konik2015predicting} was to determine the energies of optically active excitons (electron-hole bound states) in the carbon nanotubes. By dimensional analysis the energies of the excitons take the form
\begin{equation}
E_{\rm exc} = f^{\Lambda_{\rm tube}}_{\rm exc}(K_{c+})\tilde\Delta_0^{4/(5-K_{c+})}v_{c+}^{(1-K_{c+})/(5-K_{c+})},
\end{equation}
where $f^{\Lambda_{\rm tube}}_{\rm exc}$ is a dimensionless scaling function that depends in part on the total charge Luttinger parameter $K_{c+}$.
However it also depends on the effective tube bandwidth
$\Lambda_{tube}$. This dependence has the following form:
\begin{equation} \label{scalingfunc}
f^{\Lambda_{\rm tube}}_{\rm exc} = f^\infty_{\rm exc}\Bigg[1+A(K_{c+}) \bigg(\frac{\tilde \Delta_0}{v_0}\bigg)^2
\bigg(\frac{v_0}{\Lambda_{\rm tube}}\bigg)^{(5-K_{c+})/2}\Bigg].
\end{equation}
The first term in this equation for the excitonic energy scaling function involves $f^\infty_{\rm exc}$, the scaling function in the absence of a cutoff. The second involves, $A(K_{c+})$, a dimensionless constant that gives the first correction to the exciton energies coming from the presence of a finite cutoff. In Ref.~\cite{konik2015predicting}, the relationship between the bandwidth cutoff of the tube~$\Lambda_{\rm tube}$ and the TSA cutoff was argued to be:\footnote{This was argued on the basis of how the cutoff modifies the normal ordering of vertex operators.}
\begin{equation}
\Lambda_{\rm tube} = \frac{e^\gamma}{4}\Lambda_{TSA}.
\end{equation}
Once this relation was established, it was possible to use the TSA to determine the experimental excitonic energies. As we will see, taking into account the correction to these energies induced by a finite cutoff was important in obtaining a good match between the TSA result and those measured in experiments, such as Refs.~\cite{sfeir2010infrared,dukovic2005structural}.

In Fig.~\ref{exc_scaling} we present the infinite bandwidth TSA-derived scaling functions, $f^{\infty}$, as a function of $K_{c+}^{-1}$ for both the excitons that form within the $i-th$ subband, $f^\infty_{ii,\rm exc}$, as well as the scaling function for the continuum $f^\infty_{\rm cont}$ (this is twice the single particle gap). The difference in energy between the continuum and the exciton is the excitonic binding energy, typically large in carbon nanotubes. 

\begin{figure}
\includegraphics[width=0.48\textwidth]{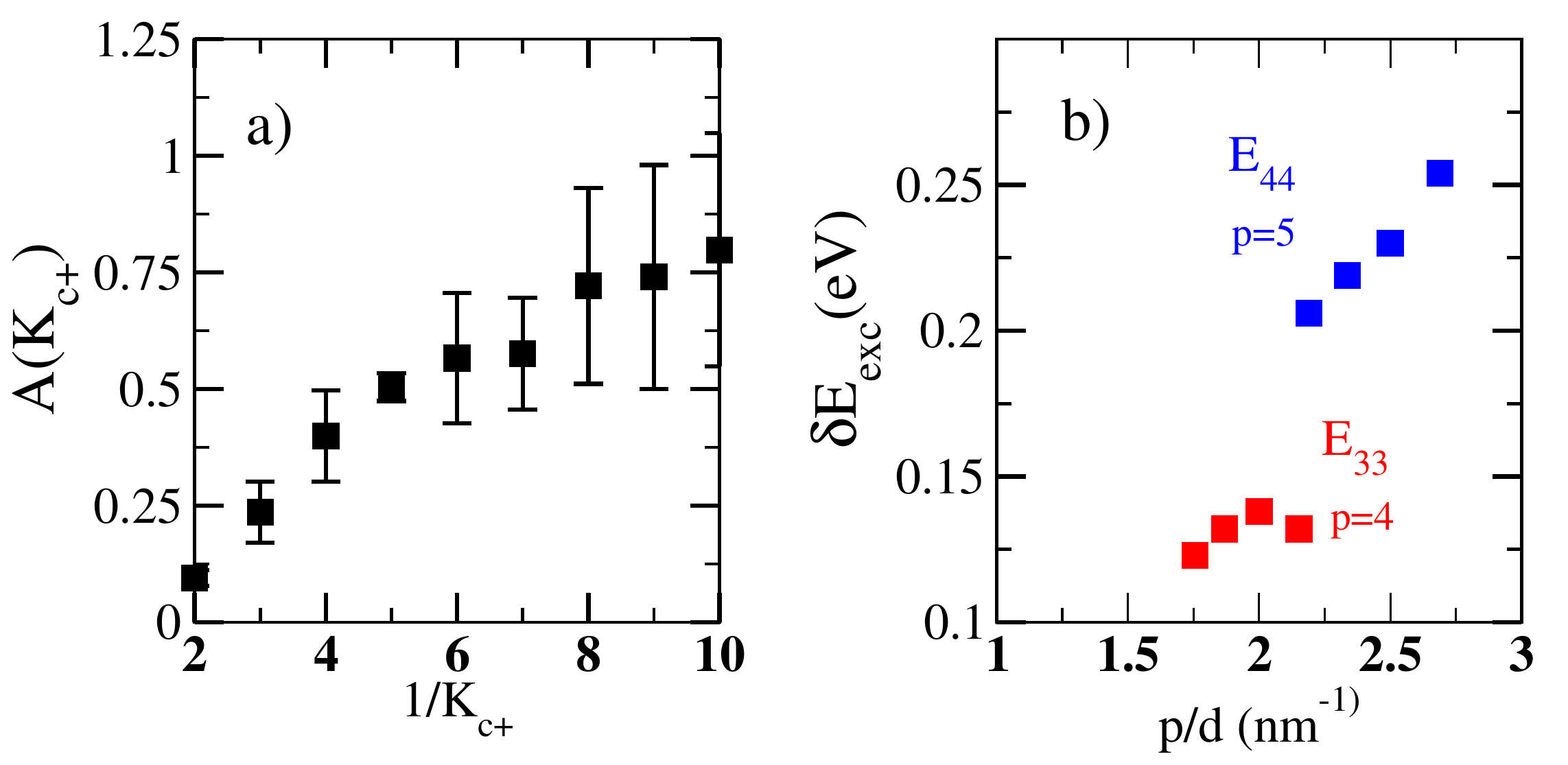}
\caption{(a) The function $A(K_{c+})$ giving the size of the finite bandwidth correction to $E_{ii,\rm exc}$. (b) The size of this correction, $\delta E_{ii,\rm exc}$, for the excitons, $E_{33,\rm exc}$ and $E_{44,\rm exc}$, of the four tubes studied in Ref.~\cite{sfeir2010infrared}. Note that here $p=i+1$ is a subband index. This figure is adapted from Ref.~\cite{konik2015predicting}.}
\label{finite_bandwidth}
\end{figure}

\begin{figure}
\includegraphics[width=0.35\textwidth]{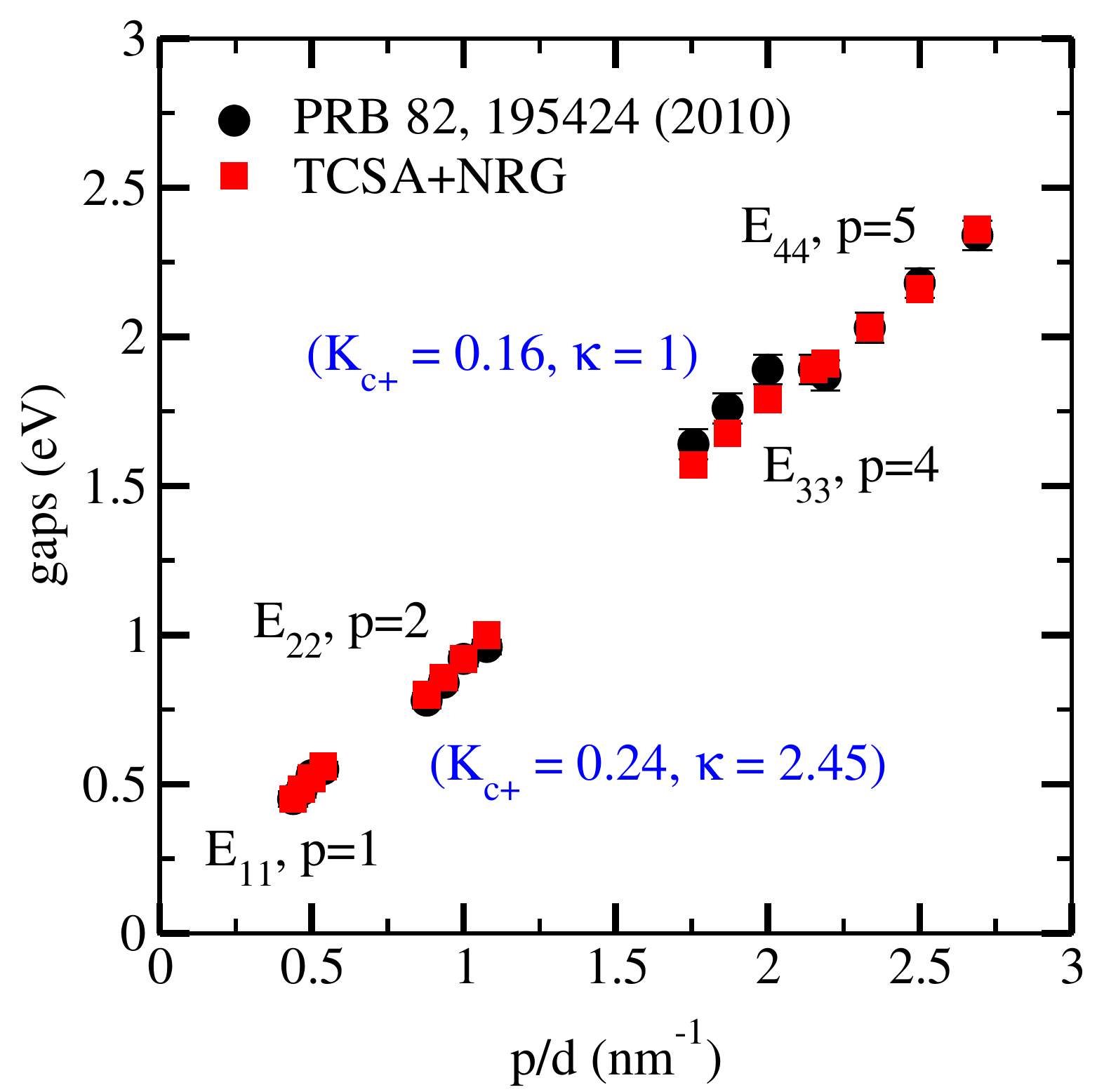}
\caption{Comparison of the measured exciton gaps in the first four subbands, $E_{ii},i=1,2,3,4$ ($p=1,2,4,5$ in the notation of Ref.~\cite{sfeir2010infrared} to that obtained using the TSA with the corrections to finite bandwidth included.}
\label{sfeir_exp_data}
\end{figure}

In the left panel of Fig.~\ref{finite_bandwidth} we plot the constant $A(K_{c+})$ as a function of $K_{c+}$, determined with the TSA.  In the right panel, we also show the corrections in energy (given in physical units of eV)  for excitons in the third and fourth subbands as measured in Ref.~\cite{sfeir2010infrared}. These corrections turned out to be substantial, on the order of 10\% of the measured excitonic energies. Without having taking into account the finite bandwidth corrections, the agreement between theory and experiment found in Ref.~\cite{konik2015predicting} would have been substantially worse.  We present in Fig.~\ref{sfeir_exp_data} the agreement between TSA and experiment for the data for all four subbands presented in Ref.~\cite{sfeir2010infrared}.

While we refer the reader to Refs.~\cite{konik2011exciton,konik2015predicting} for details of the TSA implementation, the techniques introduced in Section \ref{othermetric} were crucial to this study. In particular, to get reasonably robust results for $f^\infty_{\rm cont}$ it was necessary to work with large systems, in order to suppress finite size corrections.  Finite size corrections were considerable for the single particle excitations because of the strong renormalization of the charge velocity by the Coulomb interaction. This meant in turn that it was vital to work at a large dimensionless cutoff in the TSA.  This resulted in Hilbert space sizes on the order of $10^6$.  It would not have been possible to deal with such large Hilbert spaces without determining which of the unperturbed states in the four boson Hilbert space actually contributed to the energies in the single particle sector.

\subsection{Non-equilibrium behavior of the Lieb-Liniger model perturbed by one-body potentials}
Another non-traditional application of the TSA has been to study one-body perturbations of the Lieb-Liniger model.  The TSA is normally focused upon the study of theories with Lorentz invariance; the Lieb-Liniger model, however, is not a relativistic theory. Moreover, the ground state of this model is not the typical vacuum state that we encounter in relativistic field theories, but is instead a state that knows of the number of particles in the theory.  This presents its own unique challenges for using the TSA.

The Lieb-Liniger model has a Hamiltonian given by
\begin{equation}
H_{LL} = \int^R_0 \!\! \rd x \Big(\partial_x\psi^\dagger (x) \partial_x \psi(x)  + c\psi^\dagger(x)\psi^\dagger(x)\psi(x)\psi(x)\Big),
\end{equation}
where $\psi^\dagger(x)$ are Bose field operators satisfying the canonical commutation relations:
\be
[\psi (x),\psi^\dagger (x')] = \delta (x-x'),
\ee
$c$ describes the strength of the interaction (repulsive if $c>0$), running from $c=0$ (free bosons) to $c=\infty$ (hardcore bosons or, equivalently, free fermions). In this Hamiltonian $\hbar=1$ and the mass of the bosons $m$ has been set to $1/2$.

We will be interested in describing one-body perturbations to the Lieb-Liniger model, i.e., perturbations that involve the density operator:
\begin{equation}
V_{\rm pert} = \int^R_0 V(x) \rho(x),
\end{equation}
where $\rho(x) = \psi^\dagger(x)\psi(x)$ is the density operator. We have studied the Lieb-Liniger under two different perturbing potentials: one where $V_{\rm para}(x) = m\omega^2x^2/2$ was a harmonic trapping potential and one where $V_{\rm cos}(x) = A\cos (2\pi n x/R)$ with $n\in\mathbb{Z}$ is a cosine potential commensurate with the system size, $R$.

In our work on the Lieb-Liniger model, we were primarily interested in studying quantum quenches.  In this context, a quantum quench is a situation where we prepare the system in its ground state corresponding to one potential, $V_{\rm pre}(x)$, which at time $t=0$ is switched to a different potential, $V_{\rm post}(x)$. Having done so the system is no longer in its ground state, but is instead in some complicated superposition of eigenstate states of the post-quench Hamiltonian: $\{|E_{i,\rm post}\rangle\}$, i.e.
\begin{equation}\label{expansion}
|E_{gs,\rm pre}\rangle = \sum_i c_i |E_{i,\rm post}\rangle.
\end{equation}
Because the post-quench ($t>0$) state of a system is not (in general) an eigenstate, it begins to execute non-trivial time evolution. In Refs.~\cite{caux2012constructing,brandino2015glimmers}, it was our goal to describe such evolution. In general our strategy is to use the TSA to determine both the ground state, $|E_{gs,\rm pre}\rangle$, with respect to the pre-quench one-body potential and then to determine, again with the TSA, a sufficient number, $M$, of post-quench eigenstates $\{|E_{i,\rm post}\rangle\}$ such that two conditions held:
\begin{enumerate}
\item $M$ was large enough so that the sum of expansion coefficients satisfied $\sum_{i=1}^M |c_i|^2 > 0.99$.\\
\item The energies of these $M$ eigenstates were estimated to be determined with a relative accuracy of $10^{-3}$.
\end{enumerate}
If we could do so, we would then be able faithfully to reproduce the time evolution of the state, given by
\begin{equation}\label{time_ev_state}
|E_{gs,\rm pre}\rangle(t) = \sum_i c_i e^{iE_{i,\rm post}t} |E_{i,\rm post}\rangle,
\end{equation}
out to times (at least) many multiples of the fundamental time scale, $t_F$, of the gas (where $t_F=1/E_F$, $E_F=k_F^2$, and $k_F=\pi N/R$). We note that this is only one possible strategy for computing time evolution following a quantum quench within the TSA framework. Recently Ref.~\cite{rakovszky2016hamiltonian} showed that one could use expansions of the time evolution operator, $e^{iH_{\rm post}t}$, in terms of Chebyshev polynomials to study quenches in the quantum Ising field theory. It would be interesting to explore this technique for quenches in the Lieb-Liniger model or other integrable field theories. 

We now will consider some of the implementation details in the first step in our strategy, the TSA determination of  $|E_{gs,\rm pre}\rangle$.  As a computational basis for the TSA, we employ the eigenstates of the unperturbed Lieb-Liniger model. These states are considerably more complicated to construct and delineate than the cases we have considered previously, all of which involved manipulations of the states of either a CFT or a free massive model.  Each eigenstate of the Lieb-Liniger model is characterized by a set of $N$ rapidities, $\{\lambda_i\}_{i=1}^N$, where $N$ is the number of particles in the system. These rapidities are the solution of a set of $N$ so-called Bethe ansatz equations:
\begin{equation}
e^{i\lambda_i R} = \prod^N_{j=1}\frac{\lambda_i-\lambda_j + ic}{\lambda_i-\lambda_j - ic},
\end{equation}
where $i=1,\ldots,N$. The different states are marked by different $N$-sets of quantum numbers $\{n_i\}^N_{i=1}$ that appear after the Bethe
equations are put in logarithmic form,
\be
2\pi n_i = \lambda_i R - \frac{1}{i}\sum_{j=1}^N\log\bigg(\frac{\lambda_i-\lambda_j + ic}{\lambda_i-\lambda_j - ic}\bigg), 
\ee
for each $i=1,\ldots,N$. These sets of $\lambda_i$'s completely characterize the state. For example, the momentum $P$ and energy $E$ of a state is given in terms of these rapidities by
\begin{equation}
P = \sum_{i=1}^N \lambda_i, \quad E=\sum_{i=1}^N \lambda_i^2.
\end{equation}
To solve the Bethe equations, we employed {\sc abacus}, a set of highly optimized C++ routines written by Jean-Sebastien Caux~\cite{CauxJMathPhys09} for this express purpose. {\sc abacus} also readily enables the computation of the matrix elements of the density operator needed by the TSA. The matrix elements can be expressed in the numerically efficient form of a matrix determinant~\cite{slavnov1989calculation,slavnov1990nonequal}.

Beyond the relative computational complexity of the states and the matrix elements for the Lieb-Liniger model, the Hilbert space one must handle is larger than in the case of relativistic field theories. This is in part a consequence of having to work with reasonably large numbers of particles, $N$, and system size, $R$, in order to limit finite size effects. For large $N$ and $R$, it proved suboptimal to truncate the Hilbert space through the introduction of an energy cutoff. Instead, we found that the Hilbert space was better classified by the number of particle-hole excitations a state contained relative to the ground state. States consisting of many particle-hole excitations typically were less important and had only negligible matrix elements involving the density operator. In truncating the Hilbert space, we never considered states with more than five particle-hole excitations. Even with this constrained Hilbert space, the number of states remains immense, but many states only weakly influence the low energy physics of the ground state. To determine which states were important, we employed the methodology described in Sec.~\ref{othermetric}: i) we first constructed the ground state using a relatively small number of low energy eigenstates of the Lieb-Liniger model;  ii) we computed the matrix element of $\rho(0)$ with respect to this ground state and all the unperturbed states; iii) we then truncated the states according to this weighting. In this way we were able to study systems with up to $N=56$ particles and obtain accurate results.

\begin{figure}
\includegraphics[width=0.40\textwidth]{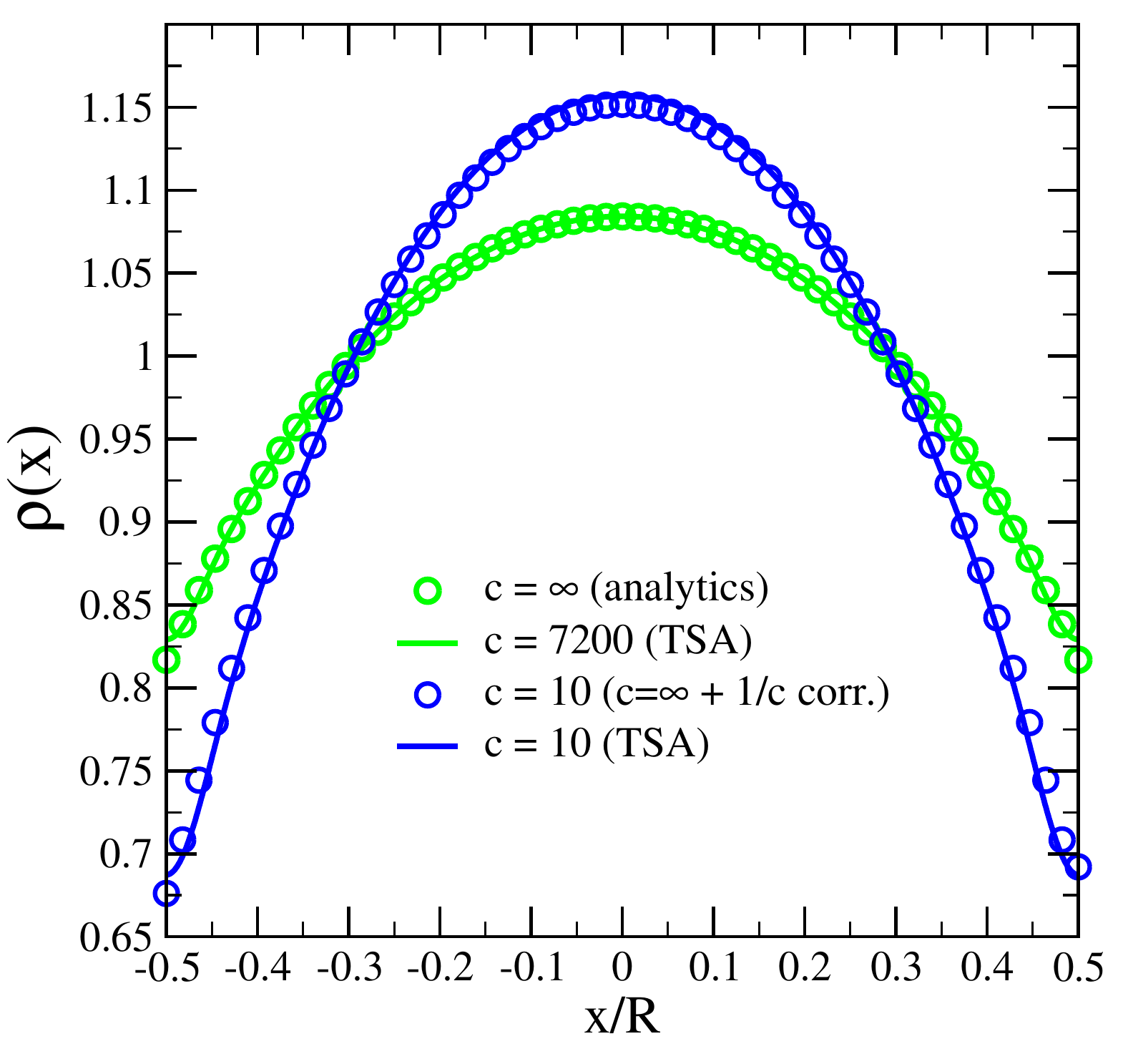}
\caption{
The density profile of the ground state of the Lieb-Liniger model in a harmonic potential $V_{\rm para}(x)= m\omega^2x^2/2$ with $\omega = 0.16$.  Here the gas has $N=56$ particles and is in a system of size $R=56$. We consider the gas at two values of its interaction parameter, $c=7200$ and $c=10$.  We plot both the analytic computation of the density using a $1/c$ expansion about the hardcore limit (for some details see Ref.~\cite{caux2012constructing}) as well as the density as computed using the TSA.  Adapted from Ref.~\cite{caux2012constructing}.}
\label{LL_density}
\end{figure}

In the first Lieb-Liniger quench studied with the TSA, see Ref.~\cite{caux2012constructing}, we initialized the system in a parabolic trap, $V_{\rm pre}(x)=V_{\rm para}(x)$, and subsequently quenched it by simply removing the trap $V_{\rm post}(x)=0$.  To demonstrate that we could successfully construct the ground state of the Lieb-Liniger model in a parabolic potential, we considered the gas in its large $c$ limit.  In this limit, the model is equivalent to free fermions perturbed by a four-fermion term whose strength is proportional to $1/c$~\cite{tonks1936complete,khodas2007dynamics,girardeau1960relationship}. In this case the density of the gas can be computed analytically. In Fig.~\ref{LL_density} we see that the density of the gas in the trap computed using the TSA compares well against the $1/c$ analytic computation.   

Having constructed with the TSA, as described above, the ground state in the parabolic potential, the subsequent post-quench time evolution was straightforward to describe. As the post-quench eigenstates are the same set of states as the computational basis, the two conditions for the construction of the post-quench eigenstates stated previously are automatically satisfied. In particular, the expansion of the pre-quench ground state in terms of the post-quench eigenbasis is completely saturated and, as we know the energies of the computational basis to arbitrary accuracy (as it is possible to solve the Bethe equations to arbitrary accuracy), we can time evolve the $t=0$ initial state to arbitrary times.

\begin{figure}
\includegraphics[width=0.45\textwidth]{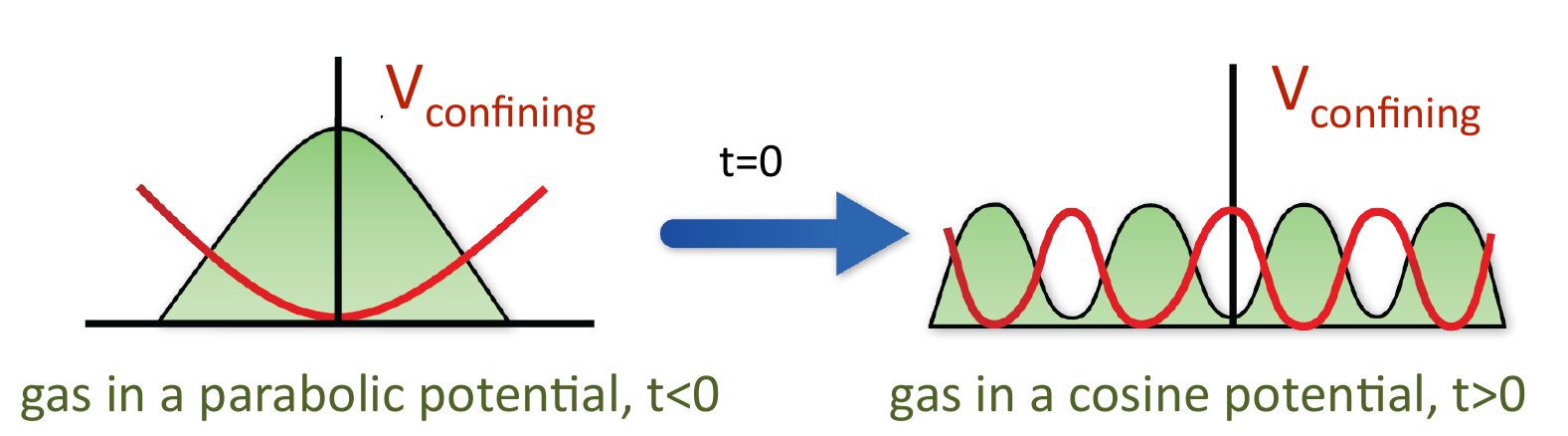}
\caption{
This sketch shows the quench of the 1D Bose gas prepared in the ground state of a parabolic potential and then released at $t=0$ into a cosine potential.  The shaded green regions represent equilibrium density profiles of the gas in the presence of the confining potentials.  Adapted from Ref.~\cite{brandino2015glimmers}.}
\label{quench_cartoon}
\end{figure}

\begin{figure}
\includegraphics[width=0.30\textwidth]{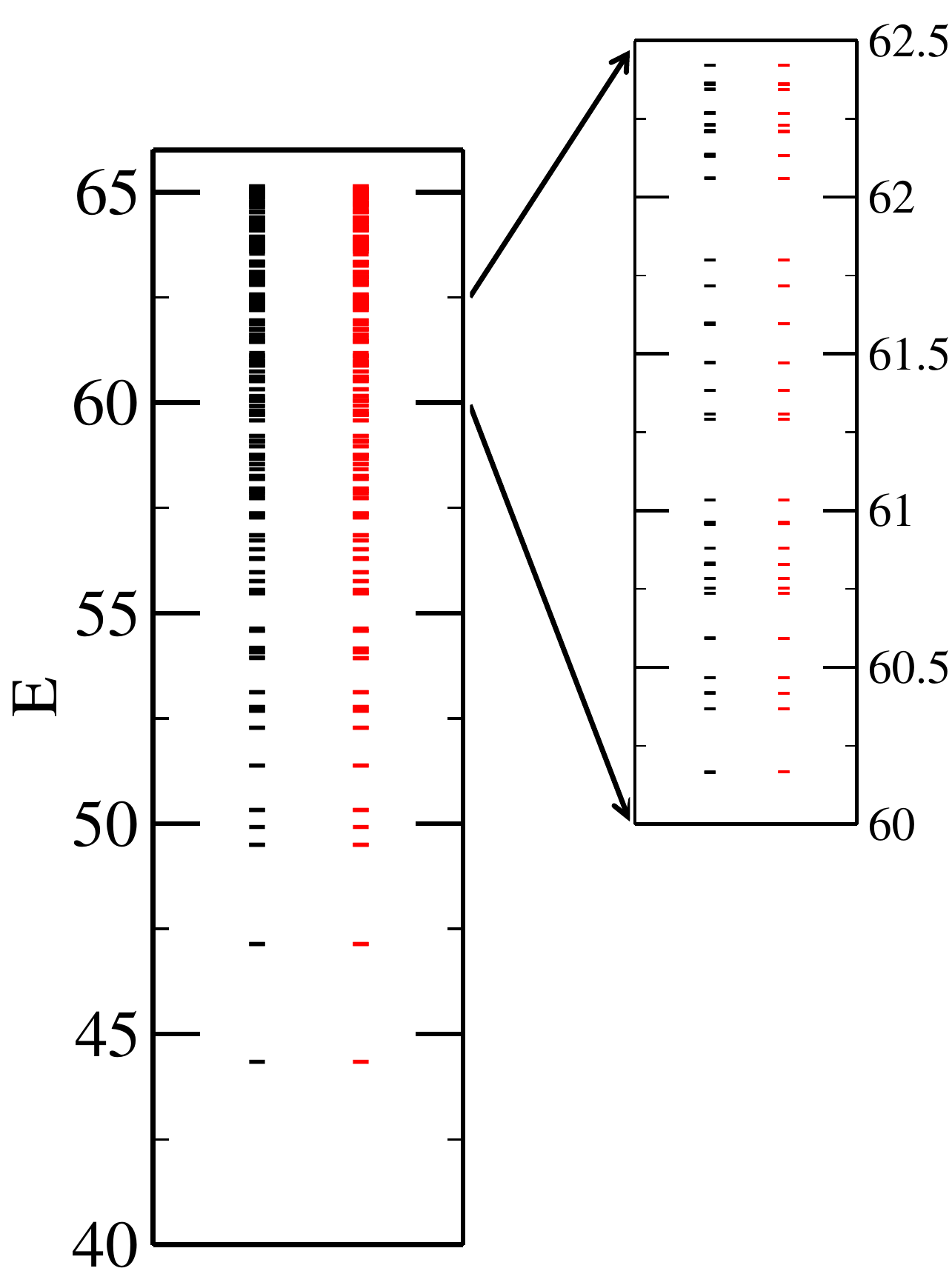}
\caption{
A plot of the energy spectra for $N=14$ particle gas with $c=7200$ in a cosine potential of amplitude $A/E_F =0.35$. The analytically computed results are given in red, while in black are the corresponding numerics. On the r.h.s., we expand a range of energy with a dense number of states so as to better exhibit agreement between the numerics and the analytics. We can determine the first 365 states (up to energies of $E=65$) with accuracy of $10^{−3}$ .  Adapted from Ref.~\cite{brandino2015glimmers}.} 
\label{cosine_energies}
\end{figure}

In our second work on quenches in the Lieb-Liniger model, Ref.~\cite{brandino2015glimmers}, we studied a more complicated quench: a quench from a parabolic potential, $V_{\rm para}(x)$, to a cosine potential, $V_{cos}(x)$ (see Fig.~\ref{quench_cartoon}). The pre-quench use of the TSA to construct the ground state in the trap was the same as in Ref.~\cite{caux2012constructing}.  But now we needed to use the TSA to construct a large number of the post-quench eigenstates of the gas in a cosine potential.  To do so we employed the sweeping mechanism discussed in Sec.~\ref{sweeping}, an adaptation necessary to compute states with finite energy density. In Fig.~\ref{cosine_energies} we plot the energies of the first 365 levels of a gas in a cosine potential in the hardcore limit; we found that the TSA was able to determine these energies to within $10^{-3}$.  

\begin{figure}
\includegraphics[width=0.35\textwidth]{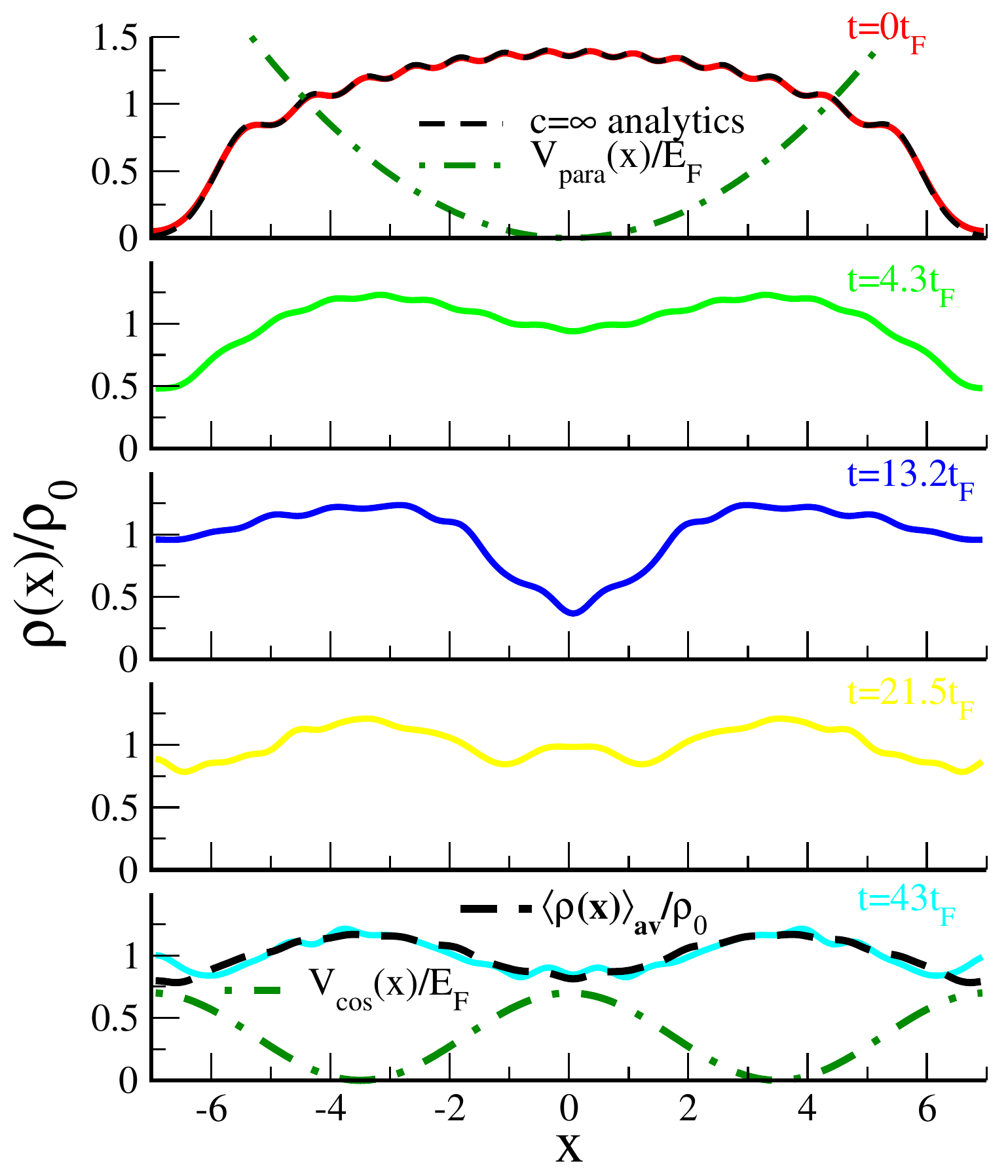}
\caption{
The density profile of $N=14$ particle gas in a system of size $R=14$ at selected times after a quench from a parabolic to a cosine potential.  This time dependence is computed after releasing the gas prepared in a parabolic potential with $m\omega^2L^2/2E_F =10.36$ (shown with a green dashed line in the $t= 0t_F$ frame) into a cosine potential $V_{cos}(x) = 0.35 E_F \cos(4\pi x/L)$ (plotted with a dashed line in the $t=43t_F$ frame). In the $t=0$ frame, we show the density profile as computed analytically in the hard-core limit.  While we only show the gas out to $t=43t_F$, we can run the time evolution as far out as $t=85t_F$ before dephasing exceeds 1\%.  We see, however, that by $t=43t_F$ the density profile of the gas has already come close to its long time average (black dashed line in the final panel).  Adapted from Ref.~\cite{brandino2015glimmers}.}
\label{time_ev}
\end{figure}

In Fig.~\ref{time_ev} we provide an example of the time evolution induced by quenching the gas from a harmonic potential to a cosine potential. In this case we consider a gas with $N=14$ particles for the same parameters as in Fig.~\ref{cosine_energies}.  While we only show the gas's evolution out to $t=43t_F$, we are able to describe time evolution out to $t=85t_F$ before 1\% of the wavefunction of the gas has dephased.\footnote{For the appropriate definition of dephasing, see Appendix A of Ref.~\cite{brandino2015glimmers}.}

Here we have focused on presenting results for the gas at large $c$ (in order to compare it to large $c$ analytic computations), but the TSA appears to work equally well down to values of $c=1$.  While our ability
to quantify this is limited, as we do not have direct checks at finite $c$, the convergence of the TSA as a function of Hilbert space truncation behaves similarly.  However, once we start to look at $c<1$ gases, we can see that the numerical performance becomes degraded at a given truncation and that, in general, more states from the unperturbed Lieb-Liniger Hilbert space are needed to describe a given perturbation of the gas.

 \subsection{Perturbed Wess-Zumino-Novikov-Witten models}
 \label{Sec:TSAWZNW}

Another class of models that have recently been studied with the TSA are perturbed WZNW models~\cite{beria2013truncated,konik2015studying,AzariaPRD16}. Such models have been the topic of extensive discussions in Secs.~\ref{Sec:NonAbelianBosonization}--\ref{Sec:ColdAtomsExamples} of this review, and we refer the reader to these for an introduction.

To study these models using the TSA, we find ourselves in a similar situation as to the study of perturbed minimal models. However, because WZNW models have central charge $c>1$, their Hilbert spaces are generically larger and more difficult to handle numerically.  Another difference between the minimal models and WZNW models is that the structure coefficients which determine the three-point functions (and are thus necessary for the determination of matrix elements) are not generally known.  The only complete classification of such structure coefficients is for the $SU(2)_k$ theories~\cite{FateevSovJNuclPhys86}. It is then not surprising that all of the examples of perturbed WZNW models treated so far involve perturbations of $SU(2)_k$.

The unperturbed space of states, $\{|s_{\rm WZNW}\rangle\}$, that forms our computational basis here is created by acting on highest weight states (formed themselves by acting with the primaries of the WZNW theory on the vacuum) with the modes of the current algebra operators [see Eq.~\fr{modes}]
\begin{equation}\label{wznw_states}
|s_{\rm WZNW}\rangle = \prod_{i=1}^M J^{a_i}_{n_i}|\Delta,\bar\Delta\rangle,
\end{equation}
where here $|\Delta,\bar\Delta\rangle$ is a highest weight state created by ${\cal O}_{\Delta,\bar\Delta}(0,0)$ (see Section~\ref{hwstates}). Like with the minimal models, the set of states of the form of Eq.~\fr{wznw_states} are not all linearly independent, so we need to project out null states using the Gram matrix from the different linear combinations of the states in Eq.~\fr{wznw_states}.

\subsubsection{The Nambu-Jona-Lasinio model}

While we refer the reader to Refs.~\cite{beria2013truncated,konik2015studying,AzariaPRD16} for further details of the implementation of the TSA in these models, we will discuss briefly the most interesting application of the TSA to a perturbed WZNW model, a study of the Nambu-Jona-Lasinio model, a model of $1+1$ dimensional QCD. In its original form, the Hamiltonian describes six fermions interacting with one another through current-current terms
\begin{eqnarray}
 {\cal H} &=&  i \Big(-R^\dagger_{j\s}\p_x R_{j\s} + L^\dagger_{j\s}\p_x L_{j\s}\Big) + 
m \Big(L^\dagger_{j\s}R_{j\s} + {\rm H.c.}\Big) \cr\cr
&& \hskip .25in +   gJ^A\bar J^A + g_f \mathcal{J}\bar{\mathcal{J}},
\label{model}
\end{eqnarray}
where $R_{j\s}, L_{j\s}$ are annihilation operators of the right- and the left moving quarks,  $j=1,2,3$ are color indices and $\s = \uparrow, \downarrow$ are flavor indices corresponding to up and down quarks (heavier quarks are neglected in this treatment). The speed of light has been set to one and a summation over repeated indices is implied henceforth. 

It is worth noting, although we will not discuss it in any detail, that the TSA has previously been applied to a $1+1$D description of QCD where the effect of the gauge field had not been reduced to an effective current-current interaction, see Ref.~\cite{katz2014solution}. However, in that case, the gauge field was coupled to a single fundamental fermion and so it is not possible to directly compare the results of this work to those obtained with the Nambu-Jona-Lasinio model. 

In Ref.~\cite{AzariaPRD16}, we treated this Hamiltonian through recourse to non-Abelian bosonization (again, see Section~\ref{Sec:NonAbelianBosonization} for an elaboration of this technique). To this end we explicitly write down $SU(3)_2$ Kac-Moody currents of right and left chirality
\be
J^A = :R^\dagger_{j\s} T^A_{jk} R_{k\s}:, \quad \bar J^A = :L^\dagger_{j\s} T^A_{jk} L_{k\s}:,
\ee
where $T^A_{jk}$ ($A= 1, \ldots, 8$) are the generators in the fundamental representation of the $SU(3)$ group, as well as the chiral $U(1)$ currents
\be
\mathcal{J} = :R^\dagger_{j\s}R_{j\s}:, \quad \bar{\mathcal{J}} = :L^\dagger_{j\s}L_{j\s}:. 
\ee
Here we stress that the $U(1)$ symmetry does not correspond to electric charge, but instead to conservation of baryon number (i.e., baryonic charge). To analyze~\fr{model} using non-Abelian bosonization, we used the fact that the Hamiltonian density of free Dirac fermions with symmetry $U(1)\times SU(N) \times SU(M)$ can be represented as a sum of a Gaussian $U(1)$ theory and WZNW models of levels $k=M$ and $k=N$ respectively~\cite{KnizhikNuclPhysB84,TsvelikBook,AffleckNuclPhysB86}. 

When $g>0$, the model~\fr{model} is asymptotically free and acquires a mass gap, $M_q = \Lambda g^{2/3}\exp(- 2\pi/3g)$ (where $\Lambda$ is the UV cut-off),  in the color sector \textit{even if the bare mass $m$ is zero}. We focus on the scenario where this dynamically generated quark mass, $M_q$, is the largest energy scale in the problem. Then, the corresponding effective Lagrangian density for low energies, $E \ll M_q$, (that is, the color singlet sector of the theory) is written in terms of the Abelian and non-Abelian Goldstone modes. It takes the form of a sigma model~\cite{GepnerNuclPhysB85,frishman2010non,TsvelikBook},
\begin{eqnarray}
{\cal L} &=& \frac{1}{8\pi}\partial^\mu\Theta\partial_\mu\Theta +  \lambda \cos \Big(2\Theta/\sqrt{6K} \Big) \cr\cr
&& + W[SU(2)_3;G]  +  m^*{\rm Tr}\Big(e^{i\Theta/\sqrt{6K}}G + {\rm H.c.}\Big), \qquad
\end{eqnarray}
for a compact boson with radius $(6K)^{1/2}$ that is coupled to an $SU(2)_3$ WZNW model [described by the Lagrangian $W[SU(2)_3;G]$, see Secs.~\ref{Sec:NonAbelianBosonization}--\ref{Sec:ColdAtomsExamples}, for the field $G$]. The effective mass $m^*$ is proportional to the bare quark mass [$m$ in~\fr{model}], $K$ is a Luttinger parameter related to the Abelian current-current coupling in the full unprojected model ($K=1+\alpha g_f$ if $g_f\ll1$ where $\alpha$ is an $O(1)$ constant), and $\lambda$ is a 't~Hooft instanton term~\cite{thooft1986how}.  

\begin{table}[!b]
\begin{center}
\begin{ruledtabular}
    \begin{tabular}{ l  r  l  l  l  l  r}
    particle species&$K=0.4$&$0.6$&$0.8$&$1.0$&$1.2$&$1.4$\\
    \hline
    nucleon&$4.5$&$4.3$&$4.5$&$4.8$&$5.1$&$5.4$\\
    isoscalar meson&$5.5$&$3.9$&$3.2$&$2.8$&$2.5$&$2.3$\\
    isovector meson&$3.6$&$3.1$&$2.9$&$2.8$&$2.7$&$2.7$\\
    isoscalar deuteron&$6.7$&$7.4$&$8.2$&$8.9$&$9.7$&$10.2$\\
    isovector deuteron&$8.2$&$8.3$&$8.7$&$9.2$&$9.7$&$10.2$\\
    \end{tabular}
\end{ruledtabular}
    \end{center}
\caption{Masses of the low-energy particles at $\lambda=0$ and zero particle 
density determined from TSA in units $M=(m^*)^{1/(2-d_{m^*})}$. We have estimated the 
error to be $0.5\,M$ and $1\,M$ in the meson and deuteron sectors, respectively, 
independent of $K$, and a relative accuracy to be one order of magnitude 
smaller.  Adapted from Ref.~\cite{AzariaPRD16}.}\label{qcd_spec}
\end{table}

In Ref.~\cite{AzariaPRD16} the masses of a large portion of the excitation spectrum in this model were determined numerically. These excitations include (for simplicity we set $K=1$ in these descriptions):
\begin{enumerate}
\item  Nucleons, Lorentz spin-1/2 bound states of three quarks carrying isospin $I=1/2$. These have the field representations
\begin{eqnarray}
n^j_{1/2} &=& C^{1/2}_{\delta,\alpha\beta\gamma}\epsilon^{abc}R_{a\alpha}R_{b\beta}L_{c\gamma} \cr\cr
&\sim& \exp\Big[i\sqrt{1/6}(2\varphi - \bar\varphi)\Big]\left[{\cal F}^{(1)}_{2/5}\bar{\cal F}^{(1/2)}_{3/20}\right], \label{H12}
\end{eqnarray}
where $j = \pm 1/2$, and $C^{1/2}_{j,\alpha\beta\gamma}$ is an appropriate coefficient projecting three isospin-1/2 representations to one of an isospin-1/2. $\varphi$ and $ \bar\varphi$ are the chiral components of the bosonic field, $\Theta = \varphi +\bar\varphi$, and  ${\cal F}^{(j)}_{h_j}, \bar {\cal F}^{({\bar j} )}_{{\bar h}_{\bar j}}$ denote the $SU(2)_3$ holomorphic and anti-holomorphic conformal blocks with isospin $j, {\bar j}=0,1/2,1,3/2$ and weights $h_j = \frac{j(j+1)}{5}$.  

Their counterparts with opposite Lorentz spin are given by similar expressions with barred quantities interchanged with their unbarred counterparts.

\item $\Delta$-baryons, Lorentz spin-3/2 particles carrying isospin $I=3/2$. These have field representations
\begin{eqnarray}
\Delta_{3/2}^{j} &=& C^{3/2}_{j,\alpha\beta\gamma}
\epsilon^{abc}R_{a\alpha}R_{b\beta}R_{c\gamma} \cr\cr
&\sim & \exp\Big(i\sqrt{3/2}\varphi\Big){\cal F}^{(3/2)}_{3/4},
\end{eqnarray}
where now $j = \pm3/2,\pm1/2$, and $C^{3/2}_{j,\alpha\beta\gamma}$ is a coefficient projecting the three isospin-1/2s onto isospin-3/2.

\item Mesons, bounds states of two quarks with Lorentz spin $0$. There are two possibilities here: we have both isosinglet mesons,
\begin{eqnarray}
M^0 &=& i ( R^\dagger_{j\alpha} L_{j \alpha} - H.c.) \cr\cr
&\sim& i\re^{-i\sqrt{1/6}\Theta}\mbox{Tr} \; G  + {\rm h.c.},
\end{eqnarray} 
and isospin-1 mesons,
\begin{eqnarray}
M^a &=& R^\dagger_{j\alpha} {\sigma^a}_{\alpha\beta} L_{j\beta} \cr\cr
&\sim & e^{-i\sqrt{1/6}\Theta}\mbox{Tr}\big[\sigma^a\big(G - G^\dagger\big)\big].
\end{eqnarray}

\item
Finally there are dibaryonic states, formed from bound states of six quarks, also known as deuterons. These can exist both with isospin 0
\begin{eqnarray}
d^0 &=& \big(R_{1\alpha}\epsilon_{\alpha\beta}L_{1\beta}\big)\big(R_{2\gamma}\epsilon_{\gamma\delta} L_{2\delta}\big)\big(R_{3\eta}\epsilon_{\eta\rho}L_{3\rho}\big) \cr\cr
&\sim & \exp\Big(i\sqrt{3/2}\Phi\Big)\mbox{Tr}\big(G + G^\dagger\big) ,
\end{eqnarray}
and with isospin 1 
\begin{eqnarray}
d^a &\sim & i\exp\Big(i\sqrt{3/2}\Phi\Big)\mbox{Tr}(\sigma^a(G-G^\dagger)),
\end{eqnarray}
where $\Phi$ is the field dual to $\Theta$. 

\end{enumerate}
The masses of these six types of excitations are given in Table~\ref{qcd_spec} for six different values of the Luttinger parameter $K$. Reference~\cite{AzariaPRD16} was the first time these masses were determined in a non-perturbative fashion. 

\subsection{Landau-Ginzburg theories in $1+1$D and higher}
\label{Sec:LandauGinzburg}

\subsubsection{Landau-Ginzburg theories in $1+1$D}
The final recent application of the TSA that we will discuss in this review is to Landau-Ginzburg theories, i.e. free bosons perturbed by polynomial interactions,
\begin{equation}
{\cal H} = \frac{1}{8\pi} \Big( (\partial_x\Theta)^2 + (\partial_x\Phi)^2 \Big) + m^2\Theta^2 + \lambda \Theta^4 + \ldots
\end{equation}
One might have thought that this canonical theory would seemingly be a target for early studies using the TSA, but in fact has only been treated in the past few years~\cite{coser2014truncated,rychkov2015hamiltonian,rychkov2016hamiltonian,bajnok2016truncated,Elias-Miro2016renormalized}. There are two reasons that make this particular application of the TSA of technical interest.  The first concerns how the zero momentum mode of the Bose field is handled. If we think of the Landau-Ginzburg theory as a perturbation of a massless free boson, we see that we immediately run into a problem: unlike the previously discussed sine-Gordon model, the boson here is non-compact. This means that the massless theory has a continuum, rather than a discrete, spectrum of highest weight states.  Thus no matter the truncation used, one will alway end up with an uncountable number of states. To deal with this difficulty several different strategies were employed.  

In the first study, Ref.~\cite{coser2014truncated} treated the zero mode of the Bose field as periodic with some large period. Thus, roughly speaking, the polynomial interaction (or at least the zero mode portion of it) was approximated by a periodic function. This gave reasonably well behaved results for small coupling constants.

In the second strategy, employed in Ref.~\cite{rychkov2015hamiltonian}, the non-interacting part of $H$ included the mass term while the $\Theta^4$ term was alone treated as the perturbation: 
\begin{eqnarray}
H_0 &=& \int^R_0 \rd x\bigg(\frac{1}{8\pi} \Big((\partial_x\Theta)^2 + (\partial_x\Phi)^2\Big) + m^2\Theta^2\bigg), \cr\cr
V_{\rm pert} &=& \lambda \int^R_0 \rd x\, \Theta^4. 
\end{eqnarray}
By using the massive non-interacting basis, the difficulties of dealing with a massless non-compact boson and its continuous spectra were avoided. The only limitation in this approach is the need to consider perturbations around the unbroken phase of the model. However, the authors of Ref.~\cite{rychkov2015hamiltonian} demonstrated that they could reach the broken phase of the $\Theta^4$ theory with sufficiently strong $\lambda$.  They were able to do so, in part, through the use of an analytic renormalization group of the type discussed in Sec.~\ref{analyticrg} of this review.

In the final strategy employed, Refs.~\cite{rychkov2016hamiltonian,bajnok2016truncated}, the zero momentum mode of the theory was singled out for a mini-superspace treatment analogous to that used in the study of the Liouville theory~\cite{seiberg1990notes,zamolodchikov1996conformal,bajnok2008sinh}. This innovation, in particular, enabled the study of the theory deep in its broken phase.  The basic idea here is to divide the Hamiltonian of theory into three pieces:
\begin{equation}
H = H_{\rm zero} + H_{\rm non-zero} + H_{\rm zero,non-zero}
\end{equation}
where $H_{\rm zero}$ is the Hamiltonian for the zero mode $a_0,a_0^\dagger$ alone:
\begin{equation}
H_{\rm zero} = \alpha\pi_0^2 + \beta a_0^2 + \gamma a_0^4,
\end{equation}
here $\pi_0$ is the momentum conjugate to $a_0$ and $\alpha,\beta,\gamma$ are constants depending on the system size, $R$, mass, $m$, and $\Theta^4$ coupling of the theory. $H_{\rm non-zero}$ involves the non-zero modes alone $\{a_{n\neq 0},a^\dagger_{n\neq 0}\}$, while $H_{\rm zero,non-zero}$ involves coupling between the two sectors.  

To proceed, one first solves the zero-mode piece of the Hamiltonian. This amounts to solving a quantum mechanics problem. One can do this by looking for solutions of the form
\begin{equation}
|\psi_0\rangle = \sum^M_{n=1} c_n (a^\dagger_0)^n|0\rangle
\end{equation}
In practice the size $M$ of the Hilbert space one searches for solutions is on the order of 500~\cite{rychkov2016hamiltonian} to a few thousand~\cite{bajnok2016truncated} states. Once one solves the zero-mode portion of the Hamiltonian, one performs a standard TSA on the full problem, but with the computational basis consisting of states of the form
\begin{equation}
|\psi_0\rangle \otimes \prod^K_{i=1}a^\dagger_{n_i<0}|0\rangle
\end{equation}
Having solved the zero-mode portion of the theory `exactly', the number of eigenstates $|\psi_0\rangle$ of $H_{\rm zero}$ one needs to include is small (less than 10~\cite{rychkov2016hamiltonian}).

The basic idea when using this approach to study the broken symmetry states is that it is the zero mode sector of the theory that is most sensitive to a negative bare mass, i.e. it is the zero mode that gets localized around a new field minima whereas the non-zero modes fluctuate around their unbroken vacuum. So, if we first solve this sector of the theory with high accuracy, the numerical effort needed to solve the remaining full theory is considerably reduced.

\subsubsection{Landau-Ginzburg theories in higher spatial dimensions}

One of the most interesting recent developments in the use of the TSA to study field theories has been the study in Ref.~\cite{hogervorst2015truncated}, where the Landau-Ginzburg model was studied in higher (albeit fractional) dimensions.\footnote{The authors here worked in fractional dimensional in order to avoid null states that appear at integer dimension for free scalar theories.} It would be tremendously exciting if the extensive body of work using the TSA to study continuum theories in 1+1 dimensions could be extended to higher dimensional field theories. Indeed, Ref.~\cite{hogervorst2015truncated} represents a promising start here: the authors were able to observe the various phases (broken, conformal, and unbroken) of the $\Theta^4$ theory and were even able to make estimates for the critical exponents of the theory.  One question that will need to be addressed in detail for TSA studies in higher dimensions will be the choice of computational basis.  One natural extension of the work of Ref.~\cite{hogervorst2015truncated} (discussed but not implemented by these authors) would be to use a massive basis to study Landau-Ginzburg theories in higher dimensions. In this vein, there has already an intriguing proposal to use light-cone quantization in infinite volume to provide a basis of states for the TSA in arbitrary dimensions~\cite{katz2016conformal}.

\subsection{Summary: the truncated space approach}
In this part of the review, we have provided a comprehensive introduction to the TSA through its application to two canonical examples: the quantum Ising model and sine-Gordon model. Following this, we discussed recent work to extend the TSA, using analytical and numerical RG methods, to ameliorate the effects of the Hilbert space truncation. Our discussions highlighted the strengths and deficiencies of the method, as well as points where one needs to take care when analyzing the results. The TSA+NRG methods were then applied to a wide variety problems: excitons in carbon nanotubes, non-equilibrium dynamics of the Lieb-Liniger model, 1+1D quantum chromodynamics, and Landau-Ginzburg models. 

Through the example applications, we saw how the TSA+NRG can be directly applied to scenarios of experimental interest (see, for example, Fig.~\ref{sfeir_exp_data} where we directly compare TSA+NRG results to experimental data), as well as those of a more theoretical interest (see, e.g., Sec.~\ref{Sec:TSAWZNW}). In Sec.~\ref{Sec:LandauGinzburg} we also presented a discussion of applications of the TSA to Landau-Ginzburg theories, an important class of problems that underpin much of our phenomenological understanding of phases of matter. Extending such studies to higher dimensions is at the forefront of current research with the TSA, see Refs.~\cite{hogervorst2015truncated,katz2016conformal}. 

In the following section, we discuss an alternative route to higher dimensional problems. By blending information from TSA analyses of continuum one-dimensional quantum systems with matrix product state technologies, we can study (discrete) arrays of continuum systems.

\section{Beyond integrability II: Matrix product states for arrays of integrable chains}
\label{Sec:CHAMPS}
\subsection{Introduction}
When there is no systematic analytical recipe to tackle a many-body problem, we must turn to explicitly numerical methods.\footnote{In contrast to approaches where analytical expressions can be written down, but still must be evaluated numerically, e.g.~\cite{CauxJMathPhys09}.}
Nevertheless even these methods must introduce some form of approximation to make progress: an exponential growth in the Hilbert space dimension with the number of degrees of freedom limits exact diagonalization treatments to systems of $\mathcal{O}(10)$ interacting objects. 
Here the largest numbers are possible only when the local Hilbert space is small and symmetries exist that reduce the Hamiltonian to block diagonal form.
An efficient, yet accurate, approximate basis for representing many-body states is therefore highly desirable.
Such a basis is now known for ground and low lying excited states of 1D quantum Hamiltonians, yielding algorithms which in many cases are \emph{numerically exact}.\footnote{This indicates that an algorithm can find the exact solution, to a specified precision.}
These matrix product states (MPS) constitute a variational basis over states with restricted entanglement, where the maximum possible entanglement between two parts of a system is controlled by the so-called \textit{bond} (or matrix) dimension.

A key development in understanding the usefulness of MPS was the introduction of the density matrix renormalization group method, by White~\cite{white1992density,white1993density}. 
This algorithm allows one to find low energy eigenstates in 1D quantum systems, though the elucidation of the relation between DMRG and MPS came somewhat later~\cite{ostlund1995thermodynamic,rommer1997class,dukelsky1998equivalence}.

The particular efficacy of MPS representations in 1D is a consequence of the behaviour of many-body entanglement, a useful measure of which is the von Neumann entanglement entropy, $S_E$.
As shown by Holzhey, Larsen and Wilczek~\cite{holzhey1994geometric}, and later by Calabrese and Cardy~\cite{calabrese2004entanglement}, $S_E$ for many-body ground states (of Hamiltonians with short range interactions) in 1D grows only logarithmically with system size in critical (gapless) systems and saturates at a scale set by the correlation length in massive (gapped) systems.
While similar behaviour is expected for low lying excited states, an arbitrary state can have much larger entanglement entropy, up to a maximum of $\sim \log N$ in a system with $N$ degrees of freedom.
As the bond dimension necessary to accurately represent a state grows with $S_E$, the worst-case logarithmic growth implies that MPS are numerically well suited to studying the low energy spectrum of 1D problems: even critical systems can be studied by a suitably controlled extrapolation of results with bond dimension~\cite{pollmann2009theory}.

A major benefit of using MPS is that they allow for robust, stable and precise algorithms; besides DMRG there are also algorithms for (real and imaginary) time evolution: \textit{time-evolving block decimation} (TEBD and variants)~\cite{vidal2004efficient} and a separate method based on the time-dependent variational principle (TDVP)~\cite{haegeman2011time}.
These techniques can often be applied directly in the thermodynamic limit, via the so-called iTEBD~\cite{vidal2007classical,orus2008infinite} and iDMRG~\cite{mcculloch2008infinite} algorithms.
They are also equally applicable to fermionic, bosonic and spin systems, and can be extended to finite temperatures~\cite{verstraete2004matrix,feiguin2005finite,karrasch2013reducing}.
Evaluation of local expectation values is highly efficient: by using matrix product operators (MPOs) and `canonical' forms of MPS these can be reduced to operations on one or few sites.
By construction matrix product states have the ability to directly access important quantum information measures, including $S_E$.

MPS have been applied to a diverse range of problems in 1D quantum systems, with studies including: 
\begin{enumerate}
\item Detailed mapping of the phase diagram of many models;
\item Examination of stripes in the $t-J$ model, motivated by high-temperature superconductors~\cite{ScalapinoPhysicaC12};
\item The calculation of dynamical correlation functions in quantum magnets~\cite{feiguin2005finite,barthel2009spectral};
\item Simulation of non-equilibrium dynamics following a `quantum quench'~\cite{daley2004time,WhitePRL04,karrasch2013dynamical};
\item Identification of topological order through the entanglement entropy~\cite{JiangNaturePhys12};
\item Describing high-energy eigenstates in many-body localized phases~\cite{KhemaniPRL16,YuPRL17}
\item Construction of Floquet eigenstates in problems with periodic time-dependent Hamiltonians~\cite{ZhangArxiv16};
\end{enumerate}
amongst many other (see, for example,~\cite{DMRGHomePage} for a list of manuscripts where MPS technology is used). 

One would like to translate these successes to strongly correlated problems in 2D and above, not least because of deficiencies in other available methods.
Relative to MPS methods, Quantum Monte Carlo~\cite{FoulkesRMP01} famously suffers from a fermionic `sign problem', requires careful treatment of statistical uncertainties, and does not provide such easy access to quantum information measures (though calculation of some generalised entropies is still possible~\cite{hastings2010measuring}).
On the other hand, dynamical mean field theory (DMFT)~\cite{georges1996dynamical} is well suited to higher dimensions (becoming exact in the infinite limit), but lacks the spatial resolution of DMRG, and still requires the solution of a complicated interacting impurity problem by some other means (such as ED, QMC or an MPS method~\cite{ganahl2015efficient}).

Unfortunately MPS algorithms are considerably less powerful in 2D than in 1D.
Why this is the case can again be understood in the context of entanglement.
Quantum information theory indicates that the entanglement of low-lying states obeys an \textit{area law}~\cite{hastings2007area,eisert2010colloquium}, with $S_E$ for a bipartite system proportional to the size (area) $\mathcal{A}$ of the interface between the two regions.\footnote{A known exception is critical fermions, for which an extra multiplicative logarithm gives $\mathcal{A} \log \mathcal{A}$ scaling~\cite{gioev2006entanglement}.}
In 2D this entails much faster growth of $S_E$ with system size and hence lower accuracy for a given bond dimension, even for gapped systems.
\begin{figure}
\includegraphics[width=0.4\textwidth]{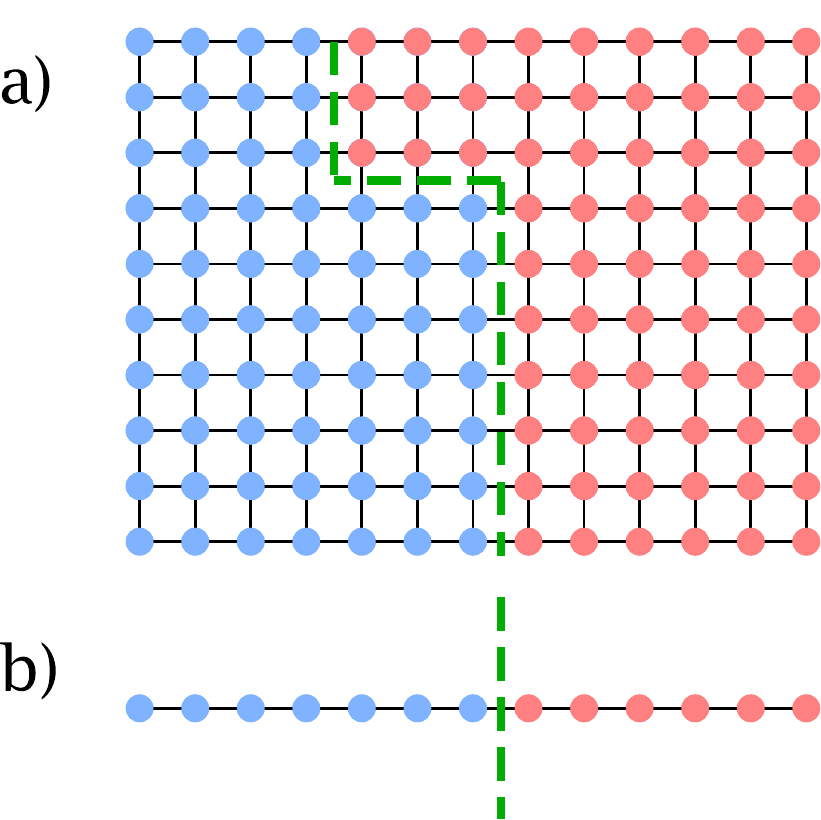}
\caption{Bipartite systems and area laws. a) The entanglement between two partitions (red and blue) of a low energy state in a gapped 2D system is proportional to the length of boundary between them. b) In 1D the boundary reduces to a finite number of points.}
\end{figure}
The above notwithstanding, MPS techniques have been fruitfully applied to two-dimensional systems.
The canonical two-dimensional approach to DMRG is to map the 2D lattice to a 1D system with long range interactions~\cite{liang1994approximate,white1996spin,stoudenmire2012studying} (see Fig.~\ref{fig:zigzag}).
This mapping can be chosen in a variety of ways, with different 1D paths through the 2D system possibly being better (in the sense of requiring smaller bond dimension for a given accuracy) depending on the structure of the state or Hamiltonian.
A few among the many applications of this method are: establishing the spin liquid ground state of the spin half Heisenberg antiferromagnet on the kagome lattice~\cite{yan2011spin,jiang2012identifying}, stripe formation in Hubbard models~\cite{white2003stripes,hager2005stripe}, and identifying toplogical orders~\cite{zaletel2013topological} (see Ref.~\cite{stoudenmire2012studying} for a more complete list).

To overcome the lingering 1D character of MPS, several generalizations have been proposed that aim to represent states in higher dimensions in a direct way.
These fall under the moniker of \textit{tensor methods} and include projected entangled pair states (PEPS)~\cite{nishino2001two,verstraete2004renormalization,verstraete2008matrix}, their thermodynamic limit variant `iPEPS'~\cite{jordan2008classical,orus2009simulation,orus2014practical} and the multiscale entanglement renormalization ansatz (MERA)~\cite{vidal2007entanglement,vidal2008class,evenbly2009algorithms}.
Both methods have proven useful for analysing 2D systems, but they also have weaknesses.
PEPS constitute a genuine variational basis for higher dimensions, and can be applied in the thermodynamic limit yet they are computationally expensive.
Furthermore the actual evaluation of expectation values within PEPS requires approximations to be made, ruining the variational nature.
MERA allows for exact evaluation of observables, however there is an even higher numerical burden than with PEPS, and it is not variational, requiring particular choices to be made which could in principle prevent convergence to the true state.

Hence, in comparison to 1D, there is still no single `best' approach to strongly correlated physics in 2D, and instead we must choose the most appropriate method based on the model of interest and the quantities we wish to measure.

The use of DMRG, MPS and tensor algorithms has been a well established field for some time, and thorough review articles on almost all aspects already exist.
The purpose of this section is not to replicate these works, but to review a specific flavour of 2D MPS that has a close connection to integrability, shares many of the advantages of 1D MPS, and is especially apt for analysing anisotropic systems~\cite{konik2009renormalization,james2013understanding,james2015quantum}.
In essence these `chain array matrix product states' (ChainAMPS) form a two-dimensional system as an array of one-dimensional quantum chains.
This may appear to negate the original benefit of MPS, namely the transformation of an exponential number of complex numbers into a linear number of finite dimensional matrices.
In fact, when an exact strongly correlated basis for the 1D chains is already known, and the entanglement between them is limited in some manner, this arrangement can be highly beneficial.

We first provide a brief overview of MPS in 1D and the canonical approach to 2D in \ref{ss:mps}, in order to provide context.
In \ref{ss:champs} we describe the reasoning behind, and the formulation of, ChainAMPS.
Finally in \ref{ss:2Dff} and \ref{ss:2Dqim} we provide example applications of ChainAMPS to two strongly correlated systems in 2D: free fermions and the quantum Ising model, covering both DMRG and TEBD algorithms.
We also point the interested reader to a software implementation of the various ideas discussed in this section, available at \url{https://bitbucket.org/chainamps}.

We conclude this introduction by pointing the interested reader to but a small sample of the extant reference literature. 
Perhaps the most comprehensive review is that due to Sch\"{o}llwock~\cite{schollwock2011density} which covers the structure of MPS and their relation to DMRG; and many other algorithms including those for time evolution; itself being an excellent source of references.
Earlier reviews by the same author~\cite{schollwock2005density} and by Hallberg~\cite{hallberg2006new}, focus on DMRG in its original (density matrix) implementation and its applications.
McCulloch gives a much more compact description of the MPS formulation of DMRG~\cite{mcculloch2007density}, but also covers the distinction between algorithms employing Abelian and Non-Abelian symmetries.
The canonical (zig-zag) extension of MPS to 2D is the subject of a thorough review by Stoudenmire and White~\cite{stoudenmire2012studying}.
Tensor methods and their application in higher dimensions are discussed in Refs.~\cite{verstraete2008matrix,cirac2009renormalization,orus2014practical}.
\subsection{A very brief guide to matrix product states}
\label{ss:mps}
\subsubsection{MPS as a variational basis with limited entanglement}
\begin{figure}
\includegraphics[width=0.45\textwidth]{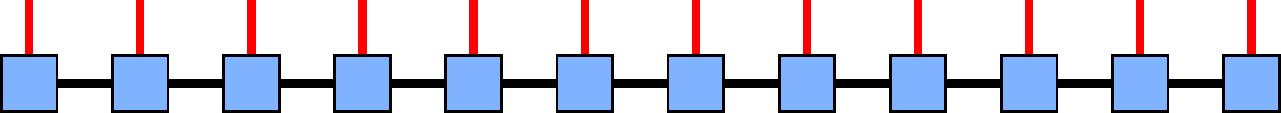}
\caption{Graphical depiction of an MPS. Each blue square represents a set of matrices, $\mathbf{M}^{\sigma_i}$, black bonds represent the (summed over) matrix indices, while the red lines indicate the physical (local) indices, $\sigma$.}
\label{fig:1dmps}
\end{figure}

Consider a many-body system consisting of $N$ local Hilbert spaces, each with a basis $\ket{\sigma}$ of dimension $d_\sigma$.
Any state of this system can be written as
\begin{align}
\ket{\Psi}&=\sum_{\boldsymbol{\sigma}} c_{\sigma_1 \sigma_2 \cdots \sigma_N} \ket{\sigma_1} \otimes \ket{\sigma_2} \otimes \cdots \otimes \ket{\sigma_N}, \nonumber \\
&=\sum_{\boldsymbol{\sigma}} c_{\boldsymbol{\sigma}} \ket{\boldsymbol{\sigma}}.
\label{eq:wf}
\end{align}
In general, for physically interesting values of $d_\sigma$ and $N$, calculating, storing or operating on the $d_\sigma^N$ complex numbers $c_{\boldsymbol{\sigma}}$ is not possible.
Instead we rewrite the $c$'s as products of matrices (hence the name \textit{matrix product state}):
\begin{align}
\ket{\Psi}&= \sum_{\boldsymbol{\sigma}} \mathbf{M}^{\sigma_1} \mathbf{M}^{\sigma_2} \cdots \mathbf{M}^{\sigma_N} \ket{\sigma_1} \otimes \ket{\sigma_2} \otimes \cdots \otimes \ket{\sigma_N}, \nonumber \\
&=\sum_{\boldsymbol{\sigma}} \prod_i \mathbf{M}^{\sigma_i} \ket{\sigma_i},
\label{eq:mps}
\end{align}
where we have dropped the explicit tensor product, and the first and last matrices are understood to be row and column vectors respectively, in order to recover a scalar (we assume open boundary conditions, for periodic boundary conditions it is necessary to take the trace instead).
A graphical depiction of this construction is given in Fig.~\ref{fig:1dmps}.

The set of $d_\sigma$ matrices $\mathbf{M}^{\sigma_i}$ can also be usefully viewed as a tensor $M^{\sigma_i}_{a_{i-1},a_i}$.
In MPS parlance the $\sigma_i$ labelling the local basis states is termed a `physical index' to distinguish it from the standard matrix indices.
Any state can be written in the form of Eq.~\fr{eq:mps}, but for a generic state the necessary matrix dimensions would grow as large as $d_\sigma^{N/2}$.
In practice the matrix dimension must be truncated at some numerically feasible value, $\chi$, termed the `bond dimension' because it applies to the bond between two partitions of the system.
This truncation implies some loss of information relative to the exact wave function: we keep up to $N d_\sigma \chi^2$ complex numbers only, instead of $d_\sigma^N$.

To best understand the nature of this approximation, we first introduce two \emph{canonical} forms of matrix.
Left canonical matrices $\mathbf{A}^{\sigma_i}$ obey
\begin{align}
\sum_{\sigma_i} \mathbf{A}^{\sigma_i \dagger} \mathbf{A}^{\sigma_i} =\mathbf{I},
\label{eq:lc}
\end{align}
while right canonical matrices $\mathbf{B}^{\sigma_i}$ obey
\begin{align}
\sum_{\sigma_i} \mathbf{B}^{\sigma_i} \mathbf{B}^{\sigma_i \dagger} =\mathbf{I}.
\label{eq:rc}
\end{align}
We can view each set of local matrices $\mathbf{M}^{\sigma_i}$ as a single matrix $M_{(\sigma_i m) n}$ or $M_{m(\sigma_i n)}$, where the parentheses indicate the collection of multiple indices into a single super-index.
This allows us to decompose them into a unitary matrix and an auxiliary matrix, for example using a QR or LQ decomposition:
\begin{align}
M_{(\sigma_i m) n}& =U_{(\sigma_i m) n'} R_{n' n}, \nonumber \\
M_{m(\sigma_i n)}& =L_{m m'} V^\dagger_{m' ((\sigma_i n)}, \nonumber
\end{align}
with $\mathbf{U}$ and $\mathbf{V}$ unitary.
Using this property we form canonical matrices
\begin{align}
U_{(\sigma_i m) n} \to A^{\sigma_i}_{m n} \quad \text{and} \quad V^\dagger_{m ((\sigma_i n)} \to B^{\sigma_i}_{m n}.
\end{align}
In the context of the full MPS, the auxiliary matrix $R$ ($L$) can be absorbed by multiplying to the right (left) with $M^{\sigma_{i+1}}$ ($M^{\sigma_{i-1}}$).
Performing these steps in sequence (for example starting at $i=1$ and QR decomposing each matrix in turn until we reach $i=N$) transforms a general MPS into one that consists entirely of left or right canonical matrices (with a multiplicative scalar that gives the normalization and an overall phase).
Generically, working inwards from the left and right leads to a mixed state:
\begin{align}
\ket{\Psi}=\sum_{\boldsymbol{\sigma}} \mathbf{A}^{\sigma_1} \cdots \mathbf{A}^{\sigma_{i-1}}   \mathbf{M}^{\sigma_i}  \mathbf{B}^{\sigma_{i+1}} \cdots \mathbf{B}^{\sigma_N} \ket{\boldsymbol{\sigma}},
\label{eq:mixed}
\end{align}
with $\mathbf{M}^{\sigma_i}$ not canonical.

We now elucidate the importance of the bond dimension by performing a \emph{singular value decomposition} (SVD) on the reshaped matrix, $M_{(\sigma_i m) n}$:
\begin{align}
M_{(\sigma_i m) n}=U_{(\sigma_i m) \ell} \Lambda_{\ell \ell'} V_{\ell' n}^\dagger.
\end{align}
According to the properties of the SVD, the matrices $\mathbf{U}$ and $\mathbf{V}$ are unitary, while $\mathbf{\Lambda}$ is a rectangular, $\dim(m) \times \dim(n)$, diagonal matrix with non-negative real entries known as singular values, $s_j$ ($j=1,\ldots,m$).  
The structure of $\mathbf{\Lambda}$ allows us to safely discard all but the first $\min(\dim(m),\dim(n))$ columns of $\mathbf{U}$ and $\mathbf{V}$ without losing any information about $\ket{\Psi}$.
Furthermore, if any of the singular values are zero, we can discard the columns of $\mathbf{U}$ and $\mathbf{V}$ they correspond to as well.
After doing so, we identify $U_{(\sigma_i m) \ell}$ as $\mathbf{A}^{\sigma_i}$ and multiply by $\mathbf{V}^\dagger \mathbf{B}^{\sigma_{i+1}}$ to form a new right canonical matrix $\tilde{\mathbf{B}}^{\sigma_{i+1}}$.
The state is now in the form of a Schmidt decomposition:
\begin{align}
\ket{\Psi}=\sum_{\boldsymbol{\sigma}} \mathbf{A}^{\sigma_1} \cdots \mathbf{A}^{\sigma_{i-1}}   \mathbf{A}^{\sigma_i} \mathbf{\Lambda} \mathbf{B}^{\sigma_{i+1}} \cdots \mathbf{B}^{\sigma_N} \ket{\boldsymbol{\sigma}},
\label{eq:schmidt}
\end{align}
with $\mathbf{\Lambda}$ a $D \times D$ diagonal matrix, $D$ being the number of nonzero singular values, or Schmidt coefficients.
Using the canonicity conditions, Eqs.~\fr{eq:lc} and~\fr{eq:rc}, the Frobenius norm of the mixed state, Eq.~\fr{eq:mixed}, is
\begin{align}
\overlap{\Psi}{\Psi}=\text{Tr}\,\mathbf{M}^{\sigma_i \dagger} \mathbf{M}^{\sigma_i}=\text{Tr}\, \mathbf{\Lambda}^2=\sum_{m=1}^D s_m^2.
\end{align}
For the state to be normalised we require
\begin{align}
\sum_{m=1}^D s_m^2 =1.
\end{align}
The effect of truncating the matrix dimensions is now apparent.
If the singular values are ordered from largest to smallest and we retain only the first $\chi < D$ of them, then this is the optimal approximation to the exact state for matrix dimension $\chi$ (in the sense of the Frobenius norm).
If we wish to keep the state normalised we must rescale the truncated singular values accordingly.
By repeated use of SVD, the truncation can be carried out on all bonds, so that all the matrices are $\chi \times \chi$ or smaller.

An important feature of Eq.~\fr{eq:schmidt} is that the distribution of Schmidt coefficients describes the entanglement across the bond $i,i+1$ because their squares are equal to the eigenvalues of the system's reduced density matrix.
The von Neumann entanglement entropy encodes this information as a single number:
\begin{align}
S_E = -\sum_{i=m}^D s_m^2 \log s_m^2,
\end{align}
which is zero for a product or separable state ($s_1=1$) and achieves its maximum value,  $S_E=\log D$, when the singular values are all equal, $s_m=1/\sqrt{D}$.
Therefore an MPS with bond dimension $\chi$ has a maximum possible entanglement entropy of $S_E=\log \chi$ for any bipartitioning of the system.

Clearly one cannot construct an MPS directly from Eq.~\fr{eq:wf}, as this would entail first calculating $c_{\boldsymbol{\sigma}}$.
Instead Eq.~\fr{eq:mps} is treated as a trial wave function in some variational scheme and optimized by iteratively sweeping back and forth through the local matrices, $\mathbf{M}^{\sigma_i}$, with canonization and truncation carried out as necessary.

The performance of these variational methods can be dramatically improved by incorporating Abelian and non-Abelian symmetries into the MPS~\cite{SierraNuclPhysB97,McCullochEPL02,schollwock2005density,hallberg2006new,mcculloch2007density,McCullochPRB08,TothPRB08,schollwock2011density,WeichselbaumAnnPhys12}, because enforcing such symmetry sectors ensures a block structure which generally leads to a heavily compressed matrix representation.
It is also possible to study infinite translationally invariant systems, using an MPS composed of a repeating unit cell of matrices, so-called \textit{uniform} MPS (uMPS)~\cite{ostlund1995thermodynamic,rommer1997class}.

When an MPS algorithm requires some initial input, this can be provided by growing an MPS iteratively from a small exactly solvable system, as with infinite volume DMRG; making an educated guess; or choosing a completely random set of matrices.
The latter is probably a bad idea as it is unlikely to respect any conservation laws.
Finally, it is sometimes possible to derive a useful MPS analytically, as with the famous AKLT state~\cite{affleck1987rigorous,fannes1989exact}.

\subsubsection{Entanglement and dimensionality}
The structure of an MPS, Eq.~\fr{eq:mps}, immediately lends itself to describing 1D lattice problems, with each of the local Hilbert spaces describing a single lattice site, but this does not guarantee a useful representation.
Instead the success of MPS in 1D is best understood through the scaling of entanglement with system size.
For a system of total length $L$, the entanglement entropy of a (contiguous) region of length $x$ scales as~\cite{holzhey1994geometric,calabrese2004entanglement},
\begin{align}
S_E & = \mathcal{A} \frac{c}{6} \log L_{\text{eff}}, \nonumber \\
L_{\text{eff}}&= \left\{
\begin{array}{cl}
 \frac{L}{\pi a}\sin\left(\frac{\pi x}{a}\right), & ~~\text{critical, or } x \le \xi   \\
 \frac{\xi}{a} & ~~\text{gapped, }  x > \xi,
\end{array}
\right.
\label{eq:1dS_E}
\end{align}
where $c$ is the central charge of the system, and $a$ is the short distance cut-off of the theory (e.g., the lattice spacing).
Equation~\fr{eq:1dS_E} indicates that $S_E$ saturates for a gapped 1D system at a scale set by the correlation length $\xi/a$ (where $a$ is the lattice constant), while for a critical 1D system the entanglement grows at worst as the $\log$ of the system size.
The factor $\mathcal{A}$ is equal to 1 or 2, depending on the number of boundary points that separate the region $x$ from the rest of the system.
In agreement with Eq.~\fr{eq:1dS_E}, explicit studies of the  density matrix eigenvalues in exactly solvable cases show that they fall off exponentially for gapped 1D systems~\cite{peschel1999densityA,peschel1999densityB,okunishi1999universal}, hence the efficacy of MPS with bond dimension $\chi$ and $S_E\le \log \chi$. 
On the other hand, the slow logarithmic growth at criticality in 1D means that useful results can be obtained through finite size or finite bond dimension scaling~\cite{pollmann2009theory,pollmann2010entanglement} even when the system is gapless.

A corresponding result does not exist in 2D and above, but the general expectation is that ground states (and to a large extent low-lying excited states) of short ranged many-body Hamiltonians should obey an \textit{area law}, such that the entanglement between two regions scales with the size, $\mathcal{A}$, of the partition between those two regions, $S_E \sim \mathcal{A}$~\cite{eisert2010colloquium}.
An exact result for \emph{critical} fermions in a $D$-dimensional (hyper)cube with side length $L$ shows that they obey an area law with logarithmic correction: $S_E \sim L^{D-1} \log L$~\cite{gioev2006entanglement}.
Therefore in 2D one expects the entanglement to grow strongly with the size of the system being studied, and the spectrum of singular values will not fall off quickly enough for an MPS with finite bond dimension to represent even a gapped system in the thermodynamic limit.
It should be noted, however, that this does not preclude MPS (and in particular DMRG) from being used to study small or highly anisotropic 2D systems.

\subsubsection{MPS algorithms}
\label{sss:mpsalgorithms}
\begin{figure}
\includegraphics[width=0.4\textwidth]{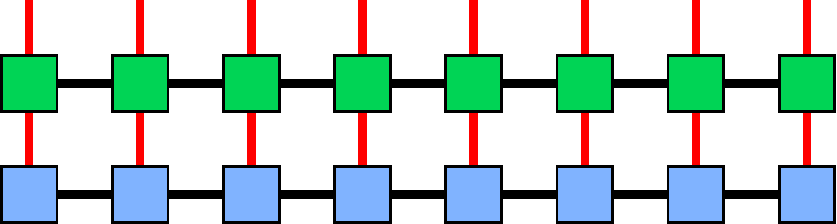}
\caption{Diagrammatic representation of a matrix product operator (green) applied to a matrix product state (blue), Fig.~\ref{fig:1dmps}.
Each tensor in the matrix product operator has two physical (e.g., external) indices shown in red and two internal indices (shown in black).
The exception is at the ends of the system where the tensors have only one internal index.
Contracting (summing over) the connected physical indices yields a new MPS.}
\label{fig:1dmpomps}
\end{figure}

MPS algorithms fall into two main categories: eigensolvers and time evolution, with some overlap because imaginary time evolution can be used to find eigenstates.
Though it was originally developed without recourse to an explicit MPS representation, DMRG is the primary example of the former, while TEBD is the seminal MPS time evolution algorithm.
Mirroring the MPS structure, an operator $\hat{\mathcal{O}}$ on the Hilbert space of the system can be described by a tensor network termed a \textit{matrix product operator} (MPO),
\begin{align}
\hat{\mathcal{O}}=\sum_{\boldsymbol{\sigma}',\boldsymbol{\sigma}} \mathbf{W}^{\sigma'_1 \sigma_1} \mathbf{W}^{\sigma'_2 \sigma_2} \cdots \mathbf{W}^{\sigma'_N \sigma_N} \ket{\boldsymbol{\sigma}'}\bra{\boldsymbol{\sigma}},
\end{align}
where the matrices (or four index tensors), $\mathbf{W}^{\sigma'_i \sigma_i}$, are labelled by two physical indices, $\sigma'_i, \sigma_i$ (in addition to their matrix indices), as opposed to the single physical index for an MPS matrix.
The action of an operator on an MPS (such as the Hamiltonian in DMRG, or the unitary time evolution operator in TEBD) is then evaluated in an iterative manner, by contracting (summing over) the corresponding physical and matrix indices of the MPO and MPS networks.
Note that the MPO has its own, separate, bond dimension, $\chi_W$.
Applying an MPO to an MPS, by contracting the matching physical indices, results in another MPS (here the $\otimes$ tells us that the matrix indices of the MPO and MPS live in different spaces, and so we sum over their tensor product)
\begin{align}
\hat{\mathcal{O}}\ket{\Psi}&=\sum_{\boldsymbol{\sigma}',\boldsymbol{\sigma}} \prod_i \mathbf{W}^{\sigma'_i \sigma_i} \otimes \mathbf{M}^{\sigma_i} \ket{\sigma'_i}\nonumber \\
&=\sum_{\boldsymbol{\sigma}',\boldsymbol{\sigma}} \sum_{\mathbf{a},\mathbf{b}} \prod_i  W^{\sigma'_i \sigma_i}_{b_{i-1} b_i} M^{\sigma_i}_{a_{i-1} a_i} \ket{\sigma'_i}\nonumber \\
&=\sum_{\boldsymbol{\sigma}'}  \sum_{\mathbf{a},\mathbf{b}} \prod_i \tilde{M}^{\sigma'_i}_{(b_{i-1}a_{i-1})(b_i a_i)} \ket{\sigma'_i}\nonumber\\
&=\sum_{\boldsymbol{\sigma}'}  \prod_i \tilde{\mathbf{M}}^{\sigma'_i} \ket{\sigma'_i},
\end{align}
with new matrices $\tilde{\mathbf{M}}^{\sigma_i}$, which have larger bond dimension, $\chi_W \times \chi$, and which will not in general be canonical.
This is shown in the diagrammatic notation of Fig.~\ref{fig:1dmps} in Fig.~\ref{fig:1dmpomps}.

\begin{figure}
\includegraphics[width=0.35\textwidth]{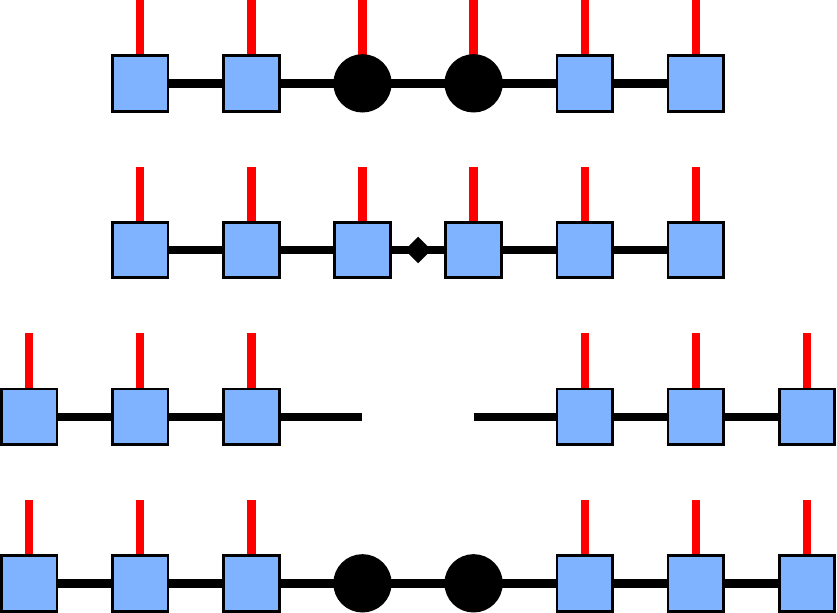}
\caption{Schematic of the DMRG ``infinite system" growth procedure (full tensor network not shown). A pair of sites is added at the centre of the chain (black circles, first panel) and the new wave function found (subject to the other matrices being held constant). An SVD is then performed on the enlarged problem, and the new reduced basis is constructed by discarding states with small Schmidt coefficients. The resulting state is the new DMRG approximation (second panel). The system is then divided (third panel) and two new sites are added (fourth panel) and the process proceeds iteratively until the required system size is achieved. Following this, a sweeping procedure is performed, see Fig.~\ref{fig:dmrgsweeps}.}
\label{fig:dmrggrowth}
\end{figure}

\begin{figure}
\includegraphics[width=0.25\textwidth]{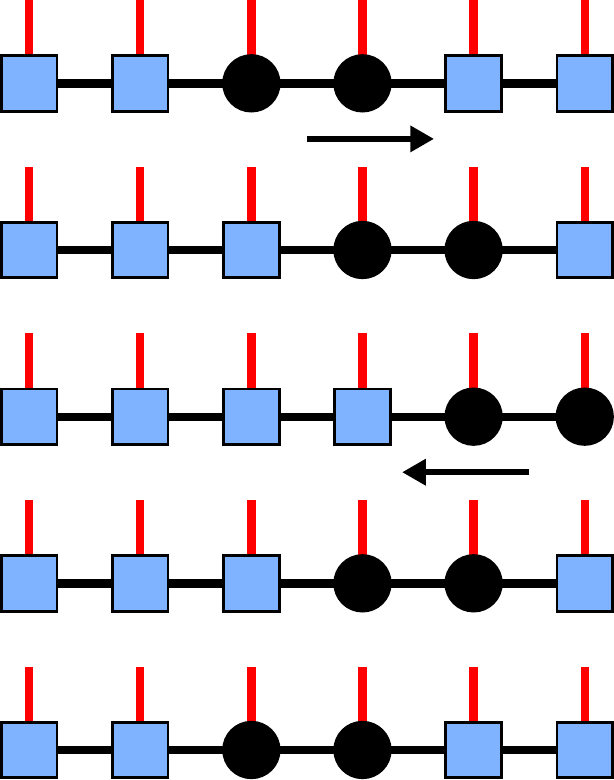}
\caption{Schematic of the DMRG finite size algorithm (full tensor network not shown). 
After the required system size is reached using the ``infinite system" DMRG algorithm, Fig.~\ref{fig:dmrggrowth}, edge effects are still present due to the small initial system size.
These are removed by repeatedly sweeping right and left through the system, in an iterative manner, locally optimizing a few sites (black circles) of the wave function at a time.}
\label{fig:dmrgsweeps}
\end{figure}

DMRG usually consists of two stages. In the \textit{infinite system algorithm} stage, an MPS approximation is grown iteratively by adding sites to the centre of a system that is initially small enough to be solved exactly.
Each growth step consists of solving an eigenvalue problem for the one or two new sites coupled to the rest of the system, followed by an SVD to compress the answer.
This process is shown schematically in Fig.~\ref{fig:dmrggrowth}, and it is stopped when the required system length is reached. 
In the second \textit{finite size sweeps} stage (Fig.~\ref{fig:dmrgsweeps}), the finite length MPS is improved by sweeping back and forth through the system, again using an eigensolver for each site (or pair of sites).
The second stage is important for removing unphysical effects produced by the edges of the system during the growth phase.

As already alluded to, DMRG traditionally takes two site or single site forms.
The former is numerically less efficient, by a factor of approximately $d_\sigma$, but the latter is liable to converge to a local, rather than the global, minimum and requires amendment in the form of a modified density matrix with noise term~\cite{white2005density}, or subspace expansion~\cite{hubig2015strictly}.


\begin{figure}
\includegraphics[width=0.45\textwidth]{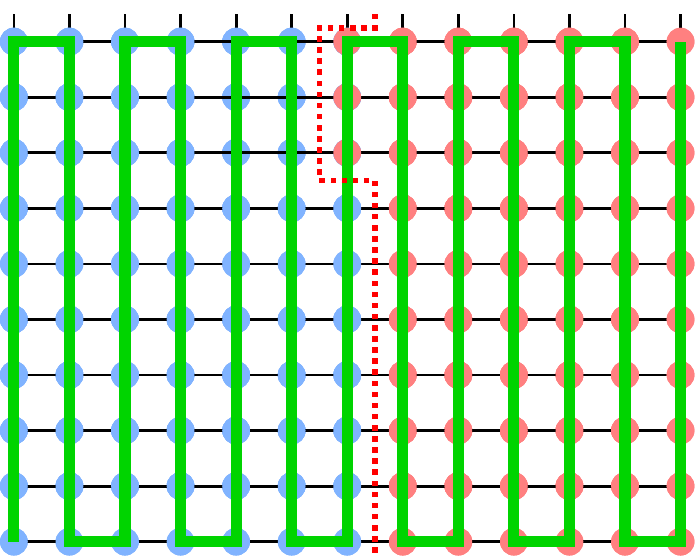}
\caption{Two-dimensional lattice and mapping to 1D. The green line shows the 1D path through the system. The red dashed line marks the boundary between two partitions (red and blue sites) at a particular step of an MPS algorithm. Note that the system is assumed to be cylindrical, with the dangling bonds at the top (black) connecting to sites at the bottom. The black nearest neighbor bonds are transformed into long-range bonds for the 1D path.}
\label{fig:zigzag}
\end{figure}

To perform DMRG on a two-dimensional lattice, the standard procedure is to map the 2D lattice Hamiltonian to a 1D Hamiltonian with long range interactions, as shown schematically in Fig.~\ref{fig:zigzag}.
Usually a 2D lattice with cylindrical geometry is considered, because the open boundary conditions along the cylinder are beneficial for MPS, while periodic boundary conditions lead to reduced finite size effects along the shorter, circumferential, direction.
Some bonds that were nearest neighbor (or short range) in the 2D system will necessarily connect distant sites of the new 1D chain, and this incurs a cost in terms of enhanced entanglement, and therefore the need for larger bond dimension.
Choosing a particular mapping can offset this to some extent (if, for example, the lattice or wave function has certain symmetries) by keeping strongly entangled bonds of the 2D lattice model nearest neighbor in the DMRG path.
The issue becomes more acute as circumference increases, requiring exponentially increasing bond dimension, and can be viewed as a manifestation of the entanglement area law.
For cylinders with small circumference this is still a numerically efficient method compared to PEPS and MERA, with most of the benefits of 1D DMRG. 

One drawback of the mapping to a 1D DMRG path is that it does not take advantage of certain useful quantum numbers, primarily the momentum in the circumferential direction, which one would expect to be particularly useful for compressing the MPS matrices.
Recently a mixed real-momentum space approach has been developed that allows the circumferential momentum to be used as a quantum number~\cite{motruk2016density}.
As expected this leads to a significantly improved algorithm, with the main additional cost being construction of a somewhat complicated matrix product operator for the Hamiltonian.

For time evolution with TEBD the underlying Trotter-Suzuki decomposition of the time evolution operator requires splitting the Hamiltonian into a sum of local (short ranged) terms. This means that the mapping cannot be used at all, with more elaborate treatments having to be developed for long-ranged Hamiltonians~\cite{zaletel2015time,haegeman2016unifying}. We will, however, avoid this issue altogether by using a method that does not require the mapping to an effective 1D chain with long-range interactions. 

\subsection{ChainAMPS}
\label{ss:champs}
\subsubsection{Large physical dimension}
A key ingredient in MPS algorithms is knowledge of a set of local Hilbert spaces of dimension $d_\sigma$ and the action on them of all local operators that appear in the Hamiltonian.
Often we choose a basis for each local Hilbert space that diagonalises the corresponding local part, $h^{\text{local}}_i$, of the full Hamiltonian, $H$, although this is not required.
The full system consists of many of these subunits coupled together,
\begin{align}
H=\sum_i \big( h^{\text{local}}_i + h^{\text{coupling}}_{i,i+1}\big).
\end{align}
In the 1D and 2D approaches described in Sec. \ref{sss:mpsalgorithms}, the local Hilbert spaces represent objects (lattice sites) with spatial dimension zero, but there is no \emph{a priori} reason why this must be so.
The ability to use larger subunits is attractive because they generally allow the use of extra conservation laws; in particular if the subunits are 1D chains with periodic boundary conditions, then chain momentum is a good quantum number.
Additionally, if the system is anisotropic in the sense that the inter-subunit entanglement is weaker than the intra-subunit entanglement (if a subunit has spatial extent we can think about the entanglement between different regions of it), then we avoid the need to perform singular value decompositions on the most strongly entangled parts of the system.
A related, heuristic, benefit is that the basis of a subunit may already reflect genuine strongly correlated physics, so that our initial variational ansatz is already closer to the true state of the total system.
Finally, by suitably coupling subunits of spatial dimension $D_s$ we can build a model in $D_s +1 $ spatial dimensions.
For example, setting $D_s=1$ we can couple multiple chains together to construct a system that is spatially 2D, \emph{but which can still be represented as a 1D MPS with short ranged interactions}, unlike the `snaking path' method described in Fig.~\ref{fig:zigzag}. 

There are two immediate concerns when using `large' objects as the subunits: (i) we must have accurate knowledge of the basis states and matrix elements of the subunits; (ii) MPS algorithms generally scale with some power $p \ge 1$ of the subunit basis size $d_\sigma$, which itself will increase exponentially with the spatial extent of a subunit. 
If the subunits are not too large, the first issue might be dealt with by numerical means, such as exact diagonalization, or even DMRG, performed on the Hamiltonian for a single subunit.
However, if $d_\sigma$ is large ED cannot be applied, and DMRG can only accurately probe the low energy sector of the subunit spectrum.
To avoid these restrictions, we will only consider subunits with local Hamiltonians that we can solve analytically, which for $D_s=1$ means integrable 1D quantum models.
As discussed in Sec. \ref{Sec:TSA} this is not too severe a restriction; although most 1D quantum models are not integrable, there is a large catalogue of models that are, and these represent a multitude of physical systems, including interacting models of spins, bosons and fermions.
Hence, in principle, many different regimes and universality classes can be explored.

The second issue with the use of large subunits is very serious, especially when we consider approaching the thermodynamic limit, where $d_\sigma \to \infty$.
Clearly we must truncate the physical basis of each subunit for practical purposes, and a simple way to implement this is via an energy cutoff $E_c$, applied to the spectra of the local Hamiltonians, $h^{\text{local}}_i$.
Under what conditions can we hope to justify this course of action, without dramatically compromising the accuracy of the method?
In Sec. \ref{Sec:TSA} on the TSA, we saw that it is possible to study integrable theories with a perturbing term using a truncated spectrum, and still compute low energy features with excellent accuracy, as long as the perturbation is relevant (in the RG sense).
Remember that the existence of the `perturbing term' does not imply that the TSA is a perturbative method.
Guided by this insight, we take the \emph{uncoupled} system of exactly solvable (by analytic or numerical means) subunits as our `integrable' system and introduce couplings, $h^{\text{coupling}}_{i,i+1}$, (perturbing terms) that are relevant.
Perhaps counterintuitively, the more relevant the perturbation the better, because this implies even less mixing between different energy sectors of the local spectra.

Alternatively, and in the spirit of most theories in many-body physics, we can view the truncated model as an effective model that captures the low energy sector of the true system of interest.
The caveat is that we must not set the cutoff so low that we throw out states with a non-negligible contribution to the low energy physics.
Any numerical study will therefore have to include a careful analysis of results at different values of $E_c$ to ensure proper convergence.
Useful quantities in this regard are the occupations of the states in the local basis and the reduced density matrix of a subunit (the latter can be found by tracing over all other subunits in the system).
If the occupations or the subunit's reduced density matrix indicate that local states near the cutoff play a significant role, then $E_c$ must be increased.

One may worry that the value of $d_\sigma$ required for accurate results will be too high for a useful algorithm.
For example a `one site' DMRG algorithm scales as $\mathcal{O}(d_\sigma \chi^3 \chi_W)$ (where $\chi_W$ is the bond dimension of the Hamiltonian MPO) with subleading terms (assuming $\chi \gg \chi_W \gtrsim d_\sigma$) that scale as $\mathcal{O}(d_\sigma^2 \chi^2 \chi_W^2)$.
What is not explicit in this analysis is that for most implementations many entries in the MPS and MPO tensors are constrained to be zero by conservation laws (i.e., the existence of `good' quantum numbers).
Knowledge of these quantum numbers allows the large tensors $M^{\sigma_i}_{a_{i-1},a_i}$ and $W^{\sigma'_i,\sigma_i}_{b_{i-1},b_i}$ (for the MPS and MPO respectively) to be stored as smaller blocks of (possibly) non zero elements, and it is the number and size of these blocks that dictate the efficiency of an algorithm.
More conserved quantities will generically produce more blocks, and for given $d_\sigma$ and $\chi$ this means that they must be reduced in size, leading to increasingly sparse tensors.
For momentum-like $\mathbb{Z}$ symmetries (or $\mathbb{Z}_{N_x}$ for a lattice of $N_x$ sites) the multitude of different values the quantum numbers can take has a similar effect.
Therefore the deciding factor in whether this approach is numerically feasible is not the value of $d_\sigma$ \emph{per se}, but the symmetries of the model of interest.
In Sec.~\ref{ss:2Dqim} we examine a model that has both $\mathbb{Z}$ and $\mathbb{Z}_2$ symmetries, and see that we can obtain accurate results near criticality, even though $\chi \lesssim d_\sigma$.
\subsubsection{Arrays of chains}
\begin{figure}
\includegraphics[width=0.45\textwidth]{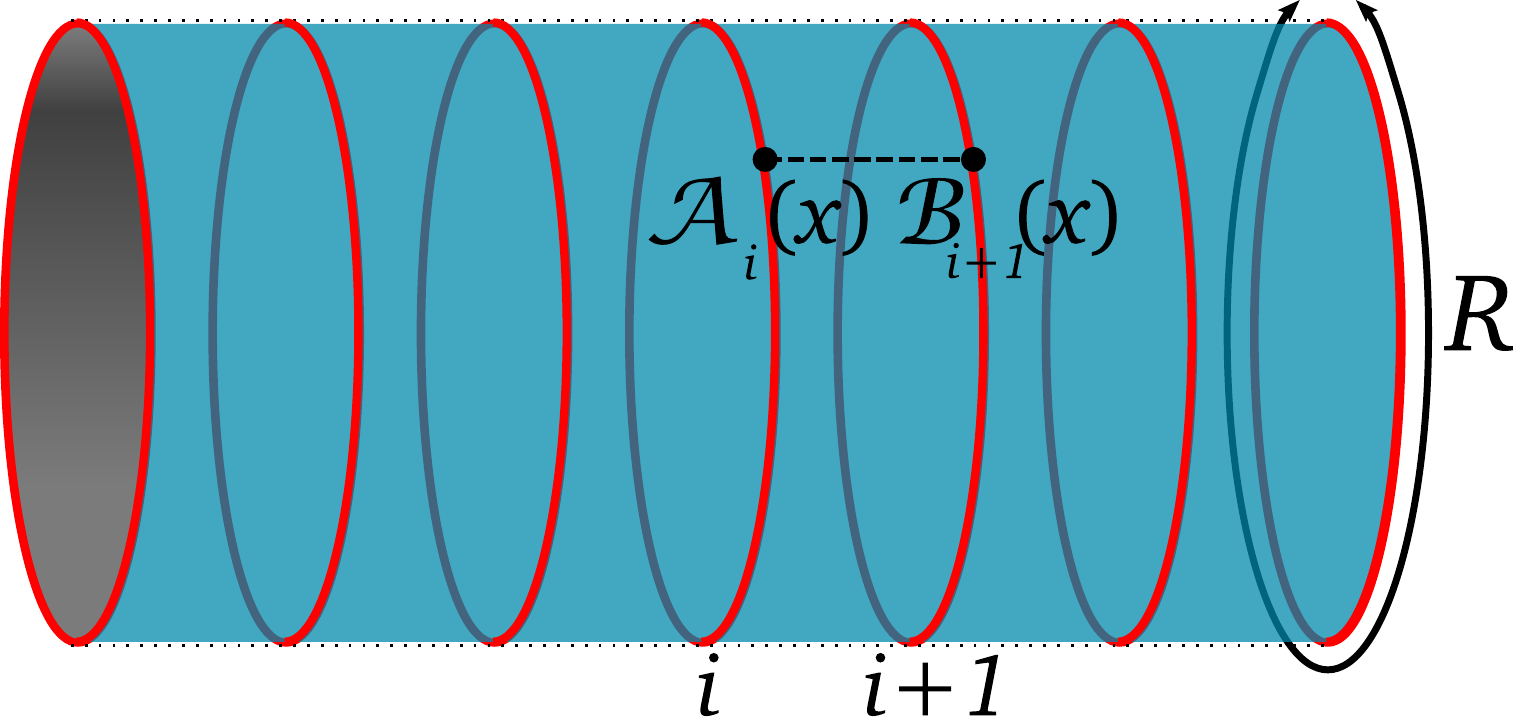}
\caption{Construction of the chain array matrix product state system. Here we use chains (red) with periodic boundary conditions, yielding a cylindrical geometry.}
\label{fig:CHAMPS}
\end{figure}
We now concentrate on the $D_s=1$ implementation of these ideas: taking an array of integrable quantum chains, we couple them together through nearest neighbor interchain interactions (for DMRG somewhat longer ranged interactions are also possible, but they are not compatible with TEBD) as depicted in Fig. \ref{fig:CHAMPS}.
Suitable integrable theories include certain spin chains, free field theories, Luttinger liquids, Lieb-Liniger and Sine-Gordon models.
By using these exactly solvable theories we avoid any issues with ED or inaccuracies that might arise from an initial DMRG calculation of the chain spectrum (especially for the excited states of the chain).

An important advantage occurs if the 1D chains we use are defined by massive continuum field theories with periodic boundary conditions: in such cases finite size effects associated with the chain length $R$ (using the same notation as Sec. \ref{Sec:TSA}) are exponentially suppressed $\sim \exp (- \Delta R)$.
Conversely, for a discrete lattice chain of $N_x$ sites we expect a slower, power law, decay of finite size terms $\sim N_x^{-\alpha}$.
It is still possible to truncate the spectrum of the continuum theory, because for finite $R$ the chain spectrum will be discrete.
Consequently we can approach the thermodynamic limit (in terms of the chain physics) using a relatively small $R$.
And because $R$ in the ChainAMPS geometry plays the role of the area in the eponymous law (see Fig. \ref{fig:CHAMPS}), its smallness \emph{will also restrict the growth of entanglement and bond dimension}.

The standard ChainAMPS system is then highly anisotropic: continuous, finite and periodic in the circumferential direction, while discrete and with open---or possibly infinite---boundary conditions along the cylinder.
If we are interested in universal physical properties this is usually not an issue, beyond the finite size effects implied by $R$ or a finite number of chains $N_y$.
Equally, we can envision studying direct manifestations of this anisotropy, such as coupled quantum wires or tubes of ultra cold atomic gas.

To be explicit, consider a chain of length $R=N_x a_x$, described by a 1D quantum Hamiltonian, $h_{\text{chain}}$, with a spectrum consisting of $d_\sigma$ states (either exactly or after truncation).
We take $N_y$ such chains, label them $i=1,\cdots,N_y$, and couple them together with nearest neighbor interactions of the form,
\begin{align}
h^{\text{coupling}}_{i,i+1}&= \int_0^R \!\! \text{d} x \: \mathcal{A}_i(x) \mathcal{B}_{i+1}(x) + \text{h.c.},
\end{align}
for continuum limit chains or
\begin{align}
h^{\text{coupling}}_{i,i+1}&= \sum_{j=0}^{N_x-1} \mathcal{A}_{i,j} \mathcal{B}_{i+1,j} +\text{h.c.},
\end{align}
for lattice chains.
Here $\mathcal{A}_i(x)$ and $\mathcal{B}_i(x)$ ($\mathcal{A}_{i,j}$ and $\mathcal{B}_{i,j}$ for the lattice case) are operators on the local Hilbert space of chain $i$.
For chain theories with periodic boundary conditions we may invoke translational invariance to find the action of the operator at position $x$ along a chain:
\begin{align}
\mathcal{A}_i(x) & = e^{-i \hat{k}_{x,i} x} \mathcal{A}_i e^{i \hat{k}_{x,i} x},
\end{align}
where $\hat{k}_{x,i}$ is the momentum operator on chain $i$.
Integrating the coupling term matrix elements over $x$ (or summing in the lattice case) then yields
\begin{align}
\bra{\sigma'_i \sigma'_{i+1}} &h^{\text{coupling}}_{i,i+1}  \ket{\sigma_i \sigma_{i+1}}\nn
=&  R\,\delta(k_{\sigma'_i}+k_{\sigma'_{i+1}}-k_{\sigma_i}-k_{\sigma_{i+1}})\nn
&\times\bra{\sigma'_i}\mathcal{A}_i\ket{\sigma_i} \bra{\sigma'_{i+1}}\mathcal{B}_{i+1}\ket{\sigma_{i+1}} +\text{H.c.},
\end{align}
and a similar result for lattice chains.
As expected, the $\delta$-function indicates that total chain momentum is conserved by the coupling, and we build this information into the Hamiltonian MPO.

With the definitions above the MPS wave function reads
\begin{align}
\ket{\Psi}&= \sum_{\boldsymbol{\sigma}} \mathbf{M}^{\sigma_1} \mathbf{M}^{\sigma_2} \cdots \mathbf{M}^{\sigma_{N_y}} \ket{\sigma_1} \otimes \ket{\sigma_2} \otimes \cdots \otimes \ket{\sigma_{N_y}},
\end{align}
where the set of $d_\sigma$ matrices $\mathbf{M}^{\sigma_i}$ parametrizes chain $i$.

A trivial example is the ground state of a system of uncoupled chains, which is just a tensor product of the individual chain ground states, $\ket{\sigma_i}=\ket{0_i}$.
In this case the bond dimension $\chi=1$ and the matrices, $\mathbf{M}^{\sigma_i}$, reduce to scalars equal to $1$ if $\sigma_i=0_i$ or $0$ otherwise,
\begin{align}
\mathbf{M}^{\sigma_i}&= \delta_{\sigma_i,0_i} \times 1,\nn
\ket{\Psi}_\text{uncoupled}&= \ket{0_1} \otimes \ket{0_2} \otimes \cdots \otimes \ket{0_{N_y}}.
\end{align}
For coupled chains the necessary bond dimension, $\chi$, will generically be $>1$ and an increasing function of the chain length, $R$, as discussed above.

\subsubsection{Infinite cylinders}
\label{sss:infinitealgorithms}
Both DMRG and TEBD can be applied to infinitely long systems (termed iDMRG and iTEBD respectively), by working with a translationally invariant MPS or uMPS, represented by a unit cell consisting of a few sites (typically one or two).
Working directly in the infinite volume limit removes the (probably unwanted) effects of the open boundary conditions used in finite MPS algorithms, and avoids the problems associated with using MPS with periodic boundary conditions.

These algorithms are very convenient when applied to ChainAMPS, because their relative efficiency---being approximately a factor of $N_y$ faster than studying a finite length system---helps to offset the effect of using large $d_\sigma$.
For uniform ChainAMPS, the unit cell consists of one or two \emph{chains} and the total system has the geometry of an infinitely long cylinder (if the individual chains have periodic boundary conditions) or strip (if the chains are open).

One issue with iDMRG is that the usual projector method for calculating excited states (see e.g.~\cite{stoudenmire2012studying}) cannot be applied in the infinite limit.
Some excited states can still be found if the values of some of their quantum numbers differ from those of the ground state, and if those quantum numbers can be written as a rational number $p/q$ where $q$ is the number of chains in the unit cell.
Otherwise we must turn to a `post matrix product state' method along the lines of a single mode approximation~\cite{haegeman2013post}.

\subsubsection{ChainAMPS summary}
At this point we have still not provided any evidence that the ChainAMPS approach works in practice.
To remedy this, in Sections \ref{ss:2Dff} and \ref{ss:2Dqim} we will cover two concrete examples.
Before doing so let us summarize the discussion thus far.
A 2D quantum system can be realized as an array of coupled quantum chains, and such a system can be written as an MPS with large physical dimension $d_\sigma$.
In order to make further progress we need a highly accurate (or exact) basis for each chain, and any matrix elements that appear in the Hamiltonian of the complete system.
This is best achieved by studying integrable chain theories.
Furthermore, it is usually necessary to truncate the spectrum of the chain so that $d_\sigma$ is not too large for numerical implementations.
This is possible if the chain spectrum is discrete (which it will be for finite length chains), in which case we can truncate by applying an energy cutoff to the spectrum of each chain's Hamiltonian.
We expect that such a truncation will not affect the physics too strongly if we are working with interchain couplings that are relevant under renormalization group transformations.
There are advantages to working with chains represented by massive, continuum limit, field theories: we can throttle the growth of entanglement by working with short chains, and still obtain results that reflect the thermodynamic limit.

Software implementing the ideas discussed in this section is available at \url{https://bitbucket.org/chainamps}.
The software includes drivers for performing DMRG on infinite and finite systems (the latter can also find excited states); for real time evolution using TEBD or iTEBD; and for performing measurements by post-processing output.
Several example continuum limit chain models, suitable for coupling together, are provided (Ising chains, free fermions and Luttinger liquids), although user defined models can also be studied.

\subsection{Free fermions in 2D}
\label{ss:2Dff}
To demonstrate some of the concepts above, and to introduce time evolution of an MPS, we first consider a trivial theory that is exactly solvable, even in 2D.
In principle we could use lattice chains for this purpose, but in order to make contact with the rest of this section, and particularly the 2D quantum Ising model we discuss next, we will take the continuum limit of the chains.

Our starting point is the theory of a free Majorana field on a ring (i.e., periodic boundary conditions) of length $R$.
This field is represented by two components $\psi$ and $\bar{\psi}$ with fermionic anticommutation relations
\be
\begin{aligned}
\{\psi(x,t),\psi(x',t) \} &=\delta(x-x'), \\
\{\psi(x,t),\bar\psi(x,t)\} &= 0.
\end{aligned}
\ee
The Lorentz invariant action is most conveniently expressed in complex coordinates, $z=t-ix$, $\bar{z}=t+ix$, and derivatives
\begin{align}
\partial_z \equiv \partial = \frac{1}{2} \big( \partial_t + i \partial_x \big) \quad \text{and} \quad \partial_{\bar z} \equiv \bar{\partial} = \frac{1}{2} \big( \partial_t - i \partial_x \big).
\end{align}
With these definitions the action of the Majorana field is
\begin{align}
S=\int \frac{\text{d}^2 z}{2\pi} \big( \psi \bar{\partial} \psi + \bar{\psi} \partial \bar{\psi}  +i \Delta \bar{\psi} \psi \big)
\label{eq:majorana_action}
\end{align}
where $\Delta > 0$ is the fermion mass (we change notations here slightly from Sec.~\ref{Sec:QuantumIsing}, setting the mass as $\Delta = m$).

The two field components have mode expansions in terms of fermion creation and annihilation operators $\{a_n,a_{n'}^\dagger\} = \delta_{n,n'}$:
\begin{align}
\psi(x,t)=&\sum_n \sqrt{\frac{\Delta}{2 \epsilon_n R}} e^{\theta_n/2}  \nonumber \\
&\qquad  \times \Big( \omega a_n e^{-i(t \epsilon_n -x p_n)}  +\omega^\ast a_n^\dagger e^{i(t \epsilon_n -x p_n)} \Big), \nonumber \\
\bar{\psi}(x,t)=&-\sum_n \sqrt{\frac{\Delta}{2 \epsilon_n R}} e^{-\theta_n/2} \nonumber\\
& \qquad \times  \Big( \omega^\ast a_n e^{-i(t \epsilon_n -x p_n)} +\omega a_n^\dagger e^{i(t \epsilon_n -x p_n)}\Big), \nonumber
\end{align}
with the parameterizations $\omega=e^{i \pi/4}$, and $\epsilon_n=\Delta \cosh \theta_n$, $p_n=\Delta \sinh \theta_n$.
The momentum and energy associated with mode $n$ are $p_n=2\pi n/R$ (for integer $n$), and $\epsilon_n =\sqrt{\Delta^2 + p_n^2}$ respectively.
Employing these expansions we obtain the Hamiltonian for chain $\ell$:
\begin{align}
H^\text{1D}_{\ell}=\sum_n \epsilon_n a^\dagger_{n,\ell} a_{n,\ell}.
\end{align}

We combine $N_y$ of these chains into an array and allow for fermions to hop between nearest neighbor chains, resulting in the Hamiltonian
\begin{align}
H_\text{free}&=\sum_{\ell} H^\text{1D}_{\ell} + H^\text{hop}_{\ell,\ell+1},\label{eq:2Dff} \\
H^\text{hop}_{\ell,\ell+1}&= -t_{\perp}\sum_{n} \frac{\Delta}{\epsilon_n} \big(a_{n,\ell}^\dagger a_{n,\ell+1} +\text{h.c.}\big),
\end{align}
with hopping parameter $t_\perp$.
If we assume our 2D system is a torus, then the Hamiltonian, $H_\text{free}$, is easily solved by Fourier transforming from chain index $\ell$ to momentum $k_m=2\pi m/N_y$ (note this momentum is transverse to that indexed by $n$ \emph{along} the chains):
\begin{gather}
a^\dagger_{n,\ell} = \frac{1}{\sqrt{N_y}}\sum_{m=1}^{N_y} e^{ik_m \ell} a^\dagger_{n,m}, \\
\{a_{k_m}, a^\dagger_{k_{m'}}\}= \delta_{k_m,k_{m'}},\\
H_\text{free}=\sum_{n,m} \Big(\epsilon_{n} -\frac{2\Delta t_\perp}{\epsilon_n} \cos k_m \Big) a^\dagger_{n,m} a_{n,m}.
\end{gather}
This diagonal Hamiltonian can be viewed as $N_y$ uncoupled 1D bands indexed by $m$, or infinitely many 1D bands, indexed by $n$.
For $\Delta > 2 t_\perp$ the ground state is the vacuum (no occupied modes).

A similar result holds for an open cylinder, with $\kappa_{m} =\pi m/(N_y+1)$, $m=1,\cdots,N_y$ and
\begin{gather}
a^\dagger_{n,\ell} = \sqrt{\frac{2}{N_y+1}}\sum_{m=1}^{N_y} \sin \big(\kappa_m \ell \big) \: \tilde{a}^\dagger_{n,m}, \\
H_\text{free}=\sum_{n,m} \Big(\epsilon_{n} -\frac{2\Delta t_\perp}{\epsilon_n} \cos \kappa_m \Big) \tilde{a}^\dagger_{n,m} \tilde{a}_{n,m}.
\label{eq:freeOBC}
\end{gather}

\begin{figure}
\includegraphics[width=0.45\textwidth]{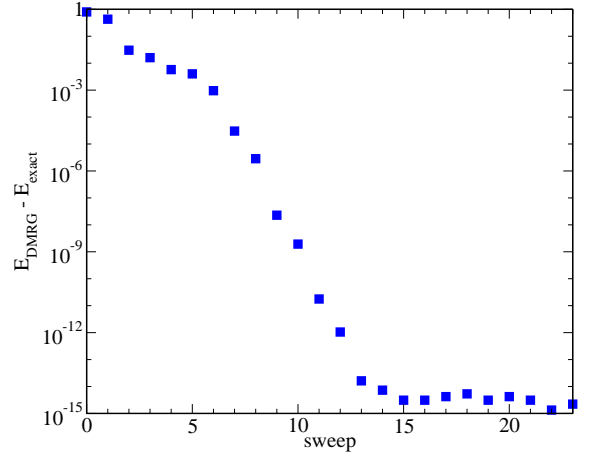}
\caption{The difference (note the logarithmic scale) between a DMRG calculation of the lowest energy two particle state of Eq.~\fr{eq:freeOBC} and the exact solution, with $t_\perp=0.2$, $R=10$, $N_y=100$, plotted against the number of finite size sweeps.
The sweep 0 value is the result of an `infinite system' DMRG growth process. Convergence tails off near the precision limit of the numerical implementation.}
\label{fig:DMRGOBC}
\end{figure}

As the energy levels of this 2D system can be calculated trivially, we can use it as a test case: apply the DMRG procedure to it, and study the implementation's convergence properties.
An example is shown in Fig. \ref{fig:DMRGOBC}.
\subsubsection{Time evolution}
\label{sss:time_evolution}
The free fermion Hamiltonian Eq.~\fr{eq:2Dff} is more interesting when we consider using it to time evolve a state that is not an eigenstate.
Such a situation occurs in a \textit{quantum quench}: a system is prepared in an eigenstate of a Hamiltonian and subsequently the Hamiltonian parameters are suddenly changed, with the state being left to evolve under the new Hamiltonian.
If the quench is instantaneous and the post quench Hamiltonian $H$ is constant in time, then the state at time $t$ is given by applying the time evolution operator $\exp(-iHt)$ to the initial (pre quench) state $\ket{\phi(0)}$:
\begin{align}
\ket{\phi(t)}=e^{-iHt} \ket{\phi(0)}.
\end{align}
For a many-body system this evolution is generally very difficult to calculate analytically, including for a free Hamiltonian unless the initial state can be written as a simple superposition of eigenstates of $H$.

TEBD and iTEBD have been extensively used in the study of quantum quenches in 1D because they are remarkably accurate and allow direct access to the time evolved wave function.
These algorithms introduce two types of error: one originates in breaking up the time evolution operator into manageable pieces by discretizing time and using Trotter-Suzuki decompositions, while the second occurs because the MPS must be compressed back to dimension $\chi$ after each time step.
Errors of the first kind can be managed by using smaller time steps, $\delta t$, and higher order, $r$, Trotter decompositions, because they scale as $\mathcal{O}\big((\delta t)^r\big)$.
The second type of error is ultimately fatal, as the entanglement can increase linearly in time after a general quench~\cite{CalabreseJStatMech05,DeChiaraJStatMech06,BarmettlerPRA08}, requiring an exponentially increasing bond dimension to preserve accuracy.
Consequently, for a fixed $\chi$, there is a maximum time up to which reliable results can be computed.

With the ChainAMPS anisotropy it is most convenient to consider quenching the parameters governing the interchain couplings.
A quench of the parameters in the 1D chain Hamiltonians is also possible, but requires that the overlaps between the initial and final chain bases are known to high accuracy (preferably analytically).

For quenches of $H_\text{free}$ in which only $t_\perp$ is changed, there is a dramatic simplification: the absence of any interactions (either on or between the chains) means that only the fermionic chain modes that are occupied in the initial, $t=0$, state participate in the evolution.
In this special case we can restrict our local (physical) bases to include only chain states that do not feature initially unoccupied modes.
Beyond this, the chain cutoff $E_c$ does not play any role.

As an example, the initial state
\begin{align}
\ket{\Phi_0}=\prod_{i=0}^{\frac{N_y}{2}-1}  \frac{1}{\sqrt{2}} \big(&\ket{\text{vac}}_{2i} \ket{n=0}_{2i+1} \nonumber \\
&+\ket{n=0}_{2i} \ket{\text{vac}}_{2i+1}\big),\label{eq:initial1D}
\end{align}
is an eigenstate when $t_\perp=0$, in which alternating pairs of chains are entangled, with a superposition of \emph{chain} vacua ($\ket{\text{vac}}_i$) and lowest excited states ($\ket{n=0}_i$).
It is not an eigenstate when $t_\perp\ne0$, but its evolution under $H_\text{free}$ does not involve any other states from the chain spectrum, and the problem reduces to a 1D model of hopping fermions.

\begin{figure}
\includegraphics[width=0.45\textwidth]{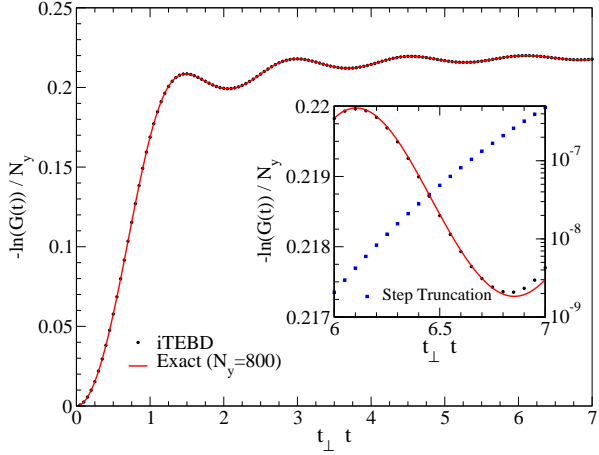}
\caption{Comparison of the logarithm of the return probability per chain for the quench starting from the one-dimensional initial state~\fr{eq:initial1D} and time-evolved with $t_\perp = 0.5$, calculated using iTEBD ($\chi$=1000) and the exact method. Inset: enlarged region showing the difference as the truncation error (right axis) increases.}
\label{fig:exact1D}
\end{figure}

A useful global measure of the quench dynamics is the Loschmidt echo (or return probability), $G(t)$, namely the absolute value squared of the overlap of the state at time $t$ with the initial state.
For a quench to a finite value of $t_\perp$ with the initial state~\fr{eq:initial1D}, it is possible to calculate $G(t)$ on a torus formed from $N_y$ chains, using a determinant method~\cite{james2015quantum}:
\begin{align}
G(t)&=\absval{\bra{\Phi_0} \exp\big[ -i H_\text{free} t\big] \ket{\Phi_0}}^2, \nonumber\\
&= \Big \vert\text{det} \big( M + (1-M) Q \exp \{-i h t\} Q \big)\Big \vert^2,\label{eq:exact_return}
\end{align}
with $N_y \times N_y$ matrices,
\begin{gather}
M=\text{diag}(0,1,0,1,0,\cdots), \nonumber \\
Q =
\sqrt{\frac{1}{2}}\left(
\begin{array}{rrrrc}
  1& 1  &   & &\\
  1 & -1  & & &\\
  &   & 1  & 1& \\
  & & 1 & -1 & \\
  & & & & \ddots 
\end{array}
\right), \quad
h =
\left(
\begin{array}{ccc}
  \Delta & -t_\perp  &   \\
 -t_\perp & \Delta  & \ddots  \\
  &  \ddots & \ddots   
\end{array}
\right).
\end{gather}
Figure~\ref{fig:exact1D} displays a comparison between iTEBD and the result of evaluating Eq.~\fr{eq:exact_return}.
While the determinant method is exact for any time $t$, it is somewhat limited in terms of which quantities can be computed, which initial states can be used, and to finite numbers of chains, $N_y$.
Conversely, with iTEBD the time evolution of the wave function (and therefore all the interesting physical information) can be calculated directly in the thermodynamic limit, $N_y \to \infty$, but only up to a time $t_\text{max}$ (dependent on the bond dimension, $\chi$) before errors become significant.

As already stated, the dynamics of this quench are effectively one-dimensional.
In fact, precisely because the model is non interacting and the individual mode occupations are all conserved, it is difficult to engineer a quench of $H_\text{free}$ in which the chain length, $R$, has a non-trivial effect.
In the next section we look to an interacting model to see truly 2D many body quantum phenomena.

\subsection{Application to the 2D quantum Ising model}
\label{ss:2Dqim}
The quantum Ising chain (or transverse field Ising model in 1D) is a paradigmatic strongly correlated system, with two massive phases separated by an order-disorder transition.
As described in detail in Sec.~\ref{Sec:QuantumIsing}, in the continuum limit the Hamiltonian of the quantum Ising chain reduces to the field theory of a free Majorana fermion with a mass $\Delta$ (\emph{c.f.} Eq.~\ref{eq:majorana_longitudinal}: we set $h=0$ as we do not consider a perturbing longitudinal field, and define $\Delta=m$ to connect with the notation in Sec.~\ref{ss:2Dff} above).
We remind the reader that the mass can be positive ($\Delta>0$), negative ($\Delta<0$) or zero, corresponding to an ordered, disordered or critical chain, respectively.  
Henceforth we shall refer to this Hamiltonian as $H_{\text{1D}}(\Delta)$.

One complication, relative to the Majorana chains considered in Sec.~\ref{ss:2Dff}, is that the Jordan-Wigner mapping from Ising spins on a ring to fermions introduces both periodic and anti-periodic boundary conditions.
While this makes the chain spectrum more intricate, as it separates into Neveu-Schwarz and Ramond sectors (see Sec.~\ref{Sec:QuantumIsing}), the chain Hamiltonian nevertheless remains non-interacting.

We form a two-dimensional quantum Ising model by coupling an array of these chains together with Ising spin-spin interactions:
\be
H_{\text{2D}}=\sum_{\ell=1}^{N_y} H_{\text{1D}}(\Delta)+J_\perp\sum_{\ell=1}^{N_y-1}\int_0^R \!\text{d}x\: \sigma^z_\ell(x) \sigma^z_{\ell+1}(x). 
\label{eq:2DQIM}
\ee
In contrast to the quantum Ising chain, this is an interacting model of fermions as the interchain coupling constitutes a genuine fermionic scattering vertex, see Eq.~\fr{eq:TFIM_ME}.
This model has two symmetries that can be easily incorporated into the MPS to produce an efficient representation.
The first is translational symmetry along the circumferential direction, leading to conservation of the total chain momentum, just as for the free model we considered previously.
The second is a $\mathbb{Z}_2$ symmetry that has its origin in the $\pi$ rotation symmetry of the spins in the 1D lattice model (i.e. spin inversion), and which leads to the overall sector (the number of chains in a Neveu-Schwarz state, modulo 2) being conserved.

\subsubsection{Static Properties}
\begin{figure}
\includegraphics[width=3.2in]{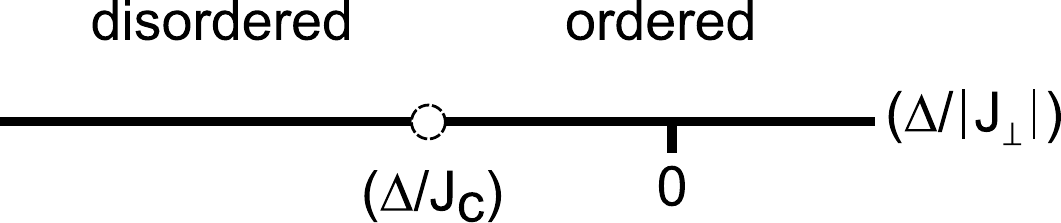}
\caption{The zero temperature phase diagram for the 2D quantum Ising model~\fr{eq:2DQIM}, in terms of the ratio $\Delta/\absval{J_\perp}$. The critical point is marked by an open circle. Note that the transition occurs for $\Delta < 0$.} 
\label{fig:2Dphasediagram}
\end{figure}
\begin{figure}
\includegraphics[width=0.45\textwidth]{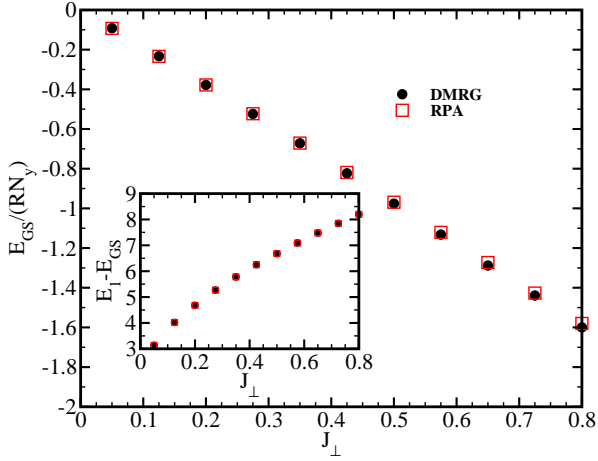}
\caption{Ground state energy of 2D quantum Ising model~\fr{eq:2DQIM} formed from an array of continuum chains, calculated using DMRG and an RPA method. Inset: Gap to the first excited state. Data from Ref.~\cite{konik2009renormalization}.} \label{fig:2DQIM_gs}
\end{figure}

We now examine the ability of the ChainAMPS construction to accurately capture the behaviour of the 2D quantum Ising model.
For chains with $\Delta>0$, the 2D system is ordered and it is possible to calculate the lowest energies of the states of the coupled chain system using a `random phase approximation' (RPA) approach, in which the ordered moment is treated self-consistently.
In Ref.~\cite{konik2009renormalization} such a calculation was compared to DMRG for the lowest lying states of the array of coupled quantum Ising chains and excellent agreement was found (see Fig. \ref{fig:2DQIM_gs}).
Calculating the energy of the ground and first excited states of the coupled chains is quite easy with DMRG, because they both have different values of the sector quantum number and therefore can be targeted by separate energy minimization runs.

\begin{figure}
\includegraphics[width=0.45\textwidth]{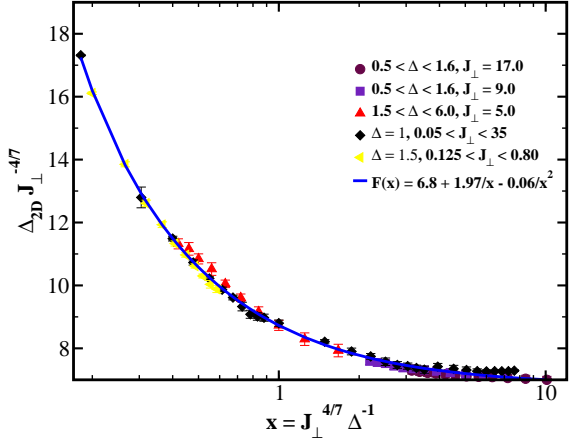}
\caption{Scaling collapse of the gap in the 2D quantum Ising model~\fr{eq:2DQIM} formed from an array of continuum chains. Also shown is a fit to the data, $F(x)$. Adapted from Ref.~\cite{konik2009renormalization}.} \label{fig:2DQIM_scaling}
\end{figure}

The DMRG results can also be used to check that the gap to the first excited state has the correct 2D quantum Ising scaling form.
Armed with the knowledge that the scaling dimension of the spin operator is $1/8$ (see, e.g., Ref.~\cite{CFTBook}), this scaling form can be discerned to be
\begin{align}
\Delta_{\text{2D}}=J_\perp^{4/7}\Phi\Big( \Delta J_\perp^{-4/7} \Big),
\end{align}
where $\Phi$ is a dimensionless scaling function.
Figure \ref{fig:2DQIM_scaling} shows the scaling collapse of the DMRG data for a wide range of $J_\perp$ and $\Delta>0$.

\begin{figure}
\includegraphics[width=0.45\textwidth]{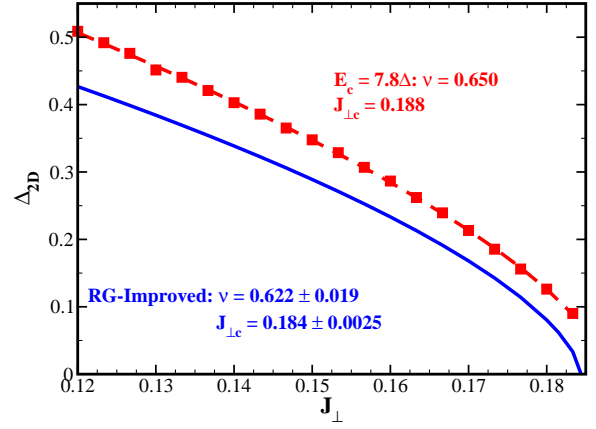}
\caption{Gap in the disordered phase of the 2D quantum Ising model~\fr{eq:2DQIM}, as calculated using DMRG on the ChainAMPS system.
Blue curve shows the results after RG improvements (cf. Sec.~\ref{Sec:TSA_Cutoff}). Adapted from Ref.~\cite{konik2009renormalization}} \label{fig:2DQIM_gap}
\end{figure}

Perhaps the most compelling argument is that the ChainAMPS system displays the correct 2D quantum critical behaviour.
As in 1D, the quantum Ising model also displays an order-disorder transition separating two gapped phases, but with properties in the universality class of the 3D \emph{classical} Ising model.
For the array of quantum Ising chains we can approach the transition to the ordered phase by starting with weakly coupled disordered ($\Delta<0$) chains and increasing $J_\perp$ (see Fig. \ref{fig:2Dphasediagram}).
In Fig. \ref{fig:2DQIM_gap} we show the gap, $\Delta_\text{2D}$, in the disordered phase as a function of $J_\perp$, calculated using DMRG for an array of 100 continuum quantum Ising chains with an energy cutoff of $E_c=7.8$.
The gap is proportional to the inverse correlation length in the system, and therefore is expected to close as,
\begin{align}
\Delta_{\text{2D}}\sim \absval{J_\perp-J_c}^\nu,\label{eq:2Dgap}
\end{align}
where $J_c$ is the critical value of the interchain Ising coupling.
Fitting the form Eq.~\fr{eq:2Dgap} to the DMRG data, one finds $\nu=0.650$ (or $0.622$ after RG improvement~\cite{konik2009renormalization}) which compares well with the value $0.630$ obtained using series expansion~\cite{hasenbusch2010finite}.\footnote{This is especially true considering extrapolations in the number of chains and the chain size, $R$, have not been performed, and that the truncation error is $\sim 10^{-6}$.}
This value should be contrasted with $\nu=1$ for the 1D quantum Ising chain (2D classical Ising universality class) \cite{IsingBook}.

For very small $R$, the model crosses over to a 1D \emph{lattice} Ising model: as the level spacing of the chain spectrum increases ($\sim R^{-1}$) only the two lowest chain eigenstates become important, leaving an effective Ising degree of freedom for each chain.
In this limit a finite size scaling analysis reveals a different critical coupling and the critical exponent $\nu=1$ as expected~\cite{james2013understanding}.

\subsubsection{Entanglement}
It is also useful to analyze the entanglement content of the model.
Far from criticality, the leading order contribution to the entanglement of a ground or low-lying state is expected to have an area law form.
By virtue of the cylindrical ChainAMPS geometry shown in Fig. \ref{fig:CHAMPS}, we can easily extract the entanglement of a bipartition formed by cutting through the cylinder between two chains.
For such a partitioning the `area' is proportional to the chain length, and therefore the entanglement entropy, $S_E$, of our 2D quantum Ising model in the disordered phase should scale linearly with $R$.

The coefficient of the area law term is non-universal, because it requires a microscopic length scale to make the contribution dimensionless.
With lattice chains, the lattice constant would provide the length scale, but for our continuum chains this is replaced by the bare correlation length on the chains, $\absval{\Delta}^{-1}$.
Taking the large $R$ limit of the various expressions defining the spectrum and matrix elements of a continuum quantum Ising chain (given in Sec.~\ref{Sec:QuantumIsing}) and performing a perturbative calculation for small $J_\perp$, we find the leading order contribution to the entanglement entropy in the disordered phase~\cite{james2013understanding}:
\begin{align}
S_E \sim -\frac{\Delta R}{8} \left( \frac{J_\perp \bar{\sigma}^2}{\Delta^2}\right)^2 \log \absval{\frac{J_\perp \bar{\sigma}^2}{\Delta^2}},
\label{eq:2DQIM_entanglement}
\end{align}
which matches our expectations for area law scaling in 2D (note that the dimensionless combination $J_\perp \bar{\sigma}^2/\Delta^2$ is the perturbative expansion parameter).

\begin{figure}
\includegraphics[width=0.45\textwidth]{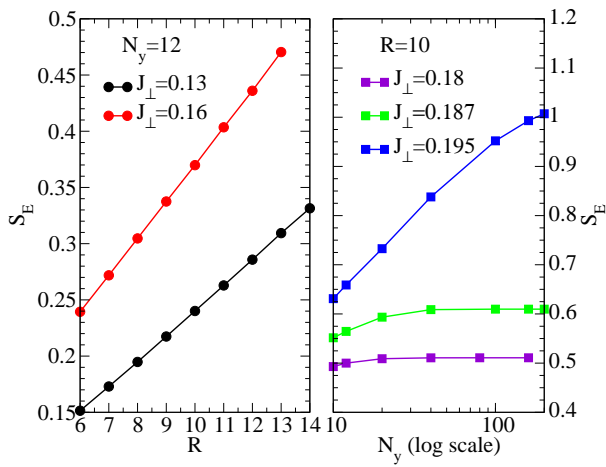}
\caption{Entanglement entropy, $S_E$, as a function of $R$ and $N_y$ for the ChainAMPS 2D quantum Ising model~\fr{eq:2DQIM} with $\Delta=-1$ and $E_c=8.0$. Deep in the disordered phase $S_E$ scales linearly with the chain length $R$ (left panel) and is independent of the number of chains $N_y$. As the critical point is approached, the correlation length increases and a $\log N_y$ dependence emerges (right panel). Data from Ref.~\cite{james2013understanding}.} \label{fig:2DQIM_area_law}
\end{figure}

The existence of area law scaling can be viewed as a statement that the dominant entanglement is short ranged, and so it makes sense that $S_E$ deep in the gapped phase (where the correlation length is very short) is independent of $N_y$, the number of chains.
Figure~\ref{fig:2DQIM_area_law} shows that this behavior persists even when $J_\perp \bar{\sigma}^2/\Delta^2$ is not strictly small, and the perturbative approach breaks down.

As the gap closes and the system approaches criticality, the correlation length increases, the entanglement becomes long-ranged, and the above arguments no longer apply.
In particular, the behavior of $S_E$ is liable to change due to the presence of logarithmic correction terms, including a $\log N_y$ piece (see the right panel of Fig. \ref{fig:2DQIM_area_law}) and a \textit{chord scaling}-like term that depends on the relative sizes of the partitions~\cite{james2013understanding}.

When examining the critical properties it can be helpful to consider not just $S_E$, but the individual eigenvalues of the system's reduced density matrix $\rho_r$ (these are just the squared Schmidt coefficients of our MPS).
Following Li and Haldane~\cite{li2008entanglement} we define a fictitious entanglement Hamiltonian, $H_\text{ES}$, in terms of the diagonalized reduced density matrix $\rho_r = \exp(-H_\text{ES})$.
The levels of the entanglement spectrum are then given by $\omega=-\log \rho_r$.

\begin{figure}
\includegraphics[width=0.45\textwidth]{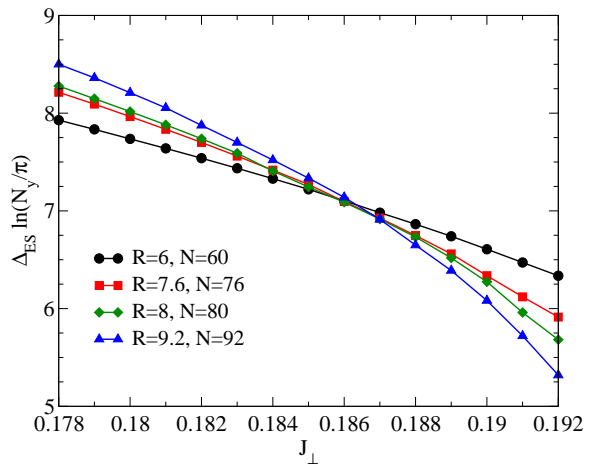}
\caption{Scaling of the entanglement gap $\Delta_\text{ES}$ in the ground state of the 2D quantum Ising model~\fr{eq:2DQIM}. The intersection of the curves provides an estimate of the critical interchain coupling $J_c$ (see text). Data from Ref.~\cite{james2013understanding}.} \label{fig:2DQIM_ES_gap_scaling}
\end{figure}

Taking the difference between the two lowest levels (corresponding to the largest Schmidt coefficients, or singular values) as the `entanglement gap' $\Delta_\text{ES}$, we can consider the scaling of this quantity with finite system size $N_y$, keeping the aspect ratio $N_y/R$ fixed (alternatively, one could use iDMRG to work in the thermodynamic limit, $N_y \to \infty$, and perform scaling with the bond dimension $\chi$).
Applying the scaling relation proposed by Calabrese and Lefevre for conformal models~\cite{calabrese2008entanglement} (and confirmed for a variety of 1D quantum critical models~\cite{pollmann2010entanglement}) to the entanglement gap,
$\Delta_\text{ES}\sim \text{const.}/\log(N_y/\pi)$, we find that curves for different system sizes cross at the same point, as shown in Fig.~\ref{fig:2DQIM_ES_gap_scaling}.
This gives an estimate of the critical coupling $J_c=0.186(2)$ which agrees very well with the value found by conventional finite size scaling of the actual spectral gap $\Delta_\text{2D}$, $J_c=0.185(2)$ and by RG improved DMRG in Ref.~\cite{konik2009renormalization}.
The former is considerably easier to obtain, because it requires a DMRG calculation of the ground state energy alone, whereas the latter also requires the first excited state to be computed.
\subsubsection{Time Evolution}
The 2D quantum Ising model~\fr{eq:2DQIM} is interacting and therefore, unlike $H_\text{free}$ in Sec.~\ref{ss:2Dff}, it is possible to see non-trivial dynamics following a quench
\be
J_\perp = 0 \to J_\perp \neq 0, 
\ee
that starts from the ground state of the uncoupled Hamiltonian.
We can discern three time scales that should feature in the dynamics on general grounds, using the quasiparticle propagation picture of Calabrese and Cardy~\cite{CalabreseJStatMech05,calabrese2006time-dependence}.
In this picture, the energy imparted by the quench\footnote{It should be noted that the initial state has an extensively high energy relative to the ground state of the post-quench Hamiltonian.} acts as a source of quasiparticle excitations.
In the initial state at time $t=0$ with $J_\perp=0$, the quasiparticles are initially entangled if they are within a distance $\sim \absval{\Delta}^{-1}$ of each other on a chain.
Once created they move along the chains with a maximum velocity $v$.
Intrachain scattering therefore begins to have an appreciable effect when quasiparticles from initially unentangled regions start to reach each other, at a time $t_\Delta =(2 v\absval{\Delta})^{-1}$.
This provides our first time scale.
The second time scale is revealed when considering the average time for quasiparticles to hop between chains, and is given by a Fermi's golden rule type argument as $t_{J_\perp}=\absval{\Delta}^{1/2}/(J_\perp R)^2$.
The third time scale occurs because the chains have a finite length $R$; two quasiparticles created at the same point and traveling along a finite chain in different directions will eventually meet again, at a time $t_R=R/(2v)=\absval{\Delta}R t_\Delta$ (for periodic chain boundary conditions).
This scale is different to $t_\text{rec}$, the time for periodic revivals or `quantum recurrences' to occur in systems with a finite number of degrees of freedom.

With ChainAMPS the revival time will depend on the number of chains, $t_\text{rec}\sim t_{J_\perp} N_y$ (roughly speaking, quasiparticles will need to hop through the entire system and back again for a revival, not just round a single chain).
Therefore if we work with infinitely long cylinders, using iTEBD, we should not see any true revivals.
Indeed there is a time window, $t_\Delta, t_{J_\perp} < t < t_R$ in which we expect to see the behavior of the thermodynamic limit of the 2D quantum Ising model.

For very small post-quench $J_\perp$, the maximum quasiparticle velocity $v$, along the chains will be given by the group velocity for excitations of the continuum quantum Ising chain, i.e. $v=1$.
As $J_\perp$ increases the additional scattering between chains will renormalize $v$.
\begin{figure}
\includegraphics[width=3in]{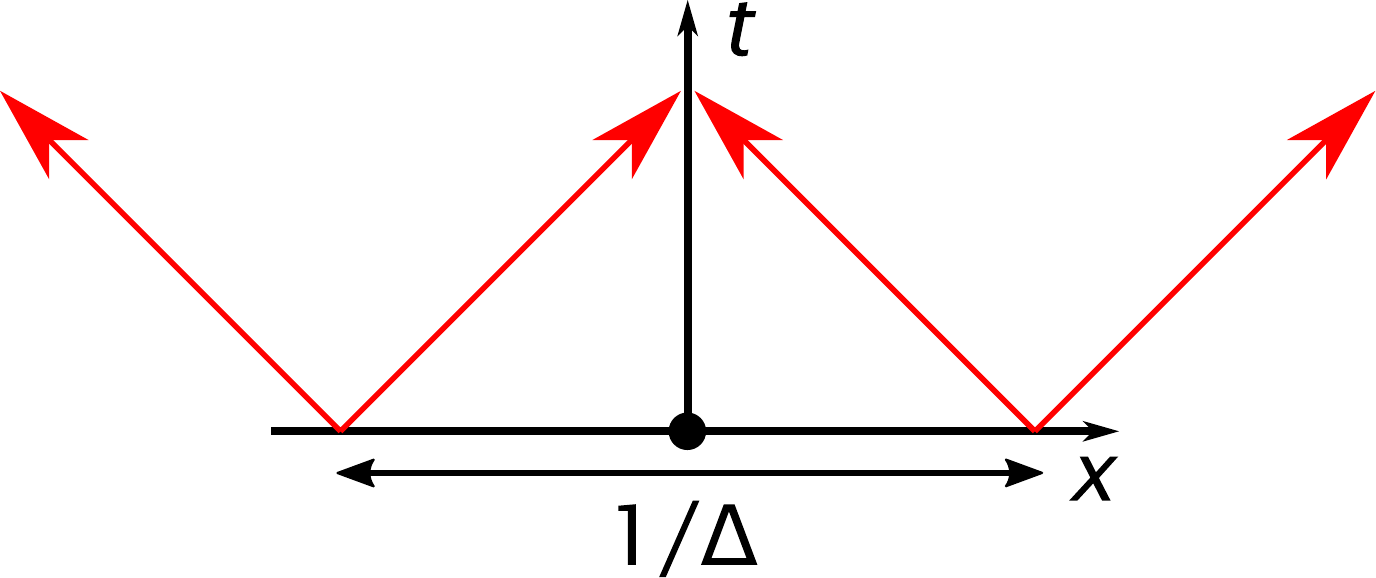}
\includegraphics[width=2.5in]{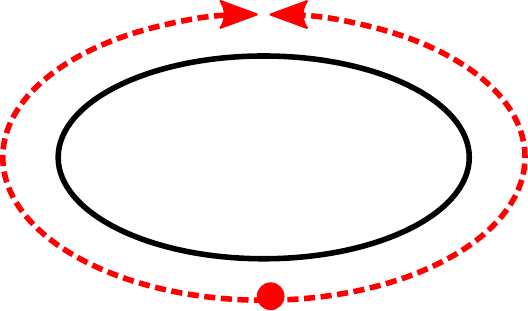}
\caption{Propagation of quasiparticles after a quantum quench. Top: Quasiparticles created by the quench must travel a distance $\sim \absval{2 \Delta}^{-1}$ to see uncorrelated quasiparticles. Bottom: Two quasiparticles created at the same point, and traveling in opposite directions, meet again on the other side of the periodic chain.}
\end{figure}

One of the questions that quenches are designed to elucidate is: how do many-body systems reach thermal equilibrium or otherwise relax?
As we consider a closed system, represented by a pure state, the time evolution is unitary and the system as a whole cannot thermalize (or, indeed, relax).
Instead we find that relaxation occurs for many local observables, for example expectation values of single site operators and short ranged correlators.
A candidate local quantity for the 2D quantum Ising model is~\expect{\sigma^z_\ell(x,t)}: the expectation of spin at position $x$ on chain $\ell$, at time $t$.
Unfortunately the symmetries of our chosen quench and the Ising chain field theories are such that this quantity is zero at all times.
We can of course still study the spin-spin correlation functions, but it is still useful to have a single chain observable that tracks the dynamics after the quench.
For this purpose we consider the mode expansions of the field theories describing the quantum Ising chains.

As explained in Sec.~\ref{Sec:QuantumIsing}, these modes are fermionic with creation and annihilation operators obeying anti-commutation relations $\{a_{i,p},a^\dagger_{i',p'}\}=\delta_{i,i'}\delta_{p,p'}$, where $p,p'$ are fermion momenta parallel to the chains, and $i,i'$ are chain indices.
Using the occupations of these modes we can calculate the occupation number (density) in position space, $n_i(x)$:
\begin{align}
R \; n_i(x)= \int_0^R \text{d} x' n_i(x')=\sum_p a^\dagger_{i,p} a_{i,p},
\end{align}
where we have invoked translational invariance around the cylinder for the first equality.
In principle calculating this quantity involves an infinite sum over all $p=2\pi n /R, n \in \mathbb{Z}$, but in practice the mode occupations fall off rapidly enough with $p$ that it is possible to obtain an accurate answer by summing a finite number of terms.
Monitoring the occupations of the individual chain modes $n_{i,p}=a^\dagger_{i,p} a_{i,p}$ also provides a good check on the effect of the chain spectrum cutoff $E_c$.
The cutoff imposes a largest possible fermion momentum along a chain through
\begin{align}
E_c > E_p-E_\text{vac}=\sqrt{\Delta^2+p^2},
\end{align}
where $E_\text{vac}$ is the appropriate chain vacuum energy.
If the occupations $n_{\ell,p}$ calculated by the ChainAMPS algorithm, are not suitably small as $p \to p_\text{max}$, then $E_c$ should be increased. 

For small $J_\perp$ we can perform a perturbative calculation of the mode occupations~\cite{james2015quantum}, following the unitary method of~\cite{kollar2011generalized} to avoid secular terms that grow in time without bound.
This calculation indicates that the mode occupations are proportional to $J_\perp^2$ at leading order, and are independent of $N_y$ (except for a boundary effect at the ends of the cylinder).

\begin{figure}
\includegraphics[width=3.4in]{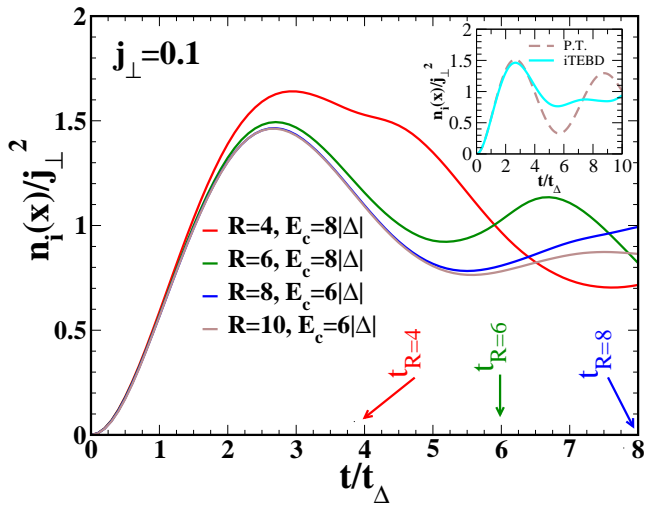}
\caption{The fermion density, $n_i(x)$ in the 2D quantum Ising ChainAMPS system, with different chain lengths $R$, after a quench from uncoupled disordered chains ($\Delta<0, J_\perp=0$) to finite coupling, $J_\perp=0.1$.
The results are scaled by the dimensionless interchain coupling $j_\perp^2=J_\perp^2 \absval{\Delta}^{-7/2}$, and collapse to a single curve up to a time $\sim t_R$, described in the text.
Inset: comparison between perturbation theory (P.T.) and the $R=10$ data, showing that the perturbation theory is no longer accurate.
Reproduced from Ref.~\cite{james2015quantum}.}
\label{fig:2DQIM_weak_quench}
\end{figure}
If we confine our discussion to disordered $\Delta<0$ chains, we can consider two types of quenches from the $J_\perp=0$ state: those in which the final coupling is less than $J_c$, and those in which it is greater.
The former are easier to perform with iTEBD and ChainAMPS, because the population of the higher energy modes remains small for a reasonable range of times.
Figure~\ref{fig:2DQIM_weak_quench} shows the results of iTEBD simulations on the ChainAMPS 2D quantum Ising model for quenches to $J_\perp=0.1<J_c$ with several different chain lengths.
Scaling the results by the dimensionless interchain coupling $j_\perp^2=J_\perp^2 \absval{\Delta}^{-7/2}$, we see that they collapse to the same curve, up to approximately $t_R=\absval{\Delta}R t_\Delta$ (the actual value will be renormalized by the interchain hopping).
In the region of collapse, the dynamics are in the 2D thermodynamic limit, outside this region the finite chain length affects the quench dynamics.

The inset of Fig.~\ref{fig:2DQIM_weak_quench} shows that the leading order perturbative result (which is appropriate for $J_\perp \ll 1$) is only accurate to short times for this quench.
The failure of the perturbative result can be linked to the growth of the Neveu-Schwarz (half-integer momentum) chain modes with time.
These modes are entirely missed by the perturbation theory at leading order, but would appear once higher order scattering processes were taken into account.

\begin{figure}
\includegraphics[width=0.45\textwidth]{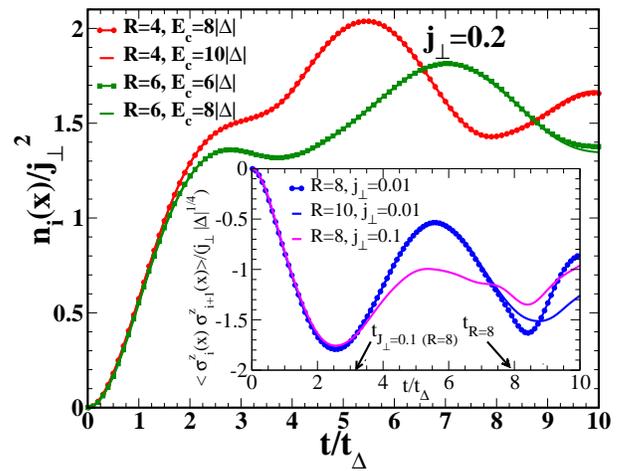}
\caption{The fermion density, $n_i(x)$ in the 2D quantum Ising ChainAMPS system, with different chain lengths $R$, after a quench through the critical coupling to $J_\perp=0.2$. 
Inset: nearest neighbor chain spin correlation function, for different antiferromagnetic post-quench $J_\perp$, showing the time scale $t_{J_\perp}$.
Reproduced from Ref.~\cite{james2015quantum}.} 
\label{fig:2DQIM_strong_quench}
\end{figure}

Quenches through the critical point can be performed, but the numerics are more challenging, and consequently the chain lengths and/or timescales that can be studied are shorter.
Figure~\ref{fig:2DQIM_strong_quench} shows results for a quench of this type, and also demonstrates that the dependence on $E_c$ becomes more significant at later times.
It is also possible to see the effect of $t_{J_\perp}$, as the approximate time at which quenches for different final $J_\perp$ begin to diverge.
This is especially evident in the inset of Fig.~\ref{fig:2DQIM_strong_quench}.

\begin{figure}
\includegraphics[width=0.45\textwidth]{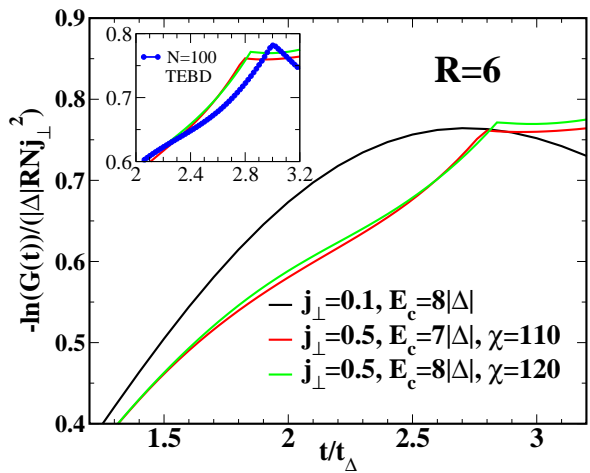}
\caption{The logarithm of the Loschmidt echo for quenches to $J_\perp=0.1$ and $0.5$. There is no apparent non-analytic behaviour in the quench to $J_\perp=0.1$, at least up to $t=10 t_\Delta$.
Inset: the difference between finite and infinite cylinders.
Reproduced from Ref.~\cite{james2015quantum}.} 
\label{fig:2DQIM_non_analytic}
\end{figure}

Quenches on systems with finite numbers of chains require approximately a factor of $N_y$ more computer time.
For reasonably large $N_y \gtrsim 10$, the results for local quantities are not significantly affected by the finite cylinder length, up to the times that can be reached by iTEBD with the ChainAMPS system (excepting boundary effects close to the ends of the cylinder).
Global quantities however, including the Loschmidt echo can be more dramatically affected.
The Loschmidt echo per unit area ($RN_y$, for this 2D model) can display non-analytic points under certain circumstances, and it has been argued that these correspond to athermal behaviour~\cite{heyl2013dynamical}.
Quenches of the coupled quantum Ising chain array to $J_\perp>J_c$ show non-analytic behaviour in the Loschmidt echo, but the qualitative nature of these non-analyticities is sensitive to boundary conditions in 1D and 2D, even for large $N_y$~\cite{james2015quantum} (numerically the non-analytic behaviour is rounded off by finite $N_y$ and $\chi$).
Figure \ref{fig:2DQIM_non_analytic} displays an example of this behaviour for a quench from the uncoupled state to $J_\perp=0.5$.
\subsection{Further directions}
In the last subsection we have attempted to establish that exactly solvable models, often thought of as a peculiarity of one dimension, can have a part to play in understanding many-body quantum physics in higher dimensions.
Here we have mainly covered the application to the 2D quantum Ising model, but it is simple to extend these ideas to other coupled quantum chains.
For example one can treat arrays of (finite length) Luttinger liquids with tunnelling terms, and even lattice systems such as Heisenberg models.
The former case admits three good $\mathbb{Z}$ quantum numbers, related to chain momentum, field winding number, and the canonical momentum conjugate to the field.
Interesting future extensions include adaptive methods to reduce the effects of the chain spectrum truncation (so that higher energy chain states can be gradually added in to the representation), and incorporating some of the ideas from Sec.~\ref{Sec:TSA_Cutoff} to reduce the effects of the cutoff on the chain spectrum.

\section{Summary}

The development of non-perturbative techniques for tackling strongly correlated quantum systems remains at the forefront of contemporary research in condensed matter theory. Whether analytical or numerical in nature, such methods provide a concrete starting point for studying problems absent a small parameter or in which the physics is not adiabatically connected to a trivial (e.g., non-interacting) point. In this review, we have covered a number of such methods for examining low-dimensional quantum systems. 

We first presented an applications-driven discussion of non-Abelian bosonization --  a formal correspondence between fermionic and bosonic theories which explicitly preserves non-Abelian symmetries through conformal embedding. We showed how this method can be applied to models with complicated symmetries, such as $SU(2)\times SU(k)$ which have applications to condensed matter systems with both spin and orbital degeneracies. We followed this by a discussion of applications to cold atom systems with large symmetries, such as $SU(N)$ or $Sp(2N)$. 

We then turned our attention to a numerical approach: the truncated space approach and its recent numerical renormalization group extensions in Secs.~\ref{Sec:TSA}--\ref{Sec:TSAapplications}. This powerful technique bootstraps exact knowledge from integrability to attack new and previously inaccessible problems, allowing us to construct low-energy (approximate) eigenstates and to compute correlation functions within these states. After introducing technical details, we presented applications of the TSA+NRG algorithm to semiconducting carbon nanotubes, the non-equilibrium dynamics of a perturbed integrable model, 2D Landau-Ginsburg theories, and perturbed WZNW models. The latter were the subject of interest in the previous section on non-Abelian bosonization.

The theme of bootstrapping integrability continued in Sec.~\ref{Sec:CHAMPS}, where we introduced matrix product states for arrays of integrable chains. This method expands the toolbox of available techniques for two-dimensional strongly correlated systems by using information from the exact solution of one-dimensional subsystems. By blending such information with matrix product state technology, the low-energy properties (including the critical point) of two-dimensional arrays of integrable one-dimensional quantum systems can be accessed. After filling out a number of technical details, we provided concrete examples where this method can be applied, including arrays of free fermions and quantum Ising chains. We showed that chain array matrix product states are useful for studying both the equilibrium properties and the non-equilibrium dynamics of 2D models that can be formed from arrays of integrable chains, and discussed some future directions for these studies.

\acknowledgments
We have greatly benefited from numerous discussions and collaborations with our colleagues in recent years. We are particularly grateful to Patrick Azaria, Valentin Bois, Edouard Boulat, Giuseppe Brandino, Bruno Bertini, Sylvain Capponi, Jean-S\'ebastien Caux, Fabian Essler, Pierre Fromholz, Andrew Green, Andrew Hallam, Fenner Harper, Curt von Keyserlingk, M\'arton Kormos, Marion Moliner, Giuseppe Mussardo, H\'eloise Nonne, Tam\'as P\'almai, Guillaume Roux, Imke Schneider, Dirk Schuricht, Vid Stojevic, G\'abor Takacs, Jesada Temaismithi, and Keisuke Totsuka for many enlightening conversations. Work by A.J.A.J. was supported by an Engineering and Physical Sciences Research Council (UK) fellowship [EP/L010623/1]. Work at Brookhaven National Laboratory (R.M.K., N.J.R., A.M.T.) was supported by the Condensed Matter Physics and Materials Science Division, under the auspices of the U.S. Department of Energy, contract number DE-AC02-98 CH 10886. P.L. is grateful to the CNRS (France) for financial support (PICS grant).

N.J.R. thanks University College London, the Simons Center for Geometry and Physics, the Aspen Center for Physics, the Georg-August-Universit\"at G\"ottingen, and the Universiteit van Amsterdam for hospitality during portions of this work. 

\section*{Glossary of Acronyms}
\begin{center}
\begin{tabular}{lcl}
{\bf 1D} & \qquad  & One dimensional \\
{\bf 2D} & & Two dimensional \\
{\bf BA} & & Bethe Ansatz \\ 
{\bf BCS} & & Bardeen-Cooper-Schrieffer \\ 
{\bf CDW} & & Charge Density Wave \\
{\bf CFT} & & Conformal Field Theory \\ 
{\bf DFT} & & Density Functional Theory\\ 
{\bf DMFT} & & Dynamical Mean Field Theory \\ 
{\bf DMRG} & & Density Matrix Renormalization Group \\
{\bf ED} & & Exact Diagonalization \\
{\bf iDMRG} & & Infinite system size DMRG \\
{\bf iTEBD} & & Infinite system size TEBD \\
{\bf iPEPS} & & Infinite system size PEPS \\
{\bf IR} & & Infrared \\
{\bf LG} & & Landau-Ginsburg (model) \\
{\bf MERA} & & Multiscale Entanglement\\ &&  Renormalization Ansatz \\
{\bf MPO} & & Matrix Product Operator \\ 
{\bf MPS} & & Matrix Product State \\
{\bf NRG} & & Numerical Renormalization Group \\
{\bf ${\cal NS}$} & & Neveu-Schwartz (sector)\\
{\bf OPE} && Operator Product Expansion \\
{\bf PEPS} & & Projected Entangled Pair States \\
{\bf QCP} & & Quantum Critical Point \\
{\bf QMC} & & Quantum Monte Carlo algorithm \\
{\bf QPT} & & Quantum Phase Transition \\
{\bf ${\cal R}$} & & Ramond (sector) \\
{\bf RG} & & Renormalization Group \\
{\bf RPA} & & Random Phase Approximation \\
{\bf SC} & & Superconductivity \\
{\bf SDW} & & Spin Density Wave \\ 
{\bf SVD} & & Singular value decomposition\\
{\bf TBA} & & Thermodynamic Bethe Ansatz \\
\end{tabular}
\begin{tabular}{lcl}
{\bf TCSA} & & Truncated Conformal Space Approach \\ 
{\bf TDVP} & & Time-dependent Variational Principle \\
{\bf TEBD} & & Time-evolving Block Decimation \\
{\bf TIM} & & Tricritical Ising Model \\
{\bf TSA} & & Truncated Space Approach \\ 
{\bf uMPS} & & `Uniform' Matrix Product State \\
{\bf UV} & & Ultraviolet \\
{\bf VEV} & & Vacuum Expectation Value \\
{\bf WZNW} & & Wess-Zumino-Novikov-Witten (model)
\end{tabular}
\end{center}

\appendix

\section{A brief recap of Abelian bosonization}
\label{App:AbelianBosonization}

There are many good introductions to Abelian bosonization: a field theoretic treatment is provided by S\'en\'echal~\cite{SenechalArxiv99} and the textbooks~\cite{GNTBook,TsvelikBook}, whilst an operator-lead constructive approach is explained in great detail in~\cite{vonDelftAnnalenderPhysik98}. The monograph by Giamarchi presents both phenomenological and constructive approaches to bosonization, and then discusses a great many applications~\cite{GiamarchiBook}. 

Over the last three decades Abelian bosonization has become a standard tool in the study of one-dimensional quantum systems. In large part this is due to its success at describing the phenomenology of strongly correlated fermionic systems: by choosing a new basis for a problem, many fermionic problems decouple into separate bosonic theories, which often can be successfully treated via integrability~\cite{GNTBook} or semi-classical approximations~\cite{GiamarchiBook}.\footnote{It is often surprisingly simple to capture the correct phenomenology, as we will see in the following. In certain cases bosonization can be combined with information from integrability to give exact results (see, e.g., Ref.~\cite{AdityaPRB12}).} In this appendix, we aim to briefly introduce bosonization through the operator correspondence, and then illustrate the so-called spin-charge separation and semi-classical treatment with a simple example.

\subsection{The bosonization identities}

At its heart, bosonization describes a formal correspondence between operators (or fields) in a fermionic theory and those in a bosonic theory. Consider left- and right-moving fermion fields $L_\s,\, R_\s$ for electrons with spin $\s=\up,\dn$ -- these may be related to the chiral bosonic fields $\varphi_\s, \bar \varphi_\s$ via the \textit{bosonization identities}
\be
\begin{split}
R_\s(x) &\sim \frac{\eta_\s}{\sqrt{2\pi}} :e^{i \varphi_\s(x)}:, \\ 
L_\s(x) &\sim \frac{\eta_\s}{\sqrt{2 \pi}} :e^{-i\bar\varphi_\s(x)}:, 
\end{split} \label{bosonizationids}
\ee
where $\eta_\s$ are Klein factors, which anticommute $\{\eta_\s, \eta_{\s'}\} = 2\delta_{\s\s'}$ to ensure the anticommutation of fermion fields of different spin species, and the bosonic fields are governed by the actions~\fr{BosonAction}. 

We have explicitly written the normal ordering of the exponential of the bosonic fields (often called vertex functions, from the high energy literature) in the above bosonization identities. This normal ordering is important in a linear theory (which is pathological in the absence of a cut-off, as there are an infinite number of electron states) and it means that the vertex functions do not multiply in the usual manner, instead they satisfy
\be
: e^A : :e^B: = : e^{A+B} : e^{ \la 0 | A B | 0\ra},
\ee
which follows from the mode-expansion of the bosonic field (see, for example, Ref.~\cite{SenechalArxiv99}). Working in Euclidean space with complex coordinate $z = \tau - ix$, this translates to the following for the vertex functions
\be
e^{i \alpha \varphi_\s(z)} e^{i \beta \varphi_{\s}(w)} = e^{i \alpha \varphi_\s(z) + i \beta \varphi_\s(w)} \left(z-w\right)^{\alpha\beta}, \label{vertexfns}
\ee 
where normal ordering of vertex functions is implicit. Notice that~\fr{vertexfns} shows that the correlation functions of the fermionic fields are reproduced by the bosonic operators
\bea
\la R^{\phantom\dagger}_\s(z) R^\dagger_{\s'}(z') \ra &=& \delta_{\s\s'} \frac{1}{2\pi}\frac{1}{z-z'},\\
&=&  \delta_{\s\s'} \frac{1}{2\pi}\frac{1}{(\tau-\tau') - i (x-x')}\, .
\eea
Furthermore, it can be proven at the level of the partition function that the free bosonic theory coincides with the free fermionic one~\cite{SenechalArxiv99,CFTBook}. An early prominent example of the fermion-boson correspondence in an interacting (1+1)-dimension system is the quantum sine-Gordon to massive Thirring model mapping discussed by Coleman in 1975~\cite{ColemanPRD75}. 

Further useful bosonization identities include the density operator
\be
:R^\dagger_\s(z) R^{\phantom\dagger}_\s(z): \sim -\frac{i}{2\pi} \p_z \varphi_\s(z), 
\ee
which can be derived via a point-splitting procedure~\cite{SenechalArxiv99}
\bea
&&:R^\dagger_\s(z) R^{\phantom\dagger}_\s(z):\nn 
&&= \lim_{\epsilon\to0} \Big[ R^\dagger_\s(z+\epsilon) R_\s(z-\epsilon) - \la R^\dagger_\s(z+\epsilon) R_\s(z-\epsilon) \ra \Big], \nn
&& = \lim_{\epsilon\to 0} \frac{-1}{4\pi \epsilon} \Big[ 1 - e^{-i(\varphi_\s(z+\epsilon) - \varphi_\s(z-\epsilon))} \Big], \nn
&& = \lim_{\epsilon\to 0} \frac{-1}{4\pi \epsilon} \Big[ 1 - 1 + 2 i \epsilon \p_z \varphi_\s(z) + O(\epsilon^2)  \Big], \nn
&& = -\frac{i}{2\pi} \p_z \varphi_\s(z). 
\eea

\subsection{Spin-charge separation} 
Let us now turn our attention to a phenomenon at the center of physics in one-dimensional quantum systems: spin-charge separation. Consider a system of interacting electrons in one-dimension; for simplicity we will consider the Hubbard (on-site) interaction. The Hamiltonian reads
\be
H_U = -t\sum_{l,\s} \Big(c^\dagger_{\s}(l)c^{\phantom\dagger}_{\s}(l+1) + {\rm H.c.}\Big)+ U \sum_l n^{\phantom\dagger}_{\up}(l)n^{\phantom\dagger}_{\dn}(l), \label{FermiHam}
\ee
where $n_{l,\s} = c^\dagger_{l,\s} c_{l,\s}$ is the number operator for electrons with spin $\s$. We focus on the case of half-filling (e.g., one-electron per site) and we proceed to bosonize the model by first linearizing the dispersion and then applying the bosonization identities~\fr{bosonizationids}. Following this, we change basis to a set of bosonic fields associated with spin ($s$) and charge ($c$) degrees of freedom: 
\be
\Phi_d = \varphi_d + \bar \varphi_d,~~\Theta_d = \varphi_d - \bar\varphi_d,~~(d = c,s)
\ee
where the chiral fields are defined as
\bea
\varphi_c &=& \varphi_\up + \varphi_\dn,~~ \varphi_s = \varphi_\up - \varphi_\dn, \\
\bar\varphi_c &=& \bar\varphi_\up + \bar\varphi_\dn,~~ \bar\varphi_s = \bar\varphi_\up - \bar\varphi_\dn\ .
\eea
Following this, we obtain the Hamiltonian density~\cite{HubbardBook}
\bea
{\cal H}_U &=& {\cal H}_c + {\cal H}_s, \label{BosonicHam}\\
{\cal H}_c &=& \frac{v_F}{16\pi} \Big[ (\p_x \Phi_c)^2 + (\p_x \Theta_c)^2 \Big] \nn
&& - \frac{g}{(2\pi)^2} \bigg\{ \cos(\Phi_c) + \frac{1}{16} \Big[ (\p_x \Theta_c)^2 - (\p_x \Phi_c)^2 \Big] \bigg\}, \nn
{\cal H}_s &=& \frac{v_F}{16\pi} \Big[ (\p_x \Phi_s)^2 + (\p_x \Theta_s)^2 \Big] \nn
&& + \frac{g}{(2\pi)^2} \bigg\{ \cos(\Phi_s) + \frac{1}{16} \Big[ (\p_x \Theta_s)^2 - (\p_x \Phi_s)^2 \Big] \bigg\}. \nonumber
\eea
We see that the theory has separated into two decoupled sectors, describing spin and charge degrees of freedom. This so-called \textit{spin-charge separation} has been observed experimentally in quasi-one-dimensional materials (see, for example, Refs.~\cite{KimPRL96,ClaessenPRL02,AuslaenderScience05,KimNatPhys06}). 

\subsection{Semi-classical treatment: a simple example}

We have seen that bosonization maps a fermionic Hamiltonian~\fr{FermiHam} to a bosonic one~\fr{BosonicHam} with decoupled spin and charge degrees of freedom. However, the bosonic theory is rather complicated: we now have non-linear interaction terms, and one might be tempted to suggest that we have in fact made our lives more difficult. Fortunately, it is often the case that a semi-classical analysis is sufficient to understand the physics.

Consider the Hamiltonian~\fr{BosonicHam}, which is the continuum limit of the Hubbard model at half-filling. Focusing on the case with repulsive interactions, a one-loop RG analysis (valid for small $g$) tells us that the fixed point Hamiltonian is~\cite{BalentsPRB96}
\bea
{\cal H} &=& \frac{v_s}{16\pi} \Big[ K_s^{-1} (\p_x \Phi_s)^2 + K_s (\p_x \Theta_s)^2 \Big] \nn
&& + \frac{v_c}{16\pi} \Big[ K_c^{-1} (\p_x \Phi_c)^2 + K_c (\p_x \Theta_c)^2 \Big] \nn
&& - \tilde g \cos (\Phi_c), 
\eea
where $K_c, K_s$ are the Luttinger parameters for the charge and spin sectors~\cite{GiamarchiBook} (which reflect the interacting nature of the underlying fermionic theory, and which change under the RG flow), $v_c, v_s$ are velocities for the charge and spin degrees of freedom, and $\tilde g \sim O(1)$ is the renormalized coupling (the coupling for the cosine of the spin boson flows to zero under the RG). The Hubbard model enjoys an $SU(2)$ spin symmetry, which fixes the Luttinger parameter in the spin sector to be $K_s = 1$.

As the RG flow is towards a strong coupling fixed point with large $\tilde g$, it is natural to treat the cosine term semi-classically: the cosine term pins the spin boson $\Phi_c$ to one of its minima. We then expand about one of the minima of the cosine potential $\Phi_c = 2n\pi + \tilde \Phi_c$ ($n\in\mathbb{Z}$), assuming that fluctuations $\tilde \Phi_c$ are small, to obtain the Hamiltonian
\bea
{\cal H} &=& \frac{\tilde v_s}{16\pi} \Big[ (\p_x \Phi_s)^2 + (\p_x \Theta_s)^2 \Big] \nn
&& + \frac{\tilde v_c}{16\pi} \Big[ K_c^{-1} (\p_x \tilde \Phi_c)^2 + K_c (\p_x \tilde \Theta_c)^2 \Big] \nn
&& + \tilde m \tilde \Phi_c^2 + O\Big(\tilde\Phi_c^4\Big) ,
\eea
where $\tilde m$ is an effective mass for the charge fluctuations and we neglect an unimportant constant. 

So, semi-classically we have a low-energy effective theory that describes a gapless spin degree of freedom and a \textit{massive} charge degree of freedom. Correlation functions can be computed: charge correlation functions will decay exponentially due to the effective mass $\tilde m$ and spin fluctuations will decay as a power law. This phenomenology is consistent with the exact solution of the Hubbard model via the Bethe ansatz~\cite{HubbardBook}. 

\section{A conformal field theory primer}
\label{App:CFT}

Conformal field theory is now a vast field in its own right and covers a huge variety of works, ranging from pure mathematics to applied physics. As a result there is a large introductory literature on the subject -- perhaps the best known is the beautiful \textit{``Big Yellow Book''} by Di Francesco, Mathieu and S\'en\'echal~\cite{CFTBook}. In this appendix, we will briefly introduce the subject of CFT and summarize some useful results. For further exposition and information, we urge the reader to consult the wider literature.

\subsection{Conformal transformations}
\label{Sec:ConfTrans}

In two dimensions, the conformal group is formed from the set of all holomorphic mappings. It is useful to consider the generators of conformal transformations
\be
\ell_n = - z^{n+1} \p_z, \quad \bar\ell_n = - \bar z^{n+1} \p_{\bar z}, \label{confgens}
\ee
which are derived from the following logic. Consider an infinitesimal holomorphic transformation 
\be
z \to z' = z + \epsilon(z), \quad \epsilon(z) = \sum_{n=-\infty}^\infty c_n z^{n+1},  \label{conftrans}
\ee
here $\epsilon(z)$ is small, and can be expressed as a Laurent series about $z=0$ (this is true by construction as we consider holomorphic transformations). A spinless field $\Psi(z,\bar z)$ (see Eq.~\fr{scalingdimconfspin}) transforms under~\fr{conftrans} as 
\be
\Psi(z,\bar z) = \Psi(z',\bar z') - \epsilon(z')\p_{z'} \Psi(z',\bar z') - \bar\epsilon(\bar z') \p_{\bar z'} \Psi(z',\bar z'),
\ee
which is easily written in terms of the generators~\fr{confgens} as 
\be
\delta \Psi = \sum_{n=-\infty}^{\infty} \Big[ c_n \ell_n \Psi(z,\bar z) + \bar c_n \bar \ell_n \Psi(z,\bar z)\Big].
\ee

It is straightforward to show that the generators obey the Witt algebra
\bea
&&[\ell_n, \ell_m] = (n-m)\ell_{n+m}, \\
&&{}[\bar\ell_n,\bar\ell_m] = (n-m)\bar\ell_{n+m}, \\
&&{}[\ell_n,\bar\ell_m] = 0.
\eea
Clearly $\ell_n$ and $\bar \ell_n$ form two infinite isomorphic algebras, each of which has an additional finite sub-algebra formed from $\ell_{-1}$, $\ell_0$, $\ell_1$. These generators correspond to: (i) $\ell_{-1} = -\p_z$ translations of the complex plane; (ii) $\ell_0 = -z\p_z$ scale transformations; (iii) $\ell_1 = -z^2\p_z$ special conformal transformations. Together (i)--(iii) form the global conformal group. 

We note that in real space ($x,y\in\mathbb{R}$), only the linear combinations
\be
\ell_n + \bar \ell_n, \quad -i(\ell_n - \bar\ell_n),
\ee 
preserve the realness of the space. For $n=0$, these combinations correspond to dilation and rotations of the real space, respectively.

\subsection{What is a CFT?}

From a purely ``computational'' point of view, a CFT can be described through the following: 
\begin{enumerate}
\item a set of primary fields $\{\phi_j \}$;
\item the conformal dimensions $\{(\Delta_j,\bar \Delta_j)\}$ of these fields;
\item the rules for fusion of these fields $\phi_i \times \phi_j \to c_{ij}^k \phi_k$;
\item its central charge.
\end{enumerate}
For the purpose of clarity in illustrating these concepts, we will focus on one particular example of a CFT: the critical Ising model.

\subsection{The critical Ising model}
This is equivalent to a model of a massless real (e.g., Majorana) fermion with the action 
\be
S = \frac{1}{2\pi} \int \rd z \rd \bar z \Big( \psi \bar\p \psi + \bar\psi \p \bar\psi\Big), \label{IsingCFT}
\ee
where $\psi,\bar\psi$ are holomorphic and anti-holomorphic Majorana fermion fields with propagators
\be
\begin{split}
\la \psi(z,\bar z) \psi(w, \bar w) \ra &= \frac{1}{z-w}, \\
\la \bar\psi(z,\bar z) \bar \psi(w,\bar w) \ra &= \frac{1}{\bar z - \bar w}, 
\end{split} \label{propagators}
\ee
and the two-dimensional plane $(x,y)$ is parameterized by the complex coordinates $z = x + i y$, $\bar z  = x - i y$. Derivatives with respect to these are denoted by 
\be
\p = \p_z = \frac12(\p_x - i \p_y), \quad \bar \p = \p_{\bar z} = \frac12(\p_x + i \p_y). 
\ee

\subsubsection{Primary fields}

Primary fields are of central importance in CFT. Under a conformal transformation  $z \to w(z),\, \bar z \to \bar w(\bar z)$ , primary fields $\phi_j(z,\bar z)$ transform as 
\be
\phi_j(z,\bar z) \to \phi_j(w,\bar w) =  \left( \frac{\p w}{\p z} \right )^{-\Delta_j} \left( \frac{\p \bar w}{\p \bar z} \right)^{-\bar \Delta_j}  \phi_j(z,\bar z),
\ee
where $(\Delta_j, \bar \Delta_j)$ are the holomorphic and antiholomorphic conformal dimensions, respectively. From these, one can define the scaling dimension $d_j$ and the conformal spin $s_j$ of the primary field $\phi_j$
\be
d _j = \Delta_j + \bar \Delta_j, \quad s_j = \Delta_j - \bar \Delta_j.  \label{scalingdimconfspin}
\ee

In terms of the Ising CFT~\fr{IsingCFT}, it follows from the form of the propagators~\fr{propagators} that the Majoranas fermions are primary fields with conformal dimensions
\be
\psi(z):\ \left(\frac12,0\right), \quad \bar \psi(\bar z):\ \left(0,\frac12\right). 
\ee
They have scaling dimension $d = 1/2$, as should be expected for fermions, and carry $s = \pm 1/2$ conformal spin. 

There are two additional operators of interest in the critical Ising theory. These are the energy operator $\varepsilon(z,\bar z) = i \bar\psi(z,\bar z)\psi(z,\bar z)$ and the spin operator $\s(z,\bar z)$, which are related to the operators $\s_i \s_{i+1}$ and $\s_i$ in the lattice two-dimensional Ising model (see, e.g., Ref.~[\onlinecite{CFTBook}]), respectively. The operators are primary fields with conformal dimensions
\bea
&&\varepsilon(z,\bar z):\ \left( \frac12,\frac12\right),\nn
&& \s(z,\bar z):\ \left(\frac{1}{16}, \frac{1}{16}\right). \nonumber
\eea
The disorder parameter $\mu(z,\bar z)$ is dual to the spin operator $\s(z,\bar z)$ and carries the same conformal dimensions. 

\subsubsection{The operator product expansion and fusion rules}
The computation of correlation functions is one of the central aims in field theory. It is typically the case that when you bring two operators together towards a single point $z\to w$ the correlation function diverges, see for example the propagators of the Majorana fields~\fr{propagators}. The operator product expansion (OPE) is a representation of this process of bringing two operators together: it expresses the product as a sum of operators which are well-behaved $z$ multiplied by functions of $z-w$ which may diverge as $z\to w$. A typical example would look like 
\be
\lim_{z\to w} \phi_i(z) \phi_j(w) \sim \sum_k c_{ij}^k \frac{\phi_k(w)}{(z-w)^{\Delta_i + \Delta_j - \Delta_k}},
\ee
where $c_{ij}^k$ are real numbers, often called the OPE coefficients and $\sim$ denotes that the relation holds only within correlation functions and keeps only the singular terms. 

A special role in the OPE is played by the stress-energy tensor of the theory. The OPE of the stress-energy tensor with a primary field is fixed
\bea
T(z)\phi(w,\bar w) \sim \frac{\Delta}{(z-w)^2} \phi(w,\bar w) + \frac{1}{z-w} \p_w \phi(w,\bar w),\nn
\bar T(\bar z)\phi(w,\bar w) \sim \frac{\bar \Delta}{(\bar z - \bar w)^2}\phi(w,\bar w) + \frac{1}{\bar z - \bar w}\p_{\bar w}\phi(w,\bar w),\nn\label{setprimary}
\eea
where $(\Delta,\bar \Delta)$ is the conformal dimension of the primary field $\phi(w,\bar w)$. It is also the case that the OPE of the stress energy tensor with itself is
\be
T(z)T(w) \sim \frac{c/2}{(z-w)^4} + \frac{2T(w)}{(z-w)^2} + \frac{\p T(w)}{(z-w)}, \label{OPEset}
\ee 
where $c$ is the central charge of the theory. 

In the context of the Ising field theory, the stress energy tensor is
\be
T(z) = -\frac12 \psi(z) \p \psi(z).
\ee
Using the OPE of the Majorana fermion fields
\be
\psi(z) \psi(w) \sim \frac{1}{z-w}, 
\ee
it is easy to see that 
\be
T(z)\psi(w) \sim \frac{1/2}{(z-w)^2}\psi(w) + \frac{1}{z-w} \p\phi(w), 
\ee
as required. Other useful OPE for the fields in the Ising theory include 
\bea
\varepsilon(z,\bar z) \varepsilon(w,\bar w) &\sim& \frac{1}{|z-w|^2},\\
\s(z,\bar z)\s(w,\bar w) &\sim& \frac{1}{|z-w|^{\frac14}} +\frac12 |z-w|^{\frac34} \varepsilon(w,\bar w). \nonumber \\ \label{opesigma}
\eea
Schematically, a quick way to characterize the above are the fusion rules:
\bea
\psi \times \psi &=& {\bf 1}, \quad \bar\psi \times \bar\psi = {\bf 1}, \nn
\varepsilon\times\varepsilon &=& {\bf 1}, \quad \s \times \s = {\bf 1 } + \varepsilon. \nonumber
\eea
We summarize the conformal information of the Ising theory and the fusion rules in Table~\ref{ConfInf}.
\begin{table}[t]
\begin{tabular}{c c c}
\begin{tabular}{|c||c|c|c|c|}
\hline
&&&& \\
& ~$\Delta$~ & ~$\bar \Delta$~ & ~$d$~ & ~$s$~ \\ 
\hline\hline
$\psi$ 		&	 $\frac12$ 	& 	0 			& 	$\frac12$ 		& 	$\frac12$ 		\\
$\bar\psi$ 		& 	0 			& $\frac12$ 		& $\frac12$ 	& 	$-\frac12$		\\ 
$\varepsilon$ 	& 	$\frac12$ 		& $\frac12$ 		& 	$1$ 		& 	0 			\\
$\s$ 			& 	$\frac{1}{16}$ 	& $\frac{1}{16}$		& $\frac{1}{8}$ 	& 	0			\\[2pt]
\hline
\end{tabular}
& \hspace{1cm} & 
\begin{tabular}{|c||c|c|c|c|}
\hline
&&&& \\
& ~$\psi$ ~ & ~$\bar\psi$ 	~ & ~$\varepsilon$~ & ~$\s$ \\ 
\hline\hline
$\psi$ 		&	{\bf 1}	&			& 			& 	$\mu$				\\
$\bar\psi$ 		& 			&	{\bf 1}	&			& 	$\mu$				\\ 
$\varepsilon$ 	&			&			&	{\bf 1}	&	$\s$					\\
$\s$ 			&	$\mu$	&	$\mu$	& 	$\s$		& 	{\bf 1}+$\varepsilon$ 	\\ 
\hline
\end{tabular}\end{tabular}
\caption{Conformal data for the Ising CFT. (Left) List of primary fields and their conformal dimensions $(\Delta,\bar \Delta)$, scaling dimensions $d$, and conformal spin $s$. (Right) The fusion rules for the primary fields of the Ising CFT, with ${\bf 1}$ being the identity field (${\bf 1} \times \phi = \phi$).}
\label{ConfInf}
\end{table}

\subsubsection{Conformal blocks}

The fusion rules immediately suggest that factorization of the multi-point correlation functions into products of holomorphic functions does not occur (unlike in the Luttinger liquid, cf. Eqs.~\fr{bosid} and the accompanying discussion). Instead, correlation functions are, generically, sums of products of holomorphic functions. This is well illustrated by the four point function of the spin operator
\be
\la \s(z_1,\bar z_1) \s(z_2,\bar z_2) \s(z_3,\bar z_3) \s(z_4,\bar z_4) \ra . 
\ee
Applying the OPE~\fr{opesigma} to both the first and second pairs, we find
\bea
&&\la \s(z_1,\bar z_1) \s(z_2,\bar z_2) \s(z_3,\bar z_3) \s(z_4,\bar z_4) \ra \nn
&& = \frac{1}{|z_1-z_2|^{\frac14} |z_3-z_4|^{\frac14}} \left( 1 + \frac{1}{4} \frac{|z_1-z_2| |z_3-z_4|}{|z_2 - z_4|^2} \right), \nn
\eea
which is \textit{a sum of products of holomorphic functions}. 

More generally, in conformal field theory multi-point correlation functions of fields will have the form
\bea
&&\la \phi(z_1,\bar z_1)\ldots \phi(z_N,\bar z_N)\ra\nn
&& ~~~= \sum_j C_j F_j(z_1,\ldots,z_N)\bar F_j(\bar z_1,\ldots,\bar z_N),
\eea
where $F_j(z_1,\ldots, z_N)$ and $\bar F_j(\bar z_1,\ldots,\bar z_N)$ are called conformal blocks, and correlation functions can be thought of as factorizing on the level of conformal blocks.

\subsection{The Hilbert space}

Before finishing our brief review of some CFT topics relevant to our discussions, it will be useful to  examine the basic structure of the Hilbert space of CFTs. 

\subsubsection{The Virasoro Algebra}

An important concept in CFT is the Virasoro algebra~\cite{VirasoroPRD70}. The algebra is formed from the mode operators of the Laurent expansion of the stress-energy tensor, traditionally denoted by $L_n,\bar L_n$~\cite{CFTBook}: 
\be
T(z) = \sum_{n=-\infty}^\infty z^{-n-2} L_n, \quad \bar T(\bar z) = \sum_{n=-\infty}^{\infty} \bar z^{-n-2} \bar L_n.
\ee
Here, the indices $-n-2$ are chosen such that the operator $L_{-n}$ transforms under $z\to z/a$ as $L_{-n} \to a^n L_{-n}$ and hence $L_n$ has scaling dimension $n$. The Laurent expansion can be inverted to give the relations~\cite{CFTBook}
\be
\begin{split}
L_n &= \frac{1}{2\pi i}\oint \rd z\, z^{n+1} T(z), \\
\bar L_n &= \frac{1}{2\pi i } \oint \rd \bar z\, \bar z^{n+1} \bar T(\bar z).
\end{split} 
\ee
Whilst the operators $\ell_n,\bar\ell_n$ in Sec.~\ref{Sec:ConfTrans} are the generators of local conformal transformations in the space of functions, the operators $L_n,\bar L_n$ generate local conformal transformations on the Hilbert space. For example, $L_0 + \bar L_0$ generate dilations of the real space $(z,\bar z) \to a(z,\bar z)$, in direct analogy with $\ell_0 + \bar\ell_0$. $L_0 + \bar L_0$ can be interpreted as the Hamiltonian of the CFT~\cite{CFTBook}. 

From the OPE of the stress-energy tensor, see Eq.~\fr{OPEset}, the Virasoro algebra~\cite{VirasoroPRD70} follows: 
\be
\begin{split}
[L_n, L_m] &= (n-m) L_{n+m} + \frac{c}{12}n(n^2-1)\delta_{n+m,0}\, , \\
[\bar L_n, \bar L_m] &= (n-m)\bar L_{n+m} + \frac{c}{12}n(n^2-1)\delta_{n+m,0}\, , \\
[L_n, \bar L_m] &= 0,
\end{split}
\ee
where $c$ is the central charge of the CFT. As with the operators $\ell_n,\bar\ell_n$, there is a subalgebra formed from $L_{-1},L_0,L_1$ which generate the global conformal group
\be
[L_{\pm 1}, L_0] = \pm L_{\pm 1}, \quad [L_1, L_{-1}] = 2 L_0.
\ee

\subsubsection{States in the Hilbert space}

Here we will briefly cover some basic terminology; the Hilbert space of a CFT can be quite complicated and may possess some intricate mathematical structure~\cite{CFTBook}. Generally, states in the CFT can be grouped into families $[\phi]$, each of which contains a single primary field $\phi$ of the theory and an infinite set of `descendant' fields. These form irreducible representations of the conformal group, with the primary field corresponding to a `highest weight state' of the representation. 

\paragraph{The vacuum.} 
We denote the vacuum state by $|0\ra$ and impose that it possesses global conformal symmetry. As a result, the vacuum state must be annihilated by the generators of the global conformal group $L_{-1}, L_0, L_1$ (and the corresponding antiholomorphic components)
\be
L_{-1} |0\ra = L_{0} |0\ra =  L_{1} |0\ra =  0. 
\ee
Furthermore, application of the stress-energy tensor to the vacuum should be well-defined in the limit $z\to 0$: 
\be
T(z) |0\ra = \sum_{n=-\infty}^\infty z^{-n-2} L_n |0\ra,
\ee 
which clearly requires 
\be
L_n |0\ra = 0,\quad \mathrm{for}~n \ge -1.
\ee
Application of the operators $L_{-n}$ with $n \ge 2$ to the vacuum generate states in the Hilbert space that form part of a representation of the Virasoro algebra. 

\paragraph{Highest weight states.} 
Let us now consider the application of the primary field $\phi(z,\bar z)$ with conformal dimensions $\Delta,\bar\Delta$ to the vacuum:
\be
|\Delta,\bar\Delta\ra = \phi(z,\bar z)|0\ra. 
\ee
The OPE of the stress-energy tensor $T(z)$ with the primary field $\phi(z,\bar z)$, see Eq.~\fr{setprimary}, fixes the commutation relation of the Virasoro operators with field
\be
[L_n, \phi(z,\bar z)] = \Delta (n+1) z^n \phi(z,\bar z) + z^{n+1} \p \phi(z,\bar z), \label{Lnphi}
\ee
where $n\ge -1$, and hence we find
\be
L_0 |\Delta,\bar \Delta \ra = \Delta |\Delta,\bar \Delta\ra, \quad
\bar L_0 |\Delta,\bar\Delta\ra = \bar\Delta | \Delta,\bar\Delta\ra. \label{L0prim}
\ee
Hence states generated by applying a primary field to the vacuum are eigenstates of the Hamiltonian. Equation~\fr{Lnphi} also implies
\be
L_n |\Delta,\bar\Delta\ra = \bar L_n |\Delta,\bar\Delta\ra = 0, \quad n > 0. \label{Lnprim}
\ee
The states $|\Delta,\bar\Delta\ra$ that satisfy Eqs.~\fr{L0prim}--\fr{Lnprim} are known as the highest weight states. 

\paragraph{Descendent states.} 
Application of the operator $L_{-m}$ ($m > 0$) to a state increases the conformal dimension (as $L_{-m}$ has scaling dimension $m$), as can be seen from the Virasoro algebra:
\be
[L_0, L_{-m}] = m L_{-m}. 
\ee
Excited states in a CFT can be obtained from application of Virasoro operators on a highest weight state:
\be
L_{-m_1}L_{-m_2}\ldots L_{-m_N} \bar L_{-\bar m_1}\bar L_{-\bar m_2}\ldots \bar L_{-\bar m_{\bar N}} |\Delta,\bar\Delta\ra 
\ee
where by convention $m_1 < m_2 < \ldots < m_N$ (and $\bar m_1 < \bar m_2 < \ldots < \bar m_{\bar N}$). This state is a simultaneous eigenstate of $L_0$ with eigenvalue \[ \Delta + \sum_{n=1}^N m_n, \] and of $\bar L_0$ with eigenvalue \[\bar \Delta + \sum_{n=1}^{\bar N} \bar m_n.\] Such states are known as descendent states (or simply `descendants') of the primary field $\phi(z,\bar z)$. 

\bibliography{review_bib}

\begin{thebibliography}{592}%
\makeatletter
\providecommand \@ifxundefined [1]{%
 \@ifx{#1\undefined}
}%
\providecommand \@ifnum [1]{%
 \ifnum #1\expandafter \@firstoftwo
 \else \expandafter \@secondoftwo
 \fi
}%
\providecommand \@ifx [1]{%
 \ifx #1\expandafter \@firstoftwo
 \else \expandafter \@secondoftwo
 \fi
}%
\providecommand \natexlab [1]{#1}%
\providecommand \enquote  [1]{``#1''}%
\providecommand \bibnamefont  [1]{#1}%
\providecommand \bibfnamefont [1]{#1}%
\providecommand \citenamefont [1]{#1}%
\providecommand \href@noop [0]{\@secondoftwo}%
\providecommand \href [0]{\begingroup \@sanitize@url \@href}%
\providecommand \@href[1]{\@@startlink{#1}\@@href}%
\providecommand \@@href[1]{\endgroup#1\@@endlink}%
\providecommand \@sanitize@url [0]{\catcode `\\12\catcode `\$12\catcode
  `\&12\catcode `\#12\catcode `\^12\catcode `\_12\catcode `\%12\relax}%
\providecommand \@@startlink[1]{}%
\providecommand \@@endlink[0]{}%
\providecommand \url  [0]{\begingroup\@sanitize@url \@url }%
\providecommand \@url [1]{\endgroup\@href {#1}{\urlprefix }}%
\providecommand \urlprefix  [0]{URL }%
\providecommand \Eprint [0]{\href }%
\providecommand \doibase [0]{http://dx.doi.org/}%
\providecommand \selectlanguage [0]{\@gobble}%
\providecommand \bibinfo  [0]{\@secondoftwo}%
\providecommand \bibfield  [0]{\@secondoftwo}%
\providecommand \translation [1]{[#1]}%
\providecommand \BibitemOpen [0]{}%
\providecommand \bibitemStop [0]{}%
\providecommand \bibitemNoStop [0]{.\EOS\space}%
\providecommand \EOS [0]{\spacefactor3000\relax}%
\providecommand \BibitemShut  [1]{\csname bibitem#1\endcsname}%
\let\auto@bib@innerbib\@empty
\bibitem [{\citenamefont {Planck}(1901)}]{PlanckAnnPhys01}%
  \BibitemOpen
  \bibfield  {author} {\bibinfo {author} {\bibfnamefont {M.}~\bibnamefont
  {Planck}},\ }\bibfield  {title} {\enquote {\bibinfo {title} {{Ueber das
  gesetz der energieverteilung im normalspectrum}},}\ }\href
  {http://dx.doi.org/10.1002/andp.19013090310} {\bibfield  {journal} {\bibinfo
  {journal} {Ann. Phys. (Berlin)}\ }\textbf {\bibinfo {volume} {309}},\
  \bibinfo {pages} {553--563} (\bibinfo {year} {1901})}\BibitemShut {NoStop}%
\bibitem [{\citenamefont {Einstein}(1905)}]{EinsteinAnnPhys05}%
  \BibitemOpen
  \bibfield  {author} {\bibinfo {author} {\bibfnamefont {A.}~\bibnamefont
  {Einstein}},\ }\bibfield  {title} {\enquote {\bibinfo {title} {{\"Uber einen
  die Erzeugung und Verwandlung des Lichtes betreffenden heuristischen
  Gesichtspunkt}},}\ }\href {\doibase 10.1002/andp.19053220607} {\bibfield
  {journal} {\bibinfo  {journal} {Ann. Phys. (Berlin)}\ }\textbf {\bibinfo
  {volume} {322}},\ \bibinfo {pages} {132--148} (\bibinfo {year}
  {1905})}\BibitemShut {NoStop}%
\bibitem [{\citenamefont {Bohr}(1913)}]{BohrPhilMag13}%
  \BibitemOpen
  \bibfield  {author} {\bibinfo {author} {\bibfnamefont {N.}~\bibnamefont
  {Bohr}},\ }\bibfield  {title} {\enquote {\bibinfo {title} {{I. On the
  constitution of atoms and molecules}},}\ }\href {\doibase
  10.1080/14786441308634955} {\bibfield  {journal} {\bibinfo  {journal} {Phil.
  Mag. Ser. 6}\ }\textbf {\bibinfo {volume} {26}},\ \bibinfo {pages} {1--25}
  (\bibinfo {year} {1913})}\BibitemShut {NoStop}%
\bibitem [{\citenamefont {Heisenberg}(1949)}]{HeisenbergBook}%
  \BibitemOpen
  \bibfield  {author} {\bibinfo {author} {\bibfnamefont {W.}~\bibnamefont
  {Heisenberg}},\ }\href {https://books.google.co.uk/books?id=NzMBh4ZxKJsC}
  {\emph {\bibinfo {title} {{The Physical Principles of the Quantum
  Theory}}}},\ Dover Books on Physics and Chemistry\ (\bibinfo  {publisher}
  {Dover Publications},\ \bibinfo {year} {1949})\BibitemShut {NoStop}%
\bibitem [{\citenamefont {Dirac}(1981)}]{DiracBook}%
  \BibitemOpen
  \bibfield  {author} {\bibinfo {author} {\bibfnamefont {P.~A.~M.}\
  \bibnamefont {Dirac}},\ }\href
  {https://books.google.co.uk/books?id=XehUpGiM6FIC} {\emph {\bibinfo {title}
  {{The Principles of Quantum Mechanics}}}},\ International series of
  monographs on physics\ (\bibinfo  {publisher} {Clarendon Press},\ \bibinfo
  {year} {1981})\BibitemShut {NoStop}%
\bibitem [{\citenamefont {Feynman}\ \emph {et~al.}(1963)\citenamefont
  {Feynman}, \citenamefont {Leighton},\ and\ \citenamefont
  {Sands}}]{FeynmanLecturesVol3}%
  \BibitemOpen
  \bibfield  {author} {\bibinfo {author} {\bibfnamefont {R.~P.}\ \bibnamefont
  {Feynman}}, \bibinfo {author} {\bibfnamefont {R.~B.}\ \bibnamefont
  {Leighton}}, \ and\ \bibinfo {author} {\bibfnamefont {M.~L.}\ \bibnamefont
  {Sands}},\ }\href {https://books.google.co.uk/books?id=\_6XvAAAAMAAJ} {\emph
  {\bibinfo {title} {{The Feynman Lectures on Physics}}}},\ \bibinfo {series}
  {The Feynman Lectures on Physics}\ No.\ \bibinfo {number} {v. 3}\ (\bibinfo
  {publisher} {Pearson/Addison-Wesley},\ \bibinfo {year} {1963})\BibitemShut
  {NoStop}%
\bibitem [{\citenamefont {Pauli}(1925{\natexlab{a}})}]{PauliZPhys25a}%
  \BibitemOpen
  \bibfield  {author} {\bibinfo {author} {\bibfnamefont {W.}~\bibnamefont
  {Pauli}},\ }\bibfield  {title} {\enquote {\bibinfo {title} {{{\"U}ber den
  Einflu{\ss} der Geschwindigkeitsabh{\"a}ngigkeit der Elektronenmasse auf den
  Zeemaneffekt}},}\ }\href {\doibase 10.1007/BF02980592} {\bibfield  {journal}
  {\bibinfo  {journal} {Z. Phys.}\ }\textbf {\bibinfo {volume} {31}},\ \bibinfo
  {pages} {373--385} (\bibinfo {year} {1925}{\natexlab{a}})}\BibitemShut
  {NoStop}%
\bibitem [{\citenamefont {Pauli}(1925{\natexlab{b}})}]{PauliZPhys25b}%
  \BibitemOpen
  \bibfield  {author} {\bibinfo {author} {\bibfnamefont {W.}~\bibnamefont
  {Pauli}},\ }\bibfield  {title} {\enquote {\bibinfo {title} {{{\"U}ber den
  Zusammenhang des Abschlusses der Elektronengruppen im Atom mit der
  Komplexstruktur der Spektren}},}\ }\href {\doibase 10.1007/BF02980631}
  {\bibfield  {journal} {\bibinfo  {journal} {Z. Phys.}\ }\textbf {\bibinfo
  {volume} {31}},\ \bibinfo {pages} {765--783} (\bibinfo {year}
  {1925}{\natexlab{b}})}\BibitemShut {NoStop}%
\bibitem [{\citenamefont {Ising}(1925)}]{IsingZPhys25}%
  \BibitemOpen
  \bibfield  {author} {\bibinfo {author} {\bibfnamefont {E.}~\bibnamefont
  {Ising}},\ }\bibfield  {title} {\enquote {\bibinfo {title} {{Beitrag zur
  Theorie des Ferromagnetismus}},}\ }\href {\doibase 10.1007/BF02980577}
  {\bibfield  {journal} {\bibinfo  {journal} {Z. Phys.}\ }\textbf {\bibinfo
  {volume} {31}},\ \bibinfo {pages} {253--258} (\bibinfo {year}
  {1925})}\BibitemShut {NoStop}%
\bibitem [{\citenamefont {Heisenberg}(1928)}]{HeisenbergZPhys28}%
  \BibitemOpen
  \bibfield  {author} {\bibinfo {author} {\bibfnamefont {W.}~\bibnamefont
  {Heisenberg}},\ }\bibfield  {title} {\enquote {\bibinfo {title} {{Zur Theorie
  des Ferromagnetismus}},}\ }\href {\doibase 10.1007/BF01328601} {\bibfield
  {journal} {\bibinfo  {journal} {Z. Phys.}\ }\textbf {\bibinfo {volume}
  {49}},\ \bibinfo {pages} {619--636} (\bibinfo {year} {1928})}\BibitemShut
  {NoStop}%
\bibitem [{\citenamefont {Bethe}(1931)}]{BetheZPhys31}%
  \BibitemOpen
  \bibfield  {author} {\bibinfo {author} {\bibfnamefont {H.}~\bibnamefont
  {Bethe}},\ }\bibfield  {title} {\enquote {\bibinfo {title} {{Zur Theorie der
  Metalle}},}\ }\href {\doibase 10.1007/BF01341708} {\bibfield  {journal}
  {\bibinfo  {journal} {Z. Phys.}\ }\textbf {\bibinfo {volume} {71}},\ \bibinfo
  {pages} {205--226} (\bibinfo {year} {1931})}\BibitemShut {NoStop}%
\bibitem [{\citenamefont {Schr\"odinger}(1926)}]{SchrodingerPR26}%
  \BibitemOpen
  \bibfield  {author} {\bibinfo {author} {\bibfnamefont {E.}~\bibnamefont
  {Schr\"odinger}},\ }\bibfield  {title} {\enquote {\bibinfo {title} {{An
  Undulatory Theory of the Mechanics of Atoms and Molecules}},}\ }\href
  {\doibase 10.1103/PhysRev.28.1049} {\bibfield  {journal} {\bibinfo  {journal}
  {Phys. Rev.}\ }\textbf {\bibinfo {volume} {28}},\ \bibinfo {pages}
  {1049--1070} (\bibinfo {year} {1926})}\BibitemShut {NoStop}%
\bibitem [{\citenamefont {Dirac}(1929)}]{DiracProcRoySocA29}%
  \BibitemOpen
  \bibfield  {author} {\bibinfo {author} {\bibfnamefont {P.~A.~M.}\
  \bibnamefont {Dirac}},\ }\bibfield  {title} {\enquote {\bibinfo {title}
  {{Quantum Mechanics of Many-Electron Systems}},}\ }\href {\doibase
  10.1098/rspa.1929.0094} {\bibfield  {journal} {\bibinfo  {journal} {Proc.
  Roy. Soc. A}\ }\textbf {\bibinfo {volume} {123}},\ \bibinfo {pages}
  {714--733} (\bibinfo {year} {1929})}\BibitemShut {NoStop}%
\bibitem [{\citenamefont {Hylleraas}(1929)}]{HylleraasZPhys29}%
  \BibitemOpen
  \bibfield  {author} {\bibinfo {author} {\bibfnamefont {E.~A.}\ \bibnamefont
  {Hylleraas}},\ }\bibfield  {title} {\enquote {\bibinfo {title} {{Neue
  Berechnung der Energie des Heliums im Grundzustande, sowie des tiefsten Terms
  von Ortho-Helium}},}\ }\href {\doibase 10.1007/BF01375457} {\bibfield
  {journal} {\bibinfo  {journal} {Z. Phys.}\ }\textbf {\bibinfo {volume}
  {54}},\ \bibinfo {pages} {347--366} (\bibinfo {year} {1929})}\BibitemShut
  {NoStop}%
\bibitem [{\citenamefont {Lamb}\ and\ \citenamefont
  {Retherford}(1947)}]{LambPR47}%
  \BibitemOpen
  \bibfield  {author} {\bibinfo {author} {\bibfnamefont {W.~E.}\ \bibnamefont
  {Lamb}}\ and\ \bibinfo {author} {\bibfnamefont {R.~C.}\ \bibnamefont
  {Retherford}},\ }\bibfield  {title} {\enquote {\bibinfo {title} {{Fine
  Structure of the Hydrogen Atom by a Microwave Method}},}\ }\href {\doibase
  10.1103/PhysRev.72.241} {\bibfield  {journal} {\bibinfo  {journal} {Phys.
  Rev.}\ }\textbf {\bibinfo {volume} {72}},\ \bibinfo {pages} {241--243}
  (\bibinfo {year} {1947})}\BibitemShut {NoStop}%
\bibitem [{\citenamefont {Gordon}(1928)}]{GordonZPhys28}%
  \BibitemOpen
  \bibfield  {author} {\bibinfo {author} {\bibfnamefont {W.}~\bibnamefont
  {Gordon}},\ }\bibfield  {title} {\enquote {\bibinfo {title} {{Die
  Energieniveaus des Wasserstoffatoms nach der Diracschen Quantentheorie des
  Elektrons}},}\ }\href {\doibase 10.1007/BF01351570} {\bibfield  {journal}
  {\bibinfo  {journal} {Z. Phys.}\ }\textbf {\bibinfo {volume} {48}},\ \bibinfo
  {pages} {11--14} (\bibinfo {year} {1928})}\BibitemShut {NoStop}%
\bibitem [{\citenamefont {Darwin}(1928)}]{DarwinProcRoySocA28}%
  \BibitemOpen
  \bibfield  {author} {\bibinfo {author} {\bibfnamefont {C.~G.}\ \bibnamefont
  {Darwin}},\ }\bibfield  {title} {\enquote {\bibinfo {title} {{The Wave
  Equations of the Electron}},}\ }\href {\doibase 10.1098/rspa.1928.0076}
  {\bibfield  {journal} {\bibinfo  {journal} {Proc. Roy. Soc. A}\ }\textbf
  {\bibinfo {volume} {118}},\ \bibinfo {pages} {654--680} (\bibinfo {year}
  {1928})}\BibitemShut {NoStop}%
\bibitem [{\citenamefont {Bethe}(1947)}]{BethePR47}%
  \BibitemOpen
  \bibfield  {author} {\bibinfo {author} {\bibfnamefont {H.~A.}\ \bibnamefont
  {Bethe}},\ }\bibfield  {title} {\enquote {\bibinfo {title} {{The
  Electromagnetic Shift of Energy Levels}},}\ }\href {\doibase
  10.1103/PhysRev.72.339} {\bibfield  {journal} {\bibinfo  {journal} {Phys.
  Rev.}\ }\textbf {\bibinfo {volume} {72}},\ \bibinfo {pages} {339--341}
  (\bibinfo {year} {1947})}\BibitemShut {NoStop}%
\bibitem [{\citenamefont {Peskin}\ and\ \citenamefont
  {Schroeder}(1995)}]{PeskinSchroeder}%
  \BibitemOpen
  \bibfield  {author} {\bibinfo {author} {\bibfnamefont {M.~E.}\ \bibnamefont
  {Peskin}}\ and\ \bibinfo {author} {\bibfnamefont {D.~V.}\ \bibnamefont
  {Schroeder}},\ }\href {https://books.google.co.uk/books?id=\_H-oPv1raioC}
  {\emph {\bibinfo {title} {{An Introduction to Quantum Field Theory}}}},\
  Advanced book classics\ (\bibinfo  {publisher} {Addison-Wesley Publishing
  Company},\ \bibinfo {year} {1995})\BibitemShut {NoStop}%
\bibitem [{\citenamefont {Zinn-Justin}(2002)}]{ZinnJustinBook}%
  \BibitemOpen
  \bibfield  {author} {\bibinfo {author} {\bibfnamefont {J.}~\bibnamefont
  {Zinn-Justin}},\ }\href {https://books.google.co.uk/books?id=N8DBpTzBCJYC}
  {\emph {\bibinfo {title} {{Quantum Field Theory and Critical Phenomena}}}},\
  International series of monographs on physics\ (\bibinfo  {publisher}
  {Clarendon Press},\ \bibinfo {year} {2002})\BibitemShut {NoStop}%
\bibitem [{\citenamefont {Srednicki}(2007)}]{SrednickiBook}%
  \BibitemOpen
  \bibfield  {author} {\bibinfo {author} {\bibfnamefont {M.}~\bibnamefont
  {Srednicki}},\ }\href {https://books.google.co.uk/books?id=5OepxIG42B4C}
  {\emph {\bibinfo {title} {{Quantum Field Theory}}}}\ (\bibinfo  {publisher}
  {Cambridge University Press},\ \bibinfo {year} {2007})\BibitemShut {NoStop}%
\bibitem [{\citenamefont {Tsvelik}(2007)}]{TsvelikBook}%
  \BibitemOpen
  \bibfield  {author} {\bibinfo {author} {\bibfnamefont {A.~M.}\ \bibnamefont
  {Tsvelik}},\ }\href
  {http://www.cambridge.org/catalogue/catalogue.asp?isbn=9780521529808} {\emph
  {\bibinfo {title} {{Quantum Field Theory in Condensed Matter Physics}}}},\
  \bibinfo {edition} {second edition}\ ed.\ (\bibinfo  {publisher} {Cambridge
  University Press},\ \bibinfo {address} {Cambridge},\ \bibinfo {year}
  {2007})\BibitemShut {NoStop}%
\bibitem [{\citenamefont {Altland}\ and\ \citenamefont
  {Simons}(2010)}]{AltlandAndSimons}%
  \BibitemOpen
  \bibfield  {author} {\bibinfo {author} {\bibfnamefont {A.}~\bibnamefont
  {Altland}}\ and\ \bibinfo {author} {\bibfnamefont {B.~D.}\ \bibnamefont
  {Simons}},\ }\href {https://books.google.co.uk/books?id=GpF0Pgo8CqAC} {\emph
  {\bibinfo {title} {{Condensed Matter Field Theory}}}},\ Cambridge books
  online\ (\bibinfo  {publisher} {Cambridge University Press},\ \bibinfo {year}
  {2010})\BibitemShut {NoStop}%
\bibitem [{\citenamefont {Mussardo}(2010)}]{MussardoBook}%
  \BibitemOpen
  \bibfield  {author} {\bibinfo {author} {\bibfnamefont {G.}~\bibnamefont
  {Mussardo}},\ }\href {https://books.google.co.uk/books?id=fakVDAAAQBAJ}
  {\emph {\bibinfo {title} {{Statistical Field Theory: An Introduction to
  Exactly Solved Models in Statistical Physics}}}},\ Oxford Graduate Texts\
  (\bibinfo  {publisher} {OUP Oxford},\ \bibinfo {year} {2010})\BibitemShut
  {NoStop}%
\bibitem [{\citenamefont {Jordan}\ and\ \citenamefont
  {Wigner}(1928)}]{JordanZPhys28}%
  \BibitemOpen
  \bibfield  {author} {\bibinfo {author} {\bibfnamefont {P.}~\bibnamefont
  {Jordan}}\ and\ \bibinfo {author} {\bibfnamefont {E.}~\bibnamefont
  {Wigner}},\ }\bibfield  {title} {\enquote {\bibinfo {title} {{{\"U}ber das
  Paulische {\"A}quivalenzverbot}},}\ }\href {\doibase 10.1007/BF01331938}
  {\bibfield  {journal} {\bibinfo  {journal} {Z. Phys.}\ }\textbf {\bibinfo
  {volume} {47}},\ \bibinfo {pages} {631--651} (\bibinfo {year}
  {1928})}\BibitemShut {NoStop}%
\bibitem [{\citenamefont {Orbach}(1958)}]{OrbachPR58}%
  \BibitemOpen
  \bibfield  {author} {\bibinfo {author} {\bibfnamefont {R.}~\bibnamefont
  {Orbach}},\ }\bibfield  {title} {\enquote {\bibinfo {title} {{Linear
  Antiferromagnetic Chain with Anisotropic Coupling}},}\ }\href {\doibase
  10.1103/PhysRev.112.309} {\bibfield  {journal} {\bibinfo  {journal} {Phys.
  Rev.}\ }\textbf {\bibinfo {volume} {112}},\ \bibinfo {pages} {309--316}
  (\bibinfo {year} {1958})}\BibitemShut {NoStop}%
\bibitem [{\citenamefont {Walker}(1959)}]{WalkerPR59}%
  \BibitemOpen
  \bibfield  {author} {\bibinfo {author} {\bibfnamefont {L.~R.}\ \bibnamefont
  {Walker}},\ }\bibfield  {title} {\enquote {\bibinfo {title}
  {{Antiferromagnetic Linear Chain}},}\ }\href {\doibase
  10.1103/PhysRev.116.1089} {\bibfield  {journal} {\bibinfo  {journal} {Phys.
  Rev.}\ }\textbf {\bibinfo {volume} {116}},\ \bibinfo {pages} {1089--1090}
  (\bibinfo {year} {1959})}\BibitemShut {NoStop}%
\bibitem [{\citenamefont {Yang}\ and\ \citenamefont
  {Yang}(1966{\natexlab{a}})}]{YangPR66a}%
  \BibitemOpen
  \bibfield  {author} {\bibinfo {author} {\bibfnamefont {C.~N.}\ \bibnamefont
  {Yang}}\ and\ \bibinfo {author} {\bibfnamefont {C.~P.}\ \bibnamefont
  {Yang}},\ }\bibfield  {title} {\enquote {\bibinfo {title} {{One-Dimensional
  Chain of Anisotropic Spin-Spin Interactions. I. Proof of Bethe's Hypothesis
  for Ground State in a Finite System}},}\ }\href {\doibase
  10.1103/PhysRev.150.321} {\bibfield  {journal} {\bibinfo  {journal} {Phys.
  Rev.}\ }\textbf {\bibinfo {volume} {150}},\ \bibinfo {pages} {321--327}
  (\bibinfo {year} {1966}{\natexlab{a}})}\BibitemShut {NoStop}%
\bibitem [{\citenamefont {Yang}\ and\ \citenamefont
  {Yang}(1966{\natexlab{b}})}]{YangPR66b}%
  \BibitemOpen
  \bibfield  {author} {\bibinfo {author} {\bibfnamefont {C.~N.}\ \bibnamefont
  {Yang}}\ and\ \bibinfo {author} {\bibfnamefont {C.~P.}\ \bibnamefont
  {Yang}},\ }\bibfield  {title} {\enquote {\bibinfo {title} {{One-Dimensional
  Chain of Anisotropic Spin-Spin Interactions. II. Properties of the
  Ground-State Energy Per Lattice Site for an Infinite System}},}\ }\href
  {\doibase 10.1103/PhysRev.150.327} {\bibfield  {journal} {\bibinfo  {journal}
  {Phys. Rev.}\ }\textbf {\bibinfo {volume} {150}},\ \bibinfo {pages}
  {327--339} (\bibinfo {year} {1966}{\natexlab{b}})}\BibitemShut {NoStop}%
\bibitem [{\citenamefont {Yang}\ and\ \citenamefont
  {Yang}(1966{\natexlab{c}})}]{YangPR66c}%
  \BibitemOpen
  \bibfield  {author} {\bibinfo {author} {\bibfnamefont {C.~N.}\ \bibnamefont
  {Yang}}\ and\ \bibinfo {author} {\bibfnamefont {C.~P.}\ \bibnamefont
  {Yang}},\ }\bibfield  {title} {\enquote {\bibinfo {title} {{One-Dimensional
  Chain of Anisotropic Spin-Spin Interactions. III. Applications}},}\ }\href
  {\doibase 10.1103/PhysRev.151.258} {\bibfield  {journal} {\bibinfo  {journal}
  {Phys. Rev.}\ }\textbf {\bibinfo {volume} {151}},\ \bibinfo {pages}
  {258--264} (\bibinfo {year} {1966}{\natexlab{c}})}\BibitemShut {NoStop}%
\bibitem [{\citenamefont {Sutherland}\ \emph {et~al.}(1967)\citenamefont
  {Sutherland}, \citenamefont {Yang},\ and\ \citenamefont
  {Yang}}]{SutherlandPRL67}%
  \BibitemOpen
  \bibfield  {author} {\bibinfo {author} {\bibfnamefont {B.}~\bibnamefont
  {Sutherland}}, \bibinfo {author} {\bibfnamefont {C.~N.}\ \bibnamefont
  {Yang}}, \ and\ \bibinfo {author} {\bibfnamefont {C.~P.}\ \bibnamefont
  {Yang}},\ }\bibfield  {title} {\enquote {\bibinfo {title} {{Exact Solution of
  a Model of Two-Dimensional Ferroelectrics in an Arbitrary External Electric
  Field}},}\ }\href {\doibase 10.1103/PhysRevLett.19.588} {\bibfield  {journal}
  {\bibinfo  {journal} {Phys. Rev. Lett.}\ }\textbf {\bibinfo {volume} {19}},\
  \bibinfo {pages} {588--591} (\bibinfo {year} {1967})}\BibitemShut {NoStop}%
\bibitem [{\citenamefont {Baxter}(1971)}]{BaxterStudApplMath71}%
  \BibitemOpen
  \bibfield  {author} {\bibinfo {author} {\bibfnamefont {R.~J.}\ \bibnamefont
  {Baxter}},\ }\bibfield  {title} {\enquote {\bibinfo {title} {{Generalized
  Ferroelectric Model on a Square Lattice}},}\ }\href {\doibase
  10.1002/sapm197150151} {\bibfield  {journal} {\bibinfo  {journal} {Stud.
  Appl. Math.}\ }\textbf {\bibinfo {volume} {50}},\ \bibinfo {pages} {51--69}
  (\bibinfo {year} {1971})}\BibitemShut {NoStop}%
\bibitem [{\citenamefont {Baxter}(1972)}]{BaxterAnnPhys72}%
  \BibitemOpen
  \bibfield  {author} {\bibinfo {author} {\bibfnamefont {R.~J.}\ \bibnamefont
  {Baxter}},\ }\bibfield  {title} {\enquote {\bibinfo {title} {{Partition
  function of the Eight-Vertex lattice model}},}\ }\href {\doibase
  http://dx.doi.org/10.1016/0003-4916(72)90335-1} {\bibfield  {journal}
  {\bibinfo  {journal} {Ann. Phys. (N.Y.)}\ }\textbf {\bibinfo {volume} {70}},\
  \bibinfo {pages} {193 -- 228} (\bibinfo {year} {1972})}\BibitemShut {NoStop}%
\bibitem [{\citenamefont {Baxter}(1973{\natexlab{a}})}]{BaxterAnnPhys73a}%
  \BibitemOpen
  \bibfield  {author} {\bibinfo {author} {\bibfnamefont {R.}~\bibnamefont
  {Baxter}},\ }\bibfield  {title} {\enquote {\bibinfo {title} {{Eight-vertex
  model in lattice statistics and one-dimensional anisotropic Heisenberg chain.
  I. Some fundamental eigenvectors}},}\ }\href {\doibase
  http://dx.doi.org/10.1016/0003-4916(73)90439-9} {\bibfield  {journal}
  {\bibinfo  {journal} {Ann. Phys. (N.Y.)}\ }\textbf {\bibinfo {volume} {76}},\
  \bibinfo {pages} {1 -- 24} (\bibinfo {year}
  {1973}{\natexlab{a}})}\BibitemShut {NoStop}%
\bibitem [{\citenamefont {Baxter}(1973{\natexlab{b}})}]{BaxterAnnPhys73b}%
  \BibitemOpen
  \bibfield  {author} {\bibinfo {author} {\bibfnamefont {R.}~\bibnamefont
  {Baxter}},\ }\bibfield  {title} {\enquote {\bibinfo {title} {{Eight-vertex
  model in lattice statistics and one-dimensional anisotropic Heisenberg chain.
  II. Equivalence to a generalized ice-type lattice model}},}\ }\href {\doibase
  http://dx.doi.org/10.1016/0003-4916(73)90440-5} {\bibfield  {journal}
  {\bibinfo  {journal} {Ann. Phys. (N.Y.)}\ }\textbf {\bibinfo {volume} {76}},\
  \bibinfo {pages} {25 -- 47} (\bibinfo {year}
  {1973}{\natexlab{b}})}\BibitemShut {NoStop}%
\bibitem [{\citenamefont {Baxter}(1973{\natexlab{c}})}]{BaxterAnnPhys73c}%
  \BibitemOpen
  \bibfield  {author} {\bibinfo {author} {\bibfnamefont {R.}~\bibnamefont
  {Baxter}},\ }\bibfield  {title} {\enquote {\bibinfo {title} {{Eight-vertex
  model in lattice statistics and one-dimensional anisotropic Heisenberg chain.
  III. Eigenvectors of the transfer matrix and hamiltonian}},}\ }\href
  {\doibase http://dx.doi.org/10.1016/0003-4916(73)90441-7} {\bibfield
  {journal} {\bibinfo  {journal} {Ann. Phys. (N.Y.)}\ }\textbf {\bibinfo
  {volume} {76}},\ \bibinfo {pages} {48 -- 71} (\bibinfo {year}
  {1973}{\natexlab{c}})}\BibitemShut {NoStop}%
\bibitem [{\citenamefont {Lieb}\ and\ \citenamefont
  {Liniger}(1963)}]{LiebPR63a}%
  \BibitemOpen
  \bibfield  {author} {\bibinfo {author} {\bibfnamefont {E.~H.}\ \bibnamefont
  {Lieb}}\ and\ \bibinfo {author} {\bibfnamefont {W.}~\bibnamefont {Liniger}},\
  }\bibfield  {title} {\enquote {\bibinfo {title} {{Exact Analysis of an
  Interacting Bose Gas. I. The General Solution and the Ground State}},}\
  }\href {\doibase 10.1103/PhysRev.130.1605} {\bibfield  {journal} {\bibinfo
  {journal} {Phys. Rev.}\ }\textbf {\bibinfo {volume} {130}},\ \bibinfo {pages}
  {1605--1616} (\bibinfo {year} {1963})}\BibitemShut {NoStop}%
\bibitem [{\citenamefont {Lieb}(1963)}]{LiebPR63b}%
  \BibitemOpen
  \bibfield  {author} {\bibinfo {author} {\bibfnamefont {E.~H.}\ \bibnamefont
  {Lieb}},\ }\bibfield  {title} {\enquote {\bibinfo {title} {{Exact Analysis of
  an Interacting Bose Gas. II. The Excitation Spectrum}},}\ }\href {\doibase
  10.1103/PhysRev.130.1616} {\bibfield  {journal} {\bibinfo  {journal} {Phys.
  Rev.}\ }\textbf {\bibinfo {volume} {130}},\ \bibinfo {pages} {1616--1624}
  (\bibinfo {year} {1963})}\BibitemShut {NoStop}%
\bibitem [{\citenamefont {Bergknoff}\ and\ \citenamefont
  {Thacker}(1979{\natexlab{a}})}]{BergknoffPRL79}%
  \BibitemOpen
  \bibfield  {author} {\bibinfo {author} {\bibfnamefont {H.}~\bibnamefont
  {Bergknoff}}\ and\ \bibinfo {author} {\bibfnamefont {H.~B.}\ \bibnamefont
  {Thacker}},\ }\bibfield  {title} {\enquote {\bibinfo {title} {{Method for
  Solving the Massive Thirring Model}},}\ }\href {\doibase
  10.1103/PhysRevLett.42.135} {\bibfield  {journal} {\bibinfo  {journal} {Phys.
  Rev. Lett.}\ }\textbf {\bibinfo {volume} {42}},\ \bibinfo {pages} {135--138}
  (\bibinfo {year} {1979}{\natexlab{a}})}\BibitemShut {NoStop}%
\bibitem [{\citenamefont {Bergknoff}\ and\ \citenamefont
  {Thacker}(1979{\natexlab{b}})}]{BergknoffPRD79}%
  \BibitemOpen
  \bibfield  {author} {\bibinfo {author} {\bibfnamefont {H.}~\bibnamefont
  {Bergknoff}}\ and\ \bibinfo {author} {\bibfnamefont {H.~B.}\ \bibnamefont
  {Thacker}},\ }\bibfield  {title} {\enquote {\bibinfo {title} {{Structure and
  solution of the massive Thirring model}},}\ }\href {\doibase
  10.1103/PhysRevD.19.3666} {\bibfield  {journal} {\bibinfo  {journal} {Phys.
  Rev. D}\ }\textbf {\bibinfo {volume} {19}},\ \bibinfo {pages} {3666--3681}
  (\bibinfo {year} {1979}{\natexlab{b}})}\BibitemShut {NoStop}%
\bibitem [{\citenamefont {Sklyanin}\ \emph {et~al.}(1979)\citenamefont
  {Sklyanin}, \citenamefont {Takhtadzhyan},\ and\ \citenamefont
  {Faddeev}}]{SklyaninTheorMathPhys79}%
  \BibitemOpen
  \bibfield  {author} {\bibinfo {author} {\bibfnamefont {E.~K.}\ \bibnamefont
  {Sklyanin}}, \bibinfo {author} {\bibfnamefont {L.~A.}\ \bibnamefont
  {Takhtadzhyan}}, \ and\ \bibinfo {author} {\bibfnamefont {L.~D.}\
  \bibnamefont {Faddeev}},\ }\bibfield  {title} {\enquote {\bibinfo {title}
  {{Quantum inverse problem method. I}},}\ }\href {\doibase 10.1007/BF01018718}
  {\bibfield  {journal} {\bibinfo  {journal} {Theor. Math. Phys.}\ }\textbf
  {\bibinfo {volume} {40}},\ \bibinfo {pages} {688--706} (\bibinfo {year}
  {1979})}\BibitemShut {NoStop}%
\bibitem [{\citenamefont {Andrei}\ and\ \citenamefont
  {Lowenstein}(1979)}]{AndreiPRL79}%
  \BibitemOpen
  \bibfield  {author} {\bibinfo {author} {\bibfnamefont {N.}~\bibnamefont
  {Andrei}}\ and\ \bibinfo {author} {\bibfnamefont {J.~H.}\ \bibnamefont
  {Lowenstein}},\ }\bibfield  {title} {\enquote {\bibinfo {title}
  {{Diagonalization of the Chiral-Invariant Gross-Neveu Hamiltonian}},}\ }\href
  {\doibase 10.1103/PhysRevLett.43.1698} {\bibfield  {journal} {\bibinfo
  {journal} {Phys. Rev. Lett.}\ }\textbf {\bibinfo {volume} {43}},\ \bibinfo
  {pages} {1698--1701} (\bibinfo {year} {1979})}\BibitemShut {NoStop}%
\bibitem [{\citenamefont {Andrei}\ and\ \citenamefont
  {Lowenstein}(1980)}]{AndreiPhysLettB80}%
  \BibitemOpen
  \bibfield  {author} {\bibinfo {author} {\bibfnamefont {N.}~\bibnamefont
  {Andrei}}\ and\ \bibinfo {author} {\bibfnamefont {J.~H.}\ \bibnamefont
  {Lowenstein}},\ }\bibfield  {title} {\enquote {\bibinfo {title} {{Derivation
  of the chiral Gross-Neveu spectrum for arbitrary SU(N) symmetry}},}\ }\href
  {\doibase http://dx.doi.org/10.1016/0370-2693(80)90061-1} {\bibfield
  {journal} {\bibinfo  {journal} {Phys. Lett. B}\ }\textbf {\bibinfo {volume}
  {90}},\ \bibinfo {pages} {106 -- 110} (\bibinfo {year} {1980})}\BibitemShut
  {NoStop}%
\bibitem [{\citenamefont {Belavin}(1979)}]{BelavinPhysLettB79}%
  \BibitemOpen
  \bibfield  {author} {\bibinfo {author} {\bibfnamefont {A.~A.}\ \bibnamefont
  {Belavin}},\ }\bibfield  {title} {\enquote {\bibinfo {title} {{Exact solution
  of the two-dimensional model with asymptotic freedom}},}\ }\href {\doibase
  http://dx.doi.org/10.1016/0370-2693(79)90033-9} {\bibfield  {journal}
  {\bibinfo  {journal} {Phys. Lett. B}\ }\textbf {\bibinfo {volume} {87}},\
  \bibinfo {pages} {117 -- 121} (\bibinfo {year} {1979})}\BibitemShut {NoStop}%
\bibitem [{\citenamefont {Hofferberth}\ \emph {et~al.}(2007)\citenamefont
  {Hofferberth}, \citenamefont {Lesanovsky}, \citenamefont {Fischer},
  \citenamefont {Schumm},\ and\ \citenamefont
  {Schmiedmayer}}]{HofferberthNature07}%
  \BibitemOpen
  \bibfield  {author} {\bibinfo {author} {\bibfnamefont {S.}~\bibnamefont
  {Hofferberth}}, \bibinfo {author} {\bibfnamefont {I.}~\bibnamefont
  {Lesanovsky}}, \bibinfo {author} {\bibfnamefont {B.}~\bibnamefont {Fischer}},
  \bibinfo {author} {\bibfnamefont {T.}~\bibnamefont {Schumm}}, \ and\ \bibinfo
  {author} {\bibfnamefont {J.}~\bibnamefont {Schmiedmayer}},\ }\bibfield
  {title} {\enquote {\bibinfo {title} {{Non-equilibrium coherence dynamics in
  one-dimensional Bose gases}},}\ }\href
  {http://dx.doi.org/10.1038/nature06149} {\bibfield  {journal} {\bibinfo
  {journal} {Nature}\ }\textbf {\bibinfo {volume} {449}},\ \bibinfo {pages}
  {324--327} (\bibinfo {year} {2007})}\BibitemShut {NoStop}%
\bibitem [{\citenamefont {Coldea}\ \emph
  {et~al.}(2010{\natexlab{a}})\citenamefont {Coldea}, \citenamefont {Tennant},
  \citenamefont {Wheeler}, \citenamefont {Wawrzynska}, \citenamefont
  {Prabhakaran}, \citenamefont {Telling}, \citenamefont {Habicht},
  \citenamefont {Smeibidl},\ and\ \citenamefont {Kiefer}}]{ColdeaScience10}%
  \BibitemOpen
  \bibfield  {author} {\bibinfo {author} {\bibfnamefont {R.}~\bibnamefont
  {Coldea}}, \bibinfo {author} {\bibfnamefont {D.~A.}\ \bibnamefont {Tennant}},
  \bibinfo {author} {\bibfnamefont {E.~M.}\ \bibnamefont {Wheeler}}, \bibinfo
  {author} {\bibfnamefont {E.}~\bibnamefont {Wawrzynska}}, \bibinfo {author}
  {\bibfnamefont {D.}~\bibnamefont {Prabhakaran}}, \bibinfo {author}
  {\bibfnamefont {M.}~\bibnamefont {Telling}}, \bibinfo {author} {\bibfnamefont
  {K.}~\bibnamefont {Habicht}}, \bibinfo {author} {\bibfnamefont
  {P.}~\bibnamefont {Smeibidl}}, \ and\ \bibinfo {author} {\bibfnamefont
  {K.}~\bibnamefont {Kiefer}},\ }\bibfield  {title} {\enquote {\bibinfo {title}
  {{Quantum Criticality in an Ising Chain: Experimental Evidence for Emergent
  E8 Symmetry}},}\ }\href {\doibase 10.1126/science.1180085} {\bibfield
  {journal} {\bibinfo  {journal} {Science}\ }\textbf {\bibinfo {volume}
  {327}},\ \bibinfo {pages} {177--180} (\bibinfo {year}
  {2010}{\natexlab{a}})}\BibitemShut {NoStop}%
\bibitem [{\citenamefont {Lake}\ \emph {et~al.}(2013)\citenamefont {Lake},
  \citenamefont {Tennant}, \citenamefont {Caux}, \citenamefont {Barthel},
  \citenamefont {Schollw\"ock}, \citenamefont {Nagler},\ and\ \citenamefont
  {Frost}}]{LakePRL13}%
  \BibitemOpen
  \bibfield  {author} {\bibinfo {author} {\bibfnamefont {B.}~\bibnamefont
  {Lake}}, \bibinfo {author} {\bibfnamefont {D.~A.}\ \bibnamefont {Tennant}},
  \bibinfo {author} {\bibfnamefont {J.-S.}\ \bibnamefont {Caux}}, \bibinfo
  {author} {\bibfnamefont {T.}~\bibnamefont {Barthel}}, \bibinfo {author}
  {\bibfnamefont {U.}~\bibnamefont {Schollw\"ock}}, \bibinfo {author}
  {\bibfnamefont {S.~E.}\ \bibnamefont {Nagler}}, \ and\ \bibinfo {author}
  {\bibfnamefont {C.~D.}\ \bibnamefont {Frost}},\ }\bibfield  {title} {\enquote
  {\bibinfo {title} {{Multispinon Continua at Zero and Finite Temperature in a
  Near-Ideal Heisenberg Chain}},}\ }\href {\doibase
  10.1103/PhysRevLett.111.137205} {\bibfield  {journal} {\bibinfo  {journal}
  {Phys. Rev. Lett.}\ }\textbf {\bibinfo {volume} {111}},\ \bibinfo {pages}
  {137205} (\bibinfo {year} {2013})}\BibitemShut {NoStop}%
\bibitem [{\citenamefont {Mourigal}\ \emph {et~al.}(2013)\citenamefont
  {Mourigal}, \citenamefont {Enderle}, \citenamefont {Klopperpieper},
  \citenamefont {Caux}, \citenamefont {Stunault},\ and\ \citenamefont
  {Ronnow}}]{MourigalNaturePhys13}%
  \BibitemOpen
  \bibfield  {author} {\bibinfo {author} {\bibfnamefont {M.}~\bibnamefont
  {Mourigal}}, \bibinfo {author} {\bibfnamefont {M.}~\bibnamefont {Enderle}},
  \bibinfo {author} {\bibfnamefont {A.}~\bibnamefont {Klopperpieper}}, \bibinfo
  {author} {\bibfnamefont {J.-S.}\ \bibnamefont {Caux}}, \bibinfo {author}
  {\bibfnamefont {A.}~\bibnamefont {Stunault}}, \ and\ \bibinfo {author}
  {\bibfnamefont {H.~M.}\ \bibnamefont {Ronnow}},\ }\bibfield  {title}
  {\enquote {\bibinfo {title} {{Fractional spinon excitations in the quantum
  Heisenberg antiferromagnetic chain}},}\ }\href
  {http://dx.doi.org/10.1038/nphys2652} {\bibfield  {journal} {\bibinfo
  {journal} {Nature Phys.}\ }\textbf {\bibinfo {volume} {9}},\ \bibinfo {pages}
  {435--441} (\bibinfo {year} {2013})}\BibitemShut {NoStop}%
\bibitem [{\citenamefont {Guan}\ \emph {et~al.}(2013)\citenamefont {Guan},
  \citenamefont {Batchelor},\ and\ \citenamefont {Lee}}]{GuanRMP13}%
  \BibitemOpen
  \bibfield  {author} {\bibinfo {author} {\bibfnamefont {X.-W.}\ \bibnamefont
  {Guan}}, \bibinfo {author} {\bibfnamefont {M.~T.}\ \bibnamefont {Batchelor}},
  \ and\ \bibinfo {author} {\bibfnamefont {C.}~\bibnamefont {Lee}},\ }\bibfield
   {title} {\enquote {\bibinfo {title} {{Fermi gases in one dimension: From
  Bethe ansatz to experiments}},}\ }\href {\doibase 10.1103/RevModPhys.85.1633}
  {\bibfield  {journal} {\bibinfo  {journal} {Rev. Mod. Phys.}\ }\textbf
  {\bibinfo {volume} {85}},\ \bibinfo {pages} {1633--1691} (\bibinfo {year}
  {2013})}\BibitemShut {NoStop}%
\bibitem [{\citenamefont {Mattis}(1974)}]{MattisJMathPhys74}%
  \BibitemOpen
  \bibfield  {author} {\bibinfo {author} {\bibfnamefont {D.~C.}\ \bibnamefont
  {Mattis}},\ }\bibfield  {title} {\enquote {\bibinfo {title} {{New
  wave-€operator identity applied to the study of persistent currents in
  1D}},}\ }\href {\doibase http://dx.doi.org/10.1063/1.1666693} {\bibfield
  {journal} {\bibinfo  {journal} {J. Math. Phys.}\ }\textbf {\bibinfo {volume}
  {15}},\ \bibinfo {pages} {609--612} (\bibinfo {year} {1974})}\BibitemShut
  {NoStop}%
\bibitem [{\citenamefont {Luther}\ and\ \citenamefont
  {Peschel}(1974)}]{LutherPRB74}%
  \BibitemOpen
  \bibfield  {author} {\bibinfo {author} {\bibfnamefont {A.}~\bibnamefont
  {Luther}}\ and\ \bibinfo {author} {\bibfnamefont {I.}~\bibnamefont
  {Peschel}},\ }\bibfield  {title} {\enquote {\bibinfo {title}
  {{Single-particle states, Kohn anomaly, and pairing fluctuations in one
  dimension}},}\ }\href {\doibase 10.1103/PhysRevB.9.2911} {\bibfield
  {journal} {\bibinfo  {journal} {Phys. Rev. B}\ }\textbf {\bibinfo {volume}
  {9}},\ \bibinfo {pages} {2911--2919} (\bibinfo {year} {1974})}\BibitemShut
  {NoStop}%
\bibitem [{\citenamefont {Coleman}(1975)}]{ColemanPRD75}%
  \BibitemOpen
  \bibfield  {author} {\bibinfo {author} {\bibfnamefont {S.}~\bibnamefont
  {Coleman}},\ }\bibfield  {title} {\enquote {\bibinfo {title} {{Quantum
  sine-Gordon equation as the massive Thirring model}},}\ }\href {\doibase
  10.1103/PhysRevD.11.2088} {\bibfield  {journal} {\bibinfo  {journal} {Phys.
  Rev. D}\ }\textbf {\bibinfo {volume} {11}},\ \bibinfo {pages} {2088--2097}
  (\bibinfo {year} {1975})}\BibitemShut {NoStop}%
\bibitem [{\citenamefont {Mandelstam}(1975)}]{MandelstamPRD75}%
  \BibitemOpen
  \bibfield  {author} {\bibinfo {author} {\bibfnamefont {S.}~\bibnamefont
  {Mandelstam}},\ }\bibfield  {title} {\enquote {\bibinfo {title} {{Soliton
  operators for the quantized sine-Gordon equation}},}\ }\href {\doibase
  10.1103/PhysRevD.11.3026} {\bibfield  {journal} {\bibinfo  {journal} {Phys.
  Rev. D}\ }\textbf {\bibinfo {volume} {11}},\ \bibinfo {pages} {3026--3030}
  (\bibinfo {year} {1975})}\BibitemShut {NoStop}%
\bibitem [{\citenamefont {Tomonaga}(1950)}]{TomonagaPTP50}%
  \BibitemOpen
  \bibfield  {author} {\bibinfo {author} {\bibfnamefont {S.-I.}\ \bibnamefont
  {Tomonaga}},\ }\bibfield  {title} {\enquote {\bibinfo {title} {{Remarks on
  Bloch's Method of Sound Waves applied to Many-Fermion Problems}},}\ }\href
  {\doibase 10.1143/ptp/5.4.544} {\bibfield  {journal} {\bibinfo  {journal}
  {Prog. Theor. Phys.}\ }\textbf {\bibinfo {volume} {5}},\ \bibinfo {pages}
  {544--569} (\bibinfo {year} {1950})}\BibitemShut {NoStop}%
\bibitem [{\citenamefont {Mattis}\ and\ \citenamefont
  {Lieb}(1965)}]{MattisJMathPhys65}%
  \BibitemOpen
  \bibfield  {author} {\bibinfo {author} {\bibfnamefont {D.~C.}\ \bibnamefont
  {Mattis}}\ and\ \bibinfo {author} {\bibfnamefont {E.~H.}\ \bibnamefont
  {Lieb}},\ }\bibfield  {title} {\enquote {\bibinfo {title} {{Exact Solution of
  a Many-Fermion System and Its Associated Boson Field}},}\ }\href {\doibase
  http://dx.doi.org/10.1063/1.1704281} {\bibfield  {journal} {\bibinfo
  {journal} {J. Math. Phys.}\ }\textbf {\bibinfo {volume} {6}},\ \bibinfo
  {pages} {304--312} (\bibinfo {year} {1965})}\BibitemShut {NoStop}%
\bibitem [{\citenamefont {Luttinger}(1963)}]{LuttingerJMathPhys63}%
  \BibitemOpen
  \bibfield  {author} {\bibinfo {author} {\bibfnamefont {J.~M.}\ \bibnamefont
  {Luttinger}},\ }\bibfield  {title} {\enquote {\bibinfo {title} {{An Exactly
  Soluble Model of a Many-€Fermion System}},}\ }\href {\doibase
  http://dx.doi.org/10.1063/1.1704046} {\bibfield  {journal} {\bibinfo
  {journal} {J. Math. Phys.}\ }\textbf {\bibinfo {volume} {4}},\ \bibinfo
  {pages} {1154--1162} (\bibinfo {year} {1963})}\BibitemShut {NoStop}%
\bibitem [{\citenamefont {Bloch}(1933)}]{BlochZPhys33}%
  \BibitemOpen
  \bibfield  {author} {\bibinfo {author} {\bibfnamefont {F.}~\bibnamefont
  {Bloch}},\ }\bibfield  {title} {\enquote {\bibinfo {title}
  {{Bremsverm{\"o}gen von Atomen mit mehreren Elektronen}},}\ }\href {\doibase
  10.1007/BF01344553} {\bibfield  {journal} {\bibinfo  {journal} {Z. Phys.}\
  }\textbf {\bibinfo {volume} {81}},\ \bibinfo {pages} {363--376} (\bibinfo
  {year} {1933})}\BibitemShut {NoStop}%
\bibitem [{\citenamefont {Bloch}(1934)}]{BlochHelvPhysActa34}%
  \BibitemOpen
  \bibfield  {author} {\bibinfo {author} {\bibfnamefont {F.}~\bibnamefont
  {Bloch}},\ }\bibfield  {title} {\enquote {\bibinfo {title} {{Inkoh\"arente
  R\"ontgenstreuung und Dichteschwankungen eines entarteten Fermigases}},}\
  }\href {http://www.e-periodica.ch/digbib/view?pid=hpa-001:1934:7#387}
  {\bibfield  {journal} {\bibinfo  {journal} {Helv. Acta. Phys.}\ }\textbf
  {\bibinfo {volume} {7}},\ \bibinfo {pages} {385--405} (\bibinfo {year}
  {1934})}\BibitemShut {NoStop}%
\bibitem [{\citenamefont {Haldane}(1981)}]{HaldaneJPhysC81}%
  \BibitemOpen
  \bibfield  {author} {\bibinfo {author} {\bibfnamefont {F.~D.~M.}\
  \bibnamefont {Haldane}},\ }\bibfield  {title} {\enquote {\bibinfo {title}
  {{'Luttinger liquid theory' of one-dimensional quantum fluids. I. Properties
  of the Luttinger model and their extension to the general 1D interacting
  spinless Fermi gas}},}\ }\href
  {http://stacks.iop.org/0022-3719/14/i=19/a=010} {\bibfield  {journal}
  {\bibinfo  {journal} {J. Phys. C}\ }\textbf {\bibinfo {volume} {14}},\
  \bibinfo {pages} {2585} (\bibinfo {year} {1981})}\BibitemShut {NoStop}%
\bibitem [{\citenamefont {Giamarchi}(2003)}]{GiamarchiBook}%
  \BibitemOpen
  \bibfield  {author} {\bibinfo {author} {\bibfnamefont {T.}~\bibnamefont
  {Giamarchi}},\ }\href
  {https://global.oup.com/academic/product/quantum-physics-in-one-dimension-9780198525004?cc=us&lang=en&}
  {\emph {\bibinfo {title} {{Quantum Physics In One Dimension}}}}\ (\bibinfo
  {publisher} {Oxford University Press},\ \bibinfo {address} {Oxford},\
  \bibinfo {year} {2003})\BibitemShut {NoStop}%
\bibitem [{\citenamefont {Gogolin}\ \emph {et~al.}(1998)\citenamefont
  {Gogolin}, \citenamefont {Nersesyan},\ and\ \citenamefont
  {Tsvelik}}]{GNTBook}%
  \BibitemOpen
  \bibfield  {author} {\bibinfo {author} {\bibfnamefont {A.~O.}\ \bibnamefont
  {Gogolin}}, \bibinfo {author} {\bibfnamefont {A.~A.}\ \bibnamefont
  {Nersesyan}}, \ and\ \bibinfo {author} {\bibfnamefont {A.~M.}\ \bibnamefont
  {Tsvelik}},\ }\href
  {http://www.cambridge.org/catalogue/catalogue.asp?isbn=9780521617192} {\emph
  {\bibinfo {title} {{Bosonization and Strongly Correlated Systems}}}}\
  (\bibinfo  {publisher} {Cambridge University Press},\ \bibinfo {address}
  {Cambridge},\ \bibinfo {year} {1998})\BibitemShut {NoStop}%
\bibitem [{\citenamefont {Fradkin}(2013)}]{FradkinBook}%
  \BibitemOpen
  \bibfield  {author} {\bibinfo {author} {\bibfnamefont {E.}~\bibnamefont
  {Fradkin}},\ }\href
  {http://admin.cambridge.org/nf/academic/subjects/physics/condensed-matter-physics-nanoscience-and-mesoscopic-physics/field-theories-condensed-matter-physics-2nd-edition}
  {\emph {\bibinfo {title} {{Field Theories of Condensed Matter Physics}}}},\
  \bibinfo {edition} {second edition}\ ed.\ (\bibinfo  {publisher} {Cambridge
  University Press},\ \bibinfo {address} {Cambridge},\ \bibinfo {year}
  {2013})\BibitemShut {NoStop}%
\bibitem [{\citenamefont {Kim}\ \emph {et~al.}(1996)\citenamefont {Kim},
  \citenamefont {Matsuura}, \citenamefont {Shen}, \citenamefont {Motoyama},
  \citenamefont {Eisaki}, \citenamefont {Uchida}, \citenamefont {Tohyama},\
  and\ \citenamefont {Maekawa}}]{KimPRL96}%
  \BibitemOpen
  \bibfield  {author} {\bibinfo {author} {\bibfnamefont {C.}~\bibnamefont
  {Kim}}, \bibinfo {author} {\bibfnamefont {A.~Y.}\ \bibnamefont {Matsuura}},
  \bibinfo {author} {\bibfnamefont {Z.-X.}\ \bibnamefont {Shen}}, \bibinfo
  {author} {\bibfnamefont {N.}~\bibnamefont {Motoyama}}, \bibinfo {author}
  {\bibfnamefont {H.}~\bibnamefont {Eisaki}}, \bibinfo {author} {\bibfnamefont
  {S.}~\bibnamefont {Uchida}}, \bibinfo {author} {\bibfnamefont
  {T.}~\bibnamefont {Tohyama}}, \ and\ \bibinfo {author} {\bibfnamefont
  {S.}~\bibnamefont {Maekawa}},\ }\bibfield  {title} {\enquote {\bibinfo
  {title} {{Observation of Spin-Charge Separation in One-Dimensional
  SrCu${\mathrm{O}}_{2}$}},}\ }\href {\doibase 10.1103/PhysRevLett.77.4054}
  {\bibfield  {journal} {\bibinfo  {journal} {Phys. Rev. Lett.}\ }\textbf
  {\bibinfo {volume} {77}},\ \bibinfo {pages} {4054--4057} (\bibinfo {year}
  {1996})}\BibitemShut {NoStop}%
\bibitem [{\citenamefont {Claessen}\ \emph {et~al.}(2002)\citenamefont
  {Claessen}, \citenamefont {Sing}, \citenamefont {Schwingenschl\"ogl},
  \citenamefont {Blaha}, \citenamefont {Dressel},\ and\ \citenamefont
  {Jacobsen}}]{ClaessenPRL02}%
  \BibitemOpen
  \bibfield  {author} {\bibinfo {author} {\bibfnamefont {R.}~\bibnamefont
  {Claessen}}, \bibinfo {author} {\bibfnamefont {M.}~\bibnamefont {Sing}},
  \bibinfo {author} {\bibfnamefont {U.}~\bibnamefont {Schwingenschl\"ogl}},
  \bibinfo {author} {\bibfnamefont {P.}~\bibnamefont {Blaha}}, \bibinfo
  {author} {\bibfnamefont {M.}~\bibnamefont {Dressel}}, \ and\ \bibinfo
  {author} {\bibfnamefont {C.~S.}\ \bibnamefont {Jacobsen}},\ }\bibfield
  {title} {\enquote {\bibinfo {title} {{Spectroscopic Signatures of Spin-Charge
  Separation in the Quasi-One-Dimensional Organic Conductor TTF-TCNQ}},}\
  }\href {\doibase 10.1103/PhysRevLett.88.096402} {\bibfield  {journal}
  {\bibinfo  {journal} {Phys. Rev. Lett.}\ }\textbf {\bibinfo {volume} {88}},\
  \bibinfo {pages} {096402} (\bibinfo {year} {2002})}\BibitemShut {NoStop}%
\bibitem [{\citenamefont {Auslaender}\ \emph {et~al.}(2005)\citenamefont
  {Auslaender}, \citenamefont {Steinberg}, \citenamefont {Yacoby},
  \citenamefont {Tserkovnyak}, \citenamefont {Halperin}, \citenamefont
  {Baldwin}, \citenamefont {Pfeiffer},\ and\ \citenamefont
  {West}}]{AuslaenderScience05}%
  \BibitemOpen
  \bibfield  {author} {\bibinfo {author} {\bibfnamefont {O.~M.}\ \bibnamefont
  {Auslaender}}, \bibinfo {author} {\bibfnamefont {H.}~\bibnamefont
  {Steinberg}}, \bibinfo {author} {\bibfnamefont {A.}~\bibnamefont {Yacoby}},
  \bibinfo {author} {\bibfnamefont {Y.}~\bibnamefont {Tserkovnyak}}, \bibinfo
  {author} {\bibfnamefont {B.~I.}\ \bibnamefont {Halperin}}, \bibinfo {author}
  {\bibfnamefont {K.~W.}\ \bibnamefont {Baldwin}}, \bibinfo {author}
  {\bibfnamefont {L.~N.}\ \bibnamefont {Pfeiffer}}, \ and\ \bibinfo {author}
  {\bibfnamefont {K.~W.}\ \bibnamefont {West}},\ }\bibfield  {title} {\enquote
  {\bibinfo {title} {{Spin-Charge Separation and Localization in One
  Dimension}},}\ }\href {\doibase 10.1126/science.1107821} {\bibfield
  {journal} {\bibinfo  {journal} {Science}\ }\textbf {\bibinfo {volume}
  {308}},\ \bibinfo {pages} {88--92} (\bibinfo {year} {2005})}\BibitemShut
  {NoStop}%
\bibitem [{\citenamefont {Kim}\ \emph {et~al.}(2006)\citenamefont {Kim},
  \citenamefont {Koh}, \citenamefont {Rotenberg}, \citenamefont {Oh},
  \citenamefont {Eisaki}, \citenamefont {Motoyama}, \citenamefont {Uchida},
  \citenamefont {Tohyama}, \citenamefont {Maekawa}, \citenamefont {Shen},\ and\
  \citenamefont {Kim}}]{KimNatPhys06}%
  \BibitemOpen
  \bibfield  {author} {\bibinfo {author} {\bibfnamefont {B.~J.}\ \bibnamefont
  {Kim}}, \bibinfo {author} {\bibfnamefont {H.}~\bibnamefont {Koh}}, \bibinfo
  {author} {\bibfnamefont {E.}~\bibnamefont {Rotenberg}}, \bibinfo {author}
  {\bibfnamefont {S.~J.}\ \bibnamefont {Oh}}, \bibinfo {author} {\bibfnamefont
  {H.}~\bibnamefont {Eisaki}}, \bibinfo {author} {\bibfnamefont
  {N.}~\bibnamefont {Motoyama}}, \bibinfo {author} {\bibfnamefont
  {S.}~\bibnamefont {Uchida}}, \bibinfo {author} {\bibfnamefont
  {T.}~\bibnamefont {Tohyama}}, \bibinfo {author} {\bibfnamefont
  {S.}~\bibnamefont {Maekawa}}, \bibinfo {author} {\bibfnamefont {Z.~X.}\
  \bibnamefont {Shen}}, \ and\ \bibinfo {author} {\bibfnamefont
  {C.}~\bibnamefont {Kim}},\ }\bibfield  {title} {\enquote {\bibinfo {title}
  {{Distinct spinon and holon dispersions in photoemission spectral functions
  from one-dimensional SrCuO2}},}\ }\href {http://dx.doi.org/10.1038/nphys316}
  {\bibfield  {journal} {\bibinfo  {journal} {Nature Phys.}\ }\textbf {\bibinfo
  {volume} {2}},\ \bibinfo {pages} {397--401} (\bibinfo {year}
  {2006})}\BibitemShut {NoStop}%
\bibitem [{\citenamefont {{Hilker}}\ \emph {et~al.}(2017)\citenamefont
  {{Hilker}}, \citenamefont {{Salomon}}, \citenamefont {{Grusdt}},
  \citenamefont {{Omran}}, \citenamefont {{Boll}}, \citenamefont {{Demler}},
  \citenamefont {{Bloch}},\ and\ \citenamefont {{Gross}}}]{HilkerArxiv17}%
  \BibitemOpen
  \bibfield  {author} {\bibinfo {author} {\bibfnamefont {T.~A.}\ \bibnamefont
  {{Hilker}}}, \bibinfo {author} {\bibfnamefont {G.}~\bibnamefont {{Salomon}}},
  \bibinfo {author} {\bibfnamefont {F.}~\bibnamefont {{Grusdt}}}, \bibinfo
  {author} {\bibfnamefont {A.}~\bibnamefont {{Omran}}}, \bibinfo {author}
  {\bibfnamefont {M.}~\bibnamefont {{Boll}}}, \bibinfo {author} {\bibfnamefont
  {E.}~\bibnamefont {{Demler}}}, \bibinfo {author} {\bibfnamefont
  {I.}~\bibnamefont {{Bloch}}}, \ and\ \bibinfo {author} {\bibfnamefont
  {C.}~\bibnamefont {{Gross}}},\ }\bibfield  {title} {\enquote {\bibinfo
  {title} {{Revealing Hidden Antiferromagnetic Correlations in Doped Hubbard
  Chains via String Correlators}},}\ }\href@noop {} {\bibfield  {journal}
  {\bibinfo  {journal} {ArXiv e-prints}\ } (\bibinfo {year} {2017})},\ \Eprint
  {http://arxiv.org/abs/1702.00642} {arXiv:1702.00642 [cond-mat.quant-gas]}
  \BibitemShut {NoStop}%
\bibitem [{\citenamefont {Noack}\ and\ \citenamefont
  {Manmana}(2005)}]{NoackAIPConfProc05}%
  \BibitemOpen
  \bibfield  {author} {\bibinfo {author} {\bibfnamefont {R.~M.}\ \bibnamefont
  {Noack}}\ and\ \bibinfo {author} {\bibfnamefont {S.~R.}\ \bibnamefont
  {Manmana}},\ }\bibfield  {title} {\enquote {\bibinfo {title}
  {{Diagonalization-€ and Numerical Renormalization-Group-€Based Methods for
  Interacting Quantum Systems}},}\ }\href {\doibase
  http://dx.doi.org/10.1063/1.2080349} {\bibfield  {journal} {\bibinfo
  {journal} {AIP Conf. Proc.}\ }\textbf {\bibinfo {volume} {789}},\ \bibinfo
  {pages} {93--163} (\bibinfo {year} {2005})}\BibitemShut {NoStop}%
\bibitem [{\citenamefont {Zhang}\ and\ \citenamefont
  {Dong}(2010)}]{ZhangEurJPhys10}%
  \BibitemOpen
  \bibfield  {author} {\bibinfo {author} {\bibfnamefont {J.~M.}\ \bibnamefont
  {Zhang}}\ and\ \bibinfo {author} {\bibfnamefont {R.~X.}\ \bibnamefont
  {Dong}},\ }\bibfield  {title} {\enquote {\bibinfo {title} {{Exact
  diagonalization: the Bose-€"Hubbard model as an example}},}\ }\href
  {http://stacks.iop.org/0143-0807/31/i=3/a=016} {\bibfield  {journal}
  {\bibinfo  {journal} {Eur. J. Phys.}\ }\textbf {\bibinfo {volume} {31}},\
  \bibinfo {pages} {591} (\bibinfo {year} {2010})}\BibitemShut {NoStop}%
\bibitem [{\citenamefont {Wilson}(1975)}]{wilson1975the}%
  \BibitemOpen
  \bibfield  {author} {\bibinfo {author} {\bibfnamefont {K.~G.}\ \bibnamefont
  {Wilson}},\ }\bibfield  {title} {\enquote {\bibinfo {title} {The
  renormalization group: Critical phenomena and the kondo problem},}\ }\href
  {\doibase 10.1103/RevModPhys.47.773} {\bibfield  {journal} {\bibinfo
  {journal} {Rev. Mod. Phys.}\ }\textbf {\bibinfo {volume} {47}},\ \bibinfo
  {pages} {773--840} (\bibinfo {year} {1975})}\BibitemShut {NoStop}%
\bibitem [{\citenamefont {Yurov}\ and\ \citenamefont
  {Zamolodchikov}(1990)}]{yurov1990truncated}%
  \BibitemOpen
  \bibfield  {author} {\bibinfo {author} {\bibfnamefont {V.~P.}\ \bibnamefont
  {Yurov}}\ and\ \bibinfo {author} {\bibfnamefont {{Al.}~B.}\ \bibnamefont
  {Zamolodchikov}},\ }\bibfield  {title} {\enquote {\bibinfo {title}
  {{Truncated conformal space approach to scaling {L}ee-{Y}ang model}},}\
  }\href {\doibase 10.1142/S0217751X9000218X} {\bibfield  {journal} {\bibinfo
  {journal} {Int. J. Mod. Phys. A}\ }\textbf {\bibinfo {volume} {05}},\
  \bibinfo {pages} {3221--3245} (\bibinfo {year} {1990})}\BibitemShut {NoStop}%
\bibitem [{\citenamefont {Yurov}\ and\ \citenamefont
  {Zamolodchikov}(1991)}]{yurov1991truncated}%
  \BibitemOpen
  \bibfield  {author} {\bibinfo {author} {\bibfnamefont {V.~P.}\ \bibnamefont
  {Yurov}}\ and\ \bibinfo {author} {\bibfnamefont {{Al.}~B.}\ \bibnamefont
  {Zamolodchikov}},\ }\bibfield  {title} {\enquote {\bibinfo {title}
  {Truncated-fermionic-space approach to the critical 2{D} {I}sing model with
  magnetic field},}\ }\href {\doibase 10.1142/S0217751X91002161} {\bibfield
  {journal} {\bibinfo  {journal} {Int. J. Mod. Phys. A}\ }\textbf {\bibinfo
  {volume} {06}},\ \bibinfo {pages} {4557--4578} (\bibinfo {year}
  {1991})}\BibitemShut {NoStop}%
\bibitem [{\citenamefont {Verstraete}\ \emph
  {et~al.}(2008{\natexlab{a}})\citenamefont {Verstraete}, \citenamefont
  {Murg},\ and\ \citenamefont {Cirac}}]{VerstraeteAdvPhys08}%
  \BibitemOpen
  \bibfield  {author} {\bibinfo {author} {\bibfnamefont {F.}~\bibnamefont
  {Verstraete}}, \bibinfo {author} {\bibfnamefont {V.}~\bibnamefont {Murg}}, \
  and\ \bibinfo {author} {\bibfnamefont {J.~I.}\ \bibnamefont {Cirac}},\
  }\bibfield  {title} {\enquote {\bibinfo {title} {{Matrix product states,
  projected entangled pair states, and variational renormalization group
  methods for quantum spin systems}},}\ }\href {\doibase
  10.1080/14789940801912366} {\bibfield  {journal} {\bibinfo  {journal} {Adv.
  Phys.}\ }\textbf {\bibinfo {volume} {57}},\ \bibinfo {pages} {143--224}
  (\bibinfo {year} {2008}{\natexlab{a}})}\BibitemShut {NoStop}%
\bibitem [{\citenamefont {Or\'us}(2014{\natexlab{a}})}]{OrusAnnPhys14}%
  \BibitemOpen
  \bibfield  {author} {\bibinfo {author} {\bibfnamefont {R.}~\bibnamefont
  {Or\'us}},\ }\bibfield  {title} {\enquote {\bibinfo {title} {{A practical
  introduction to tensor networks: Matrix product states and projected
  entangled pair states}},}\ }\href {\doibase
  http://dx.doi.org/10.1016/j.aop.2014.06.013} {\bibfield  {journal} {\bibinfo
  {journal} {Ann. Phys. (N.Y.)}\ }\textbf {\bibinfo {volume} {349}},\ \bibinfo
  {pages} {117 -- 158} (\bibinfo {year} {2014}{\natexlab{a}})}\BibitemShut
  {NoStop}%
\bibitem [{\citenamefont {White}(1992)}]{white1992density}%
  \BibitemOpen
  \bibfield  {author} {\bibinfo {author} {\bibfnamefont {S.~R.}\ \bibnamefont
  {White}},\ }\bibfield  {title} {\enquote {\bibinfo {title} {{{Density matrix
  formulation for quantum renormalization groups}}},}\ }\href {\doibase
  10.1103/PhysRevLett.69.2863} {\bibfield  {journal} {\bibinfo  {journal}
  {Phys. Rev. Lett.}\ }\textbf {\bibinfo {volume} {69}},\ \bibinfo {pages}
  {2863--2866} (\bibinfo {year} {1992})}\BibitemShut {NoStop}%
\bibitem [{\citenamefont {White}(1993{\natexlab{a}})}]{WhitePRB93}%
  \BibitemOpen
  \bibfield  {author} {\bibinfo {author} {\bibfnamefont {S.~R.}\ \bibnamefont
  {White}},\ }\bibfield  {title} {\enquote {\bibinfo {title} {{Density-matrix
  algorithms for quantum renormalization groups}},}\ }\href {\doibase
  10.1103/PhysRevB.48.10345} {\bibfield  {journal} {\bibinfo  {journal} {Phys.
  Rev. B}\ }\textbf {\bibinfo {volume} {48}},\ \bibinfo {pages} {10345--10356}
  (\bibinfo {year} {1993}{\natexlab{a}})}\BibitemShut {NoStop}%
\bibitem [{\citenamefont {{U. Schollw\"ock}}(2011)}]{SchollwockAnnPhys11}%
  \BibitemOpen
  \bibfield  {author} {\bibinfo {author} {\bibnamefont {{U. Schollw\"ock}}},\
  }\bibfield  {title} {\enquote {\bibinfo {title} {{The density-matrix
  renormalization group in the age of matrix product states}},}\ }\href
  {\doibase http://dx.doi.org/10.1016/j.aop.2010.09.012} {\bibfield  {journal}
  {\bibinfo  {journal} {Ann. Phys. (N.Y.)}\ }\textbf {\bibinfo {volume}
  {326}},\ \bibinfo {pages} {96 -- 192} (\bibinfo {year} {2011})},\ \bibinfo
  {note} {january 2011 Special Issue}\BibitemShut {NoStop}%
\bibitem [{\citenamefont {Foulkes}\ \emph {et~al.}(2001)\citenamefont
  {Foulkes}, \citenamefont {Mitas}, \citenamefont {Needs},\ and\ \citenamefont
  {Rajagopal}}]{FoulkesRMP01}%
  \BibitemOpen
  \bibfield  {author} {\bibinfo {author} {\bibfnamefont {W.~M.~C.}\
  \bibnamefont {Foulkes}}, \bibinfo {author} {\bibfnamefont {L.}~\bibnamefont
  {Mitas}}, \bibinfo {author} {\bibfnamefont {R.~J.}\ \bibnamefont {Needs}}, \
  and\ \bibinfo {author} {\bibfnamefont {G.}~\bibnamefont {Rajagopal}},\
  }\bibfield  {title} {\enquote {\bibinfo {title} {{Quantum Monte Carlo
  simulations of solids}},}\ }\href {\doibase 10.1103/RevModPhys.73.33}
  {\bibfield  {journal} {\bibinfo  {journal} {Rev. Mod. Phys.}\ }\textbf
  {\bibinfo {volume} {73}},\ \bibinfo {pages} {33--83} (\bibinfo {year}
  {2001})}\BibitemShut {NoStop}%
\bibitem [{\citenamefont {Hochkeppel}\ \emph {et~al.}(2009)\citenamefont
  {Hochkeppel}, \citenamefont {Lang}, \citenamefont {Br{\"u}nger},
  \citenamefont {Assaad},\ and\ \citenamefont {Hanke}}]{HochkeppelChapter09}%
  \BibitemOpen
  \bibfield  {author} {\bibinfo {author} {\bibfnamefont {S.}~\bibnamefont
  {Hochkeppel}}, \bibinfo {author} {\bibfnamefont {T.~C.}\ \bibnamefont
  {Lang}}, \bibinfo {author} {\bibfnamefont {C.}~\bibnamefont {Br{\"u}nger}},
  \bibinfo {author} {\bibfnamefont {F.~F.}\ \bibnamefont {Assaad}}, \ and\
  \bibinfo {author} {\bibfnamefont {W.}~\bibnamefont {Hanke}},\ }\enquote
  {\bibinfo {title} {{Quantum Monte Carlo Studies of Strongly Correlated
  Electron Systems}},}\ in\ \href {\doibase 10.1007/978-3-540-69182-2_51}
  {\emph {\bibinfo {booktitle} {High Performance Computing in Science and
  Engineering, Garching/Munich 2007}}},\ \bibinfo {editor} {edited by\ \bibinfo
  {editor} {\bibfnamefont {S.}~\bibnamefont {Wagner}}, \bibinfo {editor}
  {\bibfnamefont {M.}~\bibnamefont {Steinmetz}}, \bibinfo {editor}
  {\bibfnamefont {A.}~\bibnamefont {Bode}}, \ and\ \bibinfo {editor}
  {\bibfnamefont {M.}~\bibnamefont {Brehm}}}\ (\bibinfo  {publisher} {Springer
  Berlin Heidelberg},\ \bibinfo {address} {Berlin, Heidelberg},\ \bibinfo
  {year} {2009})\ pp.\ \bibinfo {pages} {669--686}\BibitemShut {NoStop}%
\bibitem [{\citenamefont {Tokura}\ and\ \citenamefont
  {Nagaosa}(2000)}]{TokuraScience00}%
  \BibitemOpen
  \bibfield  {author} {\bibinfo {author} {\bibfnamefont {Y.}~\bibnamefont
  {Tokura}}\ and\ \bibinfo {author} {\bibfnamefont {N.}~\bibnamefont
  {Nagaosa}},\ }\bibfield  {title} {\enquote {\bibinfo {title} {{Orbital
  Physics in Transition-Metal Oxides}},}\ }\href {\doibase
  10.1126/science.288.5465.462} {\bibfield  {journal} {\bibinfo  {journal}
  {Science}\ }\textbf {\bibinfo {volume} {288}},\ \bibinfo {pages} {462--468}
  (\bibinfo {year} {2000})}\BibitemShut {NoStop}%
\bibitem [{\citenamefont {Wang}\ \emph {et~al.}(2012)\citenamefont {Wang},
  \citenamefont {Kalantar-Zadeh}, \citenamefont {Kis}, \citenamefont
  {Coleman},\ and\ \citenamefont {Strano}}]{WangNatureNano12}%
  \BibitemOpen
  \bibfield  {author} {\bibinfo {author} {\bibfnamefont {Q.~H.}\ \bibnamefont
  {Wang}}, \bibinfo {author} {\bibfnamefont {K.}~\bibnamefont
  {Kalantar-Zadeh}}, \bibinfo {author} {\bibfnamefont {A.}~\bibnamefont {Kis}},
  \bibinfo {author} {\bibfnamefont {J.~N.}\ \bibnamefont {Coleman}}, \ and\
  \bibinfo {author} {\bibfnamefont {M.~S.}\ \bibnamefont {Strano}},\ }\bibfield
   {title} {\enquote {\bibinfo {title} {{Electronics and optoelectronics of
  two-dimensional transition metal dichalcogenides}},}\ }\href
  {http://dx.doi.org/10.1038/nnano.2012.193} {\bibfield  {journal} {\bibinfo
  {journal} {Nature Nano.}\ }\textbf {\bibinfo {volume} {7}},\ \bibinfo {pages}
  {699--712} (\bibinfo {year} {2012})}\BibitemShut {NoStop}%
\bibitem [{\citenamefont {Stewart}(1984)}]{StewartRMP84}%
  \BibitemOpen
  \bibfield  {author} {\bibinfo {author} {\bibfnamefont {G.~R.}\ \bibnamefont
  {Stewart}},\ }\bibfield  {title} {\enquote {\bibinfo {title} {{Heavy-fermion
  systems}},}\ }\href {\doibase 10.1103/RevModPhys.56.755} {\bibfield
  {journal} {\bibinfo  {journal} {Rev. Mod. Phys.}\ }\textbf {\bibinfo {volume}
  {56}},\ \bibinfo {pages} {755--787} (\bibinfo {year} {1984})}\BibitemShut
  {NoStop}%
\bibitem [{\citenamefont {Coleman}(2007)}]{ColemanArticle07}%
  \BibitemOpen
  \bibfield  {author} {\bibinfo {author} {\bibfnamefont {P.}~\bibnamefont
  {Coleman}},\ }\enquote {\bibinfo {title} {{Heavy Fermions: Electrons at the
  Edge of Magnetism}},}\ in\ \href {\doibase 10.1002/9780470022184.hmm105}
  {\emph {\bibinfo {booktitle} {Handbook of Magnetism and Advanced Magnetic
  Materials}}}\ (\bibinfo  {publisher} {John Wiley \& Sons, Ltd},\ \bibinfo
  {year} {2007})\BibitemShut {NoStop}%
\bibitem [{\citenamefont {Wu}\ \emph {et~al.}(2003)\citenamefont {Wu},
  \citenamefont {Hu},\ and\ \citenamefont {Zhang}}]{WuPRL03}%
  \BibitemOpen
  \bibfield  {author} {\bibinfo {author} {\bibfnamefont {C.}~\bibnamefont
  {Wu}}, \bibinfo {author} {\bibfnamefont {J.-p.}\ \bibnamefont {Hu}}, \ and\
  \bibinfo {author} {\bibfnamefont {S.-c.}\ \bibnamefont {Zhang}},\ }\bibfield
  {title} {\enquote {\bibinfo {title} {{Exact SO(5) Symmetry in the Spin-$3/2$
  Fermionic System}},}\ }\href {\doibase 10.1103/PhysRevLett.91.186402}
  {\bibfield  {journal} {\bibinfo  {journal} {Phys. Rev. Lett.}\ }\textbf
  {\bibinfo {volume} {91}},\ \bibinfo {pages} {186402} (\bibinfo {year}
  {2003})}\BibitemShut {NoStop}%
\bibitem [{\citenamefont {Wu}(2006)}]{WuModPhysLettB06}%
  \BibitemOpen
  \bibfield  {author} {\bibinfo {author} {\bibfnamefont {C.}~\bibnamefont
  {Wu}},\ }\bibfield  {title} {\enquote {\bibinfo {title} {Hidden symmetry and
  quantum phases in spin-3/2 cold atomic systems},}\ }\href {\doibase
  10.1142/S0217984906012213} {\bibfield  {journal} {\bibinfo  {journal} {Mod.
  Phys. Lett. B}\ }\textbf {\bibinfo {volume} {20}},\ \bibinfo {pages}
  {1707--1738} (\bibinfo {year} {2006})}\BibitemShut {NoStop}%
\bibitem [{\citenamefont {Cazalilla}\ \emph {et~al.}(2009)\citenamefont
  {Cazalilla}, \citenamefont {Ho},\ and\ \citenamefont
  {Ueda}}]{CazalillaNewJPhys09}%
  \BibitemOpen
  \bibfield  {author} {\bibinfo {author} {\bibfnamefont {M.~A.}\ \bibnamefont
  {Cazalilla}}, \bibinfo {author} {\bibfnamefont {A.~F.}\ \bibnamefont {Ho}}, \
  and\ \bibinfo {author} {\bibfnamefont {M.}~\bibnamefont {Ueda}},\ }\bibfield
  {title} {\enquote {\bibinfo {title} {{Ultracold gases of ytterbium:
  ferromagnetism and Mott states in an SU(6) Fermi system}},}\ }\href
  {http://stacks.iop.org/1367-2630/11/i=10/a=103033} {\bibfield  {journal}
  {\bibinfo  {journal} {New J. Phys.}\ }\textbf {\bibinfo {volume} {11}},\
  \bibinfo {pages} {103033} (\bibinfo {year} {2009})}\BibitemShut {NoStop}%
\bibitem [{\citenamefont {Gorshkov}\ \emph {et~al.}(2010)\citenamefont
  {Gorshkov}, \citenamefont {Hermele}, \citenamefont {Gurarie}, \citenamefont
  {Xu}, \citenamefont {Julienne}, \citenamefont {Ye}, \citenamefont {Zoller},
  \citenamefont {Demler}, \citenamefont {Lukin},\ and\ \citenamefont
  {Rey}}]{GorshkovNaturePhys10}%
  \BibitemOpen
  \bibfield  {author} {\bibinfo {author} {\bibfnamefont {A.~V.}\ \bibnamefont
  {Gorshkov}}, \bibinfo {author} {\bibfnamefont {M.}~\bibnamefont {Hermele}},
  \bibinfo {author} {\bibfnamefont {V.}~\bibnamefont {Gurarie}}, \bibinfo
  {author} {\bibfnamefont {C.}~\bibnamefont {Xu}}, \bibinfo {author}
  {\bibfnamefont {P.~S.}\ \bibnamefont {Julienne}}, \bibinfo {author}
  {\bibfnamefont {J.}~\bibnamefont {Ye}}, \bibinfo {author} {\bibfnamefont
  {P.}~\bibnamefont {Zoller}}, \bibinfo {author} {\bibfnamefont
  {E.}~\bibnamefont {Demler}}, \bibinfo {author} {\bibfnamefont {M.~D.}\
  \bibnamefont {Lukin}}, \ and\ \bibinfo {author} {\bibfnamefont {A.~M.}\
  \bibnamefont {Rey}},\ }\bibfield  {title} {\enquote {\bibinfo {title}
  {{Two-orbital SU(N) magnetism with ultracold alkaline-earth atoms}},}\ }\href
  {http://dx.doi.org/10.1038/nphys1535} {\bibfield  {journal} {\bibinfo
  {journal} {Nature Phys.}\ }\textbf {\bibinfo {volume} {6}},\ \bibinfo {pages}
  {289--295} (\bibinfo {year} {2010})}\BibitemShut {NoStop}%
\bibitem [{\citenamefont {Cazalilla}\ and\ \citenamefont
  {Rey}(2014)}]{CazalillaRPP14}%
  \BibitemOpen
  \bibfield  {author} {\bibinfo {author} {\bibfnamefont {M.~A.}\ \bibnamefont
  {Cazalilla}}\ and\ \bibinfo {author} {\bibfnamefont {A.~M.}\ \bibnamefont
  {Rey}},\ }\bibfield  {title} {\enquote {\bibinfo {title} {{Ultracold Fermi
  gases with emergent SU(N) symmetry}},}\ }\href
  {http://stacks.iop.org/0034-4885/77/i=12/a=124401} {\bibfield  {journal}
  {\bibinfo  {journal} {Rep. Prog. Phys.}\ }\textbf {\bibinfo {volume} {77}},\
  \bibinfo {pages} {124401} (\bibinfo {year} {2014})}\BibitemShut {NoStop}%
\bibitem [{\citenamefont {Capponi}\ \emph {et~al.}(2016)\citenamefont
  {Capponi}, \citenamefont {Lecheminant},\ and\ \citenamefont
  {Totsuka}}]{CapponiAnnPhys16}%
  \BibitemOpen
  \bibfield  {author} {\bibinfo {author} {\bibfnamefont {S.}~\bibnamefont
  {Capponi}}, \bibinfo {author} {\bibfnamefont {P.}~\bibnamefont
  {Lecheminant}}, \ and\ \bibinfo {author} {\bibfnamefont {K.}~\bibnamefont
  {Totsuka}},\ }\bibfield  {title} {\enquote {\bibinfo {title} {{Phases of
  one-dimensional SU(N) cold atomic Fermi gases: €"From molecular Luttinger
  liquids to topological phases}},}\ }\href {\doibase
  http://dx.doi.org/10.1016/j.aop.2016.01.011} {\bibfield  {journal} {\bibinfo
  {journal} {Ann. Phys. (N.Y.)}\ }\textbf {\bibinfo {volume} {367}},\ \bibinfo
  {pages} {50 -- 95} (\bibinfo {year} {2016})}\BibitemShut {NoStop}%
\bibitem [{\citenamefont {Gogolin}\ and\ \citenamefont
  {Eisert}(2016)}]{GogolinReview15}%
  \BibitemOpen
  \bibfield  {author} {\bibinfo {author} {\bibfnamefont {C.}~\bibnamefont
  {Gogolin}}\ and\ \bibinfo {author} {\bibfnamefont {J.}~\bibnamefont
  {Eisert}},\ }\bibfield  {title} {\enquote {\bibinfo {title} {{Equilibration,
  thermalisation, and the emergence of statistical mechanics in closed quantum
  systems}},}\ }\href {http://stacks.iop.org/0034-4885/79/i=5/a=056001}
  {\bibfield  {journal} {\bibinfo  {journal} {Rep. Prog. Phys.}\ }\textbf
  {\bibinfo {volume} {79}},\ \bibinfo {pages} {056001} (\bibinfo {year}
  {2016})}\BibitemShut {NoStop}%
\bibitem [{\citenamefont {D'Alessio}\ \emph {et~al.}(2016)\citenamefont
  {D'Alessio}, \citenamefont {Kafri}, \citenamefont {Polkovnikov},\ and\
  \citenamefont {Rigol}}]{DAlessioReview15}%
  \BibitemOpen
  \bibfield  {author} {\bibinfo {author} {\bibfnamefont {L.}~\bibnamefont
  {D'Alessio}}, \bibinfo {author} {\bibfnamefont {Y.}~\bibnamefont {Kafri}},
  \bibinfo {author} {\bibfnamefont {A.}~\bibnamefont {Polkovnikov}}, \ and\
  \bibinfo {author} {\bibfnamefont {M.}~\bibnamefont {Rigol}},\ }\bibfield
  {title} {\enquote {\bibinfo {title} {{From quantum chaos and eigenstate
  thermalization to statistical mechanics and thermodynamics}},}\ }\href
  {\doibase 10.1080/00018732.2016.1198134} {\bibfield  {journal} {\bibinfo
  {journal} {Adv. Phys.}\ }\textbf {\bibinfo {volume} {65}},\ \bibinfo {pages}
  {239--362} (\bibinfo {year} {2016})}\BibitemShut {NoStop}%
\bibitem [{\citenamefont {Essler}\ and\ \citenamefont
  {Fagotti}(2016)}]{EsslerReview16}%
  \BibitemOpen
  \bibfield  {author} {\bibinfo {author} {\bibfnamefont {F.~H.~L.}\
  \bibnamefont {Essler}}\ and\ \bibinfo {author} {\bibfnamefont
  {M.}~\bibnamefont {Fagotti}},\ }\bibfield  {title} {\enquote {\bibinfo
  {title} {{Quench dynamics and relaxation in isolated integrable quantum spin
  chains}},}\ }\href {http://stacks.iop.org/1742-5468/2016/i=6/a=064002}
  {\bibfield  {journal} {\bibinfo  {journal} {J. Stat. Mech.}\ }\textbf
  {\bibinfo {volume} {2016}},\ \bibinfo {pages} {P064002} (\bibinfo {year}
  {2016})}\BibitemShut {NoStop}%
\bibitem [{\citenamefont {Calabrese}\ and\ \citenamefont
  {Cardy}(2016)}]{CalabreseReview16}%
  \BibitemOpen
  \bibfield  {author} {\bibinfo {author} {\bibfnamefont {P.}~\bibnamefont
  {Calabrese}}\ and\ \bibinfo {author} {\bibfnamefont {J.}~\bibnamefont
  {Cardy}},\ }\bibfield  {title} {\enquote {\bibinfo {title} {{Quantum quenches
  in 1+1 dimensional conformal field theories}},}\ }\href
  {http://stacks.iop.org/1742-5468/2016/i=6/a=064003} {\bibfield  {journal}
  {\bibinfo  {journal} {J. Stat. Mech.}\ }\textbf {\bibinfo {volume} {2016}},\
  \bibinfo {pages} {P064003} (\bibinfo {year} {2016})}\BibitemShut {NoStop}%
\bibitem [{\citenamefont {Cazalilla}\ and\ \citenamefont
  {Chung}(2016)}]{CazalillaReview16}%
  \BibitemOpen
  \bibfield  {author} {\bibinfo {author} {\bibfnamefont {M.~A.}\ \bibnamefont
  {Cazalilla}}\ and\ \bibinfo {author} {\bibfnamefont {M.-C.}\ \bibnamefont
  {Chung}},\ }\bibfield  {title} {\enquote {\bibinfo {title} {{Quantum quenches
  in the Luttinger model and its close relatives}},}\ }\href
  {http://stacks.iop.org/1742-5468/2016/i=6/a=064004} {\bibfield  {journal}
  {\bibinfo  {journal} {J. Stat. Mech.}\ }\textbf {\bibinfo {volume} {2016}},\
  \bibinfo {pages} {P064004} (\bibinfo {year} {2016})}\BibitemShut {NoStop}%
\bibitem [{\citenamefont {Bernard}\ and\ \citenamefont
  {Doyon}(2016)}]{BernardReview16}%
  \BibitemOpen
  \bibfield  {author} {\bibinfo {author} {\bibfnamefont {D.}~\bibnamefont
  {Bernard}}\ and\ \bibinfo {author} {\bibfnamefont {B.}~\bibnamefont
  {Doyon}},\ }\bibfield  {title} {\enquote {\bibinfo {title} {{Conformal field
  theory out of equilibrium: a review}},}\ }\href
  {http://stacks.iop.org/1742-5468/2016/i=6/a=064005} {\bibfield  {journal}
  {\bibinfo  {journal} {J. Stat. Mech.}\ }\textbf {\bibinfo {volume} {2016}},\
  \bibinfo {pages} {P064005} (\bibinfo {year} {2016})}\BibitemShut {NoStop}%
\bibitem [{\citenamefont {Caux}(2016)}]{CauxReview16}%
  \BibitemOpen
  \bibfield  {author} {\bibinfo {author} {\bibfnamefont {J.-S.}\ \bibnamefont
  {Caux}},\ }\bibfield  {title} {\enquote {\bibinfo {title} {{The {Q}uench
  {A}ction}},}\ }\href {http://stacks.iop.org/1742-5468/2016/i=6/a=064006}
  {\bibfield  {journal} {\bibinfo  {journal} {J. Stat. Mech.}\ }\textbf
  {\bibinfo {volume} {2016}},\ \bibinfo {pages} {P064006} (\bibinfo {year}
  {2016})}\BibitemShut {NoStop}%
\bibitem [{\citenamefont {Vidmar}\ and\ \citenamefont
  {Rigol}(2016)}]{VidmarReview16}%
  \BibitemOpen
  \bibfield  {author} {\bibinfo {author} {\bibfnamefont {L.}~\bibnamefont
  {Vidmar}}\ and\ \bibinfo {author} {\bibfnamefont {M.}~\bibnamefont {Rigol}},\
  }\bibfield  {title} {\enquote {\bibinfo {title} {{Generalized Gibbs ensemble
  in integrable lattice models}},}\ }\href
  {http://stacks.iop.org/1742-5468/2016/i=6/a=064007} {\bibfield  {journal}
  {\bibinfo  {journal} {J. Stat. Mech.}\ }\textbf {\bibinfo {volume} {2016}},\
  \bibinfo {pages} {P064007} (\bibinfo {year} {2016})}\BibitemShut {NoStop}%
\bibitem [{\citenamefont {{Langen}}\ \emph {et~al.}(2016)\citenamefont
  {{Langen}}, \citenamefont {{Gasenzer}},\ and\ \citenamefont
  {{Schmiedmayer}}}]{LangenReview16}%
  \BibitemOpen
  \bibfield  {author} {\bibinfo {author} {\bibfnamefont {T.}~\bibnamefont
  {{Langen}}}, \bibinfo {author} {\bibfnamefont {T.}~\bibnamefont
  {{Gasenzer}}}, \ and\ \bibinfo {author} {\bibfnamefont {J.}~\bibnamefont
  {{Schmiedmayer}}},\ }\bibfield  {title} {\enquote {\bibinfo {title}
  {{Prethermalization and universal dynamics in near-integrable quantum
  systems}},}\ }\href {http://stacks.iop.org/1742-5468/2016/i=6/a=064009}
  {\bibfield  {journal} {\bibinfo  {journal} {J. Stat. Mech.}\ }\textbf
  {\bibinfo {volume} {2016}},\ \bibinfo {pages} {P064009} (\bibinfo {year}
  {2016})}\BibitemShut {NoStop}%
\bibitem [{\citenamefont {Ilievski}\ \emph {et~al.}(2016)\citenamefont
  {Ilievski}, \citenamefont {Medenjak}, \citenamefont {Prosen},\ and\
  \citenamefont {Zadnik}}]{ProsenReview16}%
  \BibitemOpen
  \bibfield  {author} {\bibinfo {author} {\bibfnamefont {E.}~\bibnamefont
  {Ilievski}}, \bibinfo {author} {\bibfnamefont {M.}~\bibnamefont {Medenjak}},
  \bibinfo {author} {\bibfnamefont {T.}~\bibnamefont {Prosen}}, \ and\ \bibinfo
  {author} {\bibfnamefont {L.}~\bibnamefont {Zadnik}},\ }\bibfield  {title}
  {\enquote {\bibinfo {title} {{Quasilocal charges in integrable lattice
  systems}},}\ }\href {http://stacks.iop.org/1742-5468/2016/i=6/a=064008}
  {\bibfield  {journal} {\bibinfo  {journal} {J. Stat. Mech.}\ }\textbf
  {\bibinfo {volume} {2016}},\ \bibinfo {pages} {P064008} (\bibinfo {year}
  {2016})}\BibitemShut {NoStop}%
\bibitem [{\citenamefont {Vasseur}\ and\ \citenamefont
  {Moore}(2016)}]{VasseurReview16}%
  \BibitemOpen
  \bibfield  {author} {\bibinfo {author} {\bibfnamefont {R.}~\bibnamefont
  {Vasseur}}\ and\ \bibinfo {author} {\bibfnamefont {J.~E.}\ \bibnamefont
  {Moore}},\ }\bibfield  {title} {\enquote {\bibinfo {title} {{Nonequilibrium
  quantum dynamics and transport: from integrability to many-body
  localization}},}\ }\href {http://stacks.iop.org/1742-5468/2016/i=6/a=064010}
  {\bibfield  {journal} {\bibinfo  {journal} {J. Stat. Mech.}\ }\textbf
  {\bibinfo {volume} {2016}},\ \bibinfo {pages} {P064010} (\bibinfo {year}
  {2016})}\BibitemShut {NoStop}%
\bibitem [{\citenamefont {{De Luca}}\ and\ \citenamefont
  {{Mussardo}}(2016)}]{DeLucaReview16}%
  \BibitemOpen
  \bibfield  {author} {\bibinfo {author} {\bibfnamefont {A.}~\bibnamefont {{De
  Luca}}}\ and\ \bibinfo {author} {\bibfnamefont {G.}~\bibnamefont
  {{Mussardo}}},\ }\bibfield  {title} {\enquote {\bibinfo {title}
  {{Equilibration properties of classical integrable field theories}},}\ }\href
  {http://stacks.iop.org/1742-5468/2016/i=6/a=064011} {\bibfield  {journal}
  {\bibinfo  {journal} {J. Stat. Mech.}\ }\textbf {\bibinfo {volume} {2016}},\
  \bibinfo {pages} {P064011} (\bibinfo {year} {2016})}\BibitemShut {NoStop}%
\bibitem [{\citenamefont {Imambekov}\ and\ \citenamefont
  {Glazman}(2009)}]{ImambekovScience2009}%
  \BibitemOpen
  \bibfield  {author} {\bibinfo {author} {\bibfnamefont {A.}~\bibnamefont
  {Imambekov}}\ and\ \bibinfo {author} {\bibfnamefont {L.~I.}\ \bibnamefont
  {Glazman}},\ }\bibfield  {title} {\enquote {\bibinfo {title} {{Universal
  Theory of Nonlinear Luttinger Liquids}},}\ }\href {\doibase
  10.1126/science.1165403} {\bibfield  {journal} {\bibinfo  {journal}
  {Science}\ }\textbf {\bibinfo {volume} {323}},\ \bibinfo {pages} {228--231}
  (\bibinfo {year} {2009})}\BibitemShut {NoStop}%
\bibitem [{\citenamefont {Imambekov}\ \emph {et~al.}(2012)\citenamefont
  {Imambekov}, \citenamefont {Schmidt},\ and\ \citenamefont
  {Glazman}}]{ImambekovRMP12}%
  \BibitemOpen
  \bibfield  {author} {\bibinfo {author} {\bibfnamefont {A.}~\bibnamefont
  {Imambekov}}, \bibinfo {author} {\bibfnamefont {T.~L.}\ \bibnamefont
  {Schmidt}}, \ and\ \bibinfo {author} {\bibfnamefont {L.~I.}\ \bibnamefont
  {Glazman}},\ }\bibfield  {title} {\enquote {\bibinfo {title}
  {{One-dimensional quantum liquids: Beyond the Luttinger liquid paradigm}},}\
  }\href {\doibase 10.1103/RevModPhys.84.1253} {\bibfield  {journal} {\bibinfo
  {journal} {Rev. Mod. Phys.}\ }\textbf {\bibinfo {volume} {84}},\ \bibinfo
  {pages} {1253--1306} (\bibinfo {year} {2012})}\BibitemShut {NoStop}%
\bibitem [{\citenamefont {Pereira}\ \emph {et~al.}(2008)\citenamefont
  {Pereira}, \citenamefont {White},\ and\ \citenamefont
  {Affleck}}]{PereiraPRL08}%
  \BibitemOpen
  \bibfield  {author} {\bibinfo {author} {\bibfnamefont {R.~G.}\ \bibnamefont
  {Pereira}}, \bibinfo {author} {\bibfnamefont {S.~R.}\ \bibnamefont {White}},
  \ and\ \bibinfo {author} {\bibfnamefont {I.}~\bibnamefont {Affleck}},\
  }\bibfield  {title} {\enquote {\bibinfo {title} {{Exact Edge Singularities
  and Dynamical Correlations in Spin-$1/2$ Chains}},}\ }\href {\doibase
  10.1103/PhysRevLett.100.027206} {\bibfield  {journal} {\bibinfo  {journal}
  {Phys. Rev. Lett.}\ }\textbf {\bibinfo {volume} {100}},\ \bibinfo {pages}
  {027206} (\bibinfo {year} {2008})}\BibitemShut {NoStop}%
\bibitem [{\citenamefont {Essler}(2010)}]{EsslerPRB10}%
  \BibitemOpen
  \bibfield  {author} {\bibinfo {author} {\bibfnamefont {F.~H.~L.}\
  \bibnamefont {Essler}},\ }\bibfield  {title} {\enquote {\bibinfo {title}
  {{Threshold singularities in the one-dimensional Hubbard model}},}\ }\href
  {\doibase 10.1103/PhysRevB.81.205120} {\bibfield  {journal} {\bibinfo
  {journal} {Phys. Rev. B}\ }\textbf {\bibinfo {volume} {81}},\ \bibinfo
  {pages} {205120} (\bibinfo {year} {2010})}\BibitemShut {NoStop}%
\bibitem [{\citenamefont {Shashi}\ \emph {et~al.}(2012)\citenamefont {Shashi},
  \citenamefont {Panfil}, \citenamefont {Caux},\ and\ \citenamefont
  {Imambekov}}]{AdityaPRB12}%
  \BibitemOpen
  \bibfield  {author} {\bibinfo {author} {\bibfnamefont {A.}~\bibnamefont
  {Shashi}}, \bibinfo {author} {\bibfnamefont {M.}~\bibnamefont {Panfil}},
  \bibinfo {author} {\bibfnamefont {J.-S.}\ \bibnamefont {Caux}}, \ and\
  \bibinfo {author} {\bibfnamefont {A.}~\bibnamefont {Imambekov}},\ }\bibfield
  {title} {\enquote {\bibinfo {title} {{Exact prefactors in static and dynamic
  correlation functions of one-dimensional quantum integrable models:
  Applications to the Calogero-Sutherland, Lieb-Liniger, and $XXZ$ models}},}\
  }\href {\doibase 10.1103/PhysRevB.85.155136} {\bibfield  {journal} {\bibinfo
  {journal} {Phys. Rev. B}\ }\textbf {\bibinfo {volume} {85}},\ \bibinfo
  {pages} {155136} (\bibinfo {year} {2012})}\BibitemShut {NoStop}%
\bibitem [{\citenamefont {Tiegel}\ \emph {et~al.}(2016)\citenamefont {Tiegel},
  \citenamefont {Veness}, \citenamefont {Dargel}, \citenamefont {Honecker},
  \citenamefont {Pruschke}, \citenamefont {McCulloch},\ and\ \citenamefont
  {Essler}}]{TiegelPRB16}%
  \BibitemOpen
  \bibfield  {author} {\bibinfo {author} {\bibfnamefont {A.~C.}\ \bibnamefont
  {Tiegel}}, \bibinfo {author} {\bibfnamefont {T.}~\bibnamefont {Veness}},
  \bibinfo {author} {\bibfnamefont {P.~E.}\ \bibnamefont {Dargel}}, \bibinfo
  {author} {\bibfnamefont {A.}~\bibnamefont {Honecker}}, \bibinfo {author}
  {\bibfnamefont {T.}~\bibnamefont {Pruschke}}, \bibinfo {author}
  {\bibfnamefont {I.~P.}\ \bibnamefont {McCulloch}}, \ and\ \bibinfo {author}
  {\bibfnamefont {F.~H.~L.}\ \bibnamefont {Essler}},\ }\bibfield  {title}
  {\enquote {\bibinfo {title} {{Optical conductivity of the Hubbard chain away
  from half filling}},}\ }\href {\doibase 10.1103/PhysRevB.93.125108}
  {\bibfield  {journal} {\bibinfo  {journal} {Phys. Rev. B}\ }\textbf {\bibinfo
  {volume} {93}},\ \bibinfo {pages} {125108} (\bibinfo {year}
  {2016})}\BibitemShut {NoStop}%
\bibitem [{\citenamefont {Veness}\ and\ \citenamefont
  {Essler}(2016)}]{VenessPRB16}%
  \BibitemOpen
  \bibfield  {author} {\bibinfo {author} {\bibfnamefont {T.}~\bibnamefont
  {Veness}}\ and\ \bibinfo {author} {\bibfnamefont {F.~H.~L.}\ \bibnamefont
  {Essler}},\ }\bibfield  {title} {\enquote {\bibinfo {title} {{Mobile impurity
  approach to the optical conductivity in the Hubbard chain}},}\ }\href
  {\doibase 10.1103/PhysRevB.93.205101} {\bibfield  {journal} {\bibinfo
  {journal} {Phys. Rev. B}\ }\textbf {\bibinfo {volume} {93}},\ \bibinfo
  {pages} {205101} (\bibinfo {year} {2016})}\BibitemShut {NoStop}%
\bibitem [{\citenamefont {Seabra}\ \emph {et~al.}(2014)\citenamefont {Seabra},
  \citenamefont {Essler}, \citenamefont {Pollmann}, \citenamefont {Schneider},\
  and\ \citenamefont {Veness}}]{SeabraPRB14}%
  \BibitemOpen
  \bibfield  {author} {\bibinfo {author} {\bibfnamefont {L.}~\bibnamefont
  {Seabra}}, \bibinfo {author} {\bibfnamefont {F.~H.~L.}\ \bibnamefont
  {Essler}}, \bibinfo {author} {\bibfnamefont {F.}~\bibnamefont {Pollmann}},
  \bibinfo {author} {\bibfnamefont {I.}~\bibnamefont {Schneider}}, \ and\
  \bibinfo {author} {\bibfnamefont {T.}~\bibnamefont {Veness}},\ }\bibfield
  {title} {\enquote {\bibinfo {title} {{Real-time dynamics in the
  one-dimensional Hubbard model}},}\ }\href {\doibase
  10.1103/PhysRevB.90.245127} {\bibfield  {journal} {\bibinfo  {journal} {Phys.
  Rev. B}\ }\textbf {\bibinfo {volume} {90}},\ \bibinfo {pages} {245127}
  (\bibinfo {year} {2014})}\BibitemShut {NoStop}%
\bibitem [{\citenamefont {Caux}(2009)}]{CauxJMathPhys09}%
  \BibitemOpen
  \bibfield  {author} {\bibinfo {author} {\bibfnamefont {J.-S.}\ \bibnamefont
  {Caux}},\ }\bibfield  {title} {\enquote {\bibinfo {title} {{Correlation
  functions of integrable models: A description of the ABACUS algorithm}},}\
  }\href {\doibase /10.1063/1.3216474} {\bibfield  {journal} {\bibinfo
  {journal} {J. Math. Phys.}\ }\textbf {\bibinfo {volume} {50}},\ \bibinfo
  {pages} {095214} (\bibinfo {year} {2009})}\BibitemShut {NoStop}%
\bibitem [{\citenamefont {Pozsgay}\ \emph {et~al.}(2012)\citenamefont
  {Pozsgay}, \citenamefont {van Gerven~Oei},\ and\ \citenamefont
  {Kormos}}]{PozsgayJPhysA12}%
  \BibitemOpen
  \bibfield  {author} {\bibinfo {author} {\bibfnamefont {B.}~\bibnamefont
  {Pozsgay}}, \bibinfo {author} {\bibfnamefont {W.-V.}\ \bibnamefont {van
  Gerven~Oei}}, \ and\ \bibinfo {author} {\bibfnamefont {M.}~\bibnamefont
  {Kormos}},\ }\bibfield  {title} {\enquote {\bibinfo {title} {{On form factors
  in nested Bethe Ansatz systems}},}\ }\href
  {http://stacks.iop.org/1751-8121/45/i=46/a=465007} {\bibfield  {journal}
  {\bibinfo  {journal} {J. Phys. A}\ }\textbf {\bibinfo {volume} {45}},\
  \bibinfo {pages} {465007} (\bibinfo {year} {2012})}\BibitemShut {NoStop}%
\bibitem [{\citenamefont {Belliard}\ \emph {et~al.}(2012)\citenamefont
  {Belliard}, \citenamefont {Pakuliak}, \citenamefont {Ragoucy},\ and\
  \citenamefont {Slavnov}}]{BelliardJStatMech12}%
  \BibitemOpen
  \bibfield  {author} {\bibinfo {author} {\bibfnamefont {S.}~\bibnamefont
  {Belliard}}, \bibinfo {author} {\bibfnamefont {S.}~\bibnamefont {Pakuliak}},
  \bibinfo {author} {\bibfnamefont {E.}~\bibnamefont {Ragoucy}}, \ and\
  \bibinfo {author} {\bibfnamefont {N.~A.}\ \bibnamefont {Slavnov}},\
  }\bibfield  {title} {\enquote {\bibinfo {title} {{The algebraic Bethe ansatz
  for scalar products in SU(3)-invariant integrable models}},}\ }\href
  {http://stacks.iop.org/1742-5468/2012/i=10/a=P10017} {\bibfield  {journal}
  {\bibinfo  {journal} {J. Stat. Mech.}\ }\textbf {\bibinfo {volume} {2012}},\
  \bibinfo {pages} {P10017} (\bibinfo {year} {2012})}\BibitemShut {NoStop}%
\bibitem [{\citenamefont {Belliard}\ \emph {et~al.}(2013)\citenamefont
  {Belliard}, \citenamefont {Pakuliak}, \citenamefont {Ragoucy},\ and\
  \citenamefont {Slavnov}}]{BelliardJStatMech13}%
  \BibitemOpen
  \bibfield  {author} {\bibinfo {author} {\bibfnamefont {S.}~\bibnamefont
  {Belliard}}, \bibinfo {author} {\bibfnamefont {S.}~\bibnamefont {Pakuliak}},
  \bibinfo {author} {\bibfnamefont {E.}~\bibnamefont {Ragoucy}}, \ and\
  \bibinfo {author} {\bibfnamefont {N.~A.}\ \bibnamefont {Slavnov}},\
  }\bibfield  {title} {\enquote {\bibinfo {title} {{Form factors in
  SU(3)-invariant integrable models}},}\ }\href
  {http://stacks.iop.org/1742-5468/2013/i=04/a=P04033} {\bibfield  {journal}
  {\bibinfo  {journal} {J. Stat. Mech.}\ }\textbf {\bibinfo {volume} {2013}},\
  \bibinfo {pages} {P04033} (\bibinfo {year} {2013})}\BibitemShut {NoStop}%
\bibitem [{\citenamefont {Pakuliak}\ \emph {et~al.}(2014)\citenamefont
  {Pakuliak}, \citenamefont {Ragoucy},\ and\ \citenamefont
  {Slavnov}}]{PakuliakNuclPhysB14}%
  \BibitemOpen
  \bibfield  {author} {\bibinfo {author} {\bibfnamefont {S.}~\bibnamefont
  {Pakuliak}}, \bibinfo {author} {\bibfnamefont {E.}~\bibnamefont {Ragoucy}}, \
  and\ \bibinfo {author} {\bibfnamefont {N.A.}\ \bibnamefont {Slavnov}},\
  }\bibfield  {title} {\enquote {\bibinfo {title} {{Form factors in quantum
  integrable models with GL(3)-invariant R-matrix}},}\ }\href {\doibase
  http://dx.doi.org/10.1016/j.nuclphysb.2014.02.014} {\bibfield  {journal}
  {\bibinfo  {journal} {Nucl. Phys. B}\ }\textbf {\bibinfo {volume} {881}},\
  \bibinfo {pages} {343 -- 368} (\bibinfo {year} {2014})}\BibitemShut {NoStop}%
\bibitem [{\citenamefont {Pakuliak}\ \emph
  {et~al.}(2015{\natexlab{a}})\citenamefont {Pakuliak}, \citenamefont
  {Ragoucy},\ and\ \citenamefont {Slavnov}}]{PakuliakNuclPhysB15}%
  \BibitemOpen
  \bibfield  {author} {\bibinfo {author} {\bibfnamefont {S.}~\bibnamefont
  {Pakuliak}}, \bibinfo {author} {\bibfnamefont {E.}~\bibnamefont {Ragoucy}}, \
  and\ \bibinfo {author} {\bibfnamefont {N.~A.}\ \bibnamefont {Slavnov}},\
  }\bibfield  {title} {\enquote {\bibinfo {title} {{Zero modes method and form
  factors in quantum integrable models}},}\ }\href {\doibase
  http://dx.doi.org/10.1016/j.nuclphysb.2015.02.006} {\bibfield  {journal}
  {\bibinfo  {journal} {Nucl. Phys. B}\ }\textbf {\bibinfo {volume} {893}},\
  \bibinfo {pages} {459 -- 481} (\bibinfo {year}
  {2015}{\natexlab{a}})}\BibitemShut {NoStop}%
\bibitem [{\citenamefont {Pakuliak}\ \emph
  {et~al.}(2015{\natexlab{b}})\citenamefont {Pakuliak}, \citenamefont
  {Ragoucy},\ and\ \citenamefont {Slavnov}}]{PakuliakJPhysA15}%
  \BibitemOpen
  \bibfield  {author} {\bibinfo {author} {\bibfnamefont {S.}~\bibnamefont
  {Pakuliak}}, \bibinfo {author} {\bibfnamefont {E.}~\bibnamefont {Ragoucy}}, \
  and\ \bibinfo {author} {\bibfnamefont {N.~A.}\ \bibnamefont {Slavnov}},\
  }\bibfield  {title} {\enquote {\bibinfo {title} {{Form factors of local
  operators in a one-dimensional two-component Bose gas}},}\ }\href
  {http://stacks.iop.org/1751-8121/48/i=43/a=435001} {\bibfield  {journal}
  {\bibinfo  {journal} {J. Phys. A}\ }\textbf {\bibinfo {volume} {48}},\
  \bibinfo {pages} {435001} (\bibinfo {year} {2015}{\natexlab{b}})}\BibitemShut
  {NoStop}%
\bibitem [{\citenamefont {Pakuliak}\ \emph
  {et~al.}(2015{\natexlab{c}})\citenamefont {Pakuliak}, \citenamefont
  {Ragoucy},\ and\ \citenamefont {Slavnov}}]{PakuliakSIGMA15}%
  \BibitemOpen
  \bibfield  {author} {\bibinfo {author} {\bibfnamefont {S.}~\bibnamefont
  {Pakuliak}}, \bibinfo {author} {\bibfnamefont {E.}~\bibnamefont {Ragoucy}}, \
  and\ \bibinfo {author} {\bibfnamefont {N.~A.}\ \bibnamefont {Slavnov}},\
  }\bibfield  {title} {\enquote {\bibinfo {title} {{GL(3)-Based Quantum
  Integrable Composite Models. II. Form Factors of Local Operators}},}\ }\href
  {\doibase /10.3842/SIGMA.2015.064} {\bibfield  {journal} {\bibinfo  {journal}
  {SIGMA}\ }\textbf {\bibinfo {volume} {11}},\ \bibinfo {pages} {064} (\bibinfo
  {year} {2015}{\natexlab{c}})}\BibitemShut {NoStop}%
\bibitem [{\citenamefont {Hutsalyuk}\ \emph {et~al.}(2016)\citenamefont
  {Hutsalyuk}, \citenamefont {Liashyk}, \citenamefont {Pakuliak}, \citenamefont
  {Ragoucy},\ and\ \citenamefont {Slavnov}}]{HutsalyukNuclPhysB16}%
  \BibitemOpen
  \bibfield  {author} {\bibinfo {author} {\bibfnamefont {A.}~\bibnamefont
  {Hutsalyuk}}, \bibinfo {author} {\bibfnamefont {A.}~\bibnamefont {Liashyk}},
  \bibinfo {author} {\bibfnamefont {S.~Z.}\ \bibnamefont {Pakuliak}}, \bibinfo
  {author} {\bibfnamefont {E.}~\bibnamefont {Ragoucy}}, \ and\ \bibinfo
  {author} {\bibfnamefont {N.~A.}\ \bibnamefont {Slavnov}},\ }\bibfield
  {title} {\enquote {\bibinfo {title} {{Form factors of the monodromy matrix
  entries in $gl(2|1)$-invariant integrable models}},}\ }\href {\doibase
  /10.1016/j.nuclphysb.2016.08.025} {\bibfield  {journal} {\bibinfo  {journal}
  {Nucl. Phys. B}\ }\textbf {\bibinfo {volume} {911}},\ \bibinfo {pages} {902
  -- 927} (\bibinfo {year} {2016})}\BibitemShut {NoStop}%
\bibitem [{\citenamefont {Dorey}\ \emph {et~al.}(2007)\citenamefont {Dorey},
  \citenamefont {Dunning},\ and\ \citenamefont {Tateo}}]{DoreyJPhysA07}%
  \BibitemOpen
  \bibfield  {author} {\bibinfo {author} {\bibfnamefont {P.}~\bibnamefont
  {Dorey}}, \bibinfo {author} {\bibfnamefont {C.}~\bibnamefont {Dunning}}, \
  and\ \bibinfo {author} {\bibfnamefont {R.}~\bibnamefont {Tateo}},\ }\bibfield
   {title} {\enquote {\bibinfo {title} {{The ODE/IM correspondence}},}\ }\href
  {http://stacks.iop.org/1751-8121/40/i=32/a=R01} {\bibfield  {journal}
  {\bibinfo  {journal} {J. Phys. A}\ }\textbf {\bibinfo {volume} {40}},\
  \bibinfo {pages} {R205} (\bibinfo {year} {2007})}\BibitemShut {NoStop}%
\bibitem [{\citenamefont {{Rychkov}}(2011)}]{RychkovArxiv11}%
  \BibitemOpen
  \bibfield  {author} {\bibinfo {author} {\bibfnamefont {S.}~\bibnamefont
  {{Rychkov}}},\ }\bibfield  {title} {\enquote {\bibinfo {title} {{Conformal
  Bootstrap in Three Dimensions?}}}\ }\href@noop {} {\bibfield  {journal}
  {\bibinfo  {journal} {ArXiv e-prints}\ } (\bibinfo {year} {2011})},\ \Eprint
  {http://arxiv.org/abs/1111.2115} {arXiv:1111.2115 [hep-th]} \BibitemShut
  {NoStop}%
\bibitem [{\citenamefont {{Simmons-Duffin}}(2016)}]{SimmonsDuffinArxiv16}%
  \BibitemOpen
  \bibfield  {author} {\bibinfo {author} {\bibfnamefont {D.}~\bibnamefont
  {{Simmons-Duffin}}},\ }\bibfield  {title} {\enquote {\bibinfo {title} {{TASI
  Lectures on the Conformal Bootstrap}},}\ }\href@noop {} {\bibfield  {journal}
  {\bibinfo  {journal} {ArXiv e-prints}\ } (\bibinfo {year} {2016})},\ \Eprint
  {http://arxiv.org/abs/1602.07982} {arXiv:1602.07982 [hep-th]} \BibitemShut
  {NoStop}%
\bibitem [{\citenamefont {Poland}\ and\ \citenamefont
  {Simmons-Duffin}(2016)}]{PolandNatPhys16}%
  \BibitemOpen
  \bibfield  {author} {\bibinfo {author} {\bibfnamefont {D.}~\bibnamefont
  {Poland}}\ and\ \bibinfo {author} {\bibfnamefont {D.}~\bibnamefont
  {Simmons-Duffin}},\ }\bibfield  {title} {\enquote {\bibinfo {title} {{The
  conformal bootstrap}},}\ }\href {http://dx.doi.org/10.1038/nphys3761}
  {\bibfield  {journal} {\bibinfo  {journal} {Nature Phys.}\ ,\ \bibinfo
  {pages} {535--539}} (\bibinfo {year} {2016})}\BibitemShut {NoStop}%
\bibitem [{\citenamefont {El-Showk}\ \emph {et~al.}(2012)\citenamefont
  {El-Showk}, \citenamefont {Paulos}, \citenamefont {Poland}, \citenamefont
  {Rychkov}, \citenamefont {Simmons-Duffin},\ and\ \citenamefont
  {Vichi}}]{ElShowkPRD12}%
  \BibitemOpen
  \bibfield  {author} {\bibinfo {author} {\bibfnamefont {S.}~\bibnamefont
  {El-Showk}}, \bibinfo {author} {\bibfnamefont {M.~F.}\ \bibnamefont
  {Paulos}}, \bibinfo {author} {\bibfnamefont {D.}~\bibnamefont {Poland}},
  \bibinfo {author} {\bibfnamefont {S.}~\bibnamefont {Rychkov}}, \bibinfo
  {author} {\bibfnamefont {D.}~\bibnamefont {Simmons-Duffin}}, \ and\ \bibinfo
  {author} {\bibfnamefont {A.}~\bibnamefont {Vichi}},\ }\bibfield  {title}
  {\enquote {\bibinfo {title} {{Solving the 3D Ising model with the conformal
  bootstrap}},}\ }\href {\doibase 10.1103/PhysRevD.86.025022} {\bibfield
  {journal} {\bibinfo  {journal} {Phys. Rev. D}\ }\textbf {\bibinfo {volume}
  {86}},\ \bibinfo {pages} {025022} (\bibinfo {year} {2012})}\BibitemShut
  {NoStop}%
\bibitem [{\citenamefont {El-Showk}\ \emph {et~al.}(2014)\citenamefont
  {El-Showk}, \citenamefont {Paulos}, \citenamefont {Poland}, \citenamefont
  {Rychkov}, \citenamefont {Simmons-Duffin},\ and\ \citenamefont
  {Vichi}}]{El-Showk2014}%
  \BibitemOpen
  \bibfield  {author} {\bibinfo {author} {\bibfnamefont {S.}~\bibnamefont
  {El-Showk}}, \bibinfo {author} {\bibfnamefont {M.~F.}\ \bibnamefont
  {Paulos}}, \bibinfo {author} {\bibfnamefont {D.}~\bibnamefont {Poland}},
  \bibinfo {author} {\bibfnamefont {S.}~\bibnamefont {Rychkov}}, \bibinfo
  {author} {\bibfnamefont {D.}~\bibnamefont {Simmons-Duffin}}, \ and\ \bibinfo
  {author} {\bibfnamefont {A.}~\bibnamefont {Vichi}},\ }\bibfield  {title}
  {\enquote {\bibinfo {title} {{Solving the 3d Ising Model with the Conformal
  Bootstrap II. $c$-Minimization and Precise Critical Exponents}},}\ }\href
  {\doibase 10.1007/s10955-014-1042-7} {\bibfield  {journal} {\bibinfo
  {journal} {J. Stat. Phys.}\ }\textbf {\bibinfo {volume} {157}},\ \bibinfo
  {pages} {869--914} (\bibinfo {year} {2014})}\BibitemShut {NoStop}%
\bibitem [{\citenamefont {Arsenault}\ \emph {et~al.}(2014)\citenamefont
  {Arsenault}, \citenamefont {Lopez-Bezanilla}, \citenamefont {von
  Lilienfeld},\ and\ \citenamefont {Millis}}]{ArsenaultPRB14}%
  \BibitemOpen
  \bibfield  {author} {\bibinfo {author} {\bibfnamefont {L.-F.}\ \bibnamefont
  {Arsenault}}, \bibinfo {author} {\bibfnamefont {A.}~\bibnamefont
  {Lopez-Bezanilla}}, \bibinfo {author} {\bibfnamefont {O.~A.}\ \bibnamefont
  {von Lilienfeld}}, \ and\ \bibinfo {author} {\bibfnamefont {A.~J.}\
  \bibnamefont {Millis}},\ }\bibfield  {title} {\enquote {\bibinfo {title}
  {{Machine learning for many-body physics: The case of the Anderson impurity
  model}},}\ }\href {\doibase 10.1103/PhysRevB.90.155136} {\bibfield  {journal}
  {\bibinfo  {journal} {Phys. Rev. B}\ }\textbf {\bibinfo {volume} {90}},\
  \bibinfo {pages} {155136} (\bibinfo {year} {2014})}\BibitemShut {NoStop}%
\bibitem [{\citenamefont {{Mehta}}\ and\ \citenamefont
  {{Schwab}}(2014)}]{MentaArxiv14}%
  \BibitemOpen
  \bibfield  {author} {\bibinfo {author} {\bibfnamefont {P.}~\bibnamefont
  {{Mehta}}}\ and\ \bibinfo {author} {\bibfnamefont {D.~J.}\ \bibnamefont
  {{Schwab}}},\ }\bibfield  {title} {\enquote {\bibinfo {title} {{An exact
  mapping between the Variational Renormalization Group and Deep Learning}},}\
  }\href@noop {} {\bibfield  {journal} {\bibinfo  {journal} {ArXiv e-prints}\ }
  (\bibinfo {year} {2014})},\ \Eprint {http://arxiv.org/abs/1410.3831}
  {arXiv:1410.3831 [stat.ML]} \BibitemShut {NoStop}%
\bibitem [{\citenamefont {{Ch'ng}}\ \emph {et~al.}(2016)\citenamefont
  {{Ch'ng}}, \citenamefont {{Carrasquilla}}, \citenamefont {{Melko}},\ and\
  \citenamefont {{Khatami}}}]{ChngArxiv16}%
  \BibitemOpen
  \bibfield  {author} {\bibinfo {author} {\bibfnamefont {K.}~\bibnamefont
  {{Ch'ng}}}, \bibinfo {author} {\bibfnamefont {J.}~\bibnamefont
  {{Carrasquilla}}}, \bibinfo {author} {\bibfnamefont {R.~G.}\ \bibnamefont
  {{Melko}}}, \ and\ \bibinfo {author} {\bibfnamefont {E.}~\bibnamefont
  {{Khatami}}},\ }\bibfield  {title} {\enquote {\bibinfo {title} {{Machine
  Learning Phases of Strongly Correlated Fermions}},}\ }\href@noop {}
  {\bibfield  {journal} {\bibinfo  {journal} {ArXiv e-prints}\ } (\bibinfo
  {year} {2016})},\ \Eprint {http://arxiv.org/abs/1609.02552} {arXiv:1609.02552
  [cond-mat.str-el]} \BibitemShut {NoStop}%
\bibitem [{\citenamefont {{Carrasquilla}}\ and\ \citenamefont
  {{Melko}}(2017)}]{CarrasquillaArxiv16}%
  \BibitemOpen
  \bibfield  {author} {\bibinfo {author} {\bibfnamefont {J.}~\bibnamefont
  {{Carrasquilla}}}\ and\ \bibinfo {author} {\bibfnamefont {R.~G.}\
  \bibnamefont {{Melko}}},\ }\bibfield  {title} {\enquote {\bibinfo {title}
  {{Machine learning phases of matter}},}\ }\href
  {http://dx.doi.org/10.1038/nphys4035} {\bibfield  {journal} {\bibinfo
  {journal} {Nature Phys.}\ }\textbf {\bibinfo {volume} {advance online
  publication}},\ \bibinfo {pages} {4035} (\bibinfo {year} {2017})}\BibitemShut
  {NoStop}%
\bibitem [{\citenamefont {{Broecker}}\ \emph {et~al.}(2016)\citenamefont
  {{Broecker}}, \citenamefont {{Carrasquilla}}, \citenamefont {{Melko}},\ and\
  \citenamefont {{Trebst}}}]{BroeckerArxiv16}%
  \BibitemOpen
  \bibfield  {author} {\bibinfo {author} {\bibfnamefont {P.}~\bibnamefont
  {{Broecker}}}, \bibinfo {author} {\bibfnamefont {J.}~\bibnamefont
  {{Carrasquilla}}}, \bibinfo {author} {\bibfnamefont {R.~G.}\ \bibnamefont
  {{Melko}}}, \ and\ \bibinfo {author} {\bibfnamefont {S.}~\bibnamefont
  {{Trebst}}},\ }\bibfield  {title} {\enquote {\bibinfo {title} {{Machine
  learning quantum phases of matter beyond the fermion sign problem}},}\
  }\href@noop {} {\bibfield  {journal} {\bibinfo  {journal} {ArXiv e-prints}\ }
  (\bibinfo {year} {2016})},\ \Eprint {http://arxiv.org/abs/1608.07848}
  {arXiv:1608.07848 [cond-mat.str-el]} \BibitemShut {NoStop}%
\bibitem [{\citenamefont {Carleo}\ and\ \citenamefont
  {Troyer}(2017)}]{CarleoArxiv16}%
  \BibitemOpen
  \bibfield  {author} {\bibinfo {author} {\bibfnamefont {G.}~\bibnamefont
  {Carleo}}\ and\ \bibinfo {author} {\bibfnamefont {M.}~\bibnamefont
  {Troyer}},\ }\bibfield  {title} {\enquote {\bibinfo {title} {Solving the
  quantum many-body problem with artificial neural networks},}\ }\href
  {\doibase 10.1126/science.aag2302} {\bibfield  {journal} {\bibinfo  {journal}
  {Science}\ }\textbf {\bibinfo {volume} {355}},\ \bibinfo {pages} {602--606}
  (\bibinfo {year} {2017})}\BibitemShut {NoStop}%
\bibitem [{\citenamefont {Kusne}\ \emph {et~al.}(2014)\citenamefont {Kusne},
  \citenamefont {Gao}, \citenamefont {Mehta}, \citenamefont {Ke}, \citenamefont
  {Nguyen}, \citenamefont {Ho}, \citenamefont {Antropov}, \citenamefont {Wang},
  \citenamefont {Kramer}, \citenamefont {Long},\ and\ \citenamefont
  {Takeuchi}}]{KusneSciRep14}%
  \BibitemOpen
  \bibfield  {author} {\bibinfo {author} {\bibfnamefont {A.~G.}\ \bibnamefont
  {Kusne}}, \bibinfo {author} {\bibfnamefont {T.}~\bibnamefont {Gao}}, \bibinfo
  {author} {\bibfnamefont {A.}~\bibnamefont {Mehta}}, \bibinfo {author}
  {\bibfnamefont {L.}~\bibnamefont {Ke}}, \bibinfo {author} {\bibfnamefont
  {M.~C.}\ \bibnamefont {Nguyen}}, \bibinfo {author} {\bibfnamefont {K.-M.}\
  \bibnamefont {Ho}}, \bibinfo {author} {\bibfnamefont {V.}~\bibnamefont
  {Antropov}}, \bibinfo {author} {\bibfnamefont {C.-Z.}\ \bibnamefont {Wang}},
  \bibinfo {author} {\bibfnamefont {M.~J.}\ \bibnamefont {Kramer}}, \bibinfo
  {author} {\bibfnamefont {C.}~\bibnamefont {Long}}, \ and\ \bibinfo {author}
  {\bibfnamefont {I.}~\bibnamefont {Takeuchi}},\ }\bibfield  {title} {\enquote
  {\bibinfo {title} {{On-the-fly machine-learning for high-throughput
  experiments: search for rare-earth-free permanent magnets}},}\ }\href
  {\doibase 10.1038/srep06367} {\bibfield  {journal} {\bibinfo  {journal} {Sci.
  Rep.}\ }\textbf {\bibinfo {volume} {4}},\ \bibinfo {pages} {6367} (\bibinfo
  {year} {2014})}\BibitemShut {NoStop}%
\bibitem [{\citenamefont {Kalinin}\ \emph {et~al.}(2015)\citenamefont
  {Kalinin}, \citenamefont {Sumpter},\ and\ \citenamefont
  {Archibald}}]{KalininNatureMat15}%
  \BibitemOpen
  \bibfield  {author} {\bibinfo {author} {\bibfnamefont {S.~V.}\ \bibnamefont
  {Kalinin}}, \bibinfo {author} {\bibfnamefont {B.~G.}\ \bibnamefont
  {Sumpter}}, \ and\ \bibinfo {author} {\bibfnamefont {R.~K.}\ \bibnamefont
  {Archibald}},\ }\bibfield  {title} {\enquote {\bibinfo {title}
  {{Big-deep-smart data in imaging for guiding materials design}},}\ }\href
  {http://dx.doi.org/10.1038/nmat4395} {\bibfield  {journal} {\bibinfo
  {journal} {Nature Mater.}\ }\textbf {\bibinfo {volume} {14}},\ \bibinfo
  {pages} {973--980} (\bibinfo {year} {2015})}\BibitemShut {NoStop}%
\bibitem [{\citenamefont {Ghiringhelli}\ \emph {et~al.}(2015)\citenamefont
  {Ghiringhelli}, \citenamefont {Vybiral}, \citenamefont {Levchenko},
  \citenamefont {Draxl},\ and\ \citenamefont {Scheffler}}]{GhiringhelliPRL15}%
  \BibitemOpen
  \bibfield  {author} {\bibinfo {author} {\bibfnamefont {L.~M.}\ \bibnamefont
  {Ghiringhelli}}, \bibinfo {author} {\bibfnamefont {J.}~\bibnamefont
  {Vybiral}}, \bibinfo {author} {\bibfnamefont {S.~V.}\ \bibnamefont
  {Levchenko}}, \bibinfo {author} {\bibfnamefont {C.}~\bibnamefont {Draxl}}, \
  and\ \bibinfo {author} {\bibfnamefont {M.}~\bibnamefont {Scheffler}},\
  }\bibfield  {title} {\enquote {\bibinfo {title} {{Big Data of Materials
  Science: Critical Role of the Descriptor}},}\ }\href {\doibase
  10.1103/PhysRevLett.114.105503} {\bibfield  {journal} {\bibinfo  {journal}
  {Phys. Rev. Lett.}\ }\textbf {\bibinfo {volume} {114}},\ \bibinfo {pages}
  {105503} (\bibinfo {year} {2015})}\BibitemShut {NoStop}%
\bibitem [{\citenamefont {Stone}(1994)}]{StoneBosonization}%
  \BibitemOpen
  \bibfield  {author} {\bibinfo {author} {\bibfnamefont {M.}~\bibnamefont
  {Stone}},\ }\href {https://books.google.com/books?id=PFx-tWFiEBcC} {\emph
  {\bibinfo {title} {{Bosonization}}}}\ (\bibinfo  {publisher} {World
  Scientific},\ \bibinfo {year} {1994})\BibitemShut {NoStop}%
\bibitem [{\citenamefont {Di~Francesco}\ \emph {et~al.}(1996)\citenamefont
  {Di~Francesco}, \citenamefont {Mathieu},\ and\ \citenamefont
  {S{\'e}n{\'e}chal}}]{CFTBook}%
  \BibitemOpen
  \bibfield  {author} {\bibinfo {author} {\bibfnamefont {P.}~\bibnamefont
  {Di~Francesco}}, \bibinfo {author} {\bibfnamefont {P.}~\bibnamefont
  {Mathieu}}, \ and\ \bibinfo {author} {\bibfnamefont {D.}~\bibnamefont
  {S{\'e}n{\'e}chal}},\ }\href {\doibase 10.1007/978-1-4612-2256-9} {\emph
  {\bibinfo {title} {{Conformal Field Theory}}}}\ (\bibinfo  {publisher}
  {Springer-Verlag},\ \bibinfo {address} {New York},\ \bibinfo {year}
  {1996})\BibitemShut {NoStop}%
\bibitem [{\citenamefont {Mudry}(2014)}]{MudryBook}%
  \BibitemOpen
  \bibfield  {author} {\bibinfo {author} {\bibfnamefont {C.}~\bibnamefont
  {Mudry}},\ }\href {https://books.google.com/books?id=Yl6ClwEACAAJ} {\emph
  {\bibinfo {title} {{Lecture Notes on Field Theory in Condensed Matter
  Physics}}}}\ (\bibinfo  {publisher} {World Scientific Publishing Company Pte
  Limited},\ \bibinfo {year} {2014})\BibitemShut {NoStop}%
\bibitem [{\citenamefont {Vidal}\ \emph {et~al.}(2003)\citenamefont {Vidal},
  \citenamefont {Latorre}, \citenamefont {Rico},\ and\ \citenamefont
  {Kitaev}}]{VidalPRL03}%
  \BibitemOpen
  \bibfield  {author} {\bibinfo {author} {\bibfnamefont {G.}~\bibnamefont
  {Vidal}}, \bibinfo {author} {\bibfnamefont {J.~I.}\ \bibnamefont {Latorre}},
  \bibinfo {author} {\bibfnamefont {E.}~\bibnamefont {Rico}}, \ and\ \bibinfo
  {author} {\bibfnamefont {A.}~\bibnamefont {Kitaev}},\ }\bibfield  {title}
  {\enquote {\bibinfo {title} {{Entanglement in Quantum Critical Phenomena}},}\
  }\href {\doibase 10.1103/PhysRevLett.90.227902} {\bibfield  {journal}
  {\bibinfo  {journal} {Phys. Rev. Lett.}\ }\textbf {\bibinfo {volume} {90}},\
  \bibinfo {pages} {227902} (\bibinfo {year} {2003})}\BibitemShut {NoStop}%
\bibitem [{\citenamefont {Witten}(1984)}]{WittenCommMathPhys84}%
  \BibitemOpen
  \bibfield  {author} {\bibinfo {author} {\bibfnamefont {E.}~\bibnamefont
  {Witten}},\ }\bibfield  {title} {\enquote {\bibinfo {title} {{Nonabelian
  bosonization in two dimensions}},}\ }\href
  {http://projecteuclid.org/euclid.cmp/1103940923} {\bibfield  {journal}
  {\bibinfo  {journal} {Comm. Math. Phys.}\ }\textbf {\bibinfo {volume} {92}},\
  \bibinfo {pages} {455--472} (\bibinfo {year} {1984})}\BibitemShut {NoStop}%
\bibitem [{\citenamefont {Knizhnik}\ and\ \citenamefont
  {Zamolodchikov}(1984)}]{KnizhikNuclPhysB84}%
  \BibitemOpen
  \bibfield  {author} {\bibinfo {author} {\bibfnamefont {V.~G.}\ \bibnamefont
  {Knizhnik}}\ and\ \bibinfo {author} {\bibfnamefont {A.~B.}\ \bibnamefont
  {Zamolodchikov}},\ }\bibfield  {title} {\enquote {\bibinfo {title} {{Current
  algebra and Wess-Zumino model in two dimensions}},}\ }\href {\doibase
  http://dx.doi.org/10.1016/0550-3213(84)90374-2} {\bibfield  {journal}
  {\bibinfo  {journal} {Nucl. Phys. B}\ }\textbf {\bibinfo {volume} {247}},\
  \bibinfo {pages} {83 -- 103} (\bibinfo {year} {1984})}\BibitemShut {NoStop}%
\bibitem [{\citenamefont {Polyakov}\ and\ \citenamefont
  {Wiegmann}(1984)}]{PolyakovPhysLettB84}%
  \BibitemOpen
  \bibfield  {author} {\bibinfo {author} {\bibfnamefont {A.~M.}\ \bibnamefont
  {Polyakov}}\ and\ \bibinfo {author} {\bibfnamefont {P.~B.}\ \bibnamefont
  {Wiegmann}},\ }\bibfield  {title} {\enquote {\bibinfo {title} {{Goldstone
  fields in two dimensions with multivalued actions}},}\ }\href {\doibase
  http://dx.doi.org/10.1016/0370-2693(84)90206-5} {\bibfield  {journal}
  {\bibinfo  {journal} {Phys. Lett. B}\ }\textbf {\bibinfo {volume} {141}},\
  \bibinfo {pages} {223 -- 228} (\bibinfo {year} {1984})}\BibitemShut {NoStop}%
\bibitem [{\citenamefont {Affleck}(1985)}]{AffleckPRL85}%
  \BibitemOpen
  \bibfield  {author} {\bibinfo {author} {\bibfnamefont {I.}~\bibnamefont
  {Affleck}},\ }\bibfield  {title} {\enquote {\bibinfo {title} {{Critical
  Behavior of Two-Dimensional Systems with Continuous Symmetries}},}\ }\href
  {\doibase 10.1103/PhysRevLett.55.1355} {\bibfield  {journal} {\bibinfo
  {journal} {Phys. Rev. Lett.}\ }\textbf {\bibinfo {volume} {55}},\ \bibinfo
  {pages} {1355--1358} (\bibinfo {year} {1985})}\BibitemShut {NoStop}%
\bibitem [{\citenamefont {Affleck}(1986{\natexlab{a}})}]{AffleckNuclPhysB86}%
  \BibitemOpen
  \bibfield  {author} {\bibinfo {author} {\bibfnamefont {I.}~\bibnamefont
  {Affleck}},\ }\bibfield  {title} {\enquote {\bibinfo {title} {{Exact critical
  exponents for quantum spin chains, non-linear $\sigma$-models at $\theta=\pi$
  and the quantum hall effect}},}\ }\href {\doibase
  http://dx.doi.org/10.1016/0550-3213(86)90167-7} {\bibfield  {journal}
  {\bibinfo  {journal} {Nucl. Phys. B}\ }\textbf {\bibinfo {volume} {265}},\
  \bibinfo {pages} {409 -- 447} (\bibinfo {year}
  {1986}{\natexlab{a}})}\BibitemShut {NoStop}%
\bibitem [{\citenamefont {Affleck}\ and\ \citenamefont
  {Haldane}(1987)}]{AffleckPRB87}%
  \BibitemOpen
  \bibfield  {author} {\bibinfo {author} {\bibfnamefont {I.}~\bibnamefont
  {Affleck}}\ and\ \bibinfo {author} {\bibfnamefont {F.~D.~M.}\ \bibnamefont
  {Haldane}},\ }\bibfield  {title} {\enquote {\bibinfo {title} {{Critical
  theory of quantum spin chains}},}\ }\href {\doibase 10.1103/PhysRevB.36.5291}
  {\bibfield  {journal} {\bibinfo  {journal} {Phys. Rev. B}\ }\textbf {\bibinfo
  {volume} {36}},\ \bibinfo {pages} {5291--5300} (\bibinfo {year}
  {1987})}\BibitemShut {NoStop}%
\bibitem [{\citenamefont {Fradkin}\ \emph {et~al.}(1989)\citenamefont
  {Fradkin}, \citenamefont {von Reichenbach},\ and\ \citenamefont
  {Schaposnik}}]{FradkinNuclPhysB89}%
  \BibitemOpen
  \bibfield  {author} {\bibinfo {author} {\bibfnamefont {E.}~\bibnamefont
  {Fradkin}}, \bibinfo {author} {\bibfnamefont {C.}~\bibnamefont {von
  Reichenbach}}, \ and\ \bibinfo {author} {\bibfnamefont {F.~A.}\ \bibnamefont
  {Schaposnik}},\ }\bibfield  {title} {\enquote {\bibinfo {title}
  {{Bosonization of the Kondo problem}},}\ }\href {\doibase
  http://dx.doi.org/10.1016/0550-3213(89)90065-5} {\bibfield  {journal}
  {\bibinfo  {journal} {Nucl. Phys. B}\ }\textbf {\bibinfo {volume} {316}},\
  \bibinfo {pages} {710 -- 734} (\bibinfo {year} {1989})}\BibitemShut {NoStop}%
\bibitem [{\citenamefont {Affleck}(1990)}]{AffleckNuclPhysB90}%
  \BibitemOpen
  \bibfield  {author} {\bibinfo {author} {\bibfnamefont {I.}~\bibnamefont
  {Affleck}},\ }\bibfield  {title} {\enquote {\bibinfo {title} {{A current
  algebra approach to the Kondo effect}},}\ }\href {\doibase
  http://dx.doi.org/10.1016/0550-3213(90)90440-O} {\bibfield  {journal}
  {\bibinfo  {journal} {Nucl. Phys. B}\ }\textbf {\bibinfo {volume} {336}},\
  \bibinfo {pages} {517 -- 532} (\bibinfo {year} {1990})}\BibitemShut {NoStop}%
\bibitem [{\citenamefont {Affleck}\ and\ \citenamefont
  {Ludwig}(1991{\natexlab{a}})}]{AffleckLudwigNuclPhysB91a}%
  \BibitemOpen
  \bibfield  {author} {\bibinfo {author} {\bibfnamefont {I.}~\bibnamefont
  {Affleck}}\ and\ \bibinfo {author} {\bibfnamefont {A.~W.~W.}\ \bibnamefont
  {Ludwig}},\ }\bibfield  {title} {\enquote {\bibinfo {title} {{The Kondo
  effect, conformal field theory and fusion rules}},}\ }\href {\doibase
  http://dx.doi.org/10.1016/0550-3213(91)90109-B} {\bibfield  {journal}
  {\bibinfo  {journal} {Nucl. Phys. B}\ }\textbf {\bibinfo {volume} {352}},\
  \bibinfo {pages} {849 -- 862} (\bibinfo {year}
  {1991}{\natexlab{a}})}\BibitemShut {NoStop}%
\bibitem [{\citenamefont {Affleck}\ and\ \citenamefont
  {Ludwig}(1991{\natexlab{b}})}]{AffleckLudwigNuclPhysB91b}%
  \BibitemOpen
  \bibfield  {author} {\bibinfo {author} {\bibfnamefont {I.}~\bibnamefont
  {Affleck}}\ and\ \bibinfo {author} {\bibfnamefont {A.~W.~W.}\ \bibnamefont
  {Ludwig}},\ }\bibfield  {title} {\enquote {\bibinfo {title} {{Critical theory
  of overscreened Kondo fixed points}},}\ }\href {\doibase
  http://dx.doi.org/10.1016/0550-3213(91)90419-X} {\bibfield  {journal}
  {\bibinfo  {journal} {Nucl. Phys. B}\ }\textbf {\bibinfo {volume} {360}},\
  \bibinfo {pages} {641 -- 696} (\bibinfo {year}
  {1991}{\natexlab{b}})}\BibitemShut {NoStop}%
\bibitem [{\citenamefont {Ludwig}\ and\ \citenamefont
  {Affleck}(1991)}]{LudwigAffleckPRL91}%
  \BibitemOpen
  \bibfield  {author} {\bibinfo {author} {\bibfnamefont {A.~W.~W.}\
  \bibnamefont {Ludwig}}\ and\ \bibinfo {author} {\bibfnamefont
  {I.}~\bibnamefont {Affleck}},\ }\bibfield  {title} {\enquote {\bibinfo
  {title} {{Exact, asymptotic, three-dimensional, space- and time-dependent,
  Green's functions in the multichannel Kondo effect}},}\ }\href {\doibase
  10.1103/PhysRevLett.67.3160} {\bibfield  {journal} {\bibinfo  {journal}
  {Phys. Rev. Lett.}\ }\textbf {\bibinfo {volume} {67}},\ \bibinfo {pages}
  {3160--3163} (\bibinfo {year} {1991})}\BibitemShut {NoStop}%
\bibitem [{\citenamefont {Affleck}(1995)}]{AffleckActaPhysPolon95}%
  \BibitemOpen
  \bibfield  {author} {\bibinfo {author} {\bibfnamefont {I.}~\bibnamefont
  {Affleck}},\ }\bibfield  {title} {\enquote {\bibinfo {title} {{Conformal
  Field Theory Approach to the Kondo Effect}},}\ }\href
  {http://www.actaphys.uj.edu.pl/fulltext?series=Reg&vol=26&page=1869}
  {\bibfield  {journal} {\bibinfo  {journal} {Acta Phys. Polon.}\ }\textbf
  {\bibinfo {volume} {26}},\ \bibinfo {pages} {1869--1932} (\bibinfo {year}
  {1995})}\BibitemShut {NoStop}%
\bibitem [{\citenamefont {Affleck}\ \emph {et~al.}(1995)\citenamefont
  {Affleck}, \citenamefont {Ludwig},\ and\ \citenamefont
  {Jones}}]{AffleckPRB95}%
  \BibitemOpen
  \bibfield  {author} {\bibinfo {author} {\bibfnamefont {I.}~\bibnamefont
  {Affleck}}, \bibinfo {author} {\bibfnamefont {A.~W.~W.}\ \bibnamefont
  {Ludwig}}, \ and\ \bibinfo {author} {\bibfnamefont {B.~A.}\ \bibnamefont
  {Jones}},\ }\bibfield  {title} {\enquote {\bibinfo {title}
  {{Conformal-field-theory approach to the two-impurity Kondo problem:
  Comparison with numerical renormalization-group results}},}\ }\href {\doibase
  10.1103/PhysRevB.52.9528} {\bibfield  {journal} {\bibinfo  {journal} {Phys.
  Rev. B}\ }\textbf {\bibinfo {volume} {52}},\ \bibinfo {pages} {9528--9546}
  (\bibinfo {year} {1995})}\BibitemShut {NoStop}%
\bibitem [{\citenamefont {Andrei}\ and\ \citenamefont
  {Orignac}(2000)}]{AndreiPRB00}%
  \BibitemOpen
  \bibfield  {author} {\bibinfo {author} {\bibfnamefont {N.}~\bibnamefont
  {Andrei}}\ and\ \bibinfo {author} {\bibfnamefont {E.}~\bibnamefont
  {Orignac}},\ }\bibfield  {title} {\enquote {\bibinfo {title} {{Low-energy
  dynamics of the one-dimensional multichannel Kondo-Heisenberg lattice}},}\
  }\href {\doibase 10.1103/PhysRevB.62.R3596} {\bibfield  {journal} {\bibinfo
  {journal} {Phys. Rev. B}\ }\textbf {\bibinfo {volume} {62}},\ \bibinfo
  {pages} {R3596--R3599} (\bibinfo {year} {2000})}\BibitemShut {NoStop}%
\bibitem [{\citenamefont {Ingersent}\ \emph {et~al.}(2005)\citenamefont
  {Ingersent}, \citenamefont {Ludwig},\ and\ \citenamefont
  {Affleck}}]{IngersentPRL05}%
  \BibitemOpen
  \bibfield  {author} {\bibinfo {author} {\bibfnamefont {K.}~\bibnamefont
  {Ingersent}}, \bibinfo {author} {\bibfnamefont {A.~W.~W.}\ \bibnamefont
  {Ludwig}}, \ and\ \bibinfo {author} {\bibfnamefont {I.}~\bibnamefont
  {Affleck}},\ }\bibfield  {title} {\enquote {\bibinfo {title} {{Kondo
  Screening in a Magnetically Frustrated Nanostructure: Exact Results on a
  Stable Non-Fermi-Liquid Phase}},}\ }\href {\doibase
  10.1103/PhysRevLett.95.257204} {\bibfield  {journal} {\bibinfo  {journal}
  {Phys. Rev. Lett.}\ }\textbf {\bibinfo {volume} {95}},\ \bibinfo {pages}
  {257204} (\bibinfo {year} {2005})}\BibitemShut {NoStop}%
\bibitem [{\citenamefont {Ferrero}\ \emph {et~al.}(2007)\citenamefont
  {Ferrero}, \citenamefont {Leo}, \citenamefont {Lecheminant},\ and\
  \citenamefont {Fabrizio}}]{FerreroJPhysCondMatt07}%
  \BibitemOpen
  \bibfield  {author} {\bibinfo {author} {\bibfnamefont {M.}~\bibnamefont
  {Ferrero}}, \bibinfo {author} {\bibfnamefont {L.~De}\ \bibnamefont {Leo}},
  \bibinfo {author} {\bibfnamefont {P.}~\bibnamefont {Lecheminant}}, \ and\
  \bibinfo {author} {\bibfnamefont {M.}~\bibnamefont {Fabrizio}},\ }\bibfield
  {title} {\enquote {\bibinfo {title} {{Strong correlations in a nutshell}},}\
  }\href {http://stacks.iop.org/0953-8984/19/i=43/a=433201} {\bibfield
  {journal} {\bibinfo  {journal} {J. Phys. Cond. Matt.}\ }\textbf {\bibinfo
  {volume} {19}},\ \bibinfo {pages} {433201} (\bibinfo {year}
  {2007})}\BibitemShut {NoStop}%
\bibitem [{\citenamefont {Fujimoto}\ and\ \citenamefont
  {Kawakami}(1994)}]{FujimotoJPSJ94}%
  \BibitemOpen
  \bibfield  {author} {\bibinfo {author} {\bibfnamefont {S.}~\bibnamefont
  {Fujimoto}}\ and\ \bibinfo {author} {\bibfnamefont {N.}~\bibnamefont
  {Kawakami}},\ }\bibfield  {title} {\enquote {\bibinfo {title} {{Bosonization
  Approach to the One-Dimensional Kondo Lattice Model}},}\ }\href {\doibase
  10.1143/JPSJ.63.4322} {\bibfield  {journal} {\bibinfo  {journal} {J. Phys.
  Soc. Jpn}\ }\textbf {\bibinfo {volume} {63}},\ \bibinfo {pages} {4322--4326}
  (\bibinfo {year} {1994})}\BibitemShut {NoStop}%
\bibitem [{\citenamefont {{Bernard}}(1995)}]{BernardArxiv95}%
  \BibitemOpen
  \bibfield  {author} {\bibinfo {author} {\bibfnamefont {D.}~\bibnamefont
  {{Bernard}}},\ }\bibfield  {title} {\enquote {\bibinfo {title} {{(Perturbed)
  Conformal Field Theory Applied To 2D Disordered Systems: An Introduction}},}\
  }\href {http://adsabs.harvard.edu/abs/1995hep.th....9137B} {\bibfield
  {journal} {\bibinfo  {journal} {ArXiv e-prints}\ } (\bibinfo {year}
  {1995})},\ \Eprint {http://arxiv.org/abs/hep-th/9509137} {hep-th/9509137}
  \BibitemShut {NoStop}%
\bibitem [{\citenamefont {Caux}\ \emph {et~al.}(1996)\citenamefont {Caux},
  \citenamefont {Kogan},\ and\ \citenamefont {Tsvelik}}]{CauxNuclPhysB96}%
  \BibitemOpen
  \bibfield  {author} {\bibinfo {author} {\bibfnamefont {J.-S.}\ \bibnamefont
  {Caux}}, \bibinfo {author} {\bibfnamefont {I.~I.}\ \bibnamefont {Kogan}}, \
  and\ \bibinfo {author} {\bibfnamefont {A.~M.}\ \bibnamefont {Tsvelik}},\
  }\bibfield  {title} {\enquote {\bibinfo {title} {{Logarithmic operators and
  hidden continuous symmetry in critical disordered models}},}\ }\href
  {\doibase http://dx.doi.org/10.1016/0550-3213(96)00118-6} {\bibfield
  {journal} {\bibinfo  {journal} {Nucl. Phys. B}\ }\textbf {\bibinfo {volume}
  {466}},\ \bibinfo {pages} {444 -- 462} (\bibinfo {year} {1996})}\BibitemShut
  {NoStop}%
\bibitem [{\citenamefont {Mudry}\ \emph {et~al.}(1996)\citenamefont {Mudry},
  \citenamefont {Chamon},\ and\ \citenamefont {Wen}}]{MudryNuclPhysB96}%
  \BibitemOpen
  \bibfield  {author} {\bibinfo {author} {\bibfnamefont {C.}~\bibnamefont
  {Mudry}}, \bibinfo {author} {\bibfnamefont {C.}~\bibnamefont {Chamon}}, \
  and\ \bibinfo {author} {\bibfnamefont {X.-G.}\ \bibnamefont {Wen}},\
  }\bibfield  {title} {\enquote {\bibinfo {title} {{Two-dimensional conformal
  field theory for disordered systems at criticality}},}\ }\href {\doibase
  http://dx.doi.org/10.1016/0550-3213(96)00128-9} {\bibfield  {journal}
  {\bibinfo  {journal} {Nucl. Phys. B}\ }\textbf {\bibinfo {volume} {466}},\
  \bibinfo {pages} {383 -- 443} (\bibinfo {year} {1996})}\BibitemShut {NoStop}%
\bibitem [{\citenamefont {Caux}\ \emph
  {et~al.}(1998{\natexlab{a}})\citenamefont {Caux}, \citenamefont {Taniguchi},\
  and\ \citenamefont {Tsvelik}}]{CauxPRL98}%
  \BibitemOpen
  \bibfield  {author} {\bibinfo {author} {\bibfnamefont {J.-S.}\ \bibnamefont
  {Caux}}, \bibinfo {author} {\bibfnamefont {N.}~\bibnamefont {Taniguchi}}, \
  and\ \bibinfo {author} {\bibfnamefont {A.~M.}\ \bibnamefont {Tsvelik}},\
  }\bibfield  {title} {\enquote {\bibinfo {title} {{Termination of Multifractal
  behavior for Critical Disordered Dirac Fermions}},}\ }\href {\doibase
  10.1103/PhysRevLett.80.1276} {\bibfield  {journal} {\bibinfo  {journal}
  {Phys. Rev. Lett.}\ }\textbf {\bibinfo {volume} {80}},\ \bibinfo {pages}
  {1276--1279} (\bibinfo {year} {1998}{\natexlab{a}})}\BibitemShut {NoStop}%
\bibitem [{\citenamefont {Caux}\ \emph
  {et~al.}(1998{\natexlab{b}})\citenamefont {Caux}, \citenamefont {Taniguchi},\
  and\ \citenamefont {Tsvelik}}]{CauxNuclPhysB98}%
  \BibitemOpen
  \bibfield  {author} {\bibinfo {author} {\bibfnamefont {J.-S.}\ \bibnamefont
  {Caux}}, \bibinfo {author} {\bibfnamefont {N.}~\bibnamefont {Taniguchi}}, \
  and\ \bibinfo {author} {\bibfnamefont {A.~M.}\ \bibnamefont {Tsvelik}},\
  }\bibfield  {title} {\enquote {\bibinfo {title} {{Disordered Dirac fermions:
  Multifractality termination and logarithmic conformal field theories}},}\
  }\href {\doibase http://dx.doi.org/10.1016/S0550-3213(98)00331-9} {\bibfield
  {journal} {\bibinfo  {journal} {Nucl. Phys. B}\ }\textbf {\bibinfo {volume}
  {525}},\ \bibinfo {pages} {671 -- 696} (\bibinfo {year}
  {1998}{\natexlab{b}})}\BibitemShut {NoStop}%
\bibitem [{\citenamefont {Caux}(1998)}]{CauxPRL98b}%
  \BibitemOpen
  \bibfield  {author} {\bibinfo {author} {\bibfnamefont {J.-S.}\ \bibnamefont
  {Caux}},\ }\bibfield  {title} {\enquote {\bibinfo {title} {{Exact
  Multifractality for Disordered $\mathit{N}$-Flavor Dirac Fermions in Two
  Dimensions}},}\ }\href {\doibase 10.1103/PhysRevLett.81.4196} {\bibfield
  {journal} {\bibinfo  {journal} {Phys. Rev. Lett.}\ }\textbf {\bibinfo
  {volume} {81}},\ \bibinfo {pages} {4196--4199} (\bibinfo {year}
  {1998})}\BibitemShut {NoStop}%
\bibitem [{\citenamefont {Bhaseen}\ \emph {et~al.}(2001)\citenamefont
  {Bhaseen}, \citenamefont {Caux}, \citenamefont {Kogan},\ and\ \citenamefont
  {Tsvelik}}]{BhaseenNuclPhysB01}%
  \BibitemOpen
  \bibfield  {author} {\bibinfo {author} {\bibfnamefont {M.~J.}\ \bibnamefont
  {Bhaseen}}, \bibinfo {author} {\bibfnamefont {J.-S.}\ \bibnamefont {Caux}},
  \bibinfo {author} {\bibfnamefont {I.~I.}\ \bibnamefont {Kogan}}, \ and\
  \bibinfo {author} {\bibfnamefont {A.~M.}\ \bibnamefont {Tsvelik}},\
  }\bibfield  {title} {\enquote {\bibinfo {title} {{Disordered Dirac fermions:
  the marriage of three different approaches}},}\ }\href {\doibase
  http://dx.doi.org/10.1016/S0550-3213(01)00432-1} {\bibfield  {journal}
  {\bibinfo  {journal} {Nucl. Phys. B}\ }\textbf {\bibinfo {volume} {618}},\
  \bibinfo {pages} {465 -- 499} (\bibinfo {year} {2001})}\BibitemShut {NoStop}%
\bibitem [{\citenamefont {Nersesyan}\ \emph {et~al.}(1994)\citenamefont
  {Nersesyan}, \citenamefont {Tsvelik},\ and\ \citenamefont
  {Wenger}}]{NersesyanPRL94}%
  \BibitemOpen
  \bibfield  {author} {\bibinfo {author} {\bibfnamefont {A.~A.}\ \bibnamefont
  {Nersesyan}}, \bibinfo {author} {\bibfnamefont {A.~M.}\ \bibnamefont
  {Tsvelik}}, \ and\ \bibinfo {author} {\bibfnamefont {F.}~\bibnamefont
  {Wenger}},\ }\bibfield  {title} {\enquote {\bibinfo {title} {{Disorder
  effects in two-dimensional $d$-wave superconductors}},}\ }\href {\doibase
  10.1103/PhysRevLett.72.2628} {\bibfield  {journal} {\bibinfo  {journal}
  {Phys. Rev. Lett.}\ }\textbf {\bibinfo {volume} {72}},\ \bibinfo {pages}
  {2628--2631} (\bibinfo {year} {1994})}\BibitemShut {NoStop}%
\bibitem [{\citenamefont {Nersesyan}\ \emph {et~al.}(1995)\citenamefont
  {Nersesyan}, \citenamefont {Tsvelik},\ and\ \citenamefont
  {Wenger}}]{NersesyanNuclPhysB95}%
  \BibitemOpen
  \bibfield  {author} {\bibinfo {author} {\bibfnamefont {A.~A.}\ \bibnamefont
  {Nersesyan}}, \bibinfo {author} {\bibfnamefont {A.~M.}\ \bibnamefont
  {Tsvelik}}, \ and\ \bibinfo {author} {\bibfnamefont {F.}~\bibnamefont
  {Wenger}},\ }\bibfield  {title} {\enquote {\bibinfo {title} {{Disorder
  effects in two-dimensional Fermi systems with conical spectrum: exact results
  for the density of states}},}\ }\href {\doibase
  http://dx.doi.org/10.1016/0550-3213(95)00002-A} {\bibfield  {journal}
  {\bibinfo  {journal} {Nucl. Phys. B}\ }\textbf {\bibinfo {volume} {438}},\
  \bibinfo {pages} {561 -- 588} (\bibinfo {year} {1995})}\BibitemShut {NoStop}%
\bibitem [{\citenamefont {Altland}\ \emph {et~al.}(2002)\citenamefont
  {Altland}, \citenamefont {Simons},\ and\ \citenamefont
  {Zirnbauer}}]{AltlandPhysRep02}%
  \BibitemOpen
  \bibfield  {author} {\bibinfo {author} {\bibfnamefont {A.}~\bibnamefont
  {Altland}}, \bibinfo {author} {\bibfnamefont {B.~D.}\ \bibnamefont {Simons}},
  \ and\ \bibinfo {author} {\bibfnamefont {M.~R.}\ \bibnamefont {Zirnbauer}},\
  }\bibfield  {title} {\enquote {\bibinfo {title} {{Theories of low-energy
  quasi-particle states in disordered d-wave superconductors}},}\ }\href
  {\doibase http://dx.doi.org/10.1016/S0370-1573(01)00065-5} {\bibfield
  {journal} {\bibinfo  {journal} {Phys. Rep.}\ }\textbf {\bibinfo {volume}
  {359}},\ \bibinfo {pages} {283 -- 354} (\bibinfo {year} {2002})}\BibitemShut
  {NoStop}%
\bibitem [{\citenamefont {Guruswamy}\ \emph {et~al.}(2000)\citenamefont
  {Guruswamy}, \citenamefont {LeClair},\ and\ \citenamefont
  {Ludwig}}]{GuruswamyNuclPhysB00}%
  \BibitemOpen
  \bibfield  {author} {\bibinfo {author} {\bibfnamefont {S.}~\bibnamefont
  {Guruswamy}}, \bibinfo {author} {\bibfnamefont {A.}~\bibnamefont {LeClair}},
  \ and\ \bibinfo {author} {\bibfnamefont {A.~W.~W.}\ \bibnamefont {Ludwig}},\
  }\bibfield  {title} {\enquote {\bibinfo {title} {{gl$(N|N)$ Super-current
  algebras for disordered Dirac fermions in two dimensions}},}\ }\href
  {\doibase http://dx.doi.org/10.1016/S0550-3213(00)00245-5} {\bibfield
  {journal} {\bibinfo  {journal} {Nucl. Phys. B}\ }\textbf {\bibinfo {volume}
  {583}},\ \bibinfo {pages} {475 -- 512} (\bibinfo {year} {2000})}\BibitemShut
  {NoStop}%
\bibitem [{\citenamefont {Ludwig}\ \emph {et~al.}(1994)\citenamefont {Ludwig},
  \citenamefont {Fisher}, \citenamefont {Shankar},\ and\ \citenamefont
  {Grinstein}}]{LudwigPRB94}%
  \BibitemOpen
  \bibfield  {author} {\bibinfo {author} {\bibfnamefont {A.~W.~W.}\
  \bibnamefont {Ludwig}}, \bibinfo {author} {\bibfnamefont {M.~P.~A.}\
  \bibnamefont {Fisher}}, \bibinfo {author} {\bibfnamefont {R.}~\bibnamefont
  {Shankar}}, \ and\ \bibinfo {author} {\bibfnamefont {G.}~\bibnamefont
  {Grinstein}},\ }\bibfield  {title} {\enquote {\bibinfo {title} {{Integer
  quantum Hall transition: An alternative approach and exact results}},}\
  }\href {\doibase 10.1103/PhysRevB.50.7526} {\bibfield  {journal} {\bibinfo
  {journal} {Phys. Rev. B}\ }\textbf {\bibinfo {volume} {50}},\ \bibinfo
  {pages} {7526--7552} (\bibinfo {year} {1994})}\BibitemShut {NoStop}%
\bibitem [{\citenamefont {Wen}(1990{\natexlab{a}})}]{WenPRL90}%
  \BibitemOpen
  \bibfield  {author} {\bibinfo {author} {\bibfnamefont {X.~G.}\ \bibnamefont
  {Wen}},\ }\bibfield  {title} {\enquote {\bibinfo {title} {{Electrodynamical
  properties of gapless edge excitations in the fractional quantum Hall
  states}},}\ }\href {\doibase 10.1103/PhysRevLett.64.2206} {\bibfield
  {journal} {\bibinfo  {journal} {Phys. Rev. Lett.}\ }\textbf {\bibinfo
  {volume} {64}},\ \bibinfo {pages} {2206--2209} (\bibinfo {year}
  {1990}{\natexlab{a}})}\BibitemShut {NoStop}%
\bibitem [{\citenamefont {Wen}(1990{\natexlab{b}})}]{WenPRB90}%
  \BibitemOpen
  \bibfield  {author} {\bibinfo {author} {\bibfnamefont {X.~G.}\ \bibnamefont
  {Wen}},\ }\bibfield  {title} {\enquote {\bibinfo {title} {{Chiral Luttinger
  liquid and the edge excitations in the fractional quantum Hall states}},}\
  }\href {\doibase 10.1103/PhysRevB.41.12838} {\bibfield  {journal} {\bibinfo
  {journal} {Phys. Rev. B}\ }\textbf {\bibinfo {volume} {41}},\ \bibinfo
  {pages} {12838--12844} (\bibinfo {year} {1990}{\natexlab{b}})}\BibitemShut
  {NoStop}%
\bibitem [{\citenamefont {Stone}(1991)}]{StoneAnnPhys91}%
  \BibitemOpen
  \bibfield  {author} {\bibinfo {author} {\bibfnamefont {M.}~\bibnamefont
  {Stone}},\ }\bibfield  {title} {\enquote {\bibinfo {title} {{Edge waves in
  the quantum Hall effect}},}\ }\href {\doibase
  http://dx.doi.org/10.1016/0003-4916(91)90177-A} {\bibfield  {journal}
  {\bibinfo  {journal} {Ann. Phys. (N.Y.)}\ }\textbf {\bibinfo {volume}
  {207}},\ \bibinfo {pages} {38 -- 52} (\bibinfo {year} {1991})}\BibitemShut
  {NoStop}%
\bibitem [{\citenamefont {Wen}(1992)}]{WenIntJModPhysB92}%
  \BibitemOpen
  \bibfield  {author} {\bibinfo {author} {\bibfnamefont {X.-G.}\ \bibnamefont
  {Wen}},\ }\bibfield  {title} {\enquote {\bibinfo {title} {{Theory of the edge
  states in fractional quantum Hall effects}},}\ }\href {\doibase
  10.1142/S0217979292000840} {\bibfield  {journal} {\bibinfo  {journal} {Int.
  J. Mod. Phys. B}\ }\textbf {\bibinfo {volume} {06}},\ \bibinfo {pages}
  {1711--1762} (\bibinfo {year} {1992})}\BibitemShut {NoStop}%
\bibitem [{\citenamefont {Wen}(1995)}]{WenAdvPhys95}%
  \BibitemOpen
  \bibfield  {author} {\bibinfo {author} {\bibfnamefont {X.-G.}\ \bibnamefont
  {Wen}},\ }\bibfield  {title} {\enquote {\bibinfo {title} {{Topological orders
  and edge excitations in fractional quantum Hall states}},}\ }\href {\doibase
  10.1080/00018739500101566} {\bibfield  {journal} {\bibinfo  {journal} {Adv.
  Phys.}\ }\textbf {\bibinfo {volume} {44}},\ \bibinfo {pages} {405--473}
  (\bibinfo {year} {1995})}\BibitemShut {NoStop}%
\bibitem [{\citenamefont {Kane}\ \emph {et~al.}(2002)\citenamefont {Kane},
  \citenamefont {Mukhopadhyay},\ and\ \citenamefont {Lubensky}}]{KanePRL02}%
  \BibitemOpen
  \bibfield  {author} {\bibinfo {author} {\bibfnamefont {C.~L.}\ \bibnamefont
  {Kane}}, \bibinfo {author} {\bibfnamefont {R.}~\bibnamefont {Mukhopadhyay}},
  \ and\ \bibinfo {author} {\bibfnamefont {T.~C.}\ \bibnamefont {Lubensky}},\
  }\bibfield  {title} {\enquote {\bibinfo {title} {{Fractional quantum Hall
  effect in an array of quantum wires}},}\ }\href
  {https://doi.org/10.1103/PhysRevLett.88.036401} {\bibfield  {journal}
  {\bibinfo  {journal} {Phys. Rev. Lett.}\ }\textbf {\bibinfo {volume} {88}},\
  \bibinfo {pages} {036401} (\bibinfo {year} {2002})}\BibitemShut {NoStop}%
\bibitem [{\citenamefont {Teo}\ and\ \citenamefont {Kane}(2014)}]{TeoPRB14}%
  \BibitemOpen
  \bibfield  {author} {\bibinfo {author} {\bibfnamefont {J.~C.~Y.}\
  \bibnamefont {Teo}}\ and\ \bibinfo {author} {\bibfnamefont {C.~L.}\
  \bibnamefont {Kane}},\ }\bibfield  {title} {\enquote {\bibinfo {title} {{From
  Luttinger liquid to non-Abelian quantum Hall states}},}\ }\href
  {https://doi.org/10.1103/PhysRevB.89.085101} {\bibfield  {journal} {\bibinfo
  {journal} {Phys. Rev. B}\ }\textbf {\bibinfo {volume} {89}},\ \bibinfo
  {pages} {085101} (\bibinfo {year} {2014})}\BibitemShut {NoStop}%
\bibitem [{\citenamefont {Meng}\ \emph {et~al.}(2015)\citenamefont {Meng},
  \citenamefont {Neupert}, \citenamefont {Greiter},\ and\ \citenamefont
  {Thomale}}]{MengPRB15}%
  \BibitemOpen
  \bibfield  {author} {\bibinfo {author} {\bibfnamefont {T.}~\bibnamefont
  {Meng}}, \bibinfo {author} {\bibfnamefont {T.}~\bibnamefont {Neupert}},
  \bibinfo {author} {\bibfnamefont {M.}~\bibnamefont {Greiter}}, \ and\
  \bibinfo {author} {\bibfnamefont {R.}~\bibnamefont {Thomale}},\ }\bibfield
  {title} {\enquote {\bibinfo {title} {{Coupled-wire construction of chiral
  spin liquids}},}\ }\href {https://doi.org/10.1103/PhysRevB.91.241106}
  {\bibfield  {journal} {\bibinfo  {journal} {Phys. Rev. B}\ }\textbf {\bibinfo
  {volume} {91}},\ \bibinfo {pages} {241106} (\bibinfo {year}
  {2015})}\BibitemShut {NoStop}%
\bibitem [{\citenamefont {Gorohovsky}\ \emph {et~al.}(2015)\citenamefont
  {Gorohovsky}, \citenamefont {Pereira},\ and\ \citenamefont
  {Sela}}]{GorohovskyPRB15}%
  \BibitemOpen
  \bibfield  {author} {\bibinfo {author} {\bibfnamefont {G.}~\bibnamefont
  {Gorohovsky}}, \bibinfo {author} {\bibfnamefont {R.~G.}\ \bibnamefont
  {Pereira}}, \ and\ \bibinfo {author} {\bibfnamefont {E.}~\bibnamefont
  {Sela}},\ }\bibfield  {title} {\enquote {\bibinfo {title} {{Chiral spin
  liquids in arrays of spin chains}},}\ }\href
  {https://doi.org/10.1103/PhysRevB.91.245139} {\bibfield  {journal} {\bibinfo
  {journal} {Phys. Rev. B}\ }\textbf {\bibinfo {volume} {91}},\ \bibinfo
  {pages} {245139} (\bibinfo {year} {2015})}\BibitemShut {NoStop}%
\bibitem [{\citenamefont {Huang}\ \emph
  {et~al.}(2016{\natexlab{a}})\citenamefont {Huang}, \citenamefont {Chen},
  \citenamefont {Gomes}, \citenamefont {Neupert}, \citenamefont {Chamon},\ and\
  \citenamefont {Mudry}}]{HuangPRB16}%
  \BibitemOpen
  \bibfield  {author} {\bibinfo {author} {\bibfnamefont {P.-H.}\ \bibnamefont
  {Huang}}, \bibinfo {author} {\bibfnamefont {J.-H.}\ \bibnamefont {Chen}},
  \bibinfo {author} {\bibfnamefont {P.~R.~S.}\ \bibnamefont {Gomes}}, \bibinfo
  {author} {\bibfnamefont {T.}~\bibnamefont {Neupert}}, \bibinfo {author}
  {\bibfnamefont {C.}~\bibnamefont {Chamon}}, \ and\ \bibinfo {author}
  {\bibfnamefont {C.}~\bibnamefont {Mudry}},\ }\bibfield  {title} {\enquote
  {\bibinfo {title} {{Non-Abelian topological spin liquids from arrays of
  quantum wires or spin chains}},}\ }\href
  {https://doi.org/10.1103/PhysRevB.93.205123} {\bibfield  {journal} {\bibinfo
  {journal} {Phys. Rev. B}\ }\textbf {\bibinfo {volume} {93}},\ \bibinfo
  {pages} {205123} (\bibinfo {year} {2016}{\natexlab{a}})}\BibitemShut
  {NoStop}%
\bibitem [{\citenamefont {Lecheminant}\ and\ \citenamefont
  {Tsvelik}(2016)}]{lecheminant2016lattice}%
  \BibitemOpen
  \bibfield  {author} {\bibinfo {author} {\bibfnamefont {P.}~\bibnamefont
  {Lecheminant}}\ and\ \bibinfo {author} {\bibfnamefont {A.~M.}\ \bibnamefont
  {Tsvelik}},\ }\bibfield  {title} {\enquote {\bibinfo {title} {{Lattice spin
  models for non-Abelian Chiral Spin Liquids}},}\ }\href
  {https://arxiv.org/abs/1608.05977} {\bibfield  {journal} {\bibinfo  {journal}
  {ArXiv e-prints}\ } (\bibinfo {year} {2016})},\ \Eprint
  {http://arxiv.org/abs/1608.05977} {arXiv:1608.05977} \BibitemShut {NoStop}%
\bibitem [{\citenamefont {Huang}\ \emph
  {et~al.}(2016{\natexlab{b}})\citenamefont {Huang}, \citenamefont {Chen},
  \citenamefont {Feiguin}, \citenamefont {Chamon},\ and\ \citenamefont
  {Mudry}}]{huang2016coupled}%
  \BibitemOpen
  \bibfield  {author} {\bibinfo {author} {\bibfnamefont {P.-H.}\ \bibnamefont
  {Huang}}, \bibinfo {author} {\bibfnamefont {J.-H.}\ \bibnamefont {Chen}},
  \bibinfo {author} {\bibfnamefont {A.~E.}\ \bibnamefont {Feiguin}}, \bibinfo
  {author} {\bibfnamefont {C.}~\bibnamefont {Chamon}}, \ and\ \bibinfo {author}
  {\bibfnamefont {C.}~\bibnamefont {Mudry}},\ }\bibfield  {title} {\enquote
  {\bibinfo {title} {{Coupled spin-1/2 ladders as microscopic models for
  non-Abelian chiral spin liquids}},}\ }\href
  {https://arxiv.org/abs/1611.02523} {\bibfield  {journal} {\bibinfo  {journal}
  {ArXiv e-prints}\ } (\bibinfo {year} {2016}{\natexlab{b}})},\ \Eprint
  {http://arxiv.org/abs/1611.02523} {arXiv:1611.02523} \BibitemShut {NoStop}%
\bibitem [{\citenamefont {Fuji}\ and\ \citenamefont
  {Lecheminant}(2016)}]{fuji2016non}%
  \BibitemOpen
  \bibfield  {author} {\bibinfo {author} {\bibfnamefont {Y.}~\bibnamefont
  {Fuji}}\ and\ \bibinfo {author} {\bibfnamefont {P.}~\bibnamefont
  {Lecheminant}},\ }\bibfield  {title} {\enquote {\bibinfo {title}
  {{Non-Abelian $SU(N-1)$-singlet fractional quantum Hall states from coupled
  wires}},}\ }\href {https://arxiv.org/abs/1611.07968} {\bibfield  {journal}
  {\bibinfo  {journal} {arXiv preprint arXiv:1611.07968}\ } (\bibinfo {year}
  {2016})},\ \Eprint {http://arxiv.org/abs/1611.07968} {arXiv:1611.07968}
  \BibitemShut {NoStop}%
\bibitem [{\citenamefont {Gepner}(1985)}]{GepnerNuclPhysB85}%
  \BibitemOpen
  \bibfield  {author} {\bibinfo {author} {\bibfnamefont {D.}~\bibnamefont
  {Gepner}},\ }\bibfield  {title} {\enquote {\bibinfo {title} {{Non-abelian
  bosonization and multiflavor QED and QCD in two dimensions}},}\ }\href
  {\doibase http://dx.doi.org/10.1016/0550-3213(85)90458-4} {\bibfield
  {journal} {\bibinfo  {journal} {Nucl. Phys. B}\ }\textbf {\bibinfo {volume}
  {252}},\ \bibinfo {pages} {481 -- 507} (\bibinfo {year} {1985})}\BibitemShut
  {NoStop}%
\bibitem [{\citenamefont {Affleck}(1986{\natexlab{b}})}]{AffleckNuclPhysB86b}%
  \BibitemOpen
  \bibfield  {author} {\bibinfo {author} {\bibfnamefont {I.}~\bibnamefont
  {Affleck}},\ }\bibfield  {title} {\enquote {\bibinfo {title} {{On the
  realization of chiral symmetry in (1+1) dimensions}},}\ }\href {\doibase
  http://dx.doi.org/10.1016/0550-3213(86)90168-9} {\bibfield  {journal}
  {\bibinfo  {journal} {Nucl. Phys. B}\ }\textbf {\bibinfo {volume} {265}},\
  \bibinfo {pages} {448 -- 468} (\bibinfo {year}
  {1986}{\natexlab{b}})}\BibitemShut {NoStop}%
\bibitem [{\citenamefont {Frishman}\ and\ \citenamefont
  {Sonnenschein}(1993)}]{FrishmanPhysRep93}%
  \BibitemOpen
  \bibfield  {author} {\bibinfo {author} {\bibfnamefont {Y.}~\bibnamefont
  {Frishman}}\ and\ \bibinfo {author} {\bibfnamefont {J.}~\bibnamefont
  {Sonnenschein}},\ }\bibfield  {title} {\enquote {\bibinfo {title}
  {{Bosonization and QCD in two dimensions}},}\ }\href {\doibase
  http://dx.doi.org/10.1016/0370-1573(93)90145-4} {\bibfield  {journal}
  {\bibinfo  {journal} {Phys. Rep.}\ }\textbf {\bibinfo {volume} {223}},\
  \bibinfo {pages} {309 -- 348} (\bibinfo {year} {1993})}\BibitemShut {NoStop}%
\bibitem [{\citenamefont {Azaria}\ \emph {et~al.}(2016)\citenamefont {Azaria},
  \citenamefont {Konik}, \citenamefont {Lecheminant}, \citenamefont {P\'almai},
  \citenamefont {Tak\'acs},\ and\ \citenamefont {Tsvelik}}]{AzariaPRD16}%
  \BibitemOpen
  \bibfield  {author} {\bibinfo {author} {\bibfnamefont {P.}~\bibnamefont
  {Azaria}}, \bibinfo {author} {\bibfnamefont {R.~M.}\ \bibnamefont {Konik}},
  \bibinfo {author} {\bibfnamefont {P.}~\bibnamefont {Lecheminant}}, \bibinfo
  {author} {\bibfnamefont {T.}~\bibnamefont {P\'almai}}, \bibinfo {author}
  {\bibfnamefont {G.}~\bibnamefont {Tak\'acs}}, \ and\ \bibinfo {author}
  {\bibfnamefont {A.~M.}\ \bibnamefont {Tsvelik}},\ }\bibfield  {title}
  {\enquote {\bibinfo {title} {{Particle formation and ordering in strongly
  correlated fermionic systems: Solving a model of quantum chromodynamics}},}\
  }\href {\doibase 10.1103/PhysRevD.94.045003} {\bibfield  {journal} {\bibinfo
  {journal} {Phys. Rev. D}\ }\textbf {\bibinfo {volume} {94}},\ \bibinfo
  {pages} {045003} (\bibinfo {year} {2016})}\BibitemShut {NoStop}%
\bibitem [{\citenamefont {Kojo}\ \emph {et~al.}(2010)\citenamefont {Kojo},
  \citenamefont {Pisarski},\ and\ \citenamefont {Tsvelik}}]{KojoPRD10}%
  \BibitemOpen
  \bibfield  {author} {\bibinfo {author} {\bibfnamefont {T.}~\bibnamefont
  {Kojo}}, \bibinfo {author} {\bibfnamefont {R.~D.}\ \bibnamefont {Pisarski}},
  \ and\ \bibinfo {author} {\bibfnamefont {A.~M.}\ \bibnamefont {Tsvelik}},\
  }\bibfield  {title} {\enquote {\bibinfo {title} {{Covering the Fermi surface
  with patches of quarkyonic chiral spirals}},}\ }\href {\doibase
  10.1103/PhysRevD.82.074015} {\bibfield  {journal} {\bibinfo  {journal} {Phys.
  Rev. D}\ }\textbf {\bibinfo {volume} {82}},\ \bibinfo {pages} {074015}
  (\bibinfo {year} {2010})}\BibitemShut {NoStop}%
\bibitem [{\citenamefont {Kac}(1968)}]{KacMathUSSAIzv68}%
  \BibitemOpen
  \bibfield  {author} {\bibinfo {author} {\bibfnamefont {V.~G.}\ \bibnamefont
  {Kac}},\ }\bibfield  {title} {\enquote {\bibinfo {title} {Simple irreducible
  graded {L}ie algebras of finite growth},}\ }\href
  {http://stacks.iop.org/0025-5726/2/i=6/a=A06} {\bibfield  {journal} {\bibinfo
   {journal} {Math. USSR Izv.}\ }\textbf {\bibinfo {volume} {2}},\ \bibinfo
  {pages} {1271} (\bibinfo {year} {1968})}\BibitemShut {NoStop}%
\bibitem [{\citenamefont {Moody}(1968)}]{MoodyJAlgebra68}%
  \BibitemOpen
  \bibfield  {author} {\bibinfo {author} {\bibfnamefont {R.~V.}\ \bibnamefont
  {Moody}},\ }\bibfield  {title} {\enquote {\bibinfo {title} {{A new class of
  Lie algebras}},}\ }\href {\doibase
  http://dx.doi.org/10.1016/0021-8693(68)90096-3} {\bibfield  {journal}
  {\bibinfo  {journal} {J. Algebra}\ }\textbf {\bibinfo {volume} {10}},\
  \bibinfo {pages} {211--230} (\bibinfo {year} {1968})}\BibitemShut {NoStop}%
\bibitem [{\citenamefont {Fulton}\ and\ \citenamefont
  {Harris}(2004)}]{FultonHarris}%
  \BibitemOpen
  \bibfield  {author} {\bibinfo {author} {\bibfnamefont {W.}~\bibnamefont
  {Fulton}}\ and\ \bibinfo {author} {\bibfnamefont {J.}~\bibnamefont
  {Harris}},\ }\href {\doibase 10.1007/978-1-4612-0979-9} {\emph {\bibinfo
  {title} {{Representation Theory}}}}\ (\bibinfo  {publisher}
  {Springer-Verlag},\ \bibinfo {year} {2004})\BibitemShut {NoStop}%
\bibitem [{\citenamefont {Kats}(1974)}]{KatsJFunctAnalAppl74}%
  \BibitemOpen
  \bibfield  {author} {\bibinfo {author} {\bibfnamefont {V.~G.}\ \bibnamefont
  {Kats}},\ }\bibfield  {title} {\enquote {\bibinfo {title}
  {{Infinite-dimensioned Lie algebras and Dedekind's $\eta$-function}},}\
  }\href {\doibase 10.1007/BF02028313} {\bibfield  {journal} {\bibinfo
  {journal} {Func. Anal. Appl.}\ }\textbf {\bibinfo {volume} {8}},\ \bibinfo
  {pages} {68--70} (\bibinfo {year} {1974})}\BibitemShut {NoStop}%
\bibitem [{\citenamefont {Lepowsky}\ and\ \citenamefont
  {Wilson}(1978)}]{LepowskyCommunMathPhys78}%
  \BibitemOpen
  \bibfield  {author} {\bibinfo {author} {\bibfnamefont {J.}~\bibnamefont
  {Lepowsky}}\ and\ \bibinfo {author} {\bibfnamefont {R.~L.}\ \bibnamefont
  {Wilson}},\ }\bibfield  {title} {\enquote {\bibinfo {title} {{Construction of
  the affine Lie algebra $A_{1}^{{}}(1)$}},}\ }\href
  {http://projecteuclid.org/euclid.cmp/1103904301} {\bibfield  {journal}
  {\bibinfo  {journal} {Comm. Math. Phys.}\ }\textbf {\bibinfo {volume} {62}},\
  \bibinfo {pages} {43--53} (\bibinfo {year} {1978})}\BibitemShut {NoStop}%
\bibitem [{\citenamefont {Frenkel}(1980)}]{FrenkelPNAS80}%
  \BibitemOpen
  \bibfield  {author} {\bibinfo {author} {\bibfnamefont {I.~B.}\ \bibnamefont
  {Frenkel}},\ }\bibfield  {title} {\enquote {\bibinfo {title} {{Spinor
  representations of affine Lie algebras}},}\ }\href
  {http://www.ncbi.nlm.nih.gov/pmc/articles/PMC350271/} {\bibfield  {journal}
  {\bibinfo  {journal} {Proc. Natl. Acad. USA}\ }\textbf {\bibinfo {volume}
  {77}},\ \bibinfo {pages} {6303--6306} (\bibinfo {year} {1980})}\BibitemShut
  {NoStop}%
\bibitem [{\citenamefont {Got\^o}\ and\ \citenamefont
  {Imamura}(1955)}]{GotoProgTheorPhys55}%
  \BibitemOpen
  \bibfield  {author} {\bibinfo {author} {\bibfnamefont {T.}~\bibnamefont
  {Got\^o}}\ and\ \bibinfo {author} {\bibfnamefont {T.}~\bibnamefont
  {Imamura}},\ }\bibfield  {title} {\enquote {\bibinfo {title} {{Note on the
  Non-Perturbation-Approach to Quantum Field Theory}},}\ }\href {\doibase
  10.1143/PTP.14.396} {\bibfield  {journal} {\bibinfo  {journal} {Prog. Theor.
  Phys.}\ }\textbf {\bibinfo {volume} {14}},\ \bibinfo {pages} {396--397}
  (\bibinfo {year} {1955})}\BibitemShut {NoStop}%
\bibitem [{\citenamefont {Schwinger}(1959)}]{SchwingerPRL59}%
  \BibitemOpen
  \bibfield  {author} {\bibinfo {author} {\bibfnamefont {J.}~\bibnamefont
  {Schwinger}},\ }\bibfield  {title} {\enquote {\bibinfo {title} {{Field Theory
  Commutators}},}\ }\href {\doibase 10.1103/PhysRevLett.3.296} {\bibfield
  {journal} {\bibinfo  {journal} {Phys. Rev. Lett.}\ }\textbf {\bibinfo
  {volume} {3}},\ \bibinfo {pages} {296--297} (\bibinfo {year}
  {1959})}\BibitemShut {NoStop}%
\bibitem [{\citenamefont {Coleman}\ \emph {et~al.}(1969)\citenamefont
  {Coleman}, \citenamefont {Gross},\ and\ \citenamefont
  {Jackiw}}]{ColemanPR69}%
  \BibitemOpen
  \bibfield  {author} {\bibinfo {author} {\bibfnamefont {S.}~\bibnamefont
  {Coleman}}, \bibinfo {author} {\bibfnamefont {D.}~\bibnamefont {Gross}}, \
  and\ \bibinfo {author} {\bibfnamefont {R.}~\bibnamefont {Jackiw}},\
  }\bibfield  {title} {\enquote {\bibinfo {title} {{Fermion Avatars of the
  Sugawara Model}},}\ }\href {\doibase 10.1103/PhysRev.180.1359} {\bibfield
  {journal} {\bibinfo  {journal} {Phys. Rev.}\ }\textbf {\bibinfo {volume}
  {180}},\ \bibinfo {pages} {1359--1366} (\bibinfo {year} {1969})}\BibitemShut
  {NoStop}%
\bibitem [{\citenamefont {Belavin}\ \emph
  {et~al.}(1984{\natexlab{a}})\citenamefont {Belavin}, \citenamefont
  {Polyakov},\ and\ \citenamefont {Zamolodchikov}}]{BPZJStatPhys84}%
  \BibitemOpen
  \bibfield  {author} {\bibinfo {author} {\bibfnamefont {A.~A.}\ \bibnamefont
  {Belavin}}, \bibinfo {author} {\bibfnamefont {A.~M.}\ \bibnamefont
  {Polyakov}}, \ and\ \bibinfo {author} {\bibfnamefont {A.~B.}\ \bibnamefont
  {Zamolodchikov}},\ }\bibfield  {title} {\enquote {\bibinfo {title} {{Infinite
  conformal symmetry of critical fluctuations in two dimensions}},}\ }\href
  {\doibase 10.1007/BF01009438} {\bibfield  {journal} {\bibinfo  {journal} {J.
  Stat. Phys.}\ }\textbf {\bibinfo {volume} {34}},\ \bibinfo {pages} {763--774}
  (\bibinfo {year} {1984}{\natexlab{a}})}\BibitemShut {NoStop}%
\bibitem [{\citenamefont {Belavin}\ \emph
  {et~al.}(1984{\natexlab{b}})\citenamefont {Belavin}, \citenamefont
  {Polyakov},\ and\ \citenamefont {Zamolodchikov}}]{BPZNuclPhysB84}%
  \BibitemOpen
  \bibfield  {author} {\bibinfo {author} {\bibfnamefont {A.~A.}\ \bibnamefont
  {Belavin}}, \bibinfo {author} {\bibfnamefont {A.~M.}\ \bibnamefont
  {Polyakov}}, \ and\ \bibinfo {author} {\bibfnamefont {A.~B.}\ \bibnamefont
  {Zamolodchikov}},\ }\bibfield  {title} {\enquote {\bibinfo {title} {{Infinite
  conformal symmetry in two-dimensional quantum field theory}},}\ }\href
  {\doibase http://dx.doi.org/10.1016/0550-3213(84)90052-X} {\bibfield
  {journal} {\bibinfo  {journal} {Nucl. Phys. B}\ }\textbf {\bibinfo {volume}
  {241}},\ \bibinfo {pages} {333 -- 380} (\bibinfo {year}
  {1984}{\natexlab{b}})}\BibitemShut {NoStop}%
\bibitem [{\citenamefont {Sugawara}(1968)}]{SugawaraPR68}%
  \BibitemOpen
  \bibfield  {author} {\bibinfo {author} {\bibfnamefont {H.}~\bibnamefont
  {Sugawara}},\ }\bibfield  {title} {\enquote {\bibinfo {title} {{A Field
  Theory of Currents}},}\ }\href {\doibase 10.1103/PhysRev.170.1659} {\bibfield
   {journal} {\bibinfo  {journal} {Phys. Rev.}\ }\textbf {\bibinfo {volume}
  {170}},\ \bibinfo {pages} {1659--1662} (\bibinfo {year} {1968})}\BibitemShut
  {NoStop}%
\bibitem [{\citenamefont {Zamolodchikov}\ and\ \citenamefont
  {Fateev}(1986)}]{FateevSovJNuclPhys86}%
  \BibitemOpen
  \bibfield  {author} {\bibinfo {author} {\bibfnamefont {A.~B.}\ \bibnamefont
  {Zamolodchikov}}\ and\ \bibinfo {author} {\bibfnamefont {V.~A.}\ \bibnamefont
  {Fateev}},\ }\bibfield  {title} {\enquote {\bibinfo {title} {{Operator
  algebra and correlation functions in the two-dimensional $SU(2)\times SU(2)$
  chiral Wess-Zumino model}},}\ }\href@noop {} {\bibfield  {journal} {\bibinfo
  {journal} {Sov. J. Nucl. Phys.}\ }\textbf {\bibinfo {volume} {43}},\ \bibinfo
  {pages} {657} (\bibinfo {year} {1986})}\BibitemShut {NoStop}%
\bibitem [{\citenamefont {Belavin}\ \emph
  {et~al.}(1984{\natexlab{c}})\citenamefont {Belavin}, \citenamefont
  {Polyakov},\ and\ \citenamefont {Zamolodchikov}}]{BelavinNuclPhysB84}%
  \BibitemOpen
  \bibfield  {author} {\bibinfo {author} {\bibfnamefont {A.~A.}\ \bibnamefont
  {Belavin}}, \bibinfo {author} {\bibfnamefont {A.~M.}\ \bibnamefont
  {Polyakov}}, \ and\ \bibinfo {author} {\bibfnamefont {A.~B.}\ \bibnamefont
  {Zamolodchikov}},\ }\bibfield  {title} {\enquote {\bibinfo {title} {{Infinite
  conformal symmetry in two-dimensional quantum field theory}},}\ }\href
  {\doibase http://dx.doi.org/10.1016/0550-3213(84)90052-X} {\bibfield
  {journal} {\bibinfo  {journal} {Nucl. Phys. B}\ }\textbf {\bibinfo {volume}
  {241}},\ \bibinfo {pages} {333 -- 380} (\bibinfo {year}
  {1984}{\natexlab{c}})}\BibitemShut {NoStop}%
\bibitem [{\citenamefont {Bl\"ote}\ \emph {et~al.}(1986)\citenamefont
  {Bl\"ote}, \citenamefont {Cardy},\ and\ \citenamefont
  {Nightingale}}]{BlotePRL86}%
  \BibitemOpen
  \bibfield  {author} {\bibinfo {author} {\bibfnamefont {H.~W.~J.}\
  \bibnamefont {Bl\"ote}}, \bibinfo {author} {\bibfnamefont {John~L.}\
  \bibnamefont {Cardy}}, \ and\ \bibinfo {author} {\bibfnamefont {M.~P.}\
  \bibnamefont {Nightingale}},\ }\bibfield  {title} {\enquote {\bibinfo {title}
  {{Conformal invariance, the central charge, and universal finite-size
  amplitudes at criticality}},}\ }\href {\doibase 10.1103/PhysRevLett.56.742}
  {\bibfield  {journal} {\bibinfo  {journal} {Phys. Rev. Lett.}\ }\textbf
  {\bibinfo {volume} {56}},\ \bibinfo {pages} {742--745} (\bibinfo {year}
  {1986})}\BibitemShut {NoStop}%
\bibitem [{\citenamefont {Affleck}(1986{\natexlab{c}})}]{AffleckPRL86}%
  \BibitemOpen
  \bibfield  {author} {\bibinfo {author} {\bibfnamefont {I.}~\bibnamefont
  {Affleck}},\ }\bibfield  {title} {\enquote {\bibinfo {title} {{Universal term
  in the free energy at a critical point and the conformal anomaly}},}\ }\href
  {\doibase 10.1103/PhysRevLett.56.746} {\bibfield  {journal} {\bibinfo
  {journal} {Phys. Rev. Lett.}\ }\textbf {\bibinfo {volume} {56}},\ \bibinfo
  {pages} {746--748} (\bibinfo {year} {1986}{\natexlab{c}})}\BibitemShut
  {NoStop}%
\bibitem [{\citenamefont {Calabrese}\ and\ \citenamefont
  {Cardy}(2009)}]{CalabreseJPhysA09}%
  \BibitemOpen
  \bibfield  {author} {\bibinfo {author} {\bibfnamefont {P.}~\bibnamefont
  {Calabrese}}\ and\ \bibinfo {author} {\bibfnamefont {J.}~\bibnamefont
  {Cardy}},\ }\bibfield  {title} {\enquote {\bibinfo {title} {{Entanglement
  entropy and conformal field theory}},}\ }\href
  {http://stacks.iop.org/1751-8121/42/i=50/a=504005} {\bibfield  {journal}
  {\bibinfo  {journal} {J. Phys. A}\ }\textbf {\bibinfo {volume} {42}},\
  \bibinfo {pages} {504005} (\bibinfo {year} {2009})}\BibitemShut {NoStop}%
\bibitem [{\citenamefont {Friedan}\ \emph {et~al.}(1984)\citenamefont
  {Friedan}, \citenamefont {Qiu},\ and\ \citenamefont
  {Shenker}}]{FriedanPRL84}%
  \BibitemOpen
  \bibfield  {author} {\bibinfo {author} {\bibfnamefont {D.}~\bibnamefont
  {Friedan}}, \bibinfo {author} {\bibfnamefont {Z.}~\bibnamefont {Qiu}}, \ and\
  \bibinfo {author} {\bibfnamefont {S.}~\bibnamefont {Shenker}},\ }\bibfield
  {title} {\enquote {\bibinfo {title} {{Conformal Invariance, Unitarity, and
  Critical Exponents in Two Dimensions}},}\ }\href {\doibase
  10.1103/PhysRevLett.52.1575} {\bibfield  {journal} {\bibinfo  {journal}
  {Phys. Rev. Lett.}\ }\textbf {\bibinfo {volume} {52}},\ \bibinfo {pages}
  {1575--1578} (\bibinfo {year} {1984})}\BibitemShut {NoStop}%
\bibitem [{\citenamefont {Polyakov}\ and\ \citenamefont
  {Wiegmann}(1983)}]{PolyakovPhysLettB83}%
  \BibitemOpen
  \bibfield  {author} {\bibinfo {author} {\bibfnamefont {A.}~\bibnamefont
  {Polyakov}}\ and\ \bibinfo {author} {\bibfnamefont {P.~B.}\ \bibnamefont
  {Wiegmann}},\ }\bibfield  {title} {\enquote {\bibinfo {title} {{Theory of
  nonabelian goldstone bosons in two dimensions}},}\ }\href {\doibase
  http://dx.doi.org/10.1016/0370-2693(83)91104-8} {\bibfield  {journal}
  {\bibinfo  {journal} {Phys. Lett. B}\ }\textbf {\bibinfo {volume} {131}},\
  \bibinfo {pages} {121 -- 126} (\bibinfo {year} {1983})}\BibitemShut {NoStop}%
\bibitem [{\citenamefont {Di~Vecchia}\ and\ \citenamefont
  {Rossi}(1984)}]{DiVecchiaCERN84}%
  \BibitemOpen
  \bibfield  {author} {\bibinfo {author} {\bibfnamefont {P.}~\bibnamefont
  {Di~Vecchia}}\ and\ \bibinfo {author} {\bibfnamefont {P.}~\bibnamefont
  {Rossi}},\ }\href
  {http://cds.cern.ch/record/150359/files/198404304.pdf?version=1} {\enquote
  {\bibinfo {title} {{On the equivalence between the Wess-Zumino action and the
  free Fermi theory in two dimensions}},}\ }\bibinfo {howpublished} {preprint
  TH-3808 CERN} (\bibinfo {year} {1984})\BibitemShut {NoStop}%
\bibitem [{\citenamefont {Witten}(1983)}]{WittenNuclPhysB83}%
  \BibitemOpen
  \bibfield  {author} {\bibinfo {author} {\bibfnamefont {E.}~\bibnamefont
  {Witten}},\ }\bibfield  {title} {\enquote {\bibinfo {title} {{Global aspects
  of current algebra}},}\ }\href {\doibase
  http://dx.doi.org/10.1016/0550-3213(83)90063-9} {\bibfield  {journal}
  {\bibinfo  {journal} {Nucl. Phys. B}\ }\textbf {\bibinfo {volume} {223}},\
  \bibinfo {pages} {422 -- 432} (\bibinfo {year} {1983})}\BibitemShut {NoStop}%
\bibitem [{\citenamefont {Lecheminant}\ and\ \citenamefont
  {Tsvelik}(2015)}]{LecheminantPRB15}%
  \BibitemOpen
  \bibfield  {author} {\bibinfo {author} {\bibfnamefont {P.}~\bibnamefont
  {Lecheminant}}\ and\ \bibinfo {author} {\bibfnamefont {A.~M.}\ \bibnamefont
  {Tsvelik}},\ }\bibfield  {title} {\enquote {\bibinfo {title} {{Two-leg
  $\mathrm{SU}(2n)$ spin ladder: A low-energy effective field theory
  approach}},}\ }\href {\doibase 10.1103/PhysRevB.91.174407} {\bibfield
  {journal} {\bibinfo  {journal} {Phys. Rev. B}\ }\textbf {\bibinfo {volume}
  {91}},\ \bibinfo {pages} {174407} (\bibinfo {year} {2015})}\BibitemShut
  {NoStop}%
\bibitem [{\citenamefont {Bois}\ \emph {et~al.}(2015)\citenamefont {Bois},
  \citenamefont {Capponi}, \citenamefont {Lecheminant}, \citenamefont
  {Moliner},\ and\ \citenamefont {Totsuka}}]{BoisPRB15}%
  \BibitemOpen
  \bibfield  {author} {\bibinfo {author} {\bibfnamefont {V.}~\bibnamefont
  {Bois}}, \bibinfo {author} {\bibfnamefont {S.}~\bibnamefont {Capponi}},
  \bibinfo {author} {\bibfnamefont {P.}~\bibnamefont {Lecheminant}}, \bibinfo
  {author} {\bibfnamefont {M.}~\bibnamefont {Moliner}}, \ and\ \bibinfo
  {author} {\bibfnamefont {K.}~\bibnamefont {Totsuka}},\ }\bibfield  {title}
  {\enquote {\bibinfo {title} {{Phase diagrams of one-dimensional half-filled
  two-orbital $\mathrm{SU}(N)$ cold fermion systems}},}\ }\href {\doibase
  10.1103/PhysRevB.91.075121} {\bibfield  {journal} {\bibinfo  {journal} {Phys.
  Rev. B}\ }\textbf {\bibinfo {volume} {91}},\ \bibinfo {pages} {075121}
  (\bibinfo {year} {2015})}\BibitemShut {NoStop}%
\bibitem [{\citenamefont {Akhanjee}\ and\ \citenamefont
  {Tsvelik}(2013)}]{AkhanjeePRB13}%
  \BibitemOpen
  \bibfield  {author} {\bibinfo {author} {\bibfnamefont {S.}~\bibnamefont
  {Akhanjee}}\ and\ \bibinfo {author} {\bibfnamefont {A.~M.}\ \bibnamefont
  {Tsvelik}},\ }\bibfield  {title} {\enquote {\bibinfo {title} {{Analytically
  tractable model of bad metals}},}\ }\href {\doibase
  10.1103/PhysRevB.87.195137} {\bibfield  {journal} {\bibinfo  {journal} {Phys.
  Rev. B}\ }\textbf {\bibinfo {volume} {87}},\ \bibinfo {pages} {195137}
  (\bibinfo {year} {2013})}\BibitemShut {NoStop}%
\bibitem [{\citenamefont {Altschuler}(1989)}]{AltschulerNuclPhysB89}%
  \BibitemOpen
  \bibfield  {author} {\bibinfo {author} {\bibfnamefont {D.}~\bibnamefont
  {Altschuler}},\ }\bibfield  {title} {\enquote {\bibinfo {title} {{Quantum
  equivalence of coset space models}},}\ }\href {\doibase
  http://dx.doi.org/10.1016/0550-3213(89)90320-9} {\bibfield  {journal}
  {\bibinfo  {journal} {Nucl. Phys. B}\ }\textbf {\bibinfo {volume} {313}},\
  \bibinfo {pages} {293 -- 307} (\bibinfo {year} {1989})}\BibitemShut {NoStop}%
\bibitem [{\citenamefont {Zamolodchikov}\ and\ \citenamefont
  {Fateev}(1985)}]{ZamolodchikovJETP85}%
  \BibitemOpen
  \bibfield  {author} {\bibinfo {author} {\bibfnamefont {A.~B.}\ \bibnamefont
  {Zamolodchikov}}\ and\ \bibinfo {author} {\bibfnamefont {V.~A.}\ \bibnamefont
  {Fateev}},\ }\bibfield  {title} {\enquote {\bibinfo {title} {{Nonlocal
  (parafermion) currents in two-dimensional conformal quantum field theory and
  self-dual critical points in $Z_N$-symmetric statistical systems}},}\ }\href
  {http://www.jetp.ac.ru/cgi-bin/e/index/e/62/2/p215?a=list} {\bibfield
  {journal} {\bibinfo  {journal} {Sov. Phys. JETP}\ }\textbf {\bibinfo {volume}
  {62}},\ \bibinfo {pages} {215--225} (\bibinfo {year} {1985})}\BibitemShut
  {NoStop}%
\bibitem [{\citenamefont {Stadler}\ \emph {et~al.}(2015)\citenamefont
  {Stadler}, \citenamefont {Yin}, \citenamefont {von Delft}, \citenamefont
  {Kotliar},\ and\ \citenamefont {Weichselbaum}}]{StadlerPRL15}%
  \BibitemOpen
  \bibfield  {author} {\bibinfo {author} {\bibfnamefont {K.~M.}\ \bibnamefont
  {Stadler}}, \bibinfo {author} {\bibfnamefont {Z.~P.}\ \bibnamefont {Yin}},
  \bibinfo {author} {\bibfnamefont {J.}~\bibnamefont {von Delft}}, \bibinfo
  {author} {\bibfnamefont {G.}~\bibnamefont {Kotliar}}, \ and\ \bibinfo
  {author} {\bibfnamefont {A.}~\bibnamefont {Weichselbaum}},\ }\bibfield
  {title} {\enquote {\bibinfo {title} {{Dynamical Mean-Field Theory Plus
  Numerical Renormalization-Group Study of Spin-Orbital Separation in a
  Three-Band Hund Metal}},}\ }\href {\doibase 10.1103/PhysRevLett.115.136401}
  {\bibfield  {journal} {\bibinfo  {journal} {Phys. Rev. Lett.}\ }\textbf
  {\bibinfo {volume} {115}},\ \bibinfo {pages} {136401} (\bibinfo {year}
  {2015})}\BibitemShut {NoStop}%
\bibitem [{\citenamefont {Yin}\ \emph {et~al.}(2012)\citenamefont {Yin},
  \citenamefont {Haule},\ and\ \citenamefont {Kotliar}}]{YinPRB12}%
  \BibitemOpen
  \bibfield  {author} {\bibinfo {author} {\bibfnamefont {Z.~P.}\ \bibnamefont
  {Yin}}, \bibinfo {author} {\bibfnamefont {K.}~\bibnamefont {Haule}}, \ and\
  \bibinfo {author} {\bibfnamefont {G.}~\bibnamefont {Kotliar}},\ }\bibfield
  {title} {\enquote {\bibinfo {title} {{Fractional power-law behavior and its
  origin in iron-chalcogenide and ruthenate superconductors: Insights from
  first-principles calculations}},}\ }\href {\doibase
  10.1103/PhysRevB.86.195141} {\bibfield  {journal} {\bibinfo  {journal} {Phys.
  Rev. B}\ }\textbf {\bibinfo {volume} {86}},\ \bibinfo {pages} {195141}
  (\bibinfo {year} {2012})}\BibitemShut {NoStop}%
\bibitem [{\citenamefont {Giovannetti}\ \emph {et~al.}(2015)\citenamefont
  {Giovannetti}, \citenamefont {de' Medici}, \citenamefont {Aichhorn},\ and\
  \citenamefont {Capone}}]{GiovanettiPRB15}%
  \BibitemOpen
  \bibfield  {author} {\bibinfo {author} {\bibfnamefont {G.}~\bibnamefont
  {Giovannetti}}, \bibinfo {author} {\bibfnamefont {L.}~\bibnamefont {de'
  Medici}}, \bibinfo {author} {\bibfnamefont {M.}~\bibnamefont {Aichhorn}}, \
  and\ \bibinfo {author} {\bibfnamefont {M.}~\bibnamefont {Capone}},\
  }\bibfield  {title} {\enquote {\bibinfo {title}
  {{$\mathrm{La}{}_{2}\mathrm{O}{}_{3}\mathrm{Fe}{}_{2}\mathrm{Se}{}_{2}$: A
  Mott insulator on the brink of orbital-selective metallization}},}\ }\href
  {\doibase 10.1103/PhysRevB.91.085124} {\bibfield  {journal} {\bibinfo
  {journal} {Phys. Rev. B}\ }\textbf {\bibinfo {volume} {91}},\ \bibinfo
  {pages} {085124} (\bibinfo {year} {2015})}\BibitemShut {NoStop}%
\bibitem [{\citenamefont {Gross}\ and\ \citenamefont
  {Neveu}(1974)}]{GrossPRD74}%
  \BibitemOpen
  \bibfield  {author} {\bibinfo {author} {\bibfnamefont {D.~J.}\ \bibnamefont
  {Gross}}\ and\ \bibinfo {author} {\bibfnamefont {A.}~\bibnamefont {Neveu}},\
  }\bibfield  {title} {\enquote {\bibinfo {title} {{Dynamical symmetry breaking
  in asymptotically free field theories}},}\ }\href {\doibase
  10.1103/PhysRevD.10.3235} {\bibfield  {journal} {\bibinfo  {journal} {Phys.
  Rev. D}\ }\textbf {\bibinfo {volume} {10}},\ \bibinfo {pages} {3235--3253}
  (\bibinfo {year} {1974})}\BibitemShut {NoStop}%
\bibitem [{\citenamefont {Konik}\ \emph
  {et~al.}(2015{\natexlab{a}})\citenamefont {Konik}, \citenamefont {P\'almai},
  \citenamefont {Tak\'acs},\ and\ \citenamefont {Tsvelik}}]{KonikNuclPhysB15}%
  \BibitemOpen
  \bibfield  {author} {\bibinfo {author} {\bibfnamefont {R.~M.}\ \bibnamefont
  {Konik}}, \bibinfo {author} {\bibfnamefont {T.}~\bibnamefont {P\'almai}},
  \bibinfo {author} {\bibfnamefont {G.}~\bibnamefont {Tak\'acs}}, \ and\
  \bibinfo {author} {\bibfnamefont {A.~M.}\ \bibnamefont {Tsvelik}},\
  }\bibfield  {title} {\enquote {\bibinfo {title} {{Studying the perturbed
  Wess-Zumino-€"Novikov-€"Witten theory using the truncated conformal spectrum
  approach}},}\ }\href {\doibase
  http://dx.doi.org/10.1016/j.nuclphysb.2015.08.016} {\bibfield  {journal}
  {\bibinfo  {journal} {Nucl. Phys. B}\ }\textbf {\bibinfo {volume} {899}},\
  \bibinfo {pages} {547 -- 569} (\bibinfo {year}
  {2015}{\natexlab{a}})}\BibitemShut {NoStop}%
\bibitem [{\citenamefont {Tsvelik}(1987{\natexlab{a}})}]{TsvelikJETP87}%
  \BibitemOpen
  \bibfield  {author} {\bibinfo {author} {\bibfnamefont {A.~M.}\ \bibnamefont
  {Tsvelik}},\ }\bibfield  {title} {\enquote {\bibinfo {title} {{Exact solution
  of a model of one-dimensional fermions with $SU(N)\times SU(M)$ symmetry}},}\
  }\href {http://www.jetp.ac.ru/cgi-bin/dn/e_066_04_0754.pdf} {\bibfield
  {journal} {\bibinfo  {journal} {Sov. Phys. JETP}\ }\textbf {\bibinfo {volume}
  {66}},\ \bibinfo {pages} {754—760} (\bibinfo {year}
  {1987}{\natexlab{a}})}\BibitemShut {NoStop}%
\bibitem [{\citenamefont {Smirnov}(1994)}]{SmirnovIntJModPhysA94}%
  \BibitemOpen
  \bibfield  {author} {\bibinfo {author} {\bibfnamefont {F.~A.}\ \bibnamefont
  {Smirnov}},\ }\bibfield  {title} {\enquote {\bibinfo {title} {{A New Set Of
  Exact Form Factors}},}\ }\href {\doibase 10.1142/S0217751X94002077}
  {\bibfield  {journal} {\bibinfo  {journal} {Int. J. Mod. Phys. A}\ }\textbf
  {\bibinfo {volume} {9}},\ \bibinfo {pages} {5121} (\bibinfo {year}
  {1994})}\BibitemShut {NoStop}%
\bibitem [{\citenamefont {Wilson}(1971)}]{WilsonPRB71}%
  \BibitemOpen
  \bibfield  {author} {\bibinfo {author} {\bibfnamefont {K.~G.}\ \bibnamefont
  {Wilson}},\ }\bibfield  {title} {\enquote {\bibinfo {title} {{Renormalization
  Group and Critical Phenomena. I. Renormalization Group and the Kadanoff
  Scaling Picture}},}\ }\href {\doibase 10.1103/PhysRevB.4.3174} {\bibfield
  {journal} {\bibinfo  {journal} {Phys. Rev. B}\ }\textbf {\bibinfo {volume}
  {4}},\ \bibinfo {pages} {3174--3183} (\bibinfo {year} {1971})}\BibitemShut
  {NoStop}%
\bibitem [{\citenamefont {Wegner}(1972{\natexlab{a}})}]{WegnerPRB72}%
  \BibitemOpen
  \bibfield  {author} {\bibinfo {author} {\bibfnamefont {F.~J.}\ \bibnamefont
  {Wegner}},\ }\bibfield  {title} {\enquote {\bibinfo {title} {{Corrections to
  Scaling Laws}},}\ }\href {\doibase 10.1103/PhysRevB.5.4529} {\bibfield
  {journal} {\bibinfo  {journal} {Phys. Rev. B}\ }\textbf {\bibinfo {volume}
  {5}},\ \bibinfo {pages} {4529--4536} (\bibinfo {year}
  {1972}{\natexlab{a}})}\BibitemShut {NoStop}%
\bibitem [{\citenamefont {Wegner}(1972{\natexlab{b}})}]{WegnerPRB72b}%
  \BibitemOpen
  \bibfield  {author} {\bibinfo {author} {\bibfnamefont {F.~J.}\ \bibnamefont
  {Wegner}},\ }\bibfield  {title} {\enquote {\bibinfo {title} {{Critical
  Exponents in Isotropic Spin Systems}},}\ }\href {\doibase
  10.1103/PhysRevB.6.1891} {\bibfield  {journal} {\bibinfo  {journal} {Phys.
  Rev. B}\ }\textbf {\bibinfo {volume} {6}},\ \bibinfo {pages} {1891--1893}
  (\bibinfo {year} {1972}{\natexlab{b}})}\BibitemShut {NoStop}%
\bibitem [{\citenamefont {Balents}\ and\ \citenamefont
  {Fisher}(1996)}]{BalentsPRB96}%
  \BibitemOpen
  \bibfield  {author} {\bibinfo {author} {\bibfnamefont {L.}~\bibnamefont
  {Balents}}\ and\ \bibinfo {author} {\bibfnamefont {M.~P.~A.}\ \bibnamefont
  {Fisher}},\ }\bibfield  {title} {\enquote {\bibinfo {title} {{Weak-coupling
  phase diagram of the two-chain Hubbard model}},}\ }\href {\doibase
  10.1103/PhysRevB.53.12133} {\bibfield  {journal} {\bibinfo  {journal} {Phys.
  Rev. B}\ }\textbf {\bibinfo {volume} {53}},\ \bibinfo {pages} {12133--12141}
  (\bibinfo {year} {1996})}\BibitemShut {NoStop}%
\bibitem [{\citenamefont {Lin}\ \emph {et~al.}(1997)\citenamefont {Lin},
  \citenamefont {Balents},\ and\ \citenamefont {Fisher}}]{LinPRB97}%
  \BibitemOpen
  \bibfield  {author} {\bibinfo {author} {\bibfnamefont {H.-H.}\ \bibnamefont
  {Lin}}, \bibinfo {author} {\bibfnamefont {L.}~\bibnamefont {Balents}}, \ and\
  \bibinfo {author} {\bibfnamefont {M.~P.~A.}\ \bibnamefont {Fisher}},\
  }\bibfield  {title} {\enquote {\bibinfo {title} {{$N$-chain Hubbard model in
  weak coupling}},}\ }\href {\doibase 10.1103/PhysRevB.56.6569} {\bibfield
  {journal} {\bibinfo  {journal} {Phys. Rev. B}\ }\textbf {\bibinfo {volume}
  {56}},\ \bibinfo {pages} {6569--6593} (\bibinfo {year} {1997})}\BibitemShut
  {NoStop}%
\bibitem [{\citenamefont {Assaraf}\ \emph {et~al.}(2004)\citenamefont
  {Assaraf}, \citenamefont {Azaria}, \citenamefont {Boulat}, \citenamefont
  {Caffarel},\ and\ \citenamefont {Lecheminant}}]{AssarafPRL04}%
  \BibitemOpen
  \bibfield  {author} {\bibinfo {author} {\bibfnamefont {R.}~\bibnamefont
  {Assaraf}}, \bibinfo {author} {\bibfnamefont {P.}~\bibnamefont {Azaria}},
  \bibinfo {author} {\bibfnamefont {E.}~\bibnamefont {Boulat}}, \bibinfo
  {author} {\bibfnamefont {M.}~\bibnamefont {Caffarel}}, \ and\ \bibinfo
  {author} {\bibfnamefont {P.}~\bibnamefont {Lecheminant}},\ }\bibfield
  {title} {\enquote {\bibinfo {title} {{Dynamical Symmetry Enlargement versus
  Spin-Charge Decoupling in the One-Dimensional SU(4) Hubbard Model}},}\ }\href
  {\doibase 10.1103/PhysRevLett.93.016407} {\bibfield  {journal} {\bibinfo
  {journal} {Phys. Rev. Lett.}\ }\textbf {\bibinfo {volume} {93}},\ \bibinfo
  {pages} {016407} (\bibinfo {year} {2004})}\BibitemShut {NoStop}%
\bibitem [{\citenamefont {Konik}\ \emph {et~al.}(2002)\citenamefont {Konik},
  \citenamefont {Saleur},\ and\ \citenamefont {Ludwig}}]{KonikPRB02}%
  \BibitemOpen
  \bibfield  {author} {\bibinfo {author} {\bibfnamefont {R.~M.}\ \bibnamefont
  {Konik}}, \bibinfo {author} {\bibfnamefont {H.}~\bibnamefont {Saleur}}, \
  and\ \bibinfo {author} {\bibfnamefont {A.~W.~W.}\ \bibnamefont {Ludwig}},\
  }\bibfield  {title} {\enquote {\bibinfo {title} {{Interplay of the scaling
  limit and the renormalization group: Implications for symmetry
  restoration}},}\ }\href {\doibase 10.1103/PhysRevB.66.075105} {\bibfield
  {journal} {\bibinfo  {journal} {Phys. Rev. B}\ }\textbf {\bibinfo {volume}
  {66}},\ \bibinfo {pages} {075105} (\bibinfo {year} {2002})}\BibitemShut
  {NoStop}%
\bibitem [{\citenamefont {{Frishman}}\ and\ \citenamefont
  {{Sonnenschein}}(2016)}]{FrishmanArxiv16}%
  \BibitemOpen
  \bibfield  {author} {\bibinfo {author} {\bibfnamefont {Y.}~\bibnamefont
  {{Frishman}}}\ and\ \bibinfo {author} {\bibfnamefont {J.}~\bibnamefont
  {{Sonnenschein}}},\ }\bibfield  {title} {\enquote {\bibinfo {title} {{On
  bound-states of the Gross Neveu model with massive fundamental fermions}},}\
  }\href@noop {} {\bibfield  {journal} {\bibinfo  {journal} {ArXiv e-prints}\ }
  (\bibinfo {year} {2016})},\ \Eprint {http://arxiv.org/abs/1603.00067}
  {arXiv:1603.00067 [hep-th]} \BibitemShut {NoStop}%
\bibitem [{\citenamefont {Fabrizio}\ and\ \citenamefont
  {Gogolin}(1994)}]{FabrizioPRB94}%
  \BibitemOpen
  \bibfield  {author} {\bibinfo {author} {\bibfnamefont {M.}~\bibnamefont
  {Fabrizio}}\ and\ \bibinfo {author} {\bibfnamefont {A.~O.}\ \bibnamefont
  {Gogolin}},\ }\bibfield  {title} {\enquote {\bibinfo {title} {{Toulouse limit
  for the overscreened four-channel Kondo problem}},}\ }\href {\doibase
  10.1103/PhysRevB.50.17732} {\bibfield  {journal} {\bibinfo  {journal} {Phys.
  Rev. B}\ }\textbf {\bibinfo {volume} {50}},\ \bibinfo {pages} {17732--17735}
  (\bibinfo {year} {1994})}\BibitemShut {NoStop}%
\bibitem [{\citenamefont {Affleck}(1988)}]{AffleckNuclPhysB88}%
  \BibitemOpen
  \bibfield  {author} {\bibinfo {author} {\bibfnamefont {I.}~\bibnamefont
  {Affleck}},\ }\bibfield  {title} {\enquote {\bibinfo {title} {{Critical
  behaviour of $SU(n)$ quantum chains and topological non-linear
  $\sigma$ƒ-models}},}\ }\href {\doibase
  http://dx.doi.org/10.1016/0550-3213(88)90117-4} {\bibfield  {journal}
  {\bibinfo  {journal} {Nucl. Phys. B}\ }\textbf {\bibinfo {volume} {305}},\
  \bibinfo {pages} {582 -- 596} (\bibinfo {year} {1988})}\BibitemShut {NoStop}%
\bibitem [{\citenamefont {Shelton}\ \emph {et~al.}(1996)\citenamefont
  {Shelton}, \citenamefont {Nersesyan},\ and\ \citenamefont
  {Tsvelik}}]{SheltonPRB96}%
  \BibitemOpen
  \bibfield  {author} {\bibinfo {author} {\bibfnamefont {D.~G.}\ \bibnamefont
  {Shelton}}, \bibinfo {author} {\bibfnamefont {A.~A.}\ \bibnamefont
  {Nersesyan}}, \ and\ \bibinfo {author} {\bibfnamefont {A.~M.}\ \bibnamefont
  {Tsvelik}},\ }\bibfield  {title} {\enquote {\bibinfo {title}
  {{Antiferromagnetic spin ladders: Crossover between spin \textit{S} =1/2 and
  \textit{S} =1 chains}},}\ }\href {\doibase 10.1103/PhysRevB.53.8521}
  {\bibfield  {journal} {\bibinfo  {journal} {Phys. Rev. B}\ }\textbf {\bibinfo
  {volume} {53}},\ \bibinfo {pages} {8521--8532} (\bibinfo {year}
  {1996})}\BibitemShut {NoStop}%
\bibitem [{\citenamefont {Zamolodchikov}\ and\ \citenamefont
  {Zamolodchikov}(1979{\natexlab{a}})}]{ZamoldchikovAnnPhys79}%
  \BibitemOpen
  \bibfield  {author} {\bibinfo {author} {\bibfnamefont {A.~B.}\ \bibnamefont
  {Zamolodchikov}}\ and\ \bibinfo {author} {\bibfnamefont {Al.~B.}\
  \bibnamefont {Zamolodchikov}},\ }\bibfield  {title} {\enquote {\bibinfo
  {title} {{Factorized S-matrices in two dimensions as the exact solutions of
  certain relativistic quantum field theory models}},}\ }\href {\doibase
  http://dx.doi.org/10.1016/0003-4916(79)90391-9} {\bibfield  {journal}
  {\bibinfo  {journal} {Ann. Phys. (N.Y.)}\ }\textbf {\bibinfo {volume}
  {120}},\ \bibinfo {pages} {253 -- 291} (\bibinfo {year}
  {1979}{\natexlab{a}})}\BibitemShut {NoStop}%
\bibitem [{\citenamefont {Wiegmann}(1985{\natexlab{a}})}]{WiegmannPhysLettB85}%
  \BibitemOpen
  \bibfield  {author} {\bibinfo {author} {\bibfnamefont {P.~B.}\ \bibnamefont
  {Wiegmann}},\ }\bibfield  {title} {\enquote {\bibinfo {title} {{Exact
  solution of the O(3) nonlinear $\sigma$ƒ-model}},}\ }\href {\doibase
  http://dx.doi.org/10.1016/0370-2693(85)91171-2} {\bibfield  {journal}
  {\bibinfo  {journal} {Phys. Lett. B}\ }\textbf {\bibinfo {volume} {152}},\
  \bibinfo {pages} {209 -- 214} (\bibinfo {year}
  {1985}{\natexlab{a}})}\BibitemShut {NoStop}%
\bibitem [{\citenamefont {Wiegmann}(1985{\natexlab{b}})}]{WiegmannJETPLett85}%
  \BibitemOpen
  \bibfield  {author} {\bibinfo {author} {\bibfnamefont {P.~B.}\ \bibnamefont
  {Wiegmann}},\ }\bibfield  {title} {\enquote {\bibinfo {title} {{Exact
  solution of O(3) nonlinear two-dimension $\sigma$-model}},}\ }\href
  {http://www.jetpletters.ac.ru/ps/1441/article_21926.shtml} {\bibfield
  {journal} {\bibinfo  {journal} {JETP Lett.}\ }\textbf {\bibinfo {volume}
  {41}},\ \bibinfo {pages} {95} (\bibinfo {year}
  {1985}{\natexlab{b}})}\BibitemShut {NoStop}%
\bibitem [{\citenamefont {Zamolodchikov}\ and\ \citenamefont
  {Zamolodchikov}(1992)}]{ZamolodchikovNuclPhysB92}%
  \BibitemOpen
  \bibfield  {author} {\bibinfo {author} {\bibfnamefont {A.~B.}\ \bibnamefont
  {Zamolodchikov}}\ and\ \bibinfo {author} {\bibfnamefont {{Al}.~B.}\
  \bibnamefont {Zamolodchikov}},\ }\bibfield  {title} {\enquote {\bibinfo
  {title} {{Massless factorized scattering and sigma models with topological
  terms}},}\ }\href {\doibase http://dx.doi.org/10.1016/0550-3213(92)90136-Y}
  {\bibfield  {journal} {\bibinfo  {journal} {Nucl. Phys. B}\ }\textbf
  {\bibinfo {volume} {379}},\ \bibinfo {pages} {602 -- 623} (\bibinfo {year}
  {1992})}\BibitemShut {NoStop}%
\bibitem [{\citenamefont {Fateev}\ and\ \citenamefont
  {Zamolodchikov}(1991)}]{FateevPhysLettB91}%
  \BibitemOpen
  \bibfield  {author} {\bibinfo {author} {\bibfnamefont {V.~A.}\ \bibnamefont
  {Fateev}}\ and\ \bibinfo {author} {\bibfnamefont {Al.~B.}\ \bibnamefont
  {Zamolodchikov}},\ }\bibfield  {title} {\enquote {\bibinfo {title}
  {{Integrable perturbations of $Z_N$ parafermion models and the O(3) sigma
  model}},}\ }\href {\doibase http://dx.doi.org/10.1016/0370-2693(91)91283-2}
  {\bibfield  {journal} {\bibinfo  {journal} {Phys. Lett. B}\ }\textbf
  {\bibinfo {volume} {271}},\ \bibinfo {pages} {91 -- 100} (\bibinfo {year}
  {1991})}\BibitemShut {NoStop}%
\bibitem [{\citenamefont {Fradkin}\ \emph {et~al.}(2015)\citenamefont
  {Fradkin}, \citenamefont {Kivelson},\ and\ \citenamefont
  {Tranquada}}]{FradkinRMP15}%
  \BibitemOpen
  \bibfield  {author} {\bibinfo {author} {\bibfnamefont {E.}~\bibnamefont
  {Fradkin}}, \bibinfo {author} {\bibfnamefont {S.~A.}\ \bibnamefont
  {Kivelson}}, \ and\ \bibinfo {author} {\bibfnamefont {J.~M.}\ \bibnamefont
  {Tranquada}},\ }\bibfield  {title} {\enquote {\bibinfo {title}
  {{\textit{Colloquium} : Theory of intertwined orders in high temperature
  superconductors}},}\ }\href {\doibase 10.1103/RevModPhys.87.457} {\bibfield
  {journal} {\bibinfo  {journal} {Rev. Mod. Phys.}\ }\textbf {\bibinfo {volume}
  {87}},\ \bibinfo {pages} {457--482} (\bibinfo {year} {2015})}\BibitemShut
  {NoStop}%
\bibitem [{\citenamefont {Levine}\ \emph {et~al.}(1983)\citenamefont {Levine},
  \citenamefont {Libby},\ and\ \citenamefont {Pruisken}}]{LevinePRL83}%
  \BibitemOpen
  \bibfield  {author} {\bibinfo {author} {\bibfnamefont {H.}~\bibnamefont
  {Levine}}, \bibinfo {author} {\bibfnamefont {S.~B.}\ \bibnamefont {Libby}}, \
  and\ \bibinfo {author} {\bibfnamefont {A.~M.~M.}\ \bibnamefont {Pruisken}},\
  }\bibfield  {title} {\enquote {\bibinfo {title} {{Electron Delocalization by
  a Magnetic Field in Two Dimensions}},}\ }\href {\doibase
  10.1103/PhysRevLett.51.1915} {\bibfield  {journal} {\bibinfo  {journal}
  {Phys. Rev. Lett.}\ }\textbf {\bibinfo {volume} {51}},\ \bibinfo {pages}
  {1915--1918} (\bibinfo {year} {1983})}\BibitemShut {NoStop}%
\bibitem [{\citenamefont {Levine}\ \emph
  {et~al.}(1984{\natexlab{a}})\citenamefont {Levine}, \citenamefont {Libby},\
  and\ \citenamefont {Pruisken}}]{LevineNuclPhysB84a}%
  \BibitemOpen
  \bibfield  {author} {\bibinfo {author} {\bibfnamefont {H.}~\bibnamefont
  {Levine}}, \bibinfo {author} {\bibfnamefont {S.~B.}\ \bibnamefont {Libby}}, \
  and\ \bibinfo {author} {\bibfnamefont {A.~M.~M.}\ \bibnamefont {Pruisken}},\
  }\bibfield  {title} {\enquote {\bibinfo {title} {{Theory of the quantized
  Hall effect (I)}},}\ }\href {\doibase
  http://dx.doi.org/10.1016/0550-3213(84)90277-3} {\bibfield  {journal}
  {\bibinfo  {journal} {Nucl. Phys. B}\ }\textbf {\bibinfo {volume} {240}},\
  \bibinfo {pages} {30--48} (\bibinfo {year} {1984}{\natexlab{a}})}\BibitemShut
  {NoStop}%
\bibitem [{\citenamefont {Levine}\ \emph
  {et~al.}(1984{\natexlab{b}})\citenamefont {Levine}, \citenamefont {Libby},\
  and\ \citenamefont {Pruisken}}]{LevineNuclPhysB84b}%
  \BibitemOpen
  \bibfield  {author} {\bibinfo {author} {\bibfnamefont {H.}~\bibnamefont
  {Levine}}, \bibinfo {author} {\bibfnamefont {S.~B.}\ \bibnamefont {Libby}}, \
  and\ \bibinfo {author} {\bibfnamefont {A.~M.~M.}\ \bibnamefont {Pruisken}},\
  }\bibfield  {title} {\enquote {\bibinfo {title} {{Theory of the quantized
  hall effect (II)}},}\ }\href {\doibase
  http://dx.doi.org/10.1016/0550-3213(84)90278-5} {\bibfield  {journal}
  {\bibinfo  {journal} {Nucl. Phys. B}\ }\textbf {\bibinfo {volume} {240}},\
  \bibinfo {pages} {49 -- 70} (\bibinfo {year}
  {1984}{\natexlab{b}})}\BibitemShut {NoStop}%
\bibitem [{\citenamefont {Levine}\ \emph
  {et~al.}(1984{\natexlab{c}})\citenamefont {Levine}, \citenamefont {Libby},\
  and\ \citenamefont {Pruisken}}]{LevineNuclPhysB84c}%
  \BibitemOpen
  \bibfield  {author} {\bibinfo {author} {\bibfnamefont {H.}~\bibnamefont
  {Levine}}, \bibinfo {author} {\bibfnamefont {S.~B.}\ \bibnamefont {Libby}}, \
  and\ \bibinfo {author} {\bibfnamefont {A.~M.~M.}\ \bibnamefont {Pruisken}},\
  }\bibfield  {title} {\enquote {\bibinfo {title} {{Theory of the quantized
  Hall effect (III)}},}\ }\href {\doibase
  http://dx.doi.org/10.1016/0550-3213(84)90279-7} {\bibfield  {journal}
  {\bibinfo  {journal} {Nucl. Phys. B}\ }\textbf {\bibinfo {volume} {240}},\
  \bibinfo {pages} {71 -- 90} (\bibinfo {year}
  {1984}{\natexlab{c}})}\BibitemShut {NoStop}%
\bibitem [{\citenamefont {Pruisken}(1984)}]{PruiskenNuclPhysB84}%
  \BibitemOpen
  \bibfield  {author} {\bibinfo {author} {\bibfnamefont {A.~M.~M.}\
  \bibnamefont {Pruisken}},\ }\bibfield  {title} {\enquote {\bibinfo {title}
  {{On localization in the theory of the quantized hall effect: A
  two-dimensional realization of the $\theta$-vacuum}},}\ }\href {\doibase
  http://dx.doi.org/10.1016/0550-3213(84)90101-9} {\bibfield  {journal}
  {\bibinfo  {journal} {Nucl. Phys. B}\ }\textbf {\bibinfo {volume} {235}},\
  \bibinfo {pages} {277 -- 298} (\bibinfo {year} {1984})}\BibitemShut {NoStop}%
\bibitem [{\citenamefont {Lecheminant}\ \emph {et~al.}(2005)\citenamefont
  {Lecheminant}, \citenamefont {Boulat},\ and\ \citenamefont
  {Azaria}}]{LecheminantPRL05}%
  \BibitemOpen
  \bibfield  {author} {\bibinfo {author} {\bibfnamefont {P.}~\bibnamefont
  {Lecheminant}}, \bibinfo {author} {\bibfnamefont {E.}~\bibnamefont {Boulat}},
  \ and\ \bibinfo {author} {\bibfnamefont {P.}~\bibnamefont {Azaria}},\
  }\bibfield  {title} {\enquote {\bibinfo {title} {{Confinement and
  Superfluidity in One-Dimensional Degenerate Fermionic Cold Atoms}},}\ }\href
  {\doibase 10.1103/PhysRevLett.95.240402} {\bibfield  {journal} {\bibinfo
  {journal} {Phys. Rev. Lett.}\ }\textbf {\bibinfo {volume} {95}},\ \bibinfo
  {pages} {240402} (\bibinfo {year} {2005})}\BibitemShut {NoStop}%
\bibitem [{\citenamefont {Lecheminant}\ \emph {et~al.}(2008)\citenamefont
  {Lecheminant}, \citenamefont {Azaria},\ and\ \citenamefont
  {Boulat}}]{LecheminantNuclPhysB08}%
  \BibitemOpen
  \bibfield  {author} {\bibinfo {author} {\bibfnamefont {P.}~\bibnamefont
  {Lecheminant}}, \bibinfo {author} {\bibfnamefont {P.}~\bibnamefont {Azaria}},
  \ and\ \bibinfo {author} {\bibfnamefont {E.}~\bibnamefont {Boulat}},\
  }\bibfield  {title} {\enquote {\bibinfo {title} {{Competing orders in
  one-dimensional half-integer fermionic cold atoms: A conformal field theory
  approach}},}\ }\href {\doibase
  http://dx.doi.org/10.1016/j.nuclphysb.2007.12.034} {\bibfield  {journal}
  {\bibinfo  {journal} {Nucl. Phys. B}\ }\textbf {\bibinfo {volume} {798}},\
  \bibinfo {pages} {443 -- 469} (\bibinfo {year} {2008})}\BibitemShut {NoStop}%
\bibitem [{\citenamefont {Ho}(1998)}]{HoPRL98}%
  \BibitemOpen
  \bibfield  {author} {\bibinfo {author} {\bibfnamefont {T.-L.}\ \bibnamefont
  {Ho}},\ }\bibfield  {title} {\enquote {\bibinfo {title} {{Spinor Bose
  Condensates in Optical Traps}},}\ }\href {\doibase
  10.1103/PhysRevLett.81.742} {\bibfield  {journal} {\bibinfo  {journal} {Phys.
  Rev. Lett.}\ }\textbf {\bibinfo {volume} {81}},\ \bibinfo {pages} {742--745}
  (\bibinfo {year} {1998})}\BibitemShut {NoStop}%
\bibitem [{\citenamefont {Ho}\ and\ \citenamefont {Yip}(1999)}]{HoPRL99}%
  \BibitemOpen
  \bibfield  {author} {\bibinfo {author} {\bibfnamefont {T.-L.}\ \bibnamefont
  {Ho}}\ and\ \bibinfo {author} {\bibfnamefont {S.}~\bibnamefont {Yip}},\
  }\bibfield  {title} {\enquote {\bibinfo {title} {{Pairing of Fermions with
  Arbitrary Spin}},}\ }\href {\doibase 10.1103/PhysRevLett.82.247} {\bibfield
  {journal} {\bibinfo  {journal} {Phys. Rev. Lett.}\ }\textbf {\bibinfo
  {volume} {82}},\ \bibinfo {pages} {247--250} (\bibinfo {year}
  {1999})}\BibitemShut {NoStop}%
\bibitem [{\citenamefont {Honerkamp}\ and\ \citenamefont
  {Hofstetter}(2004{\natexlab{a}})}]{HonerkampPRL04}%
  \BibitemOpen
  \bibfield  {author} {\bibinfo {author} {\bibfnamefont {C.}~\bibnamefont
  {Honerkamp}}\ and\ \bibinfo {author} {\bibfnamefont {W.}~\bibnamefont
  {Hofstetter}},\ }\bibfield  {title} {\enquote {\bibinfo {title} {{Ultracold
  Fermions and the $\mathrm{SU}(N)$ Hubbard Model}},}\ }\href {\doibase
  10.1103/PhysRevLett.92.170403} {\bibfield  {journal} {\bibinfo  {journal}
  {Phys. Rev. Lett.}\ }\textbf {\bibinfo {volume} {92}},\ \bibinfo {pages}
  {170403} (\bibinfo {year} {2004}{\natexlab{a}})}\BibitemShut {NoStop}%
\bibitem [{\citenamefont {Honerkamp}\ and\ \citenamefont
  {Hofstetter}(2004{\natexlab{b}})}]{HonerkampPRB04}%
  \BibitemOpen
  \bibfield  {author} {\bibinfo {author} {\bibfnamefont {C.}~\bibnamefont
  {Honerkamp}}\ and\ \bibinfo {author} {\bibfnamefont {W.}~\bibnamefont
  {Hofstetter}},\ }\bibfield  {title} {\enquote {\bibinfo {title} {{BCS pairing
  in Fermi systems with $N$ different hyperfine states}},}\ }\href {\doibase
  10.1103/PhysRevB.70.094521} {\bibfield  {journal} {\bibinfo  {journal} {Phys.
  Rev. B}\ }\textbf {\bibinfo {volume} {70}},\ \bibinfo {pages} {094521}
  (\bibinfo {year} {2004}{\natexlab{b}})}\BibitemShut {NoStop}%
\bibitem [{\citenamefont {Paananen}\ \emph {et~al.}(2006)\citenamefont
  {Paananen}, \citenamefont {Martikainen},\ and\ \citenamefont
  {T\"orm\"a}}]{PaananenPRA06}%
  \BibitemOpen
  \bibfield  {author} {\bibinfo {author} {\bibfnamefont {T.}~\bibnamefont
  {Paananen}}, \bibinfo {author} {\bibfnamefont {J.-P.}\ \bibnamefont
  {Martikainen}}, \ and\ \bibinfo {author} {\bibfnamefont {P.}~\bibnamefont
  {T\"orm\"a}},\ }\bibfield  {title} {\enquote {\bibinfo {title} {{Pairing in a
  three-component Fermi gas}},}\ }\href {\doibase 10.1103/PhysRevA.73.053606}
  {\bibfield  {journal} {\bibinfo  {journal} {Phys. Rev. A}\ }\textbf {\bibinfo
  {volume} {73}},\ \bibinfo {pages} {053606} (\bibinfo {year}
  {2006})}\BibitemShut {NoStop}%
\bibitem [{\citenamefont {He}\ \emph {et~al.}(2006)\citenamefont {He},
  \citenamefont {Jin},\ and\ \citenamefont {Zhuang}}]{HePRA06}%
  \BibitemOpen
  \bibfield  {author} {\bibinfo {author} {\bibfnamefont {L.}~\bibnamefont
  {He}}, \bibinfo {author} {\bibfnamefont {M.}~\bibnamefont {Jin}}, \ and\
  \bibinfo {author} {\bibfnamefont {P.}~\bibnamefont {Zhuang}},\ }\bibfield
  {title} {\enquote {\bibinfo {title} {{Superfluidity in a three-flavor Fermi
  gas with $\mathrm{SU}(3)$ symmetry}},}\ }\href {\doibase
  10.1103/PhysRevA.74.033604} {\bibfield  {journal} {\bibinfo  {journal} {Phys.
  Rev. A}\ }\textbf {\bibinfo {volume} {74}},\ \bibinfo {pages} {033604}
  (\bibinfo {year} {2006})}\BibitemShut {NoStop}%
\bibitem [{\citenamefont {Paananen}\ \emph {et~al.}(2007)\citenamefont
  {Paananen}, \citenamefont {T\"orm\"a},\ and\ \citenamefont
  {Martikainen}}]{PaananenPRA07}%
  \BibitemOpen
  \bibfield  {author} {\bibinfo {author} {\bibfnamefont {T.}~\bibnamefont
  {Paananen}}, \bibinfo {author} {\bibfnamefont {P.}~\bibnamefont {T\"orm\"a}},
  \ and\ \bibinfo {author} {\bibfnamefont {J.-P.}\ \bibnamefont
  {Martikainen}},\ }\bibfield  {title} {\enquote {\bibinfo {title}
  {{Coexistence and shell structures of several superfluids in trapped
  three-component Fermi mixtures}},}\ }\href {\doibase
  10.1103/PhysRevA.75.023622} {\bibfield  {journal} {\bibinfo  {journal} {Phys.
  Rev. A}\ }\textbf {\bibinfo {volume} {75}},\ \bibinfo {pages} {023622}
  (\bibinfo {year} {2007})}\BibitemShut {NoStop}%
\bibitem [{\citenamefont {Cherng}\ \emph {et~al.}(2007)\citenamefont {Cherng},
  \citenamefont {Refael},\ and\ \citenamefont {Demler}}]{CherngPRL07}%
  \BibitemOpen
  \bibfield  {author} {\bibinfo {author} {\bibfnamefont {R.~W.}\ \bibnamefont
  {Cherng}}, \bibinfo {author} {\bibfnamefont {G.}~\bibnamefont {Refael}}, \
  and\ \bibinfo {author} {\bibfnamefont {E.}~\bibnamefont {Demler}},\
  }\bibfield  {title} {\enquote {\bibinfo {title} {{Superfluidity and Magnetism
  in Multicomponent Ultracold Fermions}},}\ }\href {\doibase
  10.1103/PhysRevLett.99.130406} {\bibfield  {journal} {\bibinfo  {journal}
  {Phys. Rev. Lett.}\ }\textbf {\bibinfo {volume} {99}},\ \bibinfo {pages}
  {130406} (\bibinfo {year} {2007})}\BibitemShut {NoStop}%
\bibitem [{\citenamefont {Read}\ and\ \citenamefont
  {Sachdev}(1991)}]{ReadPRL91}%
  \BibitemOpen
  \bibfield  {author} {\bibinfo {author} {\bibfnamefont {N.}~\bibnamefont
  {Read}}\ and\ \bibinfo {author} {\bibfnamefont {Subir}\ \bibnamefont
  {Sachdev}},\ }\bibfield  {title} {\enquote {\bibinfo {title} {Large-
  \textit{N} expansion for frustrated quantum antiferromagnets},}\ }\href
  {\doibase 10.1103/PhysRevLett.66.1773} {\bibfield  {journal} {\bibinfo
  {journal} {Phys. Rev. Lett.}\ }\textbf {\bibinfo {volume} {66}},\ \bibinfo
  {pages} {1773--1776} (\bibinfo {year} {1991})}\BibitemShut {NoStop}%
\bibitem [{\citenamefont {Bulmash}\ \emph {et~al.}(2017)\citenamefont
  {Bulmash}, \citenamefont {Jian},\ and\ \citenamefont {Qi}}]{BulmashArxiv16}%
  \BibitemOpen
  \bibfield  {author} {\bibinfo {author} {\bibfnamefont {D.}~\bibnamefont
  {Bulmash}}, \bibinfo {author} {\bibfnamefont {C.-M.}\ \bibnamefont {Jian}}, \
  and\ \bibinfo {author} {\bibfnamefont {X.-L.}\ \bibnamefont {Qi}},\
  }\bibfield  {title} {\enquote {\bibinfo {title} {{Strongly Interacting Phases
  of Metallic Wires in Strong Magnetic Field}},}\ }\href
  {https://arxiv.org/abs/1702.08528} {\bibfield  {journal} {\bibinfo  {journal}
  {ArXiv e-prints}\ } (\bibinfo {year} {2017})},\ \Eprint
  {http://arxiv.org/abs/1702.08528} {arXiv:1702.08528 [cond-mat]} \BibitemShut
  {NoStop}%
\bibitem [{\citenamefont {Flint}\ \emph {et~al.}(2008)\citenamefont {Flint},
  \citenamefont {Dzero},\ and\ \citenamefont {Coleman}}]{FlintNatPhys08}%
  \BibitemOpen
  \bibfield  {author} {\bibinfo {author} {\bibfnamefont {R.}~\bibnamefont
  {Flint}}, \bibinfo {author} {\bibfnamefont {M.}~\bibnamefont {Dzero}}, \ and\
  \bibinfo {author} {\bibfnamefont {P.}~\bibnamefont {Coleman}},\ }\bibfield
  {title} {\enquote {\bibinfo {title} {{Heavy electrons and the symplectic
  symmetry of spin}},}\ }\href {http://dx.doi.org/10.1038/nphys1024} {\bibfield
   {journal} {\bibinfo  {journal} {Nature Phys.}\ }\textbf {\bibinfo {volume}
  {4}},\ \bibinfo {pages} {643--648} (\bibinfo {year} {2008})}\BibitemShut
  {NoStop}%
\bibitem [{\citenamefont {Ogievetsky}\ \emph {et~al.}(1987)\citenamefont
  {Ogievetsky}, \citenamefont {Reshetikhin},\ and\ \citenamefont
  {Wiegmann}}]{OgievetskyNuclPhysB87}%
  \BibitemOpen
  \bibfield  {author} {\bibinfo {author} {\bibfnamefont {E.}~\bibnamefont
  {Ogievetsky}}, \bibinfo {author} {\bibfnamefont {N.}~\bibnamefont
  {Reshetikhin}}, \ and\ \bibinfo {author} {\bibfnamefont {P.}~\bibnamefont
  {Wiegmann}},\ }\bibfield  {title} {\enquote {\bibinfo {title} {{The principal
  chiral field in two dimensions on classical lie algebras: The Bethe-ansatz
  solution and factorized theory of scattering}},}\ }\href {\doibase
  http://dx.doi.org/10.1016/0550-3213(87)90138-6} {\bibfield  {journal}
  {\bibinfo  {journal} {Nucl. Phys. B}\ }\textbf {\bibinfo {volume} {280}},\
  \bibinfo {pages} {45 -- 96} (\bibinfo {year} {1987})}\BibitemShut {NoStop}%
\bibitem [{\citenamefont {Tsvelik}(2014)}]{TsvelikPRL14}%
  \BibitemOpen
  \bibfield  {author} {\bibinfo {author} {\bibfnamefont {A.~M.}\ \bibnamefont
  {Tsvelik}},\ }\bibfield  {title} {\enquote {\bibinfo {title} {{Integrable
  Model with Parafermion Zero Energy Modes}},}\ }\href {\doibase
  10.1103/PhysRevLett.113.066401} {\bibfield  {journal} {\bibinfo  {journal}
  {Phys. Rev. Lett.}\ }\textbf {\bibinfo {volume} {113}},\ \bibinfo {pages}
  {066401} (\bibinfo {year} {2014})}\BibitemShut {NoStop}%
\bibitem [{\citenamefont {Tsvelik}(1988)}]{TsvelikJETP88}%
  \BibitemOpen
  \bibfield  {author} {\bibinfo {author} {\bibfnamefont {A.~M.}\ \bibnamefont
  {Tsvelik}},\ }\bibfield  {title} {\enquote {\bibinfo {title} {{Exact solution
  of two-dimensional $Z_N$-invariant self-dual models near the critical
  point}},}\ }\href {http://www.jetp.ac.ru/cgi-bin/e/index/e/67/7/p1436?a=list}
  {\bibfield  {journal} {\bibinfo  {journal} {Sov. Phys. JETP}\ }\textbf
  {\bibinfo {volume} {68}},\ \bibinfo {pages} {1436} (\bibinfo {year}
  {1988})}\BibitemShut {NoStop}%
\bibitem [{\citenamefont {Fateev}(1991)}]{FateevIntJModPhysA91}%
  \BibitemOpen
  \bibfield  {author} {\bibinfo {author} {\bibfnamefont {V.~A.}\ \bibnamefont
  {Fateev}},\ }\bibfield  {title} {\enquote {\bibinfo {title} {{Integrable
  deformations in $Z_N$-symmetrical models of the conformal quantum field
  theory}},}\ }\href {http://dx.doi.org/10.1142/S0217751X91001052} {\bibfield
  {journal} {\bibinfo  {journal} {Int. J. Mod. Phys. A}\ }\textbf {\bibinfo
  {volume} {6}},\ \bibinfo {pages} {2109} (\bibinfo {year} {1991})}\BibitemShut
  {NoStop}%
\bibitem [{\citenamefont {McCoy}\ and\ \citenamefont {Wu}(2014)}]{IsingBook}%
  \BibitemOpen
  \bibfield  {author} {\bibinfo {author} {\bibfnamefont {B.~M.}\ \bibnamefont
  {McCoy}}\ and\ \bibinfo {author} {\bibfnamefont {T.~T.}\ \bibnamefont {Wu}},\
  }\href {https://books.google.com/books?id=nB\_-AgAAQBAJ} {\emph {\bibinfo
  {title} {{The Two-Dimensional Ising Model: Second Edition}}}},\ Dover books
  on physics\ (\bibinfo  {publisher} {Dover Publications},\ \bibinfo {year}
  {2014})\BibitemShut {NoStop}%
\bibitem [{\citenamefont {Kitaev}(2003)}]{KitaevAnnPhys03}%
  \BibitemOpen
  \bibfield  {author} {\bibinfo {author} {\bibfnamefont {A.~Yu.}\ \bibnamefont
  {Kitaev}},\ }\bibfield  {title} {\enquote {\bibinfo {title} {{Fault-tolerant
  quantum computation by anyons}},}\ }\href
  {http://www.sciencedirect.com/science/article/pii/S0003491602000180}
  {\bibfield  {journal} {\bibinfo  {journal} {Ann. Phys. (N.Y.)}\ }\textbf
  {\bibinfo {volume} {303}},\ \bibinfo {pages} {2--30} (\bibinfo {year}
  {2003})}\BibitemShut {NoStop}%
\bibitem [{\citenamefont {Freedman}\ \emph {et~al.}(2002)\citenamefont
  {Freedman}, \citenamefont {Kitaev}, \citenamefont {Larsen},\ and\
  \citenamefont {Wang}}]{FreedmanBAMS03}%
  \BibitemOpen
  \bibfield  {author} {\bibinfo {author} {\bibfnamefont {M.~H.}\ \bibnamefont
  {Freedman}}, \bibinfo {author} {\bibfnamefont {A.}~\bibnamefont {Kitaev}},
  \bibinfo {author} {\bibfnamefont {M.~J.}\ \bibnamefont {Larsen}}, \ and\
  \bibinfo {author} {\bibfnamefont {Z.}~\bibnamefont {Wang}},\ }\bibfield
  {title} {\enquote {\bibinfo {title} {{Topological quantum computation}},}\
  }\href {\doibase http://dx.doi.org/10.1090/S0273-0979-02-00964-3} {\bibfield
  {journal} {\bibinfo  {journal} {Bull. Amer. Math. Soc.}\ }\textbf {\bibinfo
  {volume} {40}},\ \bibinfo {pages} {31--38} (\bibinfo {year}
  {2002})}\BibitemShut {NoStop}%
\bibitem [{\citenamefont {Bonderson}\ \emph {et~al.}(2008)\citenamefont
  {Bonderson}, \citenamefont {Freedman},\ and\ \citenamefont
  {Nayak}}]{BondersonPRL08}%
  \BibitemOpen
  \bibfield  {author} {\bibinfo {author} {\bibfnamefont {P.}~\bibnamefont
  {Bonderson}}, \bibinfo {author} {\bibfnamefont {M.}~\bibnamefont {Freedman}},
  \ and\ \bibinfo {author} {\bibfnamefont {C.}~\bibnamefont {Nayak}},\
  }\bibfield  {title} {\enquote {\bibinfo {title} {{Measurement-Only
  Topological Quantum Computation}},}\ }\href {\doibase
  10.1103/PhysRevLett.101.010501} {\bibfield  {journal} {\bibinfo  {journal}
  {Phys. Rev. Lett.}\ }\textbf {\bibinfo {volume} {101}},\ \bibinfo {pages}
  {010501} (\bibinfo {year} {2008})}\BibitemShut {NoStop}%
\bibitem [{\citenamefont {Nayak}\ \emph {et~al.}(2008)\citenamefont {Nayak},
  \citenamefont {Simon}, \citenamefont {Stern}, \citenamefont {Freedman},\ and\
  \citenamefont {Das~Sarma}}]{NayakRMP08}%
  \BibitemOpen
  \bibfield  {author} {\bibinfo {author} {\bibfnamefont {C.}~\bibnamefont
  {Nayak}}, \bibinfo {author} {\bibfnamefont {S.~H.}\ \bibnamefont {Simon}},
  \bibinfo {author} {\bibfnamefont {A.}~\bibnamefont {Stern}}, \bibinfo
  {author} {\bibfnamefont {M.}~\bibnamefont {Freedman}}, \ and\ \bibinfo
  {author} {\bibfnamefont {S.}~\bibnamefont {Das~Sarma}},\ }\bibfield  {title}
  {\enquote {\bibinfo {title} {{Non-Abelian anyons and topological quantum
  computation}},}\ }\href {\doibase 10.1103/RevModPhys.80.1083} {\bibfield
  {journal} {\bibinfo  {journal} {Rev. Mod. Phys.}\ }\textbf {\bibinfo {volume}
  {80}},\ \bibinfo {pages} {1083--1159} (\bibinfo {year} {2008})}\BibitemShut
  {NoStop}%
\bibitem [{\citenamefont {Moore}\ and\ \citenamefont
  {Read}(1991)}]{MooreNuclPhysB91}%
  \BibitemOpen
  \bibfield  {author} {\bibinfo {author} {\bibfnamefont {G.}~\bibnamefont
  {Moore}}\ and\ \bibinfo {author} {\bibfnamefont {N.}~\bibnamefont {Read}},\
  }\bibfield  {title} {\enquote {\bibinfo {title} {{Nonabelions in the
  fractional quantum hall effect}},}\ }\href {\doibase
  http://dx.doi.org/10.1016/0550-3213(91)90407-O} {\bibfield  {journal}
  {\bibinfo  {journal} {Nucl. Phys. B}\ }\textbf {\bibinfo {volume} {360}},\
  \bibinfo {pages} {362 -- 396} (\bibinfo {year} {1991})}\BibitemShut {NoStop}%
\bibitem [{\citenamefont {Bais}(1980)}]{BaisNuclPhysB80}%
  \BibitemOpen
  \bibfield  {author} {\bibinfo {author} {\bibfnamefont {F.~A.}\ \bibnamefont
  {Bais}},\ }\bibfield  {title} {\enquote {\bibinfo {title} {{Flux
  metamorphosis}},}\ }\href {\doibase
  http://dx.doi.org/10.1016/0550-3213(80)90474-5} {\bibfield  {journal}
  {\bibinfo  {journal} {Nucl. Phys. B}\ }\textbf {\bibinfo {volume} {170}},\
  \bibinfo {pages} {32 -- 43} (\bibinfo {year} {1980})}\BibitemShut {NoStop}%
\bibitem [{\citenamefont {Goldin}\ \emph {et~al.}(1981)\citenamefont {Goldin},
  \citenamefont {Menikoff},\ and\ \citenamefont {Sharp}}]{GoldinJMathPhys81}%
  \BibitemOpen
  \bibfield  {author} {\bibinfo {author} {\bibfnamefont {G.~A.}\ \bibnamefont
  {Goldin}}, \bibinfo {author} {\bibfnamefont {R.}~\bibnamefont {Menikoff}}, \
  and\ \bibinfo {author} {\bibfnamefont {D.~H.}\ \bibnamefont {Sharp}},\
  }\bibfield  {title} {\enquote {\bibinfo {title} {{Representations of a local
  current algebra in nonsimply connected space and the Aharonov-Bohm
  effect}},}\ }\href {\doibase http://dx.doi.org/10.1063/1.525110} {\bibfield
  {journal} {\bibinfo  {journal} {J. Math. Phys.}\ }\textbf {\bibinfo {volume}
  {22}},\ \bibinfo {pages} {1664--1668} (\bibinfo {year} {1981})}\BibitemShut
  {NoStop}%
\bibitem [{\citenamefont {Goldin}\ \emph {et~al.}(1985)\citenamefont {Goldin},
  \citenamefont {Menikoff},\ and\ \citenamefont {Sharp}}]{GoldinPRL85}%
  \BibitemOpen
  \bibfield  {author} {\bibinfo {author} {\bibfnamefont {G.~A.}\ \bibnamefont
  {Goldin}}, \bibinfo {author} {\bibfnamefont {R.}~\bibnamefont {Menikoff}}, \
  and\ \bibinfo {author} {\bibfnamefont {D.~H.}\ \bibnamefont {Sharp}},\
  }\bibfield  {title} {\enquote {\bibinfo {title} {{Comments on "General Theory
  for Quantum Statistics in Two Dimensions"}},}\ }\href {\doibase
  10.1103/PhysRevLett.54.603} {\bibfield  {journal} {\bibinfo  {journal} {Phys.
  Rev. Lett.}\ }\textbf {\bibinfo {volume} {54}},\ \bibinfo {pages} {603--603}
  (\bibinfo {year} {1985})}\BibitemShut {NoStop}%
\bibitem [{\citenamefont {Mourik}\ \emph {et~al.}(2012)\citenamefont {Mourik},
  \citenamefont {Zuo}, \citenamefont {Frolov}, \citenamefont {Plissard},
  \citenamefont {Bakkers},\ and\ \citenamefont
  {Kouwenhoven}}]{MourikScience12}%
  \BibitemOpen
  \bibfield  {author} {\bibinfo {author} {\bibfnamefont {V.}~\bibnamefont
  {Mourik}}, \bibinfo {author} {\bibfnamefont {K.}~\bibnamefont {Zuo}},
  \bibinfo {author} {\bibfnamefont {S.~M.}\ \bibnamefont {Frolov}}, \bibinfo
  {author} {\bibfnamefont {S.~R.}\ \bibnamefont {Plissard}}, \bibinfo {author}
  {\bibfnamefont {E.~P. A.~M.}\ \bibnamefont {Bakkers}}, \ and\ \bibinfo
  {author} {\bibfnamefont {L.~P.}\ \bibnamefont {Kouwenhoven}},\ }\bibfield
  {title} {\enquote {\bibinfo {title} {{Signatures of Majorana Fermions in
  Hybrid Superconductor-Semiconductor Nanowire Devices}},}\ }\href {\doibase
  10.1126/science.1222360} {\bibfield  {journal} {\bibinfo  {journal}
  {Science}\ }\textbf {\bibinfo {volume} {336}},\ \bibinfo {pages} {1003--1007}
  (\bibinfo {year} {2012})}\BibitemShut {NoStop}%
\bibitem [{\citenamefont {Willett}(2013)}]{WillettRepProgPhys13}%
  \BibitemOpen
  \bibfield  {author} {\bibinfo {author} {\bibfnamefont {R.~L.}\ \bibnamefont
  {Willett}},\ }\bibfield  {title} {\enquote {\bibinfo {title} {{The quantum
  Hall effect at 5/2 filling factor}},}\ }\href
  {http://stacks.iop.org/0034-4885/76/i=7/a=076501} {\bibfield  {journal}
  {\bibinfo  {journal} {Rep. Prog. Phys.}\ }\textbf {\bibinfo {volume} {76}},\
  \bibinfo {pages} {076501} (\bibinfo {year} {2013})}\BibitemShut {NoStop}%
\bibitem [{\citenamefont {Green}(1953)}]{GreenPR53}%
  \BibitemOpen
  \bibfield  {author} {\bibinfo {author} {\bibfnamefont {H.~S.}\ \bibnamefont
  {Green}},\ }\bibfield  {title} {\enquote {\bibinfo {title} {{A Generalized
  Method of Field Quantization}},}\ }\href {\doibase 10.1103/PhysRev.90.270}
  {\bibfield  {journal} {\bibinfo  {journal} {Phys. Rev.}\ }\textbf {\bibinfo
  {volume} {90}},\ \bibinfo {pages} {270--273} (\bibinfo {year}
  {1953})}\BibitemShut {NoStop}%
\bibitem [{\citenamefont {Fradkin}\ and\ \citenamefont
  {Kadanoff}(1980)}]{FradkinNuclPhysB80}%
  \BibitemOpen
  \bibfield  {author} {\bibinfo {author} {\bibfnamefont {E.}~\bibnamefont
  {Fradkin}}\ and\ \bibinfo {author} {\bibfnamefont {L.~P.}\ \bibnamefont
  {Kadanoff}},\ }\bibfield  {title} {\enquote {\bibinfo {title} {{Disorder
  variables and para-fermions in two-dimensional statistical mechanics}},}\
  }\href {\doibase http://dx.doi.org/10.1016/0550-3213(80)90472-1} {\bibfield
  {journal} {\bibinfo  {journal} {Nucl. Phys. B}\ }\textbf {\bibinfo {volume}
  {170}},\ \bibinfo {pages} {1 -- 15} (\bibinfo {year} {1980})}\BibitemShut
  {NoStop}%
\bibitem [{\citenamefont {Alicea}\ and\ \citenamefont
  {Fendley}(2016)}]{AliceaAnnRevCMP16}%
  \BibitemOpen
  \bibfield  {author} {\bibinfo {author} {\bibfnamefont {J.}~\bibnamefont
  {Alicea}}\ and\ \bibinfo {author} {\bibfnamefont {P.}~\bibnamefont
  {Fendley}},\ }\bibfield  {title} {\enquote {\bibinfo {title} {{Topological
  Phases with Parafermions: Theory and Blueprints}},}\ }\href
  {http://www.annualreviews.org/doi/10.1146/annurev-conmatphys-031115-011336}
  {\bibfield  {journal} {\bibinfo  {journal} {Ann. Rev. Cond. Mat. Phys.}\
  }\textbf {\bibinfo {volume} {7}},\ \bibinfo {pages} {119--139} (\bibinfo
  {year} {2016})}\BibitemShut {NoStop}%
\bibitem [{\citenamefont {Nersesyan}\ and\ \citenamefont
  {Tsvelik}(2011)}]{NersesyanEPL11}%
  \BibitemOpen
  \bibfield  {author} {\bibinfo {author} {\bibfnamefont {A.~A.}\ \bibnamefont
  {Nersesyan}}\ and\ \bibinfo {author} {\bibfnamefont {A.~M.}\ \bibnamefont
  {Tsvelik}},\ }\bibfield  {title} {\enquote {\bibinfo {title} {{Zero-energy
  Majorana modes in spin ladders and a possible realization of the Kitaev
  model}},}\ }\href {http://stacks.iop.org/0295-5075/96/i=1/a=17002} {\bibfield
   {journal} {\bibinfo  {journal} {Europhys. Lett.}\ }\textbf {\bibinfo
  {volume} {96}},\ \bibinfo {pages} {17002} (\bibinfo {year}
  {2011})}\BibitemShut {NoStop}%
\bibitem [{\citenamefont {Mong}\ \emph {et~al.}(2014)\citenamefont {Mong},
  \citenamefont {Clarke}, \citenamefont {Alicea}, \citenamefont {Lindner},
  \citenamefont {Fendley}, \citenamefont {Nayak}, \citenamefont {Oreg},
  \citenamefont {Stern}, \citenamefont {Berg}, \citenamefont {Shtengel},\ and\
  \citenamefont {Fisher}}]{MongPRX14}%
  \BibitemOpen
  \bibfield  {author} {\bibinfo {author} {\bibfnamefont {R.~S.~K.}\
  \bibnamefont {Mong}}, \bibinfo {author} {\bibfnamefont {D.~J.}\ \bibnamefont
  {Clarke}}, \bibinfo {author} {\bibfnamefont {J.}~\bibnamefont {Alicea}},
  \bibinfo {author} {\bibfnamefont {N.~H.}\ \bibnamefont {Lindner}}, \bibinfo
  {author} {\bibfnamefont {P.}~\bibnamefont {Fendley}}, \bibinfo {author}
  {\bibfnamefont {C.}~\bibnamefont {Nayak}}, \bibinfo {author} {\bibfnamefont
  {Y.}~\bibnamefont {Oreg}}, \bibinfo {author} {\bibfnamefont {A.}~\bibnamefont
  {Stern}}, \bibinfo {author} {\bibfnamefont {E.}~\bibnamefont {Berg}},
  \bibinfo {author} {\bibfnamefont {K.}~\bibnamefont {Shtengel}}, \ and\
  \bibinfo {author} {\bibfnamefont {M.~P.~A.}\ \bibnamefont {Fisher}},\
  }\bibfield  {title} {\enquote {\bibinfo {title} {{Universal Topological
  Quantum Computation from a Superconductor-Abelian Quantum Hall
  Heterostructure}},}\ }\href {\doibase 10.1103/PhysRevX.4.011036} {\bibfield
  {journal} {\bibinfo  {journal} {Phys. Rev. X}\ }\textbf {\bibinfo {volume}
  {4}},\ \bibinfo {pages} {011036} (\bibinfo {year} {2014})}\BibitemShut
  {NoStop}%
\bibitem [{\citenamefont {Zhuang}\ \emph {et~al.}(2015)\citenamefont {Zhuang},
  \citenamefont {Changlani}, \citenamefont {Tubman},\ and\ \citenamefont
  {Hughes}}]{ZhuangPRB15}%
  \BibitemOpen
  \bibfield  {author} {\bibinfo {author} {\bibfnamefont {Y.}~\bibnamefont
  {Zhuang}}, \bibinfo {author} {\bibfnamefont {H.~J.}\ \bibnamefont
  {Changlani}}, \bibinfo {author} {\bibfnamefont {N.~M.}\ \bibnamefont
  {Tubman}}, \ and\ \bibinfo {author} {\bibfnamefont {T.~L.}\ \bibnamefont
  {Hughes}},\ }\bibfield  {title} {\enquote {\bibinfo {title} {{Phase diagram
  of the ${Z}_{3}$ parafermionic chain with chiral interactions}},}\ }\href
  {\doibase 10.1103/PhysRevB.92.035154} {\bibfield  {journal} {\bibinfo
  {journal} {Phys. Rev. B}\ }\textbf {\bibinfo {volume} {92}},\ \bibinfo
  {pages} {035154} (\bibinfo {year} {2015})}\BibitemShut {NoStop}%
\bibitem [{\citenamefont {Trebst}\ \emph {et~al.}(2008)\citenamefont {Trebst},
  \citenamefont {Ardonne}, \citenamefont {Feiguin}, \citenamefont {Huse},
  \citenamefont {Ludwig},\ and\ \citenamefont {Troyer}}]{TrebstPRL08}%
  \BibitemOpen
  \bibfield  {author} {\bibinfo {author} {\bibfnamefont {S.}~\bibnamefont
  {Trebst}}, \bibinfo {author} {\bibfnamefont {E.}~\bibnamefont {Ardonne}},
  \bibinfo {author} {\bibfnamefont {A.}~\bibnamefont {Feiguin}}, \bibinfo
  {author} {\bibfnamefont {D.~A.}\ \bibnamefont {Huse}}, \bibinfo {author}
  {\bibfnamefont {A.~W.~W.}\ \bibnamefont {Ludwig}}, \ and\ \bibinfo {author}
  {\bibfnamefont {M.}~\bibnamefont {Troyer}},\ }\bibfield  {title} {\enquote
  {\bibinfo {title} {{Collective States of Interacting Fibonacci Anyons}},}\
  }\href {\doibase 10.1103/PhysRevLett.101.050401} {\bibfield  {journal}
  {\bibinfo  {journal} {Phys. Rev. Lett.}\ }\textbf {\bibinfo {volume} {101}},\
  \bibinfo {pages} {050401} (\bibinfo {year} {2008})}\BibitemShut {NoStop}%
\bibitem [{\citenamefont {Gils}\ \emph {et~al.}(2013)\citenamefont {Gils},
  \citenamefont {Ardonne}, \citenamefont {Trebst}, \citenamefont {Huse},
  \citenamefont {Ludwig}, \citenamefont {Troyer},\ and\ \citenamefont
  {Wang}}]{GilsPRB13}%
  \BibitemOpen
  \bibfield  {author} {\bibinfo {author} {\bibfnamefont {C.}~\bibnamefont
  {Gils}}, \bibinfo {author} {\bibfnamefont {E.}~\bibnamefont {Ardonne}},
  \bibinfo {author} {\bibfnamefont {S.}~\bibnamefont {Trebst}}, \bibinfo
  {author} {\bibfnamefont {D.~A.}\ \bibnamefont {Huse}}, \bibinfo {author}
  {\bibfnamefont {A.~W.~W.}\ \bibnamefont {Ludwig}}, \bibinfo {author}
  {\bibfnamefont {M.}~\bibnamefont {Troyer}}, \ and\ \bibinfo {author}
  {\bibfnamefont {Z.}~\bibnamefont {Wang}},\ }\bibfield  {title} {\enquote
  {\bibinfo {title} {{Anyonic quantum spin chains: Spin-1 generalizations and
  topological stability}},}\ }\href {\doibase 10.1103/PhysRevB.87.235120}
  {\bibfield  {journal} {\bibinfo  {journal} {Phys. Rev. B}\ }\textbf {\bibinfo
  {volume} {87}},\ \bibinfo {pages} {235120} (\bibinfo {year}
  {2013})}\BibitemShut {NoStop}%
\bibitem [{\citenamefont {Finch}\ and\ \citenamefont
  {Frahm}(2013)}]{FinchNJP13}%
  \BibitemOpen
  \bibfield  {author} {\bibinfo {author} {\bibfnamefont {P.~E.}\ \bibnamefont
  {Finch}}\ and\ \bibinfo {author} {\bibfnamefont {H.}~\bibnamefont {Frahm}},\
  }\bibfield  {title} {\enquote {\bibinfo {title} {{The D ( D 3 )-anyon chain:
  integrable boundary conditions and excitation spectra}},}\ }\href
  {http://stacks.iop.org/1367-2630/15/i=5/a=053035} {\bibfield  {journal}
  {\bibinfo  {journal} {New J. Phys.}\ }\textbf {\bibinfo {volume} {15}},\
  \bibinfo {pages} {053035} (\bibinfo {year} {2013})}\BibitemShut {NoStop}%
\bibitem [{\citenamefont {Jermyn}\ \emph {et~al.}(2014)\citenamefont {Jermyn},
  \citenamefont {Mong}, \citenamefont {Alicea},\ and\ \citenamefont
  {Fendley}}]{JermynPRB14}%
  \BibitemOpen
  \bibfield  {author} {\bibinfo {author} {\bibfnamefont {A.~S.}\ \bibnamefont
  {Jermyn}}, \bibinfo {author} {\bibfnamefont {R.~S.~K.}\ \bibnamefont {Mong}},
  \bibinfo {author} {\bibfnamefont {J.}~\bibnamefont {Alicea}}, \ and\ \bibinfo
  {author} {\bibfnamefont {P.}~\bibnamefont {Fendley}},\ }\bibfield  {title}
  {\enquote {\bibinfo {title} {{Stability of zero modes in parafermion
  chains}},}\ }\href {\doibase 10.1103/PhysRevB.90.165106} {\bibfield
  {journal} {\bibinfo  {journal} {Phys. Rev. B}\ }\textbf {\bibinfo {volume}
  {90}},\ \bibinfo {pages} {165106} (\bibinfo {year} {2014})}\BibitemShut
  {NoStop}%
\bibitem [{\citenamefont {Fendley}(2014)}]{FendleyJPhysA14}%
  \BibitemOpen
  \bibfield  {author} {\bibinfo {author} {\bibfnamefont {P.}~\bibnamefont
  {Fendley}},\ }\bibfield  {title} {\enquote {\bibinfo {title} {{Free
  parafermions}},}\ }\href {http://stacks.iop.org/1751-8121/47/i=7/a=075001}
  {\bibfield  {journal} {\bibinfo  {journal} {J. Phys. A}\ }\textbf {\bibinfo
  {volume} {47}},\ \bibinfo {pages} {075001} (\bibinfo {year}
  {2014})}\BibitemShut {NoStop}%
\bibitem [{\citenamefont {Castro-Alvaredo}\ \emph {et~al.}(2000)\citenamefont
  {Castro-Alvaredo}, \citenamefont {Fring}, \citenamefont {Korff},\ and\
  \citenamefont {Miramontes}}]{CastroAlvaredoNuclPhysB00}%
  \BibitemOpen
  \bibfield  {author} {\bibinfo {author} {\bibfnamefont {O.~A.}\ \bibnamefont
  {Castro-Alvaredo}}, \bibinfo {author} {\bibfnamefont {A.}~\bibnamefont
  {Fring}}, \bibinfo {author} {\bibfnamefont {C.}~\bibnamefont {Korff}}, \ and\
  \bibinfo {author} {\bibfnamefont {J.~L.}\ \bibnamefont {Miramontes}},\
  }\bibfield  {title} {\enquote {\bibinfo {title} {{Thermodynamic Bethe ansatz
  of the homogeneous sine-Gordon models}},}\ }\href {\doibase
  http://dx.doi.org/10.1016/S0550-3213(00)00162-0} {\bibfield  {journal}
  {\bibinfo  {journal} {Nucl. Phys. B}\ }\textbf {\bibinfo {volume} {575}},\
  \bibinfo {pages} {535 -- 560} (\bibinfo {year} {2000})}\BibitemShut {NoStop}%
\bibitem [{\citenamefont {Jerez}\ \emph {et~al.}(1998)\citenamefont {Jerez},
  \citenamefont {Andrei},\ and\ \citenamefont {Zar\'and}}]{JerezPRB98}%
  \BibitemOpen
  \bibfield  {author} {\bibinfo {author} {\bibfnamefont {A.}~\bibnamefont
  {Jerez}}, \bibinfo {author} {\bibfnamefont {N.}~\bibnamefont {Andrei}}, \
  and\ \bibinfo {author} {\bibfnamefont {G.}~\bibnamefont {Zar\'and}},\
  }\bibfield  {title} {\enquote {\bibinfo {title} {{Solution of the
  multichannel Coqblin-Schrieffer impurity model and application to multilevel
  systems}},}\ }\href {\doibase 10.1103/PhysRevB.58.3814} {\bibfield  {journal}
  {\bibinfo  {journal} {Phys. Rev. B}\ }\textbf {\bibinfo {volume} {58}},\
  \bibinfo {pages} {3814--3841} (\bibinfo {year} {1998})}\BibitemShut {NoStop}%
\bibitem [{\citenamefont {Lecheminant}\ and\ \citenamefont
  {Orignac}(2002)}]{LecheminantPRB02}%
  \BibitemOpen
  \bibfield  {author} {\bibinfo {author} {\bibfnamefont {P.}~\bibnamefont
  {Lecheminant}}\ and\ \bibinfo {author} {\bibfnamefont {E.}~\bibnamefont
  {Orignac}},\ }\bibfield  {title} {\enquote {\bibinfo {title} {{Magnetization
  and dimerization profiles of the cut two-leg spin ladder}},}\ }\href
  {\doibase 10.1103/PhysRevB.65.174406} {\bibfield  {journal} {\bibinfo
  {journal} {Phys. Rev. B}\ }\textbf {\bibinfo {volume} {65}},\ \bibinfo
  {pages} {174406} (\bibinfo {year} {2002})}\BibitemShut {NoStop}%
\bibitem [{\citenamefont {Tsvelik}(1995)}]{TsvelikPRB95}%
  \BibitemOpen
  \bibfield  {author} {\bibinfo {author} {\bibfnamefont {A.~M.}\ \bibnamefont
  {Tsvelik}},\ }\bibfield  {title} {\enquote {\bibinfo {title} {{Toulouse limit
  of the multichannel Kondo model}},}\ }\href {\doibase
  10.1103/PhysRevB.52.4366} {\bibfield  {journal} {\bibinfo  {journal} {Phys.
  Rev. B}\ }\textbf {\bibinfo {volume} {52}},\ \bibinfo {pages} {4366--4370}
  (\bibinfo {year} {1995})}\BibitemShut {NoStop}%
\bibitem [{\citenamefont {Tsvelik}(1987{\natexlab{b}})}]{TsvelikJETP87b}%
  \BibitemOpen
  \bibfield  {author} {\bibinfo {author} {\bibfnamefont {A.~M.}\ \bibnamefont
  {Tsvelik}},\ }\bibfield  {title} {\enquote {\bibinfo {title}
  {{1+1-dimensional sigma model at finite temperatures}},}\ }\href
  {http://www.jetp.ac.ru/cgi-bin/e/index/e/66/2/p221?a=list} {\bibfield
  {journal} {\bibinfo  {journal} {Sov. Phys. JETP}\ }\textbf {\bibinfo {volume}
  {66}},\ \bibinfo {pages} {221} (\bibinfo {year}
  {1987}{\natexlab{b}})}\BibitemShut {NoStop}%
\bibitem [{\citenamefont {Takahashi}(1971)}]{TakahashiProgTheorPhys71}%
  \BibitemOpen
  \bibfield  {author} {\bibinfo {author} {\bibfnamefont {M.}~\bibnamefont
  {Takahashi}},\ }\bibfield  {title} {\enquote {\bibinfo {title}
  {{One-Dimensional Heisenberg Model at Finite Temperature}},}\ }\href
  {\doibase 10.1143/PTP.46.401} {\bibfield  {journal} {\bibinfo  {journal}
  {Prog. Theor. Phys.}\ }\textbf {\bibinfo {volume} {46}},\ \bibinfo {pages}
  {401--415} (\bibinfo {year} {1971})}\BibitemShut {NoStop}%
\bibitem [{\citenamefont {Gaudin}(1971)}]{GaudinPRL71}%
  \BibitemOpen
  \bibfield  {author} {\bibinfo {author} {\bibfnamefont {M.}~\bibnamefont
  {Gaudin}},\ }\bibfield  {title} {\enquote {\bibinfo {title} {{Thermodynamics
  of the Heisenberg-Ising Ring for $\Delta>\sim 1$}},}\ }\href {\doibase
  10.1103/PhysRevLett.26.1301} {\bibfield  {journal} {\bibinfo  {journal}
  {Phys. Rev. Lett.}\ }\textbf {\bibinfo {volume} {26}},\ \bibinfo {pages}
  {1301--1304} (\bibinfo {year} {1971})}\BibitemShut {NoStop}%
\bibitem [{\citenamefont {Takahashi}\ and\ \citenamefont
  {Suzuki}(1972)}]{TakahashiProgTheorPhys72}%
  \BibitemOpen
  \bibfield  {author} {\bibinfo {author} {\bibfnamefont {M.}~\bibnamefont
  {Takahashi}}\ and\ \bibinfo {author} {\bibfnamefont {M.}~\bibnamefont
  {Suzuki}},\ }\bibfield  {title} {\enquote {\bibinfo {title} {{One-Dimensional
  Anisotropic Heisenberg Model at Finite Temperatures}},}\ }\href {\doibase
  10.1143/PTP.48.2187} {\bibfield  {journal} {\bibinfo  {journal} {Prog. Theor.
  Phys.}\ }\textbf {\bibinfo {volume} {48}},\ \bibinfo {pages} {2187--2209}
  (\bibinfo {year} {1972})}\BibitemShut {NoStop}%
\bibitem [{\citenamefont {Bazhanov}\ and\ \citenamefont
  {Reshetikhin}(1990)}]{BazhanovPTPS90}%
  \BibitemOpen
  \bibfield  {author} {\bibinfo {author} {\bibfnamefont {V.~V.}\ \bibnamefont
  {Bazhanov}}\ and\ \bibinfo {author} {\bibfnamefont {N.~Yu.}\ \bibnamefont
  {Reshetikhin}},\ }\bibfield  {title} {\enquote {\bibinfo {title} {{Scattering
  Amplitudes in Offcritical Models and RSOS Integrable Models}},}\ }\href
  {\doibase 10.1143/PTPS.102.301} {\bibfield  {journal} {\bibinfo  {journal}
  {Prog. Theor. Phys. Suppl.}\ }\textbf {\bibinfo {volume} {102}},\ \bibinfo
  {pages} {301--318} (\bibinfo {year} {1990})}\BibitemShut {NoStop}%
\bibitem [{\citenamefont {Reshetikhin}\ and\ \citenamefont
  {Smirnov}(1990)}]{ReshetikhinCMP90}%
  \BibitemOpen
  \bibfield  {author} {\bibinfo {author} {\bibfnamefont {N.}~\bibnamefont
  {Reshetikhin}}\ and\ \bibinfo {author} {\bibfnamefont {F.}~\bibnamefont
  {Smirnov}},\ }\bibfield  {title} {\enquote {\bibinfo {title} {{Hidden quantum
  group symmetry and integrable perturbations of conformal field theories}},}\
  }\href {http://projecteuclid.org/euclid.cmp/1104200706} {\bibfield  {journal}
  {\bibinfo  {journal} {Comm. Math. Phys.}\ }\textbf {\bibinfo {volume}
  {131}},\ \bibinfo {pages} {157--177} (\bibinfo {year} {1990})}\BibitemShut
  {NoStop}%
\bibitem [{\citenamefont {Lewenstein}\ \emph {et~al.}(2012)\citenamefont
  {Lewenstein}, \citenamefont {Sanpera},\ and\ \citenamefont
  {Ahufinger}}]{LewensteinBook}%
  \BibitemOpen
  \bibfield  {author} {\bibinfo {author} {\bibfnamefont {M.}~\bibnamefont
  {Lewenstein}}, \bibinfo {author} {\bibfnamefont {A.}~\bibnamefont {Sanpera}},
  \ and\ \bibinfo {author} {\bibfnamefont {V.}~\bibnamefont {Ahufinger}},\
  }\href {https://books.google.com/books?id=Wpl91RDxV5IC} {\emph {\bibinfo
  {title} {{Ultracold Atoms in Optical Lattices: Simulating quantum many-body
  systems}}}}\ (\bibinfo  {publisher} {OUP Oxford},\ \bibinfo {year}
  {2012})\BibitemShut {NoStop}%
\bibitem [{\citenamefont {Bloch}\ \emph {et~al.}(2008)\citenamefont {Bloch},
  \citenamefont {Dalibard},\ and\ \citenamefont {Zwerger}}]{BlochRMP08}%
  \BibitemOpen
  \bibfield  {author} {\bibinfo {author} {\bibfnamefont {I.}~\bibnamefont
  {Bloch}}, \bibinfo {author} {\bibfnamefont {J.}~\bibnamefont {Dalibard}}, \
  and\ \bibinfo {author} {\bibfnamefont {W.}~\bibnamefont {Zwerger}},\
  }\bibfield  {title} {\enquote {\bibinfo {title} {{Many-body physics with
  ultracold gases}},}\ }\href {\doibase 10.1103/RevModPhys.80.885} {\bibfield
  {journal} {\bibinfo  {journal} {Rev. Mod. Phys.}\ }\textbf {\bibinfo {volume}
  {80}},\ \bibinfo {pages} {885--964} (\bibinfo {year} {2008})}\BibitemShut
  {NoStop}%
\bibitem [{\citenamefont {Bloch}\ \emph {et~al.}(2012)\citenamefont {Bloch},
  \citenamefont {Dalibard},\ and\ \citenamefont {Nascimbene}}]{BlochNatPhys12}%
  \BibitemOpen
  \bibfield  {author} {\bibinfo {author} {\bibfnamefont {I.}~\bibnamefont
  {Bloch}}, \bibinfo {author} {\bibfnamefont {J.}~\bibnamefont {Dalibard}}, \
  and\ \bibinfo {author} {\bibfnamefont {S.}~\bibnamefont {Nascimbene}},\
  }\bibfield  {title} {\enquote {\bibinfo {title} {{Quantum simulations with
  ultracold quantum gases}},}\ }\href {http://dx.doi.org/10.1038/nphys2259}
  {\bibfield  {journal} {\bibinfo  {journal} {Nature Phys.}\ }\textbf {\bibinfo
  {volume} {8}},\ \bibinfo {pages} {267--276} (\bibinfo {year}
  {2012})}\BibitemShut {NoStop}%
\bibitem [{\citenamefont {Rapp}\ \emph {et~al.}(2007)\citenamefont {Rapp},
  \citenamefont {Zar\'and}, \citenamefont {Honerkamp},\ and\ \citenamefont
  {Hofstetter}}]{RappPRL07}%
  \BibitemOpen
  \bibfield  {author} {\bibinfo {author} {\bibfnamefont {\'A.}\ \bibnamefont
  {Rapp}}, \bibinfo {author} {\bibfnamefont {G.}~\bibnamefont {Zar\'and}},
  \bibinfo {author} {\bibfnamefont {C.}~\bibnamefont {Honerkamp}}, \ and\
  \bibinfo {author} {\bibfnamefont {W.}~\bibnamefont {Hofstetter}},\ }\bibfield
   {title} {\enquote {\bibinfo {title} {{Color Superfluidity and ``Baryon''
  Formation in Ultracold Fermions}},}\ }\href {\doibase
  10.1103/PhysRevLett.98.160405} {\bibfield  {journal} {\bibinfo  {journal}
  {Phys. Rev. Lett.}\ }\textbf {\bibinfo {volume} {98}},\ \bibinfo {pages}
  {160405} (\bibinfo {year} {2007})}\BibitemShut {NoStop}%
\bibitem [{\citenamefont {Hermele}\ and\ \citenamefont
  {Gurarie}(2011)}]{HermelePRB11}%
  \BibitemOpen
  \bibfield  {author} {\bibinfo {author} {\bibfnamefont {M.}~\bibnamefont
  {Hermele}}\ and\ \bibinfo {author} {\bibfnamefont {V.}~\bibnamefont
  {Gurarie}},\ }\bibfield  {title} {\enquote {\bibinfo {title} {{Topological
  liquids and valence cluster states in two-dimensional SU$(N)$ magnets}},}\
  }\href {\doibase 10.1103/PhysRevB.84.174441} {\bibfield  {journal} {\bibinfo
  {journal} {Phys. Rev. B}\ }\textbf {\bibinfo {volume} {84}},\ \bibinfo
  {pages} {174441} (\bibinfo {year} {2011})}\BibitemShut {NoStop}%
\bibitem [{\citenamefont {Nataf}\ \emph {et~al.}(2016)\citenamefont {Nataf},
  \citenamefont {Lajk\'o}, \citenamefont {Wietek}, \citenamefont {Penc},
  \citenamefont {Mila},\ and\ \citenamefont {L\"auchli}}]{NatafPRL16}%
  \BibitemOpen
  \bibfield  {author} {\bibinfo {author} {\bibfnamefont {P.}~\bibnamefont
  {Nataf}}, \bibinfo {author} {\bibfnamefont {M.}~\bibnamefont {Lajk\'o}},
  \bibinfo {author} {\bibfnamefont {A.}~\bibnamefont {Wietek}}, \bibinfo
  {author} {\bibfnamefont {K.}~\bibnamefont {Penc}}, \bibinfo {author}
  {\bibfnamefont {F.}~\bibnamefont {Mila}}, \ and\ \bibinfo {author}
  {\bibfnamefont {A.~M.}\ \bibnamefont {L\"auchli}},\ }\bibfield  {title}
  {\enquote {\bibinfo {title} {{Chiral Spin Liquids in Triangular-Lattice
  $\mathrm{SU}(N)$ Fermionic Mott Insulators with Artificial Gauge Fields}},}\
  }\href {\doibase 10.1103/PhysRevLett.117.167202} {\bibfield  {journal}
  {\bibinfo  {journal} {Phys. Rev. Lett.}\ }\textbf {\bibinfo {volume} {117}},\
  \bibinfo {pages} {167202} (\bibinfo {year} {2016})}\BibitemShut {NoStop}%
\bibitem [{\citenamefont {Ohmi}\ and\ \citenamefont
  {Machida}(1998)}]{OhmiJPSJ98}%
  \BibitemOpen
  \bibfield  {author} {\bibinfo {author} {\bibfnamefont {T.}~\bibnamefont
  {Ohmi}}\ and\ \bibinfo {author} {\bibfnamefont {K.}~\bibnamefont {Machida}},\
  }\bibfield  {title} {\enquote {\bibinfo {title} {{Bose-Einstein Condensation
  with Internal Degrees of Freedom in Alkali Atom Gases}},}\ }\href {\doibase
  10.1143/JPSJ.67.1822} {\bibfield  {journal} {\bibinfo  {journal} {J. Phys.
  Soc. Jpn.}\ }\textbf {\bibinfo {volume} {67}},\ \bibinfo {pages} {1822--1825}
  (\bibinfo {year} {1998})}\BibitemShut {NoStop}%
\bibitem [{\citenamefont {Zhang}\ \emph {et~al.}(2014)\citenamefont {Zhang},
  \citenamefont {Bishof}, \citenamefont {Bromley}, \citenamefont {Kraus},
  \citenamefont {Safronova}, \citenamefont {Zoller}, \citenamefont {Rey},\ and\
  \citenamefont {Ye}}]{ZhangScience14}%
  \BibitemOpen
  \bibfield  {author} {\bibinfo {author} {\bibfnamefont {X.}~\bibnamefont
  {Zhang}}, \bibinfo {author} {\bibfnamefont {M.}~\bibnamefont {Bishof}},
  \bibinfo {author} {\bibfnamefont {S.~L.}\ \bibnamefont {Bromley}}, \bibinfo
  {author} {\bibfnamefont {C.~V.}\ \bibnamefont {Kraus}}, \bibinfo {author}
  {\bibfnamefont {M.~S.}\ \bibnamefont {Safronova}}, \bibinfo {author}
  {\bibfnamefont {P.}~\bibnamefont {Zoller}}, \bibinfo {author} {\bibfnamefont
  {A.~M.}\ \bibnamefont {Rey}}, \ and\ \bibinfo {author} {\bibfnamefont
  {J.}~\bibnamefont {Ye}},\ }\bibfield  {title} {\enquote {\bibinfo {title}
  {{Spectroscopic observation of SU(N)-symmetric interactions in Sr orbital
  magnetism}},}\ }\href {\doibase 10.1126/science.1254978} {\bibfield
  {journal} {\bibinfo  {journal} {Science}\ }\textbf {\bibinfo {volume}
  {345}},\ \bibinfo {pages} {1467--1473} (\bibinfo {year} {2014})}\BibitemShut
  {NoStop}%
\bibitem [{\citenamefont {Scazza}\ \emph {et~al.}(2014)\citenamefont {Scazza},
  \citenamefont {Hofrichter}, \citenamefont {Hofer}, \citenamefont {De~Groot},
  \citenamefont {Bloch},\ and\ \citenamefont {Folling}}]{ScazzaNatPhys14}%
  \BibitemOpen
  \bibfield  {author} {\bibinfo {author} {\bibfnamefont {F.}~\bibnamefont
  {Scazza}}, \bibinfo {author} {\bibfnamefont {C.}~\bibnamefont {Hofrichter}},
  \bibinfo {author} {\bibfnamefont {M.}~\bibnamefont {Hofer}}, \bibinfo
  {author} {\bibfnamefont {P.~C.}\ \bibnamefont {De~Groot}}, \bibinfo {author}
  {\bibfnamefont {I.}~\bibnamefont {Bloch}}, \ and\ \bibinfo {author}
  {\bibfnamefont {S.}~\bibnamefont {Folling}},\ }\bibfield  {title} {\enquote
  {\bibinfo {title} {{Observation of two-orbital spin-exchange interactions
  with ultracold SU(N)-symmetric fermions}},}\ }\href
  {http://dx.doi.org/10.1038/nphys3061} {\bibfield  {journal} {\bibinfo
  {journal} {Nature Phys.}\ }\textbf {\bibinfo {volume} {10}},\ \bibinfo
  {pages} {779--784} (\bibinfo {year} {2014})}\BibitemShut {NoStop}%
\bibitem [{\citenamefont {Cappellini}\ \emph {et~al.}(2014)\citenamefont
  {Cappellini}, \citenamefont {Mancini}, \citenamefont {Pagano}, \citenamefont
  {Lombardi}, \citenamefont {Livi}, \citenamefont {Siciliani~de Cumis},
  \citenamefont {Cancio}, \citenamefont {Pizzocaro}, \citenamefont {Calonico},
  \citenamefont {Levi}, \citenamefont {Sias}, \citenamefont {Catani},
  \citenamefont {Inguscio},\ and\ \citenamefont {Fallani}}]{CappelliniPRL14}%
  \BibitemOpen
  \bibfield  {author} {\bibinfo {author} {\bibfnamefont {G.}~\bibnamefont
  {Cappellini}}, \bibinfo {author} {\bibfnamefont {M.}~\bibnamefont {Mancini}},
  \bibinfo {author} {\bibfnamefont {G.}~\bibnamefont {Pagano}}, \bibinfo
  {author} {\bibfnamefont {P.}~\bibnamefont {Lombardi}}, \bibinfo {author}
  {\bibfnamefont {L.}~\bibnamefont {Livi}}, \bibinfo {author} {\bibfnamefont
  {M.}~\bibnamefont {Siciliani~de Cumis}}, \bibinfo {author} {\bibfnamefont
  {P.}~\bibnamefont {Cancio}}, \bibinfo {author} {\bibfnamefont
  {M.}~\bibnamefont {Pizzocaro}}, \bibinfo {author} {\bibfnamefont
  {D.}~\bibnamefont {Calonico}}, \bibinfo {author} {\bibfnamefont
  {F.}~\bibnamefont {Levi}}, \bibinfo {author} {\bibfnamefont {C.}~\bibnamefont
  {Sias}}, \bibinfo {author} {\bibfnamefont {J.}~\bibnamefont {Catani}},
  \bibinfo {author} {\bibfnamefont {M.}~\bibnamefont {Inguscio}}, \ and\
  \bibinfo {author} {\bibfnamefont {L.}~\bibnamefont {Fallani}},\ }\bibfield
  {title} {\enquote {\bibinfo {title} {{Direct Observation of Coherent
  Interorbital Spin-Exchange Dynamics}},}\ }\href {\doibase
  10.1103/PhysRevLett.113.120402} {\bibfield  {journal} {\bibinfo  {journal}
  {Phys. Rev. Lett.}\ }\textbf {\bibinfo {volume} {113}},\ \bibinfo {pages}
  {120402} (\bibinfo {year} {2014})}\BibitemShut {NoStop}%
\bibitem [{\citenamefont {DeSalvo}\ \emph {et~al.}(2010)\citenamefont
  {DeSalvo}, \citenamefont {Yan}, \citenamefont {Mickelson}, \citenamefont
  {Martinez~de Escobar},\ and\ \citenamefont {Killian}}]{DeSalvoPRL10}%
  \BibitemOpen
  \bibfield  {author} {\bibinfo {author} {\bibfnamefont {B.~J.}\ \bibnamefont
  {DeSalvo}}, \bibinfo {author} {\bibfnamefont {M.}~\bibnamefont {Yan}},
  \bibinfo {author} {\bibfnamefont {P.~G.}\ \bibnamefont {Mickelson}}, \bibinfo
  {author} {\bibfnamefont {Y.~N.}\ \bibnamefont {Martinez~de Escobar}}, \ and\
  \bibinfo {author} {\bibfnamefont {T.~C.}\ \bibnamefont {Killian}},\
  }\bibfield  {title} {\enquote {\bibinfo {title} {{Degenerate Fermi Gas of
  $^{87}\mathrm{Sr}$}},}\ }\href {\doibase 10.1103/PhysRevLett.105.030402}
  {\bibfield  {journal} {\bibinfo  {journal} {Phys. Rev. Lett.}\ }\textbf
  {\bibinfo {volume} {105}},\ \bibinfo {pages} {030402} (\bibinfo {year}
  {2010})}\BibitemShut {NoStop}%
\bibitem [{\citenamefont {Tey}\ \emph {et~al.}(2010)\citenamefont {Tey},
  \citenamefont {Stellmer}, \citenamefont {Grimm},\ and\ \citenamefont
  {Schreck}}]{TeyPRA10}%
  \BibitemOpen
  \bibfield  {author} {\bibinfo {author} {\bibfnamefont {M.~K.}\ \bibnamefont
  {Tey}}, \bibinfo {author} {\bibfnamefont {S.}~\bibnamefont {Stellmer}},
  \bibinfo {author} {\bibfnamefont {R.}~\bibnamefont {Grimm}}, \ and\ \bibinfo
  {author} {\bibfnamefont {F.}~\bibnamefont {Schreck}},\ }\bibfield  {title}
  {\enquote {\bibinfo {title} {{Double-degenerate Bose-Fermi mixture of
  strontium}},}\ }\href {\doibase 10.1103/PhysRevA.82.011608} {\bibfield
  {journal} {\bibinfo  {journal} {Phys. Rev. A}\ }\textbf {\bibinfo {volume}
  {82}},\ \bibinfo {pages} {011608} (\bibinfo {year} {2010})}\BibitemShut
  {NoStop}%
\bibitem [{\citenamefont {Fukuhara}\ \emph {et~al.}(2007)\citenamefont
  {Fukuhara}, \citenamefont {Takasu}, \citenamefont {Kumakura},\ and\
  \citenamefont {Takahashi}}]{FukuharaPRL07}%
  \BibitemOpen
  \bibfield  {author} {\bibinfo {author} {\bibfnamefont {T.}~\bibnamefont
  {Fukuhara}}, \bibinfo {author} {\bibfnamefont {Y.}~\bibnamefont {Takasu}},
  \bibinfo {author} {\bibfnamefont {M.}~\bibnamefont {Kumakura}}, \ and\
  \bibinfo {author} {\bibfnamefont {Y.}~\bibnamefont {Takahashi}},\ }\bibfield
  {title} {\enquote {\bibinfo {title} {{Degenerate Fermi Gases of
  Ytterbium}},}\ }\href {\doibase 10.1103/PhysRevLett.98.030401} {\bibfield
  {journal} {\bibinfo  {journal} {Phys. Rev. Lett.}\ }\textbf {\bibinfo
  {volume} {98}},\ \bibinfo {pages} {030401} (\bibinfo {year}
  {2007})}\BibitemShut {NoStop}%
\bibitem [{\citenamefont {Taie}\ \emph {et~al.}(2010)\citenamefont {Taie},
  \citenamefont {Takasu}, \citenamefont {Sugawa}, \citenamefont {Yamazaki},
  \citenamefont {Tsujimoto}, \citenamefont {Murakami},\ and\ \citenamefont
  {Takahashi}}]{TaiePRL10}%
  \BibitemOpen
  \bibfield  {author} {\bibinfo {author} {\bibfnamefont {S.}~\bibnamefont
  {Taie}}, \bibinfo {author} {\bibfnamefont {Y.}~\bibnamefont {Takasu}},
  \bibinfo {author} {\bibfnamefont {S.}~\bibnamefont {Sugawa}}, \bibinfo
  {author} {\bibfnamefont {R.}~\bibnamefont {Yamazaki}}, \bibinfo {author}
  {\bibfnamefont {T.}~\bibnamefont {Tsujimoto}}, \bibinfo {author}
  {\bibfnamefont {R.}~\bibnamefont {Murakami}}, \ and\ \bibinfo {author}
  {\bibfnamefont {Y.}~\bibnamefont {Takahashi}},\ }\bibfield  {title} {\enquote
  {\bibinfo {title} {{Realization of a
  $\mathrm{SU}(2)\ifmmode\times\else\texttimes\fi{}\mathrm{SU}(6)$ System of
  Fermions in a Cold Atomic Gas}},}\ }\href {\doibase
  10.1103/PhysRevLett.105.190401} {\bibfield  {journal} {\bibinfo  {journal}
  {Phys. Rev. Lett.}\ }\textbf {\bibinfo {volume} {105}},\ \bibinfo {pages}
  {190401} (\bibinfo {year} {2010})}\BibitemShut {NoStop}%
\bibitem [{\citenamefont {Taie}\ \emph {et~al.}(2012)\citenamefont {Taie},
  \citenamefont {Yamazaki}, \citenamefont {Sugawa},\ and\ \citenamefont
  {Takahashi}}]{TaieNatPhys12}%
  \BibitemOpen
  \bibfield  {author} {\bibinfo {author} {\bibfnamefont {S.}~\bibnamefont
  {Taie}}, \bibinfo {author} {\bibfnamefont {R.}~\bibnamefont {Yamazaki}},
  \bibinfo {author} {\bibfnamefont {S.}~\bibnamefont {Sugawa}}, \ and\ \bibinfo
  {author} {\bibfnamefont {Y.}~\bibnamefont {Takahashi}},\ }\bibfield  {title}
  {\enquote {\bibinfo {title} {{An SU(6) Mott insulator of an atomic Fermi gas
  realized by large-spin Pomeranchuk cooling}},}\ }\href
  {http://dx.doi.org/10.1038/nphys2430} {\bibfield  {journal} {\bibinfo
  {journal} {Nature Phys.}\ }\textbf {\bibinfo {volume} {8}},\ \bibinfo {pages}
  {825--830} (\bibinfo {year} {2012})}\BibitemShut {NoStop}%
\bibitem [{\citenamefont {Pagano}\ \emph {et~al.}(2014)\citenamefont {Pagano},
  \citenamefont {Mancini}, \citenamefont {Cappellini}, \citenamefont
  {Lombardi}, \citenamefont {Schafer}, \citenamefont {Hu}, \citenamefont {Liu},
  \citenamefont {Catani}, \citenamefont {Sias}, \citenamefont {Inguscio},\ and\
  \citenamefont {Fallani}}]{PaganoNatPhys14}%
  \BibitemOpen
  \bibfield  {author} {\bibinfo {author} {\bibfnamefont {G.}~\bibnamefont
  {Pagano}}, \bibinfo {author} {\bibfnamefont {M.}~\bibnamefont {Mancini}},
  \bibinfo {author} {\bibfnamefont {G.}~\bibnamefont {Cappellini}}, \bibinfo
  {author} {\bibfnamefont {P.}~\bibnamefont {Lombardi}}, \bibinfo {author}
  {\bibfnamefont {F.}~\bibnamefont {Schafer}}, \bibinfo {author} {\bibfnamefont
  {H.}~\bibnamefont {Hu}}, \bibinfo {author} {\bibfnamefont {X.-J.}\
  \bibnamefont {Liu}}, \bibinfo {author} {\bibfnamefont {J.}~\bibnamefont
  {Catani}}, \bibinfo {author} {\bibfnamefont {C.}~\bibnamefont {Sias}},
  \bibinfo {author} {\bibfnamefont {M.}~\bibnamefont {Inguscio}}, \ and\
  \bibinfo {author} {\bibfnamefont {L.}~\bibnamefont {Fallani}},\ }\bibfield
  {title} {\enquote {\bibinfo {title} {{A one-dimensional liquid of fermions
  with tunable spin}},}\ }\href {http://dx.doi.org/10.1038/nphys2878}
  {\bibfield  {journal} {\bibinfo  {journal} {Nature Phys.}\ }\textbf {\bibinfo
  {volume} {10}},\ \bibinfo {pages} {198--201} (\bibinfo {year}
  {2014})}\BibitemShut {NoStop}%
\bibitem [{\citenamefont {Wu}\ and\ \citenamefont {Zhang}(2005)}]{WuPRB05}%
  \BibitemOpen
  \bibfield  {author} {\bibinfo {author} {\bibfnamefont {C.}~\bibnamefont
  {Wu}}\ and\ \bibinfo {author} {\bibfnamefont {S.-C.}\ \bibnamefont {Zhang}},\
  }\bibfield  {title} {\enquote {\bibinfo {title} {{Sufficient condition for
  absence of the sign problem in the fermionic quantum Monte Carlo
  algorithm}},}\ }\href {\doibase 10.1103/PhysRevB.71.155115} {\bibfield
  {journal} {\bibinfo  {journal} {Phys. Rev. B}\ }\textbf {\bibinfo {volume}
  {71}},\ \bibinfo {pages} {155115} (\bibinfo {year} {2005})}\BibitemShut
  {NoStop}%
\bibitem [{\citenamefont {Wu}(2005)}]{WuPRL05}%
  \BibitemOpen
  \bibfield  {author} {\bibinfo {author} {\bibfnamefont {C.}~\bibnamefont
  {Wu}},\ }\bibfield  {title} {\enquote {\bibinfo {title} {{Competing Orders in
  One-Dimensional Spin-$3/2$ Fermionic Systems}},}\ }\href {\doibase
  10.1103/PhysRevLett.95.266404} {\bibfield  {journal} {\bibinfo  {journal}
  {Phys. Rev. Lett.}\ }\textbf {\bibinfo {volume} {95}},\ \bibinfo {pages}
  {266404} (\bibinfo {year} {2005})}\BibitemShut {NoStop}%
\bibitem [{\citenamefont {Azaria}\ \emph {et~al.}(1999)\citenamefont {Azaria},
  \citenamefont {Gogolin}, \citenamefont {Lecheminant},\ and\ \citenamefont
  {Nersesyan}}]{AzariaPRL99}%
  \BibitemOpen
  \bibfield  {author} {\bibinfo {author} {\bibfnamefont {P.}~\bibnamefont
  {Azaria}}, \bibinfo {author} {\bibfnamefont {A.~O.}\ \bibnamefont {Gogolin}},
  \bibinfo {author} {\bibfnamefont {P.}~\bibnamefont {Lecheminant}}, \ and\
  \bibinfo {author} {\bibfnamefont {A.~A.}\ \bibnamefont {Nersesyan}},\
  }\bibfield  {title} {\enquote {\bibinfo {title} {{One-Dimensional SU(4)
  Spin-Orbital Model: A Low-Energy Effective Theory}},}\ }\href {\doibase
  10.1103/PhysRevLett.83.624} {\bibfield  {journal} {\bibinfo  {journal} {Phys.
  Rev. Lett.}\ }\textbf {\bibinfo {volume} {83}},\ \bibinfo {pages} {624--627}
  (\bibinfo {year} {1999})}\BibitemShut {NoStop}%
\bibitem [{\citenamefont {Controzzi}\ and\ \citenamefont
  {Tsvelik}(2006)}]{ControzziPRL06}%
  \BibitemOpen
  \bibfield  {author} {\bibinfo {author} {\bibfnamefont {D.}~\bibnamefont
  {Controzzi}}\ and\ \bibinfo {author} {\bibfnamefont {A.~M.}\ \bibnamefont
  {Tsvelik}},\ }\bibfield  {title} {\enquote {\bibinfo {title} {{Exactly
  Solvable Model for Isospin $S=3/2$ Fermionic Atoms on an Optical Lattice}},}\
  }\href {\doibase 10.1103/PhysRevLett.96.097205} {\bibfield  {journal}
  {\bibinfo  {journal} {Phys. Rev. Lett.}\ }\textbf {\bibinfo {volume} {96}},\
  \bibinfo {pages} {097205} (\bibinfo {year} {2006})}\BibitemShut {NoStop}%
\bibitem [{\citenamefont {Capponi}\ \emph {et~al.}(2007)\citenamefont
  {Capponi}, \citenamefont {Roux}, \citenamefont {Azaria}, \citenamefont
  {Boulat},\ and\ \citenamefont {Lecheminant}}]{CapponiPRB07}%
  \BibitemOpen
  \bibfield  {author} {\bibinfo {author} {\bibfnamefont {S.}~\bibnamefont
  {Capponi}}, \bibinfo {author} {\bibfnamefont {G.}~\bibnamefont {Roux}},
  \bibinfo {author} {\bibfnamefont {P.}~\bibnamefont {Azaria}}, \bibinfo
  {author} {\bibfnamefont {E.}~\bibnamefont {Boulat}}, \ and\ \bibinfo {author}
  {\bibfnamefont {P.}~\bibnamefont {Lecheminant}},\ }\bibfield  {title}
  {\enquote {\bibinfo {title} {{Confinement versus deconfinement of Cooper
  pairs in one-dimensional spin-$3?2$ fermionic cold atoms}},}\ }\href
  {\doibase 10.1103/PhysRevB.75.100503} {\bibfield  {journal} {\bibinfo
  {journal} {Phys. Rev. B}\ }\textbf {\bibinfo {volume} {75}},\ \bibinfo
  {pages} {100503} (\bibinfo {year} {2007})}\BibitemShut {NoStop}%
\bibitem [{\citenamefont {Capponi}\ \emph {et~al.}(2008)\citenamefont
  {Capponi}, \citenamefont {Roux}, \citenamefont {Lecheminant}, \citenamefont
  {Azaria}, \citenamefont {Boulat},\ and\ \citenamefont
  {White}}]{CapponiPRA08}%
  \BibitemOpen
  \bibfield  {author} {\bibinfo {author} {\bibfnamefont {S.}~\bibnamefont
  {Capponi}}, \bibinfo {author} {\bibfnamefont {G.}~\bibnamefont {Roux}},
  \bibinfo {author} {\bibfnamefont {P.}~\bibnamefont {Lecheminant}}, \bibinfo
  {author} {\bibfnamefont {P.}~\bibnamefont {Azaria}}, \bibinfo {author}
  {\bibfnamefont {E.}~\bibnamefont {Boulat}}, \ and\ \bibinfo {author}
  {\bibfnamefont {S.~R.}\ \bibnamefont {White}},\ }\bibfield  {title} {\enquote
  {\bibinfo {title} {{Molecular superfluid phase in systems of one-dimensional
  multicomponent fermionic cold atoms}},}\ }\href {\doibase
  10.1103/PhysRevA.77.013624} {\bibfield  {journal} {\bibinfo  {journal} {Phys.
  Rev. A}\ }\textbf {\bibinfo {volume} {77}},\ \bibinfo {pages} {013624}
  (\bibinfo {year} {2008})}\BibitemShut {NoStop}%
\bibitem [{\citenamefont {Roux}\ \emph {et~al.}(2009)\citenamefont {Roux},
  \citenamefont {Capponi}, \citenamefont {Lecheminant},\ and\ \citenamefont
  {Azaria}}]{RouxEPJB09}%
  \BibitemOpen
  \bibfield  {author} {\bibinfo {author} {\bibfnamefont {G.}~\bibnamefont
  {Roux}}, \bibinfo {author} {\bibfnamefont {S.}~\bibnamefont {Capponi}},
  \bibinfo {author} {\bibfnamefont {P.}~\bibnamefont {Lecheminant}}, \ and\
  \bibinfo {author} {\bibfnamefont {P.}~\bibnamefont {Azaria}},\ }\bibfield
  {title} {\enquote {\bibinfo {title} {{Spin 3/2 fermions with attractive
  interactions in a one-dimensional optical lattice: phase diagrams,
  entanglement entropy, and the effect of the trap}},}\ }\href {\doibase
  10.1140/epjb/e2008-00374-7} {\bibfield  {journal} {\bibinfo  {journal} {Eur.
  Phys. J. B}\ }\textbf {\bibinfo {volume} {68}},\ \bibinfo {pages} {293--308}
  (\bibinfo {year} {2009})}\BibitemShut {NoStop}%
\bibitem [{\citenamefont {Burovski}\ \emph {et~al.}(2009)\citenamefont
  {Burovski}, \citenamefont {Orso},\ and\ \citenamefont
  {Jolicoeur}}]{BurovskiPRL09}%
  \BibitemOpen
  \bibfield  {author} {\bibinfo {author} {\bibfnamefont {E.}~\bibnamefont
  {Burovski}}, \bibinfo {author} {\bibfnamefont {G.}~\bibnamefont {Orso}}, \
  and\ \bibinfo {author} {\bibfnamefont {T.}~\bibnamefont {Jolicoeur}},\
  }\bibfield  {title} {\enquote {\bibinfo {title} {{Multiparticle Composites in
  Density-Imbalanced Quantum Fluids}},}\ }\href {\doibase
  10.1103/PhysRevLett.103.215301} {\bibfield  {journal} {\bibinfo  {journal}
  {Phys. Rev. Lett.}\ }\textbf {\bibinfo {volume} {103}},\ \bibinfo {pages}
  {215301} (\bibinfo {year} {2009})}\BibitemShut {NoStop}%
\bibitem [{\citenamefont {Orso}\ \emph {et~al.}(2010)\citenamefont {Orso},
  \citenamefont {Burovski},\ and\ \citenamefont {Jolicoeur}}]{OrsoPRL10}%
  \BibitemOpen
  \bibfield  {author} {\bibinfo {author} {\bibfnamefont {G.}~\bibnamefont
  {Orso}}, \bibinfo {author} {\bibfnamefont {E.}~\bibnamefont {Burovski}}, \
  and\ \bibinfo {author} {\bibfnamefont {T.}~\bibnamefont {Jolicoeur}},\
  }\bibfield  {title} {\enquote {\bibinfo {title} {{Luttinger Liquid of Trimers
  in Fermi Gases with Unequal Masses}},}\ }\href {\doibase
  10.1103/PhysRevLett.104.065301} {\bibfield  {journal} {\bibinfo  {journal}
  {Phys. Rev. Lett.}\ }\textbf {\bibinfo {volume} {104}},\ \bibinfo {pages}
  {065301} (\bibinfo {year} {2010})}\BibitemShut {NoStop}%
\bibitem [{\citenamefont {Roux}\ \emph {et~al.}(2011)\citenamefont {Roux},
  \citenamefont {Burovski},\ and\ \citenamefont {Jolicoeur}}]{RouxPRA11}%
  \BibitemOpen
  \bibfield  {author} {\bibinfo {author} {\bibfnamefont {G.}~\bibnamefont
  {Roux}}, \bibinfo {author} {\bibfnamefont {E.}~\bibnamefont {Burovski}}, \
  and\ \bibinfo {author} {\bibfnamefont {T.}~\bibnamefont {Jolicoeur}},\
  }\bibfield  {title} {\enquote {\bibinfo {title} {{Multimer formation in
  one-dimensional two-component gases and trimer phase in the asymmetric
  attractive Hubbard model}},}\ }\href {\doibase 10.1103/PhysRevA.83.053618}
  {\bibfield  {journal} {\bibinfo  {journal} {Phys. Rev. A}\ }\textbf {\bibinfo
  {volume} {83}},\ \bibinfo {pages} {053618} (\bibinfo {year}
  {2011})}\BibitemShut {NoStop}%
\bibitem [{\citenamefont {Dalmonte}\ \emph {et~al.}(2011)\citenamefont
  {Dalmonte}, \citenamefont {Zoller},\ and\ \citenamefont
  {Pupillo}}]{DalmontePRL11}%
  \BibitemOpen
  \bibfield  {author} {\bibinfo {author} {\bibfnamefont {M.}~\bibnamefont
  {Dalmonte}}, \bibinfo {author} {\bibfnamefont {P.}~\bibnamefont {Zoller}}, \
  and\ \bibinfo {author} {\bibfnamefont {G.}~\bibnamefont {Pupillo}},\
  }\bibfield  {title} {\enquote {\bibinfo {title} {{Trimer Liquids and Crystals
  of Polar Molecules in Coupled Wires}},}\ }\href {\doibase
  10.1103/PhysRevLett.107.163202} {\bibfield  {journal} {\bibinfo  {journal}
  {Phys. Rev. Lett.}\ }\textbf {\bibinfo {volume} {107}},\ \bibinfo {pages}
  {163202} (\bibinfo {year} {2011})}\BibitemShut {NoStop}%
\bibitem [{\citenamefont {{Azaria}}(2017)}]{AzariaArxiv16}%
  \BibitemOpen
  \bibfield  {author} {\bibinfo {author} {\bibfnamefont {P.}~\bibnamefont
  {{Azaria}}},\ }\bibfield  {title} {\enquote {\bibinfo {title} {{Bound-States
  Dynamics in One-Dimensional Multi-Species Fermionic Systems}},}\ }\href@noop
  {} {\bibfield  {journal} {\bibinfo  {journal} {Phys. Rev. B}\ }\textbf
  {\bibinfo {volume} {95}},\ \bibinfo {pages} {125106} (\bibinfo {year}
  {2017})}\BibitemShut {NoStop}%
\bibitem [{\citenamefont {Boulat}\ \emph {et~al.}(2009)\citenamefont {Boulat},
  \citenamefont {Azaria},\ and\ \citenamefont
  {Lecheminant}}]{BoulatNuclPhysB09}%
  \BibitemOpen
  \bibfield  {author} {\bibinfo {author} {\bibfnamefont {E.}~\bibnamefont
  {Boulat}}, \bibinfo {author} {\bibfnamefont {P.}~\bibnamefont {Azaria}}, \
  and\ \bibinfo {author} {\bibfnamefont {P.}~\bibnamefont {Lecheminant}},\
  }\bibfield  {title} {\enquote {\bibinfo {title} {{Duality approach to
  one-dimensional degenerate electronic systems}},}\ }\href {\doibase
  http://dx.doi.org/10.1016/j.nuclphysb.2009.06.020} {\bibfield  {journal}
  {\bibinfo  {journal} {Nucl. Phys. B}\ }\textbf {\bibinfo {volume} {822}},\
  \bibinfo {pages} {367 -- 407} (\bibinfo {year} {2009})}\BibitemShut {NoStop}%
\bibitem [{\citenamefont {Nonne}\ \emph {et~al.}(2011)\citenamefont {Nonne},
  \citenamefont {Lecheminant}, \citenamefont {Capponi}, \citenamefont {Roux},\
  and\ \citenamefont {Boulat}}]{NonnePRB11}%
  \BibitemOpen
  \bibfield  {author} {\bibinfo {author} {\bibfnamefont {H.}~\bibnamefont
  {Nonne}}, \bibinfo {author} {\bibfnamefont {P.}~\bibnamefont {Lecheminant}},
  \bibinfo {author} {\bibfnamefont {S.}~\bibnamefont {Capponi}}, \bibinfo
  {author} {\bibfnamefont {G.}~\bibnamefont {Roux}}, \ and\ \bibinfo {author}
  {\bibfnamefont {E.}~\bibnamefont {Boulat}},\ }\bibfield  {title} {\enquote
  {\bibinfo {title} {{Competing orders in one-dimensional half-filled
  multicomponent fermionic cold atoms: The Haldane-charge conjecture}},}\
  }\href {\doibase 10.1103/PhysRevB.84.125123} {\bibfield  {journal} {\bibinfo
  {journal} {Phys. Rev. B}\ }\textbf {\bibinfo {volume} {84}},\ \bibinfo
  {pages} {125123} (\bibinfo {year} {2011})}\BibitemShut {NoStop}%
\bibitem [{\citenamefont {Karowski}\ and\ \citenamefont
  {Thun}(1981)}]{KarowskiNuclPhysB81}%
  \BibitemOpen
  \bibfield  {author} {\bibinfo {author} {\bibfnamefont {M.}~\bibnamefont
  {Karowski}}\ and\ \bibinfo {author} {\bibfnamefont {H.~J.}\ \bibnamefont
  {Thun}},\ }\bibfield  {title} {\enquote {\bibinfo {title} {{Complete S-matrix
  of the O(2N) Gross-Neveu model}},}\ }\href {\doibase
  http://dx.doi.org/10.1016/0550-3213(81)90484-3} {\bibfield  {journal}
  {\bibinfo  {journal} {Nuclear Physics B}\ }\textbf {\bibinfo {volume}
  {190}},\ \bibinfo {pages} {61 -- 92} (\bibinfo {year} {1981})}\BibitemShut
  {NoStop}%
\bibitem [{\citenamefont {{Haldane}}(1983)}]{HaldanePhysLettA83}%
  \BibitemOpen
  \bibfield  {author} {\bibinfo {author} {\bibfnamefont {F.~D.~M.}\
  \bibnamefont {{Haldane}}},\ }\bibfield  {title} {\enquote {\bibinfo {title}
  {{Continuum dynamics of the 1-D Heisenberg antiferromagnet: Identification
  with the O(3) nonlinear sigma model}},}\ }\href {\doibase
  10.1016/0375-9601(83)90631-X} {\bibfield  {journal} {\bibinfo  {journal}
  {Physics Letters A}\ }\textbf {\bibinfo {volume} {93}},\ \bibinfo {pages}
  {464--468} (\bibinfo {year} {1983})}\BibitemShut {NoStop}%
\bibitem [{\citenamefont {Haldane}(1983)}]{HaldanePRL83}%
  \BibitemOpen
  \bibfield  {author} {\bibinfo {author} {\bibfnamefont {F.~D.~M.}\
  \bibnamefont {Haldane}},\ }\bibfield  {title} {\enquote {\bibinfo {title}
  {{Nonlinear Field Theory of Large-Spin Heisenberg Antiferromagnets:
  Semiclassically Quantized Solitons of the One-Dimensional Easy-Axis N\'eel
  State}},}\ }\href {\doibase 10.1103/PhysRevLett.50.1153} {\bibfield
  {journal} {\bibinfo  {journal} {Phys. Rev. Lett.}\ }\textbf {\bibinfo
  {volume} {50}},\ \bibinfo {pages} {1153--1156} (\bibinfo {year}
  {1983})}\BibitemShut {NoStop}%
\bibitem [{\citenamefont {Nonne}\ \emph
  {et~al.}(2010{\natexlab{a}})\citenamefont {Nonne}, \citenamefont
  {Lecheminant}, \citenamefont {Capponi}, \citenamefont {Roux},\ and\
  \citenamefont {Boulat}}]{NonnePRB10}%
  \BibitemOpen
  \bibfield  {author} {\bibinfo {author} {\bibfnamefont {H.}~\bibnamefont
  {Nonne}}, \bibinfo {author} {\bibfnamefont {P.}~\bibnamefont {Lecheminant}},
  \bibinfo {author} {\bibfnamefont {S.}~\bibnamefont {Capponi}}, \bibinfo
  {author} {\bibfnamefont {G.}~\bibnamefont {Roux}}, \ and\ \bibinfo {author}
  {\bibfnamefont {E.}~\bibnamefont {Boulat}},\ }\bibfield  {title} {\enquote
  {\bibinfo {title} {{Haldane charge conjecture in one-dimensional
  multicomponent fermionic cold atoms}},}\ }\href {\doibase
  10.1103/PhysRevB.81.020408} {\bibfield  {journal} {\bibinfo  {journal} {Phys.
  Rev. B}\ }\textbf {\bibinfo {volume} {81}},\ \bibinfo {pages} {020408}
  (\bibinfo {year} {2010}{\natexlab{a}})}\BibitemShut {NoStop}%
\bibitem [{\citenamefont {Nonne}\ \emph
  {et~al.}(2010{\natexlab{b}})\citenamefont {Nonne}, \citenamefont {Boulat},
  \citenamefont {Capponi},\ and\ \citenamefont {Lecheminant}}]{NonnePRB10a}%
  \BibitemOpen
  \bibfield  {author} {\bibinfo {author} {\bibfnamefont {H.}~\bibnamefont
  {Nonne}}, \bibinfo {author} {\bibfnamefont {E.}~\bibnamefont {Boulat}},
  \bibinfo {author} {\bibfnamefont {S.}~\bibnamefont {Capponi}}, \ and\
  \bibinfo {author} {\bibfnamefont {P.}~\bibnamefont {Lecheminant}},\
  }\bibfield  {title} {\enquote {\bibinfo {title} {{Competing orders in the
  generalized Hund chain model at half filling}},}\ }\href {\doibase
  10.1103/PhysRevB.82.155134} {\bibfield  {journal} {\bibinfo  {journal} {Phys.
  Rev. B}\ }\textbf {\bibinfo {volume} {82}},\ \bibinfo {pages} {155134}
  (\bibinfo {year} {2010}{\natexlab{b}})}\BibitemShut {NoStop}%
\bibitem [{\citenamefont {Yang}(1989)}]{YangPRL89}%
  \BibitemOpen
  \bibfield  {author} {\bibinfo {author} {\bibfnamefont {C.~N.}\ \bibnamefont
  {Yang}},\ }\bibfield  {title} {\enquote {\bibinfo {title}
  {{\textit{\ensuremath{\eta}} pairing and off-diagonal long-range order in a
  Hubbard model}},}\ }\href {\doibase 10.1103/PhysRevLett.63.2144} {\bibfield
  {journal} {\bibinfo  {journal} {Phys. Rev. Lett.}\ }\textbf {\bibinfo
  {volume} {63}},\ \bibinfo {pages} {2144--2147} (\bibinfo {year}
  {1989})}\BibitemShut {NoStop}%
\bibitem [{\citenamefont {Anderson}(1958)}]{AndersonPR58}%
  \BibitemOpen
  \bibfield  {author} {\bibinfo {author} {\bibfnamefont {P.~W.}\ \bibnamefont
  {Anderson}},\ }\bibfield  {title} {\enquote {\bibinfo {title} {{Random-Phase
  Approximation in the Theory of Superconductivity}},}\ }\href {\doibase
  10.1103/PhysRev.112.1900} {\bibfield  {journal} {\bibinfo  {journal} {Phys.
  Rev.}\ }\textbf {\bibinfo {volume} {112}},\ \bibinfo {pages} {1900--1916}
  (\bibinfo {year} {1958})}\BibitemShut {NoStop}%
\bibitem [{\citenamefont {Lieb}\ and\ \citenamefont {Wu}(1968)}]{LiebPRL68}%
  \BibitemOpen
  \bibfield  {author} {\bibinfo {author} {\bibfnamefont {E.~H.}\ \bibnamefont
  {Lieb}}\ and\ \bibinfo {author} {\bibfnamefont {F.~Y.}\ \bibnamefont {Wu}},\
  }\bibfield  {title} {\enquote {\bibinfo {title} {{Absence of Mott Transition
  in an Exact Solution of the Short-Range, One-Band Model in One Dimension}},}\
  }\href {\doibase 10.1103/PhysRevLett.20.1445} {\bibfield  {journal} {\bibinfo
   {journal} {Phys. Rev. Lett.}\ }\textbf {\bibinfo {volume} {20}},\ \bibinfo
  {pages} {1445--1448} (\bibinfo {year} {1968})}\BibitemShut {NoStop}%
\bibitem [{\citenamefont {Essler}\ \emph {et~al.}(2005)\citenamefont {Essler},
  \citenamefont {Frahm}, \citenamefont {G\"ohmann}, \citenamefont {Kl\"umper},\
  and\ \citenamefont {Korepin}}]{HubbardBook}%
  \BibitemOpen
  \bibfield  {author} {\bibinfo {author} {\bibfnamefont {F.~H.~L.}\
  \bibnamefont {Essler}}, \bibinfo {author} {\bibfnamefont {H.}~\bibnamefont
  {Frahm}}, \bibinfo {author} {\bibfnamefont {F.}~\bibnamefont {G\"ohmann}},
  \bibinfo {author} {\bibfnamefont {A.}~\bibnamefont {Kl\"umper}}, \ and\
  \bibinfo {author} {\bibfnamefont {V.~E.}\ \bibnamefont {Korepin}},\ }\href
  {http://www.cambridge.org/catalogue/catalogue.asp?isbn=9780521802628} {\emph
  {\bibinfo {title} {{The One-Dimensional Hubbard Model}}}}\ (\bibinfo
  {publisher} {Cambridge University Press},\ \bibinfo {year}
  {2005})\BibitemShut {NoStop}%
\bibitem [{\citenamefont {Assaraf}\ \emph {et~al.}(1999)\citenamefont
  {Assaraf}, \citenamefont {Azaria}, \citenamefont {Caffarel},\ and\
  \citenamefont {Lecheminant}}]{AssarafPRB99}%
  \BibitemOpen
  \bibfield  {author} {\bibinfo {author} {\bibfnamefont {R.}~\bibnamefont
  {Assaraf}}, \bibinfo {author} {\bibfnamefont {P.}~\bibnamefont {Azaria}},
  \bibinfo {author} {\bibfnamefont {M.}~\bibnamefont {Caffarel}}, \ and\
  \bibinfo {author} {\bibfnamefont {P.}~\bibnamefont {Lecheminant}},\
  }\bibfield  {title} {\enquote {\bibinfo {title} {{Metal-insulator transition
  in the one-dimensional $\mathrm{SU}(N)$ Hubbard model}},}\ }\href {\doibase
  10.1103/PhysRevB.60.2299} {\bibfield  {journal} {\bibinfo  {journal} {Phys.
  Rev. B}\ }\textbf {\bibinfo {volume} {60}},\ \bibinfo {pages} {2299--2318}
  (\bibinfo {year} {1999})}\BibitemShut {NoStop}%
\bibitem [{\citenamefont {Manmana}\ \emph {et~al.}(2011)\citenamefont
  {Manmana}, \citenamefont {Hazzard}, \citenamefont {Chen}, \citenamefont
  {Feiguin},\ and\ \citenamefont {Rey}}]{ManmanaPRA11}%
  \BibitemOpen
  \bibfield  {author} {\bibinfo {author} {\bibfnamefont {S.~R.}\ \bibnamefont
  {Manmana}}, \bibinfo {author} {\bibfnamefont {K.~R.~A.}\ \bibnamefont
  {Hazzard}}, \bibinfo {author} {\bibfnamefont {G.}~\bibnamefont {Chen}},
  \bibinfo {author} {\bibfnamefont {A.~E.}\ \bibnamefont {Feiguin}}, \ and\
  \bibinfo {author} {\bibfnamefont {A.~M.}\ \bibnamefont {Rey}},\ }\bibfield
  {title} {\enquote {\bibinfo {title} {{SU$(N)$ magnetism in chains of
  ultracold alkaline-earth-metal atoms: Mott transitions and quantum
  correlations}},}\ }\href {\doibase 10.1103/PhysRevA.84.043601} {\bibfield
  {journal} {\bibinfo  {journal} {Phys. Rev. A}\ }\textbf {\bibinfo {volume}
  {84}},\ \bibinfo {pages} {043601} (\bibinfo {year} {2011})}\BibitemShut
  {NoStop}%
\bibitem [{\citenamefont {Buchta}\ \emph {et~al.}(2007)\citenamefont {Buchta},
  \citenamefont {Legeza}, \citenamefont {Szirmai},\ and\ \citenamefont
  {S\'olyom}}]{BuchtaPRB07}%
  \BibitemOpen
  \bibfield  {author} {\bibinfo {author} {\bibfnamefont {K.}~\bibnamefont
  {Buchta}}, \bibinfo {author} {\bibfnamefont {\"O.}\ \bibnamefont {Legeza}},
  \bibinfo {author} {\bibfnamefont {E.}~\bibnamefont {Szirmai}}, \ and\
  \bibinfo {author} {\bibfnamefont {J.}~\bibnamefont {S\'olyom}},\ }\bibfield
  {title} {\enquote {\bibinfo {title} {{Mott transition and dimerization in the
  one-dimensional $\mathrm{SU}(n)$ Hubbard model}},}\ }\href {\doibase
  10.1103/PhysRevB.75.155108} {\bibfield  {journal} {\bibinfo  {journal} {Phys.
  Rev. B}\ }\textbf {\bibinfo {volume} {75}},\ \bibinfo {pages} {155108}
  (\bibinfo {year} {2007})}\BibitemShut {NoStop}%
\bibitem [{\citenamefont {Sutherland}(1975)}]{SutherlandPRB75}%
  \BibitemOpen
  \bibfield  {author} {\bibinfo {author} {\bibfnamefont {B.}~\bibnamefont
  {Sutherland}},\ }\bibfield  {title} {\enquote {\bibinfo {title} {{Model for a
  multicomponent quantum system}},}\ }\href {\doibase 10.1103/PhysRevB.12.3795}
  {\bibfield  {journal} {\bibinfo  {journal} {Phys. Rev. B}\ }\textbf {\bibinfo
  {volume} {12}},\ \bibinfo {pages} {3795--3805} (\bibinfo {year}
  {1975})}\BibitemShut {NoStop}%
\bibitem [{\citenamefont {Affleck}\ \emph {et~al.}(1989)\citenamefont
  {Affleck}, \citenamefont {Gepner}, \citenamefont {Schulz},\ and\
  \citenamefont {Ziman}}]{AffleckJPhysA89}%
  \BibitemOpen
  \bibfield  {author} {\bibinfo {author} {\bibfnamefont {I.}~\bibnamefont
  {Affleck}}, \bibinfo {author} {\bibfnamefont {D.}~\bibnamefont {Gepner}},
  \bibinfo {author} {\bibfnamefont {H.~J.}\ \bibnamefont {Schulz}}, \ and\
  \bibinfo {author} {\bibfnamefont {T.}~\bibnamefont {Ziman}},\ }\bibfield
  {title} {\enquote {\bibinfo {title} {{Critical behaviour of spin-s Heisenberg
  antiferromagnetic chains: analytic and numerical results}},}\ }\href
  {http://stacks.iop.org/0305-4470/22/i=5/a=015} {\bibfield  {journal}
  {\bibinfo  {journal} {J. Phys. A}\ }\textbf {\bibinfo {volume} {22}},\
  \bibinfo {pages} {511} (\bibinfo {year} {1989})}\BibitemShut {NoStop}%
\bibitem [{\citenamefont {Majumdar}\ and\ \citenamefont
  {Mukherjee}(2002)}]{MajumdarJPhysA02}%
  \BibitemOpen
  \bibfield  {author} {\bibinfo {author} {\bibfnamefont {K.}~\bibnamefont
  {Majumdar}}\ and\ \bibinfo {author} {\bibfnamefont {M.}~\bibnamefont
  {Mukherjee}},\ }\bibfield  {title} {\enquote {\bibinfo {title} {{Logarithmic
  corrections to finite-size spectrum of SU(N) symmetric quantum chains}},}\
  }\href {http://stacks.iop.org/0305-4470/35/i=38/a=101} {\bibfield  {journal}
  {\bibinfo  {journal} {J. Phys. A}\ }\textbf {\bibinfo {volume} {35}},\
  \bibinfo {pages} {L543} (\bibinfo {year} {2002})}\BibitemShut {NoStop}%
\bibitem [{\citenamefont {Johannesson}(1986)}]{JohannesonNuclPhysB86}%
  \BibitemOpen
  \bibfield  {author} {\bibinfo {author} {\bibfnamefont {H.}~\bibnamefont
  {Johannesson}},\ }\bibfield  {title} {\enquote {\bibinfo {title} {{The
  structure of low-lying excitations in a new integrable quantum chain
  model}},}\ }\href {\doibase http://dx.doi.org/10.1016/0550-3213(86)90554-7}
  {\bibfield  {journal} {\bibinfo  {journal} {Nucl. Phys. B}\ }\textbf
  {\bibinfo {volume} {270}},\ \bibinfo {pages} {235 -- 272} (\bibinfo {year}
  {1986})}\BibitemShut {NoStop}%
\bibitem [{\citenamefont {{Bouwknegt}}\ and\ \citenamefont
  {{Schoutens}}(1996)}]{BouwknegtNuclPhysB96}%
  \BibitemOpen
  \bibfield  {author} {\bibinfo {author} {\bibfnamefont {P.}~\bibnamefont
  {{Bouwknegt}}}\ and\ \bibinfo {author} {\bibfnamefont {K.}~\bibnamefont
  {{Schoutens}}},\ }\bibfield  {title} {\enquote {\bibinfo {title} {{The
  $\overline{SU}$(n)$_{1}$ WZW models Spinon decomposition and yangian
  structure}},}\ }\href {\doibase 10.1016/S0550-3213(96)00565-2} {\bibfield
  {journal} {\bibinfo  {journal} {Nuclear Physics B}\ }\textbf {\bibinfo
  {volume} {482}},\ \bibinfo {pages} {345--372} (\bibinfo {year} {1996})},\
  \Eprint {http://arxiv.org/abs/hep-th/9607064} {hep-th/9607064} \BibitemShut
  {NoStop}%
\bibitem [{\citenamefont {Schuricht}\ and\ \citenamefont
  {Greiter}(2006)}]{SchurichtPRB06}%
  \BibitemOpen
  \bibfield  {author} {\bibinfo {author} {\bibfnamefont {D.}~\bibnamefont
  {Schuricht}}\ and\ \bibinfo {author} {\bibfnamefont {M.}~\bibnamefont
  {Greiter}},\ }\bibfield  {title} {\enquote {\bibinfo {title} {{Coloron
  excitations of the SU(3) Haldane-Shastry model}},}\ }\href {\doibase
  10.1103/PhysRevB.73.235105} {\bibfield  {journal} {\bibinfo  {journal} {Phys.
  Rev. B}\ }\textbf {\bibinfo {volume} {73}},\ \bibinfo {pages} {235105}
  (\bibinfo {year} {2006})}\BibitemShut {NoStop}%
\bibitem [{\citenamefont {Szirmai}\ \emph {et~al.}(2008)\citenamefont
  {Szirmai}, \citenamefont {Legeza},\ and\ \citenamefont
  {S\'olyom}}]{SzirmaiPRB08}%
  \BibitemOpen
  \bibfield  {author} {\bibinfo {author} {\bibfnamefont {E.}~\bibnamefont
  {Szirmai}}, \bibinfo {author} {\bibfnamefont {\"O.}\ \bibnamefont {Legeza}},
  \ and\ \bibinfo {author} {\bibfnamefont {J.}~\bibnamefont {S\'olyom}},\
  }\bibfield  {title} {\enquote {\bibinfo {title} {{Spatially nonuniform phases
  in the one-dimensional $\mathrm{SU}(n)$ Hubbard model for commensurate
  fillings}},}\ }\href {\doibase 10.1103/PhysRevB.77.045106} {\bibfield
  {journal} {\bibinfo  {journal} {Phys. Rev. B}\ }\textbf {\bibinfo {volume}
  {77}},\ \bibinfo {pages} {045106} (\bibinfo {year} {2008})}\BibitemShut
  {NoStop}%
\bibitem [{\citenamefont {{Affleck}}\ \emph {et~al.}(1991)\citenamefont
  {{Affleck}}, \citenamefont {{Arovas}}, \citenamefont {{Marston}},\ and\
  \citenamefont {{Rabson}}}]{AffleckNuclPhysB91}%
  \BibitemOpen
  \bibfield  {author} {\bibinfo {author} {\bibfnamefont {I.}~\bibnamefont
  {{Affleck}}}, \bibinfo {author} {\bibfnamefont {D.~P.}\ \bibnamefont
  {{Arovas}}}, \bibinfo {author} {\bibfnamefont {J.~B.}\ \bibnamefont
  {{Marston}}}, \ and\ \bibinfo {author} {\bibfnamefont {D.~A.}\ \bibnamefont
  {{Rabson}}},\ }\bibfield  {title} {\enquote {\bibinfo {title} {{SU(2 n)
  quantum antiferromagnets with exact C-breaking ground states}},}\ }\href
  {\doibase 10.1016/0550-3213(91)90027-U} {\bibfield  {journal} {\bibinfo
  {journal} {Nucl. Phys. B}\ }\textbf {\bibinfo {volume} {366}},\ \bibinfo
  {pages} {467--506} (\bibinfo {year} {1991})}\BibitemShut {NoStop}%
\bibitem [{\citenamefont {Onufriev}\ and\ \citenamefont
  {Marston}(1999)}]{OnufrievPRB99}%
  \BibitemOpen
  \bibfield  {author} {\bibinfo {author} {\bibfnamefont {A.~V.}\ \bibnamefont
  {Onufriev}}\ and\ \bibinfo {author} {\bibfnamefont {J.~B.}\ \bibnamefont
  {Marston}},\ }\bibfield  {title} {\enquote {\bibinfo {title} {{Enlarged
  symmetry and coherence in arrays of quantum dots}},}\ }\href {\doibase
  10.1103/PhysRevB.59.12573} {\bibfield  {journal} {\bibinfo  {journal} {Phys.
  Rev. B}\ }\textbf {\bibinfo {volume} {59}},\ \bibinfo {pages} {12573--12578}
  (\bibinfo {year} {1999})}\BibitemShut {NoStop}%
\bibitem [{\citenamefont {Dufour}\ \emph {et~al.}(2015)\citenamefont {Dufour},
  \citenamefont {Nataf},\ and\ \citenamefont {Mila}}]{DufourPRB15}%
  \BibitemOpen
  \bibfield  {author} {\bibinfo {author} {\bibfnamefont {J.}~\bibnamefont
  {Dufour}}, \bibinfo {author} {\bibfnamefont {P.}~\bibnamefont {Nataf}}, \
  and\ \bibinfo {author} {\bibfnamefont {F.}~\bibnamefont {Mila}},\ }\bibfield
  {title} {\enquote {\bibinfo {title} {{Variational Monte Carlo investigation
  of $\mathrm{SU}(N)$ Heisenberg chains}},}\ }\href {\doibase
  10.1103/PhysRevB.91.174427} {\bibfield  {journal} {\bibinfo  {journal} {Phys.
  Rev. B}\ }\textbf {\bibinfo {volume} {91}},\ \bibinfo {pages} {174427}
  (\bibinfo {year} {2015})}\BibitemShut {NoStop}%
\bibitem [{\citenamefont {Szirmai}(2013)}]{SzirmaiPRB13}%
  \BibitemOpen
  \bibfield  {author} {\bibinfo {author} {\bibfnamefont {E.}~\bibnamefont
  {Szirmai}},\ }\bibfield  {title} {\enquote {\bibinfo {title} {{Two-orbital
  physics of high-spin fermionic alkaline-earth atoms confined in a
  one-dimensional chain}},}\ }\href {\doibase 10.1103/PhysRevB.88.195432}
  {\bibfield  {journal} {\bibinfo  {journal} {Phys. Rev. B}\ }\textbf {\bibinfo
  {volume} {88}},\ \bibinfo {pages} {195432} (\bibinfo {year}
  {2013})}\BibitemShut {NoStop}%
\bibitem [{\citenamefont {Bois}\ \emph {et~al.}(2016)\citenamefont {Bois},
  \citenamefont {Fromholz},\ and\ \citenamefont {Lecheminant}}]{BoisPRB16}%
  \BibitemOpen
  \bibfield  {author} {\bibinfo {author} {\bibfnamefont {V.}~\bibnamefont
  {Bois}}, \bibinfo {author} {\bibfnamefont {P.}~\bibnamefont {Fromholz}}, \
  and\ \bibinfo {author} {\bibfnamefont {P.}~\bibnamefont {Lecheminant}},\
  }\bibfield  {title} {\enquote {\bibinfo {title} {{One-dimensional two-orbital
  SU($N$) ultracold fermionic quantum gases at incommensurate filling: A
  low-energy approach}},}\ }\href {\doibase 10.1103/PhysRevB.93.134415}
  {\bibfield  {journal} {\bibinfo  {journal} {Phys. Rev. B}\ }\textbf {\bibinfo
  {volume} {93}},\ \bibinfo {pages} {134415} (\bibinfo {year}
  {2016})}\BibitemShut {NoStop}%
\bibitem [{\citenamefont {{Szirmai}}(2016)}]{SzirmaiArxiv16}%
  \BibitemOpen
  \bibfield  {author} {\bibinfo {author} {\bibfnamefont {E.}~\bibnamefont
  {{Szirmai}}},\ }\bibfield  {title} {\enquote {\bibinfo {title} {{Orbital-FFLO
  state in a chain of high spin ultracold atoms}},}\ }\href@noop {} {\bibfield
  {journal} {\bibinfo  {journal} {ArXiv e-prints}\ } (\bibinfo {year}
  {2016})},\ \Eprint {http://arxiv.org/abs/1603.08505} {arXiv:1603.08505
  [cond-mat.quant-gas]} \BibitemShut {NoStop}%
\bibitem [{\citenamefont {Nonne}\ \emph {et~al.}(2013)\citenamefont {Nonne},
  \citenamefont {Moliner}, \citenamefont {Capponi}, \citenamefont
  {Lecheminant},\ and\ \citenamefont {Totsuka}}]{NonneEPL13}%
  \BibitemOpen
  \bibfield  {author} {\bibinfo {author} {\bibfnamefont {H.}~\bibnamefont
  {Nonne}}, \bibinfo {author} {\bibfnamefont {M.}~\bibnamefont {Moliner}},
  \bibinfo {author} {\bibfnamefont {S.}~\bibnamefont {Capponi}}, \bibinfo
  {author} {\bibfnamefont {P.}~\bibnamefont {Lecheminant}}, \ and\ \bibinfo
  {author} {\bibfnamefont {K.}~\bibnamefont {Totsuka}},\ }\bibfield  {title}
  {\enquote {\bibinfo {title} {{Symmetry-protected topological phases of
  alkaline-earth cold fermionic atoms in one dimension}},}\ }\href {\doibase
  10.1209/0295-5075/102/37008} {\bibfield  {journal} {\bibinfo  {journal}
  {Europhys. Lett.}\ }\textbf {\bibinfo {volume} {102}},\ \bibinfo {pages}
  {37008} (\bibinfo {year} {2013})}\BibitemShut {NoStop}%
\bibitem [{\citenamefont {Duivenvoorden}\ and\ \citenamefont
  {Quella}(2013)}]{DuivenvoordenPRB13}%
  \BibitemOpen
  \bibfield  {author} {\bibinfo {author} {\bibfnamefont {K.}~\bibnamefont
  {Duivenvoorden}}\ and\ \bibinfo {author} {\bibfnamefont {T.}~\bibnamefont
  {Quella}},\ }\bibfield  {title} {\enquote {\bibinfo {title} {{Topological
  phases of spin chains}},}\ }\href {\doibase 10.1103/PhysRevB.87.125145}
  {\bibfield  {journal} {\bibinfo  {journal} {Phys. Rev. B}\ }\textbf {\bibinfo
  {volume} {87}},\ \bibinfo {pages} {125145} (\bibinfo {year}
  {2013})}\BibitemShut {NoStop}%
\bibitem [{\citenamefont {{Tanimoto}}\ and\ \citenamefont
  {{Totsuka}}(2015)}]{TanimotoArxiv15}%
  \BibitemOpen
  \bibfield  {author} {\bibinfo {author} {\bibfnamefont {K.}~\bibnamefont
  {{Tanimoto}}}\ and\ \bibinfo {author} {\bibfnamefont {K.}~\bibnamefont
  {{Totsuka}}},\ }\bibfield  {title} {\enquote {\bibinfo {title}
  {{Symmetry-protected topological order in SU(N) Heisenberg magnets --quantum
  entanglement and non-local order parameters}},}\ }\href@noop {} {\bibfield
  {journal} {\bibinfo  {journal} {ArXiv e-prints}\ } (\bibinfo {year}
  {2015})},\ \Eprint {http://arxiv.org/abs/1508.07601} {arXiv:1508.07601
  [cond-mat.str-el]} \BibitemShut {NoStop}%
\bibitem [{\citenamefont {{Roy}}\ and\ \citenamefont
  {{Quella}}(2015)}]{RoyArxiv15}%
  \BibitemOpen
  \bibfield  {author} {\bibinfo {author} {\bibfnamefont {A.}~\bibnamefont
  {{Roy}}}\ and\ \bibinfo {author} {\bibfnamefont {T.}~\bibnamefont
  {{Quella}}},\ }\bibfield  {title} {\enquote {\bibinfo {title} {{Chiral
  Haldane phases of SU(N) quantum spin chains in the adjoint
  representation}},}\ }\href@noop {} {\bibfield  {journal} {\bibinfo  {journal}
  {ArXiv e-prints}\ } (\bibinfo {year} {2015})},\ \Eprint
  {http://arxiv.org/abs/1512.05229} {arXiv:1512.05229 [cond-mat.str-el]}
  \BibitemShut {NoStop}%
\bibitem [{\citenamefont {Affleck}\ and\ \citenamefont
  {Lieb}(1986)}]{AffleckLettMathPhys86}%
  \BibitemOpen
  \bibfield  {author} {\bibinfo {author} {\bibfnamefont {I.}~\bibnamefont
  {Affleck}}\ and\ \bibinfo {author} {\bibfnamefont {E.~H.}\ \bibnamefont
  {Lieb}},\ }\bibfield  {title} {\enquote {\bibinfo {title} {{A proof of part
  of Haldane's conjecture on spin chains}},}\ }\href {\doibase
  10.1007/BF00400304} {\bibfield  {journal} {\bibinfo  {journal} {Lett. Math.
  Phys.}\ }\textbf {\bibinfo {volume} {12}},\ \bibinfo {pages} {57--69}
  (\bibinfo {year} {1986})}\BibitemShut {NoStop}%
\bibitem [{\citenamefont {Rachel}\ \emph {et~al.}(2009)\citenamefont {Rachel},
  \citenamefont {Thomale}, \citenamefont {F\"uhringer}, \citenamefont
  {Schmitteckert},\ and\ \citenamefont {Greiter}}]{RachelPRB09}%
  \BibitemOpen
  \bibfield  {author} {\bibinfo {author} {\bibfnamefont {S.}~\bibnamefont
  {Rachel}}, \bibinfo {author} {\bibfnamefont {R.}~\bibnamefont {Thomale}},
  \bibinfo {author} {\bibfnamefont {M.}~\bibnamefont {F\"uhringer}}, \bibinfo
  {author} {\bibfnamefont {P.}~\bibnamefont {Schmitteckert}}, \ and\ \bibinfo
  {author} {\bibfnamefont {M.}~\bibnamefont {Greiter}},\ }\bibfield  {title}
  {\enquote {\bibinfo {title} {{Spinon confinement and the Haldane gap in
  $\text{SU}(n)$ spin chains}},}\ }\href {\doibase 10.1103/PhysRevB.80.180420}
  {\bibfield  {journal} {\bibinfo  {journal} {Phys. Rev. B}\ }\textbf {\bibinfo
  {volume} {80}},\ \bibinfo {pages} {180420} (\bibinfo {year}
  {2009})}\BibitemShut {NoStop}%
\bibitem [{\citenamefont {Read}\ and\ \citenamefont
  {Sachdev}(1989)}]{ReadNuclPhysB89}%
  \BibitemOpen
  \bibfield  {author} {\bibinfo {author} {\bibfnamefont {N.}~\bibnamefont
  {Read}}\ and\ \bibinfo {author} {\bibfnamefont {S.}~\bibnamefont {Sachdev}},\
  }\bibfield  {title} {\enquote {\bibinfo {title} {{Some features of the phase
  diagram of the square lattice SU(N) antiferromagnet}},}\ }\href {\doibase
  http://dx.doi.org/10.1016/0550-3213(89)90061-8} {\bibfield  {journal}
  {\bibinfo  {journal} {Nucl. Phys. B}\ }\textbf {\bibinfo {volume} {316}},\
  \bibinfo {pages} {609 -- 640} (\bibinfo {year} {1989})}\BibitemShut {NoStop}%
\bibitem [{\citenamefont {Takhtajan}(1982)}]{TakhtajanPhysLettA82}%
  \BibitemOpen
  \bibfield  {author} {\bibinfo {author} {\bibfnamefont {L.~A.}\ \bibnamefont
  {Takhtajan}},\ }\bibfield  {title} {\enquote {\bibinfo {title} {{The picture
  of low-lying excitations in the isotropic Heisenberg chain of arbitrary
  spins}},}\ }\href {\doibase http://dx.doi.org/10.1016/0375-9601(82)90764-2}
  {\bibfield  {journal} {\bibinfo  {journal} {Phys. Lett. A}\ }\textbf
  {\bibinfo {volume} {87}},\ \bibinfo {pages} {479 -- 482} (\bibinfo {year}
  {1982})}\BibitemShut {NoStop}%
\bibitem [{\citenamefont {Babujian}(1982)}]{BabujianPhysLettA82}%
  \BibitemOpen
  \bibfield  {author} {\bibinfo {author} {\bibfnamefont {H.~M.}\ \bibnamefont
  {Babujian}},\ }\bibfield  {title} {\enquote {\bibinfo {title} {{Exact
  solution of the one-dimensional isotropic Heisenberg chain with arbitrary
  spins S}},}\ }\href {\doibase http://dx.doi.org/10.1016/0375-9601(82)90403-0}
  {\bibfield  {journal} {\bibinfo  {journal} {Phys. Lett. A}\ }\textbf
  {\bibinfo {volume} {90}},\ \bibinfo {pages} {479 -- 482} (\bibinfo {year}
  {1982})}\BibitemShut {NoStop}%
\bibitem [{\citenamefont {Tsvelik}(1990)}]{TsvelikPRB90}%
  \BibitemOpen
  \bibfield  {author} {\bibinfo {author} {\bibfnamefont {A.~M.}\ \bibnamefont
  {Tsvelik}},\ }\bibfield  {title} {\enquote {\bibinfo {title} {{Field-theory
  treatment of the Heisenberg spin-1 chain}},}\ }\href {\doibase
  10.1103/PhysRevB.42.10499} {\bibfield  {journal} {\bibinfo  {journal} {Phys.
  Rev. B}\ }\textbf {\bibinfo {volume} {42}},\ \bibinfo {pages} {10499--10504}
  (\bibinfo {year} {1990})}\BibitemShut {NoStop}%
\bibitem [{\citenamefont {Ward}\ \emph {et~al.}(2013)\citenamefont {Ward},
  \citenamefont {Bouillot}, \citenamefont {Ryll}, \citenamefont {Kiefer},
  \citenamefont {Kr\"amer}, \citenamefont {R\"uegg}, \citenamefont {Kollath},\
  and\ \citenamefont {Giamarchi}}]{WardJPhysCondMatt13}%
  \BibitemOpen
  \bibfield  {author} {\bibinfo {author} {\bibfnamefont {S.}~\bibnamefont
  {Ward}}, \bibinfo {author} {\bibfnamefont {P.}~\bibnamefont {Bouillot}},
  \bibinfo {author} {\bibfnamefont {H.}~\bibnamefont {Ryll}}, \bibinfo {author}
  {\bibfnamefont {K.}~\bibnamefont {Kiefer}}, \bibinfo {author} {\bibfnamefont
  {K.~W.}\ \bibnamefont {Kr\"amer}}, \bibinfo {author} {\bibfnamefont {{Ch.}}\
  \bibnamefont {R\"uegg}}, \bibinfo {author} {\bibfnamefont {C.}~\bibnamefont
  {Kollath}}, \ and\ \bibinfo {author} {\bibfnamefont {T.}~\bibnamefont
  {Giamarchi}},\ }\bibfield  {title} {\enquote {\bibinfo {title} {{Spin ladders
  and quantum simulators for Tomonaga-Luttinger liquids}},}\ }\href
  {http://stacks.iop.org/0953-8984/25/i=1/a=014004} {\bibfield  {journal}
  {\bibinfo  {journal} {J. Phys.: Cond. Matt.}\ }\textbf {\bibinfo {volume}
  {25}},\ \bibinfo {pages} {014004} (\bibinfo {year} {2013})}\BibitemShut
  {NoStop}%
\bibitem [{\citenamefont {Lake}\ \emph {et~al.}(2010)\citenamefont {Lake},
  \citenamefont {Tsvelik}, \citenamefont {Notbohm}, \citenamefont {Tennant},
  \citenamefont {Perring}, \citenamefont {Reehuis}, \citenamefont {Sekar},
  \citenamefont {Krabbes},\ and\ \citenamefont {Buchner}}]{LakeNatPhys10}%
  \BibitemOpen
  \bibfield  {author} {\bibinfo {author} {\bibfnamefont {B.}~\bibnamefont
  {Lake}}, \bibinfo {author} {\bibfnamefont {A.~M.}\ \bibnamefont {Tsvelik}},
  \bibinfo {author} {\bibfnamefont {S.}~\bibnamefont {Notbohm}}, \bibinfo
  {author} {\bibfnamefont {D.~A.}\ \bibnamefont {Tennant}}, \bibinfo {author}
  {\bibfnamefont {T.~G.}\ \bibnamefont {Perring}}, \bibinfo {author}
  {\bibfnamefont {M.}~\bibnamefont {Reehuis}}, \bibinfo {author} {\bibfnamefont
  {C.}~\bibnamefont {Sekar}}, \bibinfo {author} {\bibfnamefont
  {G.}~\bibnamefont {Krabbes}}, \ and\ \bibinfo {author} {\bibfnamefont
  {B.}~\bibnamefont {Buchner}},\ }\bibfield  {title} {\enquote {\bibinfo
  {title} {{Confinement of fractional quantum number particles in a
  condensed-matter system}},}\ }\href {http://dx.doi.org/10.1038/nphys1462}
  {\bibfield  {journal} {\bibinfo  {journal} {Nature Phys.}\ }\textbf {\bibinfo
  {volume} {6}},\ \bibinfo {pages} {50--55} (\bibinfo {year}
  {2010})}\BibitemShut {NoStop}%
\bibitem [{\citenamefont {{Lecheminant}}(2015)}]{LecheminantNuclPhysB15}%
  \BibitemOpen
  \bibfield  {author} {\bibinfo {author} {\bibfnamefont {P.}~\bibnamefont
  {{Lecheminant}}},\ }\bibfield  {title} {\enquote {\bibinfo {title} {{Massless
  renormalization group flow in SU(N)$_k$ perturbed conformal field theory}},}\
  }\href {\doibase 10.1016/j.nuclphysb.2015.11.004} {\bibfield  {journal}
  {\bibinfo  {journal} {Nucl. Phys. B}\ }\textbf {\bibinfo {volume} {901}},\
  \bibinfo {pages} {510--525} (\bibinfo {year} {2015})}\BibitemShut {NoStop}%
\bibitem [{\citenamefont {van~den Bossche}\ \emph {et~al.}(2001)\citenamefont
  {van~den Bossche}, \citenamefont {Azaria}, \citenamefont {Lecheminant},\ and\
  \citenamefont {Mila}}]{vandenBosschePRL01}%
  \BibitemOpen
  \bibfield  {author} {\bibinfo {author} {\bibfnamefont {M.}~\bibnamefont
  {van~den Bossche}}, \bibinfo {author} {\bibfnamefont {P.}~\bibnamefont
  {Azaria}}, \bibinfo {author} {\bibfnamefont {P.}~\bibnamefont {Lecheminant}},
  \ and\ \bibinfo {author} {\bibfnamefont {F.}~\bibnamefont {Mila}},\
  }\bibfield  {title} {\enquote {\bibinfo {title} {{Spontaneous Plaquette
  Formation in the SU(4) Spin-Orbital Ladder}},}\ }\href {\doibase
  10.1103/PhysRevLett.86.4124} {\bibfield  {journal} {\bibinfo  {journal}
  {Phys. Rev. Lett.}\ }\textbf {\bibinfo {volume} {86}},\ \bibinfo {pages}
  {4124--4127} (\bibinfo {year} {2001})}\BibitemShut {NoStop}%
\bibitem [{\citenamefont {Weichselbaum}\ \emph {et~al.}()\citenamefont
  {Weichselbaum}, \citenamefont {Capponi}, \citenamefont {Lecheminant},
  \citenamefont {Tsvelik},\ and\ \citenamefont
  {La\"uchli}}]{SU3ladderpreprint}%
  \BibitemOpen
  \bibfield  {author} {\bibinfo {author} {\bibfnamefont {A.}~\bibnamefont
  {Weichselbaum}}, \bibinfo {author} {\bibfnamefont {S.}~\bibnamefont
  {Capponi}}, \bibinfo {author} {\bibfnamefont {P.}~\bibnamefont
  {Lecheminant}}, \bibinfo {author} {\bibfnamefont {A.~M.}\ \bibnamefont
  {Tsvelik}}, \ and\ \bibinfo {author} {\bibfnamefont {A.}~\bibnamefont
  {La\"uchli}},\ }\href@noop {} {}\bibinfo {note} {{in
  preparation}}\BibitemShut {NoStop}%
\bibitem [{\citenamefont {Delfino}\ \emph {et~al.}(1996)\citenamefont
  {Delfino}, \citenamefont {Mussardo},\ and\ \citenamefont
  {Simonetti}}]{delfino1996non}%
  \BibitemOpen
  \bibfield  {author} {\bibinfo {author} {\bibfnamefont {G.}~\bibnamefont
  {Delfino}}, \bibinfo {author} {\bibfnamefont {G.}~\bibnamefont {Mussardo}}, \
  and\ \bibinfo {author} {\bibfnamefont {P.}~\bibnamefont {Simonetti}},\
  }\bibfield  {title} {\enquote {\bibinfo {title} {Non-integrable quantum field
  theories as perturbations of certain integrable models},}\ }\href {\doibase
  http://dx.doi.org/10.1016/0550-3213(96)00265-9} {\bibfield  {journal}
  {\bibinfo  {journal} {Nucl. Phys. B}\ }\textbf {\bibinfo {volume} {473}},\
  \bibinfo {pages} {469 -- 508} (\bibinfo {year} {1996})}\BibitemShut {NoStop}%
\bibitem [{\citenamefont {Dorey}\ and\ \citenamefont
  {Tateo}(1998)}]{dorey1998excited}%
  \BibitemOpen
  \bibfield  {author} {\bibinfo {author} {\bibfnamefont {P.}~\bibnamefont
  {Dorey}}\ and\ \bibinfo {author} {\bibfnamefont {R.}~\bibnamefont {Tateo}},\
  }\bibfield  {title} {\enquote {\bibinfo {title} {{Excited states in some
  simple perturbed conformal field theories}},}\ }\href {\doibase
  http://dx.doi.org/10.1016/S0550-3213(97)00838-9} {\bibfield  {journal}
  {\bibinfo  {journal} {Nucl. Phys. B}\ }\textbf {\bibinfo {volume} {515}},\
  \bibinfo {pages} {575 -- 623} (\bibinfo {year} {1998})}\BibitemShut {NoStop}%
\bibitem [{\citenamefont {Klassen}\ and\ \citenamefont
  {Melzer}(1992{\natexlab{a}})}]{klassen1992spectral}%
  \BibitemOpen
  \bibfield  {author} {\bibinfo {author} {\bibfnamefont {T.~R.}\ \bibnamefont
  {Klassen}}\ and\ \bibinfo {author} {\bibfnamefont {E.}~\bibnamefont
  {Melzer}},\ }\bibfield  {title} {\enquote {\bibinfo {title} {{Spectral flow
  between conformal field theories in 1 + 1 dimensions}},}\ }\href {\doibase
  http://dx.doi.org/10.1016/0550-3213(92)90422-8} {\bibfield  {journal}
  {\bibinfo  {journal} {Nucl. Phys. B}\ }\textbf {\bibinfo {volume} {370}},\
  \bibinfo {pages} {511 -- 550} (\bibinfo {year}
  {1992}{\natexlab{a}})}\BibitemShut {NoStop}%
\bibitem [{\citenamefont {Klassen}\ and\ \citenamefont
  {Melzer}(1991{\natexlab{a}})}]{klassen1991relation}%
  \BibitemOpen
  \bibfield  {author} {\bibinfo {author} {\bibfnamefont {T.~R.}\ \bibnamefont
  {Klassen}}\ and\ \bibinfo {author} {\bibfnamefont {E.}~\bibnamefont
  {Melzer}},\ }\bibfield  {title} {\enquote {\bibinfo {title} {{On the relation
  between scattering amplitudes and finite-size mass corrections in QFT}},}\
  }\href {\doibase http://dx.doi.org/10.1016/0550-3213(91)90566-G} {\bibfield
  {journal} {\bibinfo  {journal} {Nucl. Phys. B}\ }\textbf {\bibinfo {volume}
  {362}},\ \bibinfo {pages} {329 -- 388} (\bibinfo {year}
  {1991}{\natexlab{a}})}\BibitemShut {NoStop}%
\bibitem [{\citenamefont {Pozsgay}\ and\ \citenamefont
  {Tak\'acs}(2008{\natexlab{a}})}]{pozsgay2008formII}%
  \BibitemOpen
  \bibfield  {author} {\bibinfo {author} {\bibfnamefont {B.}~\bibnamefont
  {Pozsgay}}\ and\ \bibinfo {author} {\bibfnamefont {G.}~\bibnamefont
  {Tak\'acs}},\ }\bibfield  {title} {\enquote {\bibinfo {title} {{Form factors
  in finite volume II: Disconnected terms and finite temperature
  correlators}},}\ }\href {\doibase
  http://dx.doi.org/10.1016/j.nuclphysb.2007.07.008} {\bibfield  {journal}
  {\bibinfo  {journal} {Nucl. Phys. B}\ }\textbf {\bibinfo {volume} {788}},\
  \bibinfo {pages} {209 -- 251} (\bibinfo {year}
  {2008}{\natexlab{a}})}\BibitemShut {NoStop}%
\bibitem [{\citenamefont {Pozsgay}\ and\ \citenamefont
  {Tak\'acs}(2008{\natexlab{b}})}]{pozsgay2008formI}%
  \BibitemOpen
  \bibfield  {author} {\bibinfo {author} {\bibfnamefont {B.}~\bibnamefont
  {Pozsgay}}\ and\ \bibinfo {author} {\bibfnamefont {G.}~\bibnamefont
  {Tak\'acs}},\ }\bibfield  {title} {\enquote {\bibinfo {title} {{Form factors
  in finite volume I: Form factor bootstrap and truncated conformal space}},}\
  }\href {\doibase http://dx.doi.org/10.1016/j.nuclphysb.2007.06.027}
  {\bibfield  {journal} {\bibinfo  {journal} {Nucl. Phys. B}\ }\textbf
  {\bibinfo {volume} {788}},\ \bibinfo {pages} {167 -- 208} (\bibinfo {year}
  {2008}{\natexlab{b}})}\BibitemShut {NoStop}%
\bibitem [{\citenamefont {Guida}\ and\ \citenamefont
  {Magnoli}(1997)}]{guida1997vacuum}%
  \BibitemOpen
  \bibfield  {author} {\bibinfo {author} {\bibfnamefont {R.}~\bibnamefont
  {Guida}}\ and\ \bibinfo {author} {\bibfnamefont {N.}~\bibnamefont
  {Magnoli}},\ }\bibfield  {title} {\enquote {\bibinfo {title} {{Vacuum
  expectation values from a variational approach}},}\ }\href {\doibase
  http://dx.doi.org/10.1016/S0370-2693(97)00983-0} {\bibfield  {journal}
  {\bibinfo  {journal} {Phys. Lett. B}\ }\textbf {\bibinfo {volume} {411}},\
  \bibinfo {pages} {127 -- 133} (\bibinfo {year} {1997})}\BibitemShut {NoStop}%
\bibitem [{\citenamefont {Klassen}\ and\ \citenamefont
  {Melzer}(1992{\natexlab{b}})}]{klassen1992kinks}%
  \BibitemOpen
  \bibfield  {author} {\bibinfo {author} {\bibfnamefont {T.~R.}\ \bibnamefont
  {Klassen}}\ and\ \bibinfo {author} {\bibfnamefont {E.}~\bibnamefont
  {Melzer}},\ }\bibfield  {title} {\enquote {\bibinfo {title} {{Kinks in finite
  volume}},}\ }\href {\doibase http://dx.doi.org/10.1016/0550-3213(92)90656-V}
  {\bibfield  {journal} {\bibinfo  {journal} {Nucl. Phys. B}\ }\textbf
  {\bibinfo {volume} {382}},\ \bibinfo {pages} {441 -- 485} (\bibinfo {year}
  {1992}{\natexlab{b}})}\BibitemShut {NoStop}%
\bibitem [{\citenamefont {Kausch}\ \emph {et~al.}(1997)\citenamefont {Kausch},
  \citenamefont {Tak\'acs},\ and\ \citenamefont {Watts}}]{kausch1997relation}%
  \BibitemOpen
  \bibfield  {author} {\bibinfo {author} {\bibfnamefont {H.}~\bibnamefont
  {Kausch}}, \bibinfo {author} {\bibfnamefont {G.}~\bibnamefont {Tak\'acs}}, \
  and\ \bibinfo {author} {\bibfnamefont {G.}~\bibnamefont {Watts}},\ }\bibfield
   {title} {\enquote {\bibinfo {title} {{On the relation between $\Phi$(1, 2)
  and $\Phi$(1, 5) perturbed minimal models and unitarity}},}\ }\href {\doibase
  http://dx.doi.org/10.1016/S0550-3213(97)00056-4} {\bibfield  {journal}
  {\bibinfo  {journal} {Nucl. Phys. B}\ }\textbf {\bibinfo {volume} {489}},\
  \bibinfo {pages} {557 -- 579} (\bibinfo {year} {1997})}\BibitemShut {NoStop}%
\bibitem [{\citenamefont {L{\"a}ssig}\ and\ \citenamefont
  {Martins}(1991)}]{lassig1991finite}%
  \BibitemOpen
  \bibfield  {author} {\bibinfo {author} {\bibfnamefont {M.}~\bibnamefont
  {L{\"a}ssig}}\ and\ \bibinfo {author} {\bibfnamefont {M.~J.}\ \bibnamefont
  {Martins}},\ }\bibfield  {title} {\enquote {\bibinfo {title} {{Finite-size
  effects in theories with factorizable S-matrices}},}\ }\href {\doibase
  http://dx.doi.org/10.1016/0550-3213(91)90371-4} {\bibfield  {journal}
  {\bibinfo  {journal} {Nucl. Phys. B}\ }\textbf {\bibinfo {volume} {354}},\
  \bibinfo {pages} {666 -- 688} (\bibinfo {year} {1991})}\BibitemShut {NoStop}%
\bibitem [{\citenamefont {Tak\'acs}(1997)}]{takacs1997new}%
  \BibitemOpen
  \bibfield  {author} {\bibinfo {author} {\bibfnamefont {G.}~\bibnamefont
  {Tak\'acs}},\ }\bibfield  {title} {\enquote {\bibinfo {title} {{A new RSOS
  restriction of the Zhiber-Mikhailov-Shabat model and $\Phi$(1,5)
  perturbations of non-unitary minimal models}},}\ }\href {\doibase
  http://dx.doi.org/10.1016/S0550-3213(97)00057-6} {\bibfield  {journal}
  {\bibinfo  {journal} {Nucl. Phys. B}\ }\textbf {\bibinfo {volume} {489}},\
  \bibinfo {pages} {532 -- 556} (\bibinfo {year} {1997})}\BibitemShut {NoStop}%
\bibitem [{\citenamefont {Ahn}\ \emph {et~al.}(2005)\citenamefont {Ahn},
  \citenamefont {Bajnok}, \citenamefont {Nepomechie}, \citenamefont {Palla},\
  and\ \citenamefont {Tak\'acs}}]{ahn2005nlie}%
  \BibitemOpen
  \bibfield  {author} {\bibinfo {author} {\bibfnamefont {C.}~\bibnamefont
  {Ahn}}, \bibinfo {author} {\bibfnamefont {Z.}~\bibnamefont {Bajnok}},
  \bibinfo {author} {\bibfnamefont {R.~I.}\ \bibnamefont {Nepomechie}},
  \bibinfo {author} {\bibfnamefont {L.}~\bibnamefont {Palla}}, \ and\ \bibinfo
  {author} {\bibfnamefont {G.}~\bibnamefont {Tak\'acs}},\ }\bibfield  {title}
  {\enquote {\bibinfo {title} {{NLIE for hole excited states in the sine-Gordon
  model with two boundaries}},}\ }\href {\doibase
  http://dx.doi.org/10.1016/j.nuclphysb.2005.03.014} {\bibfield  {journal}
  {\bibinfo  {journal} {Nucl. Phys. B}\ }\textbf {\bibinfo {volume} {714}},\
  \bibinfo {pages} {307 -- 335} (\bibinfo {year} {2005})}\BibitemShut {NoStop}%
\bibitem [{\citenamefont {Koubek}\ \emph {et~al.}(1992)\citenamefont {Koubek},
  \citenamefont {Martins},\ and\ \citenamefont
  {Mussardo}}]{koubek1992scattering}%
  \BibitemOpen
  \bibfield  {author} {\bibinfo {author} {\bibfnamefont {A.}~\bibnamefont
  {Koubek}}, \bibinfo {author} {\bibfnamefont {M.~J.}\ \bibnamefont {Martins}},
  \ and\ \bibinfo {author} {\bibfnamefont {G.}~\bibnamefont {Mussardo}},\
  }\bibfield  {title} {\enquote {\bibinfo {title} {{Scattering matrices for
  $\Phi$(1,2) perturbed conformal minimal models in absence of kink states}},}\
  }\href {\doibase http://dx.doi.org/10.1016/0550-3213(92)90215-W} {\bibfield
  {journal} {\bibinfo  {journal} {Nucl. Phys. B}\ }\textbf {\bibinfo {volume}
  {368}},\ \bibinfo {pages} {591 -- 610} (\bibinfo {year} {1992})}\BibitemShut
  {NoStop}%
\bibitem [{\citenamefont {Lepori}\ \emph {et~al.}(2008)\citenamefont {Lepori},
  \citenamefont {Mussardo},\ and\ \citenamefont
  {T{\'o}th}}]{lepori2008particle}%
  \BibitemOpen
  \bibfield  {author} {\bibinfo {author} {\bibfnamefont {L.}~\bibnamefont
  {Lepori}}, \bibinfo {author} {\bibfnamefont {G.}~\bibnamefont {Mussardo}}, \
  and\ \bibinfo {author} {\bibfnamefont {G.~Zs.}\ \bibnamefont {T{\'o}th}},\
  }\bibfield  {title} {\enquote {\bibinfo {title} {{The particle spectrum of
  the tricritical Ising model with spin reversal symmetric perturbations}},}\
  }\href {http://stacks.iop.org/1742-5468/2008/i=09/a=P09004} {\bibfield
  {journal} {\bibinfo  {journal} {J. Stat. Mech.}\ }\textbf {\bibinfo {volume}
  {2008}},\ \bibinfo {pages} {P09004} (\bibinfo {year} {2008})}\BibitemShut
  {NoStop}%
\bibitem [{\citenamefont {Martins}(1991{\natexlab{a}})}]{martins1991scaling}%
  \BibitemOpen
  \bibfield  {author} {\bibinfo {author} {\bibfnamefont {M.~J.}\ \bibnamefont
  {Martins}},\ }\bibfield  {title} {\enquote {\bibinfo {title} {{Scaling
  non-unitary models with degenerate ground state. Thermodynamic Bethe ansatz
  approach}},}\ }\href {\doibase
  http://dx.doi.org/10.1016/0370-2693(91)91899-7} {\bibfield  {journal}
  {\bibinfo  {journal} {Phys. Lett. B}\ }\textbf {\bibinfo {volume} {257}},\
  \bibinfo {pages} {317 -- 321} (\bibinfo {year}
  {1991}{\natexlab{a}})}\BibitemShut {NoStop}%
\bibitem [{\citenamefont {Colomo}\ \emph {et~al.}(1992)\citenamefont {Colomo},
  \citenamefont {Mussardo},\ and\ \citenamefont {Koubek}}]{colomo1992s}%
  \BibitemOpen
  \bibfield  {author} {\bibinfo {author} {\bibfnamefont {F.}~\bibnamefont
  {Colomo}}, \bibinfo {author} {\bibfnamefont {G.}~\bibnamefont {Mussardo}}, \
  and\ \bibinfo {author} {\bibfnamefont {A.}~\bibnamefont {Koubek}},\
  }\bibfield  {title} {\enquote {\bibinfo {title} {{On the S matrix of the
  subleading magnetic deformation of the tricritical ising model in two
  dimensions}},}\ }\href {\doibase 10.1142/S0217751X92002416} {\bibfield
  {journal} {\bibinfo  {journal} {Int. J. Mod. Phys. A}\ }\textbf {\bibinfo
  {volume} {07}},\ \bibinfo {pages} {5281--5305} (\bibinfo {year}
  {1992})}\BibitemShut {NoStop}%
\bibitem [{\citenamefont
  {Martins}(1991{\natexlab{b}})}]{martins1991constructing}%
  \BibitemOpen
  \bibfield  {author} {\bibinfo {author} {\bibfnamefont {M.~J.}\ \bibnamefont
  {Martins}},\ }\bibfield  {title} {\enquote {\bibinfo {title} {{Constructing
  an S-matrix from the truncated conformal approach data}},}\ }\href {\doibase
  http://dx.doi.org/10.1016/0370-2693(91)90639-8} {\bibfield  {journal}
  {\bibinfo  {journal} {Phys. Lett. B}\ }\textbf {\bibinfo {volume} {262}},\
  \bibinfo {pages} {39 -- 42} (\bibinfo {year}
  {1991}{\natexlab{b}})}\BibitemShut {NoStop}%
\bibitem [{\citenamefont {Lencs{\'e}s}\ and\ \citenamefont
  {Tak{\'a}cs}(2014)}]{lencses2014excited}%
  \BibitemOpen
  \bibfield  {author} {\bibinfo {author} {\bibfnamefont {M.}~\bibnamefont
  {Lencs{\'e}s}}\ and\ \bibinfo {author} {\bibfnamefont {G.}~\bibnamefont
  {Tak{\'a}cs}},\ }\bibfield  {title} {\enquote {\bibinfo {title} {{Excited
  state TBA and renormalized TCSA in the scaling Potts model}},}\ }\href
  {\doibase 10.1007/JHEP09(2014)052} {\bibfield  {journal} {\bibinfo  {journal}
  {JHEP}\ }\textbf {\bibinfo {volume} {2014}},\ \bibinfo {pages} {52} (\bibinfo
  {year} {2014})}\BibitemShut {NoStop}%
\bibitem [{\citenamefont {Mussardo}\ and\ \citenamefont
  {Tak\'acs}(2009)}]{mussardo2009effective}%
  \BibitemOpen
  \bibfield  {author} {\bibinfo {author} {\bibfnamefont {G.}~\bibnamefont
  {Mussardo}}\ and\ \bibinfo {author} {\bibfnamefont {G}~\bibnamefont
  {Tak\'acs}},\ }\bibfield  {title} {\enquote {\bibinfo {title} {Effective
  potentials and kink spectra in non-integrable perturbed conformal field
  theories},}\ }\href {http://stacks.iop.org/1751-8121/42/i=30/a=304022}
  {\bibfield  {journal} {\bibinfo  {journal} {J. Phys. A}\ }\textbf {\bibinfo
  {volume} {42}},\ \bibinfo {pages} {304022} (\bibinfo {year}
  {2009})}\BibitemShut {NoStop}%
\bibitem [{\citenamefont {Ellem}\ and\ \citenamefont
  {Bazhanov}(1998)}]{ellem1998thermodynamic}%
  \BibitemOpen
  \bibfield  {author} {\bibinfo {author} {\bibfnamefont {R.~M.}\ \bibnamefont
  {Ellem}}\ and\ \bibinfo {author} {\bibfnamefont {V.~V.}\ \bibnamefont
  {Bazhanov}},\ }\bibfield  {title} {\enquote {\bibinfo {title} {{Thermodynamic
  Bethe ansatz for the subleading magnetic perturbation of the tricritical
  Ising model}},}\ }\href {\doibase
  http://dx.doi.org/10.1016/S0550-3213(97)00748-7} {\bibfield  {journal}
  {\bibinfo  {journal} {Nucl. Phys. B}\ }\textbf {\bibinfo {volume} {512}},\
  \bibinfo {pages} {563 -- 580} (\bibinfo {year} {1998})}\BibitemShut {NoStop}%
\bibitem [{\citenamefont {Ellem}\ and\ \citenamefont
  {Bazhanov}(2002)}]{ellem2002excited}%
  \BibitemOpen
  \bibfield  {author} {\bibinfo {author} {\bibfnamefont {R.~M.}\ \bibnamefont
  {Ellem}}\ and\ \bibinfo {author} {\bibfnamefont {V.~V.}\ \bibnamefont
  {Bazhanov}},\ }\bibfield  {title} {\enquote {\bibinfo {title} {{Excited state
  TBA for the $\varphi_{2,1}$ perturbed $\mathcal{M}_{3,5}$ model}},}\ }\href
  {\doibase http://dx.doi.org/10.1016/S0550-3213(02)00843-X} {\bibfield
  {journal} {\bibinfo  {journal} {Nucl. Phys. B}\ }\textbf {\bibinfo {volume}
  {647}},\ \bibinfo {pages} {404 -- 432} (\bibinfo {year} {2002})}\BibitemShut
  {NoStop}%
\bibitem [{\citenamefont {Lencs{\'e}s}\ and\ \citenamefont
  {Tak{\'a}cs}(2015)}]{lencses2015confinement}%
  \BibitemOpen
  \bibfield  {author} {\bibinfo {author} {\bibfnamefont {M.}~\bibnamefont
  {Lencs{\'e}s}}\ and\ \bibinfo {author} {\bibfnamefont {G.}~\bibnamefont
  {Tak{\'a}cs}},\ }\bibfield  {title} {\enquote {\bibinfo {title} {{Confinement
  in the q-state Potts model: an RG-TCSA study}},}\ }\href {\doibase
  10.1007/JHEP09(2015)146} {\bibfield  {journal} {\bibinfo  {journal} {JHEP}\
  }\textbf {\bibinfo {volume} {2015}},\ \bibinfo {pages} {146} (\bibinfo {year}
  {2015})}\BibitemShut {NoStop}%
\bibitem [{\citenamefont {T\'oth}(2007)}]{toth2007study}%
  \BibitemOpen
  \bibfield  {author} {\bibinfo {author} {\bibfnamefont {G.~Zs.}\ \bibnamefont
  {T\'oth}},\ }\bibfield  {title} {\enquote {\bibinfo {title} {{A study of
  truncation effects in boundary flows of the Ising model on a strip}},}\
  }\href {http://stacks.iop.org/1742-5468/2007/i=04/a=P04005} {\bibfield
  {journal} {\bibinfo  {journal} {J. Stat. Mech.}\ }\textbf {\bibinfo {volume}
  {2007}},\ \bibinfo {pages} {P04005} (\bibinfo {year} {2007})}\BibitemShut
  {NoStop}%
\bibitem [{\citenamefont {L\"assig}(1991)}]{lassig1991exact}%
  \BibitemOpen
  \bibfield  {author} {\bibinfo {author} {\bibfnamefont {M.}~\bibnamefont
  {L\"assig}},\ }\bibfield  {title} {\enquote {\bibinfo {title} {Exact
  universal amplitude ratios in two-dimensional systems near criticality},}\
  }\href {\doibase 10.1103/PhysRevLett.67.3737} {\bibfield  {journal} {\bibinfo
   {journal} {Phys. Rev. Lett.}\ }\textbf {\bibinfo {volume} {67}},\ \bibinfo
  {pages} {3737--3740} (\bibinfo {year} {1991})}\BibitemShut {NoStop}%
\bibitem [{\citenamefont {Mossa}\ and\ \citenamefont
  {Mussardo}(2008)}]{mossa2008analytic}%
  \BibitemOpen
  \bibfield  {author} {\bibinfo {author} {\bibfnamefont {A.}~\bibnamefont
  {Mossa}}\ and\ \bibinfo {author} {\bibfnamefont {G.}~\bibnamefont
  {Mussardo}},\ }\bibfield  {title} {\enquote {\bibinfo {title} {{Analytic
  properties of the free energy: the tricritical Ising model}},}\ }\href
  {http://stacks.iop.org/1742-5468/2008/i=03/a=P03010} {\bibfield  {journal}
  {\bibinfo  {journal} {J. Stat. Mech.}\ }\textbf {\bibinfo {volume} {2008}},\
  \bibinfo {pages} {P03010} (\bibinfo {year} {2008})}\BibitemShut {NoStop}%
\bibitem [{\citenamefont {Pozsgay}\ \emph {et~al.}(2014)\citenamefont
  {Pozsgay}, \citenamefont {Sz{\'e}cs{\'e}nyi},\ and\ \citenamefont
  {Tak{\'a}cs}}]{pozsgay2014exact}%
  \BibitemOpen
  \bibfield  {author} {\bibinfo {author} {\bibfnamefont {B.}~\bibnamefont
  {Pozsgay}}, \bibinfo {author} {\bibfnamefont {I.~M.}\ \bibnamefont
  {Sz{\'e}cs{\'e}nyi}}, \ and\ \bibinfo {author} {\bibfnamefont
  {G.}~\bibnamefont {Tak{\'a}cs}},\ }\bibfield  {title} {\enquote {\bibinfo
  {title} {Exact finite volume expectation values of local operators in excited
  states},}\ }\href {https://arxiv.org/abs/1412.8436} {\bibfield  {journal}
  {\bibinfo  {journal} {arXiv preprint arXiv:1412.8436}\ } (\bibinfo {year}
  {2014})}\BibitemShut {NoStop}%
\bibitem [{\citenamefont {Pozsgay}\ \emph {et~al.}(2015)\citenamefont
  {Pozsgay}, \citenamefont {Sz{\'e}cs{\'e}nyi},\ and\ \citenamefont
  {Tak{\'a}cs}}]{pozsgay2015exact}%
  \BibitemOpen
  \bibfield  {author} {\bibinfo {author} {\bibfnamefont {B.}~\bibnamefont
  {Pozsgay}}, \bibinfo {author} {\bibfnamefont {I.~M.}\ \bibnamefont
  {Sz{\'e}cs{\'e}nyi}}, \ and\ \bibinfo {author} {\bibfnamefont
  {G.}~\bibnamefont {Tak{\'a}cs}},\ }\bibfield  {title} {\enquote {\bibinfo
  {title} {{Exact finite volume expectation values of local operators in
  excited states}},}\ }\href {\doibase 10.1007/JHEP04(2015)023} {\bibfield
  {journal} {\bibinfo  {journal} {JHEP}\ }\textbf {\bibinfo {volume} {2015}},\
  \bibinfo {pages} {23} (\bibinfo {year} {2015})}\BibitemShut {NoStop}%
\bibitem [{\citenamefont {Dorey}\ and\ \citenamefont
  {Tateo}(1996)}]{dorey1996excited}%
  \BibitemOpen
  \bibfield  {author} {\bibinfo {author} {\bibfnamefont {P.}~\bibnamefont
  {Dorey}}\ and\ \bibinfo {author} {\bibfnamefont {R.}~\bibnamefont {Tateo}},\
  }\bibfield  {title} {\enquote {\bibinfo {title} {{Excited states by analytic
  continuation of TBA equations}},}\ }\href {\doibase
  http://dx.doi.org/10.1016/S0550-3213(96)00516-0} {\bibfield  {journal}
  {\bibinfo  {journal} {Nucl. Phys. B}\ }\textbf {\bibinfo {volume} {482}},\
  \bibinfo {pages} {639 -- 659} (\bibinfo {year} {1996})}\BibitemShut {NoStop}%
\bibitem [{\citenamefont {Bazhanov}\ \emph {et~al.}(1997)\citenamefont
  {Bazhanov}, \citenamefont {Lukyanov},\ and\ \citenamefont
  {Zamolodchikov}}]{bazhanov1997quantum}%
  \BibitemOpen
  \bibfield  {author} {\bibinfo {author} {\bibfnamefont {V.~V.}\ \bibnamefont
  {Bazhanov}}, \bibinfo {author} {\bibfnamefont {S.~L.}\ \bibnamefont
  {Lukyanov}}, \ and\ \bibinfo {author} {\bibfnamefont {A.~B.}\ \bibnamefont
  {Zamolodchikov}},\ }\bibfield  {title} {\enquote {\bibinfo {title} {{Quantum
  field theories in finite volume: Excited state energies}},}\ }\href {\doibase
  http://dx.doi.org/10.1016/S0550-3213(97)00022-9} {\bibfield  {journal}
  {\bibinfo  {journal} {Nucl. Phys. B}\ }\textbf {\bibinfo {volume} {489}},\
  \bibinfo {pages} {487 -- 531} (\bibinfo {year} {1997})}\BibitemShut {NoStop}%
\bibitem [{\citenamefont {von Gehlen}(1991)}]{von1991critical}%
  \BibitemOpen
  \bibfield  {author} {\bibinfo {author} {\bibfnamefont {G.}~\bibnamefont {von
  Gehlen}},\ }\bibfield  {title} {\enquote {\bibinfo {title} {Critical and
  off-critical conformal analysis of the ising quantum chain in an imaginary
  field},}\ }\href {http://stacks.iop.org/0305-4470/24/i=22/a=021} {\bibfield
  {journal} {\bibinfo  {journal} {J. Phys. A}\ }\textbf {\bibinfo {volume}
  {24}},\ \bibinfo {pages} {5371} (\bibinfo {year} {1991})}\BibitemShut
  {NoStop}%
\bibitem [{\citenamefont {Zamolodchikov}(2002)}]{zamolodchikov2002scaling}%
  \BibitemOpen
  \bibfield  {author} {\bibinfo {author} {\bibfnamefont {A.}~\bibnamefont
  {Zamolodchikov}},\ }\bibfield  {title} {\enquote {\bibinfo {title} {{Scaling
  Lee-Yang Model on a Sphere I. Partition Function}},}\ }\href
  {http://stacks.iop.org/1126-6708/2002/i=07/a=029} {\bibfield  {journal}
  {\bibinfo  {journal} {JHEP}\ }\textbf {\bibinfo {volume} {2002}},\ \bibinfo
  {pages} {029} (\bibinfo {year} {2002})}\BibitemShut {NoStop}%
\bibitem [{\citenamefont {Delfino}\ \emph {et~al.}(2006)\citenamefont
  {Delfino}, \citenamefont {Grinza},\ and\ \citenamefont
  {Mussardo}}]{delfino2006decay}%
  \BibitemOpen
  \bibfield  {author} {\bibinfo {author} {\bibfnamefont {G.}~\bibnamefont
  {Delfino}}, \bibinfo {author} {\bibfnamefont {P.}~\bibnamefont {Grinza}}, \
  and\ \bibinfo {author} {\bibfnamefont {G.}~\bibnamefont {Mussardo}},\
  }\bibfield  {title} {\enquote {\bibinfo {title} {{Decay of particles above
  threshold in the Ising field theory with magnetic field}},}\ }\href {\doibase
  http://dx.doi.org/10.1016/j.nuclphysb.2005.12.024} {\bibfield  {journal}
  {\bibinfo  {journal} {Nucl. Phys. B}\ }\textbf {\bibinfo {volume} {737}},\
  \bibinfo {pages} {291 -- 303} (\bibinfo {year} {2006})}\BibitemShut {NoStop}%
\bibitem [{\citenamefont {Fioravanti}\ \emph
  {et~al.}(2000{\natexlab{a}})\citenamefont {Fioravanti}, \citenamefont
  {Mussardo},\ and\ \citenamefont {Simon}}]{fioravanti2000universal}%
  \BibitemOpen
  \bibfield  {author} {\bibinfo {author} {\bibfnamefont {D.}~\bibnamefont
  {Fioravanti}}, \bibinfo {author} {\bibfnamefont {G.}~\bibnamefont
  {Mussardo}}, \ and\ \bibinfo {author} {\bibfnamefont {P.}~\bibnamefont
  {Simon}},\ }\bibfield  {title} {\enquote {\bibinfo {title} {Universal ratios
  in the 2d tricritical ising model},}\ }\href {\doibase
  10.1103/PhysRevLett.85.126} {\bibfield  {journal} {\bibinfo  {journal} {Phys.
  Rev. Lett.}\ }\textbf {\bibinfo {volume} {85}},\ \bibinfo {pages} {126--129}
  (\bibinfo {year} {2000}{\natexlab{a}})}\BibitemShut {NoStop}%
\bibitem [{\citenamefont {L{\"a}ssig}\ \emph {et~al.}(1991)\citenamefont
  {L{\"a}ssig}, \citenamefont {Mussardo},\ and\ \citenamefont
  {Cardy}}]{lassig1991scaling}%
  \BibitemOpen
  \bibfield  {author} {\bibinfo {author} {\bibfnamefont {M.}~\bibnamefont
  {L{\"a}ssig}}, \bibinfo {author} {\bibfnamefont {G.}~\bibnamefont
  {Mussardo}}, \ and\ \bibinfo {author} {\bibfnamefont {J.~L.}\ \bibnamefont
  {Cardy}},\ }\bibfield  {title} {\enquote {\bibinfo {title} {{The scaling
  region of the tricritical Ising model in two dimensions}},}\ }\href {\doibase
  http://dx.doi.org/10.1016/0550-3213(91)90206-D} {\bibfield  {journal}
  {\bibinfo  {journal} {Nucl. Phys. B}\ }\textbf {\bibinfo {volume} {348}},\
  \bibinfo {pages} {591 -- 618} (\bibinfo {year} {1991})}\BibitemShut {NoStop}%
\bibitem [{\citenamefont {Feverati}\ \emph {et~al.}(1999)\citenamefont
  {Feverati}, \citenamefont {Ravanini},\ and\ \citenamefont
  {Tak\'acs}}]{feverati1999non}%
  \BibitemOpen
  \bibfield  {author} {\bibinfo {author} {\bibfnamefont {G.}~\bibnamefont
  {Feverati}}, \bibinfo {author} {\bibfnamefont {F.}~\bibnamefont {Ravanini}},
  \ and\ \bibinfo {author} {\bibfnamefont {G.}~\bibnamefont {Tak\'acs}},\
  }\bibfield  {title} {\enquote {\bibinfo {title} {{Non-linear integral
  equation and finite volume spectrum of sine-Gordon theory}},}\ }\href
  {\doibase http://dx.doi.org/10.1016/S0550-3213(98)00747-0} {\bibfield
  {journal} {\bibinfo  {journal} {Nucl. Phys. B}\ }\textbf {\bibinfo {volume}
  {540}},\ \bibinfo {pages} {543 -- 586} (\bibinfo {year} {1999})}\BibitemShut
  {NoStop}%
\bibitem [{\citenamefont {Feverati}\ \emph
  {et~al.}(1998{\natexlab{a}})\citenamefont {Feverati}, \citenamefont
  {Ravanini},\ and\ \citenamefont {Tak\'acs}}]{feverati1998truncated}%
  \BibitemOpen
  \bibfield  {author} {\bibinfo {author} {\bibfnamefont {G.}~\bibnamefont
  {Feverati}}, \bibinfo {author} {\bibfnamefont {F.}~\bibnamefont {Ravanini}},
  \ and\ \bibinfo {author} {\bibfnamefont {G.}~\bibnamefont {Tak\'acs}},\
  }\bibfield  {title} {\enquote {\bibinfo {title} {{Truncated conformal space
  at c=1, nonlinear integral equation and quantization rules for multi-soliton
  states}},}\ }\href {\doibase http://dx.doi.org/10.1016/S0370-2693(98)00543-7}
  {\bibfield  {journal} {\bibinfo  {journal} {Phys. Lett. B}\ }\textbf
  {\bibinfo {volume} {430}},\ \bibinfo {pages} {264 -- 273} (\bibinfo {year}
  {1998}{\natexlab{a}})}\BibitemShut {NoStop}%
\bibitem [{\citenamefont {Feverati}\ \emph
  {et~al.}(1998{\natexlab{b}})\citenamefont {Feverati}, \citenamefont
  {Ravanini},\ and\ \citenamefont {Tak\'acs}}]{feverati1998scaling}%
  \BibitemOpen
  \bibfield  {author} {\bibinfo {author} {\bibfnamefont {G.}~\bibnamefont
  {Feverati}}, \bibinfo {author} {\bibfnamefont {F.}~\bibnamefont {Ravanini}},
  \ and\ \bibinfo {author} {\bibfnamefont {G.}~\bibnamefont {Tak\'acs}},\
  }\bibfield  {title} {\enquote {\bibinfo {title} {{Scaling functions in the
  odd charge sector of sine-Gordon/massive Thirring theory}},}\ }\href
  {\doibase http://dx.doi.org/10.1016/S0370-2693(98)01406-3} {\bibfield
  {journal} {\bibinfo  {journal} {Phys. Lett. B}\ }\textbf {\bibinfo {volume}
  {444}},\ \bibinfo {pages} {442 -- 450} (\bibinfo {year}
  {1998}{\natexlab{b}})}\BibitemShut {NoStop}%
\bibitem [{\citenamefont {Bajnok}\ \emph {et~al.}(2004)\citenamefont {Bajnok},
  \citenamefont {Dunning}, \citenamefont {Palla}, \citenamefont {Tak\'acs},\
  and\ \citenamefont {W\'agner}}]{bajnok2004susy}%
  \BibitemOpen
  \bibfield  {author} {\bibinfo {author} {\bibfnamefont {Z.}~\bibnamefont
  {Bajnok}}, \bibinfo {author} {\bibfnamefont {C.}~\bibnamefont {Dunning}},
  \bibinfo {author} {\bibfnamefont {L.}~\bibnamefont {Palla}}, \bibinfo
  {author} {\bibfnamefont {G.}~\bibnamefont {Tak\'acs}}, \ and\ \bibinfo
  {author} {\bibfnamefont {F.}~\bibnamefont {W\'agner}},\ }\bibfield  {title}
  {\enquote {\bibinfo {title} {{SUSY sine-Gordon theory as a perturbed
  conformal field theory and finite size effects}},}\ }\href {\doibase
  http://dx.doi.org/10.1016/j.nuclphysb.2003.11.036} {\bibfield  {journal}
  {\bibinfo  {journal} {Nucl. Phys. B}\ }\textbf {\bibinfo {volume} {679}},\
  \bibinfo {pages} {521 -- 544} (\bibinfo {year} {2004})}\BibitemShut {NoStop}%
\bibitem [{\citenamefont {Bajnok}\ \emph
  {et~al.}(2002{\natexlab{a}})\citenamefont {Bajnok}, \citenamefont {Palla},\
  and\ \citenamefont {Tak\'acs}}]{bajnok2002finite}%
  \BibitemOpen
  \bibfield  {author} {\bibinfo {author} {\bibfnamefont {Z.}~\bibnamefont
  {Bajnok}}, \bibinfo {author} {\bibfnamefont {L.}~\bibnamefont {Palla}}, \
  and\ \bibinfo {author} {\bibfnamefont {G.}~\bibnamefont {Tak\'acs}},\
  }\bibfield  {title} {\enquote {\bibinfo {title} {{Finite size effects in
  boundary sine-Gordon theory}},}\ }\href {\doibase
  http://dx.doi.org/10.1016/S0550-3213(01)00616-2} {\bibfield  {journal}
  {\bibinfo  {journal} {Nucl. Phys. B}\ }\textbf {\bibinfo {volume} {622}},\
  \bibinfo {pages} {565 -- 592} (\bibinfo {year}
  {2002}{\natexlab{a}})}\BibitemShut {NoStop}%
\bibitem [{\citenamefont {T\'oth}(2004)}]{toth2004nonperturbative}%
  \BibitemOpen
  \bibfield  {author} {\bibinfo {author} {\bibfnamefont {G.~Zs.}\ \bibnamefont
  {T\'oth}},\ }\bibfield  {title} {\enquote {\bibinfo {title} {{A
  nonperturbative study of phase transitions in the multi-frequency sine-Gordon
  model}},}\ }\href {http://stacks.iop.org/0305-4470/37/i=41/a=003} {\bibfield
  {journal} {\bibinfo  {journal} {J. Phys. A}\ }\textbf {\bibinfo {volume}
  {37}},\ \bibinfo {pages} {9631} (\bibinfo {year} {2004})}\BibitemShut
  {NoStop}%
\bibitem [{\citenamefont {Feh\'er}\ and\ \citenamefont
  {Tak\'acs}(2011)}]{feher2011sine}%
  \BibitemOpen
  \bibfield  {author} {\bibinfo {author} {\bibfnamefont {G.}~\bibnamefont
  {Feh\'er}}\ and\ \bibinfo {author} {\bibfnamefont {G.}~\bibnamefont
  {Tak\'acs}},\ }\bibfield  {title} {\enquote {\bibinfo {title}
  {{Sine--€"Gordon form factors in finite volume}},}\ }\href {\doibase
  http://dx.doi.org/10.1016/j.nuclphysb.2011.06.020} {\bibfield  {journal}
  {\bibinfo  {journal} {Nucl. Phys. B}\ }\textbf {\bibinfo {volume} {852}},\
  \bibinfo {pages} {441 -- 467} (\bibinfo {year} {2011})}\BibitemShut {NoStop}%
\bibitem [{\citenamefont {Feh\'er}\ \emph {et~al.}(2012)\citenamefont
  {Feh\'er}, \citenamefont {P\'almai},\ and\ \citenamefont
  {Tak\'acs}}]{feher2012sine}%
  \BibitemOpen
  \bibfield  {author} {\bibinfo {author} {\bibfnamefont {G.~Z.}\ \bibnamefont
  {Feh\'er}}, \bibinfo {author} {\bibfnamefont {T.}~\bibnamefont {P\'almai}}, \
  and\ \bibinfo {author} {\bibfnamefont {G.}~\bibnamefont {Tak\'acs}},\
  }\bibfield  {title} {\enquote {\bibinfo {title} {{Sine-Gordon multisoliton
  form factors in finite volume}},}\ }\href {\doibase
  10.1103/PhysRevD.85.085005} {\bibfield  {journal} {\bibinfo  {journal} {Phys.
  Rev. D}\ }\textbf {\bibinfo {volume} {85}},\ \bibinfo {pages} {085005}
  (\bibinfo {year} {2012})}\BibitemShut {NoStop}%
\bibitem [{\citenamefont {P\'almai}\ and\ \citenamefont
  {Tak\'acs}(2013)}]{palmai2013diagonal}%
  \BibitemOpen
  \bibfield  {author} {\bibinfo {author} {\bibfnamefont {T.}~\bibnamefont
  {P\'almai}}\ and\ \bibinfo {author} {\bibfnamefont {G.}~\bibnamefont
  {Tak\'acs}},\ }\bibfield  {title} {\enquote {\bibinfo {title} {Diagonal
  multisoliton matrix elements in finite volume},}\ }\href {\doibase
  10.1103/PhysRevD.87.045010} {\bibfield  {journal} {\bibinfo  {journal} {Phys.
  Rev. D}\ }\textbf {\bibinfo {volume} {87}},\ \bibinfo {pages} {045010}
  (\bibinfo {year} {2013})}\BibitemShut {NoStop}%
\bibitem [{\citenamefont {Tak{\'a}cs}(2011)}]{takacs2011determining}%
  \BibitemOpen
  \bibfield  {author} {\bibinfo {author} {\bibfnamefont {G.}~\bibnamefont
  {Tak{\'a}cs}},\ }\bibfield  {title} {\enquote {\bibinfo {title} {{Determining
  matrix elements and resonance widths from finite volume: the dangerous
  $\mu$-terms}},}\ }\href {\doibase 10.1007/JHEP11(2011)113} {\bibfield
  {journal} {\bibinfo  {journal} {JHEP}\ }\textbf {\bibinfo {volume} {2011}},\
  \bibinfo {pages} {113} (\bibinfo {year} {2011})}\BibitemShut {NoStop}%
\bibitem [{\citenamefont {Palmai}(2015)}]{palmai2015edge}%
  \BibitemOpen
  \bibfield  {author} {\bibinfo {author} {\bibfnamefont {T.}~\bibnamefont
  {Palmai}},\ }\bibfield  {title} {\enquote {\bibinfo {title} {Edge exponents
  in work statistics out of equilibrium and dynamical phase transitions from
  scattering theory in one-dimensional gapped systems},}\ }\href {\doibase
  10.1103/PhysRevB.92.235433} {\bibfield  {journal} {\bibinfo  {journal} {Phys.
  Rev. B}\ }\textbf {\bibinfo {volume} {92}},\ \bibinfo {pages} {235433}
  (\bibinfo {year} {2015})}\BibitemShut {NoStop}%
\bibitem [{\citenamefont {Buccheri}\ and\ \citenamefont
  {Tak{\'a}cs}(2014)}]{buccheri2014finite}%
  \BibitemOpen
  \bibfield  {author} {\bibinfo {author} {\bibfnamefont {F.}~\bibnamefont
  {Buccheri}}\ and\ \bibinfo {author} {\bibfnamefont {G.}~\bibnamefont
  {Tak{\'a}cs}},\ }\bibfield  {title} {\enquote {\bibinfo {title} {Finite
  temperature one-point functions in non-diagonal integrable field theories:
  the sine-gordon model},}\ }\href {\doibase 10.1007/JHEP03(2014)026}
  {\bibfield  {journal} {\bibinfo  {journal} {JHEP}\ }\textbf {\bibinfo
  {volume} {2014}},\ \bibinfo {pages} {26} (\bibinfo {year}
  {2014})}\BibitemShut {NoStop}%
\bibitem [{\citenamefont {Bajnok}\ \emph
  {et~al.}(2001{\natexlab{a}})\citenamefont {Bajnok}, \citenamefont {Palla},
  \citenamefont {Tak\'acs},\ and\ \citenamefont
  {W\'agner}}]{bajnok2001nonperturbative}%
  \BibitemOpen
  \bibfield  {author} {\bibinfo {author} {\bibfnamefont {Z.}~\bibnamefont
  {Bajnok}}, \bibinfo {author} {\bibfnamefont {L.}~\bibnamefont {Palla}},
  \bibinfo {author} {\bibfnamefont {G.}~\bibnamefont {Tak\'acs}}, \ and\
  \bibinfo {author} {\bibfnamefont {F.}~\bibnamefont {W\'agner}},\ }\bibfield
  {title} {\enquote {\bibinfo {title} {{Nonperturbative study of the
  two-frequency sine-Gordon model}},}\ }\href {\doibase
  http://dx.doi.org/10.1016/S0550-3213(01)00067-0} {\bibfield  {journal}
  {\bibinfo  {journal} {Nucl. Phys. B}\ }\textbf {\bibinfo {volume} {601}},\
  \bibinfo {pages} {503 -- 538} (\bibinfo {year}
  {2001}{\natexlab{a}})}\BibitemShut {NoStop}%
\bibitem [{\citenamefont {Bajnok}\ \emph {et~al.}(2000)\citenamefont {Bajnok},
  \citenamefont {Palla}, \citenamefont {Tak\'acs},\ and\ \citenamefont
  {W\'agner}}]{bajnok2000k}%
  \BibitemOpen
  \bibfield  {author} {\bibinfo {author} {\bibfnamefont {Z.}~\bibnamefont
  {Bajnok}}, \bibinfo {author} {\bibfnamefont {L.}~\bibnamefont {Palla}},
  \bibinfo {author} {\bibfnamefont {G.}~\bibnamefont {Tak\'acs}}, \ and\
  \bibinfo {author} {\bibfnamefont {F.}~\bibnamefont {W\'agner}},\ }\bibfield
  {title} {\enquote {\bibinfo {title} {{The $k$-folded sine-Gordon model in
  finite volume}},}\ }\href {\doibase
  http://dx.doi.org/10.1016/S0550-3213(00)00441-7} {\bibfield  {journal}
  {\bibinfo  {journal} {Nucl. Phys. B}\ }\textbf {\bibinfo {volume} {587}},\
  \bibinfo {pages} {585 -- 618} (\bibinfo {year} {2000})}\BibitemShut {NoStop}%
\bibitem [{\citenamefont {Tak\'acs}\ and\ \citenamefont
  {W\'agner}(2006)}]{takacs2006double}%
  \BibitemOpen
  \bibfield  {author} {\bibinfo {author} {\bibfnamefont {G.}~\bibnamefont
  {Tak\'acs}}\ and\ \bibinfo {author} {\bibfnamefont {F.}~\bibnamefont
  {W\'agner}},\ }\bibfield  {title} {\enquote {\bibinfo {title} {{Double
  sine-Gordon model revisited}},}\ }\href {\doibase
  http://dx.doi.org/10.1016/j.nuclphysb.2006.02.004} {\bibfield  {journal}
  {\bibinfo  {journal} {Nucl. Phys. B}\ }\textbf {\bibinfo {volume} {741}},\
  \bibinfo {pages} {353 -- 367} (\bibinfo {year} {2006})}\BibitemShut {NoStop}%
\bibitem [{\citenamefont {Pozsgay}\ and\ \citenamefont
  {Tak\'acs}(2006)}]{pozsgay2006characterization}%
  \BibitemOpen
  \bibfield  {author} {\bibinfo {author} {\bibfnamefont {B.}~\bibnamefont
  {Pozsgay}}\ and\ \bibinfo {author} {\bibfnamefont {G.}~\bibnamefont
  {Tak\'acs}},\ }\bibfield  {title} {\enquote {\bibinfo {title}
  {Characterization of resonances using finite size effects},}\ }\href
  {\doibase http://dx.doi.org/10.1016/j.nuclphysb.2006.05.007} {\bibfield
  {journal} {\bibinfo  {journal} {Nucl. Phys. B}\ }\textbf {\bibinfo {volume}
  {748}},\ \bibinfo {pages} {485 -- 523} (\bibinfo {year} {2006})}\BibitemShut
  {NoStop}%
\bibitem [{\citenamefont {Tak\'acs}(2010)}]{takacs2010form}%
  \BibitemOpen
  \bibfield  {author} {\bibinfo {author} {\bibfnamefont {G.}~\bibnamefont
  {Tak\'acs}},\ }\bibfield  {title} {\enquote {\bibinfo {title} {Form factor
  perturbation theory from finite volume},}\ }\href {\doibase
  http://dx.doi.org/10.1016/j.nuclphysb.2009.10.001} {\bibfield  {journal}
  {\bibinfo  {journal} {Nucl. Phys. B}\ }\textbf {\bibinfo {volume} {825}},\
  \bibinfo {pages} {466 -- 481} (\bibinfo {year} {2010})}\BibitemShut {NoStop}%
\bibitem [{\citenamefont {Dorey}\ \emph {et~al.}(1998)\citenamefont {Dorey},
  \citenamefont {Pocklington}, \citenamefont {Tateo},\ and\ \citenamefont
  {Watts}}]{dorey1998tba}%
  \BibitemOpen
  \bibfield  {author} {\bibinfo {author} {\bibfnamefont {P.}~\bibnamefont
  {Dorey}}, \bibinfo {author} {\bibfnamefont {A.~J.}\ \bibnamefont
  {Pocklington}}, \bibinfo {author} {\bibfnamefont {R.}~\bibnamefont {Tateo}},
  \ and\ \bibinfo {author} {\bibfnamefont {G.}~\bibnamefont {Watts}},\
  }\bibfield  {title} {\enquote {\bibinfo {title} {{{TBA} and {TCSA} with
  boundaries and excited states}},}\ }\href {\doibase
  http://dx.doi.org/10.1016/S0550-3213(98)00339-3} {\bibfield  {journal}
  {\bibinfo  {journal} {Nucl. Phys. B}\ }\textbf {\bibinfo {volume} {525}},\
  \bibinfo {pages} {641 -- 663} (\bibinfo {year} {1998})}\BibitemShut {NoStop}%
\bibitem [{\citenamefont {Dorey}\ \emph {et~al.}(2000)\citenamefont {Dorey},
  \citenamefont {Runkel}, \citenamefont {Tateo},\ and\ \citenamefont
  {Watts}}]{dorey2000g}%
  \BibitemOpen
  \bibfield  {author} {\bibinfo {author} {\bibfnamefont {P.}~\bibnamefont
  {Dorey}}, \bibinfo {author} {\bibfnamefont {I.}~\bibnamefont {Runkel}},
  \bibinfo {author} {\bibfnamefont {R.}~\bibnamefont {Tateo}}, \ and\ \bibinfo
  {author} {\bibfnamefont {G.}~\bibnamefont {Watts}},\ }\bibfield  {title}
  {\enquote {\bibinfo {title} {{g -function flow in perturbed boundary
  conformal field theories}},}\ }\href {\doibase
  http://dx.doi.org/10.1016/S0550-3213(99)00772-5} {\bibfield  {journal}
  {\bibinfo  {journal} {Nucl. Phys. B}\ }\textbf {\bibinfo {volume} {578}},\
  \bibinfo {pages} {85 -- 122} (\bibinfo {year} {2000})}\BibitemShut {NoStop}%
\bibitem [{\citenamefont {Dorey}\ \emph {et~al.}(2001)\citenamefont {Dorey},
  \citenamefont {Pillin}, \citenamefont {Tateo},\ and\ \citenamefont
  {Watts}}]{dorey2001one}%
  \BibitemOpen
  \bibfield  {author} {\bibinfo {author} {\bibfnamefont {P.~E.}\ \bibnamefont
  {Dorey}}, \bibinfo {author} {\bibfnamefont {M.}~\bibnamefont {Pillin}},
  \bibinfo {author} {\bibfnamefont {R.}~\bibnamefont {Tateo}}, \ and\ \bibinfo
  {author} {\bibfnamefont {G.~M.~T.}\ \bibnamefont {Watts}},\ }\bibfield
  {title} {\enquote {\bibinfo {title} {{One-point functions in perturbed
  boundary conformal field theories}},}\ }\href {\doibase
  http://dx.doi.org/10.1016/S0550-3213(00)00622-2} {\bibfield  {journal}
  {\bibinfo  {journal} {Nucl. Phys. B}\ }\textbf {\bibinfo {volume} {594}},\
  \bibinfo {pages} {625 -- 659} (\bibinfo {year} {2001})}\BibitemShut {NoStop}%
\bibitem [{\citenamefont {Bajnok}\ \emph
  {et~al.}(2001{\natexlab{b}})\citenamefont {Bajnok}, \citenamefont {Palla},\
  and\ \citenamefont {Tak{\'a}cs}}]{bajnok2001boundary}%
  \BibitemOpen
  \bibfield  {author} {\bibinfo {author} {\bibfnamefont {Z}~\bibnamefont
  {Bajnok}}, \bibinfo {author} {\bibfnamefont {L}~\bibnamefont {Palla}}, \ and\
  \bibinfo {author} {\bibfnamefont {G}~\bibnamefont {Tak{\'a}cs}},\ }\bibfield
  {title} {\enquote {\bibinfo {title} {Boundary states and finite size effects
  in sine-gordon model with neumann boundary condition},}\ }\href@noop {}
  {\bibfield  {journal} {\bibinfo  {journal} {Nuclear Physics B}\ }\textbf
  {\bibinfo {volume} {614}},\ \bibinfo {pages} {405--448} (\bibinfo {year}
  {2001}{\natexlab{b}})}\BibitemShut {NoStop}%
\bibitem [{\citenamefont {Kormos}\ and\ \citenamefont
  {Pozsgay}(2010)}]{kormos2010one}%
  \BibitemOpen
  \bibfield  {author} {\bibinfo {author} {\bibfnamefont {M.}~\bibnamefont
  {Kormos}}\ and\ \bibinfo {author} {\bibfnamefont {B.}~\bibnamefont
  {Pozsgay}},\ }\bibfield  {title} {\enquote {\bibinfo {title} {{One-point
  functions in massive integrable QFT with boundaries}},}\ }\href {\doibase
  10.1007/JHEP04(2010)112} {\bibfield  {journal} {\bibinfo  {journal} {JHEP}\
  }\textbf {\bibinfo {volume} {2010}},\ \bibinfo {pages} {112} (\bibinfo {year}
  {2010})}\BibitemShut {NoStop}%
\bibitem [{\citenamefont {Kormos}\ \emph {et~al.}(2009)\citenamefont {Kormos},
  \citenamefont {Runkel},\ and\ \citenamefont {Watts}}]{kormos2009defect}%
  \BibitemOpen
  \bibfield  {author} {\bibinfo {author} {\bibfnamefont {M.}~\bibnamefont
  {Kormos}}, \bibinfo {author} {\bibfnamefont {I.}~\bibnamefont {Runkel}}, \
  and\ \bibinfo {author} {\bibfnamefont {G.~M.~T.}\ \bibnamefont {Watts}},\
  }\bibfield  {title} {\enquote {\bibinfo {title} {Defect flows in minimal
  models},}\ }\href {http://stacks.iop.org/1126-6708/2009/i=11/a=057}
  {\bibfield  {journal} {\bibinfo  {journal} {JHEP}\ }\textbf {\bibinfo
  {volume} {2009}},\ \bibinfo {pages} {057} (\bibinfo {year}
  {2009})}\BibitemShut {NoStop}%
\bibitem [{\citenamefont {Kormos}\ and\ \citenamefont
  {Tak\'acs}(2008)}]{kormos2008boundary}%
  \BibitemOpen
  \bibfield  {author} {\bibinfo {author} {\bibfnamefont {M.}~\bibnamefont
  {Kormos}}\ and\ \bibinfo {author} {\bibfnamefont {G.}~\bibnamefont
  {Tak\'acs}},\ }\bibfield  {title} {\enquote {\bibinfo {title} {{Boundary form
  factors in finite volume}},}\ }\href {\doibase
  http://dx.doi.org/10.1016/j.nuclphysb.2008.05.003} {\bibfield  {journal}
  {\bibinfo  {journal} {Nucl. Phys. B}\ }\textbf {\bibinfo {volume} {803}},\
  \bibinfo {pages} {277 -- 298} (\bibinfo {year} {2008})}\BibitemShut {NoStop}%
\bibitem [{\citenamefont {{Dorey}}\ \emph {et~al.}(2000)\citenamefont
  {{Dorey}}, \citenamefont {{Pillin}}, \citenamefont {{Pocklington}},
  \citenamefont {{Runkel}}, \citenamefont {{Tateo}},\ and\ \citenamefont
  {{Watts}}}]{dorey2000finite}%
  \BibitemOpen
  \bibfield  {author} {\bibinfo {author} {\bibfnamefont {P.}~\bibnamefont
  {{Dorey}}}, \bibinfo {author} {\bibfnamefont {M.}~\bibnamefont {{Pillin}}},
  \bibinfo {author} {\bibfnamefont {A.}~\bibnamefont {{Pocklington}}}, \bibinfo
  {author} {\bibfnamefont {I.}~\bibnamefont {{Runkel}}}, \bibinfo {author}
  {\bibfnamefont {R.}~\bibnamefont {{Tateo}}}, \ and\ \bibinfo {author}
  {\bibfnamefont {G.~M.~T.}\ \bibnamefont {{Watts}}},\ }\bibfield  {title}
  {\enquote {\bibinfo {title} {{Finite size effects in perturbed boundary
  conformal field theories}},}\ }\href@noop {} {\bibfield  {journal} {\bibinfo
  {journal} {ArXiv High Energy Physics - Theory e-prints}\ } (\bibinfo {year}
  {2000})},\ \Eprint {http://arxiv.org/abs/hep-th/0010278} {hep-th/0010278}
  \BibitemShut {NoStop}%
\bibitem [{\citenamefont {Kormos}(2006)}]{kormos2006boundary}%
  \BibitemOpen
  \bibfield  {author} {\bibinfo {author} {\bibfnamefont {M.}~\bibnamefont
  {Kormos}},\ }\bibfield  {title} {\enquote {\bibinfo {title} {{Boundary
  renormalisation group flows of unitary superconformal minimal models}},}\
  }\href {\doibase http://dx.doi.org/10.1016/j.nuclphysb.2006.03.018}
  {\bibfield  {journal} {\bibinfo  {journal} {Nucl. Phys. B}\ }\textbf
  {\bibinfo {volume} {744}},\ \bibinfo {pages} {358 -- 379} (\bibinfo {year}
  {2006})}\BibitemShut {NoStop}%
\bibitem [{\citenamefont {Lencs\'es}\ and\ \citenamefont
  {Tak\'acs}(2011)}]{lencses2011breather}%
  \BibitemOpen
  \bibfield  {author} {\bibinfo {author} {\bibfnamefont {M.}~\bibnamefont
  {Lencs\'es}}\ and\ \bibinfo {author} {\bibfnamefont {G.}~\bibnamefont
  {Tak\'acs}},\ }\bibfield  {title} {\enquote {\bibinfo {title} {{Breather
  boundary form factors in sine-Gordon theory}},}\ }\href {\doibase
  http://dx.doi.org/10.1016/j.nuclphysb.2011.07.010} {\bibfield  {journal}
  {\bibinfo  {journal} {Nucl. Phys. B}\ }\textbf {\bibinfo {volume} {852}},\
  \bibinfo {pages} {615 -- 633} (\bibinfo {year} {2011})}\BibitemShut {NoStop}%
\bibitem [{\citenamefont {Bajnok}\ \emph
  {et~al.}(2002{\natexlab{b}})\citenamefont {Bajnok}, \citenamefont {Palla},\
  and\ \citenamefont {Tak{\'a}cs}}]{bajnok2002spectrum}%
  \BibitemOpen
  \bibfield  {author} {\bibinfo {author} {\bibfnamefont {Z.}~\bibnamefont
  {Bajnok}}, \bibinfo {author} {\bibfnamefont {L.}~\bibnamefont {Palla}}, \
  and\ \bibinfo {author} {\bibfnamefont {G.}~\bibnamefont {Tak{\'a}cs}},\
  }\bibfield  {title} {\enquote {\bibinfo {title} {The spectrum of boundary
  sine-gordon theory},}\ }in\ \href@noop {} {\emph {\bibinfo {booktitle}
  {Statistical Field Theories}}}\ (\bibinfo  {publisher} {Springer},\ \bibinfo
  {year} {2002})\ pp.\ \bibinfo {pages} {195--204}\BibitemShut {NoStop}%
\bibitem [{\citenamefont {Tak{\'a}cs}\ and\ \citenamefont
  {Watts}(2012)}]{takacs2012excited}%
  \BibitemOpen
  \bibfield  {author} {\bibinfo {author} {\bibfnamefont {G.}~\bibnamefont
  {Tak{\'a}cs}}\ and\ \bibinfo {author} {\bibfnamefont {G.~M.~T.}\ \bibnamefont
  {Watts}},\ }\bibfield  {title} {\enquote {\bibinfo {title} {{Excited state
  g-functions from the truncated conformal space}},}\ }\href {\doibase
  10.1007/JHEP02(2012)082} {\bibfield  {journal} {\bibinfo  {journal} {JHEP}\
  }\textbf {\bibinfo {volume} {2012}},\ \bibinfo {pages} {82} (\bibinfo {year}
  {2012})}\BibitemShut {NoStop}%
\bibitem [{\citenamefont {Kormos}(2007)}]{kormos2007boundary}%
  \BibitemOpen
  \bibfield  {author} {\bibinfo {author} {\bibfnamefont {M.}~\bibnamefont
  {Kormos}},\ }\bibfield  {title} {\enquote {\bibinfo {title} {Boundary
  renormalisation group flows of the supersymmetric lee--€"yang model and its
  extensions},}\ }\href {\doibase
  http://dx.doi.org/10.1016/j.nuclphysb.2007.02.028} {\bibfield  {journal}
  {\bibinfo  {journal} {Nucl. Phys. B}\ }\textbf {\bibinfo {volume} {772}},\
  \bibinfo {pages} {227 -- 248} (\bibinfo {year} {2007})}\BibitemShut {NoStop}%
\bibitem [{\citenamefont {Konik}\ \emph
  {et~al.}(2015{\natexlab{b}})\citenamefont {Konik}, \citenamefont {Sfeir},\
  and\ \citenamefont {Misewich}}]{konik2015predicting}%
  \BibitemOpen
  \bibfield  {author} {\bibinfo {author} {\bibfnamefont {R.~M.}\ \bibnamefont
  {Konik}}, \bibinfo {author} {\bibfnamefont {M.~Y.}\ \bibnamefont {Sfeir}}, \
  and\ \bibinfo {author} {\bibfnamefont {J.~A.}\ \bibnamefont {Misewich}},\
  }\bibfield  {title} {\enquote {\bibinfo {title} {Predicting excitonic gaps of
  semiconducting single-walled carbon nanotubes from a field theoretic
  analysis},}\ }\href {\doibase 10.1103/PhysRevB.91.075417} {\bibfield
  {journal} {\bibinfo  {journal} {Phys. Rev. B}\ }\textbf {\bibinfo {volume}
  {91}},\ \bibinfo {pages} {075417} (\bibinfo {year}
  {2015}{\natexlab{b}})}\BibitemShut {NoStop}%
\bibitem [{\citenamefont {Konik}(2011)}]{konik2011exciton}%
  \BibitemOpen
  \bibfield  {author} {\bibinfo {author} {\bibfnamefont {R.~M.}\ \bibnamefont
  {Konik}},\ }\bibfield  {title} {\enquote {\bibinfo {title} {{Exciton
  Hierarchies in Gapped Carbon Nanotubes}},}\ }\href {\doibase
  10.1103/PhysRevLett.106.136805} {\bibfield  {journal} {\bibinfo  {journal}
  {Phys. Rev. Lett.}\ }\textbf {\bibinfo {volume} {106}},\ \bibinfo {pages}
  {136805} (\bibinfo {year} {2011})}\BibitemShut {NoStop}%
\bibitem [{\citenamefont {Beria}\ \emph {et~al.}(2013)\citenamefont {Beria},
  \citenamefont {Brandino}, \citenamefont {Lepori}, \citenamefont {Konik},\
  and\ \citenamefont {Sierra}}]{beria2013truncated}%
  \BibitemOpen
  \bibfield  {author} {\bibinfo {author} {\bibfnamefont {M.}~\bibnamefont
  {Beria}}, \bibinfo {author} {\bibfnamefont {G.~P.}\ \bibnamefont {Brandino}},
  \bibinfo {author} {\bibfnamefont {L.}~\bibnamefont {Lepori}}, \bibinfo
  {author} {\bibfnamefont {R.~M.}\ \bibnamefont {Konik}}, \ and\ \bibinfo
  {author} {\bibfnamefont {G.}~\bibnamefont {Sierra}},\ }\bibfield  {title}
  {\enquote {\bibinfo {title} {Truncated conformal space approach for perturbed
  wess--€"zumino--€"witten models},}\ }\href {\doibase
  http://dx.doi.org/10.1016/j.nuclphysb.2013.10.005} {\bibfield  {journal}
  {\bibinfo  {journal} {Nucl. Phys. B}\ }\textbf {\bibinfo {volume} {877}},\
  \bibinfo {pages} {457 -- 483} (\bibinfo {year} {2013})}\BibitemShut {NoStop}%
\bibitem [{\citenamefont {Konik}\ \emph
  {et~al.}(2015{\natexlab{c}})\citenamefont {Konik}, \citenamefont {P\'almai},
  \citenamefont {Tak\'acs},\ and\ \citenamefont {Tsvelik}}]{konik2015studying}%
  \BibitemOpen
  \bibfield  {author} {\bibinfo {author} {\bibfnamefont {R.~M.}\ \bibnamefont
  {Konik}}, \bibinfo {author} {\bibfnamefont {T.}~\bibnamefont {P\'almai}},
  \bibinfo {author} {\bibfnamefont {G.}~\bibnamefont {Tak\'acs}}, \ and\
  \bibinfo {author} {\bibfnamefont {A.~M.}\ \bibnamefont {Tsvelik}},\
  }\bibfield  {title} {\enquote {\bibinfo {title} {{Studying the perturbed
  Wess--Zumino--Novikov--Witten theory using the truncated conformal spectrum
  approach}},}\ }\href {\doibase
  http://dx.doi.org/10.1016/j.nuclphysb.2015.08.016} {\bibfield  {journal}
  {\bibinfo  {journal} {Nucl. Phys. B}\ }\textbf {\bibinfo {volume} {899}},\
  \bibinfo {pages} {547 -- 569} (\bibinfo {year}
  {2015}{\natexlab{c}})}\BibitemShut {NoStop}%
\bibitem [{\citenamefont {Konik}\ and\ \citenamefont
  {Adamov}(2007)}]{konik2007numerical}%
  \BibitemOpen
  \bibfield  {author} {\bibinfo {author} {\bibfnamefont {R.~M.}\ \bibnamefont
  {Konik}}\ and\ \bibinfo {author} {\bibfnamefont {Y.}~\bibnamefont {Adamov}},\
  }\bibfield  {title} {\enquote {\bibinfo {title} {Numerical renormalization
  group for continuum one-dimensional systems},}\ }\href {\doibase
  10.1103/PhysRevLett.98.147205} {\bibfield  {journal} {\bibinfo  {journal}
  {Phys. Rev. Lett.}\ }\textbf {\bibinfo {volume} {98}},\ \bibinfo {pages}
  {147205} (\bibinfo {year} {2007})}\BibitemShut {NoStop}%
\bibitem [{\citenamefont {Feverati}\ \emph {et~al.}(2008)\citenamefont
  {Feverati}, \citenamefont {Graham}, \citenamefont {Pearce}, \citenamefont
  {T\'oth},\ and\ \citenamefont {Watts}}]{feverati2008renormalization}%
  \BibitemOpen
  \bibfield  {author} {\bibinfo {author} {\bibfnamefont {G.}~\bibnamefont
  {Feverati}}, \bibinfo {author} {\bibfnamefont {K.}~\bibnamefont {Graham}},
  \bibinfo {author} {\bibfnamefont {P.~A.}\ \bibnamefont {Pearce}}, \bibinfo
  {author} {\bibfnamefont {G.~Zs.}\ \bibnamefont {T\'oth}}, \ and\ \bibinfo
  {author} {\bibfnamefont {G.~M.~T.}\ \bibnamefont {Watts}},\ }\bibfield
  {title} {\enquote {\bibinfo {title} {{A renormalization group for the
  truncated conformal space approach}},}\ }\href
  {http://stacks.iop.org/1742-5468/2008/i=03/a=P03011} {\bibfield  {journal}
  {\bibinfo  {journal} {J. Stat. Mech.}\ }\textbf {\bibinfo {volume} {2008}},\
  \bibinfo {pages} {P03011} (\bibinfo {year} {2008})}\BibitemShut {NoStop}%
\bibitem [{\citenamefont {Watts}(2012)}]{watts2012renormalisation}%
  \BibitemOpen
  \bibfield  {author} {\bibinfo {author} {\bibfnamefont {G.~M.~T.}\
  \bibnamefont {Watts}},\ }\bibfield  {title} {\enquote {\bibinfo {title} {{On
  the renormalisation group for the boundary truncated conformal space
  approach}},}\ }\href {\doibase
  http://dx.doi.org/10.1016/j.nuclphysb.2012.01.012} {\bibfield  {journal}
  {\bibinfo  {journal} {Nucl. Phys. B}\ }\textbf {\bibinfo {volume} {859}},\
  \bibinfo {pages} {177 -- 206} (\bibinfo {year} {2012})}\BibitemShut {NoStop}%
\bibitem [{\citenamefont {{Giokas}}\ and\ \citenamefont
  {{Watts}}(2011)}]{giokas2011renormalisation}%
  \BibitemOpen
  \bibfield  {author} {\bibinfo {author} {\bibfnamefont {P.}~\bibnamefont
  {{Giokas}}}\ and\ \bibinfo {author} {\bibfnamefont {G.}~\bibnamefont
  {{Watts}}},\ }\bibfield  {title} {\enquote {\bibinfo {title} {{The
  renormalisation group for the truncated conformal space approach on the
  cylinder}},}\ }\href@noop {} {\bibfield  {journal} {\bibinfo  {journal}
  {ArXiv e-prints}\ } (\bibinfo {year} {2011})},\ \Eprint
  {http://arxiv.org/abs/1106.2448} {arXiv:1106.2448 [hep-th]} \BibitemShut
  {NoStop}%
\bibitem [{\citenamefont {Rychkov}\ and\ \citenamefont
  {Vitale}(2015)}]{rychkov2015hamiltonian}%
  \BibitemOpen
  \bibfield  {author} {\bibinfo {author} {\bibfnamefont {Slava}\ \bibnamefont
  {Rychkov}}\ and\ \bibinfo {author} {\bibfnamefont {Lorenzo~G.}\ \bibnamefont
  {Vitale}},\ }\bibfield  {title} {\enquote {\bibinfo {title} {Hamiltonian
  truncation study of the ${\ensuremath{\varphi}}^{4}$ theory in two
  dimensions},}\ }\href {\doibase 10.1103/PhysRevD.91.085011} {\bibfield
  {journal} {\bibinfo  {journal} {Phys. Rev. D}\ }\textbf {\bibinfo {volume}
  {91}},\ \bibinfo {pages} {085011} (\bibinfo {year} {2015})}\BibitemShut
  {NoStop}%
\bibitem [{\citenamefont {Coser}\ \emph {et~al.}(2014)\citenamefont {Coser},
  \citenamefont {Beria}, \citenamefont {Brandino}, \citenamefont {Konik},\ and\
  \citenamefont {Mussardo}}]{coser2014truncated}%
  \BibitemOpen
  \bibfield  {author} {\bibinfo {author} {\bibfnamefont {A.}~\bibnamefont
  {Coser}}, \bibinfo {author} {\bibfnamefont {M.}~\bibnamefont {Beria}},
  \bibinfo {author} {\bibfnamefont {G.~P.}\ \bibnamefont {Brandino}}, \bibinfo
  {author} {\bibfnamefont {R.~M.}\ \bibnamefont {Konik}}, \ and\ \bibinfo
  {author} {\bibfnamefont {G.}~\bibnamefont {Mussardo}},\ }\bibfield  {title}
  {\enquote {\bibinfo {title} {Truncated conformal space approach for 2d
  landau--ginzburg theories},}\ }\href {\doibase
  10.1088/1742-5468/2014/12/P12010} {\bibfield  {journal} {\bibinfo  {journal}
  {J. Stat. Mech.}\ }\textbf {\bibinfo {volume} {2014}},\ \bibinfo {pages}
  {P12010} (\bibinfo {year} {2014})}\BibitemShut {NoStop}%
\bibitem [{\citenamefont {Rychkov}\ and\ \citenamefont
  {Vitale}(2016)}]{rychkov2016hamiltonian}%
  \BibitemOpen
  \bibfield  {author} {\bibinfo {author} {\bibfnamefont {S.}~\bibnamefont
  {Rychkov}}\ and\ \bibinfo {author} {\bibfnamefont {L.~G.}\ \bibnamefont
  {Vitale}},\ }\bibfield  {title} {\enquote {\bibinfo {title} {{Hamiltonian
  truncation study of the ${\ensuremath{\phi}}^{4}$ theory in two dimensions.
  II. The ${\mathbb{Z}}_{2}$-broken phase and the Chang duality}},}\ }\href
  {\doibase 10.1103/PhysRevD.93.065014} {\bibfield  {journal} {\bibinfo
  {journal} {Phys. Rev. D}\ }\textbf {\bibinfo {volume} {93}},\ \bibinfo
  {pages} {065014} (\bibinfo {year} {2016})}\BibitemShut {NoStop}%
\bibitem [{\citenamefont {Bajnok}\ and\ \citenamefont
  {Lajer}(2016)}]{bajnok2016truncated}%
  \BibitemOpen
  \bibfield  {author} {\bibinfo {author} {\bibfnamefont {Z.}~\bibnamefont
  {Bajnok}}\ and\ \bibinfo {author} {\bibfnamefont {M.}~\bibnamefont {Lajer}},\
  }\bibfield  {title} {\enquote {\bibinfo {title} {{Truncated Hilbert space
  approach to the 2d $\phi^4$ theory}},}\ }\href {\doibase
  10.1007/JHEP10(2016)050} {\bibfield  {journal} {\bibinfo  {journal} {JHEP}\
  }\textbf {\bibinfo {volume} {2016}},\ \bibinfo {pages} {50} (\bibinfo {year}
  {2016})}\BibitemShut {NoStop}%
\bibitem [{\citenamefont {Elias-Mir{\'o}}\ \emph {et~al.}(2016)\citenamefont
  {Elias-Mir{\'o}}, \citenamefont {Montull},\ and\ \citenamefont
  {Riembau}}]{Elias-Miro2016renormalized}%
  \BibitemOpen
  \bibfield  {author} {\bibinfo {author} {\bibfnamefont {J.}~\bibnamefont
  {Elias-Mir{\'o}}}, \bibinfo {author} {\bibfnamefont {M.}~\bibnamefont
  {Montull}}, \ and\ \bibinfo {author} {\bibfnamefont {M.}~\bibnamefont
  {Riembau}},\ }\bibfield  {title} {\enquote {\bibinfo {title} {{The
  renormalized Hamiltonian truncation method in the large $E_T$ expansion}},}\
  }\href {\doibase 10.1007/JHEP04(2016)144} {\bibfield  {journal} {\bibinfo
  {journal} {JHEP}\ }\textbf {\bibinfo {volume} {2016}},\ \bibinfo {pages}
  {144} (\bibinfo {year} {2016})}\BibitemShut {NoStop}%
\bibitem [{\citenamefont {Wu}\ \emph {et~al.}(2014)\citenamefont {Wu},
  \citenamefont {Estienne}, \citenamefont {Regnault},\ and\ \citenamefont
  {Bernevig}}]{wu2014braiding}%
  \BibitemOpen
  \bibfield  {author} {\bibinfo {author} {\bibfnamefont {Y.-L.}\ \bibnamefont
  {Wu}}, \bibinfo {author} {\bibfnamefont {B.}~\bibnamefont {Estienne}},
  \bibinfo {author} {\bibfnamefont {N.}~\bibnamefont {Regnault}}, \ and\
  \bibinfo {author} {\bibfnamefont {B.~A.}\ \bibnamefont {Bernevig}},\
  }\bibfield  {title} {\enquote {\bibinfo {title} {{Braiding Non-Abelian
  Quasiholes in Fractional Quantum Hall States}},}\ }\href {\doibase
  10.1103/PhysRevLett.113.116801} {\bibfield  {journal} {\bibinfo  {journal}
  {Phys. Rev. Lett.}\ }\textbf {\bibinfo {volume} {113}},\ \bibinfo {pages}
  {116801} (\bibinfo {year} {2014})}\BibitemShut {NoStop}%
\bibitem [{\citenamefont {Wu}\ \emph {et~al.}(2015)\citenamefont {Wu},
  \citenamefont {Estienne}, \citenamefont {Regnault},\ and\ \citenamefont
  {Bernevig}}]{wu2015matrix}%
  \BibitemOpen
  \bibfield  {author} {\bibinfo {author} {\bibfnamefont {Y.-L.}\ \bibnamefont
  {Wu}}, \bibinfo {author} {\bibfnamefont {B.}~\bibnamefont {Estienne}},
  \bibinfo {author} {\bibfnamefont {N.}~\bibnamefont {Regnault}}, \ and\
  \bibinfo {author} {\bibfnamefont {B.~A.}\ \bibnamefont {Bernevig}},\
  }\bibfield  {title} {\enquote {\bibinfo {title} {{Matrix product state
  representation of non-Abelian quasiholes}},}\ }\href {\doibase
  10.1103/PhysRevB.92.045109} {\bibfield  {journal} {\bibinfo  {journal} {Phys.
  Rev. B}\ }\textbf {\bibinfo {volume} {92}},\ \bibinfo {pages} {045109}
  (\bibinfo {year} {2015})}\BibitemShut {NoStop}%
\bibitem [{\citenamefont {Zaletel}\ and\ \citenamefont
  {Mong}(2012)}]{zaletel2012exact}%
  \BibitemOpen
  \bibfield  {author} {\bibinfo {author} {\bibfnamefont {Michael~P.}\
  \bibnamefont {Zaletel}}\ and\ \bibinfo {author} {\bibfnamefont {Roger S.~K.}\
  \bibnamefont {Mong}},\ }\bibfield  {title} {\enquote {\bibinfo {title} {Exact
  matrix product states for quantum hall wave functions},}\ }\href {\doibase
  10.1103/PhysRevB.86.245305} {\bibfield  {journal} {\bibinfo  {journal} {Phys.
  Rev. B}\ }\textbf {\bibinfo {volume} {86}},\ \bibinfo {pages} {245305}
  (\bibinfo {year} {2012})}\BibitemShut {NoStop}%
\bibitem [{\citenamefont {P\'almai}(2014)}]{palmai2014excited}%
  \BibitemOpen
  \bibfield  {author} {\bibinfo {author} {\bibfnamefont {T.}~\bibnamefont
  {P\'almai}},\ }\bibfield  {title} {\enquote {\bibinfo {title} {{Excited state
  entanglement in one-dimensional quantum critical systems: Extensivity and the
  role of microscopic details}},}\ }\href {\doibase 10.1103/PhysRevB.90.161404}
  {\bibfield  {journal} {\bibinfo  {journal} {Phys. Rev. B}\ }\textbf {\bibinfo
  {volume} {90}},\ \bibinfo {pages} {161404} (\bibinfo {year}
  {2014})}\BibitemShut {NoStop}%
\bibitem [{\citenamefont {Palmai}(2016)}]{palmai2016entanglement}%
  \BibitemOpen
  \bibfield  {author} {\bibinfo {author} {\bibfnamefont {T.}~\bibnamefont
  {Palmai}},\ }\bibfield  {title} {\enquote {\bibinfo {title} {Entanglement
  entropy from the truncated conformal space},}\ }\href {\doibase
  http://dx.doi.org/10.1016/j.physletb.2016.06.012} {\bibfield  {journal}
  {\bibinfo  {journal} {Phys. Lett. B}\ }\textbf {\bibinfo {volume} {759}},\
  \bibinfo {pages} {439 -- 445} (\bibinfo {year} {2016})}\BibitemShut {NoStop}%
\bibitem [{\citenamefont {{Katz}}\ \emph {et~al.}(2014)\citenamefont {{Katz}},
  \citenamefont {{Marques Tavares}},\ and\ \citenamefont
  {{Xu}}}]{katz2014solution}%
  \BibitemOpen
  \bibfield  {author} {\bibinfo {author} {\bibfnamefont {E.}~\bibnamefont
  {{Katz}}}, \bibinfo {author} {\bibfnamefont {G.}~\bibnamefont {{Marques
  Tavares}}}, \ and\ \bibinfo {author} {\bibfnamefont {Y.}~\bibnamefont
  {{Xu}}},\ }\bibfield  {title} {\enquote {\bibinfo {title} {{A solution of 2D
  QCD at Finite $N$ using a conformal basis}},}\ }\href@noop {} {\bibfield
  {journal} {\bibinfo  {journal} {ArXiv e-prints}\ } (\bibinfo {year}
  {2014})},\ \Eprint {http://arxiv.org/abs/1405.6727} {arXiv:1405.6727
  [hep-th]} \BibitemShut {NoStop}%
\bibitem [{\citenamefont {Rakovszky}\ \emph {et~al.}(2016)\citenamefont
  {Rakovszky}, \citenamefont {Mesty\'an}, \citenamefont {Collura},
  \citenamefont {Kormos},\ and\ \citenamefont
  {Tak\'acs}}]{rakovszky2016hamiltonian}%
  \BibitemOpen
  \bibfield  {author} {\bibinfo {author} {\bibfnamefont {T.}~\bibnamefont
  {Rakovszky}}, \bibinfo {author} {\bibfnamefont {M.}~\bibnamefont
  {Mesty\'an}}, \bibinfo {author} {\bibfnamefont {M.}~\bibnamefont {Collura}},
  \bibinfo {author} {\bibfnamefont {M.}~\bibnamefont {Kormos}}, \ and\ \bibinfo
  {author} {\bibfnamefont {G.}~\bibnamefont {Tak\'acs}},\ }\bibfield  {title}
  {\enquote {\bibinfo {title} {{Hamiltonian truncation approach to quenches in
  the Ising field theory}},}\ }\href {\doibase
  http://dx.doi.org/10.1016/j.nuclphysb.2016.08.024} {\bibfield  {journal}
  {\bibinfo  {journal} {Nucl. Phys. B}\ }\textbf {\bibinfo {volume} {911}},\
  \bibinfo {pages} {805 -- 845} (\bibinfo {year} {2016})}\BibitemShut {NoStop}%
\bibitem [{\citenamefont {Caux}\ and\ \citenamefont
  {Konik}(2012)}]{caux2012constructing}%
  \BibitemOpen
  \bibfield  {author} {\bibinfo {author} {\bibfnamefont {J.-S.}\ \bibnamefont
  {Caux}}\ and\ \bibinfo {author} {\bibfnamefont {R.~M.}\ \bibnamefont
  {Konik}},\ }\bibfield  {title} {\enquote {\bibinfo {title} {{Constructing the
  Generalized Gibbs Ensemble after a Quantum Quench}},}\ }\href {\doibase
  10.1103/PhysRevLett.109.175301} {\bibfield  {journal} {\bibinfo  {journal}
  {Phys. Rev. Lett.}\ }\textbf {\bibinfo {volume} {109}},\ \bibinfo {pages}
  {175301} (\bibinfo {year} {2012})}\BibitemShut {NoStop}%
\bibitem [{\citenamefont {Brandino}\ \emph {et~al.}(2015)\citenamefont
  {Brandino}, \citenamefont {Caux},\ and\ \citenamefont
  {Konik}}]{brandino2015glimmers}%
  \BibitemOpen
  \bibfield  {author} {\bibinfo {author} {\bibfnamefont {G.~P.}\ \bibnamefont
  {Brandino}}, \bibinfo {author} {\bibfnamefont {J.-S.}\ \bibnamefont {Caux}},
  \ and\ \bibinfo {author} {\bibfnamefont {R.~M.}\ \bibnamefont {Konik}},\
  }\bibfield  {title} {\enquote {\bibinfo {title} {{Glimmers of a Quantum KAM
  Theorem: Insights from Quantum Quenches in One-Dimensional Bose Gases}},}\
  }\href {\doibase 10.1103/PhysRevX.5.041043} {\bibfield  {journal} {\bibinfo
  {journal} {Phys. Rev. X}\ }\textbf {\bibinfo {volume} {5}},\ \bibinfo {pages}
  {041043} (\bibinfo {year} {2015})}\BibitemShut {NoStop}%
\bibitem [{\citenamefont {Hogervorst}\ \emph {et~al.}(2015)\citenamefont
  {Hogervorst}, \citenamefont {Rychkov},\ and\ \citenamefont {van
  Rees}}]{hogervorst2015truncated}%
  \BibitemOpen
  \bibfield  {author} {\bibinfo {author} {\bibfnamefont {M.}~\bibnamefont
  {Hogervorst}}, \bibinfo {author} {\bibfnamefont {S.}~\bibnamefont {Rychkov}},
  \ and\ \bibinfo {author} {\bibfnamefont {B.~C.}\ \bibnamefont {van Rees}},\
  }\bibfield  {title} {\enquote {\bibinfo {title} {{Truncated conformal space
  approach in $d$ dimensions: A cheap alternative to lattice field theory?}}}\
  }\href {\doibase 10.1103/PhysRevD.91.025005} {\bibfield  {journal} {\bibinfo
  {journal} {Phys. Rev. D}\ }\textbf {\bibinfo {volume} {91}},\ \bibinfo
  {pages} {025005} (\bibinfo {year} {2015})}\BibitemShut {NoStop}%
\bibitem [{\citenamefont {Katz}\ \emph {et~al.}(2016)\citenamefont {Katz},
  \citenamefont {Khandker},\ and\ \citenamefont {Walters}}]{katz2016conformal}%
  \BibitemOpen
  \bibfield  {author} {\bibinfo {author} {\bibfnamefont {E.}~\bibnamefont
  {Katz}}, \bibinfo {author} {\bibfnamefont {Z.~U.}\ \bibnamefont {Khandker}},
  \ and\ \bibinfo {author} {\bibfnamefont {M.~T.}\ \bibnamefont {Walters}},\
  }\bibfield  {title} {\enquote {\bibinfo {title} {A conformal truncation
  framework for infinite-volume dynamics},}\ }\href {\doibase
  10.1007/JHEP07(2016)140} {\bibfield  {journal} {\bibinfo  {journal} {JHEP}\
  }\textbf {\bibinfo {volume} {2016}},\ \bibinfo {pages} {140} (\bibinfo {year}
  {2016})}\BibitemShut {NoStop}%
\bibitem [{\citenamefont {Smirnov}(1992)}]{smirnov1992form}%
  \BibitemOpen
  \bibfield  {author} {\bibinfo {author} {\bibfnamefont {F.~A.}\ \bibnamefont
  {Smirnov}},\ }\href@noop {} {\emph {\bibinfo {title} {Form factors in
  completely integrable models of quantum field theory}}},\ Vol.~\bibinfo
  {volume} {14}\ (\bibinfo  {publisher} {World Scientific},\ \bibinfo {year}
  {1992})\BibitemShut {NoStop}%
\bibitem [{\citenamefont {Fonseca}\ and\ \citenamefont
  {Zamolodchikov}(2003)}]{fonseca2003ising}%
  \BibitemOpen
  \bibfield  {author} {\bibinfo {author} {\bibfnamefont {P.}~\bibnamefont
  {Fonseca}}\ and\ \bibinfo {author} {\bibfnamefont {A.}~\bibnamefont
  {Zamolodchikov}},\ }\bibfield  {title} {\enquote {\bibinfo {title} {{Ising
  Field Theory in a Magnetic Field: Analytic Properties of the Free Energy}},}\
  }\href {\doibase 10.1023/A:1022147532606} {\bibfield  {journal} {\bibinfo
  {journal} {J. Stat. Phys.}\ }\textbf {\bibinfo {volume} {110}},\ \bibinfo
  {pages} {527--590} (\bibinfo {year} {2003})}\BibitemShut {NoStop}%
\bibitem [{\citenamefont {Onsager}(1944)}]{onsager1944crystal}%
  \BibitemOpen
  \bibfield  {author} {\bibinfo {author} {\bibfnamefont {L.}~\bibnamefont
  {Onsager}},\ }\bibfield  {title} {\enquote {\bibinfo {title} {Crystal
  statistics. i. a two-dimensional model with an order-disorder transition},}\
  }\href {\doibase 10.1103/PhysRev.65.117} {\bibfield  {journal} {\bibinfo
  {journal} {Phys. Rev.}\ }\textbf {\bibinfo {volume} {65}},\ \bibinfo {pages}
  {117--149} (\bibinfo {year} {1944})}\BibitemShut {NoStop}%
\bibitem [{\citenamefont {Bugrii}(2001)}]{bugrii2001correlation}%
  \BibitemOpen
  \bibfield  {author} {\bibinfo {author} {\bibfnamefont {A.~I.}\ \bibnamefont
  {Bugrii}},\ }\bibfield  {title} {\enquote {\bibinfo {title} {{Correlation
  Function of the Two-Dimensional Ising Model on a Finite Lattice: I}},}\
  }\href {\doibase 10.1023/A:1010320126700} {\bibfield  {journal} {\bibinfo
  {journal} {Theor. Math. Phys.}\ }\textbf {\bibinfo {volume} {127}},\ \bibinfo
  {pages} {528--548} (\bibinfo {year} {2001})}\BibitemShut {NoStop}%
\bibitem [{\citenamefont {Sachdev}(1996)}]{sachdev1996universal}%
  \BibitemOpen
  \bibfield  {author} {\bibinfo {author} {\bibfnamefont {S.}~\bibnamefont
  {Sachdev}},\ }\bibfield  {title} {\enquote {\bibinfo {title} {{Universal,
  finite-temperature, crossover functions of the quantum transition in the
  Ising chain in a transverse field}},}\ }\href {\doibase
  http://dx.doi.org/10.1016/0550-3213(95)00657-5} {\bibfield  {journal}
  {\bibinfo  {journal} {Nucl. Phys. B}\ }\textbf {\bibinfo {volume} {464}},\
  \bibinfo {pages} {576 -- 595} (\bibinfo {year} {1996})}\BibitemShut {NoStop}%
\bibitem [{\citenamefont {McCoy}\ and\ \citenamefont
  {Wu}(1978)}]{mccoy1978two}%
  \BibitemOpen
  \bibfield  {author} {\bibinfo {author} {\bibfnamefont {B.~M.}\ \bibnamefont
  {McCoy}}\ and\ \bibinfo {author} {\bibfnamefont {T.~T.}\ \bibnamefont {Wu}},\
  }\bibfield  {title} {\enquote {\bibinfo {title} {{Two-dimensional Ising field
  theory in a magnetic field: Breakup of the cut in the two-point function}},}\
  }\href {\doibase 10.1103/PhysRevD.18.1259} {\bibfield  {journal} {\bibinfo
  {journal} {Phys. Rev. D}\ }\textbf {\bibinfo {volume} {18}},\ \bibinfo
  {pages} {1259--1267} (\bibinfo {year} {1978})}\BibitemShut {NoStop}%
\bibitem [{\citenamefont {Rutkevich}(2010)}]{rutkevich2009twokink}%
  \BibitemOpen
  \bibfield  {author} {\bibinfo {author} {\bibfnamefont {S.~B.}\ \bibnamefont
  {Rutkevich}},\ }\bibfield  {title} {\enquote {\bibinfo {title} {{Two-kink
  bound states in the magnetically perturbed Potts field theory at $T <
  T_c$}},}\ }\href {http://stacks.iop.org/1751-8121/43/i=23/a=235004}
  {\bibfield  {journal} {\bibinfo  {journal} {J. Phys. A}\ }\textbf {\bibinfo
  {volume} {43}},\ \bibinfo {pages} {235004} (\bibinfo {year}
  {2010})}\BibitemShut {NoStop}%
\bibitem [{\citenamefont {Rutkevich}(2015)}]{rutkevich2015baryon}%
  \BibitemOpen
  \bibfield  {author} {\bibinfo {author} {\bibfnamefont {S.~B.}\ \bibnamefont
  {Rutkevich}},\ }\bibfield  {title} {\enquote {\bibinfo {title} {{Baryon
  masses in the three-state Potts field theory in a weak magnetic field}},}\
  }\href {http://stacks.iop.org/1742-5468/2015/i=1/a=P01010} {\bibfield
  {journal} {\bibinfo  {journal} {J. Stat. Mech.}\ }\textbf {\bibinfo {volume}
  {2015}},\ \bibinfo {pages} {P01010} (\bibinfo {year} {2015})}\BibitemShut
  {NoStop}%
\bibitem [{\citenamefont {{Zamolodchikov}}(1989)}]{zamolodchikov1989integrals}%
  \BibitemOpen
  \bibfield  {author} {\bibinfo {author} {\bibfnamefont {A.~B.}\ \bibnamefont
  {{Zamolodchikov}}},\ }\bibfield  {title} {\enquote {\bibinfo {title}
  {{Integrals of Motion and S-Matrix of the (scaled) T = T$_{c}$ Ising Model
  with Magnetic Field}},}\ }\href {\doibase 10.1142/S0217751X8900176X}
  {\bibfield  {journal} {\bibinfo  {journal} {Int. J. Mod. Phys. A}\ }\textbf
  {\bibinfo {volume} {4}},\ \bibinfo {pages} {4235--4248} (\bibinfo {year}
  {1989})}\BibitemShut {NoStop}%
\bibitem [{\citenamefont {Fateev}(1994)}]{fateev1994exact}%
  \BibitemOpen
  \bibfield  {author} {\bibinfo {author} {\bibfnamefont {V.~A.}\ \bibnamefont
  {Fateev}},\ }\bibfield  {title} {\enquote {\bibinfo {title} {{The exact
  relations between the coupling constants and the masses of particles for the
  integrable perturbed conformal field theories}},}\ }\href {\doibase
  http://dx.doi.org/10.1016/0370-2693(94)00078-6} {\bibfield  {journal}
  {\bibinfo  {journal} {Phys. Lett. B}\ }\textbf {\bibinfo {volume} {324}},\
  \bibinfo {pages} {45 -- 51} (\bibinfo {year} {1994})}\BibitemShut {NoStop}%
\bibitem [{\citenamefont {Coldea}\ \emph
  {et~al.}(2010{\natexlab{b}})\citenamefont {Coldea}, \citenamefont {Tennant},
  \citenamefont {Wheeler}, \citenamefont {Wawrzynska}, \citenamefont
  {Prabhakaran}, \citenamefont {Telling}, \citenamefont {Habicht},
  \citenamefont {Smeibidl},\ and\ \citenamefont {Kiefer}}]{coldea2010quantum}%
  \BibitemOpen
  \bibfield  {author} {\bibinfo {author} {\bibfnamefont {R.}~\bibnamefont
  {Coldea}}, \bibinfo {author} {\bibfnamefont {D.~A.}\ \bibnamefont {Tennant}},
  \bibinfo {author} {\bibfnamefont {E.~M.}\ \bibnamefont {Wheeler}}, \bibinfo
  {author} {\bibfnamefont {E.}~\bibnamefont {Wawrzynska}}, \bibinfo {author}
  {\bibfnamefont {D.}~\bibnamefont {Prabhakaran}}, \bibinfo {author}
  {\bibfnamefont {M.}~\bibnamefont {Telling}}, \bibinfo {author} {\bibfnamefont
  {K.}~\bibnamefont {Habicht}}, \bibinfo {author} {\bibfnamefont
  {P.}~\bibnamefont {Smeibidl}}, \ and\ \bibinfo {author} {\bibfnamefont
  {K.}~\bibnamefont {Kiefer}},\ }\bibfield  {title} {\enquote {\bibinfo {title}
  {{Quantum Criticality in an Ising Chain: Experimental Evidence for Emergent
  E8 Symmetry}},}\ }\href {\doibase 10.1126/science.1180085} {\bibfield
  {journal} {\bibinfo  {journal} {Science}\ }\textbf {\bibinfo {volume}
  {327}},\ \bibinfo {pages} {177--180} (\bibinfo {year}
  {2010}{\natexlab{b}})}\BibitemShut {NoStop}%
\bibitem [{\citenamefont {Morris}\ \emph {et~al.}(2014)\citenamefont {Morris},
  \citenamefont {Vald\'es~Aguilar}, \citenamefont {Ghosh}, \citenamefont
  {Koohpayeh}, \citenamefont {Krizan}, \citenamefont {Cava}, \citenamefont
  {Tchernyshyov}, \citenamefont {McQueen},\ and\ \citenamefont
  {Armitage}}]{morris2014hierarchy}%
  \BibitemOpen
  \bibfield  {author} {\bibinfo {author} {\bibfnamefont {C.~M.}\ \bibnamefont
  {Morris}}, \bibinfo {author} {\bibfnamefont {R.}~\bibnamefont
  {Vald\'es~Aguilar}}, \bibinfo {author} {\bibfnamefont {A.}~\bibnamefont
  {Ghosh}}, \bibinfo {author} {\bibfnamefont {S.~M.}\ \bibnamefont
  {Koohpayeh}}, \bibinfo {author} {\bibfnamefont {J.}~\bibnamefont {Krizan}},
  \bibinfo {author} {\bibfnamefont {R.~J.}\ \bibnamefont {Cava}}, \bibinfo
  {author} {\bibfnamefont {O.}~\bibnamefont {Tchernyshyov}}, \bibinfo {author}
  {\bibfnamefont {T.~M.}\ \bibnamefont {McQueen}}, \ and\ \bibinfo {author}
  {\bibfnamefont {N.~P.}\ \bibnamefont {Armitage}},\ }\bibfield  {title}
  {\enquote {\bibinfo {title} {{Hierarchy of Bound States in the
  One-Dimensional Ferromagnetic Ising Chain
  ${\mathrm{CoNb}}_{2}{\mathrm{O}}_{6}$ Investigated by High-Resolution
  Time-Domain Terahertz Spectroscopy}},}\ }\href {\doibase
  10.1103/PhysRevLett.112.137403} {\bibfield  {journal} {\bibinfo  {journal}
  {Phys. Rev. Lett.}\ }\textbf {\bibinfo {volume} {112}},\ \bibinfo {pages}
  {137403} (\bibinfo {year} {2014})}\BibitemShut {NoStop}%
\bibitem [{\citenamefont {Brandino}\ and\ \citenamefont
  {Konik}(2017)}]{TruSpace}%
  \BibitemOpen
  \bibfield  {author} {\bibinfo {author} {\bibfnamefont {G.~P.}\ \bibnamefont
  {Brandino}}\ and\ \bibinfo {author} {\bibfnamefont {R.~M.}\ \bibnamefont
  {Konik}},\ }\href@noop {} {\enquote {\bibinfo {title} {Truspace: The
  truncated conformal spectrum approach for perturbations of the conformal
  minimal models},}\ }\bibinfo {howpublished}
  {\url{https://bitbucket.org/truspacedevelopers/truspace}} (\bibinfo {year}
  {2017})\BibitemShut {NoStop}%
\bibitem [{\citenamefont {L\'assig}\ and\ \citenamefont
  {Mussardo}(1991)}]{lassig1991hilbert}%
  \BibitemOpen
  \bibfield  {author} {\bibinfo {author} {\bibfnamefont {M.}~\bibnamefont
  {L\'assig}}\ and\ \bibinfo {author} {\bibfnamefont {G.}~\bibnamefont
  {Mussardo}},\ }\bibfield  {title} {\enquote {\bibinfo {title} {Hilbert space
  and structure constants of descendant fields in two-dimensional conformal
  theories},}\ }\href {\doibase http://dx.doi.org/10.1016/0010-4655(91)90009-A}
  {\bibfield  {journal} {\bibinfo  {journal} {Comp. Phys. Comm.}\ }\textbf
  {\bibinfo {volume} {66}},\ \bibinfo {pages} {71 -- 88} (\bibinfo {year}
  {1991})}\BibitemShut {NoStop}%
\bibitem [{\citenamefont {Fioravanti}\ \emph
  {et~al.}(2000{\natexlab{b}})\citenamefont {Fioravanti}, \citenamefont
  {Mussardo},\ and\ \citenamefont {Simon}}]{fioravanti2000universal1}%
  \BibitemOpen
  \bibfield  {author} {\bibinfo {author} {\bibfnamefont {D.}~\bibnamefont
  {Fioravanti}}, \bibinfo {author} {\bibfnamefont {G.}~\bibnamefont
  {Mussardo}}, \ and\ \bibinfo {author} {\bibfnamefont {P.}~\bibnamefont
  {Simon}},\ }\bibfield  {title} {\enquote {\bibinfo {title} {{Universal
  amplitude ratios of the renormalization group: Two-dimensional tricritical
  Ising model}},}\ }\href {\doibase 10.1103/PhysRevE.63.016103} {\bibfield
  {journal} {\bibinfo  {journal} {Phys. Rev. E}\ }\textbf {\bibinfo {volume}
  {63}},\ \bibinfo {pages} {016103} (\bibinfo {year}
  {2000}{\natexlab{b}})}\BibitemShut {NoStop}%
\bibitem [{\citenamefont {Mussardo}\ \emph {et~al.}(1987)\citenamefont
  {Mussardo}, \citenamefont {Sotkov},\ and\ \citenamefont
  {Stanishkov}}]{mussardo1987ramond}%
  \BibitemOpen
  \bibfield  {author} {\bibinfo {author} {\bibfnamefont {G.}~\bibnamefont
  {Mussardo}}, \bibinfo {author} {\bibfnamefont {G.}~\bibnamefont {Sotkov}}, \
  and\ \bibinfo {author} {\bibfnamefont {M.}~\bibnamefont {Stanishkov}},\
  }\bibfield  {title} {\enquote {\bibinfo {title} {Ramond sector of the
  supersymmetric minimal models},}\ }\href {\doibase
  http://dx.doi.org/10.1016/0370-2693(87)90038-4} {\bibfield  {journal}
  {\bibinfo  {journal} {Phys. Lett. B}\ }\textbf {\bibinfo {volume} {195}},\
  \bibinfo {pages} {397 -- 406} (\bibinfo {year} {1987})}\BibitemShut {NoStop}%
\bibitem [{\citenamefont {Christe}\ and\ \citenamefont
  {Mussardo}(1990)}]{christe1990integrable}%
  \BibitemOpen
  \bibfield  {author} {\bibinfo {author} {\bibfnamefont {P.}~\bibnamefont
  {Christe}}\ and\ \bibinfo {author} {\bibfnamefont {G.}~\bibnamefont
  {Mussardo}},\ }\bibfield  {title} {\enquote {\bibinfo {title} {{Integrable
  systems away from critically: The Toda field theory and S-matrix of the
  tricritical Ising model}},}\ }\href {\doibase
  http://dx.doi.org/10.1016/0550-3213(90)90119-X} {\bibfield  {journal}
  {\bibinfo  {journal} {Nucl. Phys. B}\ }\textbf {\bibinfo {volume} {330}},\
  \bibinfo {pages} {465 -- 487} (\bibinfo {year} {1990})}\BibitemShut {NoStop}%
\bibitem [{\citenamefont {Fendley}(1990)}]{fendley1990second}%
  \BibitemOpen
  \bibfield  {author} {\bibinfo {author} {\bibfnamefont {P.}~\bibnamefont
  {Fendley}},\ }\bibfield  {title} {\enquote {\bibinfo {title} {{A second
  supersymmetric S-matrix for the perturbed tricritical Ising model}},}\ }\href
  {\doibase http://dx.doi.org/10.1016/0370-2693(90)91160-D} {\bibfield
  {journal} {\bibinfo  {journal} {Phys. Lett. B}\ }\textbf {\bibinfo {volume}
  {250}},\ \bibinfo {pages} {96 -- 101} (\bibinfo {year} {1990})}\BibitemShut
  {NoStop}%
\bibitem [{\citenamefont {Henkel}(1990)}]{henkel1990mass}%
  \BibitemOpen
  \bibfield  {author} {\bibinfo {author} {\bibfnamefont {M.}~\bibnamefont
  {Henkel}},\ }\bibfield  {title} {\enquote {\bibinfo {title} {{Mass spectrum
  of the two-dimensional tricritical Ising model in an external magnetic
  field}},}\ }\href {\doibase http://dx.doi.org/10.1016/0370-2693(90)91902-N}
  {\bibfield  {journal} {\bibinfo  {journal} {Phys. Lett. B}\ }\textbf
  {\bibinfo {volume} {247}},\ \bibinfo {pages} {567 -- 570} (\bibinfo {year}
  {1990})}\BibitemShut {NoStop}%
\bibitem [{\citenamefont
  {Zamolodchikov}(1991{\natexlab{a}})}]{zamolodchikov1991tricritical}%
  \BibitemOpen
  \bibfield  {author} {\bibinfo {author} {\bibfnamefont {Al.~B.}\ \bibnamefont
  {Zamolodchikov}},\ }\bibfield  {title} {\enquote {\bibinfo {title} {{From
  tricritical Ising to critical Ising by thermodynamic Bethe ansatz}},}\ }\href
  {\doibase http://dx.doi.org/10.1016/0550-3213(91)90423-U} {\bibfield
  {journal} {\bibinfo  {journal} {Nucl. Phys. B}\ }\textbf {\bibinfo {volume}
  {358}},\ \bibinfo {pages} {524 -- 546} (\bibinfo {year}
  {1991}{\natexlab{a}})}\BibitemShut {NoStop}%
\bibitem [{\citenamefont
  {Zamolodchikov}(1991{\natexlab{b}})}]{zamolodchikov1991thermodynamic}%
  \BibitemOpen
  \bibfield  {author} {\bibinfo {author} {\bibfnamefont {Al.~B.}\ \bibnamefont
  {Zamolodchikov}},\ }\bibfield  {title} {\enquote {\bibinfo {title} {{On the
  thermodynamic Bethe ansatz equations for reflectionless ADE scattering
  theories}},}\ }\href {\doibase
  http://dx.doi.org/10.1016/0370-2693(91)91737-G} {\bibfield  {journal}
  {\bibinfo  {journal} {Phys. Lett. B}\ }\textbf {\bibinfo {volume} {253}},\
  \bibinfo {pages} {391 -- 394} (\bibinfo {year}
  {1991}{\natexlab{b}})}\BibitemShut {NoStop}%
\bibitem [{\citenamefont {Acerbi}\ \emph {et~al.}(1996)\citenamefont {Acerbi},
  \citenamefont {Valleriani},\ and\ \citenamefont {Mussardo}}]{acerbi1996form}%
  \BibitemOpen
  \bibfield  {author} {\bibinfo {author} {\bibfnamefont {C.}~\bibnamefont
  {Acerbi}}, \bibinfo {author} {\bibfnamefont {A.}~\bibnamefont {Valleriani}},
  \ and\ \bibinfo {author} {\bibfnamefont {G.}~\bibnamefont {Mussardo}},\
  }\bibfield  {title} {\enquote {\bibinfo {title} {Form factors and correlation
  functions of the stress-energy tensor in massive deformation of the minimal
  models $({E}_n)_1 \otimes ({E}_n)_1/({E}_n)_2$},}\ }\href {\doibase
  10.1142/S0217751X96002443} {\bibfield  {journal} {\bibinfo  {journal} {Int.
  J. Mod. Phys. A}\ }\textbf {\bibinfo {volume} {11}},\ \bibinfo {pages}
  {5327--5364} (\bibinfo {year} {1996})}\BibitemShut {NoStop}%
\bibitem [{\citenamefont {Guida}\ and\ \citenamefont
  {Magnoli}(1998)}]{guida1998tricritical}%
  \BibitemOpen
  \bibfield  {author} {\bibinfo {author} {\bibfnamefont {R.}~\bibnamefont
  {Guida}}\ and\ \bibinfo {author} {\bibfnamefont {N.}~\bibnamefont
  {Magnoli}},\ }\bibfield  {title} {\enquote {\bibinfo {title} {Tricritical
  ising model near criticality},}\ }\href {\doibase 10.1142/S0217751X98000512}
  {\bibfield  {journal} {\bibinfo  {journal} {Int. J. Mod. Phys. A}\ }\textbf
  {\bibinfo {volume} {13}},\ \bibinfo {pages} {1145--1157} (\bibinfo {year}
  {1998})}\BibitemShut {NoStop}%
\bibitem [{\citenamefont {Blume}\ \emph {et~al.}(1971)\citenamefont {Blume},
  \citenamefont {Emery},\ and\ \citenamefont {Griffiths}}]{blume1971ising}%
  \BibitemOpen
  \bibfield  {author} {\bibinfo {author} {\bibfnamefont {M.}~\bibnamefont
  {Blume}}, \bibinfo {author} {\bibfnamefont {V.~J.}\ \bibnamefont {Emery}}, \
  and\ \bibinfo {author} {\bibfnamefont {R.~B.}\ \bibnamefont {Griffiths}},\
  }\bibfield  {title} {\enquote {\bibinfo {title} {{Ising Model for the
  $\ensuremath{\lambda}$ Transition and Phase Separation in
  ${\mathrm{He}}^{3}$-${\mathrm{He}}^{4}$ Mixtures}},}\ }\href {\doibase
  10.1103/PhysRevA.4.1071} {\bibfield  {journal} {\bibinfo  {journal} {Phys.
  Rev. A}\ }\textbf {\bibinfo {volume} {4}},\ \bibinfo {pages} {1071--1077}
  (\bibinfo {year} {1971})}\BibitemShut {NoStop}%
\bibitem [{\citenamefont {Zamolodchikov}(1986)}]{zamolodchikov1986conformal}%
  \BibitemOpen
  \bibfield  {author} {\bibinfo {author} {\bibfnamefont {A.~B.}\ \bibnamefont
  {Zamolodchikov}},\ }\bibfield  {title} {\enquote {\bibinfo {title}
  {{Conformal Symmetry and Multicritical Points in Two-Dimensional Quantum
  Field Theory}},}\ }\href@noop {} {\bibfield  {journal} {\bibinfo  {journal}
  {Sov. J. Nucl. Phys.}\ }\textbf {\bibinfo {volume} {44}},\ \bibinfo {pages}
  {529--533} (\bibinfo {year} {1986})},\ \bibinfo {note} {[Yad. Fiz.
  \textbf{44}, 821(1986)]}\BibitemShut {NoStop}%
\bibitem [{\citenamefont {Gefen}\ \emph {et~al.}(1981)\citenamefont {Gefen},
  \citenamefont {Imry},\ and\ \citenamefont {Mukamel}}]{gefen1981phase}%
  \BibitemOpen
  \bibfield  {author} {\bibinfo {author} {\bibfnamefont {Y.}~\bibnamefont
  {Gefen}}, \bibinfo {author} {\bibfnamefont {Y.}~\bibnamefont {Imry}}, \ and\
  \bibinfo {author} {\bibfnamefont {D.}~\bibnamefont {Mukamel}},\ }\bibfield
  {title} {\enquote {\bibinfo {title} {{Phase diagram of spin-1 quantum Ising
  models: Applications to systems of weakly coupled classical Ising chains}},}\
  }\href {\doibase 10.1103/PhysRevB.23.6099} {\bibfield  {journal} {\bibinfo
  {journal} {Phys. Rev. B}\ }\textbf {\bibinfo {volume} {23}},\ \bibinfo
  {pages} {6099--6105} (\bibinfo {year} {1981})}\BibitemShut {NoStop}%
\bibitem [{\citenamefont {Alcaraz}\ \emph {et~al.}(1985)\citenamefont
  {Alcaraz}, \citenamefont {Drugowich~de Fel\'{\i}cio}, \citenamefont
  {K\"oberle},\ and\ \citenamefont {Stilck}}]{alcaraz1985hamiltonian}%
  \BibitemOpen
  \bibfield  {author} {\bibinfo {author} {\bibfnamefont {F.~C.}\ \bibnamefont
  {Alcaraz}}, \bibinfo {author} {\bibfnamefont {J.~R.}\ \bibnamefont
  {Drugowich~de Fel\'{\i}cio}}, \bibinfo {author} {\bibfnamefont
  {R.}~\bibnamefont {K\"oberle}}, \ and\ \bibinfo {author} {\bibfnamefont
  {J.~F.}\ \bibnamefont {Stilck}},\ }\bibfield  {title} {\enquote {\bibinfo
  {title} {{Hamiltonian studies of the Blume-Emery-Griffiths model}},}\ }\href
  {\doibase 10.1103/PhysRevB.32.7469} {\bibfield  {journal} {\bibinfo
  {journal} {Phys. Rev. B}\ }\textbf {\bibinfo {volume} {32}},\ \bibinfo
  {pages} {7469--7475} (\bibinfo {year} {1985})}\BibitemShut {NoStop}%
\bibitem [{\citenamefont {Balbao}\ and\ \citenamefont
  {de~Felicio}(1987)}]{balbao1987operator}%
  \BibitemOpen
  \bibfield  {author} {\bibinfo {author} {\bibfnamefont {D.~B.}\ \bibnamefont
  {Balbao}}\ and\ \bibinfo {author} {\bibfnamefont {J.~R.~Drugowich}\
  \bibnamefont {de~Felicio}},\ }\bibfield  {title} {\enquote {\bibinfo {title}
  {{Operator content of the Blume-Capel quantum chain}},}\ }\href
  {http://stacks.iop.org/0305-4470/20/i=4/a=005} {\bibfield  {journal}
  {\bibinfo  {journal} {J. Phys. A}\ }\textbf {\bibinfo {volume} {20}},\
  \bibinfo {pages} {L207} (\bibinfo {year} {1987})}\BibitemShut {NoStop}%
\bibitem [{\citenamefont {von Gehlen}(1990)}]{von1990off}%
  \BibitemOpen
  \bibfield  {author} {\bibinfo {author} {\bibfnamefont {G.}~\bibnamefont {von
  Gehlen}},\ }\bibfield  {title} {\enquote {\bibinfo {title} {{Off-criticality
  behaviour of the Blume-Capel quantum chain as a check of Zamolodchikov's
  conjecture}},}\ }\href {\doibase
  http://dx.doi.org/10.1016/0550-3213(90)90130-6} {\bibfield  {journal}
  {\bibinfo  {journal} {Nucl. Phys. B}\ }\textbf {\bibinfo {volume} {330}},\
  \bibinfo {pages} {741 -- 756} (\bibinfo {year} {1990})}\BibitemShut {NoStop}%
\bibitem [{\citenamefont {von Gehlen}\ \emph {et~al.}(1986)\citenamefont {von
  Gehlen}, \citenamefont {Rittenberg},\ and\ \citenamefont
  {Ruegg}}]{von1986conformal}%
  \BibitemOpen
  \bibfield  {author} {\bibinfo {author} {\bibfnamefont {G.}~\bibnamefont {von
  Gehlen}}, \bibinfo {author} {\bibfnamefont {V.}~\bibnamefont {Rittenberg}}, \
  and\ \bibinfo {author} {\bibfnamefont {H.}~\bibnamefont {Ruegg}},\ }\bibfield
   {title} {\enquote {\bibinfo {title} {{Conformal invariance and finite
  one-dimensional quantum chains}},}\ }\href
  {http://stacks.iop.org/0305-4470/19/i=1/a=014} {\bibfield  {journal}
  {\bibinfo  {journal} {J. Phys. A}\ }\textbf {\bibinfo {volume} {19}},\
  \bibinfo {pages} {107} (\bibinfo {year} {1986})}\BibitemShut {NoStop}%
\bibitem [{\citenamefont {Alba}\ \emph {et~al.}(2011)\citenamefont {Alba},
  \citenamefont {Fateev}, \citenamefont {Litvinov},\ and\ \citenamefont
  {Tarnopolskiy}}]{alba2011on}%
  \BibitemOpen
  \bibfield  {author} {\bibinfo {author} {\bibfnamefont {V.~A.}\ \bibnamefont
  {Alba}}, \bibinfo {author} {\bibfnamefont {V.~A.}\ \bibnamefont {Fateev}},
  \bibinfo {author} {\bibfnamefont {A.~V.}\ \bibnamefont {Litvinov}}, \ and\
  \bibinfo {author} {\bibfnamefont {G.~M.}\ \bibnamefont {Tarnopolskiy}},\
  }\bibfield  {title} {\enquote {\bibinfo {title} {{On Combinatorial Expansion
  of the Conformal Blocks Arising from {AGT} Conjecture}},}\ }\href {\doibase
  10.1007/s11005-011-0503-z} {\bibfield  {journal} {\bibinfo  {journal} {Lett.
  Math. Phys.}\ }\textbf {\bibinfo {volume} {98}},\ \bibinfo {pages} {33}
  (\bibinfo {year} {2011})}\BibitemShut {NoStop}%
\bibitem [{\citenamefont {Dotsenko}\ and\ \citenamefont
  {Fateev}(1985)}]{dotsenko1985four}%
  \BibitemOpen
  \bibfield  {author} {\bibinfo {author} {\bibfnamefont {Vl.~S.}\ \bibnamefont
  {Dotsenko}}\ and\ \bibinfo {author} {\bibfnamefont {V.~A.}\ \bibnamefont
  {Fateev}},\ }\bibfield  {title} {\enquote {\bibinfo {title} {{Four-point
  correlation functions and the operator algebra in 2D conformal invariant
  theories with central charge $c<1$}},}\ }\href {\doibase
  http://dx.doi.org/10.1016/S0550-3213(85)80004-3} {\bibfield  {journal}
  {\bibinfo  {journal} {Nucl. Phys. B}\ }\textbf {\bibinfo {volume} {251}},\
  \bibinfo {pages} {691 -- 734} (\bibinfo {year} {1985})}\BibitemShut {NoStop}%
\bibitem [{\citenamefont {Zamolodchikov}(1989)}]{zamolodchikov1989integrable}%
  \BibitemOpen
  \bibfield  {author} {\bibinfo {author} {\bibfnamefont {A.~B.}\ \bibnamefont
  {Zamolodchikov}},\ }\bibfield  {title} {\enquote {\bibinfo {title}
  {{Integrable Field Theory from Conformal Field Theory}},}\ }in\ \href
  {\doibase http://dx.doi.org/10.1016/B978-0-12-385342-4.50022-6} {\emph
  {\bibinfo {booktitle} {{Advanced Studies in Pure Mathematics 19: Integrable
  Systems in Quantum Field Theory and Statistical Mechanics}}}},\ \bibinfo
  {editor} {edited by\ \bibinfo {editor} {\bibfnamefont {M.}~\bibnamefont
  {Jimbo}}, \bibinfo {editor} {\bibfnamefont {T.}~\bibnamefont {Miwa}}, \ and\
  \bibinfo {editor} {\bibfnamefont {A.}~\bibnamefont {Tsuchiya}}}\ (\bibinfo
  {publisher} {Academic Press},\ \bibinfo {address} {San Diego},\ \bibinfo
  {year} {1989})\ pp.\ \bibinfo {pages} {641 -- 674}\BibitemShut {NoStop}%
\bibitem [{\citenamefont {Fateev}\ and\ \citenamefont
  {Zamolodchikov}(1990)}]{fateev1990conformal}%
  \BibitemOpen
  \bibfield  {author} {\bibinfo {author} {\bibfnamefont {V.~A.}\ \bibnamefont
  {Fateev}}\ and\ \bibinfo {author} {\bibfnamefont {A.~B.}\ \bibnamefont
  {Zamolodchikov}},\ }\bibfield  {title} {\enquote {\bibinfo {title} {Conformal
  field theory and purely elastic {S}-matrices},}\ }\href {\doibase
  10.1142/S0217751X90000477} {\bibfield  {journal} {\bibinfo  {journal} {Int.
  J. Mod. Phys. A}\ }\textbf {\bibinfo {volume} {05}},\ \bibinfo {pages}
  {1025--1048} (\bibinfo {year} {1990})}\BibitemShut {NoStop}%
\bibitem [{\citenamefont {Essler}\ and\ \citenamefont
  {Konik}(2012)}]{essler2005review}%
  \BibitemOpen
  \bibfield  {author} {\bibinfo {author} {\bibfnamefont {F.~H.~L.}\
  \bibnamefont {Essler}}\ and\ \bibinfo {author} {\bibfnamefont {R.~M.}\
  \bibnamefont {Konik}},\ }\enquote {\bibinfo {title} {Application of massive
  integrable quantum field theories to problems in condensed matter physics},}\
  in\ \href {\doibase 10.1142/9789812775344_0020} {\emph {\bibinfo {booktitle}
  {From Fields to Strings: Circumnavigating Theoretical Physics}}}\ (\bibinfo
  {publisher} {World Scientific},\ \bibinfo {year} {2012})\ pp.\ \bibinfo
  {pages} {684--830}\BibitemShut {NoStop}%
\bibitem [{\citenamefont {Pozsgay}(2008)}]{pozsgay2008Luscher}%
  \BibitemOpen
  \bibfield  {author} {\bibinfo {author} {\bibfnamefont {B.}~\bibnamefont
  {Pozsgay}},\ }\bibfield  {title} {\enquote {\bibinfo {title} {{L{\"u}scher's
  $\mu$-term and finite volume bootstrap principle for scattering states and
  form factors}},}\ }\href {\doibase
  http://dx.doi.org/10.1016/j.nuclphysb.2008.04.021} {\bibfield  {journal}
  {\bibinfo  {journal} {Nucl. Phys. B}\ }\textbf {\bibinfo {volume} {802}},\
  \bibinfo {pages} {435 -- 457} (\bibinfo {year} {2008})}\BibitemShut {NoStop}%
\bibitem [{\citenamefont {Gritsev}\ \emph {et~al.}(2007)\citenamefont
  {Gritsev}, \citenamefont {Polkovnikov},\ and\ \citenamefont
  {Demler}}]{gritsev2007linear}%
  \BibitemOpen
  \bibfield  {author} {\bibinfo {author} {\bibfnamefont {V.}~\bibnamefont
  {Gritsev}}, \bibinfo {author} {\bibfnamefont {A.}~\bibnamefont
  {Polkovnikov}}, \ and\ \bibinfo {author} {\bibfnamefont {E.}~\bibnamefont
  {Demler}},\ }\bibfield  {title} {\enquote {\bibinfo {title} {Linear response
  theory for a pair of coupled one-dimensional condensates of interacting
  atoms},}\ }\href {\doibase 10.1103/PhysRevB.75.174511} {\bibfield  {journal}
  {\bibinfo  {journal} {Phys. Rev. B}\ }\textbf {\bibinfo {volume} {75}},\
  \bibinfo {pages} {174511} (\bibinfo {year} {2007})}\BibitemShut {NoStop}%
\bibitem [{\citenamefont {Zamolodchikov}\ and\ \citenamefont
  {Zamolodchikov}(1979{\natexlab{b}})}]{zamolodchikov1979factorized}%
  \BibitemOpen
  \bibfield  {author} {\bibinfo {author} {\bibfnamefont {A.~B.}\ \bibnamefont
  {Zamolodchikov}}\ and\ \bibinfo {author} {\bibfnamefont {A.~B.}\ \bibnamefont
  {Zamolodchikov}},\ }\bibfield  {title} {\enquote {\bibinfo {title}
  {{Factorized S-matrices in two dimensions as the exact solutions of certain
  relativistic quantum field theory models}},}\ }\href {\doibase
  http://dx.doi.org/10.1016/0003-4916(79)90391-9} {\bibfield  {journal}
  {\bibinfo  {journal} {Ann. Phys. (N.Y.)}\ }\textbf {\bibinfo {volume}
  {120}},\ \bibinfo {pages} {253 -- 291} (\bibinfo {year}
  {1979}{\natexlab{b}})}\BibitemShut {NoStop}%
\bibitem [{\citenamefont {Izergin}\ and\ \citenamefont
  {Korepin}(1981)}]{izergin1981lattice}%
  \BibitemOpen
  \bibfield  {author} {\bibinfo {author} {\bibfnamefont {A.~G.}\ \bibnamefont
  {Izergin}}\ and\ \bibinfo {author} {\bibfnamefont {V.~E.}\ \bibnamefont
  {Korepin}},\ }\bibfield  {title} {\enquote {\bibinfo {title} {{The lattice
  quantum Sine-Gordon model}},}\ }\href {\doibase 10.1007/BF00420699}
  {\bibfield  {journal} {\bibinfo  {journal} {Lett. Math. Phys.}\ }\textbf
  {\bibinfo {volume} {5}},\ \bibinfo {pages} {199--205} (\bibinfo {year}
  {1981})}\BibitemShut {NoStop}%
\bibitem [{\citenamefont {Zamolodhchikov}(1995)}]{zamolodchikov1995mass}%
  \BibitemOpen
  \bibfield  {author} {\bibinfo {author} {\bibfnamefont {Al.~B.}\ \bibnamefont
  {Zamolodhchikov}},\ }\bibfield  {title} {\enquote {\bibinfo {title} {Mass
  scale in the sine-{G}ordon model and its reductions},}\ }\href {\doibase
  10.1142/S0217751X9500053X} {\bibfield  {journal} {\bibinfo  {journal} {Int.
  J. Mod. Phys. A}\ }\textbf {\bibinfo {volume} {10}},\ \bibinfo {pages}
  {1125--1150} (\bibinfo {year} {1995})}\BibitemShut {NoStop}%
\bibitem [{\citenamefont {Baxter}(1982)}]{baxter1982exactly}%
  \BibitemOpen
  \bibfield  {author} {\bibinfo {author} {\bibfnamefont {R.~J.}\ \bibnamefont
  {Baxter}},\ }\href@noop {} {\emph {\bibinfo {title} {Exactly solved models in
  statistical mechanics}}}\ (\bibinfo  {publisher} {Elsevier},\ \bibinfo {year}
  {1982})\BibitemShut {NoStop}%
\bibitem [{\citenamefont
  {Zamolodchikov}(1990)}]{zamolodchikov1990thermodynamic}%
  \BibitemOpen
  \bibfield  {author} {\bibinfo {author} {\bibfnamefont {Al.~B.}\ \bibnamefont
  {Zamolodchikov}},\ }\bibfield  {title} {\enquote {\bibinfo {title}
  {{Thermodynamic Bethe ansatz in relativistic models: Scaling 3-state potts
  and Lee-Yang models}},}\ }\href {\doibase
  http://dx.doi.org/10.1016/0550-3213(90)90333-9} {\bibfield  {journal}
  {\bibinfo  {journal} {Nucl. Phys. B}\ }\textbf {\bibinfo {volume} {342}},\
  \bibinfo {pages} {695 -- 720} (\bibinfo {year} {1990})}\BibitemShut {NoStop}%
\bibitem [{\citenamefont {Klassen}\ and\ \citenamefont
  {Melzer}(1991{\natexlab{b}})}]{klassen1991on}%
  \BibitemOpen
  \bibfield  {author} {\bibinfo {author} {\bibfnamefont {T.~R.}\ \bibnamefont
  {Klassen}}\ and\ \bibinfo {author} {\bibfnamefont {E.}~\bibnamefont
  {Melzer}},\ }\bibfield  {title} {\enquote {\bibinfo {title} {{On the relation
  between scattering amplitudes and finite-size mass corrections in QFT}},}\
  }\href {\doibase http://dx.doi.org/10.1016/0550-3213(91)90566-G} {\bibfield
  {journal} {\bibinfo  {journal} {Nucl. Phys. B}\ }\textbf {\bibinfo {volume}
  {362}},\ \bibinfo {pages} {329 -- 388} (\bibinfo {year}
  {1991}{\natexlab{b}})}\BibitemShut {NoStop}%
\bibitem [{\citenamefont {Bulla}\ \emph {et~al.}(2008)\citenamefont {Bulla},
  \citenamefont {Costi},\ and\ \citenamefont {Pruschke}}]{bulla2008numerical}%
  \BibitemOpen
  \bibfield  {author} {\bibinfo {author} {\bibfnamefont {R.}~\bibnamefont
  {Bulla}}, \bibinfo {author} {\bibfnamefont {T.~A.}\ \bibnamefont {Costi}}, \
  and\ \bibinfo {author} {\bibfnamefont {T.}~\bibnamefont {Pruschke}},\
  }\bibfield  {title} {\enquote {\bibinfo {title} {Numerical renormalization
  group method for quantum impurity systems},}\ }\href {\doibase
  10.1103/RevModPhys.80.395} {\bibfield  {journal} {\bibinfo  {journal} {Rev.
  Mod. Phys.}\ }\textbf {\bibinfo {volume} {80}},\ \bibinfo {pages} {395--450}
  (\bibinfo {year} {2008})}\BibitemShut {NoStop}%
\bibitem [{\citenamefont {Sz{\'e}cs{\'e}nyi}\ \emph {et~al.}(2013)\citenamefont
  {Sz{\'e}cs{\'e}nyi}, \citenamefont {Tak{\'a}cs},\ and\ \citenamefont
  {Watts}}]{szecsenyi2013one}%
  \BibitemOpen
  \bibfield  {author} {\bibinfo {author} {\bibfnamefont {I.~M.}\ \bibnamefont
  {Sz{\'e}cs{\'e}nyi}}, \bibinfo {author} {\bibfnamefont {G.}~\bibnamefont
  {Tak{\'a}cs}}, \ and\ \bibinfo {author} {\bibfnamefont {G.~M.~T.}\
  \bibnamefont {Watts}},\ }\bibfield  {title} {\enquote {\bibinfo {title}
  {{One-point functions in finite volume/temperature: a case study}},}\ }\href
  {\doibase 10.1007/JHEP08(2013)094} {\bibfield  {journal} {\bibinfo  {journal}
  {JHEP}\ }\textbf {\bibinfo {volume} {2013}},\ \bibinfo {pages} {94} (\bibinfo
  {year} {2013})}\BibitemShut {NoStop}%
\bibitem [{\citenamefont {Brandino}\ \emph {et~al.}(2010)\citenamefont
  {Brandino}, \citenamefont {Konik},\ and\ \citenamefont
  {Mussardo}}]{brandino2010energy}%
  \BibitemOpen
  \bibfield  {author} {\bibinfo {author} {\bibfnamefont {G.~P.}\ \bibnamefont
  {Brandino}}, \bibinfo {author} {\bibfnamefont {R.~M.}\ \bibnamefont {Konik}},
  \ and\ \bibinfo {author} {\bibfnamefont {G.}~\bibnamefont {Mussardo}},\
  }\bibfield  {title} {\enquote {\bibinfo {title} {Energy level distribution of
  perturbed conformal field theories},}\ }\href {\doibase
  10.1088/1742-5468/2010/07/P07013} {\bibfield  {journal} {\bibinfo  {journal}
  {J. Stat. Mech.}\ }\textbf {\bibinfo {volume} {2010}},\ \bibinfo {pages}
  {P07013} (\bibinfo {year} {2010})}\BibitemShut {NoStop}%
\bibitem [{\citenamefont {Levitov}\ and\ \citenamefont
  {Tsvelik}(2003)}]{levitov2003narrow}%
  \BibitemOpen
  \bibfield  {author} {\bibinfo {author} {\bibfnamefont {L.~S.}\ \bibnamefont
  {Levitov}}\ and\ \bibinfo {author} {\bibfnamefont {A.~M.}\ \bibnamefont
  {Tsvelik}},\ }\bibfield  {title} {\enquote {\bibinfo {title} {{Narrow-Gap
  Luttinger Liquid in Carbon Nanotubes}},}\ }\href {\doibase
  10.1103/PhysRevLett.90.016401} {\bibfield  {journal} {\bibinfo  {journal}
  {Phys. Rev. Lett.}\ }\textbf {\bibinfo {volume} {90}},\ \bibinfo {pages}
  {016401} (\bibinfo {year} {2003})}\BibitemShut {NoStop}%
\bibitem [{\citenamefont {Egger}\ and\ \citenamefont
  {Gogolin}(1998)}]{egger1998correlated}%
  \BibitemOpen
  \bibfield  {author} {\bibinfo {author} {\bibfnamefont {R.}~\bibnamefont
  {Egger}}\ and\ \bibinfo {author} {\bibfnamefont {A.~O.}\ \bibnamefont
  {Gogolin}},\ }\bibfield  {title} {\enquote {\bibinfo {title} {{Correlated
  transport and non-Fermi-liquid behavior in single-wall carbon nanotubes}},}\
  }\href {\doibase 10.1007/s100510050315} {\bibfield  {journal} {\bibinfo
  {journal} {EPJ B}\ }\textbf {\bibinfo {volume} {3}},\ \bibinfo {pages}
  {281--300} (\bibinfo {year} {1998})}\BibitemShut {NoStop}%
\bibitem [{\citenamefont {DeGottardi}\ \emph {et~al.}(2009)\citenamefont
  {DeGottardi}, \citenamefont {Wei},\ and\ \citenamefont
  {Vishveshwara}}]{degottardi2009transverse}%
  \BibitemOpen
  \bibfield  {author} {\bibinfo {author} {\bibfnamefont {W.}~\bibnamefont
  {DeGottardi}}, \bibinfo {author} {\bibfnamefont {T.-C.}\ \bibnamefont {Wei}},
  \ and\ \bibinfo {author} {\bibfnamefont {S.}~\bibnamefont {Vishveshwara}},\
  }\bibfield  {title} {\enquote {\bibinfo {title} {{Transverse-field-induced
  effects in carbon nanotubes}},}\ }\href {\doibase 10.1103/PhysRevB.79.205421}
  {\bibfield  {journal} {\bibinfo  {journal} {Phys. Rev. B}\ }\textbf {\bibinfo
  {volume} {79}},\ \bibinfo {pages} {205421} (\bibinfo {year}
  {2009})}\BibitemShut {NoStop}%
\bibitem [{\citenamefont {Sfeir}\ \emph {et~al.}(2010)\citenamefont {Sfeir},
  \citenamefont {Misewich}, \citenamefont {Rosenblatt}, \citenamefont {Wu},
  \citenamefont {Voisin}, \citenamefont {Yan}, \citenamefont {Berciaud},
  \citenamefont {Heinz}, \citenamefont {Chandra}, \citenamefont {Caldwell},
  \citenamefont {Shan}, \citenamefont {Hone},\ and\ \citenamefont
  {Carr}}]{sfeir2010infrared}%
  \BibitemOpen
  \bibfield  {author} {\bibinfo {author} {\bibfnamefont {M.~Y.}\ \bibnamefont
  {Sfeir}}, \bibinfo {author} {\bibfnamefont {J.~A.}\ \bibnamefont {Misewich}},
  \bibinfo {author} {\bibfnamefont {S.}~\bibnamefont {Rosenblatt}}, \bibinfo
  {author} {\bibfnamefont {Y.}~\bibnamefont {Wu}}, \bibinfo {author}
  {\bibfnamefont {C.}~\bibnamefont {Voisin}}, \bibinfo {author} {\bibfnamefont
  {H.}~\bibnamefont {Yan}}, \bibinfo {author} {\bibfnamefont {S.}~\bibnamefont
  {Berciaud}}, \bibinfo {author} {\bibfnamefont {T.~F.}\ \bibnamefont {Heinz}},
  \bibinfo {author} {\bibfnamefont {B.}~\bibnamefont {Chandra}}, \bibinfo
  {author} {\bibfnamefont {R.}~\bibnamefont {Caldwell}}, \bibinfo {author}
  {\bibfnamefont {Y.}~\bibnamefont {Shan}}, \bibinfo {author} {\bibfnamefont
  {J.}~\bibnamefont {Hone}}, \ and\ \bibinfo {author} {\bibfnamefont {G.~L.}\
  \bibnamefont {Carr}},\ }\bibfield  {title} {\enquote {\bibinfo {title}
  {{Infrared spectra of individual semiconducting single-walled carbon
  nanotubes: Testing the scaling of transition energies for large diameter
  nanotubes}},}\ }\href {\doibase 10.1103/PhysRevB.82.195424} {\bibfield
  {journal} {\bibinfo  {journal} {Phys. Rev. B}\ }\textbf {\bibinfo {volume}
  {82}},\ \bibinfo {pages} {195424} (\bibinfo {year} {2010})}\BibitemShut
  {NoStop}%
\bibitem [{\citenamefont {Dukovic}\ \emph {et~al.}(2005)\citenamefont
  {Dukovic}, \citenamefont {Wang}, \citenamefont {Song}, \citenamefont {Sfeir},
  \citenamefont {Heinz},\ and\ \citenamefont {Brus}}]{dukovic2005structural}%
  \BibitemOpen
  \bibfield  {author} {\bibinfo {author} {\bibfnamefont {G.}~\bibnamefont
  {Dukovic}}, \bibinfo {author} {\bibfnamefont {F.}~\bibnamefont {Wang}},
  \bibinfo {author} {\bibfnamefont {D.}~\bibnamefont {Song}}, \bibinfo {author}
  {\bibfnamefont {M.~Y.}\ \bibnamefont {Sfeir}}, \bibinfo {author}
  {\bibfnamefont {T.~F.}\ \bibnamefont {Heinz}}, \ and\ \bibinfo {author}
  {\bibfnamefont {L.~E.}\ \bibnamefont {Brus}},\ }\bibfield  {title} {\enquote
  {\bibinfo {title} {{Structural Dependence of Excitonic Optical Transitions
  and Band-Gap Energies in Carbon Nanotubes}},}\ }\href {\doibase
  10.1021/nl0518122} {\bibfield  {journal} {\bibinfo  {journal} {Nano Lett.}\
  }\textbf {\bibinfo {volume} {5}},\ \bibinfo {pages} {2314--2318} (\bibinfo
  {year} {2005})},\ \bibinfo {note} {pMID: 16277475}\BibitemShut {NoStop}%
\bibitem [{\citenamefont {Slavnov}(1989)}]{slavnov1989calculation}%
  \BibitemOpen
  \bibfield  {author} {\bibinfo {author} {\bibfnamefont {N.~A.}\ \bibnamefont
  {Slavnov}},\ }\bibfield  {title} {\enquote {\bibinfo {title} {{Calculation of
  scalar products of wave functions and form factors in the framework of the
  algebraic Bethe ansatz}},}\ }\href {\doibase 10.1007/BF01016531} {\bibfield
  {journal} {\bibinfo  {journal} {Theor. Math. Phys.}\ }\textbf {\bibinfo
  {volume} {79}},\ \bibinfo {pages} {502--508} (\bibinfo {year}
  {1989})}\BibitemShut {NoStop}%
\bibitem [{\citenamefont {Slavnov}(1990)}]{slavnov1990nonequal}%
  \BibitemOpen
  \bibfield  {author} {\bibinfo {author} {\bibfnamefont {N.~A.}\ \bibnamefont
  {Slavnov}},\ }\bibfield  {title} {\enquote {\bibinfo {title} {{Nonequal-time
  current correlation function in a one-dimensional Bose gas}},}\ }\href
  {\doibase 10.1007/BF01029221} {\bibfield  {journal} {\bibinfo  {journal}
  {Theor. Math. Phys.}\ }\textbf {\bibinfo {volume} {82}},\ \bibinfo {pages}
  {273--282} (\bibinfo {year} {1990})}\BibitemShut {NoStop}%
\bibitem [{\citenamefont {Tonks}(1936)}]{tonks1936complete}%
  \BibitemOpen
  \bibfield  {author} {\bibinfo {author} {\bibfnamefont {L.}~\bibnamefont
  {Tonks}},\ }\bibfield  {title} {\enquote {\bibinfo {title} {{The Complete
  Equation of State of One, Two and Three-Dimensional Gases of Hard Elastic
  Spheres}},}\ }\href {\doibase 10.1103/PhysRev.50.955} {\bibfield  {journal}
  {\bibinfo  {journal} {Phys. Rev.}\ }\textbf {\bibinfo {volume} {50}},\
  \bibinfo {pages} {955--963} (\bibinfo {year} {1936})}\BibitemShut {NoStop}%
\bibitem [{\citenamefont {Khodas}\ \emph {et~al.}(2007)\citenamefont {Khodas},
  \citenamefont {Pustilnik}, \citenamefont {Kamenev},\ and\ \citenamefont
  {Glazman}}]{khodas2007dynamics}%
  \BibitemOpen
  \bibfield  {author} {\bibinfo {author} {\bibfnamefont {M.}~\bibnamefont
  {Khodas}}, \bibinfo {author} {\bibfnamefont {M.}~\bibnamefont {Pustilnik}},
  \bibinfo {author} {\bibfnamefont {A.}~\bibnamefont {Kamenev}}, \ and\
  \bibinfo {author} {\bibfnamefont {L.~I.}\ \bibnamefont {Glazman}},\
  }\bibfield  {title} {\enquote {\bibinfo {title} {{Dynamics of Excitations in
  a One-Dimensional Bose Liquid}},}\ }\href {\doibase
  10.1103/PhysRevLett.99.110405} {\bibfield  {journal} {\bibinfo  {journal}
  {Phys. Rev. Lett.}\ }\textbf {\bibinfo {volume} {99}},\ \bibinfo {pages}
  {110405} (\bibinfo {year} {2007})}\BibitemShut {NoStop}%
\bibitem [{\citenamefont {Girardeau}(1960)}]{girardeau1960relationship}%
  \BibitemOpen
  \bibfield  {author} {\bibinfo {author} {\bibfnamefont {M.}~\bibnamefont
  {Girardeau}},\ }\bibfield  {title} {\enquote {\bibinfo {title} {{Relationship
  between Systems of Impenetrable Bosons and Fermions in One Dimension}},}\
  }\href {\doibase 10.1063/1.1703687} {\bibfield  {journal} {\bibinfo
  {journal} {J. Math. Phys.}\ }\textbf {\bibinfo {volume} {1}},\ \bibinfo
  {pages} {516--523} (\bibinfo {year} {1960})}\BibitemShut {NoStop}%
\bibitem [{\citenamefont {Frishman}\ and\ \citenamefont
  {Sonnenschein}(2010)}]{frishman2010non}%
  \BibitemOpen
  \bibfield  {author} {\bibinfo {author} {\bibfnamefont {Y.}~\bibnamefont
  {Frishman}}\ and\ \bibinfo {author} {\bibfnamefont {J.}~\bibnamefont
  {Sonnenschein}},\ }\href
  {http://www.cambridge.org/catalogue/catalogue.asp?isbn=9780521662659} {\emph
  {\bibinfo {title} {{Non-Perturbative Field Theory: From Two Dimensional
  Conformal Field Theory to QCD in Four Dimensions}}}}\ (\bibinfo  {publisher}
  {Cambridge University Press},\ \bibinfo {year} {2010})\BibitemShut {NoStop}%
\bibitem [{\citenamefont {'t~Hooft}(1986)}]{thooft1986how}%
  \BibitemOpen
  \bibfield  {author} {\bibinfo {author} {\bibfnamefont {G.}~\bibnamefont
  {'t~Hooft}},\ }\bibfield  {title} {\enquote {\bibinfo {title} {{How
  instantons solve the U(1) problem}},}\ }\href {\doibase
  http://dx.doi.org/10.1016/0370-1573(86)90117-1} {\bibfield  {journal}
  {\bibinfo  {journal} {Phys. Rep.}\ }\textbf {\bibinfo {volume} {142}},\
  \bibinfo {pages} {357 -- 387} (\bibinfo {year} {1986})}\BibitemShut {NoStop}%
\bibitem [{\citenamefont {Seiberg}(1990)}]{seiberg1990notes}%
  \BibitemOpen
  \bibfield  {author} {\bibinfo {author} {\bibfnamefont {N.}~\bibnamefont
  {Seiberg}},\ }\bibfield  {title} {\enquote {\bibinfo {title} {{Notes on
  quantum Liouville theory and quantum gravity}},}\ }\bibfield  {booktitle}
  {\emph {\bibinfo {booktitle} {{Cargese Study Institute: Random Surfaces,
  Quantum Gravity and Strings Cargese, France, May 27-June 2, 1990}}},\ }\href
  {\doibase 10.1143/PTPS.102.319} {\bibfield  {journal} {\bibinfo  {journal}
  {Prog. Theor. Phys. Suppl.}\ }\textbf {\bibinfo {volume} {102}},\ \bibinfo
  {pages} {319--349} (\bibinfo {year} {1990})}\BibitemShut {NoStop}%
\bibitem [{\citenamefont {Zamolodchikov}\ and\ \citenamefont
  {Zamolodchikov}(1996)}]{zamolodchikov1996conformal}%
  \BibitemOpen
  \bibfield  {author} {\bibinfo {author} {\bibfnamefont {A.}~\bibnamefont
  {Zamolodchikov}}\ and\ \bibinfo {author} {\bibfnamefont {Al.}\ \bibnamefont
  {Zamolodchikov}},\ }\bibfield  {title} {\enquote {\bibinfo {title}
  {{Conformal bootstrap in Liouville field theory}},}\ }\href {\doibase
  http://dx.doi.org/10.1016/0550-3213(96)00351-3} {\bibfield  {journal}
  {\bibinfo  {journal} {Nucl. Phys. B}\ }\textbf {\bibinfo {volume} {477}},\
  \bibinfo {pages} {577 -- 605} (\bibinfo {year} {1996})}\BibitemShut {NoStop}%
\bibitem [{\citenamefont {Bajnok}\ \emph {et~al.}(2008)\citenamefont {Bajnok},
  \citenamefont {Rim},\ and\ \citenamefont {Zamolodchikov}}]{bajnok2008sinh}%
  \BibitemOpen
  \bibfield  {author} {\bibinfo {author} {\bibfnamefont {Z.}~\bibnamefont
  {Bajnok}}, \bibinfo {author} {\bibfnamefont {C.}~\bibnamefont {Rim}}, \ and\
  \bibinfo {author} {\bibfnamefont {Al.}\ \bibnamefont {Zamolodchikov}},\
  }\bibfield  {title} {\enquote {\bibinfo {title} {{Sinh-Gordon boundary
  \{TBA\} and boundary Liouville reflection amplitude}},}\ }\href {\doibase
  http://dx.doi.org/10.1016/j.nuclphysb.2007.12.023} {\bibfield  {journal}
  {\bibinfo  {journal} {Nucl. Phys. B}\ }\textbf {\bibinfo {volume} {796}},\
  \bibinfo {pages} {622 -- 650} (\bibinfo {year} {2008})}\BibitemShut {NoStop}%
\bibitem [{\citenamefont {White}(1993{\natexlab{b}})}]{white1993density}%
  \BibitemOpen
  \bibfield  {author} {\bibinfo {author} {\bibfnamefont {S.~R.}\ \bibnamefont
  {White}},\ }\bibfield  {title} {\enquote {\bibinfo {title} {{{Density-matrix
  algorithms for quantum renormalization groups}}},}\ }\href {\doibase
  10.1103/PhysRevB.48.10345} {\bibfield  {journal} {\bibinfo  {journal} {Phys.
  Rev. B}\ }\textbf {\bibinfo {volume} {48}},\ \bibinfo {pages} {10345--10356}
  (\bibinfo {year} {1993}{\natexlab{b}})}\BibitemShut {NoStop}%
\bibitem [{\citenamefont {\"Ostlund}\ and\ \citenamefont
  {Rommer}(1995)}]{ostlund1995thermodynamic}%
  \BibitemOpen
  \bibfield  {author} {\bibinfo {author} {\bibfnamefont {S.}~\bibnamefont
  {\"Ostlund}}\ and\ \bibinfo {author} {\bibfnamefont {S.}~\bibnamefont
  {Rommer}},\ }\bibfield  {title} {\enquote {\bibinfo {title} {{Thermodynamic
  Limit of Density Matrix Renormalization}},}\ }\href {\doibase
  10.1103/PhysRevLett.75.3537} {\bibfield  {journal} {\bibinfo  {journal}
  {Phys. Rev. Lett.}\ }\textbf {\bibinfo {volume} {75}},\ \bibinfo {pages}
  {3537--3540} (\bibinfo {year} {1995})}\BibitemShut {NoStop}%
\bibitem [{\citenamefont {Rommer}\ and\ \citenamefont
  {\"Ostlund}(1997)}]{rommer1997class}%
  \BibitemOpen
  \bibfield  {author} {\bibinfo {author} {\bibfnamefont {S.}~\bibnamefont
  {Rommer}}\ and\ \bibinfo {author} {\bibfnamefont {S.}~\bibnamefont
  {\"Ostlund}},\ }\bibfield  {title} {\enquote {\bibinfo {title} {{Class of
  ansatz wave functions for one-dimensional spin systems and their relation to
  the density matrix renormalization group}},}\ }\href {\doibase
  10.1103/PhysRevB.55.2164} {\bibfield  {journal} {\bibinfo  {journal} {Phys.
  Rev. B}\ }\textbf {\bibinfo {volume} {55}},\ \bibinfo {pages} {2164--2181}
  (\bibinfo {year} {1997})}\BibitemShut {NoStop}%
\bibitem [{\citenamefont {Dukelsky}\ \emph {et~al.}(1998)\citenamefont
  {Dukelsky}, \citenamefont {Mart\'i­n-Delgado}, \citenamefont {Nishino},\ and\
  \citenamefont {Sierra}}]{dukelsky1998equivalence}%
  \BibitemOpen
  \bibfield  {author} {\bibinfo {author} {\bibfnamefont {J.}~\bibnamefont
  {Dukelsky}}, \bibinfo {author} {\bibfnamefont {M.~A.}\ \bibnamefont
  {Mart\'i­n-Delgado}}, \bibinfo {author} {\bibfnamefont {T.}~\bibnamefont
  {Nishino}}, \ and\ \bibinfo {author} {\bibfnamefont {G.}~\bibnamefont
  {Sierra}},\ }\bibfield  {title} {\enquote {\bibinfo {title} {{Equivalence of
  the variational matrix product method and the density matrix renormalization
  group applied to spin chains}},}\ }\href
  {http://stacks.iop.org/0295-5075/43/i=4/a=457} {\bibfield  {journal}
  {\bibinfo  {journal} {EPL}\ }\textbf {\bibinfo {volume} {43}},\ \bibinfo
  {pages} {457} (\bibinfo {year} {1998})}\BibitemShut {NoStop}%
\bibitem [{\citenamefont {Holzhey}\ \emph {et~al.}(1994)\citenamefont
  {Holzhey}, \citenamefont {Larsen},\ and\ \citenamefont
  {Wilczek}}]{holzhey1994geometric}%
  \BibitemOpen
  \bibfield  {author} {\bibinfo {author} {\bibfnamefont {C.}~\bibnamefont
  {Holzhey}}, \bibinfo {author} {\bibfnamefont {F.}~\bibnamefont {Larsen}}, \
  and\ \bibinfo {author} {\bibfnamefont {F.}~\bibnamefont {Wilczek}},\
  }\bibfield  {title} {\enquote {\bibinfo {title} {{Geometric and renormalized
  entropy in conformal field theory}},}\ }\href {\doibase
  http://dx.doi.org/10.1016/0550-3213(94)90402-2} {\bibfield  {journal}
  {\bibinfo  {journal} {Nucl. Phys. B}\ }\textbf {\bibinfo {volume} {424}},\
  \bibinfo {pages} {443 -- 467} (\bibinfo {year} {1994})}\BibitemShut {NoStop}%
\bibitem [{\citenamefont {Calabrese}\ and\ \citenamefont
  {Cardy}(2004)}]{calabrese2004entanglement}%
  \BibitemOpen
  \bibfield  {author} {\bibinfo {author} {\bibfnamefont {P.}~\bibnamefont
  {Calabrese}}\ and\ \bibinfo {author} {\bibfnamefont {J.}~\bibnamefont
  {Cardy}},\ }\bibfield  {title} {\enquote {\bibinfo {title} {{Entanglement
  entropy and quantum field theory}},}\ }\href
  {http://stacks.iop.org/1742-5468/2004/i=06/a=P06002} {\bibfield  {journal}
  {\bibinfo  {journal} {J. Stat. Mech.}\ }\textbf {\bibinfo {volume} {2004}},\
  \bibinfo {pages} {P06002} (\bibinfo {year} {2004})}\BibitemShut {NoStop}%
\bibitem [{\citenamefont {Pollmann}\ \emph {et~al.}(2009)\citenamefont
  {Pollmann}, \citenamefont {Mukerjee}, \citenamefont {Turner},\ and\
  \citenamefont {Moore}}]{pollmann2009theory}%
  \BibitemOpen
  \bibfield  {author} {\bibinfo {author} {\bibfnamefont {F.}~\bibnamefont
  {Pollmann}}, \bibinfo {author} {\bibfnamefont {S.}~\bibnamefont {Mukerjee}},
  \bibinfo {author} {\bibfnamefont {A.~M.}\ \bibnamefont {Turner}}, \ and\
  \bibinfo {author} {\bibfnamefont {J.~E.}\ \bibnamefont {Moore}},\ }\bibfield
  {title} {\enquote {\bibinfo {title} {{Theory of Finite-Entanglement Scaling
  at One-Dimensional Quantum Critical Points}},}\ }\href {\doibase
  10.1103/PhysRevLett.102.255701} {\bibfield  {journal} {\bibinfo  {journal}
  {Phys. Rev. Lett.}\ }\textbf {\bibinfo {volume} {102}},\ \bibinfo {pages}
  {255701} (\bibinfo {year} {2009})}\BibitemShut {NoStop}%
\bibitem [{\citenamefont {Vidal}(2004)}]{vidal2004efficient}%
  \BibitemOpen
  \bibfield  {author} {\bibinfo {author} {\bibfnamefont {G.}~\bibnamefont
  {Vidal}},\ }\bibfield  {title} {\enquote {\bibinfo {title} {{{Efficient
  Simulation of One-Dimensional Quantum Many-Body Systems}}},}\ }\href
  {\doibase 10.1103/PhysRevLett.93.040502} {\bibfield  {journal} {\bibinfo
  {journal} {Phys. Rev. Lett.}\ }\textbf {\bibinfo {volume} {93}},\ \bibinfo
  {pages} {040502} (\bibinfo {year} {2004})}\BibitemShut {NoStop}%
\bibitem [{\citenamefont {Haegeman}\ \emph {et~al.}(2011)\citenamefont
  {Haegeman}, \citenamefont {Cirac}, \citenamefont {Osborne}, \citenamefont
  {Pi\ifmmode~\check{z}\else \v{z}\fi{}orn}, \citenamefont {Verschelde},\ and\
  \citenamefont {Verstraete}}]{haegeman2011time}%
  \BibitemOpen
  \bibfield  {author} {\bibinfo {author} {\bibfnamefont {J.}~\bibnamefont
  {Haegeman}}, \bibinfo {author} {\bibfnamefont {J.~I.}\ \bibnamefont {Cirac}},
  \bibinfo {author} {\bibfnamefont {T.~J.}\ \bibnamefont {Osborne}}, \bibinfo
  {author} {\bibfnamefont {I.}~\bibnamefont {Pi\ifmmode~\check{z}\else
  \v{z}\fi{}orn}}, \bibinfo {author} {\bibfnamefont {H.}~\bibnamefont
  {Verschelde}}, \ and\ \bibinfo {author} {\bibfnamefont {F.}~\bibnamefont
  {Verstraete}},\ }\bibfield  {title} {\enquote {\bibinfo {title}
  {{Time-Dependent Variational Principle for Quantum Lattices}},}\ }\href
  {\doibase 10.1103/PhysRevLett.107.070601} {\bibfield  {journal} {\bibinfo
  {journal} {Phys. Rev. Lett.}\ }\textbf {\bibinfo {volume} {107}},\ \bibinfo
  {pages} {070601} (\bibinfo {year} {2011})}\BibitemShut {NoStop}%
\bibitem [{\citenamefont {Vidal}(2007{\natexlab{a}})}]{vidal2007classical}%
  \BibitemOpen
  \bibfield  {author} {\bibinfo {author} {\bibfnamefont {G.}~\bibnamefont
  {Vidal}},\ }\bibfield  {title} {\enquote {\bibinfo {title} {{{Classical
  Simulation of Infinite-Size Quantum Lattice Systems in One Spatial
  Dimension}}},}\ }\href {\doibase 10.1103/PhysRevLett.98.070201} {\bibfield
  {journal} {\bibinfo  {journal} {Phys. Rev. Lett.}\ }\textbf {\bibinfo
  {volume} {98}},\ \bibinfo {pages} {070201} (\bibinfo {year}
  {2007}{\natexlab{a}})}\BibitemShut {NoStop}%
\bibitem [{\citenamefont {Or{\'{u}}s}\ and\ \citenamefont
  {Vidal}(2008)}]{orus2008infinite}%
  \BibitemOpen
  \bibfield  {author} {\bibinfo {author} {\bibfnamefont {R.}~\bibnamefont
  {Or{\'{u}}s}}\ and\ \bibinfo {author} {\bibfnamefont {G.}~\bibnamefont
  {Vidal}},\ }\bibfield  {title} {\enquote {\bibinfo {title} {{{Infinite
  time-evolving block decimation algorithm beyond unitary evolution}}},}\
  }\href {\doibase 10.1103/PhysRevB.78.155117} {\bibfield  {journal} {\bibinfo
  {journal} {Phys. Rev. B}\ }\textbf {\bibinfo {volume} {78}},\ \bibinfo
  {pages} {155117} (\bibinfo {year} {2008})}\BibitemShut {NoStop}%
\bibitem [{\citenamefont {{McCulloch}}(2008)}]{mcculloch2008infinite}%
  \BibitemOpen
  \bibfield  {author} {\bibinfo {author} {\bibfnamefont {I.~P.}\ \bibnamefont
  {{McCulloch}}},\ }\bibfield  {title} {\enquote {\bibinfo {title} {{Infinite
  size density matrix renormalization group, revisited}},}\ }\href@noop {}
  {\bibfield  {journal} {\bibinfo  {journal} {ArXiv e-prints}\ } (\bibinfo
  {year} {2008})},\ \Eprint {http://arxiv.org/abs/0804.2509} {arXiv:0804.2509
  [cond-mat.str-el]} \BibitemShut {NoStop}%
\bibitem [{\citenamefont {Verstraete}\ \emph {et~al.}(2004)\citenamefont
  {Verstraete}, \citenamefont {Garc\'{\i}a-Ripoll},\ and\ \citenamefont
  {Cirac}}]{verstraete2004matrix}%
  \BibitemOpen
  \bibfield  {author} {\bibinfo {author} {\bibfnamefont {F.}~\bibnamefont
  {Verstraete}}, \bibinfo {author} {\bibfnamefont {J.~J.}\ \bibnamefont
  {Garc\'{\i}a-Ripoll}}, \ and\ \bibinfo {author} {\bibfnamefont {J.~I.}\
  \bibnamefont {Cirac}},\ }\bibfield  {title} {\enquote {\bibinfo {title}
  {{Matrix Product Density Operators: Simulation of Finite-Temperature and
  Dissipative Systems}},}\ }\href {\doibase 10.1103/PhysRevLett.93.207204}
  {\bibfield  {journal} {\bibinfo  {journal} {Phys. Rev. Lett.}\ }\textbf
  {\bibinfo {volume} {93}},\ \bibinfo {pages} {207204} (\bibinfo {year}
  {2004})}\BibitemShut {NoStop}%
\bibitem [{\citenamefont {Feiguin}\ and\ \citenamefont
  {White}(2005)}]{feiguin2005finite}%
  \BibitemOpen
  \bibfield  {author} {\bibinfo {author} {\bibfnamefont {A.~E.}\ \bibnamefont
  {Feiguin}}\ and\ \bibinfo {author} {\bibfnamefont {S.~R.}\ \bibnamefont
  {White}},\ }\bibfield  {title} {\enquote {\bibinfo {title}
  {{Finite-temperature density matrix renormalization using an enlarged Hilbert
  space}},}\ }\href {\doibase 10.1103/PhysRevB.72.220401} {\bibfield  {journal}
  {\bibinfo  {journal} {Phys. Rev. B}\ }\textbf {\bibinfo {volume} {72}},\
  \bibinfo {pages} {220401} (\bibinfo {year} {2005})}\BibitemShut {NoStop}%
\bibitem [{\citenamefont {Karrasch}\ \emph {et~al.}(2013)\citenamefont
  {Karrasch}, \citenamefont {Bardarson},\ and\ \citenamefont
  {Moore}}]{karrasch2013reducing}%
  \BibitemOpen
  \bibfield  {author} {\bibinfo {author} {\bibfnamefont {C.}~\bibnamefont
  {Karrasch}}, \bibinfo {author} {\bibfnamefont {J.~H.}\ \bibnamefont
  {Bardarson}}, \ and\ \bibinfo {author} {\bibfnamefont {J.~E.}\ \bibnamefont
  {Moore}},\ }\bibfield  {title} {\enquote {\bibinfo {title} {{Reducing the
  numerical effort of finite-temperature density matrix renormalization group
  calculations}},}\ }\href {http://stacks.iop.org/1367-2630/15/i=8/a=083031}
  {\bibfield  {journal} {\bibinfo  {journal} {New J. Phys.}\ }\textbf {\bibinfo
  {volume} {15}},\ \bibinfo {pages} {083031} (\bibinfo {year}
  {2013})}\BibitemShut {NoStop}%
\bibitem [{\citenamefont {Scalapino}\ and\ \citenamefont
  {White}(2012)}]{ScalapinoPhysicaC12}%
  \BibitemOpen
  \bibfield  {author} {\bibinfo {author} {\bibfnamefont {D.~J.}\ \bibnamefont
  {Scalapino}}\ and\ \bibinfo {author} {\bibfnamefont {S.~R.}\ \bibnamefont
  {White}},\ }\bibfield  {title} {\enquote {\bibinfo {title} {{Stripe
  structures in the $t-t'-J$ model}},}\ }\href {\doibase
  http://dx.doi.org/10.1016/j.physc.2012.04.004} {\bibfield  {journal}
  {\bibinfo  {journal} {Physica C}\ }\textbf {\bibinfo {volume} {481}},\
  \bibinfo {pages} {146 -- 152} (\bibinfo {year} {2012})}\BibitemShut {NoStop}%
\bibitem [{\citenamefont {Barthel}\ \emph {et~al.}(2009)\citenamefont
  {Barthel}, \citenamefont {Schollw\"ock},\ and\ \citenamefont
  {White}}]{barthel2009spectral}%
  \BibitemOpen
  \bibfield  {author} {\bibinfo {author} {\bibfnamefont {T.}~\bibnamefont
  {Barthel}}, \bibinfo {author} {\bibfnamefont {U.}~\bibnamefont
  {Schollw\"ock}}, \ and\ \bibinfo {author} {\bibfnamefont {S.~R.}\
  \bibnamefont {White}},\ }\bibfield  {title} {\enquote {\bibinfo {title}
  {{Spectral functions in one-dimensional quantum systems at finite temperature
  using the density matrix renormalization group}},}\ }\href {\doibase
  10.1103/PhysRevB.79.245101} {\bibfield  {journal} {\bibinfo  {journal} {Phys.
  Rev. B}\ }\textbf {\bibinfo {volume} {79}},\ \bibinfo {pages} {245101}
  (\bibinfo {year} {2009})}\BibitemShut {NoStop}%
\bibitem [{\citenamefont {Daley}\ \emph {et~al.}(2004)\citenamefont {Daley},
  \citenamefont {Kollath}, \citenamefont {Schollw\"ock},\ and\ \citenamefont
  {Vidal}}]{daley2004time}%
  \BibitemOpen
  \bibfield  {author} {\bibinfo {author} {\bibfnamefont {A.~J.}\ \bibnamefont
  {Daley}}, \bibinfo {author} {\bibfnamefont {C.}~\bibnamefont {Kollath}},
  \bibinfo {author} {\bibfnamefont {U.}~\bibnamefont {Schollw\"ock}}, \ and\
  \bibinfo {author} {\bibfnamefont {G.}~\bibnamefont {Vidal}},\ }\bibfield
  {title} {\enquote {\bibinfo {title} {{Time-dependent density-matrix
  renormalization-group using adaptive effective Hilbert spaces}},}\ }\href
  {http://stacks.iop.org/1742-5468/2004/i=04/a=P04005} {\bibfield  {journal}
  {\bibinfo  {journal} {J. Stat. Mech.}\ }\textbf {\bibinfo {volume} {2004}},\
  \bibinfo {pages} {P04005} (\bibinfo {year} {2004})}\BibitemShut {NoStop}%
\bibitem [{\citenamefont {White}\ and\ \citenamefont
  {Feiguin}(2004)}]{WhitePRL04}%
  \BibitemOpen
  \bibfield  {author} {\bibinfo {author} {\bibfnamefont {S.~R.}\ \bibnamefont
  {White}}\ and\ \bibinfo {author} {\bibfnamefont {A.~E.}\ \bibnamefont
  {Feiguin}},\ }\bibfield  {title} {\enquote {\bibinfo {title} {Real-time
  evolution using the density matrix renormalization group},}\ }\href {\doibase
  10.1103/PhysRevLett.93.076401} {\bibfield  {journal} {\bibinfo  {journal}
  {Phys. Rev. Lett.}\ }\textbf {\bibinfo {volume} {93}},\ \bibinfo {pages}
  {076401} (\bibinfo {year} {2004})}\BibitemShut {NoStop}%
\bibitem [{\citenamefont {Karrasch}\ and\ \citenamefont
  {Schuricht}(2013)}]{karrasch2013dynamical}%
  \BibitemOpen
  \bibfield  {author} {\bibinfo {author} {\bibfnamefont {C.}~\bibnamefont
  {Karrasch}}\ and\ \bibinfo {author} {\bibfnamefont {D.}~\bibnamefont
  {Schuricht}},\ }\bibfield  {title} {\enquote {\bibinfo {title} {{{Dynamical
  phase transitions after quenches in nonintegrable models}}},}\ }\href
  {\doibase 10.1103/PhysRevB.87.195104} {\bibfield  {journal} {\bibinfo
  {journal} {Phys. Rev. B}\ }\textbf {\bibinfo {volume} {87}},\ \bibinfo
  {pages} {1--9} (\bibinfo {year} {2013})}\BibitemShut {NoStop}%
\bibitem [{\citenamefont {Jiang}\ \emph
  {et~al.}(2012{\natexlab{a}})\citenamefont {Jiang}, \citenamefont {Wang},\
  and\ \citenamefont {Balents}}]{JiangNaturePhys12}%
  \BibitemOpen
  \bibfield  {author} {\bibinfo {author} {\bibfnamefont {H.-C.}\ \bibnamefont
  {Jiang}}, \bibinfo {author} {\bibfnamefont {Z.}~\bibnamefont {Wang}}, \ and\
  \bibinfo {author} {\bibfnamefont {L.}~\bibnamefont {Balents}},\ }\bibfield
  {title} {\enquote {\bibinfo {title} {Identifying topological order by
  entanglement entropy},}\ }\href {http://dx.doi.org/10.1038/nphys2465}
  {\bibfield  {journal} {\bibinfo  {journal} {Nature Phys.}\ }\textbf {\bibinfo
  {volume} {8}},\ \bibinfo {pages} {902--905} (\bibinfo {year}
  {2012}{\natexlab{a}})}\BibitemShut {NoStop}%
\bibitem [{\citenamefont {Khemani}\ \emph {et~al.}(2016)\citenamefont
  {Khemani}, \citenamefont {Pollmann},\ and\ \citenamefont
  {Sondhi}}]{KhemaniPRL16}%
  \BibitemOpen
  \bibfield  {author} {\bibinfo {author} {\bibfnamefont {V.}~\bibnamefont
  {Khemani}}, \bibinfo {author} {\bibfnamefont {F.}~\bibnamefont {Pollmann}}, \
  and\ \bibinfo {author} {\bibfnamefont {S.~L.}\ \bibnamefont {Sondhi}},\
  }\bibfield  {title} {\enquote {\bibinfo {title} {{Obtaining Highly Excited
  Eigenstates of Many-Body Localized Hamiltonians by the Density Matrix
  Renormalization Group Approach}},}\ }\href {\doibase
  10.1103/PhysRevLett.116.247204} {\bibfield  {journal} {\bibinfo  {journal}
  {Phys. Rev. Lett.}\ }\textbf {\bibinfo {volume} {116}},\ \bibinfo {pages}
  {247204} (\bibinfo {year} {2016})}\BibitemShut {NoStop}%
\bibitem [{\citenamefont {Yu}\ \emph {et~al.}(2017)\citenamefont {Yu},
  \citenamefont {Pekker},\ and\ \citenamefont {Clark}}]{YuPRL17}%
  \BibitemOpen
  \bibfield  {author} {\bibinfo {author} {\bibfnamefont {X.}~\bibnamefont
  {Yu}}, \bibinfo {author} {\bibfnamefont {D.}~\bibnamefont {Pekker}}, \ and\
  \bibinfo {author} {\bibfnamefont {B.~K.}\ \bibnamefont {Clark}},\ }\bibfield
  {title} {\enquote {\bibinfo {title} {{Finding Matrix Product State
  Representations of Highly Excited Eigenstates of Many-Body Localized
  Hamiltonians}},}\ }\href {\doibase 10.1103/PhysRevLett.118.017201} {\bibfield
   {journal} {\bibinfo  {journal} {Phys. Rev. Lett.}\ }\textbf {\bibinfo
  {volume} {118}},\ \bibinfo {pages} {017201} (\bibinfo {year}
  {2017})}\BibitemShut {NoStop}%
\bibitem [{\citenamefont {{Zhang}}\ \emph {et~al.}(2016)\citenamefont
  {{Zhang}}, \citenamefont {{Pollmann}}, \citenamefont {{Sondhi}},\ and\
  \citenamefont {{Moessner}}}]{ZhangArxiv16}%
  \BibitemOpen
  \bibfield  {author} {\bibinfo {author} {\bibfnamefont {C.}~\bibnamefont
  {{Zhang}}}, \bibinfo {author} {\bibfnamefont {F.}~\bibnamefont {{Pollmann}}},
  \bibinfo {author} {\bibfnamefont {S.~L.}\ \bibnamefont {{Sondhi}}}, \ and\
  \bibinfo {author} {\bibfnamefont {R.}~\bibnamefont {{Moessner}}},\ }\bibfield
   {title} {\enquote {\bibinfo {title} {{Density-Matrix Renormalization Group
  study of Many-Body Localization in Floquet Eigenstates}},}\ }\href@noop {}
  {\bibfield  {journal} {\bibinfo  {journal} {ArXiv e-prints}\ } (\bibinfo
  {year} {2016})},\ \Eprint {http://arxiv.org/abs/1608.06411} {arXiv:1608.06411
  [cond-mat.str-el]} \BibitemShut {NoStop}%
\bibitem [{DMR(2017)}]{DMRGHomePage}%
  \BibitemOpen
  \href@noop {} {\enquote {\bibinfo {title} {{Unofficial DMRG Home Page}},}\
  }\bibinfo {howpublished}
  {\url{http://quattro.phys.sci.kobe-u.ac.jp/dmrg.html}} (\bibinfo {year}
  {2017})\BibitemShut {NoStop}%
\bibitem [{\citenamefont {Hastings}\ \emph {et~al.}(2010)\citenamefont
  {Hastings}, \citenamefont {Gonz\'alez}, \citenamefont {Kallin},\ and\
  \citenamefont {Melko}}]{hastings2010measuring}%
  \BibitemOpen
  \bibfield  {author} {\bibinfo {author} {\bibfnamefont {M.~B.}\ \bibnamefont
  {Hastings}}, \bibinfo {author} {\bibfnamefont {I.}~\bibnamefont
  {Gonz\'alez}}, \bibinfo {author} {\bibfnamefont {A.~B.}\ \bibnamefont
  {Kallin}}, \ and\ \bibinfo {author} {\bibfnamefont {R.~G.}\ \bibnamefont
  {Melko}},\ }\bibfield  {title} {\enquote {\bibinfo {title} {{Measuring Renyi
  Entanglement Entropy in Quantum Monte Carlo Simulations}},}\ }\href {\doibase
  10.1103/PhysRevLett.104.157201} {\bibfield  {journal} {\bibinfo  {journal}
  {Phys. Rev. Lett.}\ }\textbf {\bibinfo {volume} {104}},\ \bibinfo {pages}
  {157201} (\bibinfo {year} {2010})}\BibitemShut {NoStop}%
\bibitem [{\citenamefont {Georges}\ \emph {et~al.}(1996)\citenamefont
  {Georges}, \citenamefont {Kotliar}, \citenamefont {Krauth},\ and\
  \citenamefont {Rozenberg}}]{georges1996dynamical}%
  \BibitemOpen
  \bibfield  {author} {\bibinfo {author} {\bibfnamefont {A.}~\bibnamefont
  {Georges}}, \bibinfo {author} {\bibfnamefont {G.}~\bibnamefont {Kotliar}},
  \bibinfo {author} {\bibfnamefont {W.}~\bibnamefont {Krauth}}, \ and\ \bibinfo
  {author} {\bibfnamefont {M.~J.}\ \bibnamefont {Rozenberg}},\ }\bibfield
  {title} {\enquote {\bibinfo {title} {{Dynamical mean-field theory of strongly
  correlated fermion systems and the limit of infinite dimensions}},}\ }\href
  {\doibase 10.1103/RevModPhys.68.13} {\bibfield  {journal} {\bibinfo
  {journal} {Rev. Mod. Phys.}\ }\textbf {\bibinfo {volume} {68}},\ \bibinfo
  {pages} {13--125} (\bibinfo {year} {1996})}\BibitemShut {NoStop}%
\bibitem [{\citenamefont {Ganahl}\ \emph {et~al.}(2015)\citenamefont {Ganahl},
  \citenamefont {Aichhorn}, \citenamefont {Evertz}, \citenamefont
  {Thunstr\"om}, \citenamefont {Held},\ and\ \citenamefont
  {Verstraete}}]{ganahl2015efficient}%
  \BibitemOpen
  \bibfield  {author} {\bibinfo {author} {\bibfnamefont {M.}~\bibnamefont
  {Ganahl}}, \bibinfo {author} {\bibfnamefont {M.}~\bibnamefont {Aichhorn}},
  \bibinfo {author} {\bibfnamefont {H.~G.}\ \bibnamefont {Evertz}}, \bibinfo
  {author} {\bibfnamefont {P.}~\bibnamefont {Thunstr\"om}}, \bibinfo {author}
  {\bibfnamefont {K.}~\bibnamefont {Held}}, \ and\ \bibinfo {author}
  {\bibfnamefont {F.}~\bibnamefont {Verstraete}},\ }\bibfield  {title}
  {\enquote {\bibinfo {title} {{Efficient DMFT impurity solver using real-time
  dynamics with matrix product states}},}\ }\href {\doibase
  10.1103/PhysRevB.92.155132} {\bibfield  {journal} {\bibinfo  {journal} {Phys.
  Rev. B}\ }\textbf {\bibinfo {volume} {92}},\ \bibinfo {pages} {155132}
  (\bibinfo {year} {2015})}\BibitemShut {NoStop}%
\bibitem [{\citenamefont {Hastings}(2007)}]{hastings2007area}%
  \BibitemOpen
  \bibfield  {author} {\bibinfo {author} {\bibfnamefont {M.~B.}\ \bibnamefont
  {Hastings}},\ }\bibfield  {title} {\enquote {\bibinfo {title} {{An area law
  for one-dimensional quantum systems}},}\ }\href
  {http://stacks.iop.org/1742-5468/2007/i=08/a=P08024} {\bibfield  {journal}
  {\bibinfo  {journal} {J. Stat. Mech.}\ }\textbf {\bibinfo {volume} {2007}},\
  \bibinfo {pages} {P08024} (\bibinfo {year} {2007})}\BibitemShut {NoStop}%
\bibitem [{\citenamefont {Eisert}\ \emph {et~al.}(2010)\citenamefont {Eisert},
  \citenamefont {Cramer},\ and\ \citenamefont {Plenio}}]{eisert2010colloquium}%
  \BibitemOpen
  \bibfield  {author} {\bibinfo {author} {\bibfnamefont {J.}~\bibnamefont
  {Eisert}}, \bibinfo {author} {\bibfnamefont {M.}~\bibnamefont {Cramer}}, \
  and\ \bibinfo {author} {\bibfnamefont {M.~B.}\ \bibnamefont {Plenio}},\
  }\bibfield  {title} {\enquote {\bibinfo {title} {{\textit{Colloquium}: Area
  laws for the entanglement entropy}},}\ }\href {\doibase
  10.1103/RevModPhys.82.277} {\bibfield  {journal} {\bibinfo  {journal} {Rev.
  Mod. Phys.}\ }\textbf {\bibinfo {volume} {82}},\ \bibinfo {pages} {277--306}
  (\bibinfo {year} {2010})}\BibitemShut {NoStop}%
\bibitem [{\citenamefont {Gioev}\ and\ \citenamefont
  {Klich}(2006)}]{gioev2006entanglement}%
  \BibitemOpen
  \bibfield  {author} {\bibinfo {author} {\bibfnamefont {D.}~\bibnamefont
  {Gioev}}\ and\ \bibinfo {author} {\bibfnamefont {I.}~\bibnamefont {Klich}},\
  }\bibfield  {title} {\enquote {\bibinfo {title} {{Entanglement Entropy of
  Fermions in Any Dimension and the Widom Conjecture}},}\ }\href {\doibase
  10.1103/PhysRevLett.96.100503} {\bibfield  {journal} {\bibinfo  {journal}
  {Phys. Rev. Lett.}\ }\textbf {\bibinfo {volume} {96}},\ \bibinfo {pages}
  {100503} (\bibinfo {year} {2006})}\BibitemShut {NoStop}%
\bibitem [{\citenamefont {Liang}\ and\ \citenamefont
  {Pang}(1994)}]{liang1994approximate}%
  \BibitemOpen
  \bibfield  {author} {\bibinfo {author} {\bibfnamefont {S.}~\bibnamefont
  {Liang}}\ and\ \bibinfo {author} {\bibfnamefont {H.}~\bibnamefont {Pang}},\
  }\bibfield  {title} {\enquote {\bibinfo {title} {{Approximate diagonalization
  using the density matrix renormalization-group method: A
  two-dimensional-systems perspective}},}\ }\href {\doibase
  10.1103/PhysRevB.49.9214} {\bibfield  {journal} {\bibinfo  {journal} {Phys.
  Rev. B}\ }\textbf {\bibinfo {volume} {49}},\ \bibinfo {pages} {9214--9217}
  (\bibinfo {year} {1994})}\BibitemShut {NoStop}%
\bibitem [{\citenamefont {White}(1996)}]{white1996spin}%
  \BibitemOpen
  \bibfield  {author} {\bibinfo {author} {\bibfnamefont {S.~R.}\ \bibnamefont
  {White}},\ }\bibfield  {title} {\enquote {\bibinfo {title} {{Spin Gaps in a
  Frustrated Heisenberg Model for ${\mathrm{CaV}}_{4}{\mathrm{O}}_{9}$}},}\
  }\href {\doibase 10.1103/PhysRevLett.77.3633} {\bibfield  {journal} {\bibinfo
   {journal} {Phys. Rev. Lett.}\ }\textbf {\bibinfo {volume} {77}},\ \bibinfo
  {pages} {3633--3636} (\bibinfo {year} {1996})}\BibitemShut {NoStop}%
\bibitem [{\citenamefont {Stoudenmire}\ and\ \citenamefont
  {White}(2012)}]{stoudenmire2012studying}%
  \BibitemOpen
  \bibfield  {author} {\bibinfo {author} {\bibfnamefont {E.~M.}\ \bibnamefont
  {Stoudenmire}}\ and\ \bibinfo {author} {\bibfnamefont {S.~R.}\ \bibnamefont
  {White}},\ }\bibfield  {title} {\enquote {\bibinfo {title} {{{Studying
  Two-Dimensional Systems with the Density Matrix Renormalization Group}}},}\
  }\href {\doibase 10.1146/annurev-conmatphys-020911-125018} {\bibfield
  {journal} {\bibinfo  {journal} {Annu. Rev. Condens. Matter Phys.}\ }\textbf
  {\bibinfo {volume} {3}},\ \bibinfo {pages} {111--128} (\bibinfo {year}
  {2012})}\BibitemShut {NoStop}%
\bibitem [{\citenamefont {Yan}\ \emph {et~al.}(2011)\citenamefont {Yan},
  \citenamefont {Huse},\ and\ \citenamefont {White}}]{yan2011spin}%
  \BibitemOpen
  \bibfield  {author} {\bibinfo {author} {\bibfnamefont {S.}~\bibnamefont
  {Yan}}, \bibinfo {author} {\bibfnamefont {D.~A.}\ \bibnamefont {Huse}}, \
  and\ \bibinfo {author} {\bibfnamefont {S.~R.}\ \bibnamefont {White}},\
  }\bibfield  {title} {\enquote {\bibinfo {title} {{Spin-Liquid Ground State of
  the S = 1/2 Kagome Heisenberg Antiferromagnet}},}\ }\href {\doibase
  10.1126/science.1201080} {\bibfield  {journal} {\bibinfo  {journal}
  {Science}\ }\textbf {\bibinfo {volume} {332}},\ \bibinfo {pages} {1173--1176}
  (\bibinfo {year} {2011})}\BibitemShut {NoStop}%
\bibitem [{\citenamefont {Jiang}\ \emph
  {et~al.}(2012{\natexlab{b}})\citenamefont {Jiang}, \citenamefont {Wang},\
  and\ \citenamefont {Balents}}]{jiang2012identifying}%
  \BibitemOpen
  \bibfield  {author} {\bibinfo {author} {\bibfnamefont {H.-C.}\ \bibnamefont
  {Jiang}}, \bibinfo {author} {\bibfnamefont {Z.}~\bibnamefont {Wang}}, \ and\
  \bibinfo {author} {\bibfnamefont {L.}~\bibnamefont {Balents}},\ }\bibfield
  {title} {\enquote {\bibinfo {title} {{Identifying topological order by
  entanglement entropy}},}\ }\href {http://dx.doi.org/10.1038/nphys2465}
  {\bibfield  {journal} {\bibinfo  {journal} {Nature Phys.}\ }\textbf {\bibinfo
  {volume} {8}},\ \bibinfo {pages} {902--905} (\bibinfo {year}
  {2012}{\natexlab{b}})}\BibitemShut {NoStop}%
\bibitem [{\citenamefont {White}\ and\ \citenamefont
  {Scalapino}(2003)}]{white2003stripes}%
  \BibitemOpen
  \bibfield  {author} {\bibinfo {author} {\bibfnamefont {S.~R.}\ \bibnamefont
  {White}}\ and\ \bibinfo {author} {\bibfnamefont {D.~J.}\ \bibnamefont
  {Scalapino}},\ }\bibfield  {title} {\enquote {\bibinfo {title} {{Stripes on a
  6-Leg Hubbard Ladder}},}\ }\href {\doibase 10.1103/PhysRevLett.91.136403}
  {\bibfield  {journal} {\bibinfo  {journal} {Phys. Rev. Lett.}\ }\textbf
  {\bibinfo {volume} {91}},\ \bibinfo {pages} {136403} (\bibinfo {year}
  {2003})}\BibitemShut {NoStop}%
\bibitem [{\citenamefont {Hager}\ \emph {et~al.}(2005)\citenamefont {Hager},
  \citenamefont {Wellein}, \citenamefont {Jeckelmann},\ and\ \citenamefont
  {Fehske}}]{hager2005stripe}%
  \BibitemOpen
  \bibfield  {author} {\bibinfo {author} {\bibfnamefont {G.}~\bibnamefont
  {Hager}}, \bibinfo {author} {\bibfnamefont {G.}~\bibnamefont {Wellein}},
  \bibinfo {author} {\bibfnamefont {E.}~\bibnamefont {Jeckelmann}}, \ and\
  \bibinfo {author} {\bibfnamefont {H.}~\bibnamefont {Fehske}},\ }\bibfield
  {title} {\enquote {\bibinfo {title} {{Stripe formation in doped Hubbard
  ladders}},}\ }\href {\doibase 10.1103/PhysRevB.71.075108} {\bibfield
  {journal} {\bibinfo  {journal} {Phys. Rev. B}\ }\textbf {\bibinfo {volume}
  {71}},\ \bibinfo {pages} {075108} (\bibinfo {year} {2005})}\BibitemShut
  {NoStop}%
\bibitem [{\citenamefont {Zaletel}\ \emph {et~al.}(2013)\citenamefont
  {Zaletel}, \citenamefont {Mong},\ and\ \citenamefont
  {Pollmann}}]{zaletel2013topological}%
  \BibitemOpen
  \bibfield  {author} {\bibinfo {author} {\bibfnamefont {M.~P.}\ \bibnamefont
  {Zaletel}}, \bibinfo {author} {\bibfnamefont {R.~S.~K.}\ \bibnamefont
  {Mong}}, \ and\ \bibinfo {author} {\bibfnamefont {F.}~\bibnamefont
  {Pollmann}},\ }\bibfield  {title} {\enquote {\bibinfo {title} {{Topological
  Characterization of Fractional Quantum Hall Ground States from Microscopic
  Hamiltonians}},}\ }\href {\doibase 10.1103/PhysRevLett.110.236801} {\bibfield
   {journal} {\bibinfo  {journal} {Phys. Rev. Lett.}\ }\textbf {\bibinfo
  {volume} {110}},\ \bibinfo {pages} {236801} (\bibinfo {year}
  {2013})}\BibitemShut {NoStop}%
\bibitem [{\citenamefont {Nishino}\ \emph {et~al.}(2001)\citenamefont
  {Nishino}, \citenamefont {Hieida}, \citenamefont {Okunishi}, \citenamefont
  {Maeshima}, \citenamefont {Akutsu},\ and\ \citenamefont
  {Gendiar}}]{nishino2001two}%
  \BibitemOpen
  \bibfield  {author} {\bibinfo {author} {\bibfnamefont {T.}~\bibnamefont
  {Nishino}}, \bibinfo {author} {\bibfnamefont {Y.}~\bibnamefont {Hieida}},
  \bibinfo {author} {\bibfnamefont {K.}~\bibnamefont {Okunishi}}, \bibinfo
  {author} {\bibfnamefont {N.}~\bibnamefont {Maeshima}}, \bibinfo {author}
  {\bibfnamefont {Y.}~\bibnamefont {Akutsu}}, \ and\ \bibinfo {author}
  {\bibfnamefont {A.}~\bibnamefont {Gendiar}},\ }\bibfield  {title} {\enquote
  {\bibinfo {title} {{Two-Dimensional Tensor Product Variational
  Formulation}},}\ }\href {\doibase 10.1143/PTP.105.409} {\bibfield  {journal}
  {\bibinfo  {journal} {Prog. Theor. Phys.}\ }\textbf {\bibinfo {volume}
  {105}},\ \bibinfo {pages} {409--417} (\bibinfo {year} {2001})}\BibitemShut
  {NoStop}%
\bibitem [{\citenamefont {{Verstraete}}\ and\ \citenamefont
  {{Cirac}}(2004)}]{verstraete2004renormalization}%
  \BibitemOpen
  \bibfield  {author} {\bibinfo {author} {\bibfnamefont {F.}~\bibnamefont
  {{Verstraete}}}\ and\ \bibinfo {author} {\bibfnamefont {J.~I.}\ \bibnamefont
  {{Cirac}}},\ }\bibfield  {title} {\enquote {\bibinfo {title}
  {{Renormalization algorithms for Quantum-Many Body Systems in two and higher
  dimensions}},}\ }\href@noop {} {\bibfield  {journal} {\bibinfo  {journal}
  {eprint arXiv:cond-mat/0407066}\ } (\bibinfo {year} {2004})},\ \Eprint
  {http://arxiv.org/abs/cond-mat/0407066} {cond-mat/0407066} \BibitemShut
  {NoStop}%
\bibitem [{\citenamefont {Verstraete}\ \emph
  {et~al.}(2008{\natexlab{b}})\citenamefont {Verstraete}, \citenamefont
  {Murg},\ and\ \citenamefont {Cirac}}]{verstraete2008matrix}%
  \BibitemOpen
  \bibfield  {author} {\bibinfo {author} {\bibfnamefont {F.}~\bibnamefont
  {Verstraete}}, \bibinfo {author} {\bibfnamefont {V.}~\bibnamefont {Murg}}, \
  and\ \bibinfo {author} {\bibfnamefont {J.~I.}\ \bibnamefont {Cirac}},\
  }\bibfield  {title} {\enquote {\bibinfo {title} {{Matrix product states,
  projected entangled pair states, and variational renormalization group
  methods for quantum spin systems}},}\ }\href {\doibase
  10.1080/14789940801912366} {\bibfield  {journal} {\bibinfo  {journal} {Adv.
  Phys.}\ }\textbf {\bibinfo {volume} {57}},\ \bibinfo {pages} {143--224}
  (\bibinfo {year} {2008}{\natexlab{b}})}\BibitemShut {NoStop}%
\bibitem [{\citenamefont {Jordan}\ \emph {et~al.}(2008)\citenamefont {Jordan},
  \citenamefont {Or\'us}, \citenamefont {Vidal}, \citenamefont {Verstraete},\
  and\ \citenamefont {Cirac}}]{jordan2008classical}%
  \BibitemOpen
  \bibfield  {author} {\bibinfo {author} {\bibfnamefont {J.}~\bibnamefont
  {Jordan}}, \bibinfo {author} {\bibfnamefont {R.}~\bibnamefont {Or\'us}},
  \bibinfo {author} {\bibfnamefont {G.}~\bibnamefont {Vidal}}, \bibinfo
  {author} {\bibfnamefont {F.}~\bibnamefont {Verstraete}}, \ and\ \bibinfo
  {author} {\bibfnamefont {J.~I.}\ \bibnamefont {Cirac}},\ }\bibfield  {title}
  {\enquote {\bibinfo {title} {{Classical Simulation of Infinite-Size Quantum
  Lattice Systems in Two Spatial Dimensions}},}\ }\href {\doibase
  10.1103/PhysRevLett.101.250602} {\bibfield  {journal} {\bibinfo  {journal}
  {Phys. Rev. Lett.}\ }\textbf {\bibinfo {volume} {101}},\ \bibinfo {pages}
  {250602} (\bibinfo {year} {2008})}\BibitemShut {NoStop}%
\bibitem [{\citenamefont {Or\'us}\ and\ \citenamefont
  {Vidal}(2009)}]{orus2009simulation}%
  \BibitemOpen
  \bibfield  {author} {\bibinfo {author} {\bibfnamefont {R.}~\bibnamefont
  {Or\'us}}\ and\ \bibinfo {author} {\bibfnamefont {G.}~\bibnamefont {Vidal}},\
  }\bibfield  {title} {\enquote {\bibinfo {title} {{Simulation of
  two-dimensional quantum systems on an infinite lattice revisited: Corner
  transfer matrix for tensor contraction}},}\ }\href {\doibase
  10.1103/PhysRevB.80.094403} {\bibfield  {journal} {\bibinfo  {journal} {Phys.
  Rev. B}\ }\textbf {\bibinfo {volume} {80}},\ \bibinfo {pages} {094403}
  (\bibinfo {year} {2009})}\BibitemShut {NoStop}%
\bibitem [{\citenamefont {Or\'us}(2014{\natexlab{b}})}]{orus2014practical}%
  \BibitemOpen
  \bibfield  {author} {\bibinfo {author} {\bibfnamefont {R.}~\bibnamefont
  {Or\'us}},\ }\bibfield  {title} {\enquote {\bibinfo {title} {{A practical
  introduction to tensor networks: Matrix product states and projected
  entangled pair states}},}\ }\href {\doibase
  http://dx.doi.org/10.1016/j.aop.2014.06.013} {\bibfield  {journal} {\bibinfo
  {journal} {Ann. Phys. (N.Y.)}\ }\textbf {\bibinfo {volume} {349}},\ \bibinfo
  {pages} {117 -- 158} (\bibinfo {year} {2014}{\natexlab{b}})}\BibitemShut
  {NoStop}%
\bibitem [{\citenamefont {Vidal}(2007{\natexlab{b}})}]{vidal2007entanglement}%
  \BibitemOpen
  \bibfield  {author} {\bibinfo {author} {\bibfnamefont {G.}~\bibnamefont
  {Vidal}},\ }\bibfield  {title} {\enquote {\bibinfo {title} {{Entanglement
  Renormalization}},}\ }\href {\doibase 10.1103/PhysRevLett.99.220405}
  {\bibfield  {journal} {\bibinfo  {journal} {Phys. Rev. Lett.}\ }\textbf
  {\bibinfo {volume} {99}},\ \bibinfo {pages} {220405} (\bibinfo {year}
  {2007}{\natexlab{b}})}\BibitemShut {NoStop}%
\bibitem [{\citenamefont {Vidal}(2008)}]{vidal2008class}%
  \BibitemOpen
  \bibfield  {author} {\bibinfo {author} {\bibfnamefont {G.}~\bibnamefont
  {Vidal}},\ }\bibfield  {title} {\enquote {\bibinfo {title} {{Class of Quantum
  Many-Body States That Can Be Efficiently Simulated}},}\ }\href {\doibase
  10.1103/PhysRevLett.101.110501} {\bibfield  {journal} {\bibinfo  {journal}
  {Phys. Rev. Lett.}\ }\textbf {\bibinfo {volume} {101}},\ \bibinfo {pages}
  {110501} (\bibinfo {year} {2008})}\BibitemShut {NoStop}%
\bibitem [{\citenamefont {Evenbly}\ and\ \citenamefont
  {Vidal}(2009)}]{evenbly2009algorithms}%
  \BibitemOpen
  \bibfield  {author} {\bibinfo {author} {\bibfnamefont {G.}~\bibnamefont
  {Evenbly}}\ and\ \bibinfo {author} {\bibfnamefont {G.}~\bibnamefont
  {Vidal}},\ }\bibfield  {title} {\enquote {\bibinfo {title} {{Algorithms for
  entanglement renormalization}},}\ }\href {\doibase
  10.1103/PhysRevB.79.144108} {\bibfield  {journal} {\bibinfo  {journal} {Phys.
  Rev. B}\ }\textbf {\bibinfo {volume} {79}},\ \bibinfo {pages} {144108}
  (\bibinfo {year} {2009})}\BibitemShut {NoStop}%
\bibitem [{\citenamefont {Konik}\ and\ \citenamefont
  {Adamov}(2009)}]{konik2009renormalization}%
  \BibitemOpen
  \bibfield  {author} {\bibinfo {author} {\bibfnamefont {R.~M.}\ \bibnamefont
  {Konik}}\ and\ \bibinfo {author} {\bibfnamefont {Y.}~\bibnamefont {Adamov}},\
  }\bibfield  {title} {\enquote {\bibinfo {title} {{Renormalization Group for
  Treating 2D Coupled Arrays of Continuum 1D Systems}},}\ }\href {\doibase
  10.1103/PhysRevLett.102.097203} {\bibfield  {journal} {\bibinfo  {journal}
  {Phys. Rev. Lett.}\ }\textbf {\bibinfo {volume} {102}},\ \bibinfo {pages}
  {097203} (\bibinfo {year} {2009})}\BibitemShut {NoStop}%
\bibitem [{\citenamefont {James}\ and\ \citenamefont
  {Konik}(2013)}]{james2013understanding}%
  \BibitemOpen
  \bibfield  {author} {\bibinfo {author} {\bibfnamefont {A.~J.~A.}\
  \bibnamefont {James}}\ and\ \bibinfo {author} {\bibfnamefont {R.~M.}\
  \bibnamefont {Konik}},\ }\bibfield  {title} {\enquote {\bibinfo {title}
  {{Understanding the entanglement entropy and spectra of 2D quantum systems
  through arrays of coupled 1D chains}},}\ }\href {\doibase
  10.1103/PhysRevB.87.241103} {\bibfield  {journal} {\bibinfo  {journal} {Phys.
  Rev. B}\ }\textbf {\bibinfo {volume} {87}},\ \bibinfo {pages} {241103}
  (\bibinfo {year} {2013})}\BibitemShut {NoStop}%
\bibitem [{\citenamefont {James}\ and\ \citenamefont
  {Konik}(2015)}]{james2015quantum}%
  \BibitemOpen
  \bibfield  {author} {\bibinfo {author} {\bibfnamefont {A.~J.~A.}\
  \bibnamefont {James}}\ and\ \bibinfo {author} {\bibfnamefont {R.~M.}\
  \bibnamefont {Konik}},\ }\bibfield  {title} {\enquote {\bibinfo {title}
  {{Quantum quenches in two spatial dimensions using chain array matrix product
  states}},}\ }\href {\doibase 10.1103/PhysRevB.92.161111} {\bibfield
  {journal} {\bibinfo  {journal} {Phys. Rev. B}\ }\textbf {\bibinfo {volume}
  {92}},\ \bibinfo {pages} {161111} (\bibinfo {year} {2015})}\BibitemShut
  {NoStop}%
\bibitem [{\citenamefont {Schollw{\"{o}}ck}(2011)}]{schollwock2011density}%
  \BibitemOpen
  \bibfield  {author} {\bibinfo {author} {\bibfnamefont {U.}~\bibnamefont
  {Schollw{\"{o}}ck}},\ }\bibfield  {title} {\enquote {\bibinfo {title} {{{The
  density-matrix renormalization group in the age of matrix product
  states}}},}\ }\href {\doibase 10.1016/j.aop.2010.09.012} {\bibfield
  {journal} {\bibinfo  {journal} {Ann. Phys. (N. Y).}\ }\textbf {\bibinfo
  {volume} {326}},\ \bibinfo {pages} {96--192} (\bibinfo {year}
  {2011})}\BibitemShut {NoStop}%
\bibitem [{\citenamefont {Schollw\"ock}(2005)}]{schollwock2005density}%
  \BibitemOpen
  \bibfield  {author} {\bibinfo {author} {\bibfnamefont {U.}~\bibnamefont
  {Schollw\"ock}},\ }\bibfield  {title} {\enquote {\bibinfo {title} {{The
  density-matrix renormalization group}},}\ }\href {\doibase
  10.1103/RevModPhys.77.259} {\bibfield  {journal} {\bibinfo  {journal} {Rev.
  Mod. Phys.}\ }\textbf {\bibinfo {volume} {77}},\ \bibinfo {pages} {259--315}
  (\bibinfo {year} {2005})}\BibitemShut {NoStop}%
\bibitem [{\citenamefont {Hallberg}(2006)}]{hallberg2006new}%
  \BibitemOpen
  \bibfield  {author} {\bibinfo {author} {\bibfnamefont {K.~A.}\ \bibnamefont
  {Hallberg}},\ }\bibfield  {title} {\enquote {\bibinfo {title} {{New trends in
  density matrix renormalization}},}\ }\href {\doibase
  10.1080/00018730600766432} {\bibfield  {journal} {\bibinfo  {journal} {Adv.
  Phys.}\ }\textbf {\bibinfo {volume} {55}},\ \bibinfo {pages} {477--526}
  (\bibinfo {year} {2006})}\BibitemShut {NoStop}%
\bibitem [{\citenamefont {McCulloch}(2007)}]{mcculloch2007density}%
  \BibitemOpen
  \bibfield  {author} {\bibinfo {author} {\bibfnamefont {I.~P.}\ \bibnamefont
  {McCulloch}},\ }\bibfield  {title} {\enquote {\bibinfo {title} {{From
  density-matrix renormalization group to matrix product states}},}\ }\href
  {http://stacks.iop.org/1742-5468/2007/i=10/a=P10014} {\bibfield  {journal}
  {\bibinfo  {journal} {J. Stat. Mech.}\ }\textbf {\bibinfo {volume} {2007}},\
  \bibinfo {pages} {P10014} (\bibinfo {year} {2007})}\BibitemShut {NoStop}%
\bibitem [{\citenamefont {Cirac}\ and\ \citenamefont
  {Verstraete}(2009)}]{cirac2009renormalization}%
  \BibitemOpen
  \bibfield  {author} {\bibinfo {author} {\bibfnamefont {J.~I.}\ \bibnamefont
  {Cirac}}\ and\ \bibinfo {author} {\bibfnamefont {F.}~\bibnamefont
  {Verstraete}},\ }\bibfield  {title} {\enquote {\bibinfo {title}
  {{Renormalization and tensor product states in spin chains and lattices}},}\
  }\href {http://stacks.iop.org/1751-8121/42/i=50/a=504004} {\bibfield
  {journal} {\bibinfo  {journal} {J. Phys. A}\ }\textbf {\bibinfo {volume}
  {42}},\ \bibinfo {pages} {504004} (\bibinfo {year} {2009})}\BibitemShut
  {NoStop}%
\bibitem [{\citenamefont {Sierra}\ and\ \citenamefont
  {Nishino}(1997)}]{SierraNuclPhysB97}%
  \BibitemOpen
  \bibfield  {author} {\bibinfo {author} {\bibfnamefont {G.}~\bibnamefont
  {Sierra}}\ and\ \bibinfo {author} {\bibfnamefont {T.}~\bibnamefont
  {Nishino}},\ }\bibfield  {title} {\enquote {\bibinfo {title} {{The density
  matrix renormalization group method applied to interaction round a face
  Hamiltonians}},}\ }\href {\doibase
  http://dx.doi.org/10.1016/S0550-3213(97)00217-4} {\bibfield  {journal}
  {\bibinfo  {journal} {Nucl. Phys. B}\ }\textbf {\bibinfo {volume} {495}},\
  \bibinfo {pages} {505 -- 532} (\bibinfo {year} {1997})}\BibitemShut {NoStop}%
\bibitem [{\citenamefont {McCulloch}\ and\ \citenamefont
  {Gul\'acsi}(2002)}]{McCullochEPL02}%
  \BibitemOpen
  \bibfield  {author} {\bibinfo {author} {\bibfnamefont {I.~P.}\ \bibnamefont
  {McCulloch}}\ and\ \bibinfo {author} {\bibfnamefont {M.}~\bibnamefont
  {Gul\'acsi}},\ }\bibfield  {title} {\enquote {\bibinfo {title} {{The
  non-Abelian density matrix renormalization group algorithm}},}\ }\href
  {http://stacks.iop.org/0295-5075/57/i=6/a=852} {\bibfield  {journal}
  {\bibinfo  {journal} {Europhys. Lett.}\ }\textbf {\bibinfo {volume} {57}},\
  \bibinfo {pages} {852} (\bibinfo {year} {2002})}\BibitemShut {NoStop}%
\bibitem [{\citenamefont {McCulloch}\ \emph {et~al.}(2008)\citenamefont
  {McCulloch}, \citenamefont {Kube}, \citenamefont {Kurz}, \citenamefont
  {Kleine}, \citenamefont {Schollw\"ock},\ and\ \citenamefont
  {Kolezhuk}}]{McCullochPRB08}%
  \BibitemOpen
  \bibfield  {author} {\bibinfo {author} {\bibfnamefont {I.~P.}\ \bibnamefont
  {McCulloch}}, \bibinfo {author} {\bibfnamefont {R.}~\bibnamefont {Kube}},
  \bibinfo {author} {\bibfnamefont {M.}~\bibnamefont {Kurz}}, \bibinfo {author}
  {\bibfnamefont {A.}~\bibnamefont {Kleine}}, \bibinfo {author} {\bibfnamefont
  {U.}~\bibnamefont {Schollw\"ock}}, \ and\ \bibinfo {author} {\bibfnamefont
  {A.~K.}\ \bibnamefont {Kolezhuk}},\ }\bibfield  {title} {\enquote {\bibinfo
  {title} {{Vector chiral order in frustrated spin chains}},}\ }\href {\doibase
  10.1103/PhysRevB.77.094404} {\bibfield  {journal} {\bibinfo  {journal} {Phys.
  Rev. B}\ }\textbf {\bibinfo {volume} {77}},\ \bibinfo {pages} {094404}
  (\bibinfo {year} {2008})}\BibitemShut {NoStop}%
\bibitem [{\citenamefont {T\'oth}\ \emph {et~al.}(2008)\citenamefont {T\'oth},
  \citenamefont {Moca}, \citenamefont {Legeza},\ and\ \citenamefont
  {Zar\'and}}]{TothPRB08}%
  \BibitemOpen
  \bibfield  {author} {\bibinfo {author} {\bibfnamefont {A.~I.}\ \bibnamefont
  {T\'oth}}, \bibinfo {author} {\bibfnamefont {C.~P.}\ \bibnamefont {Moca}},
  \bibinfo {author} {\bibfnamefont {\"O.}\ \bibnamefont {Legeza}}, \ and\
  \bibinfo {author} {\bibfnamefont {G.}~\bibnamefont {Zar\'and}},\ }\bibfield
  {title} {\enquote {\bibinfo {title} {{Density matrix numerical
  renormalization group for non-Abelian symmetries}},}\ }\href {\doibase
  10.1103/PhysRevB.78.245109} {\bibfield  {journal} {\bibinfo  {journal} {Phys.
  Rev. B}\ }\textbf {\bibinfo {volume} {78}},\ \bibinfo {pages} {245109}
  (\bibinfo {year} {2008})}\BibitemShut {NoStop}%
\bibitem [{\citenamefont {Weichselbaum}(2012)}]{WeichselbaumAnnPhys12}%
  \BibitemOpen
  \bibfield  {author} {\bibinfo {author} {\bibfnamefont {A.}~\bibnamefont
  {Weichselbaum}},\ }\bibfield  {title} {\enquote {\bibinfo {title}
  {{Non-abelian symmetries in tensor networks: A quantum symmetry space
  approach}},}\ }\href {\doibase http://dx.doi.org/10.1016/j.aop.2012.07.009}
  {\bibfield  {journal} {\bibinfo  {journal} {Ann. Phys. (N.Y.)}\ }\textbf
  {\bibinfo {volume} {327}},\ \bibinfo {pages} {2972 -- 3047} (\bibinfo {year}
  {2012})}\BibitemShut {NoStop}%
\bibitem [{\citenamefont {Affleck}\ \emph {et~al.}(1987)\citenamefont
  {Affleck}, \citenamefont {Kennedy}, \citenamefont {Lieb},\ and\ \citenamefont
  {Tasaki}}]{affleck1987rigorous}%
  \BibitemOpen
  \bibfield  {author} {\bibinfo {author} {\bibfnamefont {I.}~\bibnamefont
  {Affleck}}, \bibinfo {author} {\bibfnamefont {T.}~\bibnamefont {Kennedy}},
  \bibinfo {author} {\bibfnamefont {E.~H.}\ \bibnamefont {Lieb}}, \ and\
  \bibinfo {author} {\bibfnamefont {H.}~\bibnamefont {Tasaki}},\ }\bibfield
  {title} {\enquote {\bibinfo {title} {{Rigorous results on valence-bond ground
  states in antiferromagnets}},}\ }\href {\doibase 10.1103/PhysRevLett.59.799}
  {\bibfield  {journal} {\bibinfo  {journal} {Phys. Rev. Lett.}\ }\textbf
  {\bibinfo {volume} {59}},\ \bibinfo {pages} {799--802} (\bibinfo {year}
  {1987})}\BibitemShut {NoStop}%
\bibitem [{\citenamefont {Fannes}\ \emph {et~al.}(1989)\citenamefont {Fannes},
  \citenamefont {Nachtergaele},\ and\ \citenamefont
  {Werner}}]{fannes1989exact}%
  \BibitemOpen
  \bibfield  {author} {\bibinfo {author} {\bibfnamefont {M.}~\bibnamefont
  {Fannes}}, \bibinfo {author} {\bibfnamefont {B.}~\bibnamefont
  {Nachtergaele}}, \ and\ \bibinfo {author} {\bibfnamefont {R.~F.}\
  \bibnamefont {Werner}},\ }\bibfield  {title} {\enquote {\bibinfo {title}
  {{Exact Antiferromagnetic Ground States of Quantum Spin Chains}},}\ }\href
  {http://stacks.iop.org/0295-5075/10/i=7/a=005} {\bibfield  {journal}
  {\bibinfo  {journal} {EPL}\ }\textbf {\bibinfo {volume} {10}},\ \bibinfo
  {pages} {633} (\bibinfo {year} {1989})}\BibitemShut {NoStop}%
\bibitem [{\citenamefont {Peschel}\ \emph {et~al.}(1999)\citenamefont
  {Peschel}, \citenamefont {Kaulke},\ and\ \citenamefont
  {Legeza}}]{peschel1999densityA}%
  \BibitemOpen
  \bibfield  {author} {\bibinfo {author} {\bibfnamefont {I.}~\bibnamefont
  {Peschel}}, \bibinfo {author} {\bibfnamefont {M.}~\bibnamefont {Kaulke}}, \
  and\ \bibinfo {author} {\bibfnamefont {\"O.}\ \bibnamefont {Legeza}},\
  }\bibfield  {title} {\enquote {\bibinfo {title} {Density-matrix spectra for
  integrable models},}\ }\href {\doibase
  10.1002/(SICI)1521-3889(199902)8:2<153::AID-ANDP153>3.0.CO;2-N} {\bibfield
  {journal} {\bibinfo  {journal} {Ann. Phys. (Berlin)}\ }\textbf {\bibinfo
  {volume} {8}},\ \bibinfo {pages} {153--164} (\bibinfo {year}
  {1999})}\BibitemShut {NoStop}%
\bibitem [{\citenamefont {Peschel}\ and\ \citenamefont
  {Chung}(1999)}]{peschel1999densityB}%
  \BibitemOpen
  \bibfield  {author} {\bibinfo {author} {\bibfnamefont {I.}~\bibnamefont
  {Peschel}}\ and\ \bibinfo {author} {\bibfnamefont {M.-C.}\ \bibnamefont
  {Chung}},\ }\bibfield  {title} {\enquote {\bibinfo {title} {Density matrices
  for a chain of oscillators},}\ }\href
  {http://stacks.iop.org/0305-4470/32/i=48/a=305} {\bibfield  {journal}
  {\bibinfo  {journal} {J. Phys. A}\ }\textbf {\bibinfo {volume} {32}},\
  \bibinfo {pages} {8419} (\bibinfo {year} {1999})}\BibitemShut {NoStop}%
\bibitem [{\citenamefont {Okunishi}\ \emph {et~al.}(1999)\citenamefont
  {Okunishi}, \citenamefont {Hieida},\ and\ \citenamefont
  {Akutsu}}]{okunishi1999universal}%
  \BibitemOpen
  \bibfield  {author} {\bibinfo {author} {\bibfnamefont {K.}~\bibnamefont
  {Okunishi}}, \bibinfo {author} {\bibfnamefont {Y.}~\bibnamefont {Hieida}}, \
  and\ \bibinfo {author} {\bibfnamefont {Y.}~\bibnamefont {Akutsu}},\
  }\bibfield  {title} {\enquote {\bibinfo {title} {Universal asymptotic
  eigenvalue distribution of density matrices and corner transfer matrices in
  the thermodynamic limit},}\ }\href {\doibase 10.1103/PhysRevE.59.R6227}
  {\bibfield  {journal} {\bibinfo  {journal} {Phys. Rev. E}\ }\textbf {\bibinfo
  {volume} {59}},\ \bibinfo {pages} {R6227--R6230} (\bibinfo {year}
  {1999})}\BibitemShut {NoStop}%
\bibitem [{\citenamefont {Pollmann}\ and\ \citenamefont
  {Moore}(2010)}]{pollmann2010entanglement}%
  \BibitemOpen
  \bibfield  {author} {\bibinfo {author} {\bibfnamefont {F.}~\bibnamefont
  {Pollmann}}\ and\ \bibinfo {author} {\bibfnamefont {J.~E.}\ \bibnamefont
  {Moore}},\ }\bibfield  {title} {\enquote {\bibinfo {title} {{Entanglement
  spectra of critical and near-critical systems in one dimension}},}\ }\href
  {http://stacks.iop.org/1367-2630/12/i=2/a=025006} {\bibfield  {journal}
  {\bibinfo  {journal} {New. J. Phys.}\ }\textbf {\bibinfo {volume} {12}},\
  \bibinfo {pages} {025006} (\bibinfo {year} {2010})}\BibitemShut {NoStop}%
\bibitem [{\citenamefont {White}(2005)}]{white2005density}%
  \BibitemOpen
  \bibfield  {author} {\bibinfo {author} {\bibfnamefont {S.~R.}\ \bibnamefont
  {White}},\ }\bibfield  {title} {\enquote {\bibinfo {title} {{Density matrix
  renormalization group algorithms with a single center site}},}\ }\href
  {\doibase 10.1103/PhysRevB.72.180403} {\bibfield  {journal} {\bibinfo
  {journal} {Phys. Rev. B}\ }\textbf {\bibinfo {volume} {72}},\ \bibinfo
  {pages} {180403} (\bibinfo {year} {2005})}\BibitemShut {NoStop}%
\bibitem [{\citenamefont {Hubig}\ \emph {et~al.}(2015)\citenamefont {Hubig},
  \citenamefont {McCulloch}, \citenamefont {Schollw\"ock},\ and\ \citenamefont
  {Wolf}}]{hubig2015strictly}%
  \BibitemOpen
  \bibfield  {author} {\bibinfo {author} {\bibfnamefont {C.}~\bibnamefont
  {Hubig}}, \bibinfo {author} {\bibfnamefont {I.~P.}\ \bibnamefont
  {McCulloch}}, \bibinfo {author} {\bibfnamefont {U.}~\bibnamefont
  {Schollw\"ock}}, \ and\ \bibinfo {author} {\bibfnamefont {F.~A.}\
  \bibnamefont {Wolf}},\ }\bibfield  {title} {\enquote {\bibinfo {title}
  {{Strictly single-site DMRG algorithm with subspace expansion}},}\ }\href
  {\doibase 10.1103/PhysRevB.91.155115} {\bibfield  {journal} {\bibinfo
  {journal} {Phys. Rev. B}\ }\textbf {\bibinfo {volume} {91}},\ \bibinfo
  {pages} {155115} (\bibinfo {year} {2015})}\BibitemShut {NoStop}%
\bibitem [{\citenamefont {Motruk}\ \emph {et~al.}(2016)\citenamefont {Motruk},
  \citenamefont {Zaletel}, \citenamefont {Mong},\ and\ \citenamefont
  {Pollmann}}]{motruk2016density}%
  \BibitemOpen
  \bibfield  {author} {\bibinfo {author} {\bibfnamefont {J.}~\bibnamefont
  {Motruk}}, \bibinfo {author} {\bibfnamefont {M.~P.}\ \bibnamefont {Zaletel}},
  \bibinfo {author} {\bibfnamefont {R.~S.~K.}\ \bibnamefont {Mong}}, \ and\
  \bibinfo {author} {\bibfnamefont {F.}~\bibnamefont {Pollmann}},\ }\bibfield
  {title} {\enquote {\bibinfo {title} {{Density matrix renormalization group on
  a cylinder in mixed real and momentum space}},}\ }\href {\doibase
  10.1103/PhysRevB.93.155139} {\bibfield  {journal} {\bibinfo  {journal} {Phys.
  Rev. B}\ }\textbf {\bibinfo {volume} {93}},\ \bibinfo {pages} {155139}
  (\bibinfo {year} {2016})}\BibitemShut {NoStop}%
\bibitem [{\citenamefont {Zaletel}\ \emph {et~al.}(2015)\citenamefont
  {Zaletel}, \citenamefont {Mong}, \citenamefont {Karrasch}, \citenamefont
  {Moore},\ and\ \citenamefont {Pollmann}}]{zaletel2015time}%
  \BibitemOpen
  \bibfield  {author} {\bibinfo {author} {\bibfnamefont {M.~P.}\ \bibnamefont
  {Zaletel}}, \bibinfo {author} {\bibfnamefont {R.~S.~K.}\ \bibnamefont
  {Mong}}, \bibinfo {author} {\bibfnamefont {C.}~\bibnamefont {Karrasch}},
  \bibinfo {author} {\bibfnamefont {J.~E.}\ \bibnamefont {Moore}}, \ and\
  \bibinfo {author} {\bibfnamefont {F.}~\bibnamefont {Pollmann}},\ }\bibfield
  {title} {\enquote {\bibinfo {title} {{Time-evolving a matrix product state
  with long-ranged interactions}},}\ }\href {\doibase
  10.1103/PhysRevB.91.165112} {\bibfield  {journal} {\bibinfo  {journal} {Phys.
  Rev. B}\ }\textbf {\bibinfo {volume} {91}},\ \bibinfo {pages} {165112}
  (\bibinfo {year} {2015})}\BibitemShut {NoStop}%
\bibitem [{\citenamefont {Haegeman}\ \emph {et~al.}(2016)\citenamefont
  {Haegeman}, \citenamefont {Lubich}, \citenamefont {Oseledets}, \citenamefont
  {Vandereycken},\ and\ \citenamefont {Verstraete}}]{haegeman2016unifying}%
  \BibitemOpen
  \bibfield  {author} {\bibinfo {author} {\bibfnamefont {J.}~\bibnamefont
  {Haegeman}}, \bibinfo {author} {\bibfnamefont {C.}~\bibnamefont {Lubich}},
  \bibinfo {author} {\bibfnamefont {I.}~\bibnamefont {Oseledets}}, \bibinfo
  {author} {\bibfnamefont {B.}~\bibnamefont {Vandereycken}}, \ and\ \bibinfo
  {author} {\bibfnamefont {F.}~\bibnamefont {Verstraete}},\ }\bibfield  {title}
  {\enquote {\bibinfo {title} {{Unifying time evolution and optimization with
  matrix product states}},}\ }\href {\doibase 10.1103/PhysRevB.94.165116}
  {\bibfield  {journal} {\bibinfo  {journal} {Phys. Rev. B}\ }\textbf {\bibinfo
  {volume} {94}},\ \bibinfo {pages} {165116} (\bibinfo {year}
  {2016})}\BibitemShut {NoStop}%
\bibitem [{\citenamefont {Haegeman}\ \emph {et~al.}(2013)\citenamefont
  {Haegeman}, \citenamefont {Osborne},\ and\ \citenamefont
  {Verstraete}}]{haegeman2013post}%
  \BibitemOpen
  \bibfield  {author} {\bibinfo {author} {\bibfnamefont {J.}~\bibnamefont
  {Haegeman}}, \bibinfo {author} {\bibfnamefont {T.~J.}\ \bibnamefont
  {Osborne}}, \ and\ \bibinfo {author} {\bibfnamefont {F.}~\bibnamefont
  {Verstraete}},\ }\bibfield  {title} {\enquote {\bibinfo {title} {{Post-matrix
  product state methods: To tangent space and beyond}},}\ }\href {\doibase
  10.1103/PhysRevB.88.075133} {\bibfield  {journal} {\bibinfo  {journal} {Phys.
  Rev. B}\ }\textbf {\bibinfo {volume} {88}},\ \bibinfo {pages} {075133}
  (\bibinfo {year} {2013})}\BibitemShut {NoStop}%
\bibitem [{\citenamefont {Calabrese}\ and\ \citenamefont
  {Cardy}(2005)}]{CalabreseJStatMech05}%
  \BibitemOpen
  \bibfield  {author} {\bibinfo {author} {\bibfnamefont {P.}~\bibnamefont
  {Calabrese}}\ and\ \bibinfo {author} {\bibfnamefont {J.}~\bibnamefont
  {Cardy}},\ }\bibfield  {title} {\enquote {\bibinfo {title} {Evolution of
  entanglement entropy in one-dimensional systems},}\ }\href
  {http://stacks.iop.org/1742-5468/2005/i=04/a=P04010} {\bibfield  {journal}
  {\bibinfo  {journal} {J. Stat. Mech.}\ }\textbf {\bibinfo {volume} {2005}},\
  \bibinfo {pages} {P04010} (\bibinfo {year} {2005})}\BibitemShut {NoStop}%
\bibitem [{\citenamefont {Chiara}\ \emph {et~al.}(2006)\citenamefont {Chiara},
  \citenamefont {Montangero}, \citenamefont {Calabrese},\ and\ \citenamefont
  {Fazio}}]{DeChiaraJStatMech06}%
  \BibitemOpen
  \bibfield  {author} {\bibinfo {author} {\bibfnamefont {G.~De}\ \bibnamefont
  {Chiara}}, \bibinfo {author} {\bibfnamefont {S.}~\bibnamefont {Montangero}},
  \bibinfo {author} {\bibfnamefont {P.}~\bibnamefont {Calabrese}}, \ and\
  \bibinfo {author} {\bibfnamefont {R.}~\bibnamefont {Fazio}},\ }\bibfield
  {title} {\enquote {\bibinfo {title} {{Entanglement entropy dynamics of
  Heisenberg chains}},}\ }\href
  {http://stacks.iop.org/1742-5468/2006/i=03/a=P03001} {\bibfield  {journal}
  {\bibinfo  {journal} {J. Stat. Mech.}\ }\textbf {\bibinfo {volume} {2006}},\
  \bibinfo {pages} {P03001} (\bibinfo {year} {2006})}\BibitemShut {NoStop}%
\bibitem [{\citenamefont {Barmettler}\ \emph {et~al.}(2008)\citenamefont
  {Barmettler}, \citenamefont {Rey}, \citenamefont {Demler}, \citenamefont
  {Lukin}, \citenamefont {Bloch},\ and\ \citenamefont
  {Gritsev}}]{BarmettlerPRA08}%
  \BibitemOpen
  \bibfield  {author} {\bibinfo {author} {\bibfnamefont {P.}~\bibnamefont
  {Barmettler}}, \bibinfo {author} {\bibfnamefont {A.~M.}\ \bibnamefont {Rey}},
  \bibinfo {author} {\bibfnamefont {E.}~\bibnamefont {Demler}}, \bibinfo
  {author} {\bibfnamefont {M.~D.}\ \bibnamefont {Lukin}}, \bibinfo {author}
  {\bibfnamefont {I.}~\bibnamefont {Bloch}}, \ and\ \bibinfo {author}
  {\bibfnamefont {V.}~\bibnamefont {Gritsev}},\ }\bibfield  {title} {\enquote
  {\bibinfo {title} {Quantum many-body dynamics of coupled double-well
  superlattices},}\ }\href {\doibase 10.1103/PhysRevA.78.012330} {\bibfield
  {journal} {\bibinfo  {journal} {Phys. Rev. A}\ }\textbf {\bibinfo {volume}
  {78}},\ \bibinfo {pages} {012330} (\bibinfo {year} {2008})}\BibitemShut
  {NoStop}%
\bibitem [{\citenamefont {Hasenbusch}(2010)}]{hasenbusch2010finite}%
  \BibitemOpen
  \bibfield  {author} {\bibinfo {author} {\bibfnamefont {M.}~\bibnamefont
  {Hasenbusch}},\ }\bibfield  {title} {\enquote {\bibinfo {title} {Finite size
  scaling study of lattice models in the three-dimensional ising universality
  class},}\ }\href@noop {} {\bibfield  {journal} {\bibinfo  {journal} {Phys.
  Rev. B}\ }\textbf {\bibinfo {volume} {82}},\ \bibinfo {pages} {174433}
  (\bibinfo {year} {2010})}\BibitemShut {NoStop}%
\bibitem [{\citenamefont {Li}\ and\ \citenamefont
  {Haldane}(2008)}]{li2008entanglement}%
  \BibitemOpen
  \bibfield  {author} {\bibinfo {author} {\bibfnamefont {H.}~\bibnamefont
  {Li}}\ and\ \bibinfo {author} {\bibfnamefont {F.~D.~M.}\ \bibnamefont
  {Haldane}},\ }\bibfield  {title} {\enquote {\bibinfo {title} {{Entanglement
  Spectrum as a Generalization of Entanglement Entropy: Identification of
  Topological Order in Non-Abelian Fractional Quantum Hall Effect States}},}\
  }\href {\doibase 10.1103/PhysRevLett.101.010504} {\bibfield  {journal}
  {\bibinfo  {journal} {Phys. Rev. Lett.}\ }\textbf {\bibinfo {volume} {101}},\
  \bibinfo {pages} {010504} (\bibinfo {year} {2008})}\BibitemShut {NoStop}%
\bibitem [{\citenamefont {Calabrese}\ and\ \citenamefont
  {Lefevre}(2008)}]{calabrese2008entanglement}%
  \BibitemOpen
  \bibfield  {author} {\bibinfo {author} {\bibfnamefont {P.}~\bibnamefont
  {Calabrese}}\ and\ \bibinfo {author} {\bibfnamefont {A.}~\bibnamefont
  {Lefevre}},\ }\bibfield  {title} {\enquote {\bibinfo {title} {Entanglement
  spectrum in one-dimensional systems},}\ }\href {\doibase
  10.1103/PhysRevA.78.032329} {\bibfield  {journal} {\bibinfo  {journal} {Phys.
  Rev. A}\ }\textbf {\bibinfo {volume} {78}},\ \bibinfo {pages} {032329}
  (\bibinfo {year} {2008})}\BibitemShut {NoStop}%
\bibitem [{\citenamefont {Calabrese}\ and\ \citenamefont
  {Cardy}(2006)}]{calabrese2006time-dependence}%
  \BibitemOpen
  \bibfield  {author} {\bibinfo {author} {\bibfnamefont {P.}~\bibnamefont
  {Calabrese}}\ and\ \bibinfo {author} {\bibfnamefont {J.}~\bibnamefont
  {Cardy}},\ }\bibfield  {title} {\enquote {\bibinfo {title} {Time dependence
  of correlation functions following a quantum quench},}\ }\href {\doibase
  10.1103/PhysRevLett.96.136801} {\bibfield  {journal} {\bibinfo  {journal}
  {Phys. Rev. Lett.}\ }\textbf {\bibinfo {volume} {96}},\ \bibinfo {pages}
  {136801} (\bibinfo {year} {2006})}\BibitemShut {NoStop}%
\bibitem [{\citenamefont {Kollar}\ \emph {et~al.}(2011)\citenamefont {Kollar},
  \citenamefont {Wolf},\ and\ \citenamefont
  {Eckstein}}]{kollar2011generalized}%
  \BibitemOpen
  \bibfield  {author} {\bibinfo {author} {\bibfnamefont {M.}~\bibnamefont
  {Kollar}}, \bibinfo {author} {\bibfnamefont {F.~A.}\ \bibnamefont {Wolf}}, \
  and\ \bibinfo {author} {\bibfnamefont {M.}~\bibnamefont {Eckstein}},\
  }\bibfield  {title} {\enquote {\bibinfo {title} {{Generalized Gibbs ensemble
  prediction of prethermalization plateaus and their relation to nonthermal
  steady states in integrable systems}},}\ }\href {\doibase
  10.1103/PhysRevB.84.054304} {\bibfield  {journal} {\bibinfo  {journal} {Phys.
  Rev. B}\ }\textbf {\bibinfo {volume} {84}},\ \bibinfo {pages} {054304}
  (\bibinfo {year} {2011})}\BibitemShut {NoStop}%
\bibitem [{\citenamefont {Heyl}\ \emph {et~al.}(2013)\citenamefont {Heyl},
  \citenamefont {Polkovnikov},\ and\ \citenamefont
  {Kehrein}}]{heyl2013dynamical}%
  \BibitemOpen
  \bibfield  {author} {\bibinfo {author} {\bibfnamefont {M.}~\bibnamefont
  {Heyl}}, \bibinfo {author} {\bibfnamefont {A.}~\bibnamefont {Polkovnikov}}, \
  and\ \bibinfo {author} {\bibfnamefont {S.}~\bibnamefont {Kehrein}},\
  }\bibfield  {title} {\enquote {\bibinfo {title} {{Dynamical Quantum Phase
  Transitions in the Transverse-Field Ising Model}},}\ }\href {\doibase
  10.1103/PhysRevLett.110.135704} {\bibfield  {journal} {\bibinfo  {journal}
  {Phys. Rev. Lett.}\ }\textbf {\bibinfo {volume} {110}},\ \bibinfo {pages}
  {135704} (\bibinfo {year} {2013})}\BibitemShut {NoStop}%
\bibitem [{\citenamefont {{S{\'e}n{\'e}chal}}(1999)}]{SenechalArxiv99}%
  \BibitemOpen
  \bibfield  {author} {\bibinfo {author} {\bibfnamefont {D.}~\bibnamefont
  {{S{\'e}n{\'e}chal}}},\ }\bibfield  {title} {\enquote {\bibinfo {title} {{An
  introduction to bosonization}},}\ }\href
  {https://arxiv.org/abs/cond-mat/9908262} {\bibfield  {journal} {\bibinfo
  {journal} {ArXiv e-prints}\ } (\bibinfo {year} {1999})},\ \Eprint
  {http://arxiv.org/abs/cond-mat/9908262} {cond-mat/9908262} \BibitemShut
  {NoStop}%
\bibitem [{\citenamefont {{von Delft}}\ and\ \citenamefont
  {{Schoeller}}(1998)}]{vonDelftAnnalenderPhysik98}%
  \BibitemOpen
  \bibfield  {author} {\bibinfo {author} {\bibfnamefont {J.}~\bibnamefont {{von
  Delft}}}\ and\ \bibinfo {author} {\bibfnamefont {H.}~\bibnamefont
  {{Schoeller}}},\ }\bibfield  {title} {\enquote {\bibinfo {title}
  {{Bosonization for beginners - refermionization for experts}},}\ }\href
  {\doibase 10.1002/(SICI)1521-3889(199811)7:4<225::AID-ANDP225>3.0.CO;2-L}
  {\bibfield  {journal} {\bibinfo  {journal} {Ann. Phys. (Berlin)}\ }\textbf
  {\bibinfo {volume} {7}},\ \bibinfo {pages} {225--305} (\bibinfo {year}
  {1998})}\BibitemShut {NoStop}%
\bibitem [{\citenamefont {Virasoro}(1970)}]{VirasoroPRD70}%
  \BibitemOpen
  \bibfield  {author} {\bibinfo {author} {\bibfnamefont {M.~A.}\ \bibnamefont
  {Virasoro}},\ }\bibfield  {title} {\enquote {\bibinfo {title} {{Subsidiary
  Conditions and Ghosts in Dual-Resonance Models}},}\ }\href {\doibase
  10.1103/PhysRevD.1.2933} {\bibfield  {journal} {\bibinfo  {journal} {Phys.
  Rev. D}\ }\textbf {\bibinfo {volume} {1}},\ \bibinfo {pages} {2933--2936}
  (\bibinfo {year} {1970})}\BibitemShut {NoStop}%
\end{thebibliography}%

\end{document}